\documentclass[letter, amsfonts, amssymb, amsmath, reprint, showkeys, twoside, superscriptaddress, aps, pra]{revtex4-1}
\usepackage[dvipsnames]{xcolor}
\usepackage{amsthm}
\usepackage[english]{babel}
\usepackage[utf8]{inputenc}
\usepackage[colorinlistoftodos, color=green!40, prependcaption]{todonotes}
\usepackage{mathtools}
\usepackage{physics}
\usepackage{graphicx}
\usepackage[left=23mm,right=13mm,top=35mm,columnsep=15pt]{geometry} 
\usepackage{adjustbox}
\usepackage{placeins}
\usepackage[T1]{fontenc}
\usepackage{csquotes}
\usepackage{tikz}
\usepackage{bm}% bold math
\usepackage{braket}
\usepackage{multirow}
\usepackage{qcircuit}
\usepackage{siunitx}
\usepackage{amsmath}
\usepackage{bbm}
\usepackage{txfonts}

\usepackage[font=small,labelfont=bf,
   justification=Justified,
   format=plain]{caption}
\usepackage{subcaption}
\usepackage[normalem]{ulem}

\PassOptionsToPackage{hyphens}{url}%\usepackage{hyperref}
\newcommand{\R}{Ref.~}
\newcommand{\eq}{Eq.~}
\newcommand{\fig}{Fig.~}
\newcommand{\C}{\mathcal{C}}
\newcommand{\E}{\mathcal{E}}
\newcommand{\B}{\mathcal{B}}
\newcommand{\floor}[1]{\lfloor #1 \rfloor}

\newcommand{\fprep}{\ensuremath{\mathcal{F}_{\rm prep}}}
\newcommand{\fmeas}{\ensuremath{\mathcal{F}_{\rm meas}}}
\newcommand{\gateG}{\ensuremath{\mathcal{G}}}
\newcommand{\sket}[1]{\ensuremath{\left.\left\vert #1 \right\rangle\right\rangle}}
\newcommand{\sbra}[1]{\ensuremath{\left\langle\left\langle #1 \right\vert\right.}}
\newcommand{\sbraket}[2]{\ensuremath{\left\langle\left\langle #1 \middle| #2 \right\rangle\right\rangle}}
\newcommand{\Norm}[1]{\big\lVert#1\big\rVert}

\newcommand{\braopket}[3]{\ensuremath{\bra{#1}#2\ket{#3}}}
\newcommand{\proj}[1]{\ketbra{#1}{#1}}
\def\Id{\mathbb{I}}
\newcommand{\trans}{\mathrm{T}}

\newcommand{\bigsket}[1]{\ensuremath{| #1 \bigr\rangle\bigr\rangle}}
\newcommand{\bigsbra}[1]{\ensuremath{\bigl\langle\bigl\langle #1 |}}
\newcommand{\Bigsket}[1]{\ensuremath{\bigl| #1 \Bigr\rangle\Bigr\rangle}}
\newcommand{\Bigsbra}[1]{\ensuremath{\Bigl\langle\Bigl\langle #1 \bigr|}}

\usepackage{tabularx}
\newcounter{protocol}

\usepackage{titlesec}
\usepackage[linktoc=all]{hyperref}
\hypersetup{
    colorlinks=true,
    linkcolor=blue,
    urlcolor=blue,
    citecolor=blue
    }
\usepackage[capitalise]{cleveref}

\usepackage{acronym}
\begin{document}

\title{A Practical Introduction to Benchmarking and Characterization \\ of Quantum Computers}

\author{Akel Hashim}
    \email{ahashim@berkeley.edu}
    \affiliation{Department of Physics, University of California at Berkeley, Berkeley, CA 94720, USA}
    \affiliation{Applied Math and Computational Research Division, Lawrence Berkeley National Lab, Berkeley, CA 94720, USA}
\author{Long B.~Nguyen}
\author{Noah Goss}
\author{Brian Marinelli}
\author{Ravi K.~Naik}
    \affiliation{Department of Physics, University of California at Berkeley, Berkeley, CA 94720, USA}
    \affiliation{Applied Math and Computational Research Division, Lawrence Berkeley National Lab, Berkeley, CA 94720, USA}
\author{Trevor Chistolini}
    \affiliation{Department of Physics, University of California at Berkeley, Berkeley, CA 94720, USA}
\author{\\Jordan Hines}
    \affiliation{Department of Physics, University of California at Berkeley, Berkeley, CA 94720, USA}
    \affiliation{Quantum Performance Laboratory, Sandia National Laboratories, Albuquerque, NM 87185 and Livermore, CA 94550}
\author{J.~P.~Marceaux}
    \affiliation{Graduate Group in Applied Science and Technology, University of California at Berkeley, Berkeley, CA 94720, USA}
    \affiliation{Quantum Performance Laboratory, Sandia National Laboratories, Albuquerque, NM 87185 and Livermore, CA 94550}
\author{Yosep Kim}
    \affiliation{Department of Physics, Korea University, Seoul 02841, Korea}
\author{Pranav Gokhale}
\author{Teague Tomesh}
    \affiliation{Infleqtion, Chicago, IL 60604}
\author{Senrui Chen}
\author{Liang Jiang}
    \affiliation{Pritzker School of Molecular Engineering, University of Chicago, IL 60637, USA}
\author{\\Samuele Ferracin}
    \thanks{Now at IBM, Markham, ON (Canada).}
    \affiliation{Keysight Technologies Canada, Kanata, ON K2K 2W5, Canada}
\author{Kenneth Rudinger}
\author{Timothy Proctor}
\author{Kevin C.~Young}
    \affiliation{Quantum Performance Laboratory, Sandia National Laboratories, Albuquerque, NM 87185 and Livermore, CA 94550}
\author{Irfan Siddiqi}
    \affiliation{Department of Physics, University of California at Berkeley, Berkeley, CA 94720, USA}
    \affiliation{Applied Math and Computational Research Division, Lawrence Berkeley National Lab, Berkeley, CA 94720, USA}
    \affiliation{Materials Sciences Division, Lawrence Berkeley National Lab, Berkeley, CA 94720, USA}
\author{Robin Blume-Kohout}
    \affiliation{Quantum Performance Laboratory, Sandia National Laboratories, Albuquerque, NM 87185 and Livermore, CA 94550}

\date{\today} % Leave empty to omit a date

\begin{abstract}
    Rapid progress in quantum technology has transformed quantum computing and quantum information science from theoretical possibilities into tangible engineering challenges.  Breakthroughs in quantum algorithms, quantum simulations, and quantum error correction are bringing useful quantum computation closer to fruition. These remarkable achievements have been facilitated by advances in quantum characterization, verification, and validation (QCVV). QCVV methods and protocols enable scientists and engineers to scrutinize, understand, and enhance the performance of quantum information-processing devices. In this tutorial, we review the fundamental principles underpinning QCVV, and introduce a diverse array of QCVV tools used by quantum researchers. We define and explain QCVV's core models and concepts --- quantum states, measurements, and processes --- and illustrate how these building blocks are leveraged to examine a target system or operation. We survey and introduce protocols ranging from simple qubit characterization to advanced benchmarking methods. Along the way, we provide illustrated examples and detailed descriptions of the protocols, highlight the advantages and disadvantages of each, and discuss their potential scalability to future large-scale quantum computers. This tutorial serves as a guidebook for researchers unfamiliar with the benchmarking and characterization of quantum computers, and also as a detailed reference for experienced practitioners.
\end{abstract}

\maketitle

% \makeatletter
% \def\l@paragraph{\@dottedtocline{4}{5.3em}{2.1em}}
% \makeatother
\tableofcontents

\section*{List of Acronyms}
\begin{acronym}\itemsep=-1pt
    \acro{ACES}{averaged circuit eigenvalue sampling}
    \acro{AGF}{average gate fidelity}
    \acro{AGI}{average gate infidelity}
    \acro{AGSI}{average gate set infidelity}
    \acro{AWG}{arbitrary waveform generator}
    \acro{BiRB}{binary randomized benchmarking}
    \acro{BME}{Bayesian mean estimate}
    \acro{CB}{cycle benchmarking}
    \acro{CER}{cycle error reconstruction}
    \acro{CP}{complete positivity}
    \acro{CPTP}{completely positive trace-preserving}
    \acro{CRB}{Clifford-group randomized benchmarking}
    \acro{CW}{continuous-wave }
    \acro{DD}{dynamical decoupling}
    \acro{DFE}{direct fidelity estimation}
    \acro{DRB}{direct randomized benchmarking}
    \acro{EPC}{error per Clifford}
    \acro{EPLG}{error per layered gate}
    \acro{GHZ}{Greenberger–Horne–Zeilinger}
    \acro{GST}{gate set tomography}
    \acro{IRB}{interleaved randomized benchmarking}
    \acro{KL}{Kullback-Leibler}
    \acro{KNR}{$k$-body noise reconstruction}
    \acro{LGST}{linear gate set tomography}
    \acro{LRB}{leakage randomized benchmarking}
    \acro{LSGST}{long sequence gate set tomography}
    \acro{MAP}{maximum a posteriori}
    \acro{MCFE}{mirror circuit fidelity estimation}
    \acro{MCM}{mid-circuit measurement}
    \acro{MLE}{maximum likelihood estimation}
    \acro{MRB}{mirror randomized benchmarking}
    \acro{PB}{purity benchmarking}
    \acro{PLS}{projected least squares}
    \acro{POVM}{positive operator-valued measure}
    \acro{PTM}{Pauli transfer matrix}
    \acro{PVM}{projection-valued measure}
    \acro{QAOA}{quantum approximate optimization algorithm}
    \acro{QCVV}{quantum characterization, verification, and validation}
    \acro{QFT}{quantum Fourier transform}
    \acro{QI}{quantum instrument}
    \acro{QMT}{quantum measurement tomography}
    \acro{QND}{quantum non-demolition}
    \acro{QPT}{quantum process tomography}
    \acro{QST}{quantum state tomography}
    \acro{QV}{quantum volume}
    \acro{RB}{randomized benchmarking}
    \acro{RMS}{root mean square}
    \acro{RPE}{robust phase estimation}
    \acro{RWA}{rotating wave approximation}
    \acro{SPB}{speckle purity benchmarking}
    \acro{sRB}{simultaneous randomized benchmarking}
    \acro{SPAM}{state preparation and measurement}
    \acro{TP}{trace preservation}
    \acro{TVD}{total variation distance}
    \acro{XEB}{cross-entropy benchmarking}
    \acro{XRB}{eXtended randomized benchmarking}
\end{acronym}

% \printglossary[type=\acronymtype]

\section{Introduction}\label{sec:intro}

Quantum computation has grown from a mere theoretical proposition \cite{shor1994algorithms} to a tangible reality, heralding a new era of science in quantum applications across diverse domains, with recent breakthroughs in quantum measurement and control \cite{2019GoogleSupremacy, 2021USTSupremacy, 2022USTSupremacy, madsen2022quantum}, fundamental science \cite{hacohen2016quantum, colless2018computation, blok2021quantum, mi2022time, morvan2022formation, xiang2024long}, computer science \cite{yamakawa2022verifiable, aaronson2022much, chen2023complexity, anshu2023nlts, aharonov2023polynomial}, quantum chemistry \cite{cao2018potential}, materials science \cite{siddiqi2021engineering}, and many others. If quantum computers of sufficient size and precision can be built, they promise to deliver computational advantages in a diverse range of applications \cite{shor1994algorithms, kitaev1995quantum, grover1996fast, shor1999polynomial, coppersmith2002approximate, harrow2009quantum, farhi2014quantum, liu2021rigorous, daley2022practical}. Much work remains to be done before these promises become reality \cite{proctor2024benchmarking}, but the progress of quantum processors over the past three decades suggests that success will eventually be achievable. That progress --- both past, and future --- is enabled and facilitated by the growing toolbox of \emph{\ac{QCVV}}.
% \emph{quantum characterization, verification, and validation} (\acs{QCVV}).

QCVV means the characterization and benchmarking of quantum computers and their constituent components. It encompasses a large set of methods, protocols, and concepts that have been developed over the past 30 years. These techniques probe the \emph{in situ} behavior of qubits, quantum logic operations, and integrated quantum processors. Having done so, they report detailed predictive models of a device's behavior (\emph{characterization}), simple figures of merit (\emph{benchmarking}), or hybrids of the two. The majority of the QCVV literature focuses on gate-based quantum computers (rather than analogue simulators, quantum annealers, or other paradigms), and this tutorial will too. QCVV of gate-based quantum computers is primarily concerned with characterizing and/or benchmarking the quantum states, quantum gates, and quantum measurements that are implemented by (multi-)qubit devices. It is possible and useful to distinguish characterization protocols from benchmarking protocols. But, in practice these two disciplines are complementary and sometimes overlap (e.g., randomized benchmarking techniques can be deployed for either purpose), and they rely and build upon the same foundational concepts.

Put simply, the goal of QCVV is to learn about as-built quantum computing devices. This usually means using data to estimate properties of mathematical models for those devices, with the goal of predicting (either qualitatively or quantitatively) their future behavior. We conceptualize QCVV methods as tools in a large toolbox. Many of those tools are protocols that can be deployed by an experimentalist or engineer to obtain specific information about the behavior of a quantum computational device (e.g., a qubit, logic operation, or integrated processor) \footnote{The QCVV toolbox \emph{also} includes other ``tools'' besides protocols. They include conceptual tools like \emph{twirling} that are used by theorists to devise new protocols, and standardizing tools like \emph{metrics} and \emph{models} that enable clear communication between practitioners. But \emph{protocols} are the heart of the field.}. In this tutorial, we introduce the most common black-box models used to describe and predict the behavior of quantum computers. Using those models, we introduce the most commonly encountered failure modes of qubits and quantum logic gates, explain how they affect quantum computations, and survey the most common metrics used to quantify their impact. Then, in the bulk of the tutorial, we survey the most common QCVV methods and protocols used to learn models for quantum computer performance. In so doing, we discuss the advantages and disadvantages of each method, the trade-offs between them, their scalability, and their relative utility in predicting the behavior of quantum computations.

\subsection{Uses of QCVV}\label{sec:qcvv}

Most QCVV protocols fall into one of four categories:
\begin{enumerate}
    \item physical device characterization,
    \item tomographic characterization,
    \item randomized benchmarks, and
    \item holistic (application-centric) benchmarking.
\end{enumerate}
Qubits and quantum computers are (as of 2025) still mostly physics experiments. When a qubit or multi-qubit device is fabricated, it cannot be treated or operated as a quantum computer until its physical properties --- e.g. the resonant frequencies and coherence times of qubits, and the nature of couplings between them --- have been determined, calibrated, and optimized. This is the domain of physical device characterization (see Sec.~\ref{sec:qubit_characterization}).

Once a quantum computational device has been calibrated, it becomes possible to treat (and model) it as a quantum computer rather than a physics experiment. Tomographic characterization is now possible. Tomographic QCVV protocols (see Sec.~\ref{sec:tomography}) aim to measure and reconstruct (or \emph{estimate}) the state of one or more qubits, or the operation (e.g., logic gate or measurement) acting on them. For example, quantum state tomography (Sec.~\ref{sec:state_tomography}) estimates the density matrix describing an initialization operation, while quantum process tomography (Sec.~\ref{sec:qpt}) estimates the superoperator describing a reversible logic gate. Tomography-based methods are widely used to characterize individual components, but are generally not scalable to large quantum systems. 

Randomized benchmarks (see Sec.~\ref{sec:randomized_benchmarks}) are intended to build relatively \emph{qualitative} assessments of quantum device performance. Randomized benchmarks probe the performance of an entire set of quantum logic gates and summarize it with $\mathcal{O}(1)$ numbers, without attempting to characterize or model each gate in detail. They report those gates' average performance over many possible input states and many possible contexts, thus providing the end user with some intuition (but few guarantees) about how well the gate will perform in different circuits. Randomized benchmarks are often used to probe just one or two qubits, but some are scalable and can be used to assess the overall performance of an entire processor. In all cases, they provide significantly less detailed and predictive information than tomographic protocols. 

Holistic benchmarks (see Sec.~\ref{sec:holistic}) are intended to measure the performance of a quantum computer on ``relevant'' tasks. Like scalable randomized benchmarks, holistic benchmarks ignore the underlying details of individual qubits and gates. Some holistic benchmarks are designed to capture the performance of a quantum computer in a single number, while others seek to predict how well a quantum computer would perform at a range of different circuit depths and widths. Most holistic benchmarks are designed (unlike randomized benchmarks) to measure the performance of a specific application or class of algorithms.

All of these methods are useful tools in the QCVV toolbox. Our goal in this tutorial --- in addition to teaching the foundational concepts and methods that underlie \emph{all} QCVV protocols --- is to enable readers to decide which QCVV method(s) to use. They are very different tools, and the best tool for a given job depends on the user's goals and needs. Each class of protocols (1) makes different assumptions, (2) seeks to gain a different amount or kind of information about the quantum device being probed, (3) is more or less scalable to large devices, and (4) is more or less amenable to rigorous certification \cite{eisert2020quantum, kliesch2021theory}. In this tutorial, we aim to teach readers about these trade-offs, and to enable scientists and engineers to make informed decisions about which methods to use in each situation.

\subsection{Structure of this Tutorial}

This tutorial is organized as follows. We introduce fundamental models of quantum devices in Sec.~\ref{sec:models}, survey common error types in Sec.~\ref{sec:errors}, and introduce the most commonly used error metrics in Sec.~\ref{sec:overview}. Once these fundamentals have been introduced, we provide a guide to designing QCVV experiments in Sec.~\ref{sec:designing}. A condensed description of preliminary qubit characterization is provided in Sec.~\ref{sec:qubit_characterization}. Then, we discuss various tomographic QCVV techniques (mostly used to validate and debug small subsystems) in Sec.~\ref{sec:tomography}. 

We then introduce randomized benchmarks and their theory in Sec.~\ref{sec:randomized_benchmarks}. We conclude this section with a detailed comparison of different benchmarking protocols. In Sec.~\ref{sec:partial_tomography}, we discuss ``partial tomography'' methods that interpolate between randomized benchmarks and full tomographic characterization. In Sec.~\ref{sec:fidelity_estimation}, we survey protocols that measure the fidelity of entire quantum circuits. Finally, in Sec.~\ref{sec:holistic}, we introduce and survey holistic benchmarks. 
\section{Models of Imperfect Quantum Computers}\label{sec:models}

Real-world quantum computers are complex integrated devices, and the purpose of \ac{QCVV} experiments is to help understand and predict their behavior. Mathematical models that capture the most salient and important features of a real-world device play an important role in this process. They are essential for \emph{predicting} future behavior (e.g., what will happen when a novel quantum program is run on the device), and highly useful for classifying and understanding the failure modes of quantum computers.

Quantum computers have many subsystems, including control hardware (lasers, arbitrary waveform generators, etc.), environmental management (vacuum chambers, cryostats, shielding, etc.), and more. But at the heart of any gate-based quantum computer is a quantum data register (e.g., an array of qubits) that serves as a physical instantiation of quantum logic and quantum algorithms. Ultimately, the quantum computer's performance can be characterized entirely in terms of this register and the accuracy with which it carries out the quantum logic operations specified by a user. Characterizing and/or benchmarking a quantum computer almost always means probing the behavior of its quantum data register, and other subsystems' behavior is only relevant inasmuch as it impacts the quantum data register.

Thus, the models that underlie characterization and benchmarking protocols must (at a minimum) describe the state of quantum data registers, the actions of quantum logic operations, and the results of measurements. When they function perfectly, relatively simple models suffice. But real quantum registers experience errors that cannot be described by the simplest models. Modeling these errors demands greater accuracy and expressiveness, which requires more complex models. The most commonly used models for qubits and quantum registers fall into three broad categories:

\begin{itemize}
    \item \emph{The Closed Quantum System Model} (Sec.~\ref{sec:hilbert_space_model}). A quantum system that does not interact with its environment evolves reversibly, and is called \emph{closed}. When modeling a closed system, its quantum state is represented by a ray or vector in a \emph{Hilbert space}, a measurement is represented by a \emph{projection-valued measure} (PVM), and an operation on the system (e.g., its dynamical evolution) is represented by a \emph{unitary operator}. 
    
    \item \emph{The Markovian Open Quantum System Model} (Sec.~\ref{sec:open_systems_model}). Real-world quantum systems experience irreversible noise when they interact with their environments, and are called \emph{open}. We make the (artificial but useful) assumption that the environment's effects are \emph{Markovian}. In this framework, an open quantum system's state is represented by a \emph{density matrix} on its Hilbert space, a terminating measurement is represented by a \emph{positive operator-valued measure} (POVM), and an operation is represented by a \emph{completely positive trace-preserving} (CPTP) map.

    \item \emph{Non-Markovian Open Quantum System Models}. The category of \emph{non-Markovian} errors includes an enormous number of effects, including time-correlated noise and coherent coupling to a persistent environment (see Sec.~\ref{sec:nm_errors} for a brief overview). Accurate modeling of quantum systems that experience significant non-Markovian errors typically requires the use of bespoke models that are out of scope for this tutorial. 
\end{itemize}

Because QCVV is primarily concerned with noise and errors, this tutorial uses the Markovian open system models extensively. Representing quantum states as density matrices is straightforward, but the standard models of operations and measurements can get complicated. To lay the necessary groundwork for explaining QCVV metrics and protocols later in this tutorial, we explore three specific topics in detail:
\begin{itemize}
    \item \emph{Representations of Quantum Operations} (Sec.~\ref{sec:rep_quant_proc}). Quantum operations (e.g., gates) are represented by CPTP linear \emph{superoperators} that map density matrices to density matrices. Several useful and distinct representations of these CPTP maps are used in QCVV.
    
    \item \emph{Models of Quantum Measurements} (Sec.~\ref{sec:measurement}). Quantum measurements that occur at the end of quantum circuits are called \emph{terminating measurements} and can be modeled by POVMs. To model the internal dynamics of a measurement, or \emph{mid-circuit measurements} that are followed by more gates, more sophisticated models of measurement are needed. Mid-circuit measurements are represented by \emph{quantum instruments}, and if a measurement is weak and perturbs the quantum system minimally, it is possible to continuously track the trajectory of the quantum system in time.
    
    \item \emph{Gate Set Models of Quantum Computers} (Sec.~\ref{sec:gate_sets}). The entire interface of a gate-based quantum computer can be described by a \emph{gate set} that combines models of (i) state preparation, (ii) measurement, and (iii) reversible logic operations. But gate set models are more (or less) than the sum of their parts, because they have \emph{gauge symmetries} that create complications for QCVV.
\end{itemize}

%%%%%%%%%%%%%%%%%%%%%%% The Closed Quantum System Model %%%%%%%%%%%%%%%%%%%%%%% 
\subsection{The Closed Quantum System Model}\label{sec:hilbert_space_model}

A quantum register is called \emph{closed} if it does not experience noise or interact with any outside systems (i.e., its \emph{environment}). This is an artificial and oversimplified paradigm, but it is simple and elegant, and it is the basis for every introductory quantum mechanics course. Perhaps more importantly, it is the foundation upon which the more complicated and flexible ``open quantum system'' model is built. We therefore begin by laying out this foundational model, emphasizing the \emph{structure} (mutually consistent mathematical models for quantum states, measurements, and operations) that will be mirrored in the theory of open quantum systems.

%%%%%%%%%%%%%%%%%%%%%%% Quantum State Vectors and Hilbert Spaces %%%%%%%%%%%%%%%%%%%%%%% 
\subsubsection{Quantum State Vectors and Hilbert Spaces}\label{sec:state_vec_hilbert_space}

We can represent the state of a closed quantum register by a \textit{state vector} $\psi$ in the $d$-dimensional complex vector space $\mathbb{C}^d$, for some integer $d>0$. This space is denoted $\mathcal{H}$ and called the register's \emph{Hilbert space} \footnote{In mathematics, a vector space is a Hilbert space if and only if (iff) it is isomorphic to its dual space. But \emph{all} finite-dimensional vector spaces are Hilbert spaces, and finite-dimensional spaces suffice to describe quantum data registers. So, the mathematical implications of ``Hilbert space'' are an unnecessary red herring for the purposes of this tutorial.}, and $d$ is the register's \emph{Hilbert space dimension}. 

If $N$ quantum systems with Hilbert space dimensions $d_1\ldots d_N$ are considered \emph{together} as a single register, the combined system's state is represented by a vector in the tensor product space $\mathbb{C}^{d_1}\otimes \mathbb{C}^{d_2}\otimes\ldots \otimes \mathbb{C}^{d_N}$, and so its Hilbert space dimension is $\prod_{i=1}^N{d_i}$. Most quantum registers are composed of $n$ \emph{qubits}. A qubit is a quantum system with $d=2$, so an $n$-qubit register has $d=2^n$. In real-world quantum computers, each qubit is \emph{encoded} into a physical system (whose Hilbert space dimension is $\gg 2$, e.g., an atom) by selecting two quantum states, labeling them ``0'' and ``1,'' and carefully confining the physical system's quantum state to the 2-dimensional subspace that they span. This is often referred to as a \emph{two-level system approximation}. Quantum registers can also be built from qu\textit{dits} with Hilbert space dimension $d>2$ (e.g., a qu\textit{trit} has $d=3$, a qu\textit{quart} has $d=4$, etc.), but this is less common.

Following Dirac's notation, we use \emph{kets} (e.g., $\ket{\psi}$) to denote state vectors. If we specify any orthonormal basis $\{\ket{0}, \ket{1}, \ldots, \ket{d-1}\}$ for $\mathbb{C}^d$, then any state $\ket{\psi}$ can be written uniquely as a linear combination of basis vectors, whose coefficients form a column vector:
\begin{equation}
    \ket{\psi} = c_0\ket{0} + c_1\ket{1} + \ldots + c_{d-1}\ket{d-1} \doteq \begin{pmatrix} c_0 \\ c_1 \\ \vdots \\ c_{d-1} \end{pmatrix} ~.
\end{equation}
We use the $\doteq$ symbol, following Sakurai (Ref. \cite{sakurai2021modern}, Eq.~1.73), to indicate ``is concretely represented by.''

In Dirac's notation, the conjugate transpose of $\ket{\psi}$ is a \emph{bra} $\bra{\psi}$, whose coefficients form a row vector,
\begin{equation}
    \bra{\psi} = c_0^*\bra{0} + c_1^*\bra{1} + \ldots + c_{d-1}^*\bra{d-1} \doteq \begin{pmatrix} c_0^* & c_1^* & \dots & c_{d-1}^* \end{pmatrix} ~,
\end{equation}
and the inner product between two state vectors $\ket{\psi}$ and $\ket{\phi}$ is denoted $\braket{\psi|\phi}$. The inner product of $\ket{\psi}$ with itself defines its \emph{norm}, and quantum state vectors are \emph{normalized}:
\begin{equation}
    \braket{\psi|\psi} = \sum_{i=0}^{d-1}{c_i^*c_i} = 1 ~.
\end{equation}

%%%%%%%%%%%%%%%%%%%%%%% Projection-Valued Measures %%%%%%%%%%%%%%%%%%%%%%% 
\subsubsection{Quantum Measurements and Projection-Valued Measures}\label{sec:pvm}

To observe and learn about a quantum system, we perform a \emph{measurement} on it. Many different measurements can be performed on a given system. Measuring yields a particular \emph{outcome}, drawn from a set of $d$ possible outcomes for that measurement. Which outcome occurs is (usually) random, and governed by a probability distribution over the possible outcomes, $\{p(i):\ i=0\ldots d-1\}$, which is determined by the system's quantum state. The entire purpose of the quantum state is to describe and determine the probabilities of various measurement outcomes, and it is sometimes said that \emph{quantum states are linear functionals on observables}.

Each outcome of a measurement on a closed quantum system is represented by a bra \footnote{This subsection intentionally presents a simplified model of quantum mechanics consistent with most undergraduate textbooks. We neglect measurements of degenerate observables, which must be modeled by projectors of rank $>1$, for simplicity's sake. This important case is fully modeled by POVMs in the next subsection.} or row vector $\bra{\lambda}$. If the measured system is described by state $\ket{\psi}$, then the probability of an outcome labeled ``$i$'' represented by $\bra{\lambda_i}$ is given by \emph{Born's Rule}:
\begin{equation}\label{eq:born}
    p(i|\psi) = | \braket{\lambda_i|\psi} |^2 ~. 
\end{equation}
A measurement is represented by a set of bras that form an orthogonal basis, $\{ \bra{\lambda_i} \}_{i=0}^{d-1} = \{ \bra{\lambda_0}, \bra{\lambda_1}, \ldots, \bra{\lambda_{d-1}} \} $. The corresponding probabilities, $p(i|\psi)$, are all non-negative and add up to 1 because $\ket{\psi}$ is normalized, and thus define a valid probability distribution.

It is clear from Eq.~\ref{eq:born} that the state vectors $\ket{\psi}$ and $e^{i\phi}\ket{\psi}$ yield exactly the same probabilities for every measurement. They are, therefore, absolutely indistinguishable by any means, and are considered to define precisely the same state. Here, $e^{i\phi}$ is called a \emph{global phase}, and represents a gauge freedom \footnote{This means that it is a variable in the \emph{model} that has no physical reality, and can be varied without changing anything observable.} of this model. However, we can rewrite Born's Rule in a way that is useful, suggestive, and eliminates the global phase:
\begin{align}
    p(i|\psi) &= \braket{\psi \ketbra{\lambda_i}{\lambda_i} \psi} ~, \\
              &= \Tr[ \ketbra{\lambda_i}\ketbra{\psi} ] ~. \label{eq:born2}
\end{align}
In this expression, both the state and the measurement outcome are represented as \emph{projectors} (i.e., projection operators) rather than vectors. The global phase freedom vanishes, because $\ketbra{\psi}$ is invariant under $\ket{\psi} \to e^{i\phi}\ket{\psi}$, and we have an expression that is \emph{linear} in both $\ketbra{\psi}$ and $\ketbra{\lambda_i}$. This linearity is extremely useful, and this form of Born's Rule motivates the way that both states and measurements are represented for open (noisy) quantum systems.

If we represent the outcomes of a measurement by projectors $\ketbra{\lambda_i}$, then the measurement itself is represented by a \emph{set} of mutually orthogonal projectors
\begin{equation}
    \left\{\Pi_0,\Pi_1,\ldots,\Pi_{d-1}\right\} \equiv 
    \left\{\ketbra{\lambda_0},\ketbra{\lambda_1},\ldots,\ketbra{\lambda_{d-1}}\right\}
\end{equation}
that satisfy \emph{completeness} and \emph{mutual orthogonality} conditions:
\begin{align}
     &\sum\nolimits_i\Pi_i = \mathbb{I} ~, \label{eq:comp_PVM}\\
     &\Pi_i\Pi_j = \delta_{ij}\Pi_i ~, \label{eq:ortho_PVM}
 \end{align}
where $\delta_{ij}$ is the Kronecker delta. This set satisfies the mathematical definition of a \emph{measure} (over the set of possible measurement outcomes), and is called a \emph{\ac{PVM}} \footnote{It is ``projection-valued'' because it assigns a projection operator, rather than a non-negative real number, to each outcome. Born's Rule, with any state $\ketbra{\psi}$, defines a linear functional that maps a projection-valued measure to a standard probability measure, which is the probability distribution over that measurement's outcomes.}.

In this model, measurements are assumed to be \emph{repeatable}. Performing a measurement on a system does not destroy it --- the system still has a state afterward, and can be measured again --- but if the same measurement is performed again, the same outcome will be observed. This requires and implies that if a quantum system is measured and the outcome corresponding to $\bra{\lambda_i}$ (or $\ketbra{\lambda_i}$) is observed, then its \emph{post-measurement state} must be $\ket{\lambda_i}$ (or $\ketbra{\lambda_i}$):
\begin{equation}
    \ket{\psi} \mapsto \ket{\psi'} = \ket{\lambda_i} ~.
\end{equation}

Observable properties of a system --- e.g., the number of electrons in a quantum dot, or an atom's angular momentum along a particular axis --- are represented in this theory by Hermitian operators (acting on the system's Hilbert space) called \emph{observables}. Observables can be measured. Measuring an observable $O$ means performing the PVM whose elements are the projectors onto $O$'s eigenvectors, and indexed by the corresponding eigenvalues of $O$. So if
\begin{equation}
    O = \sum_i{o_i\ketbra{\lambda_i}} ~,
\end{equation}
then measuring $O$ on a system in state $\ket{\psi}$ yields value $o_i$ with probability $p(i) = |\braket{\psi|\lambda_i}|^2$. The \emph{expectation value} of $O$ is thus
\begin{equation}
    \braket{O} = \sum_i{p(i)o_i} = \bra{\psi}O\ket{\psi} ~.
\end{equation}

%%%%%%%%%%%%%%%%%%%%%%% Qubit state vectors %%%%%%%%%%%%%%%%%%%%%%% 
\subsubsection{Qubit State Vectors}\label{sec:statevector_formalism}

A qubit is a physical quantum system whose state vector $\ket{\psi}$ is restricted to a 2-dimensional subspace $\mathbb{C}^2$ of its Hilbert space. Its state space is the span of two orthogonal \emph{computational basis states}, denoted $\ket{0}$ and $\ket{1}$. They are usually eigenstates of the system's Hamiltonian, and may correspond to an atom's ground and excited states, a photon's horizontal and vertical polarization states, an electron's spin-up and spin-down states, or many other possibilities. Regardless of the physical origin, quantum computation is performed by encoding, manipulating, and measuring these states and/or superpositions of them.

The states $\{ \ket{0}, \ket{1}\}$ form a basis for $\mathcal{H}$, so an arbitrary qubit state can be written as a linear combination of them with complex coefficients $\alpha$ and $\beta$
\begin{equation}\label{eq:state_vector_alpha_beta}
    \ket{\psi} = \alpha\ket{0} + \beta\ket{1} \doteq \begin{pmatrix} \alpha \\ \beta \end{pmatrix} ~,
\end{equation}
where the computational basis states themselves are represented by unit column vectors,
\begin{equation}
    \ket{0} \doteq \begin{pmatrix} 1 \\ 0 \end{pmatrix} ~, \
    \ket{1} \doteq \begin{pmatrix} 0 \\ 1 \end{pmatrix} ~.
\end{equation}
Because states must be normalized, $\braket{\psi|\psi} = |\alpha|^2 + |\beta|^2 = 1$. 

Qubit states can be represented as linear combinations of any set of orthonormal basis vectors. Among the most commonly encountered bases are the eigenvectors of the ubiquitous \emph{Pauli operators}. The Pauli operators are represented in the computational basis by $2\times 2$ matrices:
\begin{eqnarray}
    \Id &\doteq \begin{pmatrix} 1 & 0 \\ 0 & 1 \end{pmatrix} ~, \\
    \sigma_x &\doteq \begin{pmatrix} 0 & 1 \\ 1 & 0 \end{pmatrix} ~,  \\
    \sigma_y &\doteq \begin{pmatrix} 0 & -i \\ i & 0 \end{pmatrix} ~, \\
    \sigma_z &\doteq \begin{pmatrix} 1 & 0 \\ 0 & -1 \end{pmatrix} ~.
\end{eqnarray}
It is common to also use the notation $\{I, X, Y, Z\}$ to denote the Pauli operators. The eigenbasis of the $\sigma_z$ operator is the computational basis $\{\ket{0},\ket{1}\}$. The eigenvectors of the $\sigma_x$ operator, denoted $\{\ket{+},\ket{-}\}$, can be written as linear combinations of $\ket{0}$ and $\ket{1}$ with real coefficients of equal magnitude $\alpha = \beta = \tfrac{1}{\sqrt{2}}$:
\begin{equation}
    \ket{+} = \frac{\ket{0} + \ket{1}}{\sqrt{2}} ~, \
    \ket{-} = \frac{\ket{0} - \ket{1}}{\sqrt{2}} ~.
\end{equation}
Similarly, we can also write $\ket{0}$ and $\ket{1}$ as linear combinations of $\ket{+}$ and $\ket{-}$:
\begin{equation}
    \ket{0} = \frac{\ket{+} + \ket{-}}{\sqrt{2}} ~, \
    \ket{1} = \frac{\ket{+} - \ket{-}}{\sqrt{2}} ~.
\end{equation}
Thus, an arbitrary state $\ket{\psi}$ can be expressed in the $\{\ket{+}, \ket{-}\}$ basis as
\begin{align}
    \ket{\psi} &= \alpha\ket{0} + \beta\ket{1} ~, \\
               &= \alpha\frac{\ket{+} + \ket{-}}{\sqrt{2}} + \beta\frac{\ket{+} - \ket{-}}{\sqrt{2}} ~, \\
               &= \frac{\alpha + \beta}{\sqrt{2}}\ket{+} + \frac{\alpha - \beta}{\sqrt{2}}\ket{-} ~.
\end{align}
This is the \emph{same} state as \eq\ref{eq:state_vector_alpha_beta}, just written using different basis states. 
%A similar expression can be derived for state vectors with imaginary coefficients of equal magnitude, denoted $\{\ket{i+}, \ket{i-}\}$.

Since two vectors that differ only by an overall phase describe the same quantum state, we can choose $\alpha$ to be real and, without loss of generality, define $\alpha \equiv \cos(\theta/2)$ and $\beta \equiv e^{i\phi} \sin(\theta/2)$ for some $\theta\in[0, \pi]$ and $\phi\in[0, 2\pi)$. Now, an arbitrary state vector $\ket{\psi}$ is written as
\begin{equation}\label{eq:psi_bloch_sphere}
    \ket{\psi} = \cos(\frac{\theta}{2}) \ket{0} + e^{i\phi} \sin(\frac{\theta}{2}) \ket{1} ~.
\end{equation}
If we compute the expectation values of the three non-identity Pauli operators $\{\sigma_x,\sigma_y,\sigma_z\}$ for this state, we find that
\begin{align}
    \braket{\sigma_x} &= \sin(\theta)\cos\phi ~, \label{eq:expect_X}\\
    \braket{\sigma_y} &= \sin(\theta)\sin\phi ~, \label{eq:expect_Y} \\
    \braket{\sigma_z} &= \cos(\theta) ~. \label{eq:expect_Z}
\end{align}
This parameterization of the state vector $\ket{\psi}$ associates each qubit state with a unique point on the surface of a unit sphere in $\mathbb{R}^3$, whose coordinates are $\mathbf{r}_\psi = \braket{\sigma_x},\braket{\sigma_y},\braket{\sigma_z}$. This is known as the \emph{Bloch sphere}, and the angles $\theta$ and $\phi$ are spherical coordinates for it (see \fig\ref{fig:Bloch_Sphere}). The Bloch sphere provides an intuitive visual representation of quantum states and the action of quantum operations on those states, because unitary dynamical evolution (see below) corresponds to rigid rotations of the Bloch sphere. Measuring a qubit whose state is $\ket{\psi}$ in the computational basis will yield a 1-bit result that is ``0'' with probability $|\alpha|^2 = \cos^2(\theta/2) = (1 + \braket{\sigma_z})/2$ and ``1'' with probability $|\beta|^2 = \sin^2(\theta/2) = (1 - \braket{\sigma_z})/2$.

%%%%%%%%%%%%%%%%%%%%%%% Dynamical Evolution of Quantum States %%%%%%%%%%%%%%%%%%%%%%% 
\subsubsection{Dynamical Evolution of Quantum States}\label{sec:time_evolution}

The most important parts of a quantum computation are the dynamical \emph{operations} --- e.g., logic gates --- performed on the quantum register after it is initialized (in some state) and before it is read out (by measuring it). A quantum operation is a controlled dynamical transformation or evolution of the register's state. Every dynamical evolution of a closed system's state $\ket{\psi}$ is represented by some unitary linear operator $U$:
\begin{equation}
    % \ket{\psi_\mathrm{final}} = U\ket{\psi_\mathrm{initial}} ~.
    U: \ket{\psi} \mapsto \ket{\psi^\prime} = U\ket{\psi} ~.
\end{equation}
This transformation preserves the state's norm, $\braket{\psi|\psi} = \braket{\psi^\prime|\psi^\prime} = 1$, and it is reversible because any unitary $U$ has a unitary inverse $U^\dagger = U^{-1}$.

We can choose to model a quantum register like a computer, with a discrete clock cycle. In this paradigm, time takes integer values. In each clock cycle, as time advances from $t-1$ to $t$, the register's state is transformed by some unitary $U_t$. If we denote the register's state at time $t$ by $\ket{\psi_t}$, then
\begin{align}
    \ket{\psi_1} & = U_1\ket{\psi_0} ~, \\
    \ket{\psi_2} & = U_2\ket{\psi_1} = U_2U_1\ket{\psi_0} ~,
\end{align}
and so on. Different gates (or \emph{circuit layers} of parallel gates on distinct parts of the register) will be represented by different unitaries $U$.

To go deeper and describe a register's detailed dynamics between clock ticks, we can use a continuous-time paradigm in which time $t$ is real-valued. The register's state obeys a differential equation called the \emph{time-dependent Schr\"odinger equation},
\begin{equation}\label{eq:schrodinger}
    i\hbar\frac{\partial}{\partial t} \ket{\psi(t)} = H(t) \ket{\psi(t)} ~,
\end{equation}
where $H(t)$ is a Hermitian operator called the \emph{Hamiltonian} of the register. It is said to \emph{generate} the register's dynamical evolution in time, which is
\begin{equation}
    \ket{\psi(t)} = U(t)\ket{\psi(0)}
\end{equation}
for some time-dependent $U(t)$ that solves Eq.~\ref{eq:schrodinger}. In the special but useful case where $H(t) = H$ is independent of time, the solution is
\begin{equation}\label{eq:unitary_op}
    U(t) = e^{-iHt/\hbar} ~.
\end{equation}
For arbitrary time-dependent $H(t)$, closed-form solutions to the Schr\"odinger equation do not generally exist, but many useful approximations and numerical techniques can be used.

%%%%%%%%%%%%%%%%%%%%%%% Open Systems Model %%%%%%%%%%%%%%%%%%%%%%% 
\subsection{The Markovian Open Quantum System Model}\label{sec:open_systems_model}

We assume that the state of a closed quantum system is known as precisely as it can be. This maximal knowledge is represented by a vector $\ket{\psi}$ in Hilbert space. But to model open systems, we need an efficient way to describe states of partial knowledge --- e.g., ``The system is described by $\ket{\psi_1}$ with probability $p_1$, and by $\ket{\psi_2}$ with probability $p_2$.'' This scenario is not the same as (or consistent with) a superposition state of the form $\alpha\ket{\psi_1} + \beta\ket{\psi_2}$. Instead, it means that in fact the system is either described by $\ket{\psi_1}$ or it is described by $\ket{\psi_2}$, but we are not sure which is true. In such a scenario, we call our description of the system a \emph{mixed state}, to distinguish it from scenarios consistent with a single unique $\ket{\psi}$ which we call a \emph{pure state}.

Mixed states are typically described by an object called a \textit{density matrix}. This new mathematical model for quantum states is richer than the vector state model used to describe pure states of closed systems. It can represent additional forms of uncertainty about the system that state vectors cannot. The theory of \textit{open quantum system} dynamics is built upon density matrices, describing their evolution over time as well as the results of measurements on them.

The density matrix representation of mixed states is nicely motivated by writing the probability of an event $i$ represented by $\Pi_i$ (Born's Rule, \eq\ref{eq:born2}) as
\begin{equation}
    p(i|\psi) = \Tr[\Pi_i \ketbra{\psi}] ~.
\end{equation}
It follows that if the system is described by $\ket{\psi_j}$ with probability $p_j$, then the probability of event $i$ is
\begin{align}\label{eq:borns_rule_rho}
    p(i) &= \sum_j{p_j p \left( i|\psi_j \right)} ~, \\
         &= \sum_j{p_j \Tr[\Pi_i \ketbra{\psi_j}]} ~, \\
         &= \Tr[ \Pi_i \left( \sum_j{p_j \ketbra{\psi_j}} \right) ] ~.
\end{align}
So, if we represent a pure state by the projector $\ketbra{\psi}$, then we can represent the mixed state corresponding to ``$\ket{\psi_j}$ with probability $p_j$'' by a probability-weighted linear combination of these projectors,
\begin{equation}\label{eq:density_operator}
    \rho = \sum_j p_j \ketbra{\psi_j} ~,
\end{equation}
so that any measurement probability is given by
\begin{equation} \label{eq:BornRuleMixed}
    p(i \vert \rho) = \Tr[\Pi_i\rho] ~.
\end{equation}
$\rho$ is called a \emph{density matrix} (a.k.a.~\emph{density operator}). It is a \emph{complete} description of the mixed state!  If two different probability distributions over pure states $\ket{\psi_j}$ have identical averages $\rho = \sum_j p_j \ketbra{\psi_j}$, then those scenarios predict precisely the same probabilities for every possible measurement on the system, and are in fact the same mixed state. Therefore, we always represent mixed states (of open systems) by density matrices. They enable us to model ignorance and uncertainty above and beyond the minimum amount mandated by quantum theory, and to model how noisy operations on a system create or change that ``classical'' uncertainty.

%%%%%%%%%%%%%%%%%%%%%%% Density Matrix Formalism %%%%%%%%%%%%%%%%%%%%%%% 
\subsubsection{Density Matrix Formalism}\label{sec:density_matrix_formalism}

\begin{figure*}[t]
    \centering
    \begin{subfigure}[b]{0.45\textwidth}
        \centering 
        \includegraphics[width=\textwidth]{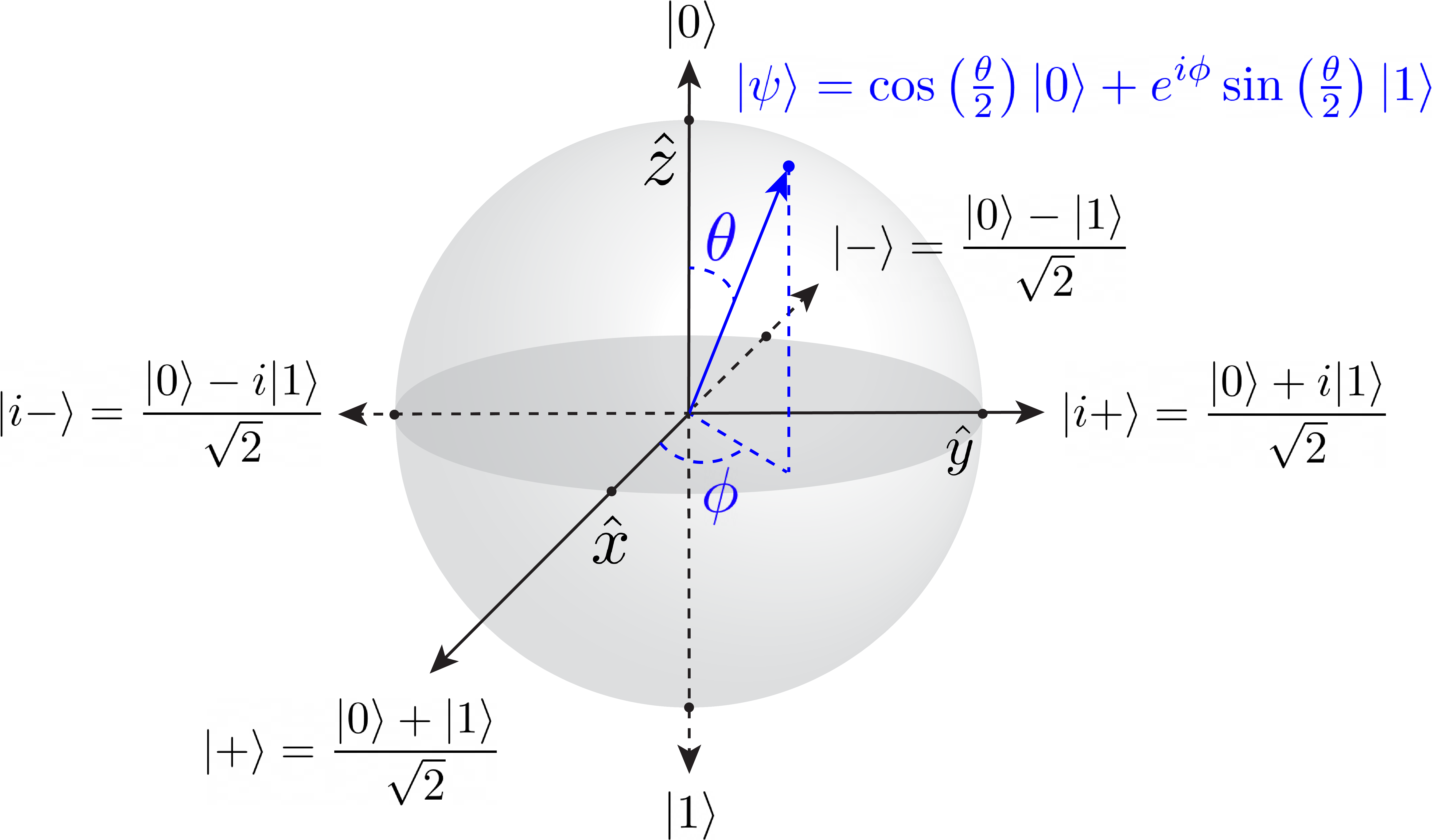}
        \caption{Bloch Sphere}
        \label{fig:Bloch_Sphere}
    \end{subfigure}
    \hspace{7pt}
    \begin{subfigure}[b]{0.25\textwidth}
        \centering 
        \includegraphics[width=\textwidth]{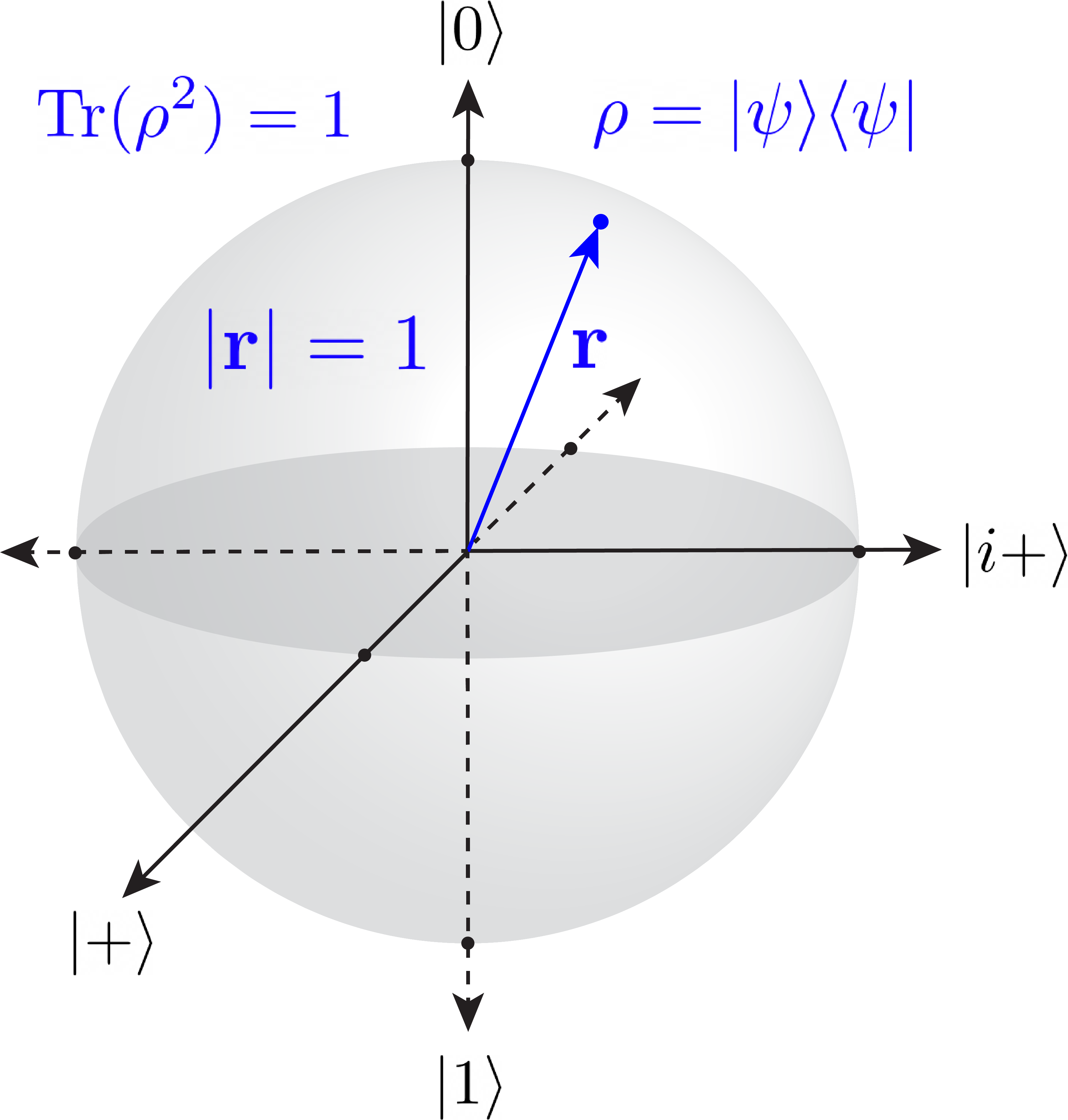}
        \caption{Pure state}
        \label{fig:Bloch_Sphere-pure}
    \end{subfigure}
    \hspace{3pt}
    \begin{subfigure}[b]{0.25\textwidth}
        \centering
        \includegraphics[width=\textwidth]{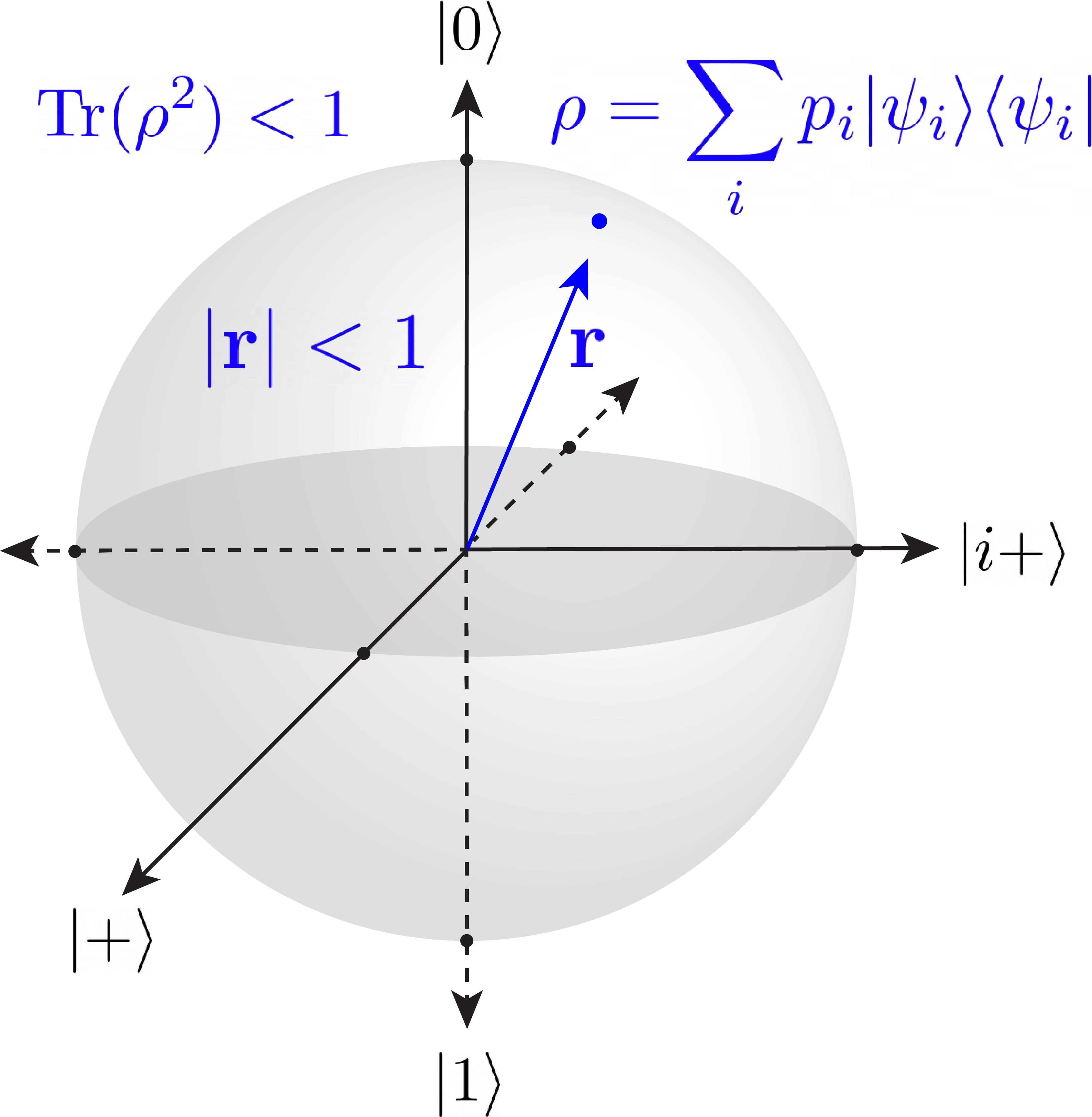}
        \caption{Mixed state}
        \label{fig:Bloch_Sphere-mixed}
    \end{subfigure}
    \caption{\textbf{Bloch Ball Representation of Qubit States.} Single-qubit quantum states, usually represented by state vectors $\ket{\psi}$ or density matrices $\rho$, can also be usefully represented by 3-dimensional real vectors called \emph{Bloch vectors} $\mathbf{r}$. Collectively, the Bloch vectors for all qubit states form the \emph{Bloch ball}. Its surface, the \emph{Bloch sphere}, contains the Bloch vectors for all pure qubit states.
    \textbf{(a)} The computational $\ket{0}$ and $\ket{1}$ states are represented by Bloch vectors at the north and south poles, respectively. Often, these correspond to a system's lowest and second-lowest energy eigenstates. The $\ket{+}$ and $\ket{-}$ states, uniform superpositions of $\ket{0}$ and $\ket{1}$ with real relative phases, are represented by Bloch vectors located along the $x$-axis. The $\ket{i+}$ and $\ket{i-}$ states, uniform superpositions of $\ket{0}$ and $\ket{1}$ with imaginary relative phases, are represented by Bloch vectors located along the $y$-axis. Any pure state can be parameterized as $\ket{\psi} = \cos\left(\tfrac{\theta}{2}\right)\ket{0} + e^{i\phi}\sin\left(\tfrac{\theta}{2}\right)\ket{1}$ (blue), where $\theta$ and $\phi$ are the polar and azimuthal angles (respectively) of its Bloch vector.
    \textbf{(b)} Every pure state's Bloch vector has length $|\mathbf{r}| = 1$.
    \textbf{(c)} The length of the Bloch vector representing a density matrix $\rho$ is $|\mathbf{r}| = \sqrt{2\Tr\rho^2 - 1}$, so Bloch vectors for mixed states have length $|\mathbf{r}| < 1$.
    In (a) -- (c), the blue dot is the projector of the end of the vector onto the surface of the Bloch sphere.
    }
    \label{fig:Bloch_sphere_all}
\end{figure*}

A density matrix $\rho$ is a linear operator that represents the state of a physical quantum system and can be used to compute probabilities for measurement outcomes via Eq.~\ref{eq:BornRuleMixed}:
\begin{equation}
    p(i \vert \rho) = \Tr[\Pi_i\rho] ~. \nonumber
\end{equation}
If a system can be described by the pure state $\ket{\psi}$, then its density matrix is the projector
\begin{equation}\label{eq:density_operator_pure}
    \rho = \ketbra{\psi} ~.
\end{equation}
Density matrices can describe at least two kinds of \emph{uncertain knowledge} that state vectors cannot. The first is represented by a probability distribution over pure states, a.k.a.~an \emph{ensemble}, in which the system's state is $\ket{\psi_j}$ with probability $p_j$ (Eq. \ref{eq:density_operator}):
\begin{equation}
    \rho = \sum_j p_j \ketbra{\psi_j} ~. \nonumber
\end{equation}
The second occurs when a system ($S$) is \emph{entangled} with a second ``reference'' system ($R$) so that they are jointly described by a pure state $\ket{\Psi}_{S,R}$ that is not equal to any tensor product $\ket{\psi}_S \otimes \ket{\phi}_R$. In this scenario, if a measurement is performed on the principal system $S$, then the probability of an outcome represented by $\Pi_i$ is
\begin{equation}
    p(i|\Psi) = \Tr[(\Pi_i\otimes\Id_R)\ketbra{\Psi}] ~.
\end{equation}
By writing $\Id_R = \sum_{j}{\ketbra{j}_R}$, we can show that this probability does not depend on all of $\ket{\Psi}$, but is determined entirely by $S$'s \emph{reduced density matrix}, $\rho_S$. Specifically,
\begin{equation}
    p(i|\Psi) = \Tr[\Pi_i \rho_S] ~,
\end{equation}
where $\rho_S$ is defined by a \emph{partial trace} over $R$,
\begin{equation}
    \rho_S = \Tr_R[\ketbra{\Psi}] = \sum_j{ \left( \Id_S \otimes \bra{j}_R\right) \ketbra{\Psi} \left(\Id_S \otimes \ket{j}_R\right)} ~.
\end{equation}
So, a density matrix can predict measurement probabilities (and thus faithfully represent a system's quantum state) both when that system is described by a distribution over pure states, and when it is known to be entangled with another system. These are both mixed states.

Every density matrix must satisfy two key properties:
\begin{enumerate}
    \item normalization: $\Tr(\rho) = 1$, and
    \item positive semidefiniteness: $\rho \ge 0$, 
\end{enumerate}
which imply three useful facts:
\begin{enumerate}
    \item[3.] $\rho$ is Hermitian: $\rho = \rho^\dagger$, 
    \item[4.] $\Tr(\rho^2) \le 1$, and
    \item[5.] $\rho = \rho^2$ iff $\Tr(\rho^2) = 1$.
\end{enumerate}
Properties (1 -- 2) enforce the basic laws of probability: Property (1) ensures that the outcome probabilities of any measurement sum to 1, and Property (2) ensures that the probability of any measurement outcome is non-negative.
Property (3) follows from Property (2), since every positive semi-definite matrix is also Hermitian. 
Property (4) is a statement about $\rho$'s \emph{purity},
\begin{equation}
    \gamma \equiv \Tr(\rho^2) ~.
\end{equation}
The maximum possible purity is $\gamma=1$, achieved uniquely when $\rho$ is a pure state (\eq\ref{eq:density_operator_pure}). For any mixed state (\eq\ref{eq:density_operator}), $\gamma < 1$, with the minimum possible purity for a $d$-dimensional state being $\gamma = 1/d$, achieved by the \emph{maximally mixed state} $\rho_\text{mix} = \Id/d$. 
Finally, Property (5) follows from Property (4) for pure states; in other words, $\rho$ is \emph{idempotent} if and only if it is pure. Equivalently, if $\rho$ is a projection operator, then it must represent a pure state.

An arbitrary single-qubit pure state $\ket{\psi} = \alpha\ket{0} + \beta\ket{1}$ is represented by the density matrix
\begin{equation}\label{eq:rho_single_qubit}
    \rho = \ketbra{\psi}{\psi} \doteq \begin{pmatrix}
            \vert \alpha \vert^2 & \alpha\beta^* \\
            \alpha^*\beta & \vert \beta \vert^2
           \end{pmatrix} ~.
\end{equation}
If we write $\ket{\psi}$ in spherical coordinates using Eq.~\ref{eq:psi_bloch_sphere}, we get:
\begin{align}
    \rho &\doteq \begin{pmatrix}
            \cos^2\left(\tfrac{\theta}{2}\right) & e^{-i\phi}\cos\left(\tfrac{\theta}{2}\right)\sin\left(\tfrac{\theta}{2}\right) \\
            e^{i\phi}\cos\left(\tfrac{\theta}{2}\right)\sin\left(\tfrac{\theta}{2}\right) & \sin^2\left(\tfrac{\theta}{2}\right)
           \end{pmatrix} ~, \label{eq:rho_spherical_param} \\
    &= \frac{1}{2}\begin{pmatrix}
            1 + \cos(\theta) & e^{-i\phi}\sin(\theta) \\
            e^{i\phi}\sin(\theta) & 1 - \cos(\theta)
           \end{pmatrix} ~, \\
    & = \frac{1}{2} \left( \mathbb{I} + \sin(\theta)\cos(\phi)\sigma_x + \sin(\theta)\sin(\phi)\sigma_y + \cos(\theta)\sigma_z \right) ~. \label{eq:rho_PauliBasis}
\end{align}
The last expression illustrates a very useful fact: operators form a \emph{vector space}. They can be added, subtracted, and scaled. Any operator can be written as a linear combination of the elements of an \emph{operator basis}. The four Pauli operators form such a basis for operators on qubits, and so we can expand $\rho$ as a linear combination of them. The space of operators on a system's Hilbert space is called its \emph{Hilbert-Schmidt space}, and is used extensively in QCVV. If the system's Hilbert space is denoted $\mathcal{H}$, then its Hilbert-Schmidt space is denoted $\mathcal{B}(\mathcal{H})$ \footnote{$\mathcal{B}(\mathcal{H})$ means ``the space of bounded operators on $\mathcal{H}$.'' Sometimes $\mathcal{L}(\mathcal{H})$, meaning ``the space of linear operators on $\mathcal{H}$,'' is used instead. These coincide when $\mathcal{H}$ is finite-dimensional.}. The inner product between two operators $A$ and $B$ in $\mathcal{B}(\mathcal{H})$ is defined by
\begin{equation}
    A\cdot B = \Tr[ A^\dagger B ] ~.
\end{equation}

Operators thus have multiple algebraic roles. They can act as transformations (on the Hilbert space $\mathcal{H}$ of state vectors), or as vectors (in the Hilbert-Schmidt space $\mathcal{B}(\mathcal{H})$ of operators). When we want to emphasize an operator's vector role, we write it in a double ket or \textit{superket}, e.g.~as $\sket{A}$ instead of $A$. To denote the Hilbert-Schmidt inner product in this notation, we define the \emph{superbra} $\sbra{A} \equiv \sket{A}^\dagger$, so that 
\begin{equation}\label{eq:hs_inner_prod}
    \sbraket{A}{B} = \Tr[A^\dagger B] ~.
\end{equation}

The inner product between a density matrix $\rho$ and an observable $P$ is therefore equal to the expectation value of $P$:
\begin{equation}
    \rho\cdot P = \sbraket{\rho}{P} = \Tr[ \rho P ] = \braket{P}_\rho ~,
\end{equation}
taking advantage of Property (3). We can expand $\rho$ in an orthogonal basis of Hermitian operators $\{P_j\}$ as $\rho = \sum_j{c_j P_j}$, where each coefficient is given by
\begin{equation}
    c_j = \frac{\sbraket{\rho}{P_j}}{\sbraket{P_j}{P_j}} = \frac{\braket{P}_\rho}{\Tr(P_j^2)} ~.
\end{equation}
As a result, Eq.~\ref{eq:rho_PauliBasis} (which expands a pure state $\rho$ in the Pauli basis) follows from Eqs.~\ref{eq:expect_X} -- \ref{eq:expect_Z} (the expectation values of the Pauli operators for that state). Therefore, any single-qubit density matrix $\rho$ can be written as
\begin{equation}\label{eq:bloch_vector}
    \rho = \frac{1}{2} \left( \mathbb{I} + \mathbf{r} \cdot \boldsymbol\sigma \right) ~,
\end{equation}
where $\mathbf{r} = r_x\mathbf{\hat{x}} + r_y\mathbf{\hat{y}} + r_z\mathbf{\hat{z}}$ and $\boldsymbol\sigma = \sigma_x\mathbf{\hat{x}} + \sigma_y\mathbf{\hat{y}} + \sigma_z\mathbf{\hat{z}}$. When $\rho = \ketbra{\psi}$ is a pure state, $\mathbf{r}$ is a unit vector. This is the \emph{Bloch sphere} representation mentioned previously, and illustrated in \fig\ref{fig:Bloch_Sphere-pure}. The Bloch sphere representation of two-level systems is extremely useful for visualizing qubit states prior to measurement and, as we will see in the next chapter, for visualizing the impact of errors on qubits.

A mixed state $\rho$ also defines a vector $\mathbf{r}$, but one with length $|\mathbf{r}|$ less than 1. The length of a mixed state's Bloch vector is determined by its purity:
\begin{equation}
    |\mathbf{r}| = \sqrt{2\Tr(\rho^2) - 1} ~. 
\end{equation}
The $\mathbf{r}$ vectors for mixed states define the \emph{Bloch ball} (the convex closure of the Bloch sphere) as shown in \fig\ref{fig:Bloch_Sphere-mixed}, with the maximally mixed state $\rho_{\mathrm{mix}} = \Id/2$ at its center ($\mathbf{r}=0$).

%%%%%%%%%%%%%%%%%%%%%%% Positive Operator-Valued Measures %%%%%%%%%%%%%%%%%%%%%%% 
\subsubsection{Positive Operator-Valued Measures}\label{sec:povm}

In closed-system quantum mechanics, a quantum state is represented by a state vector and a measurement is represented by a set of orthogonal projectors (a PVM, as described in Sec.~\ref{sec:pvm}). In open quantum systems, we need to model additional uncertainty. This requires richer representations not just of states (as density matrices), but of measurements as well. If $S$ is an open quantum system, then it is possible to perform \emph{indirect} measurements on $S$ by (1) coupling $S$ to another system $R$ (\fig\ref{fig:stinespring_rep}), and then (2) performing a PVM on $S$ and $R$ jointly (\fig\ref{fig:povm}). This enables and allows a substantially richer class of measurements called a \emph{\ac{POVM}} \footnote{Technically, a POVM is a measure (like a probability distribution) over possible events, but which is ``operator-valued,'' meaning that instead of assigning a \emph{probability} to each event, it assigns a \emph{positive semi-definite operator} to each event, whose inner product with the system's state $\rho$ defines the event's probability.}.

A POVM is a set of positive semi-definite operators $\{E_i\}$ that satisfies the completeness relation:
\begin{equation}\label{eq:povm_comp_rel}
    \sum_i E_i = \mathbb{I} ~.
\end{equation}
Each $E_i$ is called an \emph{effect}, and represents one possible outcome ``$i$'' of the measurement. When the POVM $\{E_i\}$ is performed on a state $\rho$, the probability of observing outcome $i$ is given by
\begin{equation}\label{eq:borns_rule_povm}
    p(i|\rho) = \Tr[E_i \rho] ~,
\end{equation}
which is the open-system version of Born's Rule.

Any projective measurement (PVM) is also a POVM. But POVMs are quite a bit more general. The effects in a POVM do not need to be orthogonal, nor rank-1, nor projectors. \emph{Any} set satisfying the conditions above is a valid, feasible POVM. Importantly, POVMs describe \emph{destructive} measurements, and do not specify or define what a system's state will be after measurement.

%%%%%%%%%%%%%%%%%%%%%%% Dynamical Evolution of Density Matrices %%%%%%%%%%%%%%%%%%%%%%% 
\subsubsection{Dynamical Evolution of Density Matrices}\label{sec:dyn_evolution}

\begin{figure}
    \centering
    \begin{subfigure}[b]{0.4\columnwidth}
        \centering 
        \begin{equation}
            {\Qcircuit @C=2.5em @R=3em {
                \lstick{\rho} & \multigate{1}{U} & \qw & \lstick{\E(\rho)} \\
                \lstick{\rho_\text{env}} & \ghost{U} & \qw & \lstick{\rho_\text{env}^\prime}
                }
            } \nonumber
        \end{equation}
        \caption{Dynamical evolution}
        \label{fig:stinespring_rep}
    \end{subfigure}
    \begin{subfigure}[b]{0.4\columnwidth}
        \centering 
        \begin{equation}
            {\Qcircuit @C=2.5em @R=1.0em {
                 \lstick{\cdots} & \multimeasureD{2}{\text{PVM}} \\
                 & \nghost{\text{PVM}} & \cw \\
                 \lstick{\cdots} & \ghost{\text{PVM}}
                }
            } \nonumber
        \end{equation}
        \caption{Joint measurement}
        \label{fig:povm}
    \end{subfigure}
    \caption{\textbf{System-Environment Representation of Quantum Operations.}
    \textbf{(a)} An open quantum system comprises a principal system of interest whose state is $\rho$, and a surrounding environment whose state is $\rho_\text{env}$. Any dynamical evolution of the principal system can be described by some joint evolution of the system and its environment, described by a unitary transformation $U$ that may produce correlation or entanglement between system and environment. After this evolution, the state of the principal system can be described by a reduced density matrix, defined as a partial trace over the environment. This density matrix is a linear function $\E(\rho)$ of the initial state $\rho$.
    \textbf{(b)} Terminating measurements of the principal system $\E(\rho)$ can be modeled by performing a PVM on the joint system [$\E(\rho) \otimes \rho_\text{env}^\prime$]. The results of this joint measurement are described by positive operator-valued measures (POVMs), whose outcome is a distribution of classical bits (denoted by the double line).
    }
    \label{fig:system_env_rep}
\end{figure}
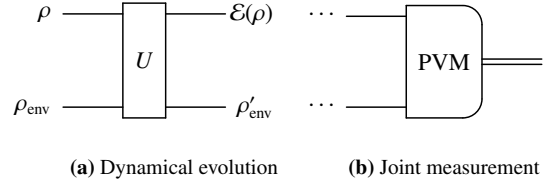

An open quantum system can interact with its environment. This possibility allows new kinds of dynamical evolution that can create or decrease uncertainty, causing the open system's state to become more (or less) mixed. In contrast, closed-system dynamics are always unitary and never change the purity of $\rho$. An open-system dynamical evolution $\rho \mapsto \E(\rho)$ can be described by a three-step process, illustrated in \fig\ref{fig:stinespring_rep}:
\begin{enumerate}
    \item The principal system of interest is described (initially) by a state $\rho$. We introduce a second system, the environment, that is assumed to be initially uncorrelated with the principal system \footnote{It is possible for initial correlation between the principal system and its environment to exist, and to be modeled. This scenario is advanced, conceptually tricky, and considered \emph{non-Markovian}. It is not often considered in QCVV, and is outside the scope of this tutorial.} and described by its own state $\rho_\text{env} = \ketbra{e_0}$.
    \item The system and environment evolve jointly, according to familiar closed-system theory, by some unitary $U$ that may induce correlations or entanglement between them.
    \item We focus on the principal system only, neglecting or ``throwing away'' the environment by performing a partial trace over its Hilbert space, to obtain a reduced density matrix for the principal system only:
\end{enumerate}
\begin{equation}\label{eq:stinespring}
    \E(\rho) = \Tr_\text{env}[ U \left( \rho \otimes \rho_\text{env} \right) U^\dagger ] ~.
\end{equation}
The dynamical map $\rho \mapsto \E(\rho)$ is called a \emph{quantum operation} (a.k.a.~a \emph{quantum channel}), and \eq\ref{eq:stinespring} is known as the \emph{system-environment}, or \emph{Stinespring}, representation of quantum operations; it is depicted graphically in Tab.~\ref{tab:op_rep}. It stems from Stinespring's \emph{dilation theorem} \cite{stinespring1955positive}, which states that every physically-allowed dynamical evolution of the principal system arises from unitary evolution on a larger system, and can thus be described by \eq\ref{eq:stinespring}. In terms of the non-square \textit{Stinespring operator} defined as $A \equiv U (\mathbb{I} \otimes \ket{e_0})$, \eq\ref{eq:stinespring} is simply
\begin{equation}\label{eq:stinespring_op}
    \E(\rho) = \Tr_\text{env}[ A \rho A^\dagger ] ~.
\end{equation}
Equation \ref{eq:stinespring_op} shows clearly that the dynamical map $\E$ acts \emph{linearly} on density matrices --- i.e., if $\rho = a\rho_1 + b\rho_2$, then $\E(\rho) = a\E(\rho_1) + b\E(\rho_2)$.

Such a linear map can represent a real, physically-realizable quantum operation if --- and only if --- it satisfies two conditions:
\begin{enumerate}
    % \item \emph{Complete Positivity} (\acs{CP}):
    \item \emph{\Ac{CP}}: Given any positive semidefinite density matrix $\rho \geq 0$, applying $\E$ to $\rho$ must yield a matrix that is also positive semidefinite, even if $\E$ only acts on a part (subsystem) of $\rho$. So $(\E\otimes\Id)[ \rho ] \geq 0$ for every $\rho \geq 0$. This is a stricter requirement than simple \textit{positivity} --- $\E[\rho] \geq 0$ for every $\rho \geq 0$ --- because a linear map can be positive yet not completely positive. A canonical example is the transpose map, $\E(\rho) = \rho^\trans$. If such a map could be experimentally applied to arbitrary states, then by applying it to a system properly entangled with another system, a negative probability (for some measurement outcome) could be produced.
    
    % \item \emph{Trace Preservation} (\acs{TP}):  
    \item \emph{\Ac{TP}}: $\Tr[\E(\rho)] = \Tr[\rho]$ for all $\rho$.
\end{enumerate}
Just like Properties (1 -- 2) of density matrices, these conditions guarantee the two essential properties of probability distributions. CP ensures that no event can have negative probability, while TP ensures that the outcome probabilities of every measurement add up to 1. Linear maps satisfying both conditions are called \emph{\ac{CPTP}} maps, and every CPTP map represents a physically realizable quantum operation.

The Stinespring representation of an operation $\E$ is not unique, because many different physical scenarios ($U$ and $\rho_{\text{env}}$) can produce identical reduced dynamics for the principal system. As a result, the Stinespring representation is rarely used in practical calculations, because more convenient representations exist. As we will see in this tutorial, quantum operations are very important in QCVV, and the QCVV literature uses several distinct representations of them for specific purposes. We examine these representations in detail in the next subsection.

%%%%%%%%%%%%%%%%%%%%%%% Models of Quantum Operations %%%%%%%%%%%%%%%%%%%%%%% 
\subsection{Representations of Quantum Operations}\label{sec:rep_quant_proc}

\begin{table*}[t]
\renewcommand{\arraystretch}{1.1}

\centering
\resizebox{1.6\columnwidth}{!}{
\begin{tabular}{c | c | l}
    \hline
    \hline
    Mathematical & Graphical & Representation \\
    \hline
    $\E(\rho) = \Tr_\text{env}[ U \left( \rho \otimes \rho_\text{env} \right) U^\dagger ]$ & \includegraphics[height=1.8cm]{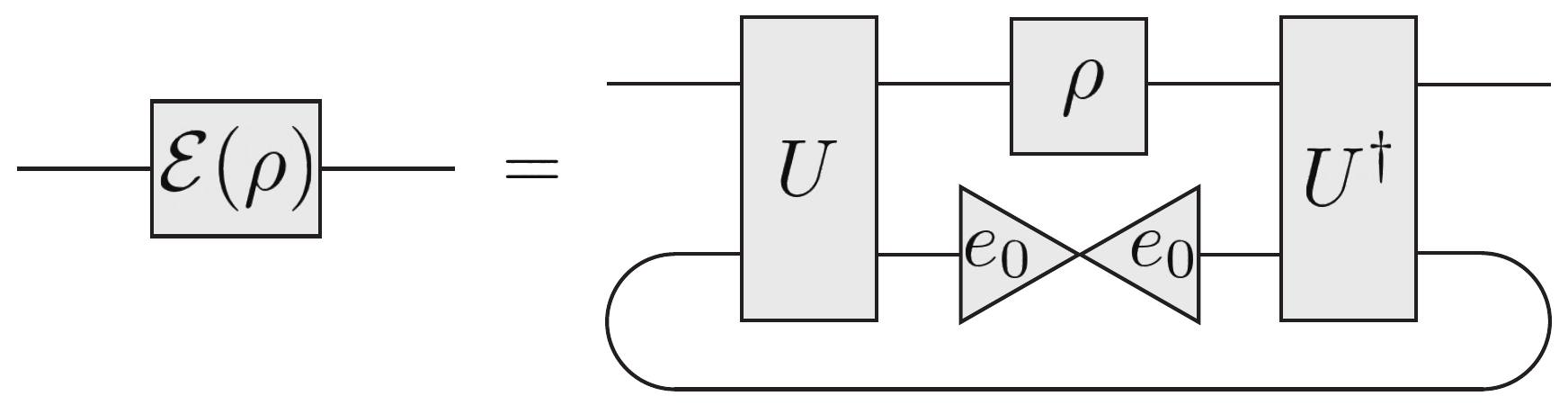} & Stinespring \\
    \hline
    $\E(\rho) = \sum_i^N K_i \rho K_i^\dagger$ & \includegraphics[height=0.75cm]{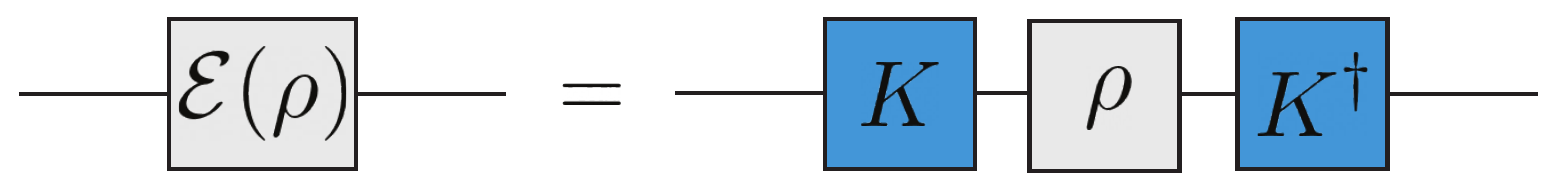} & Kraus \\
    \hline
    $\sket{\E(\rho)} = \Lambda\sket{\rho}$ & \includegraphics[height=1.35cm]{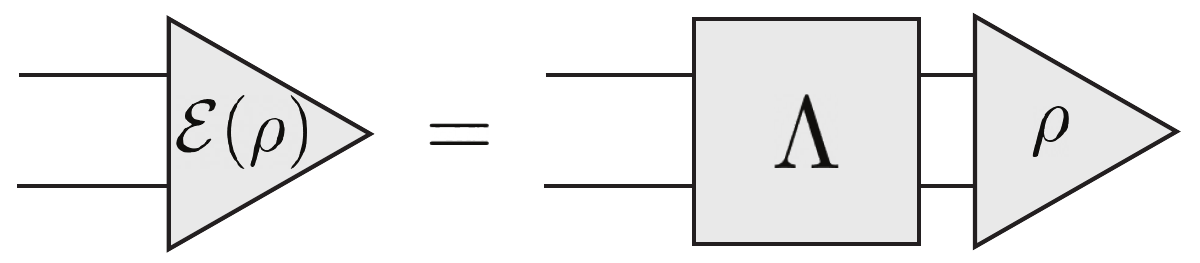} & Transfer matrix \\
    \hline
    $ \E(\rho) = \sum_{jk}{ \chi_{jk} P_j \rho P_k }$ & \includegraphics[height=1.85cm]{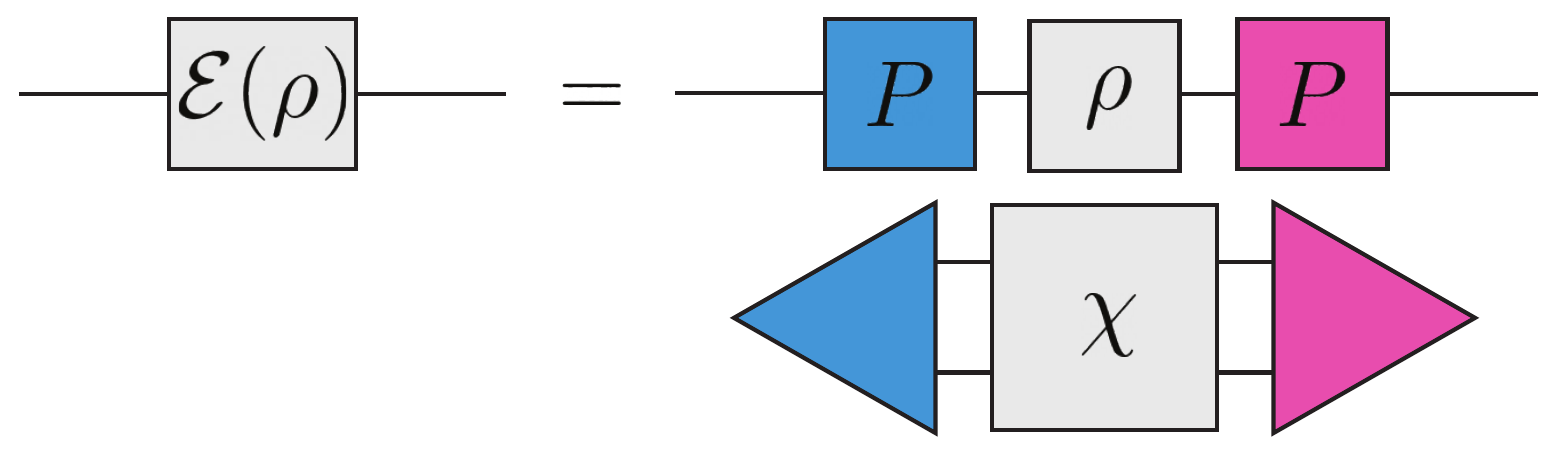} & Chi (process) matrix \\
    \hline
    $\C = \sum_k \mathbf{vec}(K_k)\mathbf{vec}(K_k)^\dagger$ & \includegraphics[height=1.25cm]{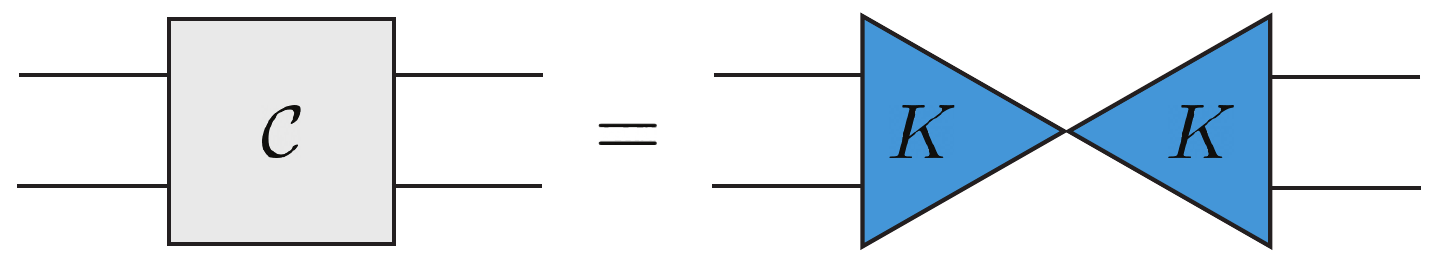} & Choi matrix \\
    \hline
    \hline
\end{tabular}}
\caption{
\textbf{Graphical Representations of Operations.}
\R\cite{wood2011tensor} introduces a useful and intuitive ``graphical calculus,'' closely related to the visual language of quantum circuits. These graphical representations can be a useful aid to intuition for students learning the mathematics of open quantum systems. This table shows the graphical representations of the Stinespring, Kraus, transfer matrix, Chi (process) matrix, and Choi matrix representations of a quantum operation. Although we do not explain here how these diagrams are interpreted and manipulated, the interested reader is referred to \R\cite{wood2011tensor} for a comprehensive discussion of graphical calculus and its use in quantum computing.
}
\label{tab:op_rep}

\end{table*}

We have already seen one way to represent a quantum operation $\E$, as a unitary transformation on a larger Hilbert space (\eq\ref{eq:stinespring}). In this section, we will construct several more, each with their own unique properties and advantages. These mathematical representations can also be represented \emph{graphically}, which the interested reader might find helpful for understanding the math of this section. We provide several examples of these graphical representations in Tab.~\ref{tab:op_rep}, but a formal introduction to this material is beyond the scope of this tutorial. For a complete introduction to the graphical calculus of open quantum systems, the reader is referred to \R\cite{wood2011tensor}.

%%%%%%%%%%%%%%%%%%%%%%% Kraus Form %%%%%%%%%%%%%%%%%%%%%%% 
\subsubsection{Kraus (Operator-Sum) Representation}\label{sec:kraus_rep}

We can construct a second representation by rewriting \eq\ref{eq:stinespring} in terms of a set of orthonormal basis states $\{ \ket{e_i} \}$ for the environment's Hilbert space, as
\begin{equation}\label{eq:Kraus_derivation}
     \mathcal{E}(\rho) = \sum_i \bra{e_i} U \left( \rho \otimes \ketbra{e_0} \right) U^\dagger \ket{e_i} ~.
\end{equation}
If we define $K_i \equiv \bra{e_i} U \ket{e_0}$, then \eq\ref{eq:Kraus_derivation} becomes
\begin{equation}\label{eq:kraus}
    \E(\rho) = \sum_i^N K_i \rho K_i^\dagger ~.
\end{equation}
The operators $\{ K_i \}$ are known as \emph{Kraus operators}, and \eq\ref{eq:kraus} is known as the \emph{Kraus} or \emph{operator-sum} representation of a quantum operation $\E$; it is depicted graphically in Tab.~\ref{tab:op_rep}. The Kraus representation is also not unique --- distinct sets $\{K_i\}$ and $\{K'_i\}$ can produce identical operations $\E$. However, it is always possible to construct a Kraus representation with $N \le d^2$ Kraus operators $K_i$ (where $d=2^n$ for $n$ qubits) that are mutually orthogonal --- i.e., $\Tr(K_i^\dagger K_j)=0$ for $i \ne j$. This representation is almost always unique (see Section 8.2.4 of Ref.~\cite{nielsen2002quantum} for an in-depth analysis).

The Kraus representation of $\E$ avoids any explicit reference to the environment's state or dynamics, describing the evolution of $\rho$ using only operators acting on the principal system. The Kraus operators do not need to be derived (as we did above) from properties of the environment. \emph{Any} Kraus representation automatically satisfies the CP condition, and a set of Kraus operators $\{K_i\}$ satisfies the TP condition (and thus describes a physically allowed quantum operation) iff it satisfies a \emph{completeness relation}:
\begin{equation}\label{eq:kraus_completeness}
    \sum_i^N K_i^\dagger K_i = \Id ~.
\end{equation}

If the Kraus operators of a single-qubit operation $\E$ are proportional to Pauli operators, then we call $\E$ a \emph{Pauli channel}. This concept can be extended to quantum operations acting on $n>1$ qubits using the $n$-qubit Pauli operators, which comprise all $4^n$ tensor products of 1-qubit Pauli operators. If we define $\mathbb{P} = \{I,X,Y,Z\}$, then the $n$-qubit Pauli group $\mathbb{P}_n$ is given by
\begin{equation}
    \mathbb{P}_n \equiv \mathbb{P}^{\otimes n} = \{I, X, Y, Z\}^{\otimes n} ~.
\end{equation}
An $n$-qubit operation $\E$ is a Pauli channel if
\begin{equation}\label{eq:kraus_pauli}
    \mathcal{E}(\rho) = \sum_{P \in \mathbb{P}_n} p_P P \rho P^\dagger
\end{equation}
for some probability distribution $\{p_P\}$. Pauli channels are useful and intuitive because they describe probabilistic (a.k.a., stochastic or random) processes. Each Pauli operator $P$ is a unitary operation that could ``happen'' to $\rho$, and if $\rho$ evolves according to a Pauli channel, then $P$ occurs with probability $p_P$.

%%%%%%%%%%%%%%%%%%%%%%% Superoperators %%%%%%%%%%%%%%%%%%%%%%% 
\subsubsection{Transfer Matrix Representation}\label{sec:superop}

The third common representation of a quantum operation $\E$ is as a \emph{linear superoperator} or \emph{transfer matrix}. We saw above that a quantum operation's action on a density matrix must be described by a \emph{linear} map $\rho \mapsto \E(\rho)$. And, as observed previously, density matrices describing a system's state can be thought of as vectors in the Hilbert-Schmidt space of $d \times d$ matrices. So, just as $\ket{\psi}$ is a vector in Hilbert space that is transformed by unitary operators $U$, $\rho$ can be viewed as a vector $\sket{\rho}$ in Hilbert-Schmidt space that is transformed by linear maps $\E$. 

We can make this action explicit by taking a $d \times d$ density matrix $\rho$ whose elements in a standard basis are
\begin{equation}
    \rho \doteq \begin{pmatrix}
                \rho_{11} & \rho_{12} & \ldots & \rho_{1d}  \\
                \rho_{21} & \rho_{22} & \ldots & \rho_{2d}  \\
                \vdots & \vdots & \ddots & \vdots           \\
                \rho_{d1} & \rho_{d2} & \ldots & \rho_{dd}
           \end{pmatrix}
\end{equation}
and \emph{vectorizing} \footnote{``Vectorization'' is used, in quantum information science and QCVV specifically, in at least two distinct ways. This one (Eq.~\ref{eq:vectorization}) maps a mathematical ``vector'' (an abstract element $\rho$ of Hilbert-Schmidt space) to a computer programming ``vector'' (a concrete list of numbers). But around Eq.~\ref{eq:vec}, we introduce another ``vectorization'' that maps mathematical vectors in $\mathcal{B}(\mathcal{H})$ to mathematical vectors in $\mathcal{H}\otimes\mathcal{H}$. Both are standard in the literature, and we apologize to the reader for our field's overloaded terminology.} it into a $d^2 \times 1$ column vector \cite{gilchrist2009vectorization}
\begin{equation}\label{eq:vectorization}
    \ket{\rho}\rangle \doteq (\rho_{11}, \rho_{21}, \ldots, \rho_{d1}, \rho_{12}, \ldots, \rho_{dd})^\trans ~.
\end{equation}
This particular representation, sometimes called ``column stacking'' because the columns of $\rho$ are stacked on top of each other, is one specific way to represent $\sket{\rho}$ concretely in a particular operator basis. To get Eq.~\ref{eq:vectorization}, we use the basis of \emph{matrix units} defined (in terms of the Hilbert space basis used to represent $\rho$) by $\{\ketbra{i}{j},\ i,j=1\ldots d\}$, because
\begin{equation}
    \rho_{ij} = \Tr[ \ketbra{j}{i} \rho ] = \sbraket{\ketbra{i}{j}}{\rho}~.
\end{equation}
But an operator $A$ can be ``vectorized'' in any operator basis. One particularly useful choice is the $n$-qubit Pauli basis. However, whereas the matrix units are orthonormal because
\begin{equation}
    \sbraket{ \ketbra{i}{j}}{\ketbra{k}{l}} = \Tr[ \ketbra{j}{i} \ketbra{k}{l} ] = \delta_{ik}\delta_{jl} ~,
\end{equation}
the Paulis are mutually orthogonal but not normalized, because 
\begin{equation}
    \sbraket{ P_i}{ P_j} = \delta_{ij}\Tr[\Id] = d\delta_{ij} ~.
\end{equation}
This can be dealt with either by using \emph{normalized} Pauli operators $\{P_i/\sqrt{d}\}$, or by computing the coefficient of each Pauli basis operator as $\sbraket{ P_i}{A}/d$ instead of $\sbraket{ P_i}{A}$.

Using this framework, a quantum operation $\E$ is just a linear transformation that maps any density matrix $\rho \in \mathcal{B}(\mathcal{H})$ to a new density matrix $\rho' \in \mathcal{B}(\mathcal{H})$ in the same Hilbert-Schmidt space \footnote{It is possible to define quantum operations that map $\mathcal{B}(\mathcal{H})\mapsto \mathcal{B}(\mathcal{H'})$, where $\mathcal{H}\neq\mathcal{H'}$, but these are used relatively rarely in QCVV and out of scope for this tutorial.}. Therefore, $\E$ can be concretely represented by a $d^2 \times d^2$ matrix $\Lambda$ that acts on vectorized states $\ket{\rho}\rangle$ by matrix multiplication:
\begin{equation}\label{eq:supop_map}
    \sket{\rho} \mapsto \sket{\E(\rho)} = \Lambda\sket{\rho} ~.
\end{equation}
A linear map on density matrices, or a matrix that acts on vectorized density matrices, is called a \emph{superoperator}. Every quantum operation can be described by a superoperator $\Lambda$. Superoperators representing quantum operations are often called \emph{transfer matrices} because $\Lambda$'s action on vectorized density matrices resembles the action of transfer operators in dynamical systems or statistical mechanics. This representation is also sometimes called the \emph{Liouville} or \emph{associative} representation, and is depicted graphically in Tab.~\ref{tab:op_rep}.

An operation $\E$'s transfer matrix $\Lambda$ can be concretely constructed by choosing an orthonormal basis $\{B_i\}$ for the vector space of $d\times d$ matrices, then defining the elements of $\Lambda$ using the Hilbert-Schmidt inner product:
\begin{equation}\label{eq:supop}
    \Lambda_{ij} \equiv \Tr \left[ B_i^\dagger \E(B_j) \right] ~.
\end{equation}
It is common to construct $\Lambda$ in the basis of matrix units, or the Pauli basis (see Sec.~\ref{sec:ptm_rep}). We usually work in the Pauli basis; transfer matrices and other objects that are explicitly represented in the basis of matrix units are identified with the subscript $c$. If $\E$ has Kraus operators $\{K_i\}$, then its transfer matrix in the basis of matrix units is
\begin{equation}\label{eq:kraus_to_supop}
    \Lambda_c = \sum_i K_i^* \otimes K_i ~.
\end{equation}

In the transfer matrix representation, the composition of quantum operations is associative. In other words, if two operations with transfer matrices $\Lambda_1$ and $\Lambda_2$ are applied in succession, then the net effect is to apply the product of the two matrices, i.e., $\Lambda_2\Lambda_1$:
\begin{equation}
    \ket{\rho'}\rangle = \ket{\E_2(\E_1(\rho))}\rangle = \Lambda_2 \ket{\E_1(\rho)}\rangle = \Lambda_2 \Lambda_1 \ket{\rho}\rangle ~.
\end{equation}
Because of this useful property, the transfer matrix representation is widely used to model errors in quantum logic operations (i.e., gates). When those errors are \emph{small}, a variation called the \emph{error generator} formalism \cite{blume2022taxonomy} --- that represents transfer matrices by their logarithms --- is useful for distinguishing and classifying small errors (see Appendix \ref{sec:error_gen} for a brief summary). Error generators can be used to classify and quantify the rates at which different types of errors occur in quantum gates \cite{mkadzik2021precision}.

%%%%%%%%%%%%%%%%%%%%%%% Pauli Transfer Matrix %%%%%%%%%%%%%%%%%%%%%%% 
\subsubsection{Pauli Transfer Matrix Representation}\label{sec:ptm_rep}

When a transfer matrix is constructed in the Pauli basis, we call it a \emph{\ac{PTM}}. This is also sometimes known as the \emph{Pauli-Liouville} representation of quantum operations. We use the symbol $\Lambda$ for all transfer matrix representations, regardless of basis, but in this tutorial $\Lambda$ will always indicate a PTM unless another basis is specified.

The PTM representation of a quantum operation $\E$ is a $4^n \times 4^n$ superoperator $\Lambda$ with entries
\begin{equation}
    \Lambda_{ij} = \sbraket{P_i}{\E(P_j)} = \frac{1}{d}\Tr \left[ P_i\E(P_j) \right] ~,
\end{equation}
where $P_i$ and $P_j$ are elements of the $n$-qubit Pauli group $\mathbb{P}_n$. PTMs act on density matrices which have also been expanded in the Pauli basis,
\begin{equation}
    \rho = \sum_{P \in  \mathbb{P}_n} \rho_P P ~,
\end{equation}
where $\rho_P = \sbraket{P}{\rho} / d$ are the expansion coefficients. By vectorizing the expansion coefficients into a single column vector,
\begin{equation}
    \sket{\rho} \doteq \begin{pmatrix} \rho_{I^{\otimes n}} & ... & \rho_{Z^{\otimes n}} \end{pmatrix}^\trans ~,
\end{equation}
the quantum map $\rho' = \E(\rho)$ can be expressed in vector form, where the PTM $\Lambda$ acts on $\sket{\rho}$ by direct matrix multiplication: $\sket{\rho'} = \Lambda \sket{\rho}$. For example, for a single qubit,
\begin{equation}
    \sket{\rho} \doteq \begin{pmatrix}
                        \rho_{I} \\
                        \rho_{X} \\
                        \rho_{Y} \\
                        \rho_{Z}
                   \end{pmatrix} ~,
\end{equation}
and the map $\rho' = \E(\rho)$ is given by
\begin{equation}\label{eq:single_qubit_ptm}
    \begin{pmatrix}
        \rho_{I}' \\
        \rho_{X}' \\
        \rho_{Y}' \\
        \rho_{Z}
    \end{pmatrix} = 
    \begin{pmatrix}
        \Lambda_{II} & \Lambda_{IX} & \Lambda_{IY} & \Lambda_{IZ} \\
        \Lambda_{XI} & \Lambda_{XX} & \Lambda_{XY} & \Lambda_{XZ} \\
        \Lambda_{YI} & \Lambda_{YX} & \Lambda_{YY} & \Lambda_{YZ} \\
        \Lambda_{ZI} & \Lambda_{ZX} & \Lambda_{ZY} & \Lambda_{ZZ}
    \end{pmatrix}
    \begin{pmatrix}
        \rho_{I} \\
        \rho_{X} \\
        \rho_{Y} \\
        \rho_{Z}
    \end{pmatrix} ~.
\end{equation}

\begin{figure}[t]
    \centering
    \includegraphics[width=\columnwidth]{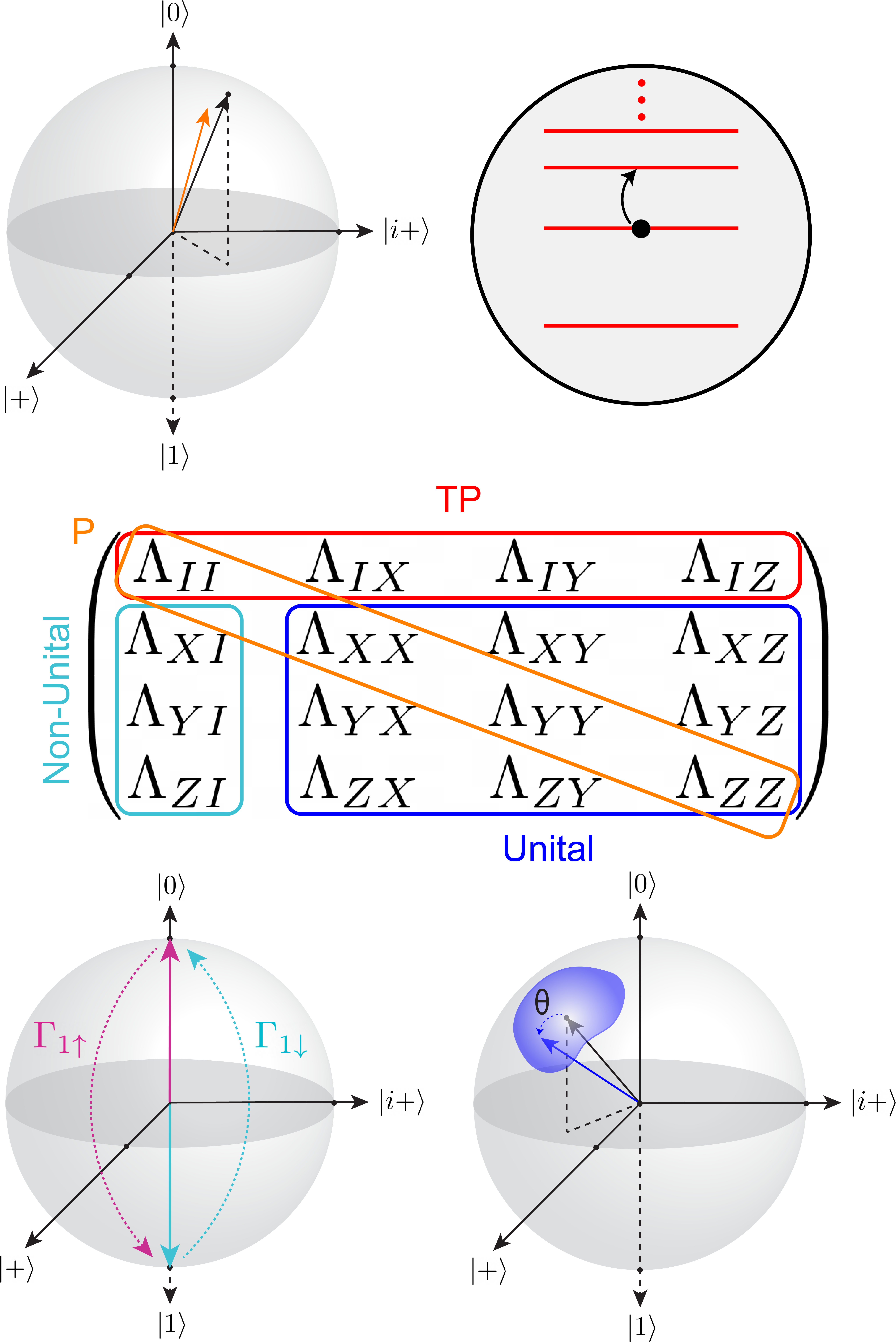}
    \caption[Pauli Transfer Matrix.]{\textbf{Pauli Transfer Matrix.} 
    We can identify four useful blocks within a PTM. The top row is typically fixed as $[1, 0, 0, 0]$ by trace preservation (TP, red), although postselected operations can be non-TP. The lower right-hand block (blue) captures unital processes, such as unitary errors. The column to the left of the unital block (cyan) indicates non-unital processes, such as $T_1$ decay, resulting in $\Lambda_{PI} \ne 0$ for $P \in \{X, Y, Z\}$. The diagonal elements indicate how well polarization (P, orange) along the various Pauli axes is preserved, and are directly impacted by stochastic Pauli noise. The spheres at each corner depict example errors captured by each block (see \fig\ref{fig:Bloch_sphere_errors_noise}).
    }
    \label{fig:ptm}
\end{figure}

The entries of a PTM are all real numbers bounded by $\Lambda_{ij} \in [-1, 1]$. Some important properties of an operation can be extracted directly from its PTM. We can isolate four (slightly overlapping) useful blocks within a PTM, as shown in \fig\ref{fig:ptm}.
\begin{itemize}
    \item $\Lambda$'s top row reveals whether it is trace-preserving. The operation is TP if and only if $\Lambda_{0j} = \delta_{0j}$ (i.e., if the first row of the PTM is $[1, 0, ..., 0]$). Every deterministic process must be TP, but postselected operations provide an example of non-TP processes \footnote{Note that some authors consider leakage, for example, to be a non-TP process, in which case the top row of the PTM captures state-dependent leakage. This is true if one only considers the qubit subspace within the full Hilbert space. However, strictly speaking, leakage is still TP, since the total probability of observing \emph{some} outcome is preserved. For example, in some platforms leakage cannot be detected, and might instead be (erroneously) measured as 0 or 1, but the total number of shots will remain the same. In other platforms leakage can be more easily measured (see, for example, \fig\ref{fig:lrb}), in which case the total probability of observing 0, 1, or 2 is preserved. Therefore, when considering only the qubit subspace in the presence of leakage, it is sometimes common to relax the TP constraint, and instead simply require that the total probability must not increase (i.e., $\Tr[\E(\rho)] \le \Tr[\rho]$).}.
    
    \item The bottom right $(d^2-1) \times (d^2-1)$ block of the PTM is called the \emph{unital} block. An operation is unital if it preserves the identity [i.e., $\E(\Id) = \Id$], and the PTM for a unital operation is restricted to this block and the $1\times 1$ block defined by $\Lambda_{00}$. Unital processes cannot increase purity (or decrease entropy). Unitary dynamics (Sec.~\ref{sec:coherent_errors}) and stochastic Pauli errors (Sec.~\ref{sec:stoch_pauli}) are unital.
    
    \item The leftmost column of $\Lambda$ is called the \emph{non-unital} block. For any unital operation, $\Lambda_{i0} = \delta_{i0}$ (i.e., the first column of the PTM is $[1, 0, ..., 0]^\trans$). If it does not take this form, its elements indicate entropy-decreasing processes like cooling, energy relaxation, or spontaneous emission (e.g., $T_1$ decay; see Sec.~\ref{sec:spont_emis_amp_damp}). 
    
    \item The diagonal elements of $\Lambda$ quantify how well \emph{polarization} is preserved along each Pauli axis (see Sec.~\ref{sec:polarization}), with $\Lambda_{PP} = 1$ if the operation preserves the component of Pauli operator $P$ in $\rho$. $\Lambda_{PP} < 1$ indicates loss of polarization or coherence along a Pauli axis. A PTM's diagonal elements $\Lambda_{PP}$ are sometimes called the \emph{survival probabilities}, \emph{Pauli fidelities}, or (when $\Lambda$ is diagonal) \emph{Pauli eigenvalues}. An operation's PTM is diagonal if and only if it is a Pauli channel. 
\end{itemize}
The TP constraint is obvious and easy to enforce in the PTM representation, by requiring that $\Lambda_{0j} = \delta_{0j}$. In contrast, the CP constraint is hard to express or evaluate in the PTM representation. The easiest way to test whether a PTM $\Lambda$ describes a CP map is to construct its Choi matrix representation (see Sec.~\ref{sec:choi_mat_rep}).

An operation's PTM can be computed directly from a transfer matrix represented in a different basis (\eq\ref{eq:supop}) by applying a unitary change of basis. Suppose, for example, $\Lambda_c$ is a transfer matrix written in the basis of matrix units (\eq\ref{eq:kraus_to_supop}). We can construct the equivalent PTM $\Lambda$ as
\begin{equation}\label{eq:supop_to_ptm}
    \Lambda =  U_{c \rightarrow P} \Lambda_c U_{c \rightarrow P}^\dagger ~,
\end{equation}
where 
\begin{equation}\label{eq:c2p_basis_transform}
    U_{c \rightarrow P} \equiv \frac{1}{\sqrt{d}}\sum_{i=0}^{d^2 - 1} \ketbra{c_i\rangle}{\langle P_i} ~,
\end{equation}
where $\{\sket{c_i}\}_{i=0}^{d^2 - 1}$ is the basis of matrix units. The factor of $1/\sqrt{d}$ normalizes the Pauli basis elements to 1 and makes $U_{c \rightarrow P}$ unitary. The inverse transformation is also possible with
\begin{equation}\label{eq:ptm_to_supop}
    \Lambda_c =  U_{P \rightarrow c} \Lambda U_{P \rightarrow c}^\dagger ~,
\end{equation}
where $U_{P \rightarrow c} = U_{c \rightarrow P}^\dagger$.

%%%%%%%%%%%%%%%%%%%%%%% Chi Matrix Representation %%%%%%%%%%%%%%%%%%%%%%% 
\subsubsection{Chi (Process) Matrix Representation}\label{sec:chi_mat_rep}

We can construct a fourth representation of $\E$ by expanding $\E$'s Kraus operators $\{K_i\}$ (\eq\ref{eq:kraus}) in a fixed operator basis $\{P_j\}$, such as the Pauli basis:
\begin{equation}
    K_i = \sum_j{ c_{ij} P_j} ~,
\end{equation}
where $c_{ij}$ are the expansion coefficients. Plugging this expansion into \eq\ref{eq:kraus} yields
\begin{equation} \label{eq:chi_rep}
    \E(\rho) = \sum_{jk}{ \chi_{jk} P_j \rho P_k } ~,
\end{equation}
where $\chi_{jk} = \sum_i{c_{ij}c^*_{ik}}$. The $d^2\times d^2$matrix of coefficients $\chi_{jk}$ is called a \emph{$\chi$ matrix}, and this is known as the \emph{$\chi$ matrix representation} of an operation $\E$; it is depicted graphically in Tab.~\ref{tab:op_rep}. Historically, an operation's $\chi$ matrix was called its \emph{process matrix}. More recently, the term ``process matrix'' has been used more broadly to describe other matrix representations of $\E$ (e.g., the Pauli transfer matrix). In this tutorial, \emph{process matrix} will always refer to the $\chi$ matrix, and $\chi$ matrices will be constructed in the Pauli basis unless otherwise specified.

The $\chi$ matrix representation is closely related to the Kraus representation, but has certain advantages. It is easy to construct mechanically, and it is unique once an operator basis $\{P_j\}$ is chosen. A map $\E$ is CP iff its $\chi$ matrix is positive semi-definite. Constructing $\E$'s $\chi$ matrix and checking whether it is positive semi-definite is the easiest way to check complete positivity. Trace preservation (TP) is also easy to check in this representation; $\E$ is TP iff $\sum_{j,k} \chi_{jk} P_j^\dagger P_k = \mathbb{I}$. This condition is equivalent to \eq\ref{eq:kraus_completeness}. Moreover, it constrains $d^2$ of the $d^4$ parameters in $\chi$, and so the $\chi$ matrix for a CPTP map has $d^2(d^2-1)$ free parameters.

Equation \ref{eq:chi_rep} can be seen as an expansion of $\E$ as a linear combination of non-CPTP linear maps known as \emph{Choi units}, denoted $X$:
\begin{align}
    \E &= \sum_{jk}{ \chi_{jk} X_{jk}} ~, \\
    X_{jk}(\rho) &\equiv P_j \rho P_k ~.
\end{align}
Each Choi unit is a superoperator acting on operators. Choi units are Hermitian ($X^\dagger_{jk} = X_{jk}$), and they form an orthogonal basis,
\begin{equation}
    \Tr[ X_{ij}^\dagger X_{kl} ] = d^2\delta_{ik}\delta_{jl} ~.
\end{equation}
This makes it easy to construct the $\chi$ matrix for any operation $\E$, since
\begin{equation}
    \chi_{jk} = \frac{1}{d^2}\Tr[ X_{jk}^\dagger \E ] = \frac{1}{d^3}\sum_{P_m \in \mathbb{P}_n}{ \Tr[ P_m P_j \E( P_m ) P_k ]} ~.
\end{equation}
For example, the $\chi$ matrix coefficients for the identity operation $\E(\rho) = \rho$ are
\begin{align}
    \chi_{jk} 
        &= \frac{1}{d^3} \sum_{P_m \in \mathbb{P}_n}{ \Tr[ P_m P_j P_m P_k ]} ~, \\
        &= \frac{1}{d^2} \Tr[P_j] \Tr[P_k] ~, \\
        &= \delta_{j0}\delta_{k0} ~,
\end{align}
using the identity $\sum_{P \in \mathbb{P}_n}{PMP} = d\Tr(M)\Id$.

Any $\chi$ matrix for a CP map can be diagonalized by a unitary change of basis. This diagonal form,
\begin{equation}
    \E(\rho) = \sum_j{ \lambda_j Q_j \rho Q^\dagger_j } ~,
\end{equation}
gives the orthogonal Kraus representation of $\E$, with $K_i = \sqrt{\lambda_i} Q_i$. It is unique up to degeneracies ($\lambda_j = \lambda_{j'}$ for some $j\neq j'$). Constructing the $\chi$ matrix in some basis and diagonalizing it is the easiest way to find the orthogonal Kraus form of a generic operation.

Both the PTM and $\chi$ matrix representations of a map $\E$ are unique. So, it is possible to compute the PTM from the $\chi$ matrix as
\begin{equation}\label{eq:chi_to_ptm}
    \Lambda_{ij} = \frac{1}{d} \sum_{kl} \chi_{kl} \Tr \left[ P_i P_k P_j P_l \right] ~,
\end{equation}
and the $\chi$ matrix from the PTM as
\begin{equation}\label{eq:ptm_to_chi}
    \chi_{ij} = \frac{1}{d^3} \sum_{kl} \Lambda_{kl} \Tr \left[ P_l P_i P_k P_j \right] ~.
\end{equation}

%%%%%%%%%%%%%%%%%%%%%%% Choi Matrix Representation %%%%%%%%%%%%%%%%%%%%%%% 
\subsubsection{Choi Matrix Representation}\label{sec:choi_mat_rep}

The final representation of a quantum operation $\E$ that we consider represents $\E$ as an unnormalized quantum state of an larger (2-copy) system. We begin with a maximally entangled state of two systems, which can be defined in terms of any orthonormal basis $\{\ket{i}\}_{i=0}^{d - 1}$ as
\begin{equation}
    \ket{\Phi_0} = \sum_i \ket{i} \otimes \ket{i} ~,
\end{equation} 
keeping in mind that this ``state'' has norm $d$. Its density matrix is
\begin{align}
    \ketbra{\Phi_0} = \sum_{i,j} \ketbra{i}{j}\otimes\ketbra{i}{j} ~,
\end{align}
which for an $n$-qubit system can also be written using Pauli operators as
\begin{equation} \label{eq:max_ent_Pauli}
    \ketbra{\Phi_0} = \frac{1}{d}\sum_{P\in\mathbb{P}_n} P \otimes P^\trans ~.
\end{equation}
We now apply $\Id\otimes\E$ to this state to get
\begin{align}
    \C &= (\mathbb{I} \otimes \E)\left[ \ketbra{\Phi_0}\right] ~, \\
    &= \sum_{i,j=0}^{d-1} \ketbra{i}{j} \otimes \E(\ketbra{i}{j}) ~, \label{eq:choi} \\
    &= \frac{1}{d}\sum_{P\in\mathbb{P}_n} P \otimes \E(P)^\trans ~.
\end{align}
$\C$ is called the \emph{Choi matrix} \cite{choi1975completely}, or sometimes the \emph{dynamical matrix} \cite{sudarshan1961stochastic}, of $\E$. The action of an operation $\E$ can be written in terms of its Choi matrix $\C$ as
\begin{equation}\label{eq:choi_evolution}
    \E(\rho) = \Tr_1 \left[ \C (\rho^\trans \otimes \mathbb{I}) \right] ~,
\end{equation}
where $\Tr_1$ denotes the partial trace over the 1st subsystem. By substituting this into \eq\ref{eq:choi}, it is straightforward to verify that $\C$ faithfully and uniquely represents $\E$.

This one-to-one correspondence between completely positive \emph{operations} $\E$ and positive semidefinite (bipartite) \emph{states} $\C$ is known as the \emph{Choi-Jamiołkowski isomorphism} (or \emph{channel-state duality}) \cite{jamiolkowski1972linear, zyczkowski2004duality}. It implies several useful properties:
\begin{enumerate}
    \item $\E$ is CP $\iff$ $\C \ge 0$,
    \item $\E$ is TP $\iff$ $\Tr_1[\C] = \mathbb{I}$,
    \item $\E$ is Hermitian-preserving (HP) $\iff$ $\C  = \C^\dagger$,
    % \item $\E$ is unital $\iff$ $\Tr_2[\C] = \mathbb{I}$,
\end{enumerate}
where $\iff$ denotes \emph{if and only if}.

An even more fundamental (and perhaps surprising) statement is that the Choi matrix is proportional to the transpose of the $\chi$ matrix in suitably chosen bases. In other words, given any basis $\{\ket{\phi_i}\}$ for the bipartite Hilbert space $\mathcal{H} \otimes \mathcal{H}$, there is a corresponding operator basis $\{\ket{P_i}\rangle\}$ for $\mathcal{B}(\mathcal{H})$ so that the $\chi$ and Choi representations of any operation $\E$ obey
\begin{equation}\label{eq:ChoiChi}
    \chi_{ji} = \frac{1}{d}\braket{\phi_i|\C|\phi_j}~.
\end{equation}
We can demonstrate this using the Pauli operator basis, and the basis of \emph{normalized} maximally entangled states (for $\mathcal{H} \otimes \mathcal{H}$) given by $\{\ket{\phi_j}\} = \left\{ (P_j\otimes\Id)\ket{\Phi_0} / \sqrt{d} \right\}_{P_j \in \mathbb{P}_n}$. Using \eq\ref{eq:choi} and \eq\ref{eq:max_ent_Pauli}, we can write
\begin{align}
    \frac{1}{d} \braket{\phi_i|\C|\phi_j} 
        &= \frac{1}{d} \Tr[ \C (P_j \otimes \Id) \ketbra{\phi_0} (P_i \otimes \Id) ] ~, \\
        &= \frac{1}{d^4} \sum_{k,l} \Tr[ (P_k \otimes \E(P_k)^\trans)(P_j\otimes\Id)(P_l\otimes P_l^\trans)(P_i\otimes\Id) ] ~, \\
        &= \frac{1}{d^4 }\sum_{k,l} \Tr[P_k P_j P_l P_i] \Tr[P_l\E(P_k)] ~, \\
        &= \frac{1}{d^3} \sum_{k,l} \Lambda_{lk} \Tr[P_l P_i P_k P_j] ~, \\
        &= \chi_{ji} ~. \label{eq:Choi_Chi}
\end{align}
where in the last line we have used the cyclic property of the trace and \eq\ref{eq:ptm_to_chi}. This equivalence is very powerful, since it implies that $\C$ and $\chi$ are essentially identical [up to a transpose, a factor of $d$, and appropriate choice of bases for $\mathcal{H} \otimes \mathcal{H}$ and $\mathcal{B}(\mathcal{H})$].

If an operation $\E$ has the Kraus representation $\E(\rho) = \sum_k{K_k \rho K_k^\dagger}$, then we can construct its Choi representation (depicted graphically in Tab.~\ref{tab:op_rep}) beginning with Eq.~\ref{eq:choi} as
\begin{align}
    \C &= \sum_{i,j=0}^{d-1} \ketbra{i}{j} \otimes \E(\ketbra{i}{j}) ~, \\
    &= \sum_{i,j,k} \ketbra{i}{j} \otimes K_k\ketbra{i}{j}K_k^\dagger ~, \\
    &= \sum_{i,j,k} \left(\ket{i}\otimes K_k\ket{i}\right) \left(\bra{j}\otimes \bra{j}K_k^\dagger\right) ~, \\
    &= \sum_k \left(\sum_i\ket{i}\otimes K_k\ket{i}\right) \left(\sum_j\bra{j}\otimes \bra{j}K_k^\dagger\right) ~, \\
    \C &= \sum_k \mathbf{vec}(K_k)\mathbf{vec}(K_k)^\dagger \label{eq:choi_kraus} ~,
\end{align}
where $\mathbf{vec}$ is a linear map between $d\times d$ matrices (e.g., $K_k$) and bipartite states (e.g., $\ket{\Phi}$) defined by
\begin{equation}\label{eq:vec}
    \mathbf{vec}(\ketbra{i}{j}) = \ket{j}\otimes\ket{i} ~.
\end{equation}
This is an explicit form of the Choi-Jamiołkowski isomorphism. It is important to note that this ``vectorization'' is distinct from the vectorization introduced in Sec.~\ref{sec:superop}. The $\textbf{vec}$ operation associates a matrix acting on Hilbert space $\mathcal{H}$ with a vector in $\mathcal{H} \otimes \mathcal{H}$ (e.g., $\ket{j} \otimes \ket{i}$), whereas the vectorization in Sec.~\ref{sec:superop} merely constructs a concrete representation of that matrix as an element of $\mathcal{B}(\mathcal{H})$ (e.g., $\ket{ \ketbra{i}{j} \rangle}$). If the bipartite state in Eq.~\ref{eq:vec} is represented concretely, then it is possible to choose a particular basis for which these two concrete representations coincide. For example, in the computational basis,
\begin{equation}
    \mathbf{vec} \begin{pmatrix} a & b \\ c & d \end{pmatrix} 
        \doteq \begin{pmatrix} a \\ c \\ b \\ d \end{pmatrix} ~.
\end{equation}

%%%%%%%%%%%%%%%%%%%%%%% Models of Quantum Measurements %%%%%%%%%%%%%%%%%%%%%%% 
\subsection{Models of Quantum Measurements}\label{sec:measurement}

If a quantum system could not be observed, its state would be meaningless. Observations of quantum systems are called \emph{measurements}. Subsections \ref{sec:pvm} and \ref{sec:povm} introduced models for \emph{terminating} measurements (PVMs and POVMs) that can be used when the quantum system is used up during the measurement (e.g., photodetection) or can be thrown away (e.g., readout that concludes a quantum computation). But if the measured quantum system persists and might be observed again post-measurement, the POVM formalism is not sufficient to predict both the measurement outcome \emph{and} the post-measurement state. In the context of quantum computing (and thus QCVV), such measurements are usually called \emph{\ac{MCM}s}. In this section, we introduce \emph{quantum instruments} that model MCMs, and we discuss continuous weak measurements that model the internal dynamics of the readout process.

%%%%%%%%%%%%%%%%%%%%%%% Quantum Instruments %%%%%%%%%%%%%%%%%%%%%%% 
\subsubsection{Quantum Instruments}\label{sec:quantum_instruments}

Quantum measurements typically change quantum states. This is called \emph{measurement back-action} \cite{hatridge2013quantum}. Moreover, consecutive quantum measurements can give rise to geometric phases contingent upon the order of measurements \cite{cho2019emergence}. The POVM formalism, which maps a quantum state into a classical probability distribution, $\rho \mapsto \{p(i|\rho)\}$, is inadequate to describe the measurement-induced state dynamics. To address this limitation, the \emph{\ac{QI}} formalism is introduced, providing an extended framework that accounts for quantum operations influenced by the measurement outcome \cite{davies1970operational, rudinger2022characterizing}. Quantum instruments can model the three-step quantum measurement procedure \cite{vonNeumann1932mathematische, zurek1991decoherence} introduced by John von Neumann:
\begin{enumerate}
	\setlength\itemsep{-0.3em}
 	\item Initialize meter state to $|m_0\rangle$.
  	\item Apply interaction between the system and meter.
	\item Read meter state.
 \end{enumerate} 
Without loss of generality, the measurement interaction transforms the system-meter state of $\rho \otimes \ketbra{m_0}{m_0}$ into
\begin{equation}\label{eq:M_int}
    \mathcal{M}(\rho\otimes|m_0\rangle\langle m_0|) = \sum_{ij} \mathcal{M}_{ij}(\rho) \otimes \ketbra{m_i}{m_j} ~,
\end{equation}
and then the meter is projected onto one of the orthonormal eigenstates $\{|m_i\rangle\}$. The meter reads $m_i$ with a probability $p(i|\rho)$,
\begin{equation}
    p(i|\rho) = \Tr[\mathcal{M}_{ii}(\rho)] ~,
\end{equation}
and the projection also ``collapses'' the principal system into the post-measurement state,
\begin{equation}\label{eq:QI_state}
    \rho_i = \frac{\mathcal{M}_{ii}(\rho)}{p(i|\rho)} ~.
\end{equation}
A QI models this process by describing the transformation of a quantum state into a composite quantum-classical state, represented as $\rho\mapsto \{[\rho_i, p(i|\rho)]\}$. A QI $\mathcal{I}$ is a CPTP map that transforms a state $\rho$ as:
\begin{equation}\label{eq:QI}
    \rho \mapsto \mathcal{I}(\rho) = \sum_{i} \mathcal{M}_{i}(\rho)\otimes \ketbra{m_i}{m_i} ~.
\end{equation}
Each of the conditional quantum operations $\{\mathcal{M}_i\}$ is CP, and if there is no loss, the overall TP condition is imposed as follows:
\begin{equation}\label{eq:QI_TP}
    \sum_i \Tr[\mathcal{M}_i(\rho)] = 1 ~. 
\end{equation}

%%%%%%%%%%%%%%%%%%%%%%% Quantum Non-Demolition Measurements %%%%%%%%%%%%%%%%%%%%%%% 
\subsubsection{Quantum Non-Demolition Measurements}\label{sec:qnd}

A \emph{\ac{QND}} measurement is a quantum measurement that extracts information about a system while disturbing its quantum state as little as possible. QND measurements still ``collapse'' quantum states into a specific eigenspace of the observable that was measured, but performing repeated QND measurements will consistently produce identical outcomes, and will not change the expectation value of the measured observable~\cite{braginsky1980quantum, braginsky1996quantum}. The reproducible and minimally-perturbing nature of QND measurements plays a crucial role in achieving high fidelity readout~\cite{volz2011measurement,dassonneville2020fast} and is integral to quantum computing protocols including syndrome measurements in quantum error correction~\cite{google2023suppressing}, qubit recycling~\cite{liu2019variational}, and algorithms for quantum machine learning~\cite{cong2019quantum}.

The QND property can be expressed in terms of the quantum instrument (QI) formalism (Eq.~\ref{eq:QI}). Consistency of outcomes across repeated QND measurements implies that the measurement probabilities satisfy 
\begin{equation}
    p(i|\rho_i) = \Tr[\mathcal{M}_i(\rho_i)] = \Tr[\mathcal{M}_i(\mathcal{M}_i(\rho_i))] = 1 ~.
\end{equation}
% Consider an observable $O$ expressed as a sum of projectors,
% \begin{equation}\label{eq:observable_O}
%     O = \sum_i o_i \ \! \Pi_i = \sum_i o_i \ketbra{\lambda_i} ~,
% \end{equation}
% Considering the observable $O$ as defined in Eq.~\ref{eq:observable_O}, 
% with eigenstates represented by $\rho_i = \ketbra{\lambda_i}{\lambda_i}$. 
Therefore, a QND measurement’s conditional operations must be projectors:
\begin{equation}
    % \mathcal{M}_i(\rho) = \ketbra{\lambda_i}{\lambda_i} \rho \ketbra{\lambda_i}{\lambda_i} ~.
    \mathcal{M}_i^2 = \mathcal{M}_i \Longrightarrow \mathcal{M}_i(\rho) = \Pi_i \rho \Pi_i ~,
\end{equation}
where the $\Pi_i$ are mutually orthogonal projectors. If the system experiences a Hamiltonian, then the post-measurement states must remain unchanged under the state evolution governed by the system Hamiltonian $H_s$, 
\begin{equation}\label{eq:qnd}
    e^{-iH_s t/\hbar} \Pi_i e^{iH_s t/\hbar} = \Pi_i ~.
\end{equation}
This holds true, for all $i$, if and only if
\begin{equation}
    [\Pi_i, H_s] = 0 ~.
\end{equation}
% This commutation relation further establishes that the measurement interaction on $O$ remains unaffected by the system Hamiltonian during the interaction, leading to high-fidelity readout. 
Interestingly, the condition given by Eq.~\ref{eq:qnd} can be eased if the system exhibits periodicity, such that  $e^{-iH_s\tau/\hbar} = e^{i\phi}\mathbb{I}$, where $\tau$ is the measurement interval. Therefore, even when $[\Pi_i, H_s] \neq 0$, it becomes feasible to conduct a QND measurement by measuring the system at intervals of $\tau$. This approach is known as a stroboscopic QND measurement \cite{caves1980measurement}.

%%%%%%%%%%%%%%%%%%%%%%% Continuous Weak Measurements %%%%%%%%%%%%%%%%%%%%%%% 
\subsubsection{Continuous Weak Measurements}\label{eq:weak_meas}

Consider a dispersive interaction \cite{siddiqi2006dispersive} between a qubit and a cavity described by the following approximated Hamiltonian:
\begin{equation}\label{eq:Hdisp}
    H_\mathrm{disp} = \hbar\chi\sigma_z \otimes a^{\dagger}a ~,
\end{equation}
where $\chi$ is the dispersive shift in frequency of the qubit, and $a^{\dagger}$ and $a$ are creation and annihilation operators for the cavity mode. We take the initial qubit state to be $\ket{\psi}$ and, for simplicity, we assume that the cavity initially contains a coherent meter state $\ket{\alpha}$ with no energy loss (more realistic models and analyses can be found in Ref.~\cite{blais2021circuit}). After the qubit-meter state $|\psi\rangle|\alpha\rangle$ evolves under the dispersive Hamiltonian for time $t$, they become entangled as
\begin{equation}
    e^{-iH_\mathrm{disp}t/\hbar}|\psi\rangle|\alpha\rangle=\Pi_0|\psi\rangle|e^{-i\alpha \chi t}\alpha\rangle+\Pi_1|\psi\rangle|e^{i\alpha \chi t}\alpha\rangle ~, 
\end{equation}
where $\Pi_i = \ketbra{i}{i}$ are the projectors onto the computational basis states. If the interaction time $t$ is sufficiently long and the amplitude $\alpha$ is large enough to satisfy $|\langle e^{i\alpha \chi t}\alpha|e^{-i\alpha \chi t}\alpha\rangle|^2=0$, the measurement is QND because the measurement operators are projectors that commute with the qubit Hamiltonian, since $H_\mathrm{disp} \propto \sigma_z$.

In practical measurements, however, the continuous readout of the meter state introduces uncertainties due to quantum and classical sources of noise. After a short interaction time, if $|\langle e^{i\alpha \chi t}\alpha|e^{-i\alpha \chi t}\alpha\rangle|^2\neq0$, the meter readout of $|\beta\rangle$ becomes uncertain, resulting in a non-projective measurement described by $\mathcal{M}(\rho)=K_\beta\rho K_\beta^{\dagger}$, where
\begin{equation}
    K_{\beta}=\langle \beta|e^{-i\alpha \chi t}\alpha\rangle\Pi_0+\langle \beta|e^{i\alpha \chi t}\alpha\rangle\Pi_1 ~.
\end{equation}
Such quantum measurements are commonly referred to as \emph{weak} measurements because, while providing some information about the system, they do not completely collapse the system to its eigenstates at once~\cite{clerk2010introduction}. 
Nevertheless, successive weak measurements consistently alter the system according to $...K_{\beta_2}K_{\beta_1}\rho K_{\beta_1}^{\dagger}K_{\beta_2}^{\dagger}...$ and guide the state toward a specific eigenstate. Intriguingly, the state trajectory can be deduced from the measurement outcomes $\{\beta_n\}$ through the quantum Bayesian approach \cite{korotkov2016quantum, murch2013observing}. From the perspective of characterizing quantum computers, continuous weak measurements can be employed to monitor system dynamics \cite{weber2014mapping, koolstra2022monitoring}, perform quantum process tomography \cite{kim2018direct}, and diagnose gate errors \cite{siva2023time}. The trade-off between information gain, state disturbance, and the reversibility of weak measurements has been extensively studied in Refs.~\cite{fuchs1996quantum, hong2022demonstration}.

%%%%%%%%%%%%%%%%%%%%%%% Gate Set Models of Quantum Computers %%%%%%%%%%%%%%%%%%%%%%% 
\subsection{Gate Set Models of Quantum Computers}\label{sec:gate_sets}

Gate-based quantum computers are devices that implement quantum circuits, which are sequences of instructions for applying logic operations to physical qubits. These instructions generally include a discrete set of quantum gates, as well as state preparation, and terminating (and possibly intermediate) measurements. In the preceding sections, we have described mathematical models of all of these operations. For many QCVV protocols, it is convenient to construct a single mathematical object called a \emph{gate set} that contains representations of all of the native instructions for a quantum device. 

Formally, a gate set is the union of three distinct sets. The first one lists the $N_\rho$ possible initial states that can be natively prepared, $\left\{ \rho^{(i)} \right\}_{i=1}^{N_\rho}$. Often, quantum computers only provide a single initialization state (e.g., $\ketbra{0}{0}$), in which case $N_{\rho}=1$. The second set is a list of the computer's $N_{\mathrm{G}}$ native operations or \textit{gates}, $\left\{ G_i \right\}_{i=1}^{N_{\mathrm{G}}}$. The third set lists the computer's $N_{\mathrm{M}}$ native measurement outcomes (POVM effects), $\left\{ E_i^{(m)} \right \}_{m=1, i=1}^{N_{\mathrm{M}}, N_{\mathrm{E}}^{(m)}}$, where $N_{\mathrm{E}}^{(m)}$ is the the number of possible outcomes of the $m$th measurement. Many quantum computers offer a single native measurement in the computational basis of $n$ qubits, in which case $N_{\mathrm{M}} = 1$ and $N_{\mathrm{E}} = 2^n$. The entire gate set is thus:
\begin{equation}\label{eq:gate_set_model}
    \mathcal{G} = \left\{
        \left\{ \rho^{(i)} \right\}_{i=1}^{N_\rho},\;
        \left\{ G_i \right\}_{i=1}^{N_{\mathrm{G}}},\;
        \left\{ E_i^{(m)} \right\}_{m=1,i=1}^{N_{\mathrm{M}},N_{\mathrm{E}}^{(m)}}
    \right\} ~.
\end{equation}
A gate set describes a \textit{specific, limited} set of operations. A quantum processor may be capable of implementing other operations that are not listed in a particular gate set. A gate set $\mathcal{G}$ can only be used to describe and predict circuits built from the operations in $\mathcal{G}$. 

In the context of gate sets, the word ``gate'' indicates an operation acting on the entire computer. The existing gate set formalism is not consistent with the alternative meaning of ``gate'' to denote an operation acting only on a subsystem (e.g., 1 or 2 qubits) of a quantum computer, which can be combined in parallel (by tensor product) with other gates on disjoint subsystems to produce a whole-computer operation called a circuit \emph{layer} or \emph{cycle}. In the gate set formalism, each layer (configuration of parallel gates) that can be performed constitutes a distinct ``gate.'' Gate sets do not generally assume any connection or correlation between the actions of, for example, an $X$ gate on qubit 1, an $X$ gate on qubit 2, and parallel $X$ gates on qubits 1 and 2. Treating each layer as an independent operation makes it possible --- by comparing and contrasting different parallel combinations of gates --- to study the effects of crosstalk on a device \cite{Nielsen_2021, PRXQuantum.2.040338}.

A gate set can be expressed in any representation that is convenient for the task at hand, but the most common convention is to use the transfer matrix representation and Hilbert-Schmidt space notation introduced in Sec.~\ref{sec:superop} to represent initialization operations as superkets $\sket{\rho}$, logic gates $G$ as transfer matrices, and POVM measurements as lists of effects $\{\sbra{E_i}\}$. 
Using these representations, a general gate set is written as 
\begin{equation}\label{eq:gate_set_model_vectorized}
    \mathcal{G} = \left\{
        \left\{ \sket{\rho^{(i)}} \right\}_{i=1}^{N_\rho},\;
        \left\{ G_i \right\}_{i=1}^{N_{\mathrm{G}}},\;
        \left\{ \sbra{E_i^{(m)}} \right\}_{m=1,i=1}^{N_{\mathrm{M}},N_{\mathrm{E}}^{(m)}}
    \right\} ~.
\end{equation}
In Sec.~\ref{sec:models:circuits}, we discuss how this representation can be conveniently used to predict quantum circuit outcomes. 

A \emph{gate set model} is a gate-set-valued function of some parameters --- i.e., a map from a list of parameters to gate sets. A \textit{fully-parameterized} gate set model assigns a free parameter to each matrix element of each operation in a gate set. For an $n$-qubit processor, a fully parameterized gate set model contains $4^n-1$ parameters per initial state, $16^n-4^n$ parameters per logic gate, and $8^n-4^n$ parameters per projective measurement. 

Reduced models can be constructed that use fewer parameters \cite{PRXQuantum.2.040338, hashim2023benchmarking, PRXQuantum.4.010325}, motivated by sparse matrix ansätze, structural properties of the processor's Hilbert space \cite{blume2022taxonomy}, or knowledge of its low-level physics. These models have been proposed as a way to overcome the exponential growth of parameters with system size, and their construction is an area of active research. They have fewer parameters, but generally rely on assumptions, such as limited or no crosstalk, symmetries, or \emph{ad hoc} ansätze like low-rank tensor networks. Physics-informed reduced models can have the additional advantage of more easily interpretable parameters, such as the intensity, frequency, or phase of a control field.

Finding the parameters of a gate set model (whether fully-parameterized or reduced) that fit and describe data from a particular device is the task of \emph{gate set tomography}, discussed in detail in Sec.~\ref{sec:gst}. 

%%%%%%%%%%%%%%%%%%%%%%% Circuits %%%%%%%%%%%%%%%%%%%%%%% 
\subsubsection{Circuits}\label{sec:models:circuits}

A gate set is a model of a quantum computer that can be used to predict the measurement outcome distribution for arbitrary quantum circuits composed of elements of the gate set. For a circuit that comprises (i) preparing native state $\rho$, (ii) applying the sequence of operations $C = (g_1, g_2, ..., g_L)$, and (iii) measuring the POVM $\{E_i\}$, the probability of measurement outcome $i$ is given by Born's rule as
\begin{equation}
    p(i|\rho,C) =  \sbra{E_i} G_{g_L} G_{g_{L-1}} ... G_{g_1} \sket{\rho} ~.
\end{equation}
This can be written in the more familiar form (see \eq\ref{eq:borns_rule_povm}),
\begin{equation}
   p(i|\rho,C) = \Tr[E_i \E_C(\rho)] ~,
\end{equation}
where $\E_C$ is the quantum operation defined by the sequence of gates $C$.

%%%%%%%%%%%%%%%%%%%%%%% Gauge Ambiguity %%%%%%%%%%%%%%%%%%%%%%% 
\subsubsection{Gauge Ambiguity}\label{sec:gauge}

A gate set model is a complete description of a Markovian quantum processor, but it is actually an \emph{over-complete} description. A gate set contains extra \emph{gauge} degrees of freedom that have no effect at all on \textit{any} observable probabilities, and therefore cannot be observed. No experiment can reveal information about a gauge parameter. A \textit{gauge transformation} on a gate set changes the gate set without changing any observable property.

A gauge transformation can be described by an arbitrary invertible $d^2 \times d^2$ matrix $M$, and transforms the gate set as follows:
\begin{align}
    \sbra{E_i^{(m)}} &\mapsto \sbra{E_i^{(m)}} M^{-1} ~, \\
    \sket{\rho} &\mapsto M \sket{\rho} ~,  \\
    G_i &\mapsto M G_i M^{-1} ~. 
\end{align}
This transformation maps the gate set $\mathcal{G}$ to a new gate set $\mathcal{G}'$ with a new set of parameters, but $\mathcal{G}$ and $\mathcal{G}'$ predict identical outcome probabilities for all possible circuits because 
\begin{multline}
    \sbra{E} M^{-1} M G_{g_L} M^{-1} ... M^{-1} M G_{g_1} M^{-1}  M \sket{\rho} = \\
    \sbra{E} G_{g_L} ... G_{g_1} \sket{\rho}
\end{multline}
for all pairs of state preparations and measurement outcomes $\{\rho, E\}$, and all sequences of gates $g_1, \ldots , g_L$. Gauge freedom implies the existence of equivalence classes of gate set models (a.k.a. \textit{gauge orbits}) that are physically indistinguishable. As an example, Appendix \ref{sec:pauli_gauge} provides a introduction to gauge ambiguities in Pauli noise learning (Sec.~\ref{sec:pnr}).

Gauge degrees of freedom within gate set models can significantly complicate comparisons between two models, because popular gate error metrics like diamond norm and fidelity are gauge-dependent (see Sec.~\ref{sec:overview}). One approach to mitigate these metrics' gauge-dependence is to employ gauge-fixing. This is most commonly done via ``gauge optimization,'' which varies over all possible gauge transformations to find a gauge that minimizes the deviation between a (noisy) gate set model and an ideal ``target'' model. The metric of deviation is somewhat arbitrary, but weighted Frobenius distance is commonly used for convenience. The need for gauge-fixing can be avoided by using strictly gauge-invariant metrics of error. Gauge transformations do not change the eigenvalues of a gate's transfer matrix, so any error metric that depends only on the transfer matrix’s spectrum is gauge-invariant.

Some work has explored alternative model constructions that circumvent the gauge problem. For example, \R\cite{DiMatteo2020operationalgauge} employs a representation of gate sets in terms of the probabilities of linear inversion gate set tomography (see Sec.~\ref{sec:gst}). This parameterization is overcomplete, and somewhat inconvenient, but completely avoids gauge freedom because every parameter in the representation is explicitly gauge-invariant. Other work \cite{KalmanGST,  Nielsen2022-jx} makes use of ``first-order gauge-invariant'' (FOGI) parameterizations that are invariant under small gauge transformations. This is an active area of research.

\section{Common Errors in Quantum Computers}\label{sec:errors}

Markovian errors in quantum computing can be broadly placed into two categories: \emph{coherent errors} and \emph{incoherent noise}. Coherent errors describe a reversible (purity-preserving) process in which an imperfect or unwanted unitary operator rotates the quantum register to the wrong state relative to the intended target state. Coherent errors can manifest from imperfections in gate calibrations, classical crosstalk signals that unintentionally drive a qubit, or unwanted coupling between qubits. Incoherent noise, on the other hand, describes irreversible processes, which are often referred to as \textit{decoherence}.

The design and analysis of quantum devices must account for various intrinsic noise sources that can lead to different types of errors within the systems~\cite{guillaud2019repetition, darmawan2021practical, nguyen2022blueprint}. In this Section, we introduce the following commonly encountered errors and noise, and illustrate their impact on single-qubit states using the Bloch sphere:
\begin{itemize}
    \item \textit{Coherent Errors} (Sec.~\ref{sec:coherent_errors}). When a unitary operation (including the idle) is implemented incorrectly but reversibly, the quantum register experiences a unitary (a.k.a.~\emph{coherent} or \emph{Hamiltonian}) error. In the case of a single qubit, the qubit's state will be rotated to an incorrect point on the Bloch sphere. Coherent errors preserve purity, and can be caused by control miscalibration or entangling Hamiltonians between neighboring qubits that produce crosstalk.

    \item \emph{Dephasing Noise} (Sec.~\ref{sec:dephasing}). Qubits in a superposition state can experience noise which leads to phase decoherence over time. For example, fluctuations in qubit frequency cause the qubit's Bloch vector to precess randomly with respect to the rotating frame, leading to random phase errors. This results in the \emph{dephasing} of superposition states, which is visualized as shrinking of the Bloch vector towards the polar axis of the Bloch sphere.
    
    \item \textit{Amplitude Damping} (Sec.~\ref{sec:spont_emis_amp_damp}). A qubit in an excited state will eventually thermalize to its ground state. This energy relaxation process is dictated by the underlying physics of the qubit --- i.e., whether it is an atom, superconducting qubit, spin qubit, etc. --- but is often termed ``spontaneous emission,'' as this is the physical pathway by which qubits thermalize for many systems. Therefore, the amplitude (or probability) of remaining in the excited state is \emph{damped} over time. Amplitude damping is an example of a \emph{non-unital} error, because it does not preserve the maximally mixed state.
    
    \item \textit{Depolarizing Noise} (Sec.~\ref{sec:dep_noise}). When incoherent noise acts isotropically about the Bloch sphere (i.e., all states have an equal probability of experiencing bit- and phase-flip errors), a qubit will eventually undergo decoherence, resulting in the complete loss of quantum information. This process is called \emph{depolarizing} noise, because it results in the depolarization of the Bloch vector toward the center of the Bloch sphere.
    
    \item \textit{Stochastic Pauli Noise} (Sec.~\ref{sec:stoch_pauli}). Noise in many systems is biased such that random bit- or phase-flips about different axes can occur with different rates. Such noise can be modeled by random (or stochastic) Pauli errors, whereby each type of Pauli error (e.g., $X$, $Y$, or $Z$) has a distinct probability of occurring. 
    
    \item \textit{Leakage} (Sec.~\ref{sec:leakage}). Qubits are defined by two computational basis states (see Sec.~\ref{sec:statevector_formalism}). \emph{Leakage} describes any process by which a qubit transitions out of the computational subspace, either via random noise or some coherent driving process. Leakage is often considered a \emph{non-Markovian} process in the context of qubit computations because it cannot be described by a CPTP map on the computational subspace, and it can produce temporal correlations across multiple gates or clock cycles.
    
    \item \textit{Non-Markovian and Unmodeled Errors} (Sec.~\ref{sec:nm_errors}). Sometimes an error in a quantum state, gate, or measurement cannot be captured by any \ac{CPTP} model. In such cases, these \emph{unmodeled errors} are typically ascribed to some \emph{non-Markovian} process in the system. Non-Markovian errors are not the focus of this tutorial, but understanding their impact on Markovian error models is important in \ac{QCVV}.
\end{itemize}

\begin{figure*}
    \centering
    \subfloat[Coherent errors]{
    \includegraphics[width=0.28\textwidth]{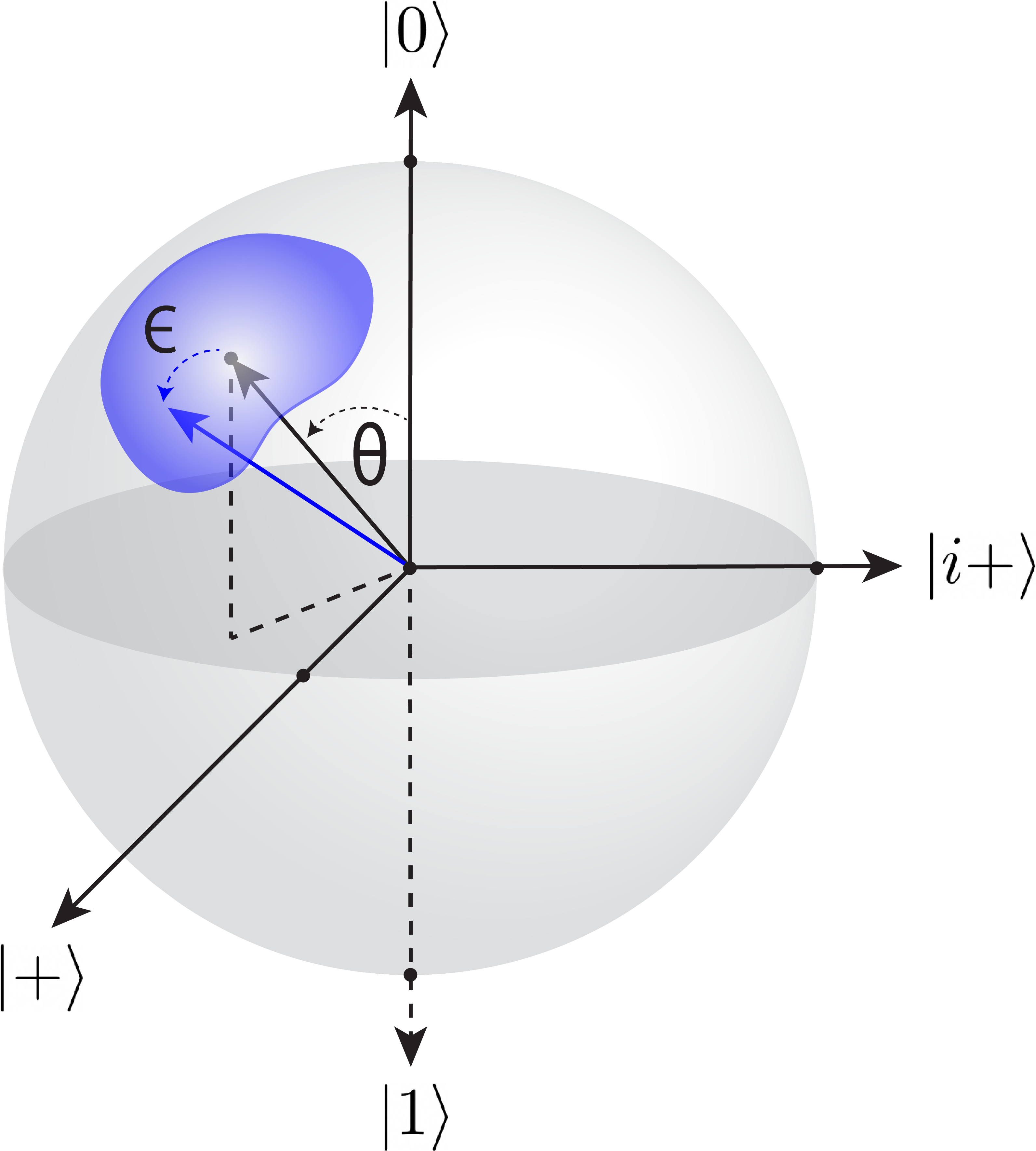}\label{fig:Coherent_errors}
        }
    \qquad
    \subfloat[Pure dephasing]{
    \includegraphics[width=0.28\textwidth]{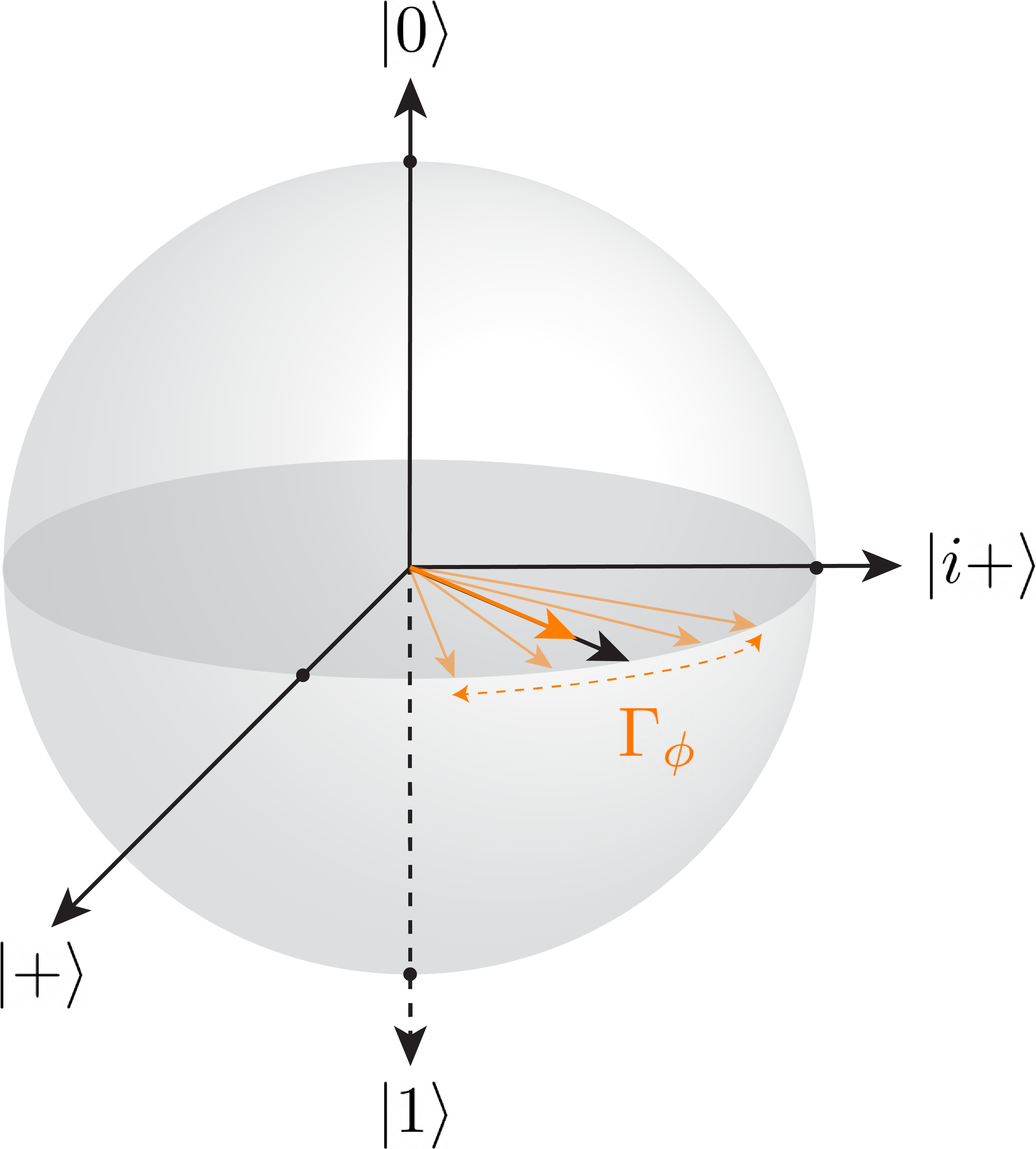}\label{fig:Dephasing_noise}
        }
    \qquad
    \subfloat[Longitudinal transitions]{
    \includegraphics[width=0.28\textwidth]{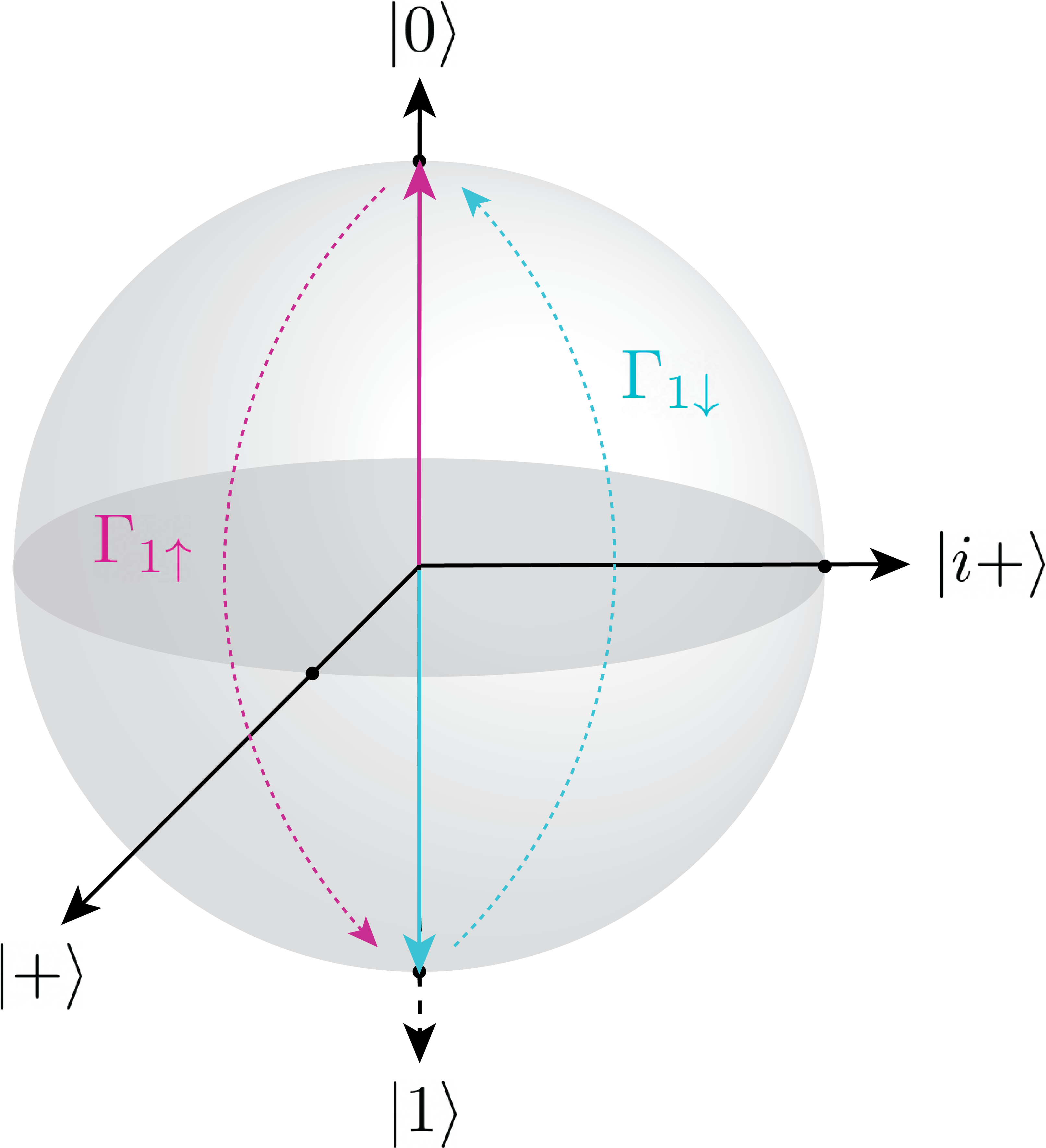}\label{fig:Relaxation_T1}
        }
    \\
    \subfloat[Depolarizing noise]{
    \includegraphics[width=0.28\textwidth]{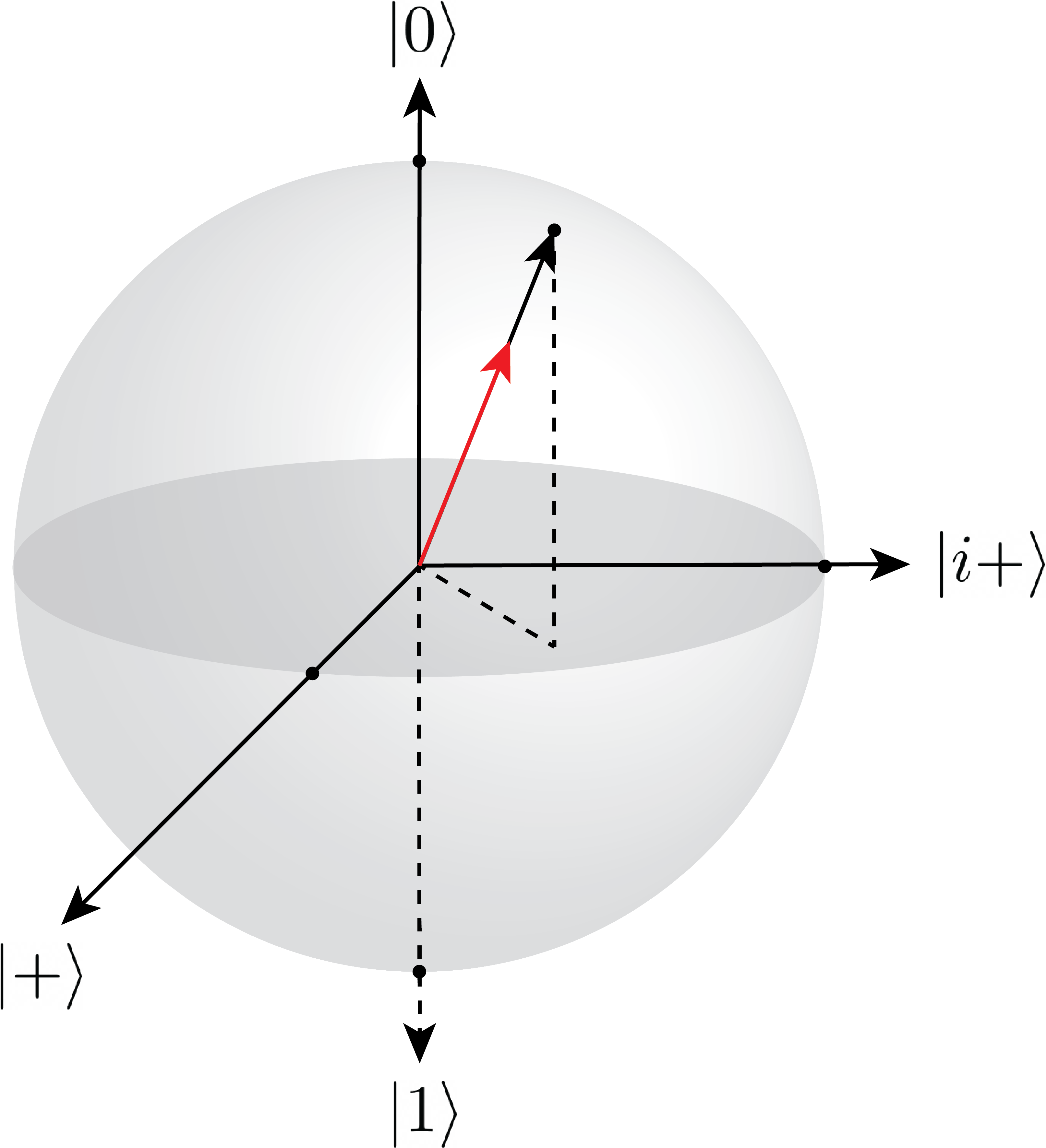}\label{fig:Depolarizing_noise}
        }
    \qquad
    \subfloat[Stochastic Pauli noise]{
    \includegraphics[width=0.28\textwidth]{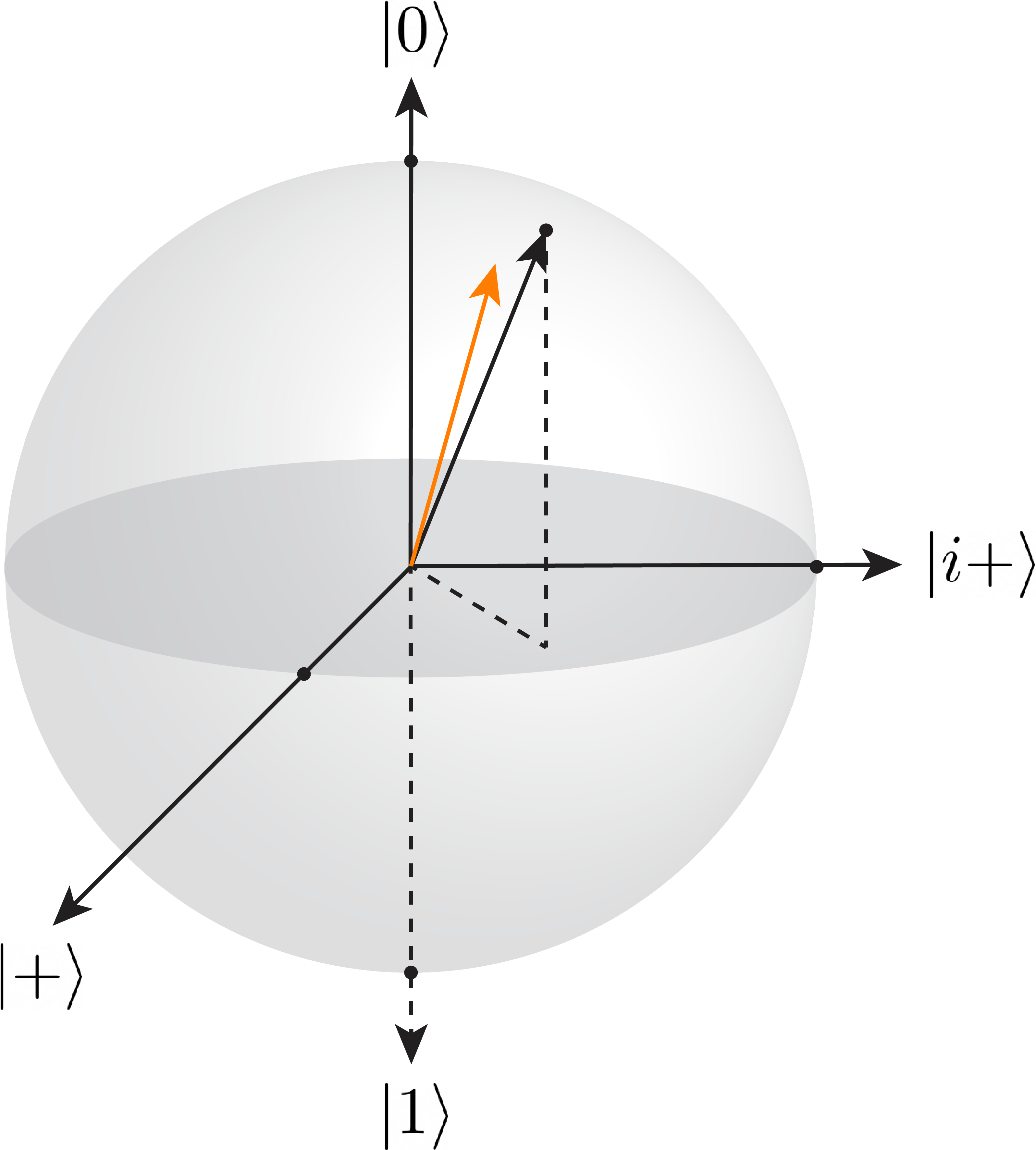}\label{fig:Stochastic_noise}
        }
    \qquad
    \subfloat[Leakage]{
    \includegraphics[width=0.28\textwidth]{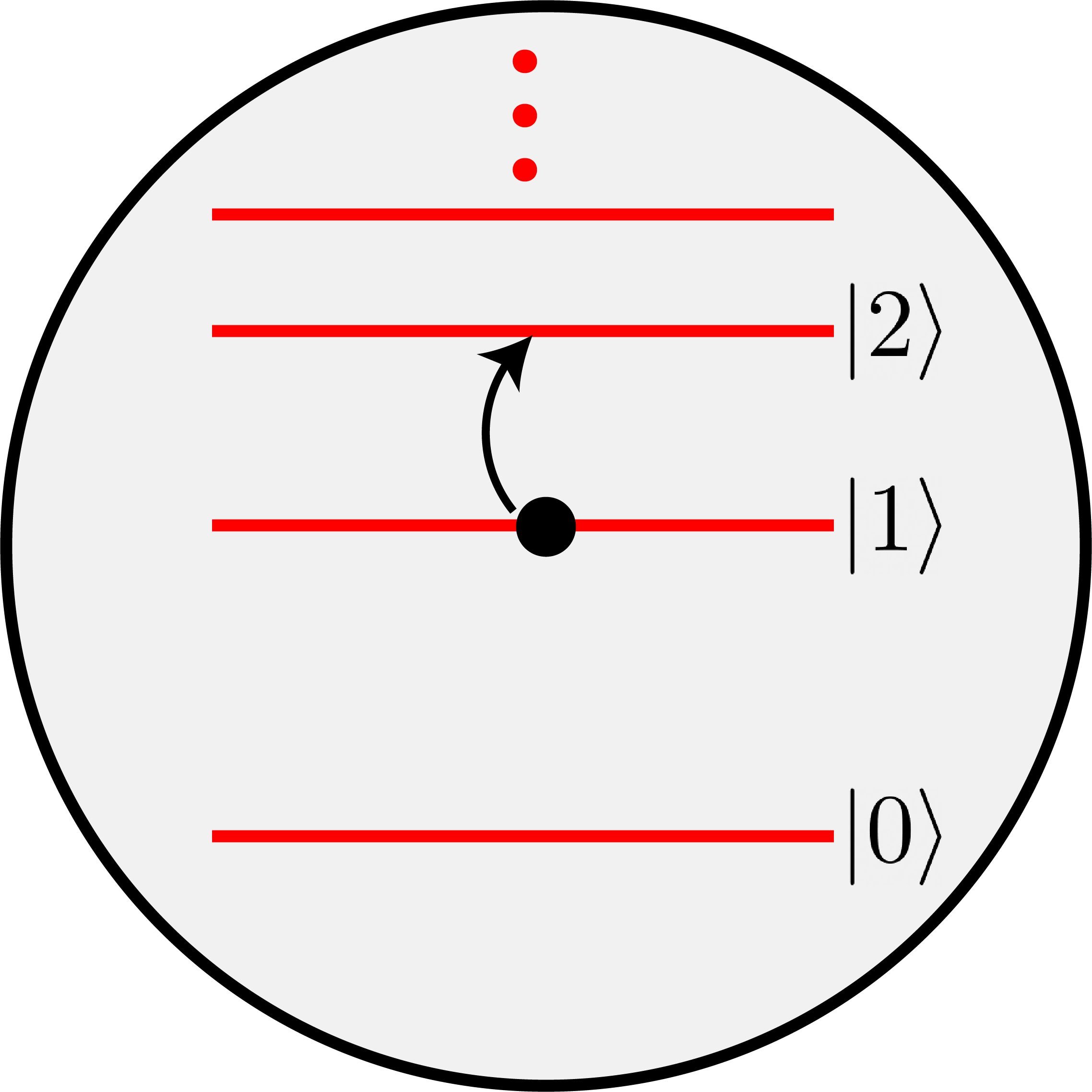}\label{fig:Leakage}
        }
    \caption{\textbf{Common Errors and Noise.}
    \textbf{(a)} Coherent errors result in an unintended rotation by an angle $\epsilon$ (blue arrow) relative to the intended target state (black arrow). The axis of rotation can act in any direction relative to the intended target state, depicted by the large blue region.
    \textbf{(b)} Dephasing noise can result from fluctuations in the qubit $\omega_{01}$ transition frequency, such that the qubit Bloch vector along the equator precesses randomly with respect to the rotating frame of the qubit (light orange arrows), shrinking the Bloch vector at a rate of $\Gamma_\phi$ (orange).
    \textbf{(c)} Longitudinal transitions result from spontaneous decay (cyan) at a rate $\Gamma_{1\downarrow}$ or spontaneous excitation (purple) at a rate $\Gamma_{1\uparrow}$.
    \textbf{(d)} Depolarizing noise acts isotropically around the Bloch sphere, shrinking the length of the Bloch vector (red) relative to a pure quantum state on the surface of the Bloch sphere (black).
    \textbf{(e)} Stochastic Pauli noise acts anisotropically around the Bloch sphere, shrinking the length of the Bloch vector and resulting in an offset (orange) relative to the intended quantum state (black).
    \textbf{(f)} Leakage describes the excitation of a qubit out of the computational basis $\{ \ket{0}, \ket{1}\}$ into higher energy levels.
    }
    \label{fig:Bloch_sphere_errors_noise}
\end{figure*}

%%%%%%%%%%%%%%%%%%%%%%% Coherent Errors %%%%%%%%%%%%%%%%%%%%%%% 
\subsection{Coherent Errors}\label{sec:coherent_errors}

Single-qubit unitary rotation operators $U_\mathbf{\hat{n}}(\theta)$ rotate a state vector $\ket{\psi}$ by an angle $\theta$ about an axis $\mathbf{\hat{n}}$. The resulting quantum state can be written as
\begin{equation}\label{eq:unitary_rotation}
    \ket{\psi'} = U_\mathbf{\hat{n}}(\theta) \ket{\psi} = e^{-i\frac{\theta}{2} \mathbf{\hat{n}} \cdot \boldsymbol{\sigma}} \ket{\psi} ~,
\end{equation}
where $\boldsymbol{\sigma}$ is the vector of Pauli operators. Rotations about the coordinate axes of the Bloch sphere are particularly common, and their representations as unitary operators are
\begin{align}
    U_\mathbf{\hat{x}}(\theta) &= R_x(\theta) \doteq \begin{pmatrix} \cos(\tfrac{\theta}{2}) & - i\sin(\tfrac{\theta}{2}) \\ - i\sin(\tfrac{\theta}{2}) & \cos(\tfrac{\theta}{2}) \end{pmatrix} ~, \\
    U_\mathbf{\hat{y}}(\theta) &= R_y(\theta) \doteq \begin{pmatrix} \cos(\tfrac{\theta}{2}) & - \sin(\tfrac{\theta}{2}) \\ \sin(\tfrac{\theta}{2}) & \cos(\tfrac{\theta}{2}) \end{pmatrix} ~, \\
    U_\mathbf{\hat{z}}(\theta) &= R_z(\theta) \doteq \begin{pmatrix} e^{-i\theta/2} & 0 \\ 0 & e^{i\theta/2} \end{pmatrix} ~,
    \label{eq:R_z_theta}
\end{align}
where we have used the fact that
\begin{equation}
    e^{-i\frac{\theta}{2} \mathbf{\hat{n}} \cdot \boldsymbol{\sigma}} = \cos(\frac{\theta}{2}) -i(\mathbf{\hat{n}} \cdot \boldsymbol{\sigma}) \sin(\frac{\theta}{2}) ~,
\end{equation}
and where the notation $R_n(\theta)$ is commonly used in place of the notation $U_\mathbf{\hat{n}}(\theta)$.

Unitary (or coherent) \emph{errors} manifest as unwanted or imperfect unitary rotations acting on qubits. This can be modeled as an ideal operator $U_\mathbf{\hat{n}}(\theta)$ followed by an erroneous operator $U_\mathbf{\hat{m}}(\epsilon)$, such that the actual final state is given by
\begin{align}\label{eq:coherent_error}
    \ket{\psi'} &= U_\mathbf{\hat{m}}(\epsilon) U_\mathbf{\hat{n}}(\theta) \ket{\psi} ~, \\
               &= e^{-i\frac{\epsilon}{2}\mathbf{\hat{m}} \cdot \boldsymbol{\sigma}} e^{-i\frac{\theta}{2}\mathbf{\hat{n}}\cdot\boldsymbol{\sigma}} \ket{\psi} ~,
\end{align}
where $\mathbf{\hat{m}}$ can be arbitrary relative to $\mathbf{\hat{n}}$. When $\mathbf{\hat{m}}=\mathbf{\hat{n}}$, as is common for certain calibration errors, the rotation axis is correct, but the rotation \textit{angle} is wrong; this is called an over- or under-rotation error. A coherent error changes where the state vector is located on the Bloch sphere relative to the intended target state, but has no effect on the length of the Bloch vector and therefore maintains the purity of the state; see \fig\ref{fig:Coherent_errors}.

In the Kraus representation, a coherent error by an angle $\theta$ is given by
\begin{equation}\label{eq:kraus_coh_err}
    \mathcal{E}(\rho) = U_\mathbf{\hat{m}}(\theta) \rho U_\mathbf{\hat{m}}(\theta)^\dagger = e^{-i\frac{\theta}{2} \mathbf{\hat{m}} \cdot \boldsymbol{\sigma}} \rho e^{i\frac{\theta}{2} \mathbf{\hat{m}} \cdot \boldsymbol{\sigma}} ~,
\end{equation}
where $K = U_\mathbf{\hat{m}}(\theta) = e^{-i\frac{\theta}{2} \mathbf{\hat{m}} \cdot \boldsymbol{\sigma}}$ is the Kraus operator.
Below, we list various superoperator representations for a coherent error about the $Z$ axis with Kraus operator $K = R_z(\theta)$ (see \eq\ref{eq:R_z_theta}):
\begin{itemize}
    \item Transfer matrix (basis of matrix units):
        \begin{align}\label{eq:supop_coh_err}
              \Lambda_c &= R_z(\theta)^* \otimes R_z(\theta) \nonumber \\
                          &= \begin{pmatrix}
                                1 & 0 & 0 & 0 \\
                                0 & e^{i\theta} & 0 & 0 \\
                                0 & 0  & e^{-i\theta} & 0 \\
                                0 & 0 & 0 & 1
                             \end{pmatrix} ~.
        \end{align}
    \item PTM:
        \begin{equation}\label{eq:ptm_coh_err}
            \Lambda = 
                \begin{pmatrix}
                    1 & 0 & 0 & 0 \\
                    0 & \cos(\theta) & -\sin(\theta) & 0 \\
                    0 & \sin(\theta) & \cos(\theta) & 0 \\
                    0 & 0 & 0 & 1
                \end{pmatrix} ~.
        \end{equation}
    \item $\chi$ matrix (Pauli basis):
        \begin{equation}\label{eq:chi_coh_err}
            \chi \doteq \frac{1}{2}
                \begin{pmatrix}
                    1 + \cos(\theta) & 0 & 0 & i\sin(\theta) \\
                    0 & 0 & 0 & 0 \\
                    0 & 0 & 0 & 0 \\
                    -i\sin(\theta) & 0 & 0 & 1 - \cos(\theta)
                \end{pmatrix} ~.
        \end{equation}
    \item Choi matrix (computational basis):
        \begin{align}\label{eq:choi_coh_err}
             \C &= \mathbf{vec}[R_z(\theta)]\mathbf{vec}[R_z(\theta)]^\dagger ~, \\
                &\doteq \begin{pmatrix}
                        1 & 0 & 0 & e^{-i\theta} \\
                        0 & 0 & 0 & 0 \\
                        0 & 0 & 0 & 0 \\
                        e^{i\theta} & 0 & 0 & 1
                    \end{pmatrix} ~.
        \end{align}
\end{itemize}

%%%%%%%%%%%%%%%%%%%%%%% Dephasing Noise %%%%%%%%%%%%%%%%%%%%%%% 
\subsection{Dephasing}\label{sec:dephasing}

\emph{Dephasing} is the loss of phase coherence in a quantum state. This manifests as the decay in the absolute value of the off-diagonal entries of the system's density matrix. In many physical qubit implementations, the $\ket{0}$ and $\ket{1}$ states are chosen to be energy eigenstates. The relative phase of these two states will evolve in time at a rate proportional to their energy difference. In order to perform coherent operations, external control fields must be resonant (or near resonant) with this transition, meaning that the oscillation frequency of the control fields should equal the phase evolution frequency of the qubit. When noise causes a complete loss of phase coherence between these two oscillators (the qubit and the control field), the qubit is said to have \textit{dephased}. This can happen if either (or both) of the oscillators suffer from fluctuations in their oscillation frequency, leading to uncertainty in their relative phase (see \fig\ref{fig:Dephasing_noise}). Qubits can experience frequency uncertainty due to changes in their energy splittings, such as magnetic field fluctuations, or coupling to other quantum systems, such as neighboring qubits, AC Stark shifts from control amplitude fluctuations, paramagnetic defects in semiconductors, or even the electromagnetic vacuum field. Control systems can similarly suffer a range of errors that lead to frequency instability, such as finite laser linewidths, acoustic noise in fiber optics, or clock jitter in arbitrary waveform generators. 

Irreversible dephasing can arise when the qubit/clock relative frequency is changing quickly compared to the characteristic control timescale. In this case, off-diagonal entries of the density matrix are seen to decay exponentially with a characteristic timescale $T_2$ (``T two''; see Sec.~\ref{sec:t2}).
$T_2$ is sometimes called the \textit{transverse relaxation time} or the \textit{intrinsic dephasing time}. In the absence of spontaneous emission effects (discussed in Sec.~\ref{sec:spont_emis_amp_damp}), the $T_2$ time is the inverse of the \textit{pure dephasing rate}, $\Gamma_\phi=1/T_2$. 

If the relative frequency is changing slowly compared to the control timescale, then frequency errors can build up coherently for some time, and the decay of quantum coherence is non-exponential (e.g., Gaussian) rather than exponential. The characteristic timescale $T_2^*$ (``T two star'') is known as the \textit{effective transverse relaxation time} or \textit{inhomogeneous dephasing time}. Because the errors are correlated in time, dynamical decoupling/refocusing tools, such as the Hahn echo~\cite{hahn1950spin}, can be used to extend the phase coherence time. The coherence decay timescale after refocusing is typically used as an estimate of the intrinsic dephasing time, and is denoted $T_{2E}$ (``T two echo''). Protocols for characterizing the $T_2^*$ and $T_{2E}$ times are introduced in Sec.~\ref{sec:t2}. See~\cite{carr1954effects, meiboom1958modified, maudsley1986modified, ahmed2013robustness} for background on more advanced dynamical decoupling schemes.

The Kraus representation of dephasing noise is
\begin{equation}\label{eq:kraus_dephasing}
  \mathcal{E}(\rho) = \left( 1 - \frac{p}{2} \right)\rho + \frac{p}{2} Z \rho Z ~,
\end{equation}
with Kraus operators $K_I = \sqrt{1 - p/2}I$ and $K_Z = \sqrt{p/2}Z$. Here, a quantum state $\rho$ under goes a phase-flip with probability $p/2$, and is unchanged with probability $1 - p/2$. Below, we list various superoperator representations for dephasing noise occurring with probability $p/2$:
\begin{itemize}
    \item Transfer matrix (basis of matrix units):
        \begin{align}\label{eq:supop_dephasing}
                \Lambda_c &= K_I^* \otimes K_I + K_Z^* \otimes K_Z \nonumber \\
                            &= \begin{pmatrix}
                                1 & 0 & 0 & 0 \\
                                0 & 1 - p & 0 & 0 \\
                                0 & 0  & 1 - p & 0 \\
                                0 & 0 & 0 & 1
                               \end{pmatrix} ~.
            \end{align}
    \item PTM:
        \begin{equation}\label{eq:ptm_dephasing}
          \Lambda = 
            \begin{pmatrix}
                1 & 0 & 0 & 0 \\
                0 & 1 - p & 0 & 0 \\
                0 & 0 & 1 - p & 0 \\
                0 & 0 & 0 & 1
          \end{pmatrix} ~.
        \end{equation}
    \item $\chi$ matrix (Pauli basis):
        \begin{equation}\label{eq:chi_dephasing}
          \chi \doteq 
            \begin{pmatrix}
                1 - p/2 & 0 & 0 & 0 \\
                0 & 0 & 0 & 0 \\
                0 & 0 & 0 & 0 \\
                0 & 0 & 0 & p/2
          \end{pmatrix} ~.
        \end{equation}
    \item Choi matrix (computational basis):
        \begin{align}\label{eq:choi_dephasing}
           \C &= \mathbf{vec}(K_I) \mathbf{vec}(K_I)^\dagger + \mathbf{vec}(K_Z) \mathbf{vec}(K_Z)^\dagger ~, \\
              &\doteq \begin{pmatrix}
                        1 & 0 & 0 & 1 - p \\
                        0 & 0 & 0 & 0 \\
                        0 & 0 & 0 & 0 \\
                        1 - p & 0 & 0 & 1
                 \end{pmatrix} ~.
        \end{align}
\end{itemize}

%%%%%%%%%%%%%%%%%%%%%%% Spontaneous Emission and Amplitude Damping %%%%%%%%%%%%%%%%%%%%%%% 
\subsection{Amplitude Damping (Spontaneous Emission)}\label{sec:spont_emis_amp_damp}

Many physical qubit species use two non-degenerate energy eigenstates to store quantum information, with the  $\ket{1}$ state often higher in energy than the $\ket{0}$ state. Because of this energy gap, the qubit can experience \emph{spontaneous emission}, a process in which an excited system decays to a lower energy state by the emission of a photon, or similar (but non-radiative) energy-loss processes. These effects lead to a loss of quantum information, and are typically modeled as an \emph{amplitude damping} error. Amplitude damping can also model the reverse process, where a qubit absorbs energy from the environment. Together, the combination of the emission and decay processes describe \textit{thermalization}. 

Strong amplitude damping on a qubit maps all points on (and within) the Bloch sphere to a single pure state, making amplitude damping the paradigmatic example of a \textit{non-unital} process --- it does not preserve the maximally mixed state. Weaker (partial) amplitude damping errors are characterized by their decay rate. If amplitude damping describes a decay from $\ket{1}$ to $\ket{0}$, we denote the decay rate $\Gamma_{1\downarrow}$; the rate of the reverse process is denoted $\Gamma_{1\uparrow}$ (see \fig\ref{fig:Relaxation_T1}). The characteristic thermalization rate, also frequently called the \textit{longitudinal} relaxation rate, is their sum: 
 \begin{equation}\label{eq:long_relaxation_rate}
     \Gamma_1 = \frac{1}{T_1} = \Gamma_{1\uparrow} + \Gamma_{1\downarrow} ~.
 \end{equation}
Here $T_1$, the ``T-one time,'' is the characteristic thermalization timescale. If the temperature of the environment is small relative to the qubit energy splitting --- i.e., $k_B T \ll \hbar\omega_{01}$ --- then the decay rate will be significantly larger than the absorption rate, $\Gamma_{1\downarrow} \gg \Gamma_{1\uparrow}$, and $T_1 \simeq 1/\Gamma_{1\downarrow}$. Energy decay will also impact the phase coherence of the qubit, since a qubit that decays to the ground state loses all information about its prior phase. A well-known bound~\cite{Slichter2010-te} on the dephasing timescale is:
\begin{equation}
    T_2 \le 2 T_1 ~.
\end{equation}
In Sec.~\ref{sec:coherence}, we introduce protocols for characterizing both timescales.

The energy decay rate can be derived from low-level physics models through the use of Fermi's golden rule:
\begin{equation}\label{eq:fermigolden}
    \Gamma_{1\downarrow} = \frac{1}{\hbar^2} |\braket{0|\hat{\mathcal{C}}|1}|^2 \mathcal{S}_\alpha(\omega_{01}) ~,
\end{equation}
where $\hat{\mathcal{C}}$ is the coupling operator to an environmental bath $\alpha$ at the qubit frequency $\omega_{01}$, which is described by the noise spectral density $\mathcal{S}_\alpha$. Careful engineering of the noise environment --- e.g., by placing qubits in cavities --- has been shown to significantly extend $T_1$ times in superconducting qubits \cite{siddiqi2021engineering, kerman2010metastable, lin2018demonstration, earnest2018realization, nguyen2019high}. Optical-frequency qubits in atomic systems have $T_1$ times typically on the order of seconds, while hyperfine atomic qubits can have radiative $T_1$ times approaching the age of the universe. This long lifetime is due to a combination of weak magnetic dipole coupling to the electromagnetic field, and the relatively small density of states available to photons at low (microwave) splittings. 

The Kraus representation of spontaneous emission (amplitude damping) is
\begin{equation}\label{eq:kraus_spont_amp}
            \mathcal{E}(\rho) = K_0 \rho K_0^\dagger + K_1 \rho K_1^\dagger ~,
\end{equation}
with Kraus operators
$$K_0 = \sqrt{I - p\sigma_+\sigma_-} = \begin{pmatrix} 1 & 0 \\ 0 & \sqrt{1 - p} \end{pmatrix}$$
and
$$K_1 = \sqrt{p} \sigma_- =  \begin{pmatrix} 0 & \sqrt{p} \\ 0 & 0 \end{pmatrix} ~,$$
where $\sigma_+ = \ketbra{1}{0}$ and $\sigma_- = \ketbra{0}{1}$. Each Kraus operator represents an event that can occur, transforming the state. $K_1$ can only occur if the system is initially in $\ket{1}$, and if it occurs then a quantum of energy is lost to the environment, leaving the system in $\ket{0}$. Otherwise $K_0$ occurs, and the amplitude for the system to be in $\ket{1}$ gets smaller.
% Spontaneous emission and amplitude damping is an example of a \emph{non-unital} process, which does not map the identity $\mathbb{I}$ back to itself: $\E(\mathbb{I}) \ne \mathbb{I}$. 
Below, we list various superoperator representations for an amplitude damping process with decay probability $p$:

\begin{itemize}
    \item Transfer matrix (basis of matrix units):
        \begin{align}\label{eq:supop_spont_amp}
            \Lambda_c &= K_0^* \otimes K_0 + K_1^* \otimes K_1 ~, \nonumber \\
                        &= \begin{pmatrix}
                            1 & 0 & 0 & p \\
                            0 & \sqrt{1 - p} & 0 & 0 \\
                            0 & 0  & \sqrt{1 - p} & 0 \\
                            0 & 0 & 0 & 1 - p
                           \end{pmatrix} ~.
        \end{align}
    \item PTM:
        \begin{equation}\label{eq:ptm_spont_amp}
          \Lambda = 
            \begin{pmatrix}
                1 & 0 & 0 & 0 \\
                0 & \sqrt{1 - p} & 0 & 0 \\
                0 & 0 & \sqrt{1 - p} & 0 \\
                p & 0 & 0 & 1 - p
            \end{pmatrix} ~.
        \end{equation}
    \item $\chi$ matrix (Pauli basis):
        \begin{equation}\label{eq:chi_spont_amp}
          \chi \doteq \frac{1}{4}
            \begin{pmatrix}
                \left(1 + \sqrt{1 - p}\right)^2 & 0 & 0 & p \\
                0 & p & -ip & 0 \\
                0 & ip & p & 0 \\
                p & 0 & 0 & \left(1 - \sqrt{1 - p}\right)^2
           \end{pmatrix} ~.
        \end{equation}
    \item Choi matrix (computational basis):
        \begin{align}\label{eq:choi_spont_amp}
          % \C &= \frac{1}{2} \bigg( \ketbra{K_0 \rangle}{\langle K_0} + \ketbra{K_1 \rangle}{\langle K_1} \bigg) \\
            \C &= \mathbf{vec}(K_0)\mathbf{vec}(K_0)^\dagger + \mathbf{vec}(K_1)\mathbf{vec}(K_1)^\dagger ~, \\
               &\doteq % \frac{1}{2}
                    \begin{pmatrix}
                        1 & 0 & 0 & \sqrt{1 - p} \\
                        0 & 0 & 0 & 0 \\
                        0 & 0 & p & 0 \\
                        \sqrt{1 - p} & 0 & 0 & 1 - p
                    \end{pmatrix} ~.
        \end{align}
\end{itemize}
In the \ac{PTM} representation, we can directly observe that amplitude damping is a non-unital process, because the first column is not $[1, 0, 0, 0]^\trans$.

%%%%%%%%%%%%%%%%%%%%%%% Depolarizing Noise %%%%%%%%%%%%%%%%%%%%%%% 
\subsection{Depolarizing Noise}\label{sec:dep_noise}

\emph{Depolarizing noise} describes the process in which a quantum state $\rho$ is replaced by a completely mixed state with some probability $p$,
\begin{align}\label{eq:kraus_depolarizing}
    \mathcal{E}(\rho) &= p \mathbb{I}/d + (1 - p)\rho ~, \\
                      &= \left( 1 - \frac{3p}{4} \right)\rho + \frac{p}{4}(X \rho X + Y \rho Y + Z \rho Z) ~.
\end{align}
Its Kraus operators are $K_I = \sqrt{1 - 3p/4}I$, $K_X = \sqrt{p/4}X$, $K_Y = \sqrt{p/4}Y$, and $K_Z = \sqrt{p/4}Z$. Depolarizing noise acts isotropically on the Bloch sphere, and Pauli X, Y, and Z errors have an equal probability of occurring for all states. Therefore, depolarizing noise scales the length of the Bloch vector by the depolarizing probability $p$; see \fig\ref{fig:Depolarizing_noise}. 
Note that depolarizing noise is often written in a more intuitive way,
\begin{equation}\label{eq:kraus_depolarizing_alt}
     \mathcal{E}(\rho) = \left( 1 - p' \right)\rho + \frac{p'}{3}(X \rho X + Y \rho Y + Z \rho Z) ~,
\end{equation}
where we take $p' = 3p/4$. In this form, we may interpret a depolarizing noise channel as one in which the qubit experiences an $X$, $Y$, and $Z$ error, each with probability $p'/3$, but remains unchanged with probability $1 - p'$. Below, we list various superoperator representations for depolarizing noise:
\begin{itemize}
    \item Transfer matrix (basis of matrix units):
        \begin{align}\label{eq:supop_depolarizing}
          \Lambda_c &= \left( 1 - \frac{3p}{4} \right) I \otimes I + \frac{p}{4}\left( X \otimes X + Y^* \otimes Y + Z \otimes Z \right) ~, \nonumber \\
                      &= \begin{pmatrix}
                            1 - p/2 & 0 & 0 & p/2 \\
                            0 & 1 - p & 0 & 0 \\
                            0 & 0  & 1 - p & 0 \\
                            p/2 & 0 & 0 &  1 - p/2
                         \end{pmatrix} ~.
        \end{align}
    \item PTM:
        \begin{equation}\label{eq:ptm_depolarizing}
          \Lambda = 
            \begin{pmatrix}
                1 & 0 & 0 & 0 \\
                0 & 1 - p & 0 & 0 \\
                0 & 0 & 1 - p & 0 \\
                0 & 0 & 0 & 1 - p
          \end{pmatrix} ~.
        \end{equation}
    \item $\chi$ matrix (Pauli basis):
       \begin{equation}\label{eq:chi_depolarizing}
          \chi \doteq \begin{pmatrix}
                1 - 3p/4 & 0 & 0 & 0 \\
                0 & p/4 & 0 & 0 \\
                0 & 0 & p/4 & 0 \\
                0 & 0 & 0 & p/4
          \end{pmatrix} ~.
        \end{equation}
    \item Choi matrix (computational basis):
        \begin{align}\label{eq:choi_depolarizing}
            \C &= \sum_{P \in \{I,X,Y,Z\}} \mathbf{vec}(K_P) \mathbf{vec}(K_P)^\dagger ~, \\
               &\doteq \begin{pmatrix}
                        1 - p/2 & 0 & 0 & 1 - p \\
                        0 & p/2 & 0 & 0 \\
                        0 & 0 & p/2 & 0 \\
                        1 - p & 0 & 0 & 1 - p/2
                   \end{pmatrix} ~.
        \end{align}
\end{itemize}

%%%%%%%%%%%%%%%%%%%%%%% Stochastic Pauli Noise %%%%%%%%%%%%%%%%%%%%%%% 
\subsection{Stochastic Pauli Noise}\label{sec:stoch_pauli}

\emph{Stochastic Pauli noise} generalizes both depolarizing and dephasing noise, allowing all three of the Pauli $X$, $Y$, and $Z$ errors to have distinct probabilities $p_X$, $p_Y$, and $p_Z$, respectively. The Kraus representation of stochastic Pauli noise is 
\begin{align}\label{eq:kraus_stoch_pauli}
    \mathcal{E}(\rho) = \; &(1- p_X - p_Y - p_Z) \rho \nonumber \\
                           &+ p_X X\rho X + p_Y Y \rho Y + p_Z Z \rho Z ~,
\end{align}
with Kraus operators $K_I = \sqrt{1 - p_X - p_Y - p_Z}I$, $K_X = \sqrt{p_X}X$, $K_Y = \sqrt{p_Y}Y$, and $K_Z = \sqrt{p_Z}Z$, subject to $p_X, p_Y, p_Z \ge 0$ and $p_X + p_Y + p_Z \le 1$. Dephasing noise is the special case of Pauli noise where $p_X = p_Y = 0$, and depolarizing noise is the special case where $p_X = p_Y = p_Z$.

Stochastic Pauli noise is unital [i.e., $\E(\Id) = \Id$], and for a single qubit it shrinks the Bloch vector anisotropically. This reduces the Bloch vector's length, but because the shrinking is anisotropic it can also change the Bloch vector's \textit{direction} relative to the intended target state in a way that depends on the relative probabilities of the Pauli errors and the location of the vector on the Bloch sphere (see \fig\ref{fig:Stochastic_noise}). The various representations of stochastic Pauli noise with probabilities $p_X$, $p_Y$, and $p_Z$ are given by the following, where $p' = p_X + p_Y + p_Z$ and $p_I = 1 - p'$:
\begin{itemize}
    \item Transfer matrix (basis of matrix units):
        \begin{align}\label{eq:supop_stoch_pauli}
          \Lambda_c &= p_I I \otimes I + p_X X \otimes X + p_Y Y^* \otimes Y + p_Z Z \otimes Z \nonumber \\ 
                    &= \begin{pmatrix}
                            p_I+ p_Z & 0 & 0 & p_X + p_Y \\
                            0 & p_I - p_Z & p_X - p_Y & 0 \\
                            0 & p_X - p_Y  & p_I - p_Z & 0 \\
                            p_X + p_Y & 0 & 0 & p_I + p_Z
                        \end{pmatrix} ~.
        \end{align}
    \item PTM:
       \begin{equation}\label{eq:ptm_stoch_pauli}
        %   \Lambda = 
            \begin{pmatrix}
                1 & 0 & 0 & 0 \\
                0 & 1 - 2(p_Y + p_Z) & 0 & 0 \\
                0 & 0 & 1 - 2(p_X + p_Z) & 0 \\
                0 & 0 & 0 & 1 - 2(p_X + p_Y)
          \end{pmatrix} ~.
        \end{equation}
    \item $\chi$ matrix (Pauli basis):
       \begin{equation}\label{eq:chi_stoch_pauli}
          \chi \doteq 
            \begin{pmatrix}
                1 - p' & 0 & 0 & 0 \\
                0 & p_X & 0 & 0 \\
                0 & 0 & p_Y & 0 \\
                0 & 0 & 0 & p_Z
          \end{pmatrix} ~.
        \end{equation}
    \item Choi matrix (computational basis):
        \begin{align}\label{eq:choi_stoch_pauli}
             \C &= \sum_{P \in \{I,X,Y,Z\}} \mathbf{vec}(K_P) \mathbf{vec}(K_P)^\dagger ~, \\
                &\doteq \begin{pmatrix}
                        p_I + p_Z & 0 & 0 & p_I - p_Z \\
                        0 & p_X + p_Y & p_X - p_Y & 0 \\
                        0 & p_X - p_Y & p_X + p_Y & 0 \\
                        p_I - p_Z & 0 & 0 & p_I + p_Z
                    \end{pmatrix} ~.
        \end{align}
\end{itemize}
The $\chi$ matrix of any Pauli stochastic noise process is diagonal in the Pauli basis, and its diagonal elements can be determined directly from the probability coefficients in the Kraus representation. A Pauli channel's PTM is \textit{also} diagonal in the Pauli basis, and so its eigenoperators are the Pauli operators (i.e., $\E(P) = \lambda_P P$ for each Pauli $P$). The diagonal elements of the PTM for a stochastic Pauli noise process are thus called \textit{Pauli eigenvalues}.

Equation \ref{eq:ptm_stoch_pauli} makes apparent an important connection between the Kraus and PTM representations for stochastic Pauli noise. Namely, a Pauli error $Q$ in the Kraus representation that occurs with probability $p_Q$ will attenuate the Pauli eigenvalue corresponding to any Pauli operator $P$ that anticommutes with $Q$ (i.e., $\{P, Q\} = 0$). This can be generalized in the following way: for a set of Pauli-Kraus operators $\{ Q \}$ with $\sum_Q p_Q = 1$, the resulting Pauli eigenvalues $\lambda_P = \Lambda_{P P}$ in the PTM representation are given as
\begin{equation}\label{eq:ptm_kraus_rel}
    \Lambda_{P P} = 1 - 2\sum_{Q\ s.t.\ \{P, Q\} = 0} p_Q ~.
\end{equation}
The connection between Pauli errors in the Kraus representation and Pauli eigenvalues in the PTM representation will be important when we discuss QCVV protocols for Pauli noise learning (Sec.~\ref{sec:pnr}).

%%%%%%%%%%%%%%%%%%%%%%% Leakage %%%%%%%%%%%%%%%%%%%%%%% 
\subsection{Leakage}\label{sec:leakage}

Leakage refers to any process in which a qubit or quantum register is excited out of its computational basis states (e.g., $\{ \ket{0}, \ket{1} \}$) to an orthogonal state.  When a qubit is encoded into the lowest energy levels of a system, the \textit{leakage states} are higher energy levels (e.g., $\ket{2}$); see \fig\ref{fig:Leakage}. Leakage can be coherent (preserving phases between the computational basis and the leakage state[s]), if a qubit is unintentionally driven at its $\ket{1} \rightarrow \ket{2}$ resonant frequency, or incoherent (no phase coherence is preserved between the computational space and the leakage state[s]), if the excitation is due to thermal noise. 

Some authors describe leakage as a \emph{non-trace-preserving} (\emph{non-\ac{TP}}) process, i.e., as a qubit process that maps $\rho \mapsto \rho'$ where $\Tr(\rho') < \Tr(\rho)$ for a qubit state $\rho$. However, the probability of observing \emph{some} outcome will always be 1, which makes the representation of leakage as non-TP problematic. In a system that can detect leakage events (e.g., many superconducting systems; see \fig\ref{fig:lrb}) every experiment will either return $\ket{0}$, $\ket{1}$, or $\ket{2}$. Conversely, in systems unable or not designed to detect leakage (e.g., atomic qubits measured via resonance fluorescence, where $\ket{0}$ is measured by a dark state and $\ket{1}$ is measured by a bright state), leakage events will be \emph{misclassified} as either $\ket{0}$ or $\ket{1}$. However, in both cases, every measurement yields \textit{some} outcome. An example of a processes which is truly non-TP is \emph{postselection}, in which some outcomes are \emph{discarded} after measurement.

There is no TP Kraus representation of leakage within the qubit subspace. But leakage can be modeled rigorously by including additional states, promoting a qubit to a ``qudit'' with a $d>2$-dimensional Hilbert space. Leakage can then be modeled by CPTP maps acting on $d \times d$ density matrices, by introducing Kraus operators $K = \sqrt{p_{ij}} \ketbra{j}{i}$ that map computational states (e.g., $\{\ket{0},\ket{1}\}$) to leakage states (e.g.~$\ket{2}$). Leakage events are often subject to selection rules between the $i$th and $j$th energy level; for example, while a transition from $\ket{1} \rightarrow \ket{2}$ is responsible for leakage out of the computational basis states, a direct transition from $\ket{0} \rightarrow \ket{2}$ might be quantum mechanically forbidden, depending on the underlying physics of the system. Transitions from the leakage states back into the computational basis states are called \textit{seepage} \cite{wood2018quantification}, and can also be modeled using qudit Kraus operators.

%%%%%%%%%%%%%%%%%%%%%%% Non-Markovian Errors %%%%%%%%%%%%%%%%%%%%%%% 
\subsection{Non-Markovian and Unmodeled Errors}\label{sec:nm_errors}

\begin{figure}%[ht]
    \centering
    \includegraphics{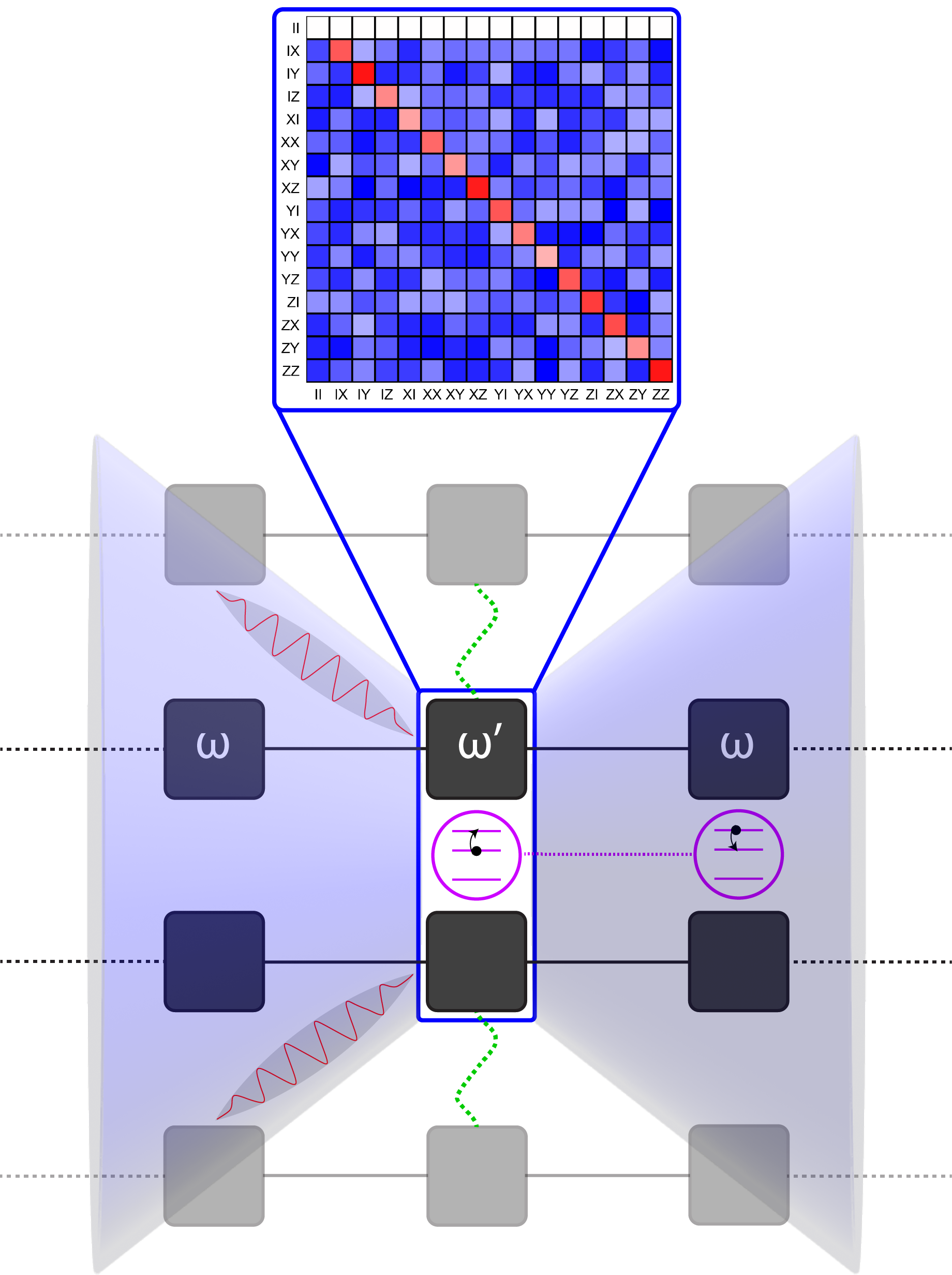}
    \caption{\textbf{Non-Markovian Errors in Gate-based Quantum Computing.}
    For a system of two active qubits (black), all Markovian errors that occur within the timescale of a given cycle of gates (blue rectangle) can be modeled by a two-qubit transfer matrix (PTM at top of the figure; in this schematic, colored cells indicate errors in the target operation). Examples of non-Markovian errors that cannot be modeled by a two-qubit transfer matrix include (but are not limited to) drift in qubit properties over the timescale of multiple layers (e.g., fluctuations in the qubit frequency $\omega$, with $\omega' = \omega + \delta\omega$), leakage to higher energy levels with a memory longer than the duration of a gate (purple), unwanted entanglement (green) with outside qubits (grey; e.g., due to static $ZZ$ coupling), and classical EM crosstalk signals (red) originating from other qubits outside of the defined system that arrive within the light cone (light blue) of the system qubits. (Figure adapted with permission from \R\cite{hashim2023benchmarking}.)}
    \label{fig:nm_errors}
\end{figure}

So far, we have only considered \textit{Markovian} errors. In the context of gate-based quantum computing, in particular for QCVV, an error is Markovian if it can be modeled by a CPTP map (i.e., a transfer matrix or process matrix). If an operation $g$ on $n$ qubits is modeled by an $n$-qubit CPTP map $G$, then $G$ can describe any Markovian errors in the implementation of $g$. But there are commonly-encountered phenomena that $G$ cannot model. For example, if the $n$ ``active’’ qubits are accompanied by one or more ``spectator’’ qubits that $G$ does not describe, then $G$ cannot account for coupling between active and spectator qubits. If applying $g$ causes a persistent effect that induces errors during subsequent operations, that also cannot be captured by a CPTP map $G$. Errors that violate the assumption of temporal locality --- including coupling to spectator degrees of freedom --- are by definition \emph{non}-Markovian (see Sec.~\ref{sec:model_violation}). Therefore, for $n$ active qubits, non-Markovian errors are deviations from ideal behavior that cannot be modeled by a context-independent $n$-qubit transfer or process matrix.

Common types of non-Markovian errors in the NISQ era include fluctuation or drift of qubit parameters (e.g., qubit transition frequency) \cite{proctor2020detecting} and leakage outside of the computational basis states \cite{ghosh2013understanding, wallman2016robust, chen2016measuring, wood2018quantification, hayes2020eliminating, babu2021state} (see \fig\ref{fig:Leakage}) with a memory longer than the timescale of the gate, correlated errors \cite{li2024direct, harrington2024synchronous} or unwanted entanglement with qubits outside of the defined $n$-qubit system (e.g., static $ZZ$ coupling in superconducting qubits \cite{mundada2019suppression, zhao2020high, ni2021scalable}), coupling to other external fluctuators (e.g., nonequilibrium quasiparticles) \cite{serniak2018hot, de2020two, berlin2021changes}, qubit heating \cite{webb2018resilient}, and $1/f$ noise \cite{burkard2009non, groszkowski2022simple}; see \fig\ref{fig:nm_errors}. However, if we were to instead enlarge our Hilbert space to include higher energy levels and more (perhaps non-local) qubits, then processes like leakage and unwanted entanglement are no longer non-Markovian. Therefore, in general, quantum non-Markovianity is highly dependent upon the definition of one's system and the timescales under consideration; i.e., whether observed dynamics are Markovian or not depends on how big a model the observer uses to describe them.

The study of quantum non-Markovianity is an active area of research \cite{diosi1998non, wolf2008assessing, piilo2008non, breuer2009measure, liu2011experimental, de2017dynamics, glick2020markovian, head2021capturing, link2022non, tserkis2022information}, and defining \emph{quantum non-Markovian processes} is the subject of much debate \cite{rivas2014quantum, breuer2016colloquium, li2018concepts, li2019non, milz2021quantum}. However, there are efforts to unify all non-Markovian processes under a common theoretical framework \cite{white2023unifying}. In this tutorial, we focus mainly on the characterization and benchmarking of \emph{Markovian} errors, and only mention non-Markovian errors in passing. However, understanding non-Markovian errors is important for many reasons, including the fact that they are an unavoidable consequence of open quantum systems, in which the system under study is in contact with an external bath or environment with uncontrolled degrees of freedom. Additionally, non-Markovian errors interfere with the characterization of Markovian errors, which is the central goal of QCVV. Finally, their impact on quantum error correction is not well understood, which is important for fault-tolerant quantum computation. 

\section{Fidelities and Error Metrics}\label{sec:overview}

The purpose of \ac{QCVV} is to discover and describe what is happening inside a quantum computer. In almost all cases, the computer is \emph{intended} to do a particular thing. We call this a \emph{target}. The models or descriptions for the target, and the thing that actually happened, can be complex and unwieldy. QCVV results are therefore often summarized by a single \emph{performance metric} that compares what \emph{actually} happened to what was \emph{intended} to happen. Many such metrics exist. In this section, we introduce and explain the most common ones.

The metrics we consider in this section compare two mathematical models, one of which (the target) describes the ideal operation of the quantum computer. So, a metric \footnote{Here, we use the term ``metric'' loosely. For example, we discuss different types of fidelity in this section, but fidelity is strictly not a metric in a mathematical sense, as it does not obey the triangle inequality.} is generally a function $f(x, \overline{x})$, where $x$ describes actual (i.e., experimental) behavior, and $\overline{x}$ is the target. The nature of $x$ and $\overline{x}$ depend on what aspect of the quantum computer's operation is being examined. In this tutorial (and generally in QCVV), we consider metrics for five aspects of a quantum computer's behavior:
\begin{enumerate}
    \item \emph{Probability Distributions} (Sec.~\ref{sec:prob_dists}). Probability distributions describe the results of running quantum circuits or experiments.
    \item \emph{Quantum States} (Sec.~\ref{sec:quantum_states}). Quantum states describe the configuration of a quantum register before it is measured.
    \item \emph{Quantum Processes} (Sec.~\ref{sec:quantum_processes}). Quantum processes describe how an operation or logic gate transforms states.
    \item \emph{Quantum Measurements} (Sec.~\ref{sec:quantum_meas}). Quantum measurements describe readout operations that extract classical data from quantum states.
    \item \emph{Quantum Processors (Gate Sets)} (Sec.~\ref{sec:quantum_gate_sets}). Quantum gate sets describe a complete set of logic operations on a quantum register.
\end{enumerate}
We discuss multiple distinct metrics that are commonly used for each kind of quantum object. These distinct metrics are not redundant; they quantify different aspects of an error, and each uniquely solves a particular problem. Understanding the differences between these metrics, and when to use each, is essential to reading and communicating QCVV results.

The metrics used in QCVV emerged organically. Most were borrowed or adapted from other fields of quantum information science, where their original purpose was \emph{not} to quantify ``error,'' but to quantify the difficulty of distinguishing two objects. As a result, their organization is somewhat haphazard. Some quantify the \emph{similarity} of two objects. These are usually called ``fidelity,'' and take the value $f(x, \overline{x}) = 1$ when $x = \overline{x}$. Others quantify the \emph{deviation} between (or \emph{distinguishability} of) two objects. These take the value $f(x, \overline{x}) = 0$ when $x = \overline{x}$, and some (but not all!) satisfy the mathematical definition of a metric. If $f(x,\overline{x})$ is a fidelity metric, then generally the corresponding \emph{infidelity} $1 - f(x, \overline{x})$ can be used as a deviation metric. Deviation metrics, including infidelities, are often interpreted as ``error rates,'' but we caution that no single ``error rate'' coincides with the probability of quantum computer failures in all contexts (which is why multiple metrics exist!).

%%%%%%%%%%%%%%%%%%%%%%% Classical Probability Distributions %%%%%%%%%%%%%%%%%%%%%%% 
\subsection{Classical Probability Distributions}\label{sec:prob_dists}

QCVV is primarily about modeling the behavior and performance of \emph{quantum} states and processes, but they cannot be observed directly. A quantum state gains tangible reality by being measured. Quantum processes act on states, which can then be measured to yield data. Thus, the necessary common denominator in \emph{any} experiment that tests a quantum state or process is the \emph{probability distribution} of a measurement's outcome. For this reason, a good metric for quantum states or quantum processes is almost always derived from a more elementary metric on probability distributions. We therefore begin our survey with metrics that compare two probability distributions. However, these metrics also appear in QCVV \emph{in their own right}, when they are used directly to evaluate the execution accuracy of a large quantum circuit.

If an experiment has a deterministic (non-random) outcome, and is intended to produce an outcome $X$, then it is very easy to test whether the experiment is working correctly. Perform a single trial, record the outcome $Y$, and ask whether $Y = X$. But the outcomes of quantum experiments are generally not deterministic. Both the target outcome and the actual outcome are \emph{random} variables, described by probability distributions $\mathbf{\overline{p}}$ and $\mathbf{p}$ over some sample space. Testing whether such an experiment is working correctly is trickier. Even in the simplest case, where the target outcome is deterministic ($\mathbf{\overline{p}}$ is supported on a single unique outcome $X$), we must extend the Boolean measure of correctness (``it works correctly'' or ``it does not'') to a real-valued probability $p(X)\in[0,1]$ describing \emph{how often} the experiment works correctly. For general $\mathbf{\overline{p}}$ and $\mathbf{p}$, quantifying the experiment's correctness becomes nontrivial. The following metrics are frequently used for this purpose.

%%%%%%%%%%%%%%%%%%%%%%% TVD %%%%%%%%%%%%%%%%%%%%%%% 
\subsubsection{Total Variation Distance}\label{sec:tvd}

The \emph{\ac{TVD}} between $\mathbf{\overline{p}}$ and $\mathbf{p}$ is:
\begin{align}\label{eq:tvd}
    d_\textrm{TV}(\mathbf{\overline{p}}, \mathbf{p}) 
        &\equiv \frac{1}{2} \sum_{k} \abs{\overline{p}_k - p_k} ~, \\
        &= \frac{1}{2} \norm{\mathbf{\overline{p}} - \mathbf{p}}_1 ~.
\end{align}
It has several important operational interpretations (practical questions to which it is the answer). The best-known interpretation of TVD involves \emph{single-shot discrimination} between distributions. The optimal probability of guessing correctly whether a single sample was drawn from distribution $\mathbf{\overline{p}}$ or $\mathbf{p}$, with both theories deemed equally probable, is $\left[1 + d_{\textrm{TV}}(\mathbf{\overline{p}}, \mathbf{p})\right] / 2$. Perhaps more importantly, if $N$ samples are drawn from distribution $\mathbf{p}$, then as $N\to\infty$ the fraction of those samples that must be \emph{changed} in order to make them consistent with $\mathbf{\overline{p}}$ is exactly $d_{\textrm{TV}}(\mathbf{\overline{p}}, \mathbf{p})$. This supports interpreting $d_{\textrm{TV}}(\mathbf{\overline{p}},\mathbf{p})$ as an \emph{error rate} --- i.e., the rate of events produced by sampling from $\mathbf{p}$ that are inconsistent with $\mathbf{\overline{p}}$. 

The TVD between any two distributions is bounded between 0 (achieved uniquely when they are equal) and 1 (achieved when the distributions have disjoint support). TVD is a metric in the strict mathematical sense (e.g., it satisfies the triangle inequality).

%%%%%%%%%%%%%%%%%%%%%%% Classical (Hellinger) Fidelity %%%%%%%%%%%%%%%%%%%%%%% 
\subsubsection{Classical (Hellinger) Fidelity}\label{sec:classical_fidelity}

TVD does not capture everything. For example, consider two different discrimination problems over the set $\{0,1\}$:
\begin{enumerate}
    \item Distinguish $\mathbf{\overline{p}}=(1,0)$ from $\mathbf{p}=(0.98,0.02)$,
    \item Distinguish $\mathbf{\overline{q}}=(0.49,51)$ from $\mathbf{q}=(0.51,0.49)$.
\end{enumerate}
Both pairs are separated by the same TVD ($0.02$), and so can be distinguished with equal probability ($0.51$) given a single sample. But this is barely better than random guessing. Distinguishing either pair with reasonable confidence requires examining $N\gg1$ samples. Rather than asking ``What's the probability of guessing correctly given \emph{one} sample?'', we should ask ``How many samples are required to guess correctly with high probability (e.g., 90\%)?''

Remarkably, the answers for the two pairs are quite different. To distinguish $\mathbf{\overline{p}}$ from $\mathbf{p}$, we guess $\mathbf{p}$ if we see any ``1'' outcomes whatsoever, and $\mathbf{\overline{p}}$ if we do not. It takes just 80 samples to ensure a 90\% probability of guessing correctly. But to distinguish $\mathbf{\overline{q}}$ from $\mathbf{q}$, we guess $\mathbf{q}$ if we see more ``0'' outcomes than ``1'' outcomes, and $\mathbf{\overline{q}}$ otherwise. The random fluctuations in the number of ``0'' and ``1'' outcomes are much larger in this case, and a whopping 4105 samples --- greater than $50\times$ more! --- are needed to ensure a 90\% probability of guessing correctly. Thus, the TVD does not \emph{regularize} well --- i.e., the TVD between $\mathbf{\overline{p}}$ and $\mathbf{p}$ does not accurately predict the TVD between the $N$-copy distributions $\mathbf{\overline{p}}^{\otimes N}$ and $\mathbf{p}^{\otimes N}$.

A metric that \emph{does} regularize well is the \emph{Bhattacharya coefficient} \cite{bhattacharyya1943measure} (a.k.a., ``statistical overlap'' \cite{fuchs1996distinguishability}):
\begin{equation}
    BC(\mathbf{\overline{p}}, \mathbf{p}) \equiv \sum_{k}{\sqrt{\overline{p}_{k} p_{k}}} ~.
\end{equation}
In quantum information science, the square of the Bhattacharya coefficient is often called \emph{classical fidelity} or \emph{Hellinger fidelity}:
\begin{equation}\label{eq:classical_fidelity}
    F(\mathbf{\overline{p}}, \mathbf{p}) \equiv \left(\sum_{k}{\sqrt{\overline{p}_{k} p_{k}}}\right)^2 ~.
\end{equation}
Classical fidelity is a measure of \emph{similarity}: $F(\mathbf{\overline{p}},\mathbf{p})=1$ iff $\mathbf{p} = \mathbf{\overline{p}}$, and $F(\mathbf{\overline{p}},\mathbf{p})=0$ iff they have disjoint support. Unlike the TVD, the fidelity between two distributions does predict the fidelity between $N$ copies of the same distributions, because
\begin{equation}
    F(\mathbf{\overline{p}}^{\otimes N}, \mathbf{p}^{\otimes N}) = F(\mathbf{\overline{p}}, \mathbf{p})^{N} ~.
\end{equation}

Classical fidelity is very closely related to the \emph{Hellinger distance},
\begin{equation}
    H(\mathbf{\overline{p}}, \mathbf{p}) \equiv \sqrt{1-\sqrt{F(\mathbf{\overline{p}}, \mathbf{p})}} ~,
\end{equation}
which is a metric in the strict mathematical sense. The Hellinger distance is related to the TVD by a bounding inequality,
\begin{equation}
    H^{2}(\mathbf{\overline{p}}, \mathbf{p}) \leq d_\textrm{TV}(\mathbf{\overline{p}}, \mathbf{p}) \leq \sqrt{2} H(\mathbf{\overline{p}}, \mathbf{p}) ~.
\end{equation}
This kind of inequality can help us understand $N$-copy distinguishability. If we rewrite it in terms of the classical fidelity, we get
\begin{equation}
    1 - \sqrt{F(\mathbf{\overline{p}}, \mathbf{p})} \leq d_\textrm{TV}(\mathbf{\overline{p}}, \mathbf{p}) \leq \sqrt{2}\sqrt{1 - \sqrt{F(\mathbf{\overline{p}}, \mathbf{p})}} ~.
\end{equation}
There is a strictly more powerful inequality \cite{fuchs1999cryptographic},
\begin{equation}\label{eq:ClassicalFvdG}
    1 - \sqrt{F(\mathbf{\overline{p}}, \mathbf{p})} \leq d_\textrm{TV}(\mathbf{\overline{p}}, \mathbf{p}) \leq \sqrt{1 - F(\mathbf{\overline{p}}, \mathbf{p})} ~,
\end{equation}
and if we apply this to the $N$-copy distributions, we get
\begin{equation}
    1 - F(\mathbf{\overline{p}}, \mathbf{p})^{N/2} \leq d_\textrm{TV} \left(\mathbf{\overline{p}}^{\otimes N}, \mathbf{p}^{\otimes N}\right) \leq \sqrt{1 - F(\mathbf{\overline{p}}, \mathbf{p})^{N}} ~.
\end{equation}
Therefore, the TVD between $\mathbf{\overline{p}}^{\otimes N}$ and $\mathbf{p}^{\otimes N}$ will be close to 1 iff~$F(\mathbf{\overline{p}},\mathbf{p})^{N/2} \ll 1$ --- which is to say, when $N \gtrsim 2/\left(-\ln[F(\mathbf{\overline{p}},\mathbf{p})]\right)$. This is the inverse of the \emph{Bhattacharya distance},
\begin{equation}
    d_{B}(\mathbf{\overline{p}}, \mathbf{p}) \equiv -\frac{1}{2}\ln[F(\mathbf{\overline{p}},\mathbf{p})] ~,
\end{equation}
which quantifies the difficulty of distinguishing very similar distributions $\mathbf{p}$ and $\mathbf{\overline{p}}$ much more accurately than the TVD. For the specific examples given in the beginning of the section, $1/d_{B}(\mathbf{\overline{p}}, \mathbf{p})\approx 99$, whereas $1/d_{B}(\mathbf{\overline{q}}, \mathbf{q})\approx 4999$ --- quite close (in both cases) to the exact number of samples required to distinguish the distributions 90\% of the time.

%%%%%%%%%%%%%%%%%%%%%%% Entropy %%%%%%%%%%%%%%%%%%%%%%% 
\subsubsection{Relative Entropy and Cross-Entropy}\label{sec:entropy}

In classical information theory, the most important and commonly used metric of deviation between distributions $\mathbf{p}$ and $\mathbf{\overline{p}}$ is neither TVD nor fidelity. Rather, it is the \emph{\ac{KL} divergence}, also known as \emph{relative entropy}:
\begin{equation}
    d_{\mathrm{KL}}(\mathbf{p} \vert\vert \mathbf{\overline{p}}) \equiv \sum_{k}{p_{k}\log\left(\frac{p_{k}}{\overline{p}_k}\right)} ~.
\end{equation}
The KL divergence quantifies \emph{deviation} (not \emph{similarity}). It is always non-negative, it is zero iff $\mathbf{p} = \mathbf{\overline{p}}$, and it is not a mathematical metric. Unlike fidelity or TVD, $d_{\mathrm{KL}}$ can be arbitrarily large. Furthermore, it is \emph{asymmetric} with respect to its two arguments, and is usually stated as ``the KL divergence \emph{from} $\mathbf{\overline{p}}$ \emph{to} $\mathbf{p}$.'' Its first argument (here $\mathbf{p}$) should represent truth or reality, while its second argument (here $\mathbf{\overline{p}}$) should represent a theory or model. 

The KL divergence is deeply rooted in statistics and information theory, and has too many operational interpretations to list here. It often quantifies the consequences of \emph{believing} that samples are being drawn from $\mathbf{\overline{p}}$ when they are actually being drawn from $\mathbf{p}$. For example, it describes the rate at which a gambler or investor will lose money if they use a suboptimal strategy, the extra bandwidth required to send a message using a code adapted for the wrong distribution of symbols, and the rate at which a skeptical observer will accumulate evidence against the theory $\mathbf{\overline{p}}$ when data are actually generated by $\mathbf{p}$.

In all of these usages, the KL divergence appears as the \emph{difference} between two quantities known as the \emph{entropy} and \emph{cross-entropy}:
\begin{align}
    H(\mathbf{p}) &\equiv -\sum_{k}{p_{k} \log(p_{k})} ~, \label{eq:entropy} \\
    H(\mathbf{p}, \mathbf{\overline{p}}) &\equiv -\sum_{k}{p_{k} \log(\overline{p}_k)} ~. \label{eq:cross-entropy}
\end{align}
The entropy of $\mathbf{p}$ (usually known as \textit{Shannon entropy} in the information theory literature \footnote{The logarithm that appears in entropic quantities can be evaluated in any base; using $\log_2$ yields \textit{bits} of entropy, while $\ln$ yields units called \textit{nats}.}) quantifies the intrinsic cost of performing a task on $\mathbf{p}$, while the cross-entropy of $\mathbf{\overline{p}}$ relative to $\mathbf{p}$ quantifies the same cost using a suboptimal strategy optimized for $\mathbf{\overline{p}}$.

Each of these quantities has many uses in its own right. The cross-entropy is particularly useful in QCVV and machine learning, because it has a rigorous mathematical meaning \emph{and} it can be estimated easily in experiments since it is strictly linear in the true distribution $\mathbf{p}$, and can thus be written as an expectation value:
\begin{equation}
    H(\mathbf{p}, \mathbf{\overline{p}}) = \left\langle \log(\mathbf{\overline{p}}) \right\rangle_{\mathbf{p}} ~. \label{eq:crossentropy}
\end{equation}
If the entropy of a candidate (model) distribution $\mathbf{\overline{p}}$ is known, then an easy way to check whether the true $\mathbf{p}$ is equal (or close) to $\mathbf{\overline{p}}$ is to estimate $H(\mathbf{p}, \mathbf{\overline{p}})$ by drawing some samples, estimating $\left\langle \log(\mathbf{\overline{p}}) \right\rangle_{\mathbf{p}}$, and comparing it to the known $H(\mathbf{\overline{p}})$.

%%%%%%%%%%%%%%%%%%%%%%% Linear cross-entropy and heavy output probability %%%%%%%%%%%%%%%%%%%%%%% 
\subsubsection{Linear Cross-Entropy and Heavy Output Probability}\label{sec:linear_xe_heavy_output}

As noted above, cross-entropy is a well-motivated metric, with important operational interpretations, that can be measured directly. However, if $\mathbf{\overline{p}}$ has very small (e.g., zero) entries, then the variance of the estimate can be very large, making it slow to converge. 

When the precise properties of cross-entropy are not important, and it is only being used as a proxy for similarity of $\mathbf{p}$ to $\mathbf{\overline{p}}$, the so-called \emph{linear cross-entropy},
\begin{equation}
    H_{\mathrm{lin}}(\mathbf{p}, \mathbf{\overline{p}}) \equiv d\sum_{k=1}^d {p_{k}\overline{p}_k} - 1 ~,
    \label{eq:lincrossentropy}
\end{equation}
where $d$ is the size of the sample space, can be used instead. It is less sensitive to arbitrary small deviations in the probabilities than the real cross-entropy, but estimates of it converge with fewer samples.

\emph{Heavy output probability} is another metric designed for ease of measurement. Given a $d$-element probability distribution $\mathbf{\overline{p}}$, the ``heavy'' outcomes are simply the ones whose probability is greater than the median --- i.e., the $d/2$ elements to which $\mathbf{\overline{p}}$ assigns the highest probabilities \footnote{Heavy output probability is not necessarily well-defined for highly degenerate distributions.}. The heavy output probability of a distribution $\mathbf{p}$ with respect to $\mathbf{\overline{p}}$ is simply the total probability assigned by $\mathbf{p}$ to $\mathbf{\overline{p}}$'s heavy outcomes. Heavy output probability is easily estimated by simply drawing samples from $\mathbf{p}$ and checking whether they are ``heavy'' for $\mathbf{\overline{p}}$. 

Linear cross-entropy \cite{boixo2018characterizing} and heavy output probability \cite{cross2019validating} are used in QCVV to test and verify distributions over enormously large sample spaces, where sampling the entire space is infeasible. Both can be estimated fairly accurately using just a few samples. However, neither has a particularly compelling interpretation. Moreover, estimating them \emph{does} require calculating elements of the reference distribution $\mathbf{\overline{p}}$, which can be difficult for distributions produced by quantum algorithms (the classical hardness of this task is partly why we are developing quantum computers in the first place!).

%%%%%%%%%%%%%%%%%%%%%%% Roles of metrics %%%%%%%%%%%%%%%%%%%%%%% 
\subsubsection{Roles of Metrics}

TVD, fidelity, and relative/cross-entropy are just a few of the many metrics, divergences, deviations, and similarity measures used in the literature to quantify similarity or distinguishability of distributions. But they are the ones that appear most frequently in the context of quantum computing and QCVV. More importantly, they are the ones from which the most commonly used properties of \emph{quantum} objects are derived. These quantities, and their quantum counterparts discussed below, are distinct, inequivalent, and generally \emph{not} interchangeable. When a QCVV practitioner is choosing  how to quantify accuracy, error, similarity, or deviation, it is important to consider the specific task at hand. In almost every circumstance, no more than one of these quantities will faithfully capture the experimental behavior of interest.

%%%%%%%%%%%%%%%%%%%%%%% Quantum States %%%%%%%%%%%%%%%%%%%%%%% 
\subsection{Quantum States}\label{sec:quantum_states}

The quantum state of a qubit, qudit, or quantum register \emph{before} it gets measured is described by a state vector $\ket{\psi}$ (Sec.~\ref{sec:state_vec_hilbert_space}) or a density matrix $\rho = \sum_{i}{p_{i}\proj{\psi_{i}}}$ (Sec.~\ref{sec:density_matrix_formalism}). Like probability distributions, quantum states assign probabilities to events. But for a quantum system, the sample space of possible events is determined not by the nature of the system, but by how an observer interacts with (measures) it. How similar or distinguishable two quantum states are thus depends on how they are measured. Because it is always easy to find measurements that \emph{fail} to distinguish between quantum states, metrics for comparing two quantum states are typically defined by maximizing distinguishability, or minimizing similarity, over all possible measurements.

Metrics on quantum states can be used to compare any two states $\rho$ and $\sigma$, but in QCVV the most common use by far is to compare a ``real'' state $\rho$ to an ``ideal'' target state $\overline{\rho}$, and thus quantify state preparation error. Since these metrics optimize over all possible measurements, they generally define \emph{upper bounds} on the probability of observing an error in a specific measurement, protocol, or algorithm that uses the ``real'' state.

%%%%%%%%%%%%%%%%%%%%%%% Trace Distance %%%%%%%%%%%%%%%%%%%%%%% 
\subsubsection{Trace Distance}\label{sec:trace_distance}

The \emph{trace distance} between two quantum states $\rho$ and $\overline{\rho}$ is a measure of their distinguishability. It varies from 0 (iff~$\rho=\overline{\rho}$) to 1 (when their supports are orthogonal). It is the maximum over all \ac{POVM} measurements $M = \{E_{m}\}$  of the TVD between $M$'s outcome distribution given $\rho$, and $M$'s outcome distribution given $\overline{\rho}$. We say that the two distributions $\Pr(m \vert \rho)$ and $\Pr(m \vert \overline{\rho})$ are \emph{induced} by the states $\rho$ and $\overline{\rho}$. They are given by
\begin{align}
    \Pr(m \vert \rho) &= \Tr[E_{m}\rho] ~, \\
    \Pr(m \vert \overline{\rho}) &= \Tr[E_{m}\overline{\rho}] ~,
\end{align}
and so the TVD between them equals
\begin{equation}
    d_\textrm{TV} = \frac12\sum_{m}{\left|\Tr[E_{m}(\rho-\overline{\rho})]\right|} ~.
\end{equation}
Helstrom \cite{helstrom1969quantum} proved that this is maximized by a 2-outcome POVM whose effects are the projectors onto the positive and negative eigenspaces of $(\rho - \overline{\rho})$, and that the trace distance between $\rho$ and $\overline{\rho}$ is given by the \emph{nuclear norm} of $(\rho - \overline{\rho})$,
\begin{equation}
    d_{\textrm{tr}}(\rho, \overline{\rho}) = \frac12\|\rho-\overline{\rho}\|_{1} = \frac12\Tr|\rho-\overline{\rho}|,
\end{equation}
where $|\rho - \overline{\rho}| = \sqrt{(\rho - \overline{\rho})^{2}}$ can be obtained by diagonalizing $(\rho - \overline{\rho})$ and replacing each of its eigenvalues $\lambda_{i}$ with its absolute value $|\lambda_{i}|$.

Trace distance is a metric in the rigorous sense, and a measure of distinguishability. It inherits essentially all the properties of the TVD. In particular, like TVD, it gives the probability of success for single-shot discrimination between $\rho$ and $\overline{\rho}$. If we are given a single quantum system, prepared either according to $\rho$ or $\overline{\rho}$ with equal prior probabilities, then the maximum achievable probability of guessing correctly how it was prepared is achieved by performing Helstrom's measurement and is equal to $[1 + d_{\textrm{tr}}(\rho, \overline{\rho})] / 2$.

The trace distance is related to the Euclidean distance between $\rho$ and $\overline{\rho}$ for the special case of single-qubit states on the Bloch sphere. To see this, we can write $\rho$ and $\overline{\rho}$ in terms of their respective Bloch vectors $\mathbf{r}$ and $\overline{\mathbf{r}}$, 
\begin{equation}
    \rho = \frac{1}{2} (\Id + \mathbf{r} \cdot \boldsymbol{\sigma}), ~~ \overline{\rho} = \frac{1}{2} (\Id + \overline{\mathbf{r}} \cdot \boldsymbol{\sigma}) ~.
\end{equation}
The trace distance between $\rho$ and $\overline{\rho}$ is
\begin{equation}
   d_{\textrm{tr}}(\rho, \overline{\rho}) = \frac{1}{2} \Tr\abs{\rho - \overline{\rho}} = \frac{1}{4} \Tr\abs{ (\mathbf{r} - \overline{\mathbf{r}}) \cdot \boldsymbol{\sigma} } ~.
\end{equation}
Because the eigenvalues of $\boldsymbol{\sigma}$ are $\pm 1$, the trace of $\abs{(\mathbf{r} - \overline{\mathbf{r}}) \cdot \boldsymbol{\sigma} } = 2 \abs{ \mathbf{r} - \overline{\mathbf{r}}}$, and thus
\begin{equation}
    d_{\textrm{tr}}(\rho, \overline{\rho}) = \frac{1}{2} \abs{ \mathbf{r} - \overline{\mathbf{r}} } ~.
\end{equation}
Therefore, the trace distance between two single-qubit states is exactly equal to one-half the Euclidean distance between their Bloch vectors.

%%%%%%%%%%%%%%%%%%%%%%% State Fidelity %%%%%%%%%%%%%%%%%%%%%%% 
\subsubsection{State Fidelity}\label{sec:state_fidelity}

The fidelity between two quantum states $\rho$ and $\overline{\rho}$ is a measure of their similarity. It varies from 1 (iff~$\rho = \overline{\rho}$) to 0 (when their supports are orthogonal). In the simple special case where both states are pure, so $\rho = \proj{\psi}$ and $\overline{\rho} = \proj{\phi}$, their fidelity is exactly equal to the \emph{transition probability},
\begin{equation}\label{eq:state_fidelity}
    F(\proj{\psi},\proj{\phi}) = |\braket{\psi | \phi}|^2 ~.
\end{equation}
If state $\ket{\psi}$ is measured in a basis containing $\bra{\phi}$, then $F$ is the probability of observing $\bra{\phi}$ and thus collapsing into $\ket{\phi}$ (and vice-versa). It is important to note that the state fidelity is sometimes defined in the literature as the square root of the transition probability ($F' = \sqrt{F} = \abs{\braket{\psi | \phi}}$). We (like most authors) prefer the definition given above because $F$ is an actual probability, but readers should be aware of (and alert for) both definitions in the literature \footnote{See, for example, the book \textit{Quantum Computation and Quantum Information} \cite{nielsen2002quantum}.}.

If one state is pure, but the other is mixed --- e.g., $\rho = \sum_i{p_i\proj{\psi_i}}$ and $\overline{\rho} = \proj{\phi}$ --- then there is still a well-defined transition probability from $\rho \to \proj{\phi}$. Schumacher \cite{schumacher1995quantum} was the first to define the fidelity as
\begin{equation}\label{eq:SemimixedStateFidelity}
    F(\rho, \proj{\phi}) = \braopket{\phi}{\rho}{\phi} = \Tr[ \rho \proj{\phi} ] ~. 
\end{equation}
This special case is very common in QCVV, where state fidelity is commonly used to quantify error when an experimentalist intended to prepare $\proj{\phi}$ but prepared $\rho$ instead. 

Defining the fidelity between two \emph{mixed} quantum states $\rho$ and $\overline{\rho}$ is a bit trickier because there is no obvious ``transition probability'' to a mixed state. There are at least two independent ways to define the fidelity between two mixed states. Remarkably, they lead to exactly the same result! Uhlmann \cite{uhlmann1976transition} and (later) Jozsa \cite{jozsa1994fidelity} sought to generalize ``transition probability'' to mixed states by considering \emph{purifications} of $\rho$ and $\overline{\rho}$ on a larger Hilbert space. If $\ket{\psi}$ and $\ket{\phi}$ are purifications of $\rho$ and $\overline{\rho}$ (respectively), then the \emph{maximum} value (over all possible purifications) of the transition probability $F(\proj{\psi},\proj{\phi}) = |\braket{\psi | \phi}|^2$ is equal to
\begin{equation}\label{eq:MixedStateFidelity}
    F(\rho,\overline{\rho}) = \left(\Tr\sqrt{\sqrt{\rho}\overline{\rho}\sqrt{\rho}}\right)^2 = \left(\Tr\sqrt{\sqrt{\overline{\rho}}\rho\sqrt{\overline{\rho}}}\right)^2 ~, 
\end{equation}
which is now widely accepted as the definition of fidelity between two mixed quantum states. Fuchs \cite{fuchs1996distinguishability} asked a different question that is a direct analogue to Helstrom's derivation of trace distance: what is the \emph{minimum} value, over all POVM measurements $M$, of the classical fidelity between $\Pr(m \vert \rho)$ and $\Pr(m \vert \overline{\rho})$? The answer turns out to be identical to Josza's fidelity (\eq\ref{eq:MixedStateFidelity}).

%%%%%%%%%%%%%%%%%%%%%%% Infidelity %%%%%%%%%%%%%%%%%%%%%%% 
\subsubsection{Infidelity} \label{sec:Infidelity}

Although \eq\ref{eq:MixedStateFidelity} is celebrated, it is \emph{very} rarely necessary in QCVV. The primary use of quantum state fidelity in QCVV is to quantify and report the \emph{error} in an experimental attempt to prepare a pure target state $\proj{\phi}$. In this situation, although the experimentally prepared state $\rho$ is mixed, the target state is pure. The far simpler formula in \eq\ref{eq:SemimixedStateFidelity} can be used instead.

Quantifying \emph{error} is usually better done by reporting \emph{infidelity} instead of fidelity:
\begin{equation}
    \epsilon_F \equiv 1 - F ~.
\end{equation}
Like trace distance or other measures of \emph{distinguishability}, infidelity ranges from 0 (when $\rho=\overline{\rho}$) to 1 (when they have disjoint support). If a target state $\rho = \proj{\phi}$ is prepared with infidelity 0, then the probability of an error resulting from that preparation is also zero, making infidelity a good metric of error. 

Many experiments in the literature report fidelity. However, the only rationale for the awkwardness of reporting $F = 0.99981 \pm 0.00007$ instead of $\epsilon_F = (1.9\pm 0.7)\times 10^{-4}$ are habit and a vague sense that ``fidelity'' plays a privileged role in the quantum information literature. This is largely a historical accident. Fault tolerance thresholds are always described by error rates, and for most QCVV purposes ``low-error'' is a more descriptive epithet than ``high-fidelity.''

%%%%%%%%%%%%%%%%%%%%%%% Contrasting Trace Distance and Infidelity %%%%%%%%%%%%%%%%%%%%%%% 
\subsubsection{Contrasting Trace Distance and Infidelity} \label{sec:tracedistance_vs_infidelity}

\begin{figure}[t]
     \centering
     \includegraphics[width=0.8\columnwidth]{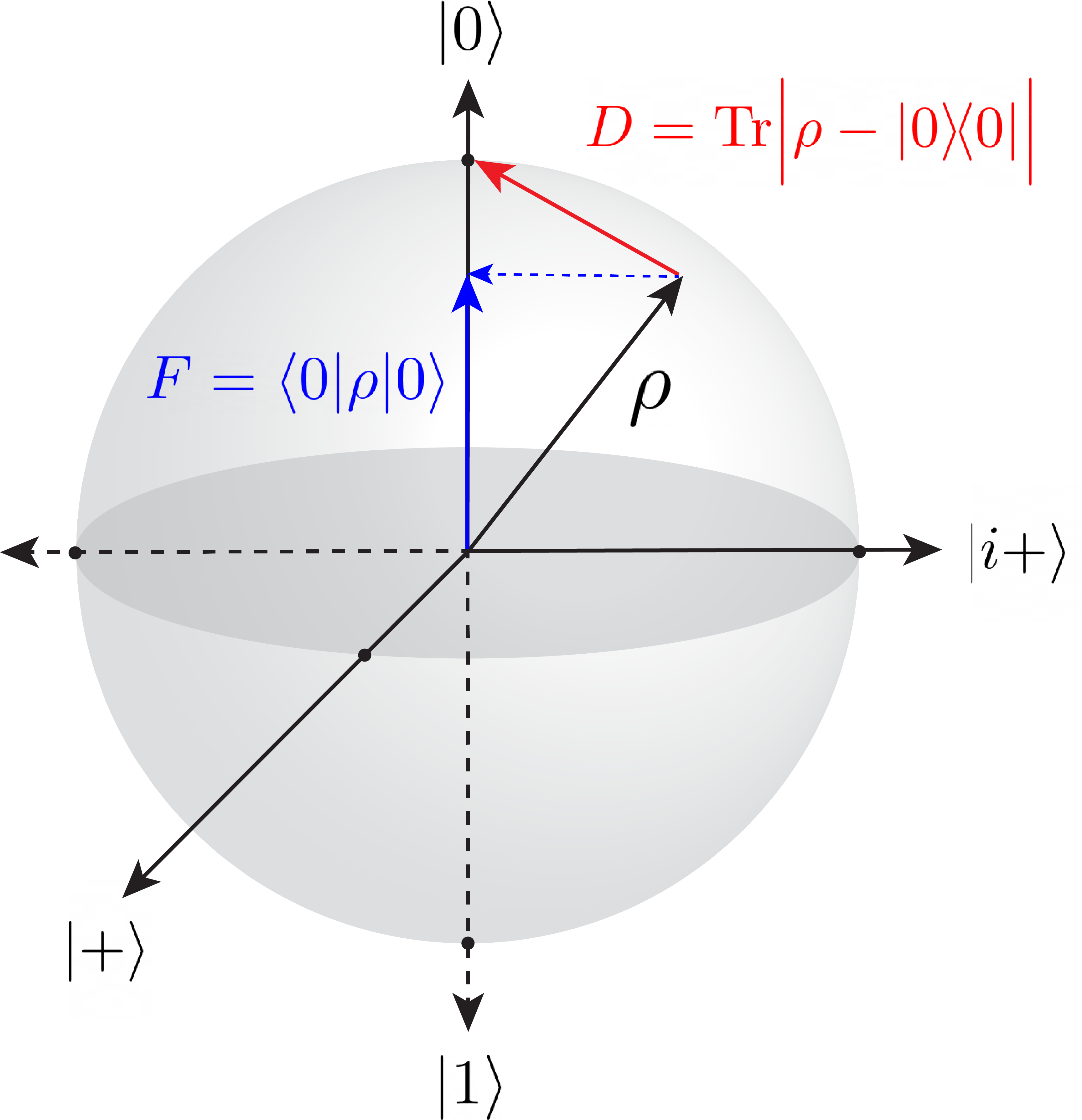}
     \caption{\textbf{Fidelity vs.~Trace Distance.} 
     Given an arbitrary state $\rho$ and an ideal target state $\ketbra{0}{0}$, the ``error'' in $\rho$ can be quantified in different ways. For example, the fidelity (blue) of $\rho$ with $\ketbra{0}{0}$ is the \emph{projection}, or \emph{overlap}, of $\rho$ with $\ket{0}$: $F = \bra{0} \rho \ket{0}$. On the other hand, the trace distance between $\rho$ and $\ketbra{0}{0}$ is $d_\textrm{tr} = \tfrac{1}{2} \Tr \big| \rho - \ketbra{0}{0} \big|$, which equals one-half the Euclidean distance between the two vectors on the Bloch sphere. In the figure above, $D$ (red) is the Euclidean distance between $\rho$ and $\ketbra{0}{0}$, thus $D = 2d_\textrm{tr} = \Tr \big| \rho - \ketbra{0}{0} \big|$. The infidelity ($1 - F$) and the trace distance between $\rho$ and $\ketbra{0}{0}$ can be very different, but neither is fundamentally better at quantifying the ``error'' in $\rho$. Each is appropriate for a particular operational scenario.}
     \label{fig:bloch_sphere_measurement}
 \end{figure}

Trace distance and infidelity are the most commonly used metrics of deviation or distinguishability for quantum states. It is worth briefly examining what makes them different, and why neither can replace the other. Both correspond directly to classical counterparts (TVD and classical fidelity) as shown by Helstrom \cite{helstrom1969quantum} and Fuchs \cite{fuchs1996distinguishability}, respectively. They inherit all the properties (and differences) of those classical counterparts. So, for example, the fidelity between $\rho$ and $\overline{\rho}$ regularizes nicely to $\rho^{\otimes N}$ and $\overline{\rho}^{\otimes N}$, whereas their trace distance does not.

But there are additional differences that appear only at the quantum level. The simplest of these have to do with the behavior of fidelity and trace distance for nearby \emph{pure} quantum states. Suppose that $\ket{\psi}$ and $\ket{\phi}$ are ``nearby'' pure states, meaning that $1 - |\braket{\psi | \phi}|^2 = \epsilon \ll 1$. Since $\ket{\psi}$ and $\ket{\phi}$ span a 2-dimensional subspace, we can consider a single-qubit system without any loss of generality (see  \fig\ref{fig:bloch_sphere_measurement}). Their fidelity is $F = |\braket{\psi | \phi}|^2 = 1 - \epsilon$, so their infidelity is $\epsilon$. The trace distance between them is $d_{\mathrm{tr}} = \frac12 \Tr \big| \proj{\psi} - \proj{\phi} \big|$, which we can compute by observing that because $\proj{\psi} - \proj{\phi}$ has trace 0, its eigenvalues are $\{+\lambda, -\lambda\}$, and $2\lambda^2 = \Tr[\left( \proj{\psi} - \proj{\phi} \right)^2] = 2 - 2(1 - \epsilon) = 2\epsilon$, so $\lambda = \sqrt{\epsilon}$ and thus $d_{\mathrm{tr}} = \sqrt{\epsilon}$. So, if the infidelity between two pure states is small (e.g., $\epsilon = 10^{-4}$), then the trace distance between them will be much larger ($\sqrt\epsilon = 10^{-2}$). It is reasonable to ask whether this behavior is generic --- i.e., is it generally true that $d_{\mathrm{tr}} \approx \sqrt{\epsilon_F}$?  It is not. This behavior is specific to \emph{pure} states that differ by a unitary operation. 

If instead we compare $\proj{\psi}$ to a mixed state $\rho = (1-\epsilon)\proj{\psi} + \epsilon\proj{\overline{\psi}}$, where $\braket{\psi | \overline{\psi}} = 0$, then it is very easy to show that $d_{\mathrm{tr}} = \epsilon_F = \epsilon$. For \emph{these} states, the two metrics coincide.

These two cases illustrate the two extremes of the \emph{Fuchs–van de Graaf inequalities} \cite{fuchs1999cryptographic}, which relate infidelity and trace distance for any pair of quantum states:
\begin{equation} \label{eq:FuchsVanDeGraaf}
    1 - \sqrt{1 - \epsilon_F} \leq d_{\mathrm{tr}} \leq \sqrt{\epsilon_F} ~.
\end{equation}
These inequalities are an \emph{exact} quantum analogue of the inequalities given in Eq. \ref{eq:ClassicalFvdG} for classical probability distributions. In the quantum case, if one state is pure, then a tighter and simpler lower bound holds:
\begin{equation}
    \epsilon_F \leq d_{\mathrm{tr}} \leq \sqrt{\epsilon_F} ~.
\end{equation}

%%%%%%%%%%%%%%%%%%%%%%% Quantum Relative Entropy %%%%%%%%%%%%%%%%%%%%%%% 
\subsubsection{Quantum Relative Entropy}\label{sec:qrelative_entropy}

\emph{Quantum relative entropy} is a measure of distinguishability between quantum states. Although it is less common in the QCVV literature, it is important in quantum information theory. It generalizes KL divergence to quantum states, and is given by
\begin{equation}
    S(\rho \vert\vert \overline{\rho} ) = \Tr[\rho \log(\rho)] -\Tr[\rho \log(\overline{\rho})] ~.
\end{equation}
Like the metrics discussed above, it is defined by maximizing the \emph{classical} KL divergence of $\Pr(m \vert \rho)$ with respect to $\Pr(m \vert \overline{\rho})$ over all POVM measurements $M = \{E_m\}$. Like the KL divergence, it is not a metric, it is asymmetric with respect to its arguments, and it diverges to infinity whenever there exists a measurement outcome such that $\Pr(m \vert \rho) > 0$ but $\Pr(m \vert \overline{\rho}) = 0$.

%%%%%%%%%%%%%%%%%%%%%%% Quantum Processes %%%%%%%%%%%%%%%%%%%%%%% 
\subsection{Quantum Processes}\label{sec:quantum_processes}

Quantum computing requires that quantum states be \emph{transformed} by precise, controlled evolution. Logic gates, circuit layers (comprising multiple gates in parallel), and quantum circuits (comprising multiple layers in sequence) describe particular \emph{unitary} transformations that are supposed to change a $d$-dimensional quantum register's state as
\begin{equation}
    \rho_{\mathrm{in}} \mapsto \rho_{\mathrm{out}} = U\rho_{\mathrm{in}}U^\dagger
\end{equation}
for some $d\times d$ unitary matrix $U$. Real-world attempts to implement unitary transformations are imperfect, so the register's evolution is not generally described by any unitary $U$, but as discussed in Sec.~\ref{sec:models}, it can often be described by a quantum process (\ac{CPTP} map) acting on $d\times d$ density matrices,
\begin{equation}
    \rho_{\mathrm{in}} \mapsto \rho_{\mathrm{out}} = G[ \rho_{\mathrm{in}} ] ~.
\end{equation}
In this tutorial, a ``quantum process'' \footnote{It is worth emphasizing that a \emph{quantum process} is not analogous to a classical \emph{stochastic process}. The classical analogue of a quantum process is a \emph{stochastic matrix} (see Sec.~\ref{sec:truthtable}), which is related to stochastic processes, but quite distinct.} means a CPTP map whose input and output spaces are the same. Such maps describe the action of gates, layers, reversible circuits, idle time, and/or imperfect unitaries. 
%In the next subsection, on quantum measurements, we will consider CPTP maps whose output space is different from their input space (POVMs and quantum instruments), but in this subsection we focus on quantum processes.
In this section, we will refer to any such operation as a \emph{gate}. We will denote its real (noisy) action by $G$, and its ideal ``target'' action by $\overline{G}$. The target action is almost always unitary, so $\overline{G}^{\, -1}$ is also a valid operation, and we can write
\begin{align}
    G &= \E\overline{G} \\
    & \Updownarrow \\ \nonumber
    \E &= G \overline{G}^{\, -1} ~,
\end{align}
and refer to $\E$ as the \emph{error process} for $G$. Most metrics for quantum processes can be used to compare two arbitrary processes, but in QCVV they are almost always used to compare a real process $G$ to its ideal target $\overline{G}$. If a metric $f(\cdot,\cdot)$ is unitarily invariant, then $f(G, \overline{G}) = f(\E, \Id)$.

Section \ref{sec:models} introduced several representations of quantum processes. The metrics we discuss here are properties of the quantum process itself, and their validity does not depend on what representation is being used. But each metric is most easily \emph{defined} (and/or computed) in a particular representation. We will make extensive use of two representations (defined in Sec.~\ref{sec:models}, but outlined again here):
\begin{itemize}
    \item The \emph{transfer matrix} representation $\Lambda_G$ of a quantum process $G$ (Sec.~\ref{sec:superop}), which is constructed by choosing an orthonormal basis $\{B_i\}$ for the vector space of $d\times d$ matrices, and using the Hilbert-Schmidt inner product to define
        \begin{equation}
            \left(\Lambda_G\right)_{i,j} \equiv \Tr\left( B_i^\dagger G[ B_j ] \right) ~.
        \end{equation}
    \item The \emph{$\chi$ matrix} (a.k.a.~``process matrix'') representation $\chi_G$ of $G$ (Sec.~\ref{sec:chi_mat_rep}) is constructed by choosing an orthonormal basis $\{B_i\}$ for the vector space of $d \times d$ matrices, and then finding a matrix of coefficients $\left(\chi_G\right)_{i,j}$ such that, for any $d\times d$ density matrix $\rho$,
        \begin{equation}
            G[ \rho ] = \sum_{i,j}{\left(\chi_G\right)_{i,j} B_i\rho B_j^\dagger} ~.
        \end{equation}
\end{itemize}

It is easy to define \textit{ad hoc} metrics of similarity or deviation between the matrix representations of $G$ and $\overline{G}$. But most have no operational meaning, and are not useful. The metrics we use in QCVV and quantum computing are chosen because they have observable meanings. Since quantum processes are (like quantum states) not directly observable, meaningful metrics of similarity or deviation between $G$ and $\overline{G}$ compare probability distributions \emph{induced} by $G$ and $\overline{G}$. Metrics specify (1) an initial state $\rho$, (2) a POVM $\{E_m\}$, and (3) a classical metric between distributions to compare
\begin{align}
    \Pr(m \vert G[\rho]) & = \Tr( E_m G[\rho] ) ~, \\
    \Pr(m \vert \overline{G}[\rho]) & = \Tr( E_m \overline{G}[\rho] ) ~.
\end{align}
This can usually be condensed into ``Choose an input state $\rho$ and compute a known quantum state metric between $G[\rho]$ and $\overline{G}[\rho]$.''

As a result, metrics for quantum processes mirror metrics for quantum states and classical distributions. The most commonly used ones are direct generalizations of TVD and classical fidelity, and inherit their properties. However, quantum processes are a richer set than states (or distributions), and display some novel behaviors. So do their metrics. In particular, \emph{more} metrics are necessary, because there is more than one sensible way to choose a fiducial state.

%%%%%%%%%%%%%%%%%%%%%%% Diamond Distance %%%%%%%%%%%%%%%%%%%%%%% 
\subsubsection{Diamond Distance}\label{sec:ddist}

Given two processes $G$ and $\overline{G}$, a simple natural question is ``How much error would be induced by substituting $G$ for $\overline{G}$?'' There is no unique answer, because ``error'' is not precisely defined in this context. A more \emph{precise} formulation is ``If an \emph{unknown} process is used just once, what is the maximum probability of guessing whether it was actually $G$ or $\overline{G}$, given equal prior probability?'' Another reasonable formulation is ``If we accidentally used $G$ in place of $\overline{G}$ in \emph{a single spot} in a quantum information processing protocol repeated $N$ times, what is the maximum fraction of the $N$ outcomes that would need to be changed to cover up the mistake?''

Both framings lead to the same answer, the \emph{diamond norm distance} (or \emph{diamond distance}) between $G$ and $\overline{G}$ \cite{kitaev1997quantum}. Derived from trace distance and TVD, the diamond distance $d_{\diamond}(G, \overline{G})$ is the maximum trace distance between $G[\rho]$ and $\overline{G}[\rho]$, maximized over all possible input states $\rho$. But remarkably, the maximum value of $\norm{G[\rho] - \overline{G}[\rho]}_{1}$ may not be attained for any \emph{local} state $\rho$ describing just the system on which $G$ or $\overline{G}$ acts. A strictly higher value --- and thus, greater probability of correctly distinguishing $G$ from $\overline{G}$ --- can be achieved by applying the unknown process to a system that is \emph{entangled} with another ``reference'' system that is not affected by the process, but can be measured jointly afterward. This counterintuitive phenomenon, akin to superdense coding \cite{BennettPRL92}, envariance \cite{ZurekPRL03}, and teleportation \cite{BennettPRL93}, is important, because quantum logic gates are often applied to qubits that are entangled with other qubits. Restricting the maximization to local states would yield a metric that does not actually capture the worst case.

The diamond distance is defined as:
\begin{align}
    d_{\diamond}(G,\overline{G}) &\equiv \frac12\norm{G - \overline{G}}_{\diamond} ~, \\
    &\equiv \frac12 \max_{\rho_{AB}} \Norm{ \big((G_A - \overline{G}_A) \otimes \Id_B \big) [\rho_{AB}]}_{1} ~, \\
    &= \max_{\rho_{AB}} d_{\mathrm{tr}} \left( (G_A\otimes\Id_B)[\rho_{AB}], (\overline{G}_A\otimes\Id_B)[\rho_{AB}] \right) ~, \label{eq:diamond_norm}
\end{align}
where $A$ indicates the system on which $G$ and $\overline{G}$ act, $B$ indicates a reference system of the same dimension, and $\Id_B$ is the identity process on the reference system. Defined this way, $d_{\diamond}\in[0,1]$, with $d_{\diamond}=0$ iff $G = \overline{G}$ and $d_{\diamond}=1$ iff they can be distinguished perfectly with a single use. In the literature, diamond distance is sometimes defined without the factor of $\frac12$.

The diamond distance is unitarily invariant, so if $G = \E \overline{G}$ and $\overline{G}$ is unitary, then $d_{\diamond}(G, \overline{G}) = d_{\diamond}(\E, \Id)$. The \emph{diamond norm error} of an error process $\E$ is the diamond distance between it and the identity,
\begin{equation}
    d_\diamond(\E) =  \frac12\Norm{\E - \Id}_\diamond = \frac12\max_{\rho_{AB}} \Norm{ \big((\E_A - \Id_A) \otimes \Id_B \big) [\rho_{AB}]}_{1} ~.
\end{equation}
When the ``diamond norm error'' of a gate is mentioned in the literature, it generally means the diamond norm error of the gate's error process (which, as noted here, is equal to the diamond distance between the gate and its target). In general, the diamond norm error of a gate is not trivial to measure, but it can be bounded by the average gate infidelity or entanglement/process infidelity of a gate (see Sec.~\ref{sec:inf_dn}), which we introduce in \ref{sec:fidelities}.

The best-known operational interpretation of $d_{\diamond}$ is the first one given above --- it is an achievable upper bound on the probability of distinguishing $G$ from $\overline{G}$ in a single-shot experiment. But the most \emph{important} role of the diamond norm in quantum computing is as an error bound for circuits that use a gate \emph{multiple} times. Aharonov \textit{et al} \cite{aharonov1998quantum} showed that the diamond norm is \emph{subadditive}. This means that if two quantum circuits (Circuit 1 and Circuit 2) are identical \emph{except} that where operations $\overline{G}_1$ and $\overline{G}_2$ appear in Circuit 1, operations $G_1$ and $G_2$ appear instead in Circuit 2, then the diamond norm distance between the processes implemented by Circuit 1 and Circuit 2 is less than or equal to $d_\diamond(G_1, \overline{G}_1) + d_\diamond(G_2, \overline{G}_2)$.

This property, not shared by any other commonly used error metrics, makes diamond distance uniquely useful. It is often a very pessimistic upper bound on the error probability of \emph{specific} circuits, because in many circuits (i) the initial state and final measurement are not chosen to maximize the observed TVD, and (ii) gates are arranged so that the errors in their implementation either cancel each other out (e.g., via dynamical decoupling \cite{viola1998dynamical, viola1999dynamical}) or add up non-constructively (e.g., via Pauli frame randomization \cite{knill2004fault, kern2005quantum, ware2021experimental} or randomized compiling \cite{wallman2016noise, hashim2021randomized}). But diamond distance provides a \emph{guaranteed} upper bound on the accumulation of error in any quantum circuit --- which can be saturated in some circumstances (e.g., error-amplifying circuits \cite{nielsen2021gate}) --- because the TVD between a circuit's ideal and experimental output distributions is bounded above by the sum of every operation's diamond norm error \cite{kitaev1997quantum, aharonov1998quantum}. So, it is sometimes used, for example, in rigorous proofs of fault tolerance \cite{aharonov1998quantum, aliferis2006quantum}.

%%%%%%%%%%%%%%%%%%%%%%% Jamiołkowski Trace Distance %%%%%%%%%%%%%%%%%%%%%%% 
\subsubsection{Jamiołkowski Trace Distance}\label{sec:ndist}

Another metric which is closely related to the diamond distance is the \emph{Jamiołkowski trace distance}. It is obtained by replacing the maximization over input states in Eq.~\ref{eq:diamond_norm} with a maximally entangled state between the system of interest and a reference of the same size:
\begin{equation}\label{eq:Jtr}
    d_{J-tr}(G, \overline{G}) \equiv \frac12 \norm{ \big(G_A\otimes\Id_B - \overline{G}_A\otimes\Id_B\big)\big[\proj{\Psi_{AB}}\big] }_{1} ~.
\end{equation}
The Jamiołkowski trace distance provides a closed-form \emph{lower} bound for $d_{\diamond}$. It is equal to the trace distance between the $\chi$ matrices of $G$ and $\overline{G}$, and proportional to the trace distance between their Choi matrices:
\begin{equation}
    d_{J-tr}(G,\overline{G}) = \frac12 \norm{ \chi_{G} - \chi_{\overline{G}} }_{1} = \frac{1}{2d} \norm{ \mathcal{C}_{G} - \mathcal{C}_{\overline{G}} }_{1}~.
\end{equation}
This relationship can be derived from \eq\ref{eq:Jtr} using the relationship between the Choi and $\chi$ representations (\eq\ref{eq:ChoiChi}) and the definition of the Choi representation (\eq\ref{eq:choi}).

% %%%%%%%%%%%%%%%%%%%%%%% Fidelities %%%%%%%%%%%%%%%%%%%%%%% 
\subsubsection{Fidelities}\label{sec:fidelities}

The most commonly encountered performance metrics for quantum gates are \emph{fidelities}. In fact, the word ``fidelity'' now transcends its technical context (like ``Xerox machine'' or ``Kleenex''), and appears in paper titles and abstracts as a generic synonym for ``quality.''  Despite this usage, it is still a precise technical term in quantum information science and quantum computing, and we urge readers to avoid unfortunate usage like ``We quantify gate fidelity using diamond norm distance.''

At least three distinct fidelity metrics appear, and are used, in the literature. The difference between them is in the initial state to which $G$ or $\overline{G}$ is applied. But every ``fidelity'' quantifies \emph{similarity}, is derived from quantum state fidelity, and inherits its properties in exactly the same way that diamond distance inherits the properties of trace distance. For every fidelity $F$, there is a corresponding \emph{infidelity} $r = 1 - F$ that quantifies discrepancy and can be used as a metric of error.\\

%%%%%%%%%%%%%%%%%%%%%%% Average Gate Fidelity %%%%%%%%%%%%%%%%%%%%%%% 
\paragraph{Average Gate Fidelity}\label{sec:ave_gate_fid}
\hfill \break

There is a simple reason for the existence of multiple definitions of fidelity for quantum processes: quantum processes can only be ``observed'' by applying them to a state. The fidelity, distinguishability, or erroneousness of a process therefore depends on \emph{context} --- i.e., on what state it acts. So, the key ingredient in any definition of fidelity for quantum processes is the \emph{output-state fidelity} for a given input state, defined in terms of the state fidelity $F(\rho,\overline{\rho})$ (\eq\ref{eq:MixedStateFidelity}) as
\begin{equation}
    F_{\rho}( G,\overline{G} ) \equiv F\left( G[\rho], \overline{G}[\rho] \right) ~.
\end{equation}
To see why $\rho$ matters, consider a flawed idle gate $G_{\Id}$ that is supposed to leave states unchanged, but actually dephases them in the $Z$ basis. If applied to a $Z$ eigenstate ($\ket{0}$, $\ket{1}$, or any mixture of them), it acts exactly like its target, so $F_{\proj{0}}(G_{\Id},\overline{G}_\Id)=1$. But if applied to an eigenstate of $X$ or $Y$, it decoheres them completely, so $F_{\proj{+}}(G_{\Id},\overline{G}_\Id)=1/2$.

The \emph{\ac{AGF}} eliminates this variation by the simple expedient of averaging $F_\rho$ over all pure states using the unique (normalized) unitarily invariant Haar measure (see Appendix~\ref{sec:haar}):
\begin{equation}\label{eq:ave_gate_fidelity}
    F_{\mathrm{avg}}(G,\overline{G}) \equiv \int{F \left( G[\proj{\psi}], \overline{G}[\proj{\psi}] \right)\mathrm{d}\psi} ~.
\end{equation}
This definition applies for any $G$ and $\overline{G}$, but $\overline{G}$ is usually unitary. If $\overline{G}[\rho] = \mathcal{U}[\rho] = U\rho U^{\dagger}$ for some unitary operator $U$, then $\overline{G}[\proj{\psi}]$ is pure, and 
\begin{align}
    F_{\mathrm{avg}}(G,\mathcal{U}) 
        &= \int{ \left( \bra{\psi} U^\dagger G[\proj{\psi}] U \ket{\psi} \right) \mathrm{d}\psi } ~, \\
        &= \int{ \left( \sbra{\proj{\psi}}\left(\mathcal{U}^{-1}G\right)\sket{\proj{\psi}} \right) \mathrm{d}\psi } ~.
\end{align}
This form of the AGF makes it clear that $F_{\mathrm{avg}}(G,\mathcal{U})$ quantifies how well the noisy process $G$ implements the desired unitary operation $U$. 

From the AGF, we can define the \emph{\ac{AGI}} $r(G, \overline{G})$:
\begin{equation}\label{eq:ave_gate_infidelity}
    r(G, \overline{G}) = 1 - F_{\mathrm{avg}}(G, \overline{G}) ~.
\end{equation}
It should be noted that while $r(G,\overline{G})$ is also commonly referred to as the \emph{average error rate} of a gate, some draw a distinction between the average error rate and average gate infidelity \cite{sanders2015bounding}.

Many QCVV benchmarking procedures are constructed such that $U = \mathbb{I}$. In this case, the average gate fidelity is
\begin{equation}\label{eq:ave_gate_fidelity_I}
    F_{\mathrm{avg}}(G) = \int \bra{\psi} G[\ketbra{\psi}] \ket{\psi} d\psi ~.
\end{equation}
Here, $F_{\mathrm{avg}}$ defines the probability that $G$ produces no detectable change in a random pure state $\rho = \ketbra{\psi}$. Note that this is not the same as ``the probability that $G$ leaves $\rho = \ketbra{\psi}$ unchanged,'' since if $G$ deterministically rotates $\ket{\psi} \mapsto \ket{\phi} \neq \ket{\psi}$, the probability of \emph{detecting} the change is only $1 - |\braket{\psi | \phi}|^2$.

The integral in Eqs.~\ref{eq:ave_gate_fidelity}--\ref{eq:ave_gate_fidelity_I} can be computed explicitly \cite{nielsen2002simple, emerson2005scalable, magesan2011gate} to yield a simple relationship between AGF and the (arguably more fundamental) entanglement fidelity \cite{horodecki1999general, nielsen2002simple} discussed below,
\begin{equation}\label{eq:FidelityRelationship}
    F_{\mathrm{avg}}(G,\mathcal{U}) = \frac{dF_e( G,\mathcal{U} ) + 1}{d+1} ~,
\end{equation}
where $d$ is the dimension of the system's Hilbert space. AGF and AGI are particularly relevant and useful in randomized benchmarking (Sec.~\ref{sec:randomized_benchmarks}) and direct fidelity estimation (Sec.~\ref{sec:direct_fid_est}), because these protocols apply a process or processes to a system initialized in randomly distributed pure --- or nearly-pure --- \emph{local} (unentangled) states.\\

%%%%%%%%%%%%%%%%%%%%%%% Entanglement Fidelity %%%%%%%%%%%%%%%%%%%%%%% 
\paragraph{Entanglement (Process) Fidelity}\label{sec:ent_fid}
\hfill \break

There is another way to eliminate the state-dependence of output-state fidelity. If we apply the unknown operation ($G$ or $\overline{G}$) to a system that is maximally entangled with a reference system, so that their joint state is a maximally entangled state $\ket{\Psi}$, then the fidelity between the resulting states,
\begin{equation}\label{eq:EntanglementFidelity}
    F_e(G,\overline{G}) \equiv F\Big( (G\otimes\Id)[\proj{\Psi}], (\overline{G}\otimes\Id)[\proj{\Psi}] \Big) ~,
\end{equation}
does not depend on \emph{which} maximally entangled state was used. This quantity is known as the \emph{entanglement fidelity} between $G$ and $\overline{G}$ \cite{schumacher1996sending, nielsen1996entanglement}. The Choi-Jamiołkowski isomorphism implies that the two states on the right hand side of \eq\ref{eq:EntanglementFidelity} are unitarily equivalent to the $\chi$ matrices $\chi_G$ and $\chi_{\overline{G}}$ (and to the trace-normalized Choi matrices $\mathcal{C}_G$ and $\mathcal{C}_{\overline{G}}$), respectively. Therefore, if $\overline{G} = \mathcal{U}$ is unitary, so that $\chi_\mathcal{U}$ is rank-1, then
\begin{equation}\label{eq:ent_fidelity_choi}
    F_e(G, \mathcal{U}) = \Tr(\chi_G \chi_\mathcal{U}) = \Tr\left( \chi_{\mathcal{U}^{-1}G} \chi_{\Id} \right) ~.
\end{equation}
So, the fidelity between $G$ and a unitary target process $\overline{G} = \mathcal{U}$ equals the fidelity between the \emph{error process} $\E = G \overline{G}^{\, -1}$ and the identity process. As discussed in the context of average gate fidelity, we often want to measure the fidelity of $G$ with the identity operation. In this case, the entanglement fidelity is sometimes written as
\begin{equation} \label{eq:SingleArgumentEntanglementFidelity}
    F_e(G) = \bra{\Psi} (G \otimes \Id)[\proj{\Psi}] \ket{\Psi} = \Tr( \chi_{G}\chi_{\Id} )~.
\end{equation}

It is sometimes \cite{nielsen2002simple} said that $F_e(G)$ quantifies how well $G$ preserves entanglement, but this is not strictly correct. $\E$ can be entanglement-breaking, yet still have nonzero entanglement fidelity. Conversely, if $G$ is a Pauli unitary, then $F_e = 0$ even though $G$ does not destroy entanglement. $F_e$ is more accurately described as the fidelity of a process \emph{when acting on (maximally) entangled states}. For this reason, entanglement \emph{infidelity} ($1 - F_e$) is usually the most appropriate metric of error for quantum computing, where a gate will often act on qubits that are entangled with other qubits. Conversion between $F_e$ and $F_{\mathrm{avg}}$ is very easy using \eq\ref{eq:FidelityRelationship} (see also Tab.~\ref{tab:table_rel_fid_dep}). Entanglement fidelity is lower (more pessimistic) than average gate fidelity, because entangled states are generically more sensitive to error than random local states.

$F_{e}$ is also commonly referred to as \emph{process fidelity}. However, this usage is not entirely reliable --- sometimes ``process fidelity'' is used to refer to other fidelity-type metrics (e.g., average gate fidelity), or as a catch-all for \emph{any} fidelity-like metric between quantum operations. Throughout this tutorial, we only use ``process fidelity'' to denote $F_{e}$, but generally use (and recommend) the term ``entanglement fidelity'' to minimize ambiguity.

Process/entanglement fidelity can also be computed in the Pauli transfer matrix representation (see Sec.~\ref{sec:ptm_rep}) \emph{if} one of the two arguments is unitary:
\begin{equation}\label{eq:process_fidelity}
    F_e(G, \mathcal{U}) = \frac{1}{d^2}\Tr[\Lambda_{G\mathcal{U}^{-1}}] = \frac{1}{d^2}\Tr[\Lambda_G \Lambda_\mathcal{U}^{-1}] ~.
\end{equation}
The derivation is simple, starting from \eq\ref{eq:ent_fidelity_choi}:
\begin{align}
    F_e(G, \mathcal{U}) &= \Tr(\chi_G \chi_\mathcal{U}) ~ \\
        &= \frac{1}{d^2} \sum_{i,j,k,l}{ \Tr( G[ \ketbra{i}{j} ] \mathcal{U}[ \ketbra{k}{l} ] \otimes \ \ketbra{i}{j} \ketbra{k}{l} ) } ~, \\
        &= \frac{1}{d^2} \sum_{i,j}{ \Tr( G[ \ketbra{i}{j} ]\mathcal{U}[ \ketbra{j}{i} ] ) } ~, \\
        &= \frac{1}{d^2} \sum_{i,j}{ \Tr( G[ \ketbra{i}{j} ](\mathcal{U}^\dagger[ \ketbra{i}{j} ])^\dagger ) } ~, \\
        &= \frac{1}{d^2} \Tr(\Lambda_{G\mathcal{U}^{-1}}) = \frac{1}{d^2} \Tr(\Lambda_{G}\Lambda_{U^{-1}}) ~.
\end{align}
Equivalently, if we write $\Lambda_G = \Lambda_\E\Lambda_\mathcal{U}$, so that $\E$ is the post-gate error process of the noisy operation $G$, then 
\begin{equation}\label{eq:process_fidelity_E}
    F_e(G,\mathcal{U}) = F_e(\E,\Id) = \frac{1}{d^2}\Tr[\Lambda_\E] ~.
\end{equation}
For the remainder of this tutorial, any reference to process fidelity is a reference to Eqs.~\ref{eq:process_fidelity} or \ref{eq:process_fidelity_E}.

%The process fidelity can also be given in terms of the $\chi$-matrix representation of a process. 
We often write $\chi$ matrices in the Pauli basis. In this basis, the $\chi$ matrix for the identity process has only one nonzero element, which is $\chi_{\Id,\Id} = 1$. It is common to enumerate the Pauli basis elements from $0 \ldots d^2-1$, starting with $\Id$, in which case this is written as $\chi_{0,0} = 1$. The process fidelity between an error process $\E$ and the identity is then given by 
\begin{equation}
    F_e(\E, \Id) = (\chi_\E)_{\Id, \Id} = (\chi_\E)_{0,0} ~.
\end{equation}

The process \emph{infidelity} of a gate is simply
\begin{equation}\label{eq:process_infidelity}
    e_F = 1 - F_e ~.
\end{equation}
It is related to average gate infidelity by a simple dimension-dependent proportionality factor \cite{horodecki1999general, nielsen2002simple} (see Tab.~\ref{tab:table_rel_fid_dep}):
\begin{equation}\label{eq:proc_ave_infidelity}
    e_F = \frac{d+1}{d} r ~.
\end{equation}
If the error process $\E$ has an orthogonal Kraus decomposition in which the first Kraus operator is the identity ($K_0 \propto \Id$), then we call the error process \emph{stochastic}, because we can model it as a probabilistic mixture of (i) no error ($K_0$) occurs, or (ii) an error occurs (see, e.g., Secs.~\ref{sec:dephasing} --  \ref{sec:stoch_pauli}). For stochastic error processes, the process infidelity is precisely the probability that an error occurs. There is then a simple intuition for the difference between process infidelity and AGI: \emph{every} error is detectable if it occurs on a maximally entangled state, but if the error occurs on a random pure state, it may go undetected (e.g., if the state is an eigenstate of the error).

As a result of this, process fidelity behaves well under composition (i.e., when two gates are combined by tensor product to describe a layer of parallel gates). From \eq\ref{eq:SingleArgumentEntanglementFidelity}, and the fact that a product of two maximally entangled states is maximally entangled, it follows that
\begin{equation}
    F_e(\E_1\otimes \E_2) = F_e(\E_1)F_e(\E_2) ~,
\end{equation}
and therefore that
\begin{equation}
    e_F(\E_1\otimes \E_2) = e_F(\E_1) + e_F(\E_2) + \mathcal{O}(\epsilon^2)
\end{equation}
if both $e_F(\E_1)$ and $e_F(\E_2)$ are $\mathcal{O}(\epsilon)$. These relationships do not hold for the AGI, because of the dimension-dependent factor. Again, this has a useful intuitive explanation: combining subsystems by tensor product increases the overall system dimension, which reduces the probability that an error will go undetected if it occurs on a random pure state.\\

%%%%%%%%%%%%%%%%%%%%%%% Worst-Case (min) Fidelity %%%%%%%%%%%%%%%%%%%%%%% 
\paragraph{Worst-Case (min) Fidelity}\label{sec:min_fid}
\hfill \break

One final fidelity for processes, rarely used but deserving mention, is the \emph{stabilized minimum fidelity} \cite{gilchrist2005distance}:
\begin{equation}
    F_{\mathrm{stab}}(G, \overline{G}) \equiv \min_{\ket{\psi}_{AB}}{ F\Big( (G_A\otimes\Id_B)[\proj{\psi}] ,(\overline{G}_A\otimes\Id_B)[\proj{\psi}] \Big) } ~.
\end{equation}
This is a fidelity-based metric (rather than a TVD-based one), but is extremized (like diamond distance) rather than averaged over input states. As with diamond distance, the minimum could also be taken over local states. But Gilchrist \textit{et al.}~observe that the resulting metric is not \emph{stable} with respect to adding unrelated ancillary systems \cite{gilchrist2005distance}, and recommend $F_{\mathrm{stab}}$ instead.

Worst-case fidelity is not commonly used in QCVV. There is, to the best of our knowledge, no \emph{good} reason for this. In many contexts, it may be better-motivated than AGF or entanglement/process fidelity. However, it suffers from sociological factors. It requires numerical computation without a nice analytic form (making it less appealing to theorists) and is strictly lower than any other fidelity metric (making it less appealing to experimentalists).\\

%%%%%%%%%%%%%%%%%%%%%%% Process Polarization %%%%%%%%%%%%%%%%%%%%%%% 
\subsubsection{Process Polarization}\label{sec:polarization}

\begin{table*}[t]
    \renewcommand{\arraystretch}{1.8}
    \centering
    \resizebox{0.65\textwidth}{!}{
        \begin{tabular}{r || c c c c c}
            & $F_\mathrm{avg}$ & $r$ & $F_e$ & $e_F$ & $f$ \\
            \hline
            \hline
            \hyperref[eq:ave_gate_fidelity]{$F_\mathrm{avg} =$} & $F_\mathrm{avg}$ & $1 - r$ & $\frac{dF_e + 1}{d + 1}$ & $1 - \frac{d}{d+1} e_F$ & $\frac{(d-1)f + 1}{d}$ \\
            \hyperref[eq:ave_gate_infidelity]{$r =$} & $1 - F_\mathrm{avg}$ & $r$ & $\frac{d}{d+1} (1 - F_e)$ & $\frac{d}{d+1} e_F$ & $\frac{d - 1}{d} (1 - f)$ \\
            \hyperref[eq:process_fidelity]{$F_e =$} & $\frac{(d+1)F_\mathrm{avg} - 1}{d}$ & $1 - \frac{d+1}{d} r$ & $F_e$ & $1 - e_F$ & $\frac{(d^2 - 1)f + 1}{d^2}$ \\
            \hyperref[eq:process_infidelity]{$e_F =$} & $\frac{d+1}{d}(1 - F_\mathrm{avg})$ & $\frac{d+1}{d} r$ & $1 - F_e$ & $e_F$ & $\frac{d^2 - 1}{d^2} (1 - f)$ \\
            \hyperref[eq:process_polarization]{$f =$} & $\frac{dF_\mathrm{avg} - 1}{d - 1}$ & $1 - \frac{d}{d-1} r$ & $\frac{d^2F_e - 1}{d^2 - 1}$ & $1 - \frac{d^2}{d^2 - 1} e_F$ & $f$ \\
        \end{tabular}
    }
    \caption{\textbf{Linear Relations between Performance Metrics.}
    Summary of the linear relationship between the average gate fidelity $F_\mathrm{avg}$, the average gate infidelity $r$, the entanglement (process) fidelity $F_e$, the entanglement (process) infidelity $e_F$, and the process polarization $f$, where $d = 2^n$ for $n$ qubits. (Table adapted from Ref.~\cite{carignan2019walk}.)}
\label{tab:table_rel_fid_dep}
\end{table*}

Equation \ref{eq:proc_ave_infidelity} shows that the process infidelity and the AGI of an error process are identical up to a constant factor. A third re-scaling of this quantity has a particularly intuitive use. This is the \emph{effective depolarizing parameter} or \emph{process polarization} of an error channel.

Many benchmarking procedures use gates in a specific way that ``twirls'' their error processes (see Sec.~\ref{sec:rb_math_twirling} and Appendix \ref{sec:twirling}), effectively replacing each noisy gate 
$G = \E \overline{G}$ with $G' = \E_{\mathrm{twirled}} \overline{G}$, where
\begin{equation}
    \E_{\mathrm{twirled}} = \int{u\E u^{-1}\mathrm{d}\mu(u)} ~.
\end{equation}
In this expression, $u$ applies a unitary transformation, and $\mathrm{d}\mu(u)$ is the normalized Haar measure over \emph{all} unitaries acting on the gate's target Hilbert space. The effect of twirling is to symmetrize and simplify the error process drastically. It replaces $\E$ with a \emph{partial depolarizing channel} $\E_{\mathrm{twirled}}$ of the form
\begin{equation}\label{eq:global_dep_noise}
    \E_{\mathrm{twirled}} = f\Id + (1-f)\mathcal{D} ~,
\end{equation}
where $\mathcal{D}$ is the depolarizing process that acts as $\mathcal{D}[\rho] = \Tr[\rho]\Id/d$, and $1-f$ is the probability of depolarization. So,
\begin{equation}
    \E_{\mathrm{twirled}}[\rho] = f \rho + (1-f)\frac{\Id}{d}
\end{equation}
for any normalized density matrix $\rho$. We call $f$ the \emph{process polarization} of $\E$, because it quantifies the amount of polarization in $\rho$ that remains after applying $G$ in a context that twirls its error process. 

The process polarization $f$ is closely related to process fidelity. Analysis of twirling (see Appendix \ref{sec:twirling}) shows that $\E_{\mathrm{twirled}}$ and $\E$ have exactly the same $\chi_{00}$, and thus the same process fidelity. It is straightforward to compute that $\chi_{00}=1$ for the identity process $\Id$, and $\chi_{00}=1/d^2$ for the depolarizing process $\mathcal{D}$. It follows that $F_e(\E_{\mathrm{twirled}}) = f + (1 - f)/d^2$, and since $F_e(\E_{\mathrm{twirled}}) = F_e(\E)$,
\begin{equation}\label{eq:process_polarization}
    f(\E) = 1 - p = \frac{d^2F_e(\E) - 1}{d^2-1} ~,
\end{equation}
where $p$ is the probability of depolarization. Process polarization can also be computed straightforwardly from the error channel's Pauli transfer matrix as
\begin{equation}\label{eq:depolarizing_parameter}
    f(\E) = \frac{\Tr[\Lambda_\E] - 1}{d^2 - 1} ~.
\end{equation}
These relationships (and others) are summarized in Tab.~\ref{tab:table_rel_fid_dep}.

A useful property of process polarization is that the polarizations of two twirled gates applied in sequence combine by simple multiplication.  If $G_1 = \E_1 \overline{G}_1$ and $G_2 = \E_2 \overline{G}_2$, and both gates are performed in a context that twirls them, then
\begin{equation}
    G'_2 G'_1 = \E_{2,\mathrm{twirled}}\E_{1,\mathrm{twirled}}\overline{G}_2\overline{G}_1
\end{equation}
and
\begin{equation}
    f( \E_{2,\mathrm{twirled}}\E_{1,\mathrm{twirled}} ) = f(\E_{2,\mathrm{twirled}})f(\E_{1,\mathrm{twirled}}) ~.
\end{equation}
Note that this simplification does not apply for \emph{parallel} composition, because twirling is system-dependent, and the tensor product of two (locally) twirled error channels is not (globally) twirled.

%%%%%%%%%%%%%%%%%%%%%%% Contrasting Diamond Distance and Infidelity %%%%%%%%%%%%%%%%%%%%%%% 
\subsubsection{Contrasting Diamond Distance and Infidelity}\label{sec:inf_dn}

As we have seen in this section, there are many different ways to quantify the ``error rate'' of a quantum process. Both the AGI $r$ (\eq\ref{eq:ave_gate_infidelity}) and entanglement/process infidelity $e_F$ (\eq\ref{eq:process_infidelity}) have the convenient interpretation of being \emph{average} error rates (i.e., the rate at which an error would be observed, averaged in some way over possible input states). On the other hand, TVD-derived error metrics such as the diamond distance $d_{\diamond}$ (\eq\ref{eq:diamond_norm}) are \emph{maximizations} over all possible POVMs and/or input states, and thus are sometimes called \emph{worst-case} error rates. 

While average error rates can be efficiently measured by Monte Carlo sampling (see, e.g., Sec.~\ref{sec:randomized_benchmarks}), estimating extremal quantities like diamond distance is harder. For example, although tomographic reconstruction methods (see Sec.~\ref{sec:tomography}) can be used to estimate the diamond distance \cite{Blume_Kohout_2017} by means of semi-definite programs \cite{watrous2009semidefinite, watrous2012simpler}, the cost of tomography grows exponentially with the number of qubits. However, an error channel's AGI or process infidelity provides a \emph{bound} on its diamond norm error \cite{wallman2014randomized, sanders2015bounding, wallman2015bounding}:
\begin{align}
    \frac{d+1}{d} r &\leq d_\diamond \leq \sqrt{d(d+1)} \sqrt{r} ~, \label{eq:dn_bounds_r} \\
    e_F &\leq d_\diamond \leq d \sqrt{e_F} ~, \label{eq:dn_bounds_ef}
\end{align}
where $d$ is the dimension of the Hilbert space. Tighter bounds can be obtained if one has knowledge of the \emph{unitarity} of a gate \cite{wallman2015bounding}, which we introduce in Sec.~\ref{sec:pb}.

To illustrate the types of errors that saturate the bounds of the diamond norm, we consider two types of single-qubit errors: (1) a coherent (unitary) error, and (2) a stochastic error. A single-qubit coherent $X$ error corresponds to a unitary rotation by some angle $\theta$,
\begin{equation}
   R_x(\theta) = \left( \begin{array}{cc} \cos(\theta / 2) & - i \sin(\theta / 2) \\  
                                i \sin(\theta / 2) & \cos(\theta / 2) \end{array} \right) ~.
\end{equation}
The PTM superoperator of this error is given as 
\begin{equation}
   \Lambda_\E = \left(
        \begin{array}{cccc}
            1 & 0 & 0 & 0 \\
            0 & 1 & 0 & 0 \\
            0 & 0 & \cos(\theta) & - \sin(\theta) \\
            0 & 0 & \sin(\theta) & \cos(\theta) \\
        \end{array}
        \right) ~.
\end{equation}
Considering the difference between the identity operation and the error,
\begin{equation}
   \mathbb{I} - \Lambda_\E = \left(
        \begin{array}{cccc}
            0 & 0 & 0 & 0 \\
            0 & 0 & 0 & 0 \\
            0 & 0 & 1 - \cos(\theta) & - \sin(\theta) \\
            0 & 0 & \sin(\theta) & 1 - \cos(\theta) \\
        \end{array}
        \right) ~,
\end{equation}
we observe that, for small $\theta$, the magnitude of the diagonal elements scales as $\abs{1 - \cos(\theta)} \approx \tfrac{1}{2}\theta^2$, and the magnitude of the off-diagonal elements scales as $\abs{\sin(\theta)} \approx \theta$. Because the diamond norm is the maximization over all possible input states and, via the trace distance, also a maximization over all POVMs, it is sensitive to the largest elements of $\mathbb{I} - \Lambda_\E$. In the case of the coherent error given above, the largest elements of $\mathbb{I} - \Lambda_\E$ are the off-diagonal elements, which scale as $\mathcal{O}(\theta)$, and thus $d_\diamond \sim \theta$. Now, consider \emph{twirling} $\Lambda_\E$ into a stochastic Pauli channel (Sec.~\ref{sec:stoch_pauli}) or depolarizing channel (Sec.~\ref{sec:dep_noise}) via Pauli or Clifford twirling, respectively (see Sec.~\ref{sec:rb_math_twirling} and Appendix \ref{sec:twirling}). In both cases, the diamond norm of $(\Lambda_\E)_\textrm{twirled}$ scales as its process infidelity, $d_\diamond \sim e_F \approx \theta^2$. In this example, we observe that the diamond norm is \emph{at least} $d_\diamond \sim e_F \approx \theta^2$ (when $\Lambda_\E$ represents a stochastic error channel), and \emph{at most} $d_\diamond \sim \sqrt{e_F} \approx \theta $ (when $\Lambda_\E$ represents a unitary channel). Thus, it is often said that the lower bound of the diamond norm is saturated by a purely stochastic noise channel, and the upper bound of the diamond norm is saturated by a purely unitary error channel \cite{wallman2015bounding, kueng2016comparing, hashim2023benchmarking}. This example is only meant to be a \emph{heuristic} --- it is not a rigorous derivation of the bounds of the diamond norm --- but it does highlight where the quadratic difference between the lower and upper bounds of the diamond norm come from, and how the diamond norm can differ by orders of magnitude in the presence of stochastic noise versus coherent errors.

%%%%%%%%%%%%%%%%%%%%%%% Quantum Measurements %%%%%%%%%%%%%%%%%%%%%%% 
\subsection{Quantum Measurements}\label{sec:quantum_meas}

Quantum measurements, or \emph{readout operations}, are the third essential logic operation in a quantum processor. Two distinct kinds of measurement operation appear in quantum circuits and quantum computing experiments, \emph{terminating measurements} that mark the end of a circuit (after which the entire processor may be re-initialized, cooled, and/or recalibrated before another circuit is run), and \emph{mid-circuit measurements} (\ac{MCM}s). Mid-circuit measurements are harder to implement, because they must (1) not disrupt other qubits that are not being measured, and (2) leave the measured qubit(s) in a usable state.

Terminating measurements have been studied and analyzed much more thoroughly than mid-circuit measurements in the QCVV community. However, the analysis of error metrics for both kinds of measurement is surprisingly rare in the literature. The metrics most commonly used (readout fidelity and \ac{QND}-ness) are relatively ad-hoc in comparison to the systematic framework that has been developed for states and processes. We outline the most commonly-used metrics below.

%%%%%%%%%%%%%%%%%%%%%%% Terminating Measurements %%%%%%%%%%%%%%%%%%%%%%% 
\subsubsection{Terminating Measurements} \label{sec:term_meas}

Terminating measurements are modeled by POVMs (see Sec.~\ref{sec:povm}), and have been an important subject of QCVV since at least 1999 \cite{Luis1999-ek}. However, the literature on error metrics for POVMs is sparse \cite{Ji2006-fp, Magesan2013-vy, Dressel2014-rb, blumoff2016implementing}.

Quantum computing experimentalists usually seek to perform \emph{orthogonal rank-1 projective} measurements. For each outcome $i$ of the measurement, there is a unique pure state $\ket{i}$ for which $p(i|i)=1$. Some of the most common metrics are specialized for this case. One example is \emph{readout fidelity}, defined as the average probability of observing outcome $i$ given state $\ket{i}$, or
\begin{equation}\label{eq:ro_fidelity}
    F_\mathrm{readout} = \frac{1}{d} \sum_{i=0}^{d-1} p(i|i) = \frac{1}{d} \sum_{i=0}^{d-1} \Tr[ E_i \proj{i} ] ~,
\end{equation}
where $d$ is the number of outcomes (and Hilbert space dimension), and $E_i$ is the POVM effect associated with measurement outcome $i$ (see \eq\ref{eq:borns_rule_povm}). Readout fidelity is used extensively as folklore (without citation) in the superconducting qubit literature \cite{mallet2009single, johnson2012heralded, heinsoo2018rapid, elder2020high}. It can be defined as the average of a more fundamental quantity that we denote \emph{effect-wise fidelity}:
\begin{equation}\label{eq:effectwise_fidelity}
	F_{i} = p(i|i) = \Tr[ E_i \ketbra{i}{i} ] ~.
\end{equation}
The $F_{i}$ are useful when they vary substantially over outcomes $i$, in which case the worst-case fidelity,
\begin{equation}\label{eq:ro_min_fidelity}
	F_{\mathrm{min}} = \min_i{ F_i } ~,
\end{equation}
is relevant. For example, the excited state readout fidelity $F_1$ of many systems is often worse than the ground state readout fidelity $F_0$ due to energy relaxation (e.g., $T_1$ decay; see Sec.~\ref{sec:spont_emis_amp_damp}). However, these effect-wise fidelities can be equalized through methods that twirl measurement noise \cite{beale2023randomized, hashim2023quasi}.

None of these quantities is directly observable, unless the experimenter has the ability to prepare perfect initial states $\proj{i}$. In real-world experiments, it is impossible to unambiguously distinguish errors in state preparation from errors in measurement. So, these errors are often grouped together and referred to as \emph{\ac{SPAM}} errors \cite{magesan2012efficient}. Directly measured quantities generally depend (in more or less complicated ways) on
\begin{equation}\label{eq:SPAM_fidelity}
	F_{\mathrm{SPAM}(i)} = p\mathrm{(i|\rho_i)} = \Tr[ E_i \rho_i ] ~.
\end{equation}

Error metrics for terminating measurements that are \emph{not} supposed to be orthogonal rank-1 projective are almost nonexistent in the literature, to the best of our knowledge. Although general non-projective POVMs are rarely implemented on purpose, non-orthogonal rank-1 measurements (e.g., SIC POVMs \cite{Renes2004-pc}) and non-rank-1 projective measurements (e.g., stabilizers \cite{Chow2014-bc}) are important use cases to which Eqs. \ref{eq:ro_fidelity} and \ref{eq:effectwise_fidelity} do not necessarily apply. General metrics can be obtained by observing that a POVM is a kind of CPTP map (a ``quantum-classical'' or q-c channel \cite{Holevo98}), so every process metric defined previously (entanglement fidelity, diamond distance, etc.) can be computed for POVMs. However, none are in common usage (although \cite{Dressel2014-rb} discusses many possible fidelities). For example, the entanglement fidelity between a POVM $\{E_i\}$ and the ideal POVM $\{\proj{i}\}$ works out to
\begin{equation} \label{eq:POVM_ent_fidelity}
	F_e\left( \{E_i\}, \{\proj{i}\} \right) = \left(\sum_i{\sqrt{F_i}}\right)^2 ~,
\end{equation}
which is not quite the same as the ubiquitous Eq. \ref{eq:ro_fidelity} (although they agree to leading order in $1-F_i$). TVD-based metrics, though well-motivated whenever the intended output distribution is nontrivial, have not seen widespread use.

%%%%%%%%%%%%%%%%%%%%%%% Mid-circuit Measurements %%%%%%%%%%%%%%%%%%%%%%% 
\subsubsection{Mid-circuit Measurements} \label{sec:MCM}

Mid-circuit measurements (\ac{MCM}s) are modeled by quantum instruments (see Sec.~\ref{sec:quantum_instruments}), and are critical for quantum error correction. Their importance to QCVV has grown rapidly in recent years. However, the literature on error metrics for MCMs is essentially limited to Ref.~\cite{McLaren2023-gf}, which focuses on the important special case of \emph{uniform stochastic instruments} and shows that entanglement fidelity and diamond distance (defined by treating the instrument as a CPTP map) are suitable metrics.

The experimental literature on characterization of MCMs (e.g., Ref.~\cite{blumoff2016implementing}) primarily reports readout fidelity as defined in \eq\ref{eq:ro_fidelity}, and another folklore metric called \emph{QND-ness} (see Sec.~\ref{sec:qnd}, and also \cite{pereira2022complete, pereira2023parallel}) which is defined as the average probability of getting the same measurement result twice in a row:
\begin{align}
    Q &= \frac{1}{d} \sum_{i=0}^{d-1}{p(i, i|i)} ~, \\
      &= \frac{1}{d} \sum_{i=0}^{d-1} \frac{\text{Tr}[ \ketbra{i}{i} \mathcal{M}_i( \ketbra{i}{i} )]}{p(i|i)} ~,
\end{align}
% \begin{align}
where the noisy mid-circuit measurement is given by a set of \ac{CP} maps $\{\mathcal{M}_i\}$. Like readout fidelity, QND-ness is practical, but impossible to measure exactly without the ability to prepare perfect input states. As noted in \cite{pereira2022complete}, QND-ness is not a very useful metric for non-rank-1 measurements, because it only measures repeatability and has no sensitivity to whether the mid-circuit measurement disrupts other observables that it should commute with.

The difference between QND-ness and readout fidelity can be illustrated by considering qubit measurements that failed to satisfy the QND requirements as described in Sec.~\ref{sec:qnd}. For instance, measurements that inherently alter a quantum state such as charge detection~\cite{nakamura1999coherent} or resonance fluorescence~\cite{astafiev2010resonance, cottet2021electron} may achieve high fidelity, but leave the system in a state outside the qubit manifold. Alternatively, state errors during measurement can lead to significant non-QND-ness, but have only minimal effect on the readout fidelity. For example, $T_1$ decay processes during dispersive readout of the excited state of superconducting qubit can be observed and corrected for by time-resolved state discrimination (this can be accomplished using weak, continuous measurement, outlined in Sec.~\ref{eq:weak_meas}), but the final state of the system may have decayed to the ground state. In such cases, QND-ness can often be improved by actively re-initializing the state $\ketbra{i}{i}$ corresponding to whatever $i$ was read out, or shortening the measurement time.

%%%%%%%%%%%%%%%%%%%%%%% Quantum Processors (Gate Sets) %%%%%%%%%%%%%%%%%%%%%%% 
\subsection{Quantum Processors (Gate Sets)}\label{sec:quantum_gate_sets}

So far, we have considered each logic operation independently, in isolation. This is consistent with the history of the field. However, it is internally \emph{inconsistent} --- and, more importantly, unrealistic. We needed measurements to define state fidelity, states to define measurement fidelity, and \emph{both} of them to define process fidelity. Every logic operation is only defined (and observable) \emph{relative} to other logic operations. This became widely recognized between 2012 and 2016 \cite{merkel2013selfconsistent, Blume-Kohout2013gst, proctor2017randomized}. This relationality creates a gauge freedom that couples all of a quantum processor's logic operations and requires them to be treated as a \emph{gate set}, rather than a set of independent operations \cite{Nielsen2021gatesettomography} (see Sec.~\ref{sec:gate_sets}).

Many modern QCVV protocols (starting with randomized benchmarking) explicitly mix together properties of all the operations in a gate set, to produce a holistic metric that quantifies the error rate not of any single operation, but of a processor's entire gate set. We outline this below.

%%%%%%%%%%%%%%%%%%%%%%% Average Gate Set (In)Fidelity %%%%%%%%%%%%%%%%%%%%%%% 
\subsubsection{Average Gate Set (In)Fidelity}\label{sec:ave_gate_fids}

Perhaps the most common error metric for gate sets is \emph{\ac{AGSI}} \cite{proctor2017randomized}. This is simply the average, over all gates in a gate set (\emph{not} including state preparation or measurement), of the AGI (\eq\ref{eq:ave_gate_infidelity}).

AGSI was originally believed to correspond accurately to the error rate observed in randomized benchmarking (RB) \cite{magesan2012efficient}, which is broadly agreed to be one useful ``error rate'' for a quantum processor. Gauge freedom turns out to complicate this relationship \cite{proctor2017randomized}, but if AGSI is evaluated in the gauge that minimizes the gate-to-gate variation of the individual gates' error channels, it does in fact correspond well to the RB error rate \cite{wallman2018randomized}.

Since 2018, randomized benchmarking protocols have proliferated (see Sec.~\ref{sec:randomized_benchmarks} for some examples), and they do not all measure the same ``error rate.'' Therefore, it is a good idea to read the defining paper for a particular RB protocol carefully before interpreting its result!  However, most RB error rates are related, at some level, to AGSI.

%%%%%%%%%%%%%%%%%%%%%%% Circuit Output Distributions %%%%%%%%%%%%%%%%%%%%%%% 
\subsubsection{Circuit Output Distributions}\label{sec:circ_output_dists}

The other class of metrics that are commonly used to evaluate the performance of processors and gate sets are actually the \emph{classical} metrics discussed in Section \ref{sec:prob_dists}, applied to the outcome distributions of quantum circuits. In particular, linear cross-entropy \cite{2019GoogleSupremacy} and heavy output probability \cite{cross2019validating} are widely used to quantify how accurately a quantum processor has executed a circuit. Other metrics (e.g., TVD \cite{hashim2021randomized, zhong2020quantum} or Hellinger fidelity \cite{Dasgupta2022-nv}) are also used, but less commonly.
%%%%%%%%%%%%%%%%%%%%%%% Robust QCVV experiment design %%%%%%%%%%%%%%%%%%%%%%%
\section{Design and Implementation of QCVV Experiments}\label{sec:designing}

Quantum computers implement quantum algorithms by preparing quantum states, applying quantum gates, and performing quantum measurements. Characterization and benchmarking experiments each provide insight into the types and rates of the errors that affect these operations, but can be broadly distinguished by what they measure. Characterization experiments are typically designed to fit the parameters of a \emph{statistical model} that attempts to capture some aspect of the data generating process. These models are often \emph{interpretable} --- their parameters have physical meaning --- and so may be used to identify the physical source of an error. Benchmarking experiments, on the other hand, are typically designed to assess performance as captured by some empirical measure of success or accuracy, such as the average success probability of circuits specified by a particular algorithm. These performance metrics do not generally permit reliable attribution of errors to physical sources, but benchmarking protocols usually scale more efficiently to many-qubit processors than detailed characterization protocols, and may be more indicative of application performance. The distinction between characterization and benchmarking protocols is often somewhat blurred in practice. 

A \ac{QCVV} \emph{protocol} can be thought of as a recipe for the design and analysis of a characterization or benchmarking experiment. For the purposes of this tutorial, a protocol takes as input a register of qubits and a set of native quantum operations, and outputs a set of (possibly randomized) quantum circuits. A protocol also specifies a data analysis procedure for fitting a model to the experimental data (in the case of characterization) or extracting a performance metric (in the case of benchmarking). Most protocols leave additional important experiment design parameters unspecified --- e.g., the order in which circuits should be run, the number of shots to be taken per circuit, and the frequency of recalibration. In this Section, we outline several principles that can inform these decisions, and discuss some experimental realities that can constrain them. To ensure clarity and reproducibility, papers that report QCVV results should state clearly the specific choices made when implementing QCVV protocols. We divide our discussion in two parts:
\begin{itemize}
    \item \textit{Principles of QCVV Experiment Design} (Sec.~\ref{sec:principles}). QCVV experiments utilize particular families of quantum circuits to probe the noisy dynamics of quantum hardware. Many of these circuit families share a common structure that helps ensure the protocol is \emph{robust} and \emph{informationally complete}. Running these circuits on real hardware often requires \emph{compilation} and \emph{scheduling} that can impact the performance and results of the protocol. 
    
    \item \textit{Identifying and Mitigating Out-of-Model Effects} (Sec.~\ref{sec:identifyingandmitigatingoutofmodeleffects}). QCVV protocols are typically designed to be accurate across a wide range of experimental conditions. However, a quantum computing system may suffer errors that were not accounted for, such as leakage or non-Markovianity, that may cause biased or nonsensical results. Models can be \emph{extended} to include these error types, or the protocols can be designed and run in a way that \emph{averages}, \emph{mitigates}, or \emph{quantifies} their effects. 
\end{itemize}

%%%%%%%%%%%%%%%%%%%%%%% Principles of QCVV experiment design %%%%%%%%%%%%%%%%%%%%%%%
\subsection{Principles of QCVV Experiment Design}\label{sec:principles}

\begin{figure}[t]
    \begin{equation*}
        {\Qcircuit @C=1.5em @R=1em {
            \lstick{\ket{0}} & \multigate{3}{\mathcal{F}_{\rm prep}} & \qw & 
                \multigate{3}{\hspace{.2cm}\mathcal{G}\hspace{.2cm}} & 
                \ustick{\times N}\qw &\multigate{3}{\mathcal{F}_{\rm meas}} 
                & \meter \\
            \lstick{\ket{0}} & \ghost{\mathcal{F}_{\rm prep}} & \qw &
                \ghost{\hspace{.2cm}\mathcal{G}\hspace{.2cm}} &\qw& \ghost{\mathcal{F}_{\rm meas}} 
                & \meter \\
            \lstick{\ket{0}} & \ghost{\mathcal{F}_{\rm prep}} & \qw &
                \ghost{\hspace{.2cm}\mathcal{G}\hspace{.2cm}} &\qw& \ghost{\mathcal{F}_{\rm meas}} 
                & \meter \\
            \lstick{\ket{0}} & \ghost{\mathcal{F}_{\rm prep}} & \qw &
                \ghost{\hspace{.2cm}\mathcal{G}\hspace{.2cm}} &\qw& \ghost{\mathcal{F}_{\rm meas}} 
                & \meter 
            \gategroup{1}{4}{4}{4}{1.5em}{\{}
            \gategroup{1}{4}{4}{4}{1.5em}{\}}
            }}
    \end{equation*}
    \caption[Typical QCVV ``Sandwich'' Structure.]{\textbf{Typical QCVV Circuit Structure.} Circuits used in QCVV often comprise a short state preparation gate sequence $\mathcal{F}_{\rm{prep}}$, an $N$-fold repeated (possibly randomized) gate sequence $\mathcal{G}$, a measurement preparation circuit $\mathcal{F}_{\rm{meas}}$, and a computational basis measurement.}
    \label{fig:circuit_structure}
\end{figure}
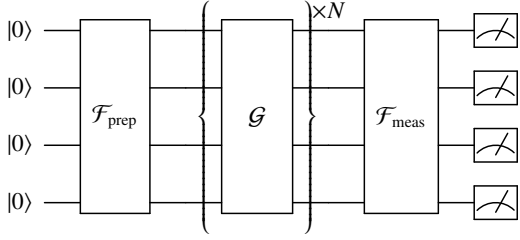

%%%%%%%%%%%%%%%%%%%%%%% Quantum Circuit Families %%%%%%%%%%%%%%%%%%%%%%%
\subsubsection{Quantum Circuit Families}\label{sec:quantumcircuitfamilies}

\begin{table*}[ht]
    \renewcommand{\arraystretch}{1.5}
    \centering
    \begin{tabular}{l|ccc}
        Experiment                                  & \;\;\fprep\;\; & \gateG & \;\;\fmeas\;\; \\
        \hline
        Rabi Oscillations (Sec.~\ref{sec:rabi})     & -- & $X_{\pi/2}$ & --\\
        Ramsey Oscillations (Sec.~\ref{sec:ramsey}) & $X_{\pi/2}$ & $I$ & $X_{\pi/2}$ \\
        Randomized Benchmarking (Sec.~\ref{sec:standardrb}) & -- & $\mathsf{Random}(\mathbb{C}_n)$ & $\mathsf{InverseClifford}$ \\
        Gate Set Tomography (Sec.~\ref{sec:gst})    & \fprep & \gateG & \fmeas  \\
        Quantum Volume (Sec.~\ref{sec:qv})          & -- & $\mathsf{Random}(\mathbb{S}_n) \circ \mathsf{Random}\left[\mathsf{SU}(4)^{\otimes \floor{n/2}} \right]$ & -- \\ 
    \end{tabular}
    \caption{\bf{Examples of ``Sandwich'' QCVV Circuit Structure.} 
    \normalfont Many QCVV circuits possess a sandwich structure, illustrated in \fig\ref{fig:circuit_structure}, wherein a state preparation operation \fprep~is applied first, followed by $N$ applications of a short (possibly random) circuit \gateG, concluding with a measurement preparation circuit \fmeas~and terminating measurement of all qubits. Here, we explicitly list the components of common QCVV experiments. Unless otherwise specified, randomized gates are generally intended to be drawn uniformly at random from the Haar distribution over the associated group (see Appendix \ref{sec:haar}) each time the randomized gate is applied within a circuit. Here, $\mathbb{C}_n$ is the set of $n$-qubit Clifford operations, $\mathbb{S}_n$ is the set of permutation operations on $n$ elements, and $\mathsf{SU}(4)^{\otimes \floor{n/2}}$ is the $\floor{n/2}$-fold Cartesian product of $\mathsf{SU}(4)$, where each factor represents an arbitrary two-qubit operation acting on a pair of qubits. The $\mathsf{InverseClifford}$ operation is the measurement preparation that, conditional on a particular realization of the random Clifford $\{ \mathsf{Random}(\mathbb{C}_n) \}$ elements, acts as the inverse operation, ideally returning any  state to its initial value at the start of the circuit. The gate \gateG~for Rabi and Ramsey experiments are sometimes implemented \emph{discretely}, as indicated in the table above, or via \emph{continuous} driving; see Sec.~\ref{sec:qubit_characterization} for more details.} 
    \label{tab:qcvv_sandwich}
\end{table*}

A QCVV experiment design should enable the effective and efficient study of the specific errors under test while remaining (ideally) agnostic to other sources of error that might be present in the system. The experiment design necessarily includes a family of circuits to run, which is often specified by the QCVV protocol. QCVV circuit families may comprise a finite set of specific circuits, as in state (Sec.~\ref{sec:state_tomography}) or process tomography (Sec.~\ref{sec:qpt}), or an ensemble of random circuits to be sampled according to some prescribed measure, as in randomized benchmarking (Sec.~\ref{sec:randomized_benchmarks}).

The circuits specified by QCVV protocols often share a common \emph{sandwich} structure, as illustrated in \fig\ref{fig:circuit_structure}. Such circuits typically comprise initialization in the all-zeros state, a short state preparation gate sequence $\mathcal{F}_{\rm{prep}}$ (sometimes called a \emph{preparation fiducial circuit}), an $N$-fold repeated (possibly randomized) gate sequence $\mathcal{G}$, a measurement preparation circuit $\mathcal{F}_{\rm{meas}}$ (sometimes called a \emph{measurement fiducial circuit}), and a concluding computational basis measurement. This structure facilitates the estimation of specific errors: the \fprep{} circuit creates a state that is sensitive to some aspect of \gateG{}'s performance, the $N$-fold repetition of \gateG{} amplifies some aspect of its errors, and \fmeas{} implements a measurement that is sensitive to the target errors. Table~\ref{tab:qcvv_sandwich} illustrates how this structure manifests in several common QCVV protocols. 

In order to serve as effective and efficient probes of error, circuits defined by a QCVV protocol should possess a few common properties:
\begin{description}
    \item[Informational completeness] Taken together, data from all specified circuits should be sufficient to enable the estimation of all desired parameters. \emph{Informational completeness} occurs when there is sufficient independent data (circuit outcome statistics) to compute the protocol's performance metric or to reconstruct the target model parameters. If the circuit ensemble is \emph{over}-complete, and the QCVV protocol reconstructs a statistical model for the data, then formal model validation can be used to assess the quality of the fit \cite{nielsen2021gate, Nielsen_2021}. \emph{Under}-complete circuit ensembles generally do not allow for the estimation of all model parameters without some additional regularization, but can be useful, for example, when doing sparse model selection, as in compressed sensing \cite{Gross2010-lb, Riofrio2017-yq}. 
    
    Formally, informational completeness of an experiment design corresponds to the associated Fisher information matrix \cite{Meyer2021-lp, Ostrove2023-xp} being full rank over the vector space of model parameters (see Sec.~\ref{sec:gauge} for a more nuanced perspective). The Fisher information is a powerful tool for analyzing an experiment design and evaluating its ability to probe the parameters of a statistical model. The Cramér-Rao bound states that the absolute precision of any estimator is bounded by the Fisher information, so circuits with a large Fisher information are therefore preferable to those with small Fisher information. Further details about the Fisher information are beyond the scope of this tutorial. The interested reader is encouraged to consult \R\cite{Meyer2021-lp} for an introduction to the fundamentals of Fisher information and its applications in quantum information processing. 
    
    \item[Amplificational completeness] Informational completeness guarantees that an experiment will have \emph{some} sensitivity to every parameter of interest.  Its sensitivity --- i.e., the precision with which parameters can be estimated --- can be increased by increasing the number $N$ of experimental shots (counts).  But in almost all cases, expected estimation error decreases relatively slowly, as $\mathcal{O}(1)/\sqrt{N}$, with the number of shots.  A different (and often better) way to make experiments more sensitive is to grow the \emph{length} ($L$) of the circuits in the experiment, ``amplifying'' certain parameters.  Many QCVV experiments define scalable families of circuits parameterized by a nominal length $L$. If such an experiment's sensitivity to \emph{all} parameters of interest grows uniformly with $L$, so that the expected estimation error decreases as $\mathcal{O}(1)/L$, then we say the experiment is ``amplificationally complete.''  Amplificational completeness is usually evaluated using Fisher information \cite{Ostrove2023-xp}.
    
    \item[Robustness] QCVV protocols are typically designed to probe particular errors or some measure of performance (e.g., fidelity; see Sec.~\ref{sec:overview}). But the errors under test are rarely the only errors present in the system. QCVV protocols are often designed to suppress other phenomena while amplifying the target errors. Dynamical decoupling \cite{viola1998dynamical, viola1999dynamical}, Pauli frame randomization \cite{knill2004fault, kern2005quantum, ware2021experimental}, randomized compiling \cite{wallman2016noise, hashim2021randomized}, and group twirling (see Sec.~\ref{sec:rb_math_twirling} and Appendix \ref{sec:twirling}) are particularly well known examples of such techniques.
    
    \item[Classical simulability] QCVV experiments probe the noise and errors in quantum processors by comparing observed circuit outcomes to those expected in a noiseless system, or those predicted by a noise model. This comparison often requires calculating the expected outcome distribution for a quantum circuit. The classical hardness of this problem is the entire reason we are building quantum computers! QCVV circuits use a number of techniques to preserve classical simulability. For comparing to ideal unitary evolution, these include: restriction to non-universal gate sets (such as Clifford circuits) \cite{Tong2024-el}, structured inversion \cite{proctor2021scalable}, and restriction to small systems. When comparing to statistical noise models, additional constraints on the model are often enforced to preserve classical simulability, such as tensor network ansätze \cite{Torlai2023-ca} or restriction to low-weight errors \cite{Evans2019-kp, Huang2021-bn, huang2020predicting}. Circuit primitives, such as group twirling, can further enhance the performance of simplified error models and improve simulability. 
\end{description}

%%%%%%%%%%%%%%%%%%%%%%% Implementation Details %%%%%%%%%%%%%%%%%%%%%%%
\subsubsection{Implementation Details}

The precision with which an experiment can measure a parameter is controlled by the experiment design (the quantum circuits to be run), the amount of data (the number of ``shots'' per circuit), and the data analysis procedure (the estimator). 

Among the more important experimental constraints is the time it takes to run a full QCVV experiment. Model-based characterization of multi-qubit systems can be among the most experimentally taxing applications of near-term quantum computers. A single run of two-qubit gate set tomography (see Sec.~\ref{sec:gst}), for instance, can take several hours on atomic-based quantum computers. During this time, the environmental degrees of freedom are likely to drift, so that data from circuits run at the beginning of data collection will reflect a different noise environment than those from the end. Periodic gate recalibration can mitigate these effects to some extent, but the recalibration rate should be consistent with the expected drift during typical operation.

The experimentalist must also make a number of choices about the specifics of the data collection procedure that will impact the precision and reliability of the results. Some of these choices include:
\begin{description}
    \item[Circuit repetitions (shots)] The precision of an estimator grows with the amount of data collected, so more repetitions are often better than fewer. However, more data requires more experiment time.  The experimentalist must balance the benefit of greater precision against the cost of time.  Estimation error decreases only as $1/\sqrt{N}$ with the number of shots $N$, so more data yield diminishing returns.
    
    \item[Maximum circuit depth] Deeper (longer) circuits can amplify errors and therefore serve as more precise probes of gate error ---  particularly coherent errors --- than short circuits. However, if the circuits are too long, then decoherence can reduce the visibility of the target errors. Furthermore, short circuits are useful for ensuring consistency of an estimator, and are often used to improve convergence of an optimizer in post-processing. Several protocols, including robust phase estimation (Sec.~\ref{sec:phase_estimation}) and gate set tomography (Sec.~\ref{sec:gst}), use logarithmically spaced circuit lengths in order to strike an appropriate balance between stability and precision. 
    
    \item[Data collection order] Drift and/or hardware recalibration can lead to time correlations in physical error rates. If data is taken in batches (all samples of a circuit are taken in sequence before moving on to the next circuit), different circuits can experience different noise environments. This can cause bias in  parameter/metric estimation and be difficult or impossible to identify \textit{post hoc}. If data collection is instead \emph{rastered} (data is taken in many passes, with each pass taking one shot of each circuit in the experiment design), drift effects will be smoothed across the dataset, and data can be analyzed for signs of time-correlated noise. Not all experimental systems can be configured to take rastered data, and for these an intermediate collection procedure may be necessary wherein all data is retaken in two or three batches. Such data can be used to probe for low-frequency drift \cite{proctor2020detecting}.
    
    \item[Circuit compilation rules] Integrated quantum processors often feature a \textit{compiler} that optimizes the scheduling of circuits to maximize the processor's performance. When running QCVV protocols, this can sometimes lead to confusing results. For instance, Ramsey experiments (Sec.~\ref{sec:ramsey}) use long idle periods or sequences of repeated idle gates. Care must be taken that the compiler does not identify this and remove the ``extra'' idles, or the expected Ramsey decay may not be observed. Similarly, it is impossible to probe certain crosstalk errors if, for example, the compiler forbids multiple two-qubit gates from acting in parallel. In some cases, programmatic ``barriers'' in the low-level quantum assembly code can enforce compilation restrictions. When characterizing and benchmarking quantum systems, users should take steps to ensure that the compiler is not altering quantum circuit instructions in ways detrimental to the QCVV protocol. 
\end{description}

In addition to these considerations, experiments must contend with control system constraints, including: the data collection rate, recalibration times, memory buffer sizes, network latency, \ac{AWG} upload times, constraints on batching versus rastering, and others. These constraints can prevent a QCVV protocol from collecting data optimally, and some care should be taken to consider the impact of these constraints on the reliability,  susceptibility to bias, and potentially increased variance of derived performance metrics.

%%%%%%%%%%%%%%%%%%%%%%% Identifying and Mitigating out-of-model effects %%%%%%%%%%%%%%%%%%%%%%%
\subsection{Identifying and Mitigating Out-of-Model Effects}
\label{sec:identifyingandmitigatingoutofmodeleffects}

As discussed above, characterization experiments are typically designed to learn all or some parameters of a statistical error model describing a quantum device. If physical errors are present in the experiment that are not captured by this model, then the model will not fit the observed data, and estimates of model parameters can be significantly biased by out-of-model effects. In this section, we briefly survey techniques that can be used to make QCVV protocols robust to out-of-model effects, or at least estimate their impact.

%%%%%%%%%%%%%%%%%%%%%%% Extending Models %%%%%%%%%%%%%%%%%%%%%%%
\subsubsection{Extending Models}\label{sec:extendingmodels}

The simplest approach to mitigating out-of-model effects is to modify the protocol to turn them into in-model effects. An example of this is leakage quantification (see Secs.~\ref{sec:leakage} and \ref{sec:lrb}) using ``blind'' randomized benchmarking (RB) \cite{andrews2019quantifying}, a form of character benchmarking \cite{helsen2019new} (see Sec.~\ref{sec:randomized_benchmarks} for details on randomized benchmarking and its various variants, such as character RB). Blind RB modifies the models used by standard RB to explicitly include a parameter describing the population of leakage levels that cannot be directly observed. Unobserved leakage causes the usual RB decay curve to become a mixture of \emph{two} exponential decays rather than a single exponential decay:
\begin{align}
    \Bar{p}(m) = A + B f^m + C g^m ~.
\end{align}
Fitting a mixture of exponentials can be difficult, especially in the presence of noise. Blind RB modifies the experiment design so that, rather than all circuits compiling to the identity, half of the circuits are chosen to compile to a bit flip operation. In the absence of leakage, the success probability of these circuits should decay at the same rate as the standard identity circuits. However, the bit flip operation does not act on the leakage state, so the success probability decays differently:
\begin{align}
    \Bar{p}_x(d) = A^\prime - B f^d + C g^d ~.
\end{align}
By adding and subtracting the decay curves for the two experiments, we get two new curves that decay as single exponentials and so are easy to fit. Blind RB estimates both the error rate per Clifford and the \emph{leakage rate} per Clifford.

Tomographic protocols (Sec.~\ref{sec:tomography}) can also be extended by explicitly growing the size of the model. Leakage can be captured, for instance, by modeling a qubit as a qu\textit{trit} (or, more generally, a qu\textit{dit}), and performing state, process, or gate set tomography on the larger model (see Appendix \ref{sec:qudit_tomography} for some examples of tomography applied to qutrits and ququarts). This can get expensive, though, and requires careful thought if the leakage levels are not coherently addressable. This also makes it hard to construct a tomographically complete set of states and measurements, and so it may be impossible to fit a full qudit model. Instead, reduced models that assume, for example, only \emph{incoherent} leakage may be more experimentally tractable.

%%%%%%%%%%%%%%%%%%%%%%% Averaging Out-of-Model Effects %%%%%%%%%%%%%%%%%%%%%%%
\subsubsection{Averaging Out-of-Model Effects}\label{sec:averagingoutofmodeleffects}

Sometimes QCVV models are \emph{known} to not capture all of the errors in a system. For instance, tomographic protocols often fit a static model to a system that is actually experiencing drift in some physical parameter. If the effect of drift is not mitigated by the experiment design, the tomographic estimate will be biased.  For instance, drift in the measurement error rate can significantly impact the RB decay curve and subsequent error rate estimates. If data are taken in order of increasing circuit length, growing measurement errors can make the decay curve steeper than it should be, causing the error per gate to be overestimated. On the other hand, if the data is taken with circuits in \emph{decreasing} order of length, then growing measurement errors over the course of the experiment will produce a shallower curve, and an underestimate of the error per gate. 

This bias effect also manifests in tomographic routines. Bias can be reduced (at the possible cost of increased variance) by removing the correlation between execution time and circuit properties. This can be done by selecting a new circuit from the experiment list for each shot, until all repetitions and circuits have been consumed. Another approach is to \emph{raster} the data --- collecting a single shot from each circuit in some order, and then repeating until enough shots have been taken for all of the circuits. If the shots from each circuit are then averaged, the drift will be distributed uniformly across the data set. But rastered data can also be analyzed directly, using methods such as the ones introduced in \R\cite{proctor2020detecting}, to yield time-dependent estimates of error rates.

%%%%%%%%%%%%%%%%%%%%%%% Mitigating Out-of-Model Effects %%%%%%%%%%%%%%%%%%%%%%%
\subsubsection{Mitigating Out-of-Model Effects}\label{sec:mitigatingoutofmodeleffects}

While the above methods can be used to average out-of-model effects, sometimes we just want to eliminate large classes of noise. Again, one particularly frustrating source of noise is drift in control parameters. This drift can lead to errors that change over the course of an experiment. In general, there are three main approaches to reducing the effects of parameter drift: recalibration, dynamical decoupling, and twirling. 

When experiments are impacted by low-frequency noise (like the ubiquitous $1/f$ noise in solid state systems), periodic recalibration can dramatically reduce the scale of the drift problem. This naturally comes at the cost of experimental time, but is often necessary in long experiments to ensure that data taken over long times is consistent. Recalibration may involve fine-tuning experimental control parameters or completely rerunning the calibration procedure \emph{ab initio}, feeding back on results from just a few carefully chosen circuits, or feeding forward data taken from spectator qubits \cite{Gupta2020-kl, Majumder2020-ut}. What method is chosen will depend on experimental capabilities, the timescale of the experiment, and the nature of the drifting error rates. 

\Ac{DD} \cite{viola1998dynamical, viola1999dynamical}  --- or dynamically corrected gates more generally --- has a long history of eliminating the impact of unknown coherent sources of errors. DD evolved from the Hahn echo in nuclear magnetic resonance \cite{hahn1950spin}, where sequential radio frequency pulses are used to reverse the effects of inhomogeneities in the local magnetic field, effectively refocusing the spins and producing a detectable echo signal. In quantum computing, DD uses sequences of gate operations to cancel coherent errors or low-frequency dephasing noise (Sec.~\ref{sec:dephasing}) whose magnitudes are unknown or drifting.

Twirling is a technique closely related to dynamical decoupling, where interleaved gate operations are used to prevent the buildup of coherent errors. Unlike DD, twirling uses random gates, and often aggregates data taken from different randomized circuit realizations (a.k.a.~``randomizations''). Thus, twirling can result in a larger experimental overhead, requiring a different circuit to be measured per randomization. This can be largely mitigated by performing the twirling directly on the control hardware on a shot-by-shot basis \cite{fruitwala2024hardware}. Twirling can dramatically reduce the complexity of an error channel, and is a key component of randomized benchmarks (Sec.~\ref{sec:randomized_benchmarks}) and methods such as randomized compiling \cite{wallman2016noise, hashim2021randomized}. Twirling is discussed in detail in Sec.~\ref{sec:rb_math_twirling} and Appendix \ref{sec:twirling}.

%%%%%%%%%%%%%%%%%%%%%%% Quantifying Out-of-Model Effects %%%%%%%%%%%%%%%%%%%%%%%
\subsubsection{Quantifying Out-of-Model Effects}\label{sec:quantifyingoutofmodeleffects}

Even if all the techniques above are deployed, there remain scenarios in which data will display clear evidence of out-of-model effects. Models will simply not fit the data. In this circumstance, several statistical tools (e.g., likelihood ratio tests) can be deployed to detect and quantify a model's failure to fit the data \cite{Nielsen_2021}. However, statistical techniques based on hypothesis testing can only determine the confidence with which a model can be rejected. They do not generally provide a measure of \emph{effect size}, i.e., how much a model's predictions deviate from actual observations. For instance, consider a coin that is modeled as fair (i.e., 50/50), but when flipped a trillion times yields 501 billion heads and 499 billion tails. The $\chi^2$ statistic --- a simple statistical model validation tool --- is approximately $4\times 10^6$, meaning that the fair coin model can be rejected with incredibly high confidence (a $p$-value of practically zero \cite{Schervish1996-in}). But, for most practical purposes, a coin that is biased by 0.1\% can be well-approximated as fair. The effect size --- $0.1\%$ --- is simply too small to matter for, say, a football game coin toss. 

Quantum tomography experiments often provide a lot of data, so the best-fit model can often be rejected by a statistical hypothesis test. This does not necessarily mean that it is a bad model.  It means that it is demonstrably not a \textit{perfect} model.  There are visible deviations from the model, which are not just statistical fluctuations, but indicate the existence of unmodeled effects. Whether the model is ``good enough'' should be evaluated not using a statistical confidence measure (which grows with the size of the dataset), but using some measure of the \emph{size} of out-of-model effects. 

One proposed way to quantify effect size in QCVV is to compute a wildcard error model \cite{blume2020wildcard}. Wildcard models have been deployed for gate set tomography (GST) experiments, where they relax GST models so that, rather than predicting a circuit's outcome probabilities, they predict a \emph{range} of outcome probabilities. One kind of wildcard model does this by adding a bit of extra ``wildcard'' error to each gate. Wildcard models state that the total variation distance (\eq\ref{eq:tvd}) between each circuit's outcome distribution and the standard GST model prediction should be no larger than the circuit's wildcard error. The per-gate wildcard error is then chosen to be minimally sufficient to make the data statistically consistent with the wildcard model predictions. The wildcard error can then be compared against the parameters of the GST model. If the wildcard is small relative to the gate error, then the GST model captures the most important noise sources, even if statistical tests indicate high confidence for rejecting the GST model. See, for example, Refs.~\cite{PRXQuantum.2.040338, hashim2023benchmarking} for how this done in practice.
\section{Qubit and Gate Characterization}\label{sec:qubit_characterization}

The first steps needed to run a quantum computer are to characterize basic qubit properties and calibrate quantum gates. Qubit characterization involves measuring the resonant frequency and coherence times (i.e., how long a qubit behaves quantum mechanically) of the qubit. These properties are important for calibrating the quantum gates used in algorithms. For example, the coherence times of a qubit place fundamental limits on the fidelity of quantum gates or measurements performed on that qubit, and additionally inform the end user of the circuit depth with which one can perform useful computations with that qubit. Moreover, many quantum gates require coherently driving qubits on-resonance; therefore, it is necessary to characterize qubit frequencies to high accuracy. In this section, we review standard methods for characterizing basic qubit properties and discuss how they can be utilized for measuring errors in quantum gates to high precision:
\begin{enumerate}
    \item \textit{Frequency-domain Spectroscopy} (Sec.~\ref{sec:spectroscopy}). The first step in probing a quantum system is to find its transition frequencies. To achieve this, one can irradiate a driving field on the qubit and sweep its frequency across a broad range. As the drive induces the qubit's transitions when it is swept across the corresponding resonant frequencies, the subsequent measurement of the qubit then reveals its energy spectrum.
    
    \item \textit{Rabi Oscillations} (Sec.~\ref{sec:rabi}). A fundamental test of qubit control is to perform \textit{Rabi oscillations}, whereby a qubit is coherently driven on-resonance between its ground and excited states. Measurement of the qubit after a varied drive duration reveals an oscillation pattern characteristic of a two-level quantum system.
    
    \item \textit{Time-domain Spectroscopy} (Sec.~\ref{sec:frequency}). 
    The coherent control of a qubit requires driving it at its resonant frequency. Therefore, accurately finding qubit frequencies is an important step in the calibration of quantum gates. We review two methods for characterizing qubit frequencies in the time domain, including \textit{Ramsey spectroscopy}, a standard interferometric experiment that is used throughout atomic, molecular, optical, and solid state physics.
    
    \item \textit{Qubit Coherence} (Sec.~\ref{sec:coherence}). The \emph{coherence times} of a qubit are characterized by two timescales: (1) thermalization (or energy relaxation), which quantifies how long a qubit will remain excited before decaying to the ground state; and (2) phase relaxation, which quantifies how long a qubit in a superposition state will maintain phase coherence. Measuring these two properties can be accomplished with simple Rabi and Ramsey experiments.
    
    \item \textit{Phase Estimation} (Sec.~\ref{sec:phase_estimation}). Standard Rabi and Ramsey experiments are typically performed in a continuous manner (see, e.g., \fig\ref{fig:rabi}). However, in a gate-based setting, one can instead perform \textit{discrete} Rabi and Ramsey experiments which are constructed out of a set of defined quantum logic gates. This is the basis for a class of methods known as \textit{phase estimation}, which can be used to perform precision measurements of small errors in the rotation angles of quantum gates.  
\end{enumerate}

%%%%%%%%%%%%%%%%%%%%%%% Frequency-domain Spectroscopy %%%%%%%%%%%%%%%%%%%%%%% 
\subsection{Frequency-domain Spectroscopy}\label{sec:spectroscopy}

\begin{figure}[t]
    \centering
    \includegraphics[width=0.9\columnwidth]{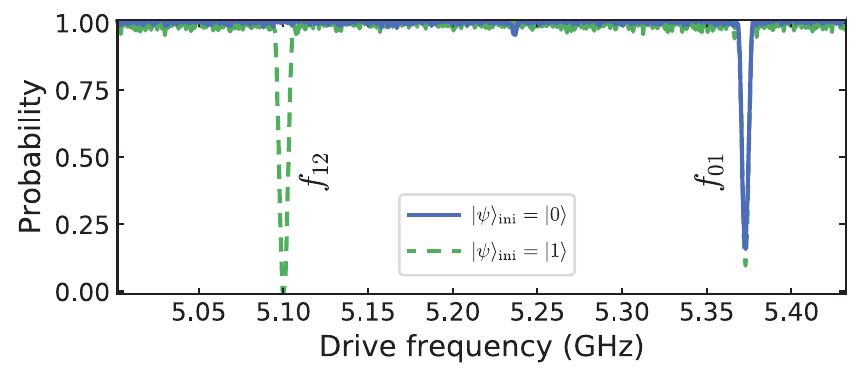}
    \caption{\textbf{Frequency-domain Spectroscopy}. By sweeping the frequency of a microwave tone driving a transmon qubit and monitoring the readout signal, we can detect the transitions of the qubit from its initial states, $|\psi\rangle_\mathrm{ini} = |0\rangle$ (solid blue line) and $|\psi\rangle_\mathrm{ini} = |1\rangle$ (dashed green line). The observed dips indicate the resonant frequencies, which correspond to the $\ket{0}\rightarrow \ket{1}$ and $\ket{1}\rightarrow \ket{2}$ transitions. (The data are reproduced with permission from Ref.~\cite{nguyen2022programmable}.)}
    \label{fig:spectroscopy}
\end{figure}

Measuring the energy spectrum of a qubit is a foundational step in quantum characterization. Coarse measurements of qubit transition frequencies can be performed using standard laboratory spectroscopy methods, such as absorption spectroscopy~\cite{demtroder1973laser}. When driven on-resonant using an external electromagnetic field (e.g., optical laser, microwave signal, etc.), the qubit will absorb some energy from the radiation field and undergo a transition from its initial state, resulting in a change in the measurement signal. Otherwise, the signal remains constant up to the noise level. Therefore, by sweeping the frequency of the external field and monitoring the reflected or transmitted signal, we can detect the transition frequency of the qubit. For example, \fig\ref{fig:spectroscopy} shows the transition spectrum from the $|0\rangle$ and $|1\rangle$ states of a superconducting transmon qubit.

If the qubit is continuously driven across a wide range of frequencies during the measurement, the technique is broadly referred to as \emph{\ac{CW} spectroscopy}. To mitigate spurious and higher-order effects from multi-photon processes, the qubit drive may be deactivated during the measurement phase, which is then termed \emph{pulsed spectroscopy}. Together, these techniques are often called \emph{frequency-domain spectroscopy}. While the broad linewidths found using frequency-domain spectroscopy typically provide sufficient frequency information to observe the coherent nature of qubits (see Sec.~\ref{sec:rabi}), it is not generally precise enough to calibrate quantum gates; instead, one must resort to time-domain techniques, such as Ramsey spectroscopy, to obtain such information (see Sec.~\ref{sec:frequency}). 

While some platforms can be probed directly, such as atomic systems, where resonance fluorescence is often used to measure qubit states, other platforms are measured using an ancilla system. For example, superconducting qubits are often measured via dispersive coupling to a readout resonator \cite{blais2021circuit}, in which the frequency of the readout resonator is dependent on the state of the coupled qubit; thus, by probing the resonant frequency of the readout resonator, one can determine what state the qubit was in. In such cases, it is necessary to first perform frequency-domain spectroscopy on the ancilla system to characterize its resonant frequency. Then, to measure the resonant frequency of the qubit, we sweep the frequency of the field driving the qubit, while also monitoring the frequency spectrum of the readout resonator. This is referred to as \emph{two-tone spectroscopy}, since the qubit and ancilla system often operate at different frequencies.

%%%%%%%%%%%%%%%%%%%%%%% Rabi Oscillations %%%%%%%%%%%%%%%%%%%%%%% 
\subsection{Rabi Oscillations}\label{sec:rabi}

\begin{figure}[th]
    \centering
    \includegraphics[width=0.9\columnwidth]{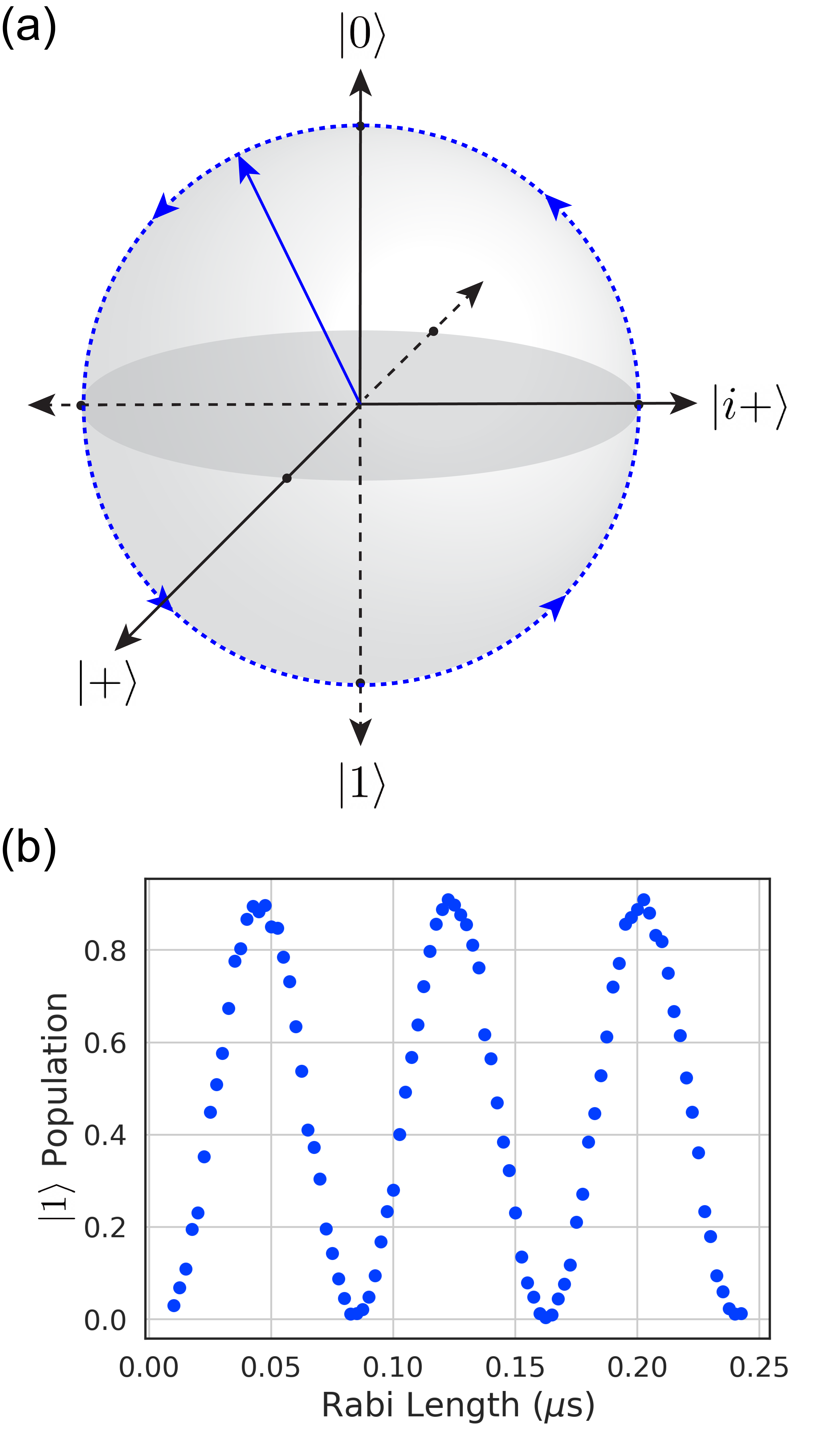}
    \caption{\textbf{Rabi Oscillations}. 
    (a) Rabi oscillations are performed by driving a qubit (represented by a blue arrow on the Bloch sphere) coherently between the ground state $\ket{0}$ and excited state $\ket{1}$. 
    (b) Rabi oscillations of a superconducting qubit. The qubit is driven on-resonance and the probability of finding the qubit in the excited state $\ket{1}$ is measured as a function of time. The offset from 1.0 of the peaks of the oscillations is due to finite readout error of the excited state and/or an off-axis rotation, in which case the oscillations would not reach full contrast between $\ket{0}$ and $\ket{1}$.}
    \label{fig:rabi}
\end{figure}

A two-level system, such as a qubit, can be rotated about the Bloch sphere (see \fig\ref{fig:Bloch_Sphere}) using a classical oscillating field
\begin{equation}
    \mathbf{E} = E_0 \left( \boldsymbol{\epsilon}_d e^{-i(\omega_d t + \phi_d)} + \text{h.c.} \right) ~,
\end{equation}
where $E_0$ is the amplitude of the field, $\omega_d$ is the driving frequency, $\phi_d$ is the phase, $\boldsymbol{\epsilon}_d$ is the complex polarization vector, and h.c.~is the Hermitian conjugate. This field is couples to the qubit's dipole moment
\begin{equation}
    \mathbf{D} = d\left( \boldsymbol{\epsilon}_s \sigma_+ + \text{h.c.} \right) ~,
\end{equation}
where $d$ is the dipole matrix element, $\boldsymbol{\epsilon}_s$ is the qubit's polarization vector, and $\sigma_+$ the qubit raising operator. The coupling Hamiltonian between the qubit and the field is then given by $H_d = \mathbf{E} \cdot \mathbf{D}$. By aligning the field with the polarization of the qubit, and by performing the \emph{\ac{RWA}} to simplify the driving scheme, we can write the Hamiltonian in the interaction picture as
\begin{equation}
\begin{aligned}\label{eqn:driverotatingframehamiltonian}
        H_I &= \frac{\hbar \delta\omega}{2}\sigma_Z + \frac{\hbar\Omega}{2}(e^{-i\phi_d}\sigma_+ + e^{i\phi_d}\sigma_-) ~,\\
            &= \frac{\hbar \delta\omega}{2}\sigma_Z + \frac{\hbar\Omega}{2}\left( \cos \phi_d\sigma_X + \sin\phi_d\sigma_Y \right) ~,
\end{aligned}
\end{equation}
where $\delta\omega = \omega_q - \omega_d$ is the qubit-drive detuning, and $\Omega = 2dE_0/\hbar$ is the \emph{Rabi frequency} of the driven system, which is intuitively proportional to the dipole matrix element $d$ and the amplitude of the field $E_0$. 

Due to the finite detuning between the drive and qubit frequencies, the qubit does not always rotate at the Rabi frequency. More generally, we can write the interaction Hamiltonian as
\begin{equation}\label{eqn:rabihamiltonian}
    H_I = \frac{\hbar \Omega '}{2}\boldsymbol{\sigma} \cdot \hat{\mathbf{n}} ~,
\end{equation}
where 
$\Omega^\prime = \sqrt{\delta\omega^2 + \Omega^2}$ is the \emph{effective Rabi frequency}, and
\begin{equation}\label{eqn:rotationaxis}
    \hat{\mathbf{n}} \equiv \frac{\delta\omega \mathbf{u}_Z  + \Omega \cos \phi_d \mathbf{u}_X + \Omega \sin \phi_d \mathbf{u}_Y}{\Omega '}
\end{equation}
is a unit vector on the Bloch sphere, whose direction is determined by the ratio between the detuning $\delta\omega$ and the on-resonance frequency $\Omega$. The Hamiltonian given by \eq\ref{eqn:rabihamiltonian} thus describes the motion of a qubit in the Bloch sphere which precesses around an effective axis along $\hat{\mathbf{n}}$ at a frequency $\Omega^\prime$. 
    
In \fig\ref{fig:rabi}, we depict Rabi oscillations around the Bloch sphere about the $\hat{\mathbf{x}}$ axis, and plot oscillations for a superconducting qubit from 0 -- 250 ns, in which we measure the excited state $\ket{1}$ population as a function of time. We find that the qubit coherently oscillates between the ground $\ket{0}$ and excited $\ket{1}$ states. The frequency of oscillation depends on the driving amplitude of the Rabi pulse, with higher amplitudes leading to faster oscillations. In general, one can Rabi-drive a qubit at a frequency that is near-resonant and still observe oscillations, although off-resonant drives will not produce full contrast between $\ket{0}$ and $\ket{1}$ (see Sec.~\ref{sec:chevron}). Therefore, the coarse measurements of frequency given by frequency-domain spectroscopy are generally sufficient to probe the coherent nature of a qubit via Rabi oscillations. Beyond being a fundamental test of qubit control, Rabi oscillations are also important for single-qubit gates, which are typically calibrated using resonant Rabi-driven pulses. This requires the precise characterization of qubit frequencies, which is the topic of the following section.

%%%%%%%%%%%%%%%%%%%%%%% Frequency Characterization %%%%%%%%%%%%%%%%%%%%%%%
\subsection{Time-domain Spectroscopy}\label{sec:frequency}

\begin{figure}[t]
    \centering
    \includegraphics[width=\columnwidth]{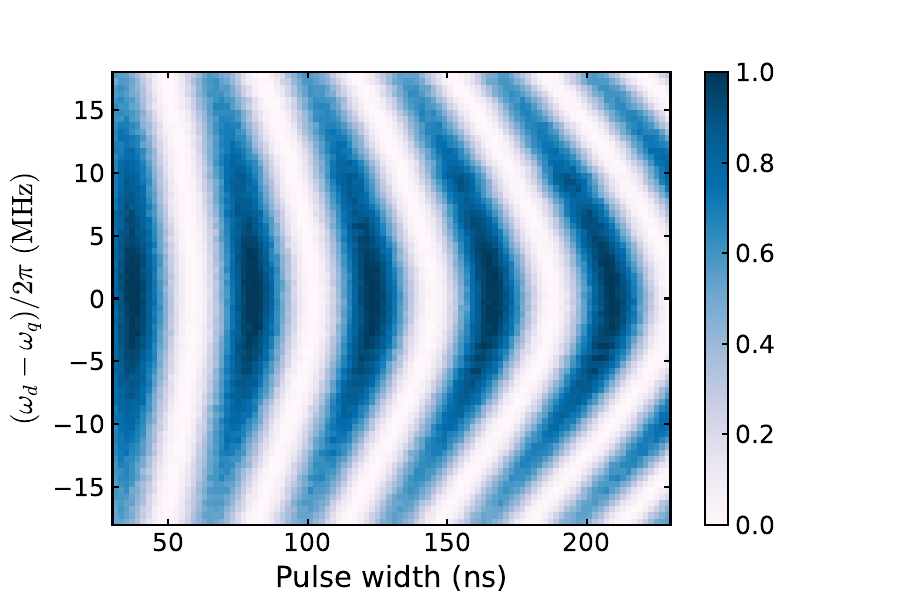}
    \caption{\textbf{Rabi Chevron}. A superconducting qubit is Rabi driven across a range of different frequency detunings, producing a chevron-shaped interferometry pattern. Each horizontal line-out in the above plot is an individual Rabi experiment, with the blue regions depicting the peaks of the oscillations and the white regions depicting the troughs. The qubit frequency can be found by extracting the drive frequency at which the period of oscillations is the largest; this corresponds to the apex of the chevron pattern.}
    \label{fig:chevron}
\end{figure}

Characterizing qubit frequencies is a fundamental component of performing high-fidelity qubit operations. Imperfect frequency calibrations or off-resonant qubit drives will result in coherent phase errors. These errors can be modeled with a small modification to an arbitrary single-qubit density matrix (\eq\ref{eq:rho_single_qubit}),
\begin{equation}\label{eq:rho_detuning}
    \rho = \begin{pmatrix}
            \vert \alpha \vert^2 & \alpha\beta^* e^{i\delta\omega t} \\
            \alpha^*\beta e^{-i\delta\omega t} & \vert \beta \vert^2
           \end{pmatrix} ~,
\end{equation}
where we have added an explicit phase term $\exp(\pm i \delta\omega t)$, and $\delta\omega = \omega_q - \omega_d$ is the detuning between the qubit frequency and the drive frequency, which determines the rotating frame. This phase term accounts for a drive frequency that is off-resonant from the qubit frequency, in which case the qubit will precess in the rotating frame. In most cases, single-qubit quantum gates are designed to drive qubits on-resonance; therefore, characterizing and correcting any off-resonant phase errors is important for gate calibration. While coarse measurements of qubit frequencies can be performed using classical frequency-domain spectroscopy, introduced in Sec.~\ref{sec:spectroscopy}, fine-tuned measurements of qubit frequencies require quantum-based protocols. In this section, we outline two fundamental characterization methods for measuring the detuning in qubit drive frequencies. The first is based on Rabi oscillations, outlined in the previous section, and the second introduces an important interferometric method known as \emph{Ramsey spectroscopy}. Together, these methods are often referred to as \emph{time-domain spectroscopy}, because they are generally implemented by driving a qubit at a given frequency for specific duration of time.

\begin{figure*}[th]
    \centering
    \includegraphics[width=2\columnwidth]{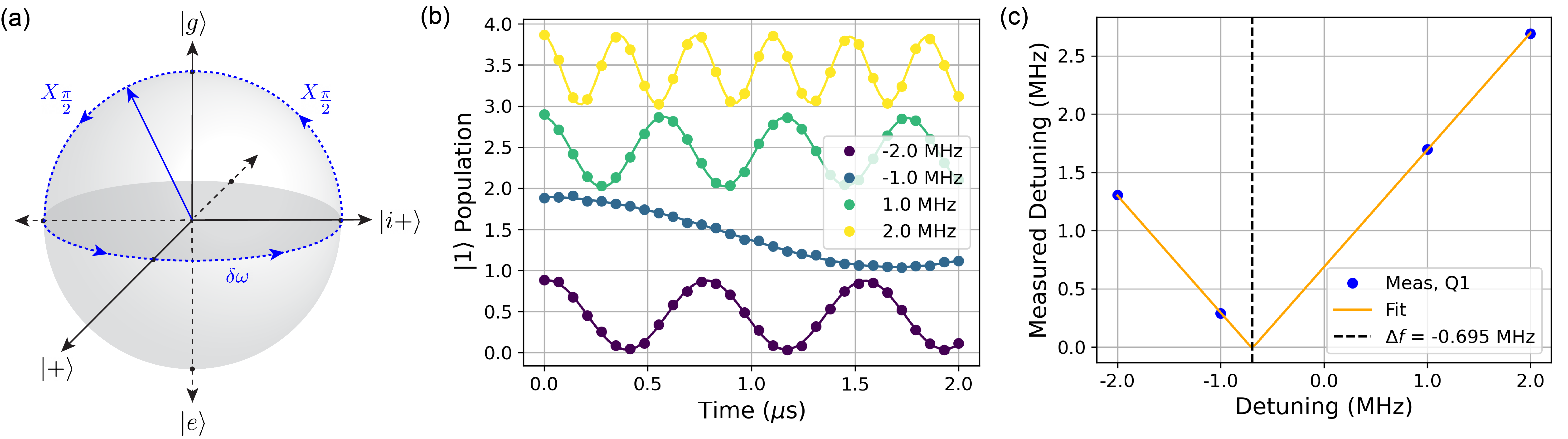}
    \caption{\textbf{Ramsey Spectroscopy}. 
    (a) Ramsey oscillations around the Bloch sphere. A qubit prepared in a superposition state with an $X_{\pi/2}$ pulse will precess along the equator at a frequency given by $\delta\omega$, where $\delta\omega = \omega_q - \omega_d$ is the detuning between the qubit frequency $\omega_q$ and the frequency of the rotating frame $\omega_d$. After some time $t$, another $X_{\pi/2}$ gate is performed and the qubit is measured in the computational basis.
    (b) Ramsey spectroscopy of a superconducting qubit (Q1) is performed for four different artificial detunings ($\Delta = \{-2, -1, 1, 2\}$ MHz). The resulting Ramsey oscillations are plotted and spaced vertically apart for visual clarity. The data are fit to an exponential cosine function, from which the frequency of oscillations can be extracted.
    (c) Measured detunings are extracted from the frequency fits from (b) and plotted as a function of the artificial detunings. The data is then fit to an absolute value function, with the vertex of the fit representing the actual frequency detuning of the qubit drive. From these data, it was found that the qubit was detuned 695 kHz from the drive frequency.}
    \label{fig:ramsey}
\end{figure*}

%%%%%%%%%%%%%%%%%%%%%%% Chevron %%%%%%%%%%%%%%%%%%%%%%% 
\subsubsection{Rabi Chevron}\label{sec:chevron}

A basic method for finding the resonant frequency of a qubit is to Rabi drive the qubit across a range of different frequencies and measure the resulting period of the Rabi oscillations. As shown in the Rabi oscillation formalism, the effective Rabi frequency $\Omega'$ increases with larger detuning $\delta\omega$, resulting in shorter oscillation periods. Moreover, the amplitude of oscillations also decreases with larger detuning. Therefore, when the Rabi drive is on-resonance with the qubit, both the amplitude and period of oscillations will be at peak value. When sweeping over a large range of detunings around the expected qubit resonance and measuring the resulting Rabi oscillations, a chevron-like pattern is produced, as shown in \fig\ref{fig:chevron}. Here, we observe that the Rabi oscillations are largest closest to the middle of the frequency sweep, and that the period and amplitude of oscillations slowly dies off at larger detunings, suggesting that the qubit drive is already near resonant with the qubit frequency. By finding the detuning at which peak oscillations occur, one is able to accurately characterize the drive frequency needed to perform resonant operations on the qubit.

%%%%%%%%%%%%%%%%%%%%%%% Ramsey Spectroscopy %%%%%%%%%%%%%%%%%%%%%%% 
\subsubsection{Ramsey Spectroscopy}\label{sec:ramsey}

Ramsey spectroscopy \cite{ramsey1950molecular}, or Ramsey interferometry, is a precise method for characterizing qubit frequencies. In a typical Ramsey experiment, a qubit is prepared in a superposition state (via a $X_{\pi/2}$ or $Y_{\pi/2}$ pulse) and allowed to evolve naturally for some amount of time $t$, after which the qubit is mapped back to the computational basis via the same gate used to prepare the state, and subsequently measured (see \fig\ref{fig:ramsey}a). Within the rotating frame of the qubit drive, defined by the frequency $f_d = \omega_d/2\pi$, any finite detuning between the qubit frequency and the drive frequency $\delta\omega$ will result in a qubit state which precesses along the equator of the Bloch sphere. Measuring the qubit in the computational basis for different times will result in a sinusoidal oscillation between $\ket{0}$ and $\ket{1}$, much like measurements of Rabi oscillations. However, in this case the oscillations are not caused by coherent driving between $\ket{0}$ and $\ket{1}$, but rather by coherent precession between $\ket{+i}$ and $\ket{-i}$ due to a frequency detuning, which is subsequently mapped back to the computation basis. 

In addition to the sinusoidal oscillations caused by any frequency detuning, a qubit in a superposition state will also experience stochastic noise along the longitudinal axis of the qubit due to interactions with the environment, causing the qubit frequency to fluctuate in time, which results in a Bloch vector that precesses both forwards and backwards in the rotating frame. This process is known as \emph{pure dephasing} (see \fig\ref{fig:Dephasing_noise} and Sec.~\ref{sec:dephasing}), which results in the depolarization of the Bloch vector towards the polar axis of the Bloch sphere. Pure dephasing will lead to the exponential decay of the Ramsey oscillations as a function of time. Thus, measurements of Ramsey spectroscopy are typically fit to an exponential cosine function, for which the probability of measuring the qubit in the excited state is given by
\begin{equation}\label{eq:ramsey_oscillations}
    P_{\ket{1}}(t) = e^{-\Gamma_2 t} \cos(\omega_m t) ~,
\end{equation}
where $\Gamma_2$ is the rate at which the qubit loses phase coherence, and $\omega_m$ is the measured frequency of oscillations. 

The dephasing rate $\Gamma_2$ places a limit on the duration of time over which Ramsey oscillations can be observed. Therefore, if $\Gamma_2$ is large and/or $\delta\omega$ is small, it may be difficult to fit the sinusoidal component of \eq\ref{eq:ramsey_oscillations} to the observed data. For this reason, it is common to drive the qubit at an intentionally large detuning, such that $\delta\omega = (\omega_q - \omega_d) + \Delta$, where $\Delta$ is an additional \emph{artificial} detuning which has been added to the natural detuning. This enables one to fit the sinusoidal component to the observed Ramsey oscillations at short timescales even if the natural detuning is small.

In \fig\ref{fig:ramsey}b, we plot the Ramsey oscillations of a superconducting qubit for four different artificial detunings, $\Delta = \{-2, -1, 1, 2\}$ MHz. We observe that $P_{\ket{1}}$ oscillates sinusoidally, with only a small exponential component visible due to the short time span of the measurements (2 $\mu$s). By extracting the measured frequency of oscillation $\omega_m$, we can compute the \emph{measured detuning} (i.e., the difference between the measured frequency and the artificial detuning, $\omega_m - \Delta$) for each artificial detuning. In \fig\ref{fig:ramsey}c, we plot the measured detuning versus the artificial detuning. By choosing artificial detunings above and below where we expect the true qubit frequency to reside, we can fit the measured detuning to an absolute value curve, with the vertex of the fit representing the detuning between the qubit frequency and the drive frequency. For the data in \fig\ref{fig:ramsey}c, we find that the qubit frequency was detuned 695 kHz below the qubit drive.

%%%%%%%%%%%%%%%%%%%%%%% Qubit Coherence %%%%%%%%%%%%%%%%%%%%%%% 
\subsection{Qubit Coherence}\label{sec:coherence}

There are two important characteristic timescales that define the \emph{coherence} of a qubit. The first timescale --- called $T_1$ --- describes how long a qubit will remain in an excited state before it decays to the ground state. The second characteristic timescale --- called $T_2$ --- describes how long a qubit can maintain phase coherence in a superposition state. The characterization of these timescales is important, as they place fundamental limits on the gate fidelities achievable for each qubit, as well as fundamental limits on the time within which one can perform useful computations on a quantum processor.

In general, different types of qubits can have drastically different coherence times. For example, while coherence times of superconducting qubits ranging from $\sim 100 \mu$ s to 1 ms are considered quite long \cite{somoroff2023millisecond}, atomic-based systems such as neutral atoms or trapped ions can exhibit drastically longer coherence times, ranging from seconds to even hours \cite{wang2021single}. However, gate times can also differ by orders of magnitude between different platforms, typically ranging from tens of nanoseconds on superconducting systems to tens of milliseconds on atomic systems. Therefore, in the context of gate-based quantum computing, one should consider the relative number of gates that can be implemented within the coherence times of qubits on quantum processor, as this determines the maximum circuit depth achievable for the processor.

%%%%%%%%%%%%%%%%%%%%%%% Energy relaxation %%%%%%%%%%%%%%%%%%%%%%% 
\subsubsection{Energy relaxation: $T_1$}\label{sec:t1}

\begin{figure}[t]
    \centering
    \includegraphics[width=\columnwidth]{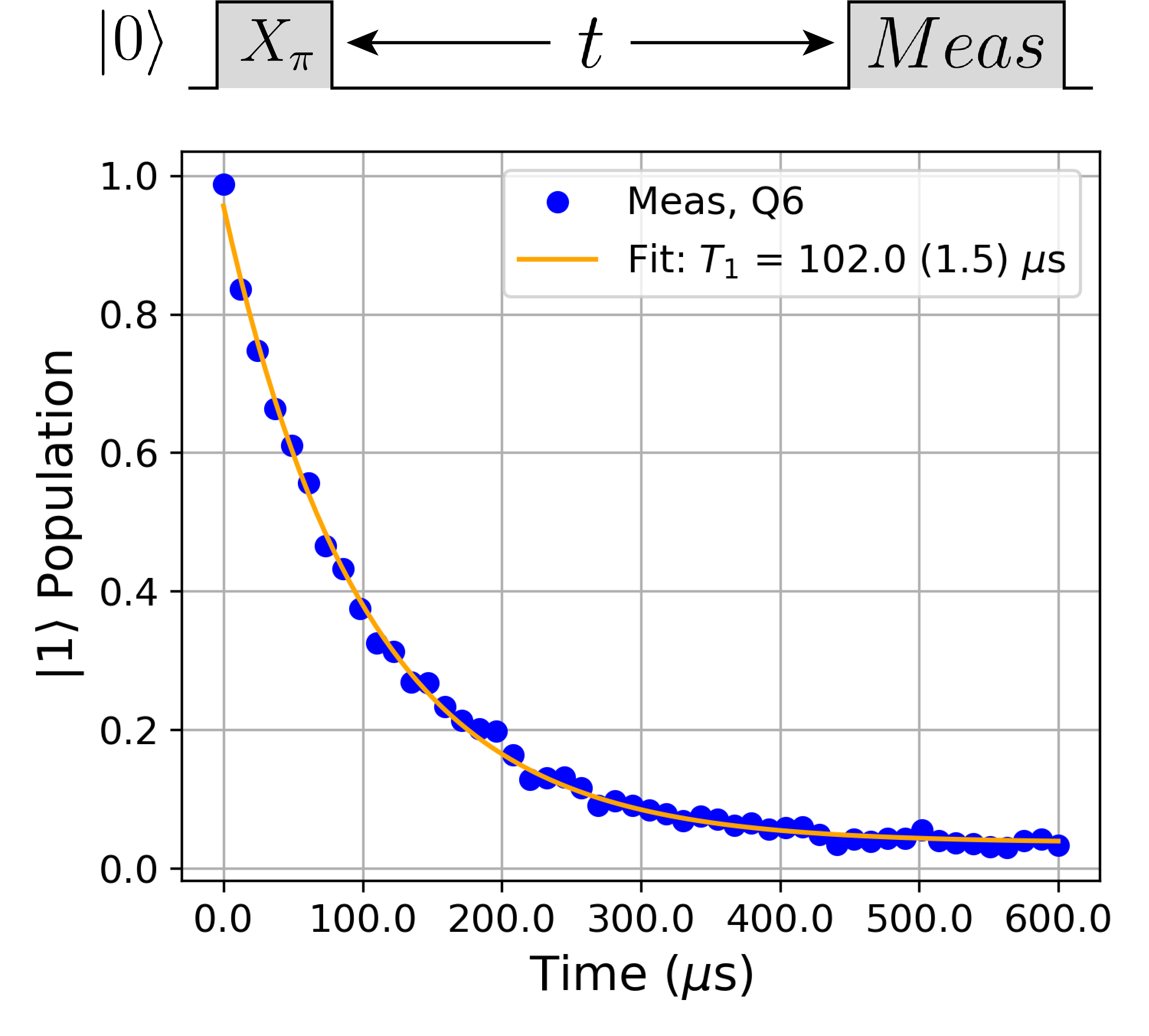}
    \caption{\textbf{Qubit $T_1$ Characterization}. Exponential decay curve for a superconducting qubit (labeled Q6) prepared in the excited state and measured after a variable amount of time. The raw data (blue points) corresponds to the ensemble probabilities of the qubit being measured in the $\ket{1}$ state after each waiting period, and orange curve is the exponential fit to the data. From this fit, we can extract a characteristic time of $T_1 = 102.0 (1.5) ~ \mu$s for this qubit.}
    \label{fig:t1}
\end{figure}

\begin{figure*}[t]
    \centering
    \includegraphics[width=2\columnwidth]{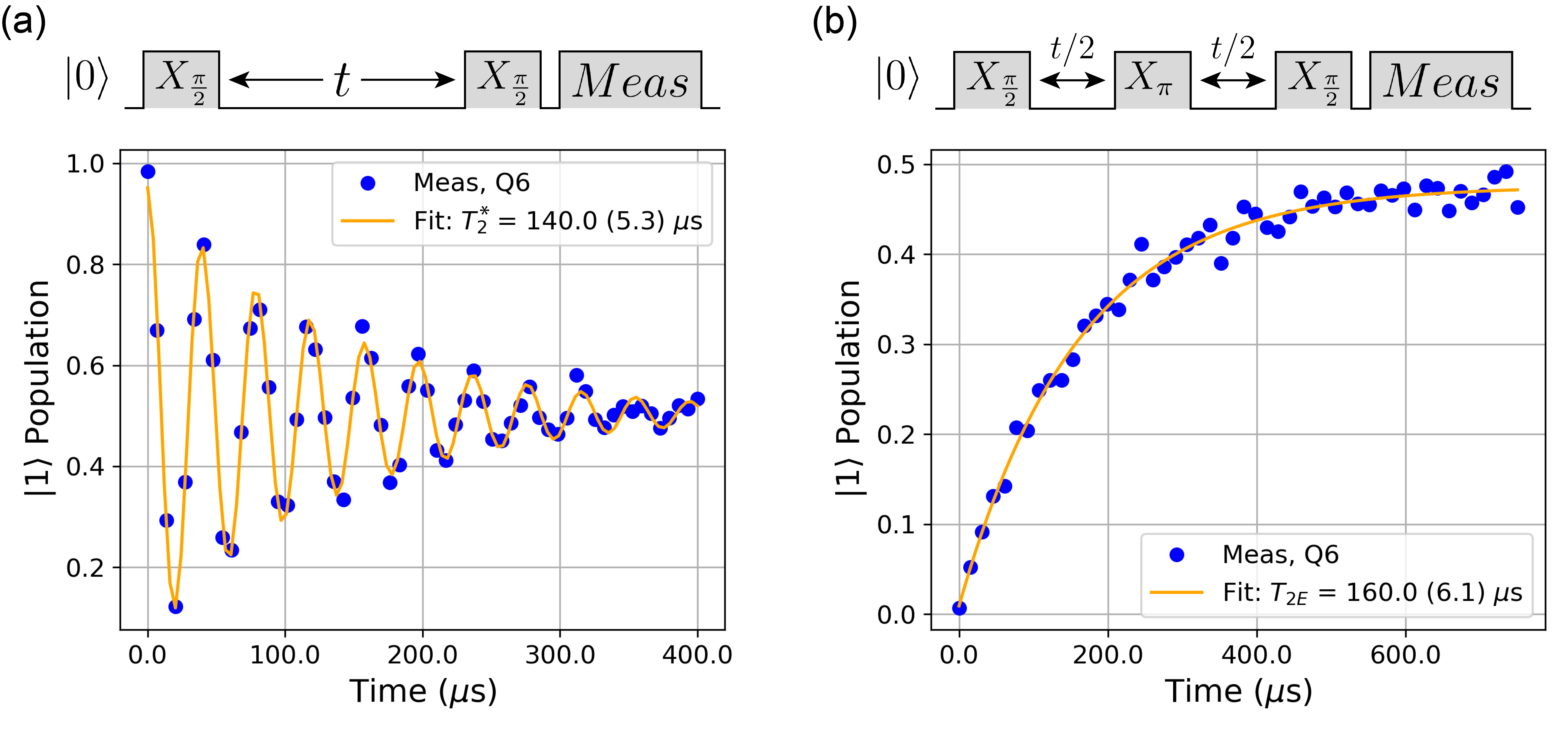}
    \caption{\textbf{Qubit $T_2$ Characterization}. 
    \textbf{(a)} Decaying sinusoidal curve for a superconducting qubit (labeled Q6) which has been prepared in a superposition state using an $X_{\pi/2}$ pulse, allowed precess along the equator for some time $t$, after which it is mapped back to the computational basis with a final $X_{\pi/2}$ pulse and subsequently measured. 
    From the fit of the data, we extract a characteristic time of $T_2^* = 140.0 (5.3) ~ \mu$s for this qubit.
    \textbf{(b)} Hahn echo experiment for Q6. The sequence is identical to the Ramsey sequence in (a), but an $X_\pi$ pulse is added to the middle of the experiment to echo away the effects of low-frequency noise. From the exponential fit, we extract a characteristic time of $T_{2E} = 160.0 (6.1) ~ \mu$s for this qubit.
    }
    \label{fig:t2}
\end{figure*}

A qubit in an excited state will eventually decay to the ground state due to energy relaxation, such as spontaneous emission (see Sec.~\ref{sec:spont_emis_amp_damp}). The characteristic timescale for thermalization --- termed $T_1$ --- is defined by the longitudinal relaxation rate $\Gamma_1$ (\eq\ref{eq:long_relaxation_rate}),
\begin{equation}\label{eq:t1}
    T_1 \equiv \frac{1}{\Gamma_1} ~.
\end{equation}
$T_1$ is the $1/e$ decay constant for energy relaxation, whereby with some probability $p(t) = 1 - \exp(-t/T_1)$ at time $t$ a qubit in the excited state will thermalize to the ground state. For an arbitrary single-qubit state, this process can be modeled in the density matrix formalism:
\begin{equation}\label{eq:rho_t1}
    \rho = \begin{pmatrix}
            1 + (\vert \alpha \vert^2 - 1)e^{-t/T_1} & \alpha\beta^* e^{i\delta\omega t} \\
            \alpha^*\beta e^{-i\delta\omega t} & \vert \beta \vert^2 e^{-t/T_1}
           \end{pmatrix} ~,
\end{equation}
where we note that as $t \longrightarrow \infty$, $\rho_{00} = 1 + (\vert \alpha \vert^2 - 1)e^{-t/T_1} \longrightarrow 1$ and $\rho_{11} = \vert \beta \vert^2 e^{-t/T_1} \longrightarrow 0$.

To measure the $T_1$ time of a qubit, the qubit is prepared in the $\ket{1}$ state using an $X_\pi$ pulse and then measured after waiting some time $t$. By repeating this process for many different times, the measured data is fit to an exponential decay function $A \exp(-t/T_1)$, from which $T_1$ can be extracted. In \fig\ref{fig:t1}, we plot the $T_1$ characterization curve for a superconducting transmon qubit, and find that it has a $T_1$ time of 102.0 (1.5) $\mu$s.

%%%%%%%%%%%%%%%%%%%%%%% Phase coherence %%%%%%%%%%%%%%%%%%%%%%% 
\subsubsection{Phase decoherence: $T_2$}\label{sec:t2}

A qubit prepared in an superposition state will eventually experience phase decoherence due to both energy relaxation and pure dephasing (see Sec.~\ref{sec:dephasing}). The characteristic timescale for phase decoherence --- termed $T_2$ --- is defined by the transverse relaxation rate $\Gamma_2$,
\begin{equation}\label{eq:t2}
    T_2 \equiv \frac{1}{\Gamma_2} = \left( \frac{\Gamma_1}{2} + \Gamma_\phi \right)^{-1} ~,
\end{equation}
where $\Gamma_\phi$ is the rate of pure dephasing. Here, we see that in the limit of no pure dephasing (i.e., $\Gamma_\phi = 0$), the timescale for phase coherence is determined by the timescale for energy relaxation, with $T_2 = 2T_1$. This reflects the fact that $T_1$ events erase all phase knowledge of the qubit state, limiting the maximum length of time that a qubit can maintain phase coherence.

$T_2$ is the $1/e$ decay constant for phase decoherence, whereby with some probability $p(t) = 1 - \exp(-t/T_2)$ at time $t$ a qubit in a superposition state will depolarize toward the polar axis. Equation \ref{eq:rho_t1} can be modified to include phase decoherence,
\begin{equation}\label{eq:rho_t2}
    \rho = \begin{pmatrix}
            1 + (\vert \alpha \vert^2 - 1)e^{-t/T_1} & \alpha\beta^* e^{i\delta\omega t} e^{-t/T_2} \\
            \alpha^*\beta e^{-i\delta\omega t} e^{-t/T_2} & \vert \beta \vert^2 e^{-t/T_1}
           \end{pmatrix} ~,
\end{equation}
where we have added $e^{-t/T_2}$ to the off-diagonal terms to account for phase decoherence as $t \longrightarrow \infty$. In the long-time limit, all terms converge to zero except $\rho_{00}$. \eq\ref{eq:rho_t2} is known as the Bloch-Redfield model of two-level systems \cite{redfield1957theory}.

The phase decoherence time $T_2$ can be measured using Ramsey spectroscopy (Sec.~\ref{sec:ramsey}). First, the qubit is prepared in a superposition state, then allowed to naturally dephase along the equator for a variable amount of time, after which the resulting state is rotated back to the computational basis and subsequently measured (see \fig\ref{fig:ramsey}a). In a Ramsey experiment, one should observe decaying oscillations between $\ket{0}$ and $\ket{1}$. By fitting the data to a decaying sinusoid (\eq\ref{eq:ramsey_oscillations}), $T_2$ can be determined directly from the exponential fit parameter, $T_2 = 1/\Gamma_2$. If both the drive-qubit detuning and the dephasing rate are small, then it can be difficult to fit the observed data to \eq\ref{eq:ramsey_oscillations}. In this case, it is convenient to add an artificial detuning to the drive (see \fig\ref{fig:ramsey}) such that the data can be accurately fit to a decaying sinusoid. The dephasing time measured using Ramsey spectroscopy is typically written as $T_2^*$ to denote that it is sensitive to inhomogeneous low-frequency noise (see the discussion in Sec.~\ref{sec:dephasing}). In \fig\ref{fig:t2}a, we plot the $T_2^*$ characterization curve for a superconducting transmon qubit, and find that it has a $T_2^*$ time of 140.0 (5.3) $\mu$s. We observe that it is less than $2T_1$ of the same qubit (see \fig\ref{fig:t1}), indicating the presence of pure dephasing.

Ramsey measurements are generally sensitive to low-frequency noise. Here, low-frequency noise is defined to be quasi-static over the timescale of an experiment, but can vary from experiment to experiment. Therefore, it is possible to ``echo'' away the effect of the noise using a Hahn echo pulse \cite{hahn1950spin}. Hahn echo experiments are identical to Ramsey experiments in the state-preparation and measurement, but halfway through the experiment an $X_\pi$ pulse is applied to the qubit. This reverses the effects of inhomogeneous broadening caused by the quasi-static noise, effectively refocusing the Bloch vector. By performing a Hahn echo experiment for different durations of time, one can fit the measurements to an exponential function whose exponential fit parameter determines the $T_2$ time of the qubit with an echo pulse --- denoted $T_{2E}$. In \fig\ref{fig:t2}b, we plot the $T_{2E}$ characterization curve for a superconducting transmon qubit, and find that it has a $T_{2E}$ time of 160.0 (6.1) $\mu$s. While this is longer than the $T_2^*$ time of the qubit, it does not saturate the $T_2 = 2T_1$ limit. This indicates the presence of not only low-frequency quasi-static noise (since $T_{2E} > T_2^*$), but also high-frequency noise which likely varies over the timescale of the experiment.

%%%%%%%%%%%%%%%%%%%%%%% Phase Estimation %%%%%%%%%%%%%%%%%%%%%%% 
\subsection{Phase Estimation}\label{sec:phase_estimation}

\begin{figure}[t]
    \centering
    \includegraphics[width=0.85\columnwidth]{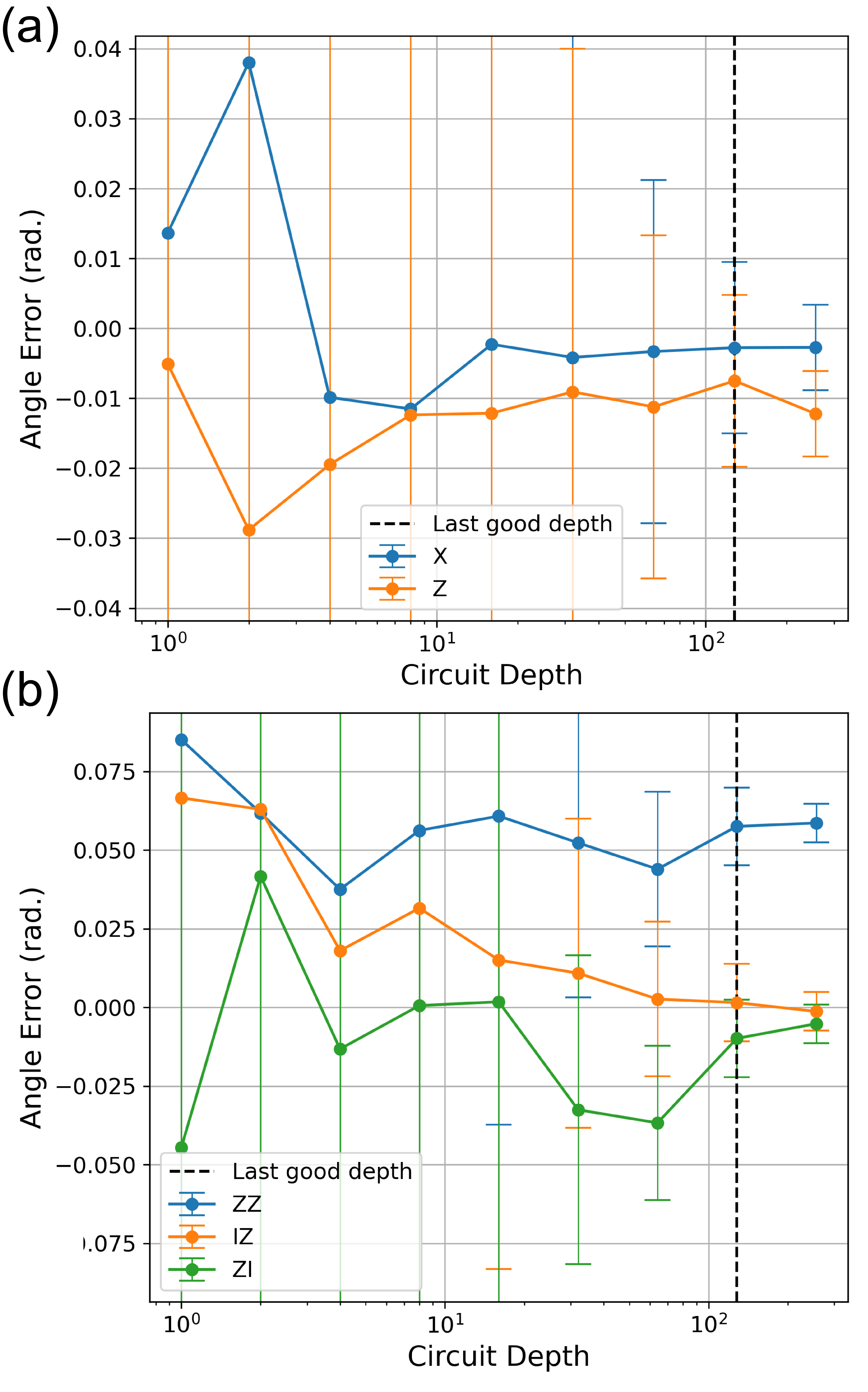}
        \caption{\textbf{Robust Phase Estimation}. 
    \textbf{(a)} RPE of an $R_x(\pi/2)$ gate. The $X$ error (i.e., over/under-rotation) and $Z$ error (i.e., axis angle error) in the gate are plotted above. Both the $X$ and $Z$ errors are equal to zero (up to uncertainty).
    \textbf{(b)} RPE of a CZ gate. Rotation errors on the $IZ$, $ZI$, and $ZZ$ Hamiltonian generators of the gate are plotted above. RPE is able to verify that the errors on the $IZ$ and $ZI$ terms are zero (up to uncertainty), but there remains an error on the $ZZ$ entangling term of approximately $0.06\pm\tfrac{\pi}{512}\approx0.06\pm .006$ radians.
    The error bars in the plot represent the $\pi/(2\cdot k_{\max})$ upper bound on the estimates' RMS error (we omit displaying the full error bar at lower circuit depths for visual clarity), and the most trusted estimate of the errors (i.e., ``last good depth'') from the RPE consistency checks are indicated by dashed vertical lines.
    }
    \label{fig:rpe}
\end{figure}

While \emph{continuous} Rabi, Ramsey, and Hahn echo experiments are useful for learning basic properties of qubits, they do not provide detailed information about the performance of quantum gates in a gate-based setting. However, by performing \emph{discrete} Rabi and Ramsey experiments composed of defined quantum logic operations (e.g., $X_{\pi/2}$ or $Y_{\pi/2}$ gates), one can learn detailed information about the underlying gates themselves. This is the goal of a set of characterization methods under the broad term \emph{phase estimation} \footnote{The term ``phase estimation'' has multiple and closely related uses, as it also refers to the similar but distinct task of estimating a Hamiltonian's eigenvalues \cite{kitaev1995quantum}.}.

As an ideal $n$-qubit quantum gate implements a unitary in $\mathsf{SU}(2^n)$, we may think of any such gate as implementing some manner of \textit{rotation} of vectors in $d = 2^n$ dimensional Hilbert space.  Implementing incorrect rotation angles is a primary source of coherent error in quantum hardware (see Sec.~\ref{sec:coherent_errors}); accurate characterization of such angles is necessary for the calibration of high-quality gates. The task of specifically estimating a gate's rotation angle is called phase estimation, and it is the task we concern ourselves with in this subsection.

While there exist multiple protocols for phase estimation, here we review a particular flavor of it called \emph{\ac{RPE}} \cite{kimmel2015robust}. RPE may be thought of in some sense as an interpolation between the aforementioned Rabi and Ramsey oscillations and the gate set tomography (GST) protocol, discussed in Section~\ref{sec:gst}. Like Rabi oscillations, RPE estimates one particular Hamiltonian parameter of a gate operation, but like GST, it uses a set of circuits with logarithmically spaced depths, allowing it to learn that parameter with Heisenberg-like accuracy.

Without loss of generality, we may consider the task of estimating the phase $\theta$ from a single-qubit gate $U=e^{-i\theta\sigma_j/2}$, with $\theta\in[-\pi,\pi]$ and where $\sigma_j$ may be taken to be any Pauli matrix \footnote{A unit-normalized linear combination of Pauli matrices also works, i.e., $\bf{\hat{n}} \cdot \boldsymbol{\sigma}$, but makes state preparation and measurement mildly more complicated.}.  If we simply apply the gate $U$ to a uniform superposition of its two eigenstates and then perform a projective measurement onto that same superposition, the probability $P_c$ that the system is projected onto that same superposition is given by
\begin{equation}\label{eq:RPE_cos_0}
    P_c = |\bra{+} U \ket{+}|^2 = \frac{1 + \cos\theta}{2} ~,
\end{equation}
where $\ket{+}$ denotes the uniform superposition of the two eigenstates of $U$ [by analogy with the standard definition of $\ket{+}=\tfrac{1}{\sqrt{2}}(\ket{0}+\ket{1})$].  Similarly, if we instead perform a projective measurement onto the $\ket{i+}$ state (where we have put a relative phase of $i$ between the two eigenstates), the probability of measuring in the $\ket{i+}$ state is
\begin{equation}\label{eq:RPE_sin_0}
    P_s = |\bra{i+} U \ket{+}|^2 = \frac{1+\sin\theta}{2} ~.
\end{equation}
By repeating both of these experiments many times to build up approximations of $P_c$ and $P_s$ --- which we will denote as $\hat{P}_c$ and $\hat{P}_s$, respectively --- we can estimate $\theta$ by $\hat{\theta}$:
\begin{equation}\label{eq:RPE_theta_0}
    \hat{\theta} = \arctan2(2\hat{P}_{s} - 1, 2\hat{P}_{c} - 1) ~,
\end{equation}
where $\arctan2$ is the arc-tangent function which accounts for branch cuts by tracking the signs of its two arguments. While one can, in principle, estimate $\theta$ in this manner, this approach suffers from two problems. First, unwanted error terms (e.g., decoherence effects, SPAM errors, etc.) can pollute $\hat{P}_c$ and $\hat{P}_s$, corrupting $\hat{\theta}$. Second, even if such errors are \emph{not} present, the accuracy of the estimate $\hat{\theta}$ is at the standard quantum limit, i.e., if $N$ repetitions (shots) of each of the two circuits are taken, then the uncertainty in $\hat{\theta}$ scales as $1/\sqrt{N}$.  For learning $\theta$ to high precision, this approach becomes very expensive.

RPE solves both of these problems, as we describe in the rest of this subsection. Instead of only using circuits with just a single repetition of $U$, RPE replaces the single instances of $U$ in Eqs.~\ref{eq:RPE_cos_0} and~\ref{eq:RPE_sin_0} with $k$ repetitions of $U$, for $k\in\{1,2,4,\ldots,k_{\max}\}$, giving an RPE experiment a total of $2(1+\log_2k_{\max})$ circuits. Thus, the target probability distributions that an RPE experiment attempts to sample from look unsurprisingly similar to Eqs.~\ref{eq:RPE_cos_0} and~\ref{eq:RPE_sin_0}. However, as we noted above, Eqs.~\ref{eq:RPE_cos_0} and~\ref{eq:RPE_sin_0} do not take into account other errors which could perturb the distributions we wish to sample from (and estimate). If we denote such additive perturbations by $\delta_{k,c}$ and $\delta_{k,s}$ for the $P_{k,c}$ and $P_{k,s}$ distributions, respectively, then we find that the \textit{actual} probability distributions an RPE experiment samples from are given by
\begin{align}\label{eq:RPE_probs}
    P_{k,c} = \frac{1 + \cos(k\theta)}{2} + \delta_{k,c} ~, \\
    P_{k,s} = \frac{1 + \sin(k\theta)}{2} + \delta_{k,s} ~.
\end{align}

As detailed in Refs.~\cite{kimmel2015robust, russo2021consistency}, built into RPE is a robustness against such additive errors.  As long as $\max_{\{i \in \{c, s\}, k\} }|\delta_{k,i}| < \sqrt{\tfrac{3}{32}} \approx 30.6\%$, RPE can still successfully estimate $\theta$. In fact, if that constraint on the additive errors is satisfied for all $k$ up to some $k_{\max}$, then the \ac{RMS} error of RPE's estimate of $\theta$ will be no greater than $\pi/(2\cdot k_{\max})$.  
Thus, RPE yields an estimate of $\theta$ that is Heisenberg-limited in its accuracy (up to decoherence), allowing $\theta$ to be estimated extraordinarily efficiently. For example, it was shown in \cite{rudinger2017experimental} that RPE could be used to learn a single-qubit gate's phase to within $4 \times 10^{-4}$ radians, with only 176 total experimental samples.

We now turn to discussing \textit{how} RPE constructs an estimate of $\theta$ from its experimental estimates $\hat{P}_{k,c}$ and $\hat{P}_{k,s}$.  For a given ``generation'' of circuits corresponding to $k$ repetitions of $U$, we can estimate $k \theta$ according to
\begin{equation}\label{eq:RPE_ktheta}
    k\hat{\theta} = \arctan2(2\hat{P}_{c,k} - 1, 2\hat{P}_{k,s} - 1) \mod 2\pi ~.
\end{equation}
Because sinusoids exhibit periodicity, we cannot learn $\theta$ from just Eq.~\ref{eq:RPE_ktheta}.  In other words, there are multiple values of $\hat{\theta}$ that satisfy Eq.~\ref{eq:RPE_ktheta}.  However, we can use successive generations of RPE data to learn $\theta$ iteratively.  For each successive generation, the angular space in which $\hat{\theta}$ can fall is cut in half (thus allowing the uncertainty to shrink by a factor of two with every generation, yielding Heisenberg-like scaling). To begin, we start with $k=1$, and compute an initial estimate of $\hat{\theta}_1$, given by Eq.~\ref{eq:RPE_theta_0}.  For each subsequent generation, we compute
\begin{equation}
    \Tilde{\theta}_k = \arctan2(2\hat{P}_{c,k} - 1, 2\hat{P}_{k,s} - 1)/k ~.
\end{equation}
This quantity, $\Tilde{\theta}_k$, is effectively the update to our previous estimate $\hat{\theta}_{k-1}$.  However, in order for the new estimate to be in the correct branch, we must add (or subtract) $2\pi/k$ to (or from) $\hat{\theta}_k$ until the resulting quantity falls within $\hat{\theta}_{k-1} \pm \pi/k$.  This resulting quantity is then taken to be $\hat{\theta}_k$.

In \fig\ref{fig:rpe}a, we plot the results of an RPE experiment performed on a single-qubit $R_x(\pi/2)$ gate implemented on a superconducting qubit and analyzed using the python package \texttt{pyRPE}~\cite{pyrpe2021repo}. As discussed in Refs.~\cite{kimmel2015robust, rudinger2017experimental}, small modifications can be made to the standard RPE routine to learn both the error in rotation angle and the axis misalignment of a gate relative to the target operation [i.e., an ideal $R_x(\pi/2)$ gate]. We plot both of these errors for the $R_x(\pi/2)$ gate in \fig\ref{fig:rpe}a, where the $X$ error refers to an over/under-rotation, and the $Z$ error refers to a phase (i.e., axis) error on the gate. The error bars in the plot represent the $\pi/(2\cdot k_{\max})$ upper bound on the phase estimates' RMS error, and the most trusted estimate of the errors are indicated from the RPE consistency checks. Both the $X$ and $Z$ errors on the $R_x(\pi/2)$ are zero, up to uncertainty. 

Additionally, RPE may be used to characterize two-qubit rotation angles, with only minor modifications required~\cite{russo2021evaluating}. We demonstrate RPE on a CZ gate that acts on pair of superconducting qubits \cite{rudinger2025heisenberg}. As, up to a global phase,
\begin{equation}
    \text{CZ} = e^{\tfrac{-i}{2}\left(\tfrac{\pi}{2}ZI+\tfrac{\pi}{2}IZ-\tfrac{\pi}{2}ZZ\right)} ~,
\end{equation}
RPE can be used to learn each of the coefficients of the $ZI$, $IZ$, and $ZZ$ terms in the Hamiltonian which generates the CZ gate. The results, summarized in Fig.~\ref{fig:rpe}(b), show that while the errors on the $IZ$ and $ZI$ phases are zero (up to uncertainty), the $ZZ$ entangling phase has an angle error of approximately 0.06 radians. Because RPE is rather inexpensive to perform, the aforementioned experiment cost fewer than 60 distinct circuits, and its error estimates can be used to (re)calibrate the gates being characterized using parameter sweeps or closed-loop optimization \cite{rudinger2025heisenberg}.

\section{Tomographic Reconstruction}\label{sec:tomography}

Tomographic reconstruction methods (a.k.a. ``tomography'') estimate all aspects of an object by probing it along several axes and combining the results. In \ac{QCVV}, tomography is used to estimate the mathematical object representing a quantum logic operation -- a quantum state, process, or measurement. Tomography protocols estimate the \emph{entire} density matrix, transfer matrix, or \ac{POVM}. This requires more experimental data and analytic effort than just estimating a few properties of the object, but yields more diagnostic power. A complete tomographic characterization reveals \emph{all} interesting properties of a logic operation, enabling debugging of quantum hardware and predicting the behavior of operations \emph{in situ}. 

Quantum tomography methods are among the oldest characterization tools. Quantum state tomography (Sec.~\ref{sec:state_tomography}) appears in the literature as early as 1968 \cite{gale1968determination}, and quantum process tomography (Sec.~\ref{sec:qpt}) dates to 1997 \cite{chuang1997prescription}. Traditional methods such as quantum state and process tomography are still widely used to this day, but they are unreliable in the presence of imperfect state preparation and measurement (\ac{SPAM}) \cite{merkel2013selfconsistent}. Self-consistent tomographic methods like gate set tomography (Sec.~\ref{sec:gst}) avoid this problem, and have superseded state and process tomography in contexts that require reliability. 

\begin{figure}[t]
    \centering
    \includegraphics[width=0.4\textwidth]{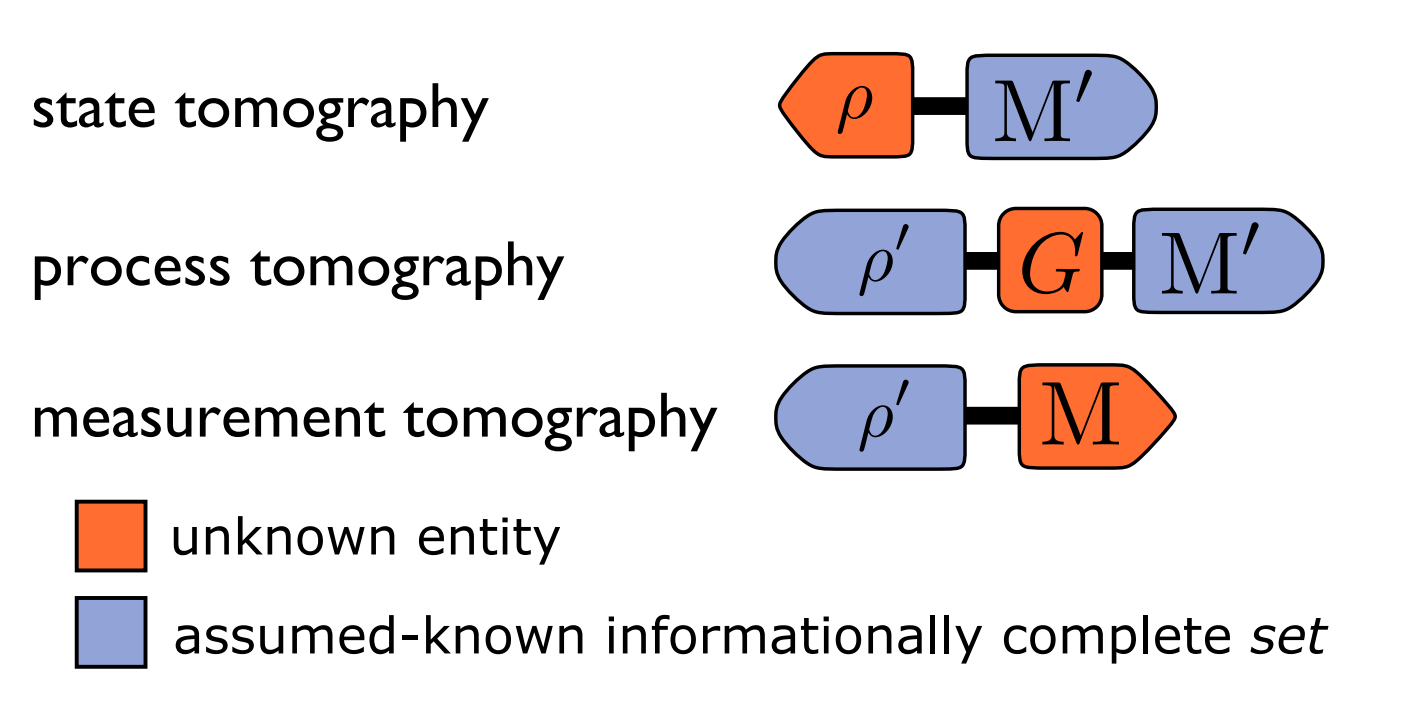}
    \caption{\textbf{Tomographic Reconstruction.} The structure of the circuits required for state, process, and measurement tomography are shown. Each of these protocols reconstructs an initially-unknown quantum operation (a state $\rho$, process $G$, or POVM $M$) by combining that operation into simple circuits with a set of \emph{known} complementary operations. The known operations form a reference frame for estimation of the unknown operation's matrix elements. If the reference frame is \emph{informationally complete}, then all matrix elements of the unknown operation can be learned. We denote reference frame operations with ``primed'' symbols ($\rho^\prime$ and $M^\prime$) to indicate that they are \emph{effective} (rather than \emph{native}) state preparations and measurements, usually implemented by applying gate operations after or before a native state preparation or measurement. These techniques are limited by systematic errors stemming from imperfect \emph{a priori} knowledge of the $\rho^\prime$ and $M^\prime$. Figure and caption reproduced with permission from \R\cite{nielsen2021gate}. }
    \label{fig:spm_tomography}
\end{figure}

In this section, we provide overviews of the following tomographic reconstruction methods:
\begin{itemize}
    \item \emph{Quantum State Tomography} (Sec.~\ref{sec:state_tomography}). Quantum state tomography is designed to reconstruct an unknown quantum state $\rho$ (i.e., density matrix) by performing an informationally complete set of measurements $\{M_j^\prime\}$ on many identical copies of $\rho$ (see \fig\ref{fig:spm_tomography}).
    
    \item \emph{Quantum Process Tomography} (Sec.~\ref{sec:qpt}). Quantum process tomography is designed to reconstruct an unknown quantum operation (e.g., a gate $G$) by applying many identical implementations of $G$ to an informationally complete set of distinct states $\{\rho_i^\prime\}$ and performing an informationally complete set of measurements $\{M_j^\prime\}$ on many identical copies of each $G[\rho'_{i}]$ (see \fig\ref{fig:spm_tomography}).
    
    \item \emph{Quantum Measurement Tomography} (Sec.~\ref{sec:meas_tomography}). Quantum measurement tomography is designed to reconstruct an unknown POVM $M$ by applying it to an informationally complete set of states $\{\rho_i^\prime\}$ (see \fig\ref{fig:spm_tomography}).
    
    \item \emph{Gate Set Tomography} (Sec.~\ref{sec:gst}). Gate set tomography is designed to self-consistently reconstruct an entire \emph{set} of quantum operations including at least one initialization (state), at least one measurement (POVM), and at least two logic gates (quantum processes) --- i.e., a \emph{gate set} --- by running a wide range of circuits composed from those operations (see \fig\ref{fig:lgst}).
\end{itemize}

%%%%%%%%%%%%%%%%%%%%%%% State Tomography %%%%%%%%%%%%%%%%%%%%%%% 
\subsection{Quantum State Tomography}\label{sec:state_tomography}

\begin{figure}[t]
    \includegraphics[width=0.5\textwidth]{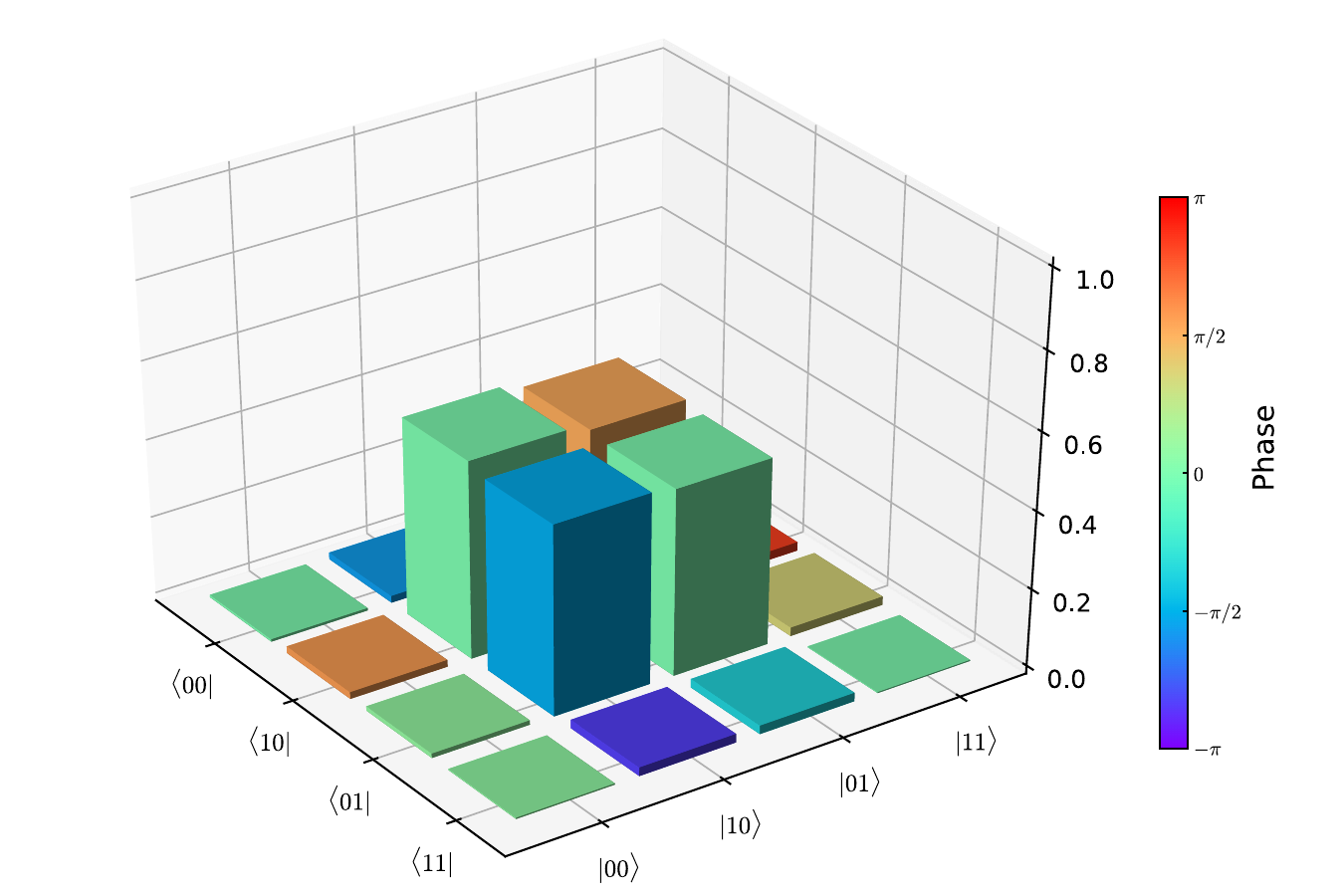}
    \caption{\textbf{Quantum State Tomography.} 
    The estimated density matrix of a two-qubit Bell state that was experimentally realized by applying a $\sqrt{\mathrm{iSWAP}}$ gate to the initial state $\ket{10}$, and then reconstructed using quantum state tomography~\cite{nguyen2022programmable}. The tomographic estimate indicated that the desired Bell state was prepared with high fidelity ($F\approx0.995$). Each bar represents a single element of the $4\times 4$ density matrix, with its height indicating the matrix element's absolute value and its color indicating complex phase.}
    \label{fig:qst}
\end{figure}

The goal of \emph{\ac{QST}} \cite{gale1968determination} is to accurately estimate the density matrix $\rho$ representing a quantum state. This cannot be done using just one copy of the unknown state $\rho$, because no single measurement will reveal $\rho$, and measuring a system disrupts its unmeasured properties. State tomography therefore requires \emph{many} ($N$) identically prepared systems. It is usually assumed, for simplicity, that these systems are identically and independently prepared, so that their joint state is $\rho^{\otimes N}$ for some unknown $\rho$. In real experiments, this assumption is only an approximation, and state tomography estimates the \emph{average reduced density matrix} of the $N$ sample systems. 

State tomography is performed by performing an \emph{informationally complete} measurement, or set of measurements, on the samples of $\rho$ (see \fig\ref{fig:spm_tomography}). Informational completeness is a property of a set of POVM effects $\{E_i\}$, and it does not matter whether those effects all come from a single POVM (i.e., $\{E_i\}$ is itself an experimentally performable POVM $M^\prime$) or a set of distinct measurements (e.g., $\{E_i\}$ is the union of the effects of several PVMs $\{M^\prime_1, M^\prime_2, \ldots\})$. A set of effects is informationally complete if (and only if) they span the vector space $\B(\mathcal{H})$ of operators. Since $\B(\mathcal{H})$ for a $d$-dimensional Hilbert space $\mathcal{H}$ is $d^2$-dimensional, a measurement or set of measurements must contain $d^2$ linearly independent effects to be informationally complete and enable state tomography. This is usually achieved by choosing at least $d+1$ \ac{PVM}s (orthogonal bases), but can in principle be achieved with a single $d^2$-element POVM. Informationally complete sets are not all created equal --- the accuracy of tomographic reconstruction is controlled by the condition number of $\{E_i\}$'s Gram matrix, and optimal accuracy is achieved by a \emph{2-design} \cite{Renes2004-pc}, such as a full set of mutually unbiased bases \cite{wootters1989optimal, adamson2010improving} or a symmetric informationally-complete POVM \cite{Renes2004-pc, stricker2022experimental}.

The basic principle of state tomography is very simple. By repeating measurements many times, we estimate the probability $p(E_i)$ of each effect. The simplest estimator $\hat{p}(E_i)$ is the number of times $E_i$ was observed divided by the number of times it could have occurred:
\begin{equation}
    \hat{p}(E_i) = \frac{n_i}{N_i} ~.
\end{equation}
By Born's rule, 
\begin{equation} \label{eq:statetomoBorn}
    p(E_i) = \Tr(E_i\rho) = \sbraket{E_i}{\rho} ~.
\end{equation} 
If the set $\{E_i\}$ spans $\B(\mathcal{H})$, then \eq\ref{eq:statetomoBorn} defines a set of linear equations that can be solved uniquely for $\rho$ (see \eq\ref{eq:statetomolinearinversion} below). This is state tomography.

To gain insight into how this is done in practice, consider the simplest case of a single-qubit state described by a $2\times 2$ density matrix (see Sec.~\ref{sec:density_matrix_formalism}). A single projective measurement is not sufficient to determine $\rho$, even if we repeat it $N \longrightarrow \infty$ times. Measuring, for example, $Z$ reveals only $p(\proj{0})$ and $p(\proj{1})$, from which we can deduce $\braket{Z}$ and $\braket{\Id}$ but not $\braket{X}$ or $\braket{Y}$. To estimate $\rho$, we also need to learn $\braket{X}$ and $\braket{Y}$, because a single-qubit $\rho$ is defined by a Bloch vector $\mathbf{r}\in\mathbb{R}^3$ in the Bloch ball (see \eq\ref{eq:bloch_vector} and \fig\ref{fig:Bloch_sphere_all}). 

We can obtain informationally complete data by dividing $N$ samples of $\rho$ into three groups, then measuring $X$ on each sample in the first group, $Y$ on the second group, and $Z$ on the third group. From the resulting count data, $\{\braket{X},\braket{Y},\braket{Z}\}$ can all be estimated. Then the density matrix can be reconstructed as
\begin{equation}
    \rho = \frac{1}{2}\left( I + \braket{X} X + \braket{Y} Y + \braket{Z} Z \right) ~.
\end{equation}
Often, the only native measurement is a $Z$-basis measurement. In this case, \emph{effective} measurements of $X$ and $Y$ are performed by (1) rotating the qubit using a $R_y(-\pi/2)$ or $R_x(\pi/2)$ operation (respectively), then (2) performing the native $Z$-basis measurement.

This informationally complete experiment can be described (as above) as a union of three distinct PVMs. But it can equally well be described as a \emph{single} POVM with 6 outcomes, $M = \left\{\frac13\proj{\psi_i}\right\}$, where $\ket{\psi_i}$ ranges over the 6 single-qubit Pauli eigenstates (i.e., the $\pm 1$ eigenstates of $X$, $Y$, and $Z$). This POVM can be implemented by drawing a uniformly random number $k$ from $1\ldots 3$ and performing the $k$th Pauli PVM (repeating both steps for each shot). State tomography experiments are sometimes described this way \cite{flammia2005minimal, filippov2011mutually}, representing a set of PVMs or POVMs as a single POVM, because it is simpler.

A straightforward extension of this single-qubit tomography protocol can be used to perform state tomography on an $n$-qubit system. To do single-qubit state tomography, we measured in 3 independent bases. For $n$-qubit state tomography, we need $3^n$ distinct ``measurement configurations'' or PVMs. Each measurement configuration corresponds to simultaneously measuring one of the 3 Paulis ($X,Y,Z$) on each of the $n$ qubits. So, for example, state tomography on $n=2$ qubits uses $9 = 3^2$ distinct configurations each corresponding to measuring one of $\{X,Y,Z\}$ on the first qubit and (independently) one of $\{X,Y,Z\}$ on the second. Each of the 9 measurements has 4 possible outcomes, which are represented by rank-1 POVM effects. For example, the 4 effects for measuring $Z$ on both qubits are $\{\proj{00},\proj{01},\proj{10},\proj{11}\}$. After performing this measurement on $N$ samples, it is straightforward to estimate each effect's probability as $\hat{p}_{ij} = n_{ij}/N$ (e.g., $\hat{p}_{00} = n_{00} / N$). Born's Rule says that $p_{ij} = \Tr( \proj{ij}\rho ) = \braket{\proj{ij}}$. So, we can transform these estimated probabilities into estimates of the expectation values of 3 Pauli observables just by writing out those Paulis as linear combinations of POVM effects,
\begin{align}
    ZZ &= \proj{00} - \proj{01} - \proj{10} + \proj{11} ~, \\
    IZ &= \proj{00} - \proj{01} + \proj{10} - \proj{11} ~, \\
    ZI &= \proj{00} + \proj{01} - \proj{10} - \proj{11} ~, \\
\end{align}
and then multiplying both sides by $\rho$ and taking the trace:
\begin{align}
    \braket{ZZ} &\simeq \frac{n_{00}}{N} - \frac{n_{01}}{N} - \frac{n_{10}}{N} + \frac{n_{11}}{N} ~, \\
    \braket{IZ} &\simeq \frac{n_{00}}{N} - \frac{n_{01}}{N} + \frac{n_{10}}{N} - \frac{n_{11}}{N} ~, \\
    \braket{ZI} &\simeq \frac{n_{00}}{N} + \frac{n_{01}}{N} - \frac{n_{10}}{N} - \frac{n_{11}}{N} ~.
\end{align}
The coefficients in this linear transformation from effect probabilities to Pauli expectation values correspond to the 2-bit Walsh-Hadamard transform (\eq\ref{eq:walsh_hadamard}), which can be defined for any $n$.

The expectation value of \emph{every} $n$-qubit Pauli operator in the $4^n$-element Pauli group $\mathbb{P}_n = \{I, X, Y, Z\}^{\otimes n}$ can be straightforwardly estimated from at least one of the $3^n$ measurement configurations defined above. Then, because the Pauli operators form a complete orthogonal basis for $\B(\mathcal{H})$, $\rho$ can be reconstructed as
\begin{equation}
    \rho = \frac{1}{2^n} \sum_{P \in \mathbb{P}_n} \braket{P} P ~.
\end{equation}
In the example of $n=2$ qubits, 
\begin{equation}
    \begin{split}
        &\rho = \frac{1}{4} ( II + \langle IX\rangle IX + \langle IY\rangle IY + \langle IZ\rangle IZ \\
        &+\langle XI\rangle XI + \langle XX\rangle XX + \langle XY\rangle XY + \langle XZ\rangle XZ \\
        &+ \langle YI\rangle YI + \langle YX\rangle YX + \langle YY\rangle YY + \langle YZ\rangle YZ \\
        &+ \langle ZI\rangle ZI + \langle ZX\rangle ZX + \langle ZY\rangle ZY + \langle ZZ\rangle ZZ ) ~.
    \end{split}
\end{equation}
\fig\ref{fig:qst} shows an experimental realization of this state tomography protocol to estimate a two-qubit Bell state.

This simple $n$-qubit tomography protocol illustrates two valuable points. First, the $n$-qubit measurements described above are PVMs -- their outcome effects are mutually orthogonal rank-1 projectors -- but they are \emph{not} measurements of individual Pauli observables. When we perform $Z$ measurements simultaneously on both qubits, that is not ``a measurement of $ZZ$.'' $ZZ$ is a Pauli operator with only two eigenvalues ($+1$ and $-1$) that each label a 2-dimensional eigenspace. A measurement of $ZZ$ is described by a rank-2 POVM with 2 outcomes that yields exactly 1 bit of information. In contrast, measuring both qubits in $Z$ yields 4 outcomes and 2 bits of information. This is a simultaneous measurement of multiple commuting Paulis, a.k.a.~a \emph{stabilizer} \cite{gottesman1997stabilizer}. The simultaneously measured observables, $ZI$ and $IZ$, generate a \emph{stabilizer group} (a maximal abelian subgroup of $\mathbb{P}_n$ containing $2^n$ commuting Pauli operators). Each of the $2^n$ Paulis in the stabilizer group can be written as a linear combination of the $2^n$ rank-1 projectors describing outcomes of the PVM measurement of the stabilizer. In this example, the Paulis $\{II, IZ, ZI, ZZ\}$ are linear combinations of $\{\proj{00},\proj{01},\proj{10},\proj{11}\}$). A good estimate of the outcome probabilities of a stabilizer measurement is sufficient to estimate \emph{all} the expectation values of the $2^n$ Paulis in the stabilizer. Note that some tomography theory papers do consider direct measurements of Pauli observables (which have 2 outcomes and reveal 1 expectation value) instead of stabilizers (which have $2^n$ outcomes and reveal $2^n-1$ expectation values). 

Second, the $3^n$ distinct stabilizer measurements are not mutually independent. Performing 9 measurements that each have 4 outcomes should reveal $27 = 9(4-1)$ independent probabilities. But these 27 distinct probabilities are not linearly independent. For example, in one measurement configuration we measure the stabilizer $\{ZI,IZ\}$, while in another we measure $\{ZI,IX\}$. Each yields 3 Pauli expectation values, but they are not linearly independent because $ZI$ appears in both stabilizers. In fact, $ZI$ appears in the stabilizers of three different measurement configurations, whereas $ZX$ only appears in one. As a result, this tomographic experiment yields more information about $ZI$ than $ZX$, and therefore higher precision in the estimate of $\braket{ZI}$ than $\braket{ZX}$. In the $n$-qubit case, a Pauli that acts as $\Id$ on $k$ of the $n$ qubits will appear in $3^{k}$ measurement configurations. As a result, this tomographic measurement protocol is both redundant (more measurement configurations than necessary) and heteroskedastic (some Pauli expectation values are estimated to much lower precision than others).

These features are unavoidable if each each qubit is measured independently. There is a 1:1 correspondence between weight-$n$ Paulis and measurement configurations, so removing even a single one of the $3^n$ measurement configurations breaks informational completeness (there is some Pauli $P \in \mathbb{P}_n$ whose expectation value cannot be estimated). However, by performing \emph{entangling} $n$-qubit measurements --- i.e., POVMs or PVMs whose effects are not tensor products of $n$ single-qubit projectors --- it is possible to construct a set of just $2^n+1$ PVMs that measure mutually unbiased bases \cite{wootters1989optimal}. This set is informationally complete, minimal, and enables more accurate tomography than the local measurement described above.

State tomography can be performed on any quantum system, e.g., $d$-dimensional systems where $d \neq 2^n$. The basic principle is the same: (1) define a set of measurements whose effects $\{E_i\}$ span $\B(\mathcal{H})$; (2) perform those measurements on $N$ samples of the unknown $\rho$; and (3) estimate $\rho$ by inverting Born's rule. The main complications are technical. The Pauli operators cannot be used, and the available operator bases for $d$-dimensional qudits are less convenient (see Appendix \ref{sec:weyl_gellman}). Mutually unbiased bases are not known (or believed) to exist unless $d = D^n$, where $D$ is prime. More details can be found in Appendix \ref{sec:qudit_state_tomography}.

%%%%%%%%%%%%%%%%%%%%%%% MLE %%%%%%%%%%%%%%%%%%%%%%% 
\subsubsection{Maximum Likelihood Estimation}\label{sec:mle}

The simple description of quantum state tomography in the previous section glosses over some key (if nonobvious) points:
\begin{enumerate}
    \item What should be done if we have measured \emph{more than $d^2$} distinct observables or POVM effects?  In this case, the equations in \eq\ref{eq:statetomoBorn} will \emph{over-constrain} $\rho$ and may have no solution.
    
    \item What should be done if solving \eq\ref{eq:statetomoBorn} yields an estimated state $\hat\rho$ that is not positive semidefinite?
\end{enumerate}
Both of these issues arise because finite-sample fluctuations (a.k.a. \emph{shot noise}) \footnote{Shot noise is not the only source of fluctuations and errors. Laboratory measurements are also subject to, for example, imperfect signal amplification, electronic noise, poor quantum efficiency, imperfect signal conversion, errors in digitization and classification, and a terrifying range of \emph{systematic} errors like drift over the duration of a tomography experiment. However, there is no systematic theoretical treatment of these noise sources. In practice, the techniques used to deal with shot noise (which \emph{does} have a solid theory) can deal with these noise sources too, although not optimally.} cause $\hat{\rho}$ to fluctuate randomly around the true $\rho$. We have ignored these fluctuations so far, implicitly assuming that the \emph{estimated} value of any observable (e.g., $\braket{Z}$) is equal to its true value. But this is not true in practice. As a result, tomography is a \emph{statistical} problem. The two issues highlighted above are solved by reformulating tomography not as a set of linear equations, but as a statistical inference problem.

The easiest way to address these issues is to treat \eq\ref{eq:statetomoBorn},
\begin{equation*}
    p(E_i) = \Tr[E_i\rho] = \sbraket{E_i}{\rho} ~,
\end{equation*}  
not as an exact linear inversion problem, but as a least-squares problem. These equations can be written in matrix form, by arranging the effect probabilities into a column vector $\vec{p} = [p(E_1), p(E_2), \ldots]^\trans$ and stacking the vectorized effects into a matrix
\begin{equation}
    T = \left(\begin{array}{c} \sbra{E_1} \\ \sbra{E_2} \\ \vdots \end{array}\right)
\end{equation}
so that
\begin{equation} \label{eq:statetomoBorn2}
    \vec{p} = T\sket{\rho} ~.
\end{equation}
Now, if these equations have a unique solution $\hat{\rho}$, it is given by
\begin{equation}\label{eq:statetomolinearinversion}
    \sket{\hat{\rho}} = T^{-1}\vec{p} ~.
\end{equation}
This is linear inversion state tomography, in a single equation.

If we have measured more than $d^2$ observable probabilities, then $T$ will not be square, and therefore not invertible. But if we admit that the estimated probabilities $\hat{p}(E_i)$ will fluctuate around the true probabilities, then we can reformulate \eq\ref{eq:statetomoBorn2} as a least-squares problem and seek the $\sket{\hat{\rho}}$ that minimizes
\begin{equation} \label{eq:statetomoLSobjective}
    \left\|\vec{p} - T\!\sket{\rho}\right\|_2^2 = \sum_i{\left[\hat{p}(E_i) - \Tr (E_i\hat{\rho}) \right]^2} ~.
\end{equation}
This actually has a closed-form solution, in terms of the \emph{Moore-Penrose pseudo-inverse} of the matrix $T$:
\begin{equation}
    \sket{\hat{\rho}} = T^{+}\vec{p} ~,
\end{equation}
where the pseudoinverse is defined as
\begin{equation}
    T^{+} \equiv (T^\trans T)^{-1}T^\trans ~.
\end{equation}
The second issue (what if $\hat{\rho} \not\geq 0$) can also be solved by reformulating inversion as an ordinary least-squares problem, by \emph{constraining} the optimization of \eq\ref{eq:statetomoLSobjective} to positive semidefinite $\rho\geq0$. This is a tractable convex optimization problem. 
It can be solved by \emph{projecting} the unconstrained linear inversion estimate onto the nearest positive semidefinite state \cite{smolin2012efficient}, an algorithm known as \emph{\ac{PLS}} \cite{guta2020fast,surawy2022projected}.

However, these \emph{least-squares tomography} estimators are ad-hoc solutions, and not optimal in any sense except simplicity. They can be seen as approximations to a statistically well-motivated approach called \emph{\ac{MLE}} \cite{hradil1997quantum, banaszek1999maximum}. MLE is a simple, widely used method for statistical inference --- i.e., the estimation of unknown parameters from data --- in which the estimated values of the unknown parameters are the ones that maximize the probability of observing the data that were actually observed. The probability of the observed data, as a function of the unknown parameters, is called the \emph{likelihood function}:
\begin{equation}
    \mathcal{L}(\mathbf{\theta}) = \mathrm{Pr}(D_{\mathrm{observed}}|\mathbf{\theta}) ~.
\end{equation}
Here, $\mathbf{\theta}$ is a vector of parameters whose values we would like to estimate, and $D_{\mathrm{observed}}$ is the actual data that have been observed. The likelihood function is a compressed, efficient representation of the information that $D_{\mathrm{observed}}$ provides about the unknown $\mathbf{\theta}$. The maximum likelihood estimate of $\mathbf{\theta}$ is
\begin{equation}
    \hat{\mathbf{\theta}}_{\mathrm{MLE}} = \mathrm{argmax}[\mathcal{L}(\mathbf{\theta})] ~.
\end{equation}

In quantum state tomography, the statistical model is Born's rule [$p(E_i) = \Tr(E_i\rho)$], and its parameters are the matrix elements of $\rho$. The observed data can be described very simply by a set of POVM effects $\{E_i\}$ and the number of times each effect has been observed, $\{n_i\}$. The likelihood function is
\begin{align}
    \mathcal{L}(\rho) &= \mathrm{Pr(D_{\mathrm{observed}}|\rho)} ~, \\
    &= \prod_{i}\Tr(E_i\rho)^{n_i} ~.
\end{align}
The maximum likelihood estimate, $\hat{\rho}_{\mathrm{MLE}}$, is simply the density matrix $\rho$ that maximizes $\mathcal{L}(\rho)$. No general closed-form solutions exist, but finding $\hat{\rho}_{\mathrm{MLE}}$ is a tractable convex optimization problem because the argmax of $\mathcal{L}$ is also the argmax of the log-likelihood function $\log \mathcal{L}(\rho)$,
\begin{equation}
    \log \mathcal{L}(\rho) = \sum_i{ n_i\log(\Tr[E_i \rho])} ~,
\end{equation}
which is concave downward. A variety of numerical algorithms can be used to find the maximum of $\log \mathcal{L}(\rho)$, \emph{constrained} to the convex subset of Hermitian matrices that satisfy (1) $\Tr(\rho)=1$ and (2) $\rho \geq 0$. The trace constraint is a straightforward linear (holonomic) constraint, but the positivity constraint is trickier. One way to enforce it is by using the nonlinear parameterization
\begin{equation}
\rho = L L^\dagger / \Tr(L L^\dagger),
\end{equation}
which guarantees both $\rho\geq 0$ and $\Tr(\rho)=1$. If $L$ is restricted to (complex) lower triangular matrices with real diagonal elements, it is the unique Cholesky factorization of $\rho$. However, in this parameterization $\log \mathcal{L}(L)$ is not necessarily convex.

In certain circumstances, $\hat{\rho}_{\mathrm{MLE}}$ coincides \emph{exactly} with the linear inversion estimate $\hat{\rho}$ from \eq\ref{eq:statetomolinearinversion}. If the $T$ matrix from \eq\ref{eq:statetomolinearinversion} is invertible (i.e., the tomographic data is informationally complete, but not \emph{overcomplete}), and we ignore the positivity constraint $\rho\geq 0$ and extend the likelihood function to all Hermitian trace-1 $\rho$ for which $\mathcal{L}(\rho) \geq 0$ (including matrices that are not positive semidefinite), then $\mathcal{L}(\rho)$ achieves its maximum uniquely at the linear inversion $\hat{\rho}$. It follows that \emph{if} $\hat{\rho} \geq 0$, then it is the MLE. This can provide significant time savings if/when $\hat{\rho} \geq 0$, because computing \eq\ref{eq:statetomolinearinversion} is often much faster than finding $\hat{\rho}_{\mathrm{MLE}}$ numerically..

If $\hat{\rho}$ is not positive, then it follows that $\hat{\rho}_{\mathrm{MLE}}$ must lie on the boundary of the set defined by $\rho \geq 0$ -- i.e., it will have at least one zero eigenvalue \cite{blume2010optimal}. In this case, it is sometimes useful to \emph{approximate} $\hat{\rho}_{\mathrm{MLE}}$ efficiently by observing that because the linear inversion $\hat{\rho}$ is the maximum of the unconstrained likelihood, $\log \mathcal{L}$ is necessarily quadratic in a neighborhood of $\hat{\rho}$. If the Hessian of $\log \mathcal{L}$ around $\hat{\rho}$ can be efficiently computed, then constrained weighted least-squares optimization (instead of generic convex optimization) algorithms can be used to find the $\rho \geq 0$ that maximizes the quadratic approximation to $\log \mathcal{L}$. This approach is sometimes (confusingly) described as ``maximum likelihood estimation'' in the literature; in fact, it is only an approximation to MLE.

%%%%%%%%%%%%%%%%%%%%%%% Bayesian %%%%%%%%%%%%%%%%%%%%%%% 
\subsubsection{Bayesian Tomography}\label{sec:BayesianTomography}

Other statistical inference approaches can be applied to tomography. MLE is merely the most common. The most well-studied alternative is \emph{Bayesian} estimation \cite{buvzek1998reconstruction, blume2010optimal, granade2016practical}. Bayesian statistical inference \cite{caves2002quantum,von2011bayesian} differs from \emph{frequentist} approaches like MLE by assuming (or asserting) that ignorance about an unknown parameter --- e.g., a quantum state --- can and should be described by a \emph{probability distribution} over it. Therefore, the standard Bayesian approach to estimating $\rho$ starts by assigning a \emph{prior probability distribution} (``prior'') $\mu_\mathrm{prior}(\rho)$ to the unknown parameter. The observed data $D$ are used to perform \emph{Bayesian update} on the prior to get a \emph{posterior} distribution by applying \emph{Bayes' Rule}:
\begin{equation}\label{eq:Bayes}
    \mu_\mathrm{post}(\rho) = \frac{\mathrm{Pr}(D|\rho)\mu_\mathrm{prior}(\rho)}{\mathrm{Pr}(D)} = \frac{\mathcal{L}(\rho) \mu_\mathrm{prior}(\rho)}{\int_\rho{\mathcal{L}(\rho)\mu_\mathrm{prior}(\rho)}} ~,
\end{equation}
where $\mathcal{L}(\rho) = \mathrm{Pr}(D|\rho)$ is the same likelihood function whose maximum defines the MLE. The most common estimator, the \emph{\ac{BME}} \cite{blume2010optimal}, is the mean of the posterior:
\begin{equation}
    \hat{\rho}_{\mathrm{BME}} = \int_\rho{ \rho \mu_{\mathrm{post}}(\rho)} ~.
\end{equation}
Other estimators are possible. However, despite common misconception, the posterior \emph{mode} --- i.e., the maximum of $\mu_{\mathrm{post}}$ --- is not a valid estimator. Because $\mu_{\mathrm{post}}$ is a continuous \emph{measure} over a continuous space, not a function, the posterior probability of any single $\rho$ is precisely zero. For this reason, the posterior cannot have a maximum at any $\rho$, and \emph{\ac{MAP}} estimators do not exist. (MAP estimators exist iff the posterior distribution is discrete, but this scenario defines quantum state \emph{discrimination} \cite{barnett2009quantum, bae2015quantum} rather than tomography.) Attempts to construct MAP estimators (e.g., \cite{siddhu2019maximum}) actually yield a variant of \emph{hedged maximum likelihood} estimation \cite{blume2010hedged}.

Bayesian tomography generally requires more computation than MLE, because sampling or integrating a posterior distribution is usually harder than maximizing a convex likelihood function. They also depend critically on the choice of prior, which is a double-edged sword --- it is harder to claim ``objective'' results, but very straightforward to take optimal advantage of ``informative'' prior information about partially-known states. Bayesian estimates are often better behaved and more accurate than frequentist ones \cite{blume2010optimal}, and they simplify adaptive tomography \cite{huszar2012adaptive, kravtsov2013experimental, granade2017practical} and uncertainty quantification \cite{christandl2012reliable, oh2019efficient}. Software implementations of efficient algorithms \cite{granade2016practical, granade2017practical, lukens2020practical} have made Bayesian tomography of states \emph{and} other objects (e.g., gate sets \cite{DiMatteo2020operationalgauge, evans2022fast}) more feasible and accessible.

%%%%%%%%%%%%%%%%%%%%%%% QPT %%%%%%%%%%%%%%%%%%%%%%% 
\subsection{Quantum Process Tomography}\label{sec:qpt}

\begin{figure*}[t]
    \includegraphics[width=\textwidth]{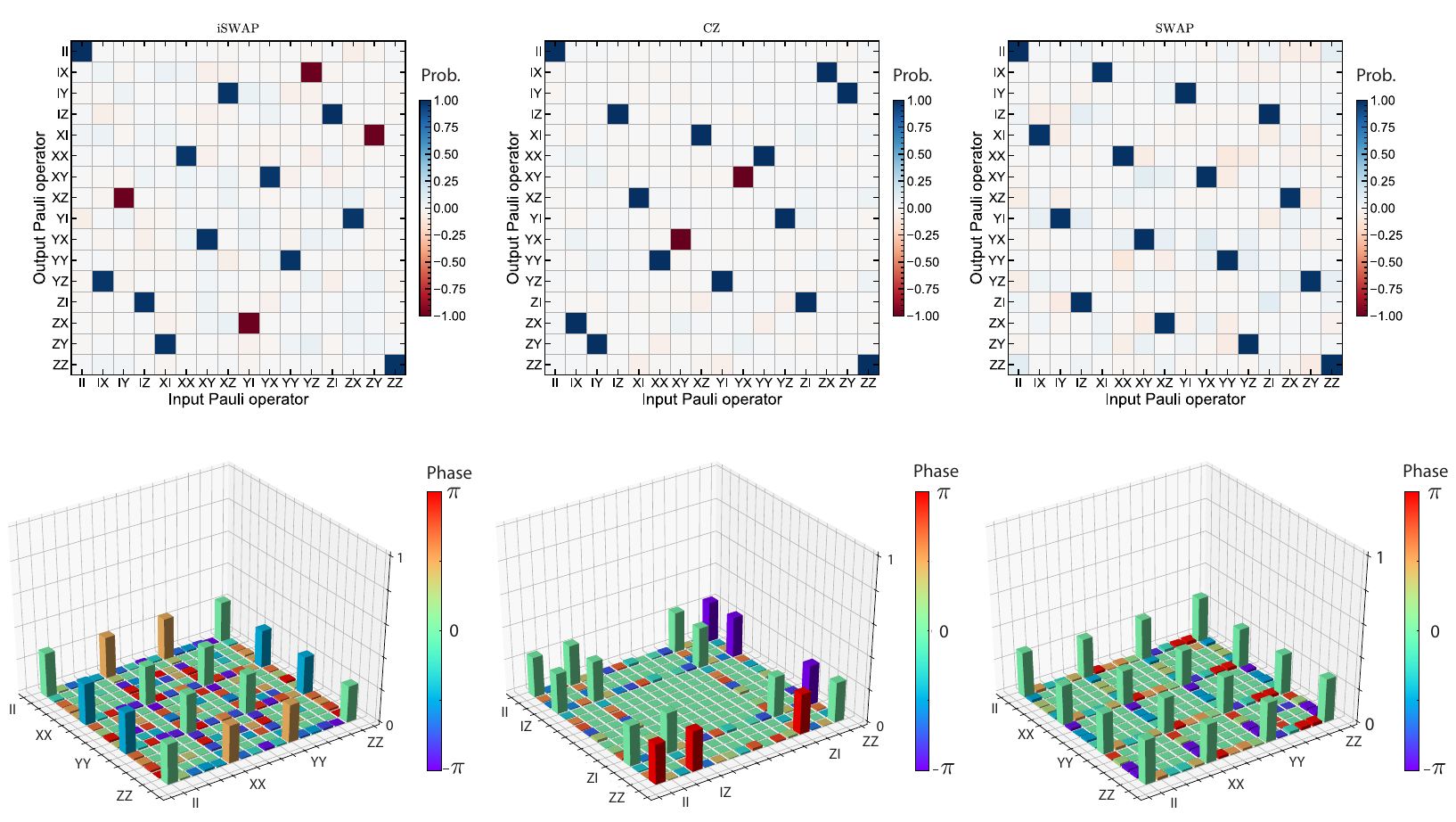}
    \caption{\textbf{Quantum Process Tomography.} 
     Top row: PTMs for experimental $\mathrm{iSWAP}$, $\mathrm{CZ}$, and $\mathrm{SWAP}$ gates reconstructed using QPT \cite{nguyen2022programmable}. Note that all of the values of a PTM are real and bounded between [-1, 1]. Bottom row: $\chi$ (process) matrices for the $\mathrm{iSWAP}$, $\mathrm{CZ}$, and $\mathrm{SWAP}$ gates. The $\chi$-matrix is a complex matrix, so each matrix element’s magnitude is represented by the height of the corresponding column, and its phase is represented by the column’s color. The process fidelities of the gates are $99.32(3)\%$, $99.72(2)\%$, and $98.93(5)\%$, respectively.
     }
    \label{fig:qpt}
\end{figure*}

Tomography can also be used to reconstruct (estimate) the \ac{CPTP} map that best describes a quantum operation (e.g., a logic gate). This is called \emph{\ac{QPT}} \cite{chuang1997prescription, poyatos1997complete}. A CPTP map is a linear map on density matrices, a.k.a.~a \emph{superoperator} acting on $\mathcal{B}(\mathcal{H})$ (see Sec.~\ref{sec:rep_quant_proc}). In QPT, a CPTP map to be estimated is generally represented either as a \emph{transfer matrix} $\Lambda$ that acts on a vectorized density matrix $\sket{\rho}$ by matrix multiplication (see Sec.~\ref{sec:superop}),
\begin{equation}
    \sket{\rho} \mapsto \Lambda \sket{\rho}~,
\end{equation}
or as a $\chi$ matrix describing
\begin{equation}
    \rho \mapsto \sum_{i,j} \chi_{i,j} P_i \rho P_j ~,
\end{equation}
where $\{P_i\}$ are a basis (often the Pauli basis) for $\mathcal{B}(\mathcal{H})$. The goal of QPT is to estimate a complete mathematical description of the transfer matrix $\Lambda$ or process matrix $\chi$. Since $\Lambda$ and $\chi$ are equivalent (see Sec.~\ref{sec:rep_quant_proc}), analyses of QPT usually just pick whichever representation is more convenient for the specific protocol being described. We will follow the same convention here.

QPT is performed by choosing an informationally complete set of input states $\{\rho_j^\prime\}$ \emph{and} an informationally complete set of measurements $\{M_i^\prime\}$. The measurements used for QPT must satisfy exactly the same criteria as those used for QST (see \fig\ref{fig:spm_tomography}), and the input states must collectively span $\B(\mathcal{H})$. Like QST, QPT is very simple in principle. Suppose that $\{\rho_j^\prime\}$ form an informationally complete set of states, so that $\{\,\bigsket{\rho_j^\prime}\,\}$ span $\B(\mathcal{H})$, and $\{E_i\}$ (the union of all the effects of the measurements $\{M_i^\prime\}$) form an informationally complete set of effects, so that $\{\,\bigsbra{E_i}\,\}$ also span $\B(\mathcal{H})$. It follows that the set of superoperators $\{\,\bigsket{\rho_j^\prime}\bigsbra{E_i}\,\}$ (for all $i,j$) span the entire space of superoperators. Now, we prepare many copies of every $\rho_j^\prime$, apply the unknown process to all of them, and then divide the (processed) copies of $\rho_j^\prime$ into groups labeled by $i$ and perform measurement $M_i^\prime$ on the $i$th group. By doing so, we can estimate every probability
\begin{equation} \label{eq:processtomoBorn}
%    p_{i,j} = \mathrm{Pr}(E_j | \rho_i^\prime) = \sbra{E_j}{\Lambda} \sket{\rho_i^\prime} = \Tr[ \Lambda \sket{\rho_i^\prime} \sbra{E_j}] ~.
    p_{i,j} = \mathrm{Pr}(E_i | \rho_j^\prime) = \Bigsbra{E_i}{\,\Lambda\,} \Bigsket{\rho_j^\prime} = \Tr[ \Lambda\, \Bigsket{\rho_j^\prime} \Bigsbra{E_i}] ~.
\end{equation}
This defines a (large!) set of linear equations that can be solved for $\Lambda$ in terms of measurable probabilities $p_{i,j}$.

QPT on a system described by a $d$-dimensional Hilbert space requires at least $d^2$ distinct input states and enough measurement configurations to perform QST (see Sec.~\ref{sec:state_tomography}). If only rank-1 PVMs (orthogonal basis measurements) are used, this requires at least $d+1$ distinct measurement configurations, for a total of $d^2(d+1)$ distinct state/measurement configurations, to estimate the $d^4-d^2$ free parameters of the unknown process. For the special case of $n$ qubits, where $d = 2^n$, this works out to at least $8^n + 4^n$ distinct state/measurement configurations. Achieving this bound requires entangling measurements. If only simultaneous single-qubit measurements are used, then $n$-qubit process tomography requires $3^n$ measurement configurations, and thus at least $12^n$ state/measurement configurations. The experimental complexity of QPT grows rapidly for $n$ qubits!

\begin{figure*}[t]
    \includegraphics[width=0.8\textwidth]{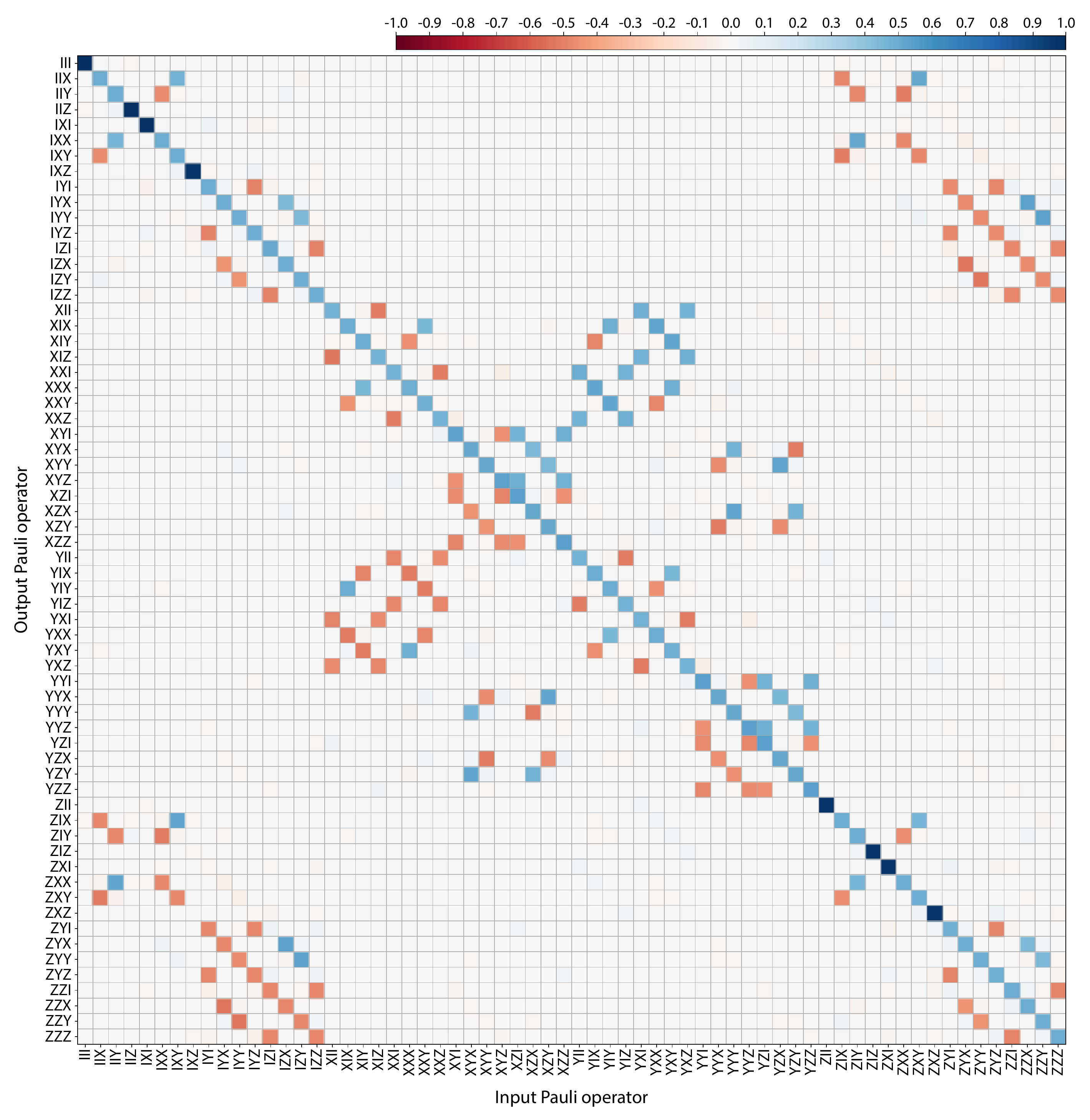}
    \caption{\textbf{Three-qubit QPT.}
    PTM of a three-qubit iToffoli gate \cite{kim2022high}. A three-qubit PTM contains $16^{3} - 4^{3} = 4032$ independent parameters, which can be estimated from a minimum of $4^3 \cdot 3^3 = 1728$ independent experiments. The process fidelity is estimated to be $97.1(8)\%$.}
    \label{fig:iToff_qpt}
\end{figure*}

In principle, analysis of QPT data is as simple as ``solve \eq\ref{eq:processtomoBorn} for $\Lambda$.'' But, as the discussion of QST in Sec.~\ref{sec:state_tomography} illustrates, there are many ways to solve these equations. All of the complications discussed in the context of QST also appear for QPT, which is very nearly isomorphic to QST because of the Choi-Jamiołkowski isomorphism. Much of the QPT literature consists of taking a new QST algorithm (e.g., MLE) and adapting it to QPT. In this tutorial, we do not attempt to explore this literature in detail. Instead, we explain one simple approach to linear-inversion QPT in detail. We assume an $n$-qubit system, but this analysis generalizes straightforwardly to qudits (see Appendix \ref{sec:qudit_qpt}) using a non-Pauli operator basis (see Appendix \ref{sec:weyl_gellman}).

Suppose that by performing QPT experiments we have estimated each of the probabilities in \eq\ref{eq:processtomoBorn} as 
\begin{equation}
    \hat{p}_{i,j} \simeq  \Bigsbra{E_i}{\,\Lambda\,} \Bigsket{\rho_j^\prime} ~.
\end{equation}
To solve for $\Lambda$, we begin by arranging these probabilities into a matrix $\mathsf{P}$ so that $\mathsf{P}_{ij} = \hat{p}_{i,j}$. Next, we construct two matrices by vectorizing the input states and effects in the Pauli operator basis,
\begin{equation}
    B = \begin{pmatrix}
        \ket{\rho_1' \rangle}, 
        \ket{\rho_2' \rangle}, 
        ..., 
        \ket{\rho_{N}' \rangle}
    \end{pmatrix}~,
\end{equation}
and
\begin{equation}
    A = \begin{pmatrix}
        \bra{\langle E_1} \\
        \bra{\langle E_2} \\
        \vdots \\
        \bra{\langle E_N} \\
    \end{pmatrix} ~.
\end{equation}
Choosing the Pauli basis makes the final $\Lambda$ a Pauli transfer matrix (PTM; see Sec.~\ref{sec:ptm_rep}). The elements of the input (B) and output (A) matrices are then
\begin{align}
    B_{ij} &= \frac{1}{\sqrt{d}}\Tr[P_i \rho_j] ~, \\
    A_{ij} &= \frac{1}{\sqrt{d}} \Tr[E_i P_j] ~.
\end{align}
Now, using $A$, $B$, and $\mathsf{P}$, we can express \eq\ref{eq:processtomoBorn} as
\begin{equation}
    \mathsf{P} = A \Lambda B ~,
\end{equation}
and extract $\Lambda$ by inverting the $A$ and $B$ matrices~\cite{chow2012universal, corcoles2013process}:
\begin{equation}
    \Lambda = A^{-1} \mathsf{P} B^{-1} ~.
\end{equation}
When the input states and/or effects are overcomplete, a Moore-Penrose pseudoinverse can be used instead \cite{blume2024easy}.

An equivalent approach, which offers complementary intuitions about QPT, is to use the data to perform quantum state tomography on $\Lambda\sket{\rho'_j}$ for each input state $\rho'_j$. Then, if we write both the input and output states as vectors in the Pauli basis, we can simply solve for the PTM $\Lambda$ using least-squares fitting. \fig\ref{fig:qpt} shows PTMs and $\chi$ matrices reconstructed this way from experimental QPT on 2-qubit $\mathrm{iSWAP}$, $\mathrm{CZ}$, and $\mathrm{SWAP}$ gates implemented on a superconducting quantum processor. A minimal set of input states $\{\proj{0},\proj{1},\proj{+},\proj{-}\}^{\otimes 2}$ was used.

Adding a third qubit increases the number of experimental configurations required from 144 to 1728, illustrating the rapid growth of QPT's experimental complexity with $n$. \fig\ref{fig:iToff_qpt} shows the PTM for a 3-qubit iToffoli gate, which contains 4032 independent parameters estimated from 1728 distinct experiments. 

As we noted at the beginning of Sec.~\ref{sec:tomography}, both QST and QPT are vulnerable to SPAM errors, which cause systematic bias in the estimate. Therefore, the process fidelities quoted in Figs.~\ref{fig:qpt} and \ref{fig:iToff_qpt} do not separate gate errors from SPAM errors. For this reason, experimental tomography of gates has moved toward tomographic reconstruction methods that characterize SPAM errors and gate errors simultaneously and self-consistently, such as gate set tomography (Sec.~\ref{sec:gst}). A variation of QPT that leverages ideas from gate set tomography to eliminate or correct SPAM bias has also been proposed \cite{blume2024easy}.

The examples above illustrated \emph{linear-inversion} QPT. But, just as density matrices reconstructed using linear-inversion QST can easily violate the positivity constraint $\rho\geq0$, superoperators reconstructed using linear-inversion QPT can  violate complete positivity (CP; see discussion in Sec.~\ref{sec:rep_quant_proc}). There are many ways to constrain a reconstructed PTM or process matrix to be CP, including MLE \cite{chow2012universal, mitchell2003diagnosis, o2004quantum} or projection algorithms \cite{knee2018quantum}. CP-constrained MLE can be done in a variety of ways (e.g., via semi-definite programs \cite{chow2012universal}), but the easiest approach to understand uses the Choi-Jamiołkowski isomorphism. In this approach, the process is parameterized by its $\chi$ matrix, which is isomorphic to a density matrix on a larger system. Now, MLE can be performed using algorithms designed for state tomography (although an additional constraint on the $\chi$ matrix, corresponding to trace preservation, must be added).

%%%%%%%%%%%%%%%%%%%%%%% Measurement Tomgoraphy %%%%%%%%%%%%%%%%%%%%%%% 
\subsection{Quantum Measurement Tomography}\label{sec:meas_tomography}

The goal of \emph{\ac{QMT}} is to estimate the parameters of a mathematical model for a quantum measurement operation. For the most common case where the measurement is modeled by a POVM $M = \{E_i\}$ (see Sec.~\ref{sec:povm}), QMT requires applying the unknown measurement to an informationally complete set of input states $\{\rho_j^\prime\}$ (see \fig\ref{fig:spm_tomography}). So, QMT can be seen as the complement to QST (Sec.~\ref{sec:state_tomography}): where unknown states are estimated by performing a range of known measurements on them, an unknown measurement is estimated by applying it to a range of known input states. 

A POVM describing an $m$-outcome measurement on a $d$-dimensional quantum system comprises $m$ $d\times d$ effects $E_i\geq 0$. Estimating such a POVM requires repeatedly applying it to an informationally complete set of at least $d^2$ input states whose density matrices span $\B(\mathcal{H})$. For an $n$-qubit system, $4^n$ linearly independent input states are required. 

The most common QMT procedure is to choose $4^n$ linearly independent input states from the $6^n$ tensor products of single-qubit Pauli eigenstates, prepare $N$ samples of each, apply the unknown measurement to all the samples, record the outcome statistics, and estimate the probabilities
\begin{equation} \label{eq:QMTprobs}
    p(i|j) = \mathrm{Pr}(E_i|\rho_j^\prime) = \Tr[E_i\rho_j^\prime] = \sbraket{E_i}{\rho_j^\prime} ~.
\end{equation}
There is no uniquely good way to select a subset of $4^n$ Pauli eigentates. Optimal accuracy is achieved when the input states form a 2-design, but even for a single qubit, achieving this optimum requires either (1) choosing non-Pauli eigenstates such as a SIC-POVM \cite{Renes2004-pc}, or (2) using all 6 Pauli eigenstates. 

Analysis of QMT data proceeds identically to QST and QMT. Equation \ref{eq:QMTprobs} is solved using the same techniques and methods --- e.g., linear inversion, least-squares, or MLE --- to find each $E_i$, and thus the entire unknown POVM $M$. Measurement tomography implies slightly different constraints than state or process tomography; each effect $E_i$ must be positive semidefinite, but the analogue of the trace or \ac{TP} constraint is that $\sum_i{E_i} = \Id$. This requires technical changes to constrained MLE algorithms, but no conceptual novelty~\cite{fiuravsek2001maximum, lundeen2009tomography}. 

Tomographic estimation of mid-circuit measurements requires a different model. POVMs model \emph{terminating} measurements, and mid-circuit measurements are modeled not by POVMs, but by \emph{quantum instruments} (\eq\ref{eq:QI}). Tomographic reconstruction of quantum instruments is a reasonably straightforward fusion of QPT and POVM tomography, and the interested reader is referred to Refs.~\cite{blumoff2016implementing, pereira2022complete, pereira2023parallel}.

\begin{figure}[t]
    \includegraphics[width=0.4\textwidth]{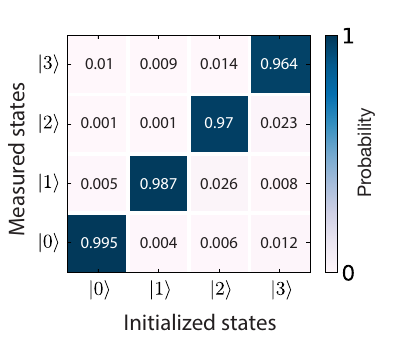}
    \caption{\textbf{Response Matrix.} Heralding and measuring the states of a qudit of dimension $D=4$ yields the probabilities $p(i|j)$ of measuring state $|i\rangle$ after preparing state $|j\rangle$. These constitute the response matrix~R. (The data are reproduced with permission from Ref.~\cite{nguyen2023empowering}.)}
    \label{fig:rmtrx}
\end{figure}

Sometimes, it is adequate or desirable to use a simpler  model for terminating measurements instead of a full POVM. A \emph{response} (or \emph{confusion}) matrix provides such a model when the unknown measurement is intended to measure a canonical basis (e.g., the computational basis). The response matrix $R$ is a $d\times d$ matrix whose elements $R_{ij} = p(E_i|\rho_j)$ are the probabilities of getting outcome $E_i \approx \proj{i}$ if state $\rho_j \approx \proj{j}$ is prepared and measured. For example, a single-qubit response matrix is given by
\begin{equation}\label{eq:response_matrix}
    R=\begin{pmatrix}
    p(0|0) &&\! p(0|1) \\
    p(1|0) &&\! p(1|1)
    \end{pmatrix} ~.
\end{equation}
$R$ can be estimated in the obvious way, by preparing each basis state as accurately as possible, applying the measurement, and recording the empirical probabilities of each outcome, as shown in Fig.~\ref{fig:rmtrx}. Response matrix estimation requires preparing only $d$ input states, whereas QMT estimation of a POVM requires $d^2$. It provides a subset of the information in a POVM. For example, it can be used to compute the readout fidelity defined in \eq\ref{eq:ro_fidelity}, which is given by $\Tr(R)$, but it does not predict the results of performing the measurement on superpositions of the canonical basis states. Like QST and QPT, QMT's experimental complexity grows exponentially with the number of qubits $n$, making it effectively infeasible for more than a few qubits. However, measurements can be characterized much more efficiently if a valid ansatz applies. 

A common and powerful ansatz is to assume that crosstalk in multi-qubit readout is negligible. 
%(This is already assumed in the response matrix model of the previous paragraph --- readout crosstalk can invalidate the response matrix model of readout \cite{beale2023randomized}). 
If correlated readout errors are negligible, then a multi-qubit response matrix $R$ or POVM $M$ can be approximated as the tensor product of each individual qubit's $R$ or $M$ \cite{bravyi2021mitigating}. This holds true for mid-circuit measurements modeled by quantum instruments as well. Crosstalk-free models of measurements can generally be characterized using a set of input states whose size scales just \emph{linearly} with the number of qubits. But because crosstalk effects are often non-negligible~\cite{blumoff2016implementing, chen2019detector, pereira2023parallel}, crosstalk-free models should be viewed with caution as a potentially-useful approximation. Full POVM characterization is usually necessary for accurate assessment and modeling. Some forms of crosstalk (and other errors in measurement) can be effectively eliminated by ``twirling'' readout error into stochastic bit flip channels \cite{beale2023randomized, hashim2023quasi}. Twirled measurements display less crosstalk, and are almost always more consistent with the response matrix model, than native measurements.

%%%%%%%%%%%%%%%%%%%%%%% GST %%%%%%%%%%%%%%%%%%%%%%% 
\subsection{Gate Set Tomography}\label{sec:gst}

State, process, and measurement tomography are powerful tools for diagnosing errors in a quantum processor. However, each of these protocols implicitly assumes the existence of a pre-calibrated \emph{reference frame} (see Fig.~\ref{fig:spm_tomography}) of perfect states and/or measurements. Errors in the operations that define such a reference frame can bias the tomographic reconstructions, and lead to incorrect models for the operations under test. \emph{\Ac{GST}} \cite{nielsen2021gate, greenbaum2015introduction} is a family of calibration-free approaches to tomography that explicitly acknowledge that all elements of a quantum computer's \emph{gate set} --- the native state preparations, measurements, and logic gates (see Sec.~\ref{sec:gate_sets}) --- are subject to errors. GST protocols are able to reconstruct self-consistent mathematical representations of a quantum computer's native gate set and the the errors afflicting it. 

Around 2012, groups at IBM \cite{Merkel2013gst} and Sandia National Laboratories \cite{Blume-Kohout2013gst} independently identified the need for calibration-free characterizations of quantum operations. IBM approached this problem using a so-called ``overkill'' tomography protocol that utilizes all circuits of depth 3 or less and fits a gate set model with MLE. Sandia's ``linear GST'' method uses similar circuits to standard QPT and fits a model with linear inversion. Variants of these early protocols are still in use to some extent, but since their introduction the family of GST protocols has evolved significantly. It now encompasses a rather broad set of experiment design and data analysis techniques for self-consistently estimating the parameters of a gate set model (see Sec.~\ref{sec:gate_sets}). In this tutorial, we limit our discussion to two essential protocols: \emph{linear GST}, mentioned above, and \emph{long-sequence GST}, which uses long, structured quantum circuits and iterative MLE. Significant extensions to these protocols \cite{PRXQuantum.4.010325, PRXQuantum.2.030328} have introduced approaches for characterizing larger processors and devices with mid-circuit measurements \cite{rudinger2022characterizing}, and other work \cite{KalmanGST} introduced an efficient Kalman filter formalism for GST that replaces the MLE analysis with a Bayesian estimator capable of streaming, real-time data processing. Experimental implementations of GST can be found in many papers, including (but not limited to) \cite{Blume-Kohout2013gst, dehollain2016optimization, Blume_Kohout_2017, PRXQuantum.2.040338, mkadzik2021precision, xue2022quantum, hashim2023benchmarking}.

We use the term ``GST'' without qualification to indicate ``long-sequence GST.''  Reference implementations of linear and long sequence GST can be found in the \texttt{pyGSTi} python package \cite{Nielsen2020pygsti}. Throughout the rest of this subsection, we use the term ``gate set'' to describe both the ensemble of logical instructions available on a given quantum computer (e.g., ``prepare $\ket{0}$,'' ``Hadamard gate on qubit 3,'' ``measure qubit 1,'' etc.), and the mathematical representations of those objects (e.g., density matrices, transfer or process matrices, and POVM elements). In discussing those mathematical objects, we follow the conventions of \eq\ref{eq:gate_set_model_vectorized} for defining a gate set $\mathcal{G}$:
% \begin{equation}
%     \mathcal{M} = \Big\{ \{ | \rho^{(i)} \rangle \rangle \}_{i = 1}^{N_\rho} , \{ G(g_j) \}_{j=1}^{N_G}, \{ \langle \langle E_i^{m} | \}_{m=1, i=1}^{N_M, N_E^{(m)}} \Big\} \tag{\ref{equ:gate_set_model}}
% \end{equation}
\begin{equation}
    \mathcal{G} = \left\{
        \left\{ \sket{\rho^{(i)}} \right\}_{i=1}^{N_\rho},\;
        \left\{ G_i \right\}_{i=1}^{N_{\mathrm{G}}},\;
        \left\{ \sbra{E_i^{(m)}} \right\}_{m=1,i=1}^{N_{\mathrm{M}},N_{\mathrm{E}}^{(m)}}
    \right\} ~. \tag{\ref{eq:gate_set_model_vectorized}}
\end{equation}
Here, $N_\rho$ is the number of native state preparations, $N_G$ is the number of native gates, $N_M$ is the number of native measurements, and $N_E^{(m)}$ is the number of outcomes for the $m$\textsuperscript{th} native measurement. In most cases, $N_\rho = N_M = 1$.

Gate set models often have many parameters, and their size grows very rapidly with the number of qubits $n$. Each operation in the gate set is a matrix whose dimension grows exponentially with $n$. Moreover, the total number of possible $n$-qubit operations can grow \emph{combinatorially}. Fitting a model with exponentially many parameters requires experimental complexity (and time) that also grows exponentially with $n$. For these reasons, standard GST protocols have so far been applied only to one- and two-qubit systems. 

%%%%%%%%%%%%%%%%%%%%%%% Linear GST %%%%%%%%%%%%%%%%%%%%%%% 
\subsubsection{Linear GST}

\begin{figure}[!ht]
    \centering
    \includegraphics[width=0.45\textwidth]{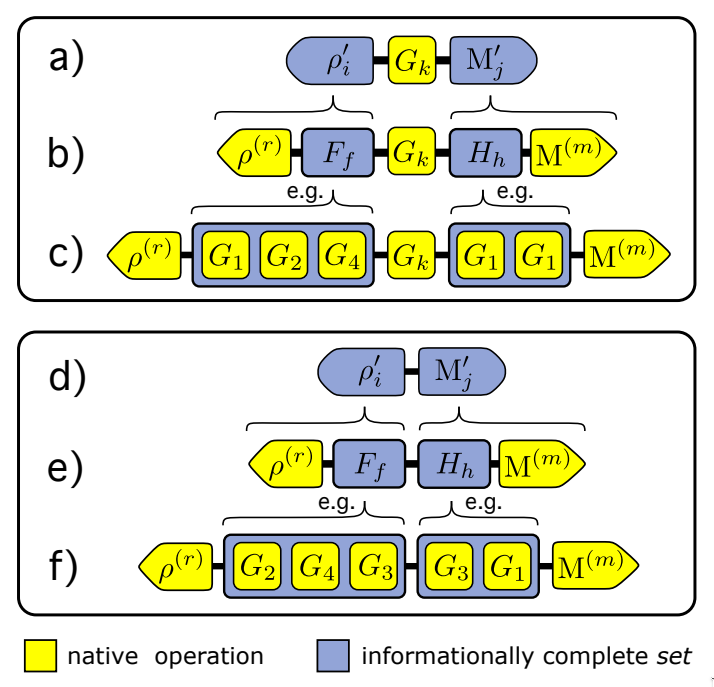}
    \caption{\textbf{Linear Gate Set Tomography.} Structures of the two types of circuits required by the LGST algorithm. 
        \textbf{Upper panel:} each native gate, $G_k$, is sandwiched between the elements of informationally complete sets of \emph{effective state preparations}, $\{ \rho_i^\prime\}$, and of \emph{effective measurements,} $\{M_j^\prime\}$. These are the same circuits that QPT requires to characterize $G_k$. 
        (a) Shows these circuits in their simplest form, with each informationally complete set displayed as a unit. 
        (b) Depicts the common case when the set of effective preparations (measurements) is implemented by following (preceding) a single native state preparation (measurement) operation with a \emph{fiducial circuit} $F_f$ ($H_h$). 
        (c) Exemplifies that the fiducial circuits are composed of native gates $G_i$ and gives the circuit entirely in terms of native operations. 
        \textbf{Lower panel:} because LGST does not assume knowledge of $\rho_i^\prime$ and $M_j^\prime$, it requires circuits that sandwich nothing between pairs of fiducials in order to be self-calibrating. The circuit diagrams in (d), (e), and (f) parallel those in (a), (b), and (c). LGST also requires the circuits that perform state (measurement) tomography on $\rho$ ($M$), but these are not explicitly shown. They are similar to (d) -- (f) (replacing $\rho^\prime$ with $\rho$ or $M^\prime$ with $M$), and are actually included as a subset of these circuits when the gate set contains only a single native state preparation (measurement) and one of the preparation (measurement) fiducial circuits is the empty (do-nothing) circuit. (Figure and caption reproduced with permission from \R\cite{nielsen2021gate}.)
    }
    \label{fig:lgst}
\end{figure}

\emph{\Ac{LGST}} is a self-consistent approach to simultaneous state, process, and measurement tomography that uses short quantum circuits and reconstructs a gate set model using linear inversion (see Fig.~\ref{fig:lgst}). Like QPT, it assembles elements of the gate set to construct informationally complete (see Sec.~\ref{sec:principles}) sets of states and measurements, which are then used to probe the errors in elementary logic operations. Most quantum computing systems can natively prepare only a single initial state (typically $\ket{000...}$) and perform measurements only in a single basis (e.g., the computational basis). So, a full, informationally-complete set of states and measurements must be constructed from these native operations by the application of short \emph{fiducial} gate sequences. For instance, measurement in the $\{\ket{+},\ket{-}\}$ basis can be performed by preceding a computational basis measurement by a Hadamard operation. Informationally \emph{over}complete fiducial sets are often used in the literature, but for simplicity of presentation, we assume exact informational completeness. See the Appendix of \R\cite{nielsen2021gate} for the general case. 

Given an informationally complete set of fiducial states $\left\{ \ket{\rho_j'\rangle} \right\}_{j = 1}^{N_\rho}$ and an informationally complete set of fiducial measurement effects $\left\{ \bra{\langle E_i'} \right\}_{i=1}^{N_E}$, the LGST protocol prescribes a set of circuits whose output distributions provide sufficient information to estimate the parameters of a gate set model. To see how the protocol works, it is convenient to start by collecting all the fiducial states and measurement effects into matrices $A$ and $B$ defined as 
\begin{equation}
    A = \begin{pmatrix}
        \bra{\langle E_1'} \\
        \bra{\langle E_2'} \\
        \vdots \\
        \bra{\langle E_N'} \\
    \end{pmatrix}
\end{equation}
and
\begin{equation}
    B = \begin{pmatrix}
        \ket{\rho_1' \rangle}, 
        \ket{\rho_2' \rangle}, 
        ..., 
        \ket{\rho_{N_\rho}' \rangle}
    \end{pmatrix} ~.
\end{equation}
Now, the parameters of a gate $G_k$ can be estimated by first estimating a matrix of probabilities $\mathsf{P}_k$ defined component-wise as
\begin{equation}
    [\mathsf{P}_k]_{i, j} = \bra{\langle E_i'} G_k \ket{\rho_j' \rangle} ~,
\end{equation}
or as a matrix equation,
\begin{equation} \label{equ:gate_meas_gst}
    \mathsf{P}_k = A G_k B ~.
\end{equation}
Measuring $\mathsf{P}_k$ is essentially standard QPT, but in the absence of a pre-calibrated reference frame, neither $A$ nor $B$ is known, so we cannot solve \eq\ref{equ:gate_meas_gst} directly for $G_k$.

Instead, LGST also measures an additional set of circuits that would correspond to QPT on the null operation. Their probabilities, which can be estimated from the circuits' observed outcome statistics, form a \emph{Gram matrix} $\tilde{\mathbbm{1}}$ whose elements are
\begin{equation}
    [\tilde{\mathbbm{1}}]_{i, j} = \langle \langle E_i' | \rho_j' \rangle \rangle ~.
\end{equation}
The Gram matrix is precisely equal to
\begin{equation}
    \tilde{\mathbbm{1}} = A B ~.
\end{equation}
Since the fiducial states and measurements are (by assumption) informationally complete, both $A$ and $B$ are square invertible matrices, and $\tilde{\mathbbm{1}}^{-1} = B^{-1} A^{-1}$. It follows that multiplying both sides of \eq\ref{equ:gate_meas_gst} by $\tilde{\mathbbm{1}}^{-1}$ yields
\begin{equation} 
    \tilde{\mathbbm{1}}^{-1} \mathsf{P}_k = B^{-1} G_k B ~,
\end{equation}
or, solving for $G_k$, 
\begin{equation} \label{eq:LGST_final}
    G_k = B \tilde{\mathbbm{1}}^{-1} \mathsf{P}_k B^{-1} ~. 
\end{equation}
This equation defines the process matrices for every gate in the gate set, in terms of physically measurable quantities $\tilde{\mathbbm{1}}$ and $\mathsf{P}_k$, \emph{up to an unknown superoperator $B$}. It turns out that $B$ is not just unknown, it is unknowable -- $B$ embodies the \emph{gauge freedom} in gate set models. So \eq\ref{eq:LGST_final} defines a complete estimate of all gates in the gate set up to gauge freedom.

The native state preparations $\left\{ \rho^{(i)} \right\}_{i=1}^{N_\rho}$ and measurements $\left\{ E_j^{(m)} \right\}_{m=1, j=1}^{N_M, N_E^{(m)}}$ can be estimated (in the same gauge) by constructing the following vectors,
\begin{align}
    [\mathbf{R}^{(l)}]_j &= \braket{\langle E_j' | \rho^{(l)} \rangle} ~, \\
    [\mathbf{Q}_l^{(m)}]_j &= \braket{\langle E_l^{(m)} | \rho_j' \rangle} ~,
\end{align}
using empirically measured circuit outcome probabilities. We write them as
\begin{align}
    \mathbf{R}^{(l)} &= A \ket{\rho^{(l)} \rangle} ~, \\
    \mathbf{Q}_l^{(m)T} &= \bra{\langle E_l^{(m)}} B ~, 
\end{align}
and use the Gram matrix identity $\tilde{\mathbbm{1}} = AB$ to write all the elements of a gate set model in terms of measurable quantities and a gauge transformation $B$ as
\begin{align}\label{eq:gk_from_pk}
    G_k &= B \tilde{\mathbbm{1}}^{-1} \mathsf{P}_k B^{-1} ~, \\
    \ket{\rho^{(l)} \rangle}  &= B \tilde{\mathbbm{1}}^{-1} \mathbf{R}^{(l)} ~, \\
    \bra{\langle E_l^{(m)}} &= \mathbf{Q}_l^{(m)T} B^{-1} ~.
\end{align}
Any invertible matrix $B$ is a valid choice, and the predicted outcome probabilities of circuits are entirely independent of $B$. Different choices of $B$ simply define different gauges (bases) in which the gate set can be written. Thus, no physical experiment can single out a ``proper'' gauge. Inconveniently, most metrics of performance (Sec.~\ref{sec:overview}) are \emph{not} gauge-invariant (independent of $B$). Therefore, it is customary and necessary to choose a convenient gauge in which to report the estimated $\hat{\mathcal{G}}$. This is discussed in Sec.~\ref{sec:gauge}; an easy starting choice is to use the $B$ defined by the \emph{intended} fiducial states.

In the above discussion, $\mathsf{P}_k$ is a matrix of circuit outcome probabilities that must be estimated from data. The maximum likelihood estimator for those probabilities is simply the observed frequency. If the circuit is run $N$ times, then finite sample fluctuations will lead to error in the estimate that scales like $\order{1/\sqrt{N}}$. Since $G_k$ is linear in $\mathsf{P}_k$ (\eq\ref{eq:gk_from_pk}), the error bars on the estimate of the $G_k$ transfer matrix also scale as $1/\sqrt{N}$. This is the so-called ``standard quantum limit'' for parameter estimation \cite{Giovannetti2004quantumlimit}, and it results here from the fact that each gate is only used once per circuit (excluding any potential uses in creating the fiducials). Long-sequence GST evades the standard quantum limit by using long circuits that amplify errors and achieve ``Heisenberg'' scaling in estimation precision.

%%%%%%%%%%%%%%%%%%%%%%% LSGST %%%%%%%%%%%%%%%%%%%%%%% 
\subsubsection{Long Sequence GST}\label{sec:lsgst}

\begin{figure}[ht]
    \centering
    \includegraphics[width=0.45\textwidth]{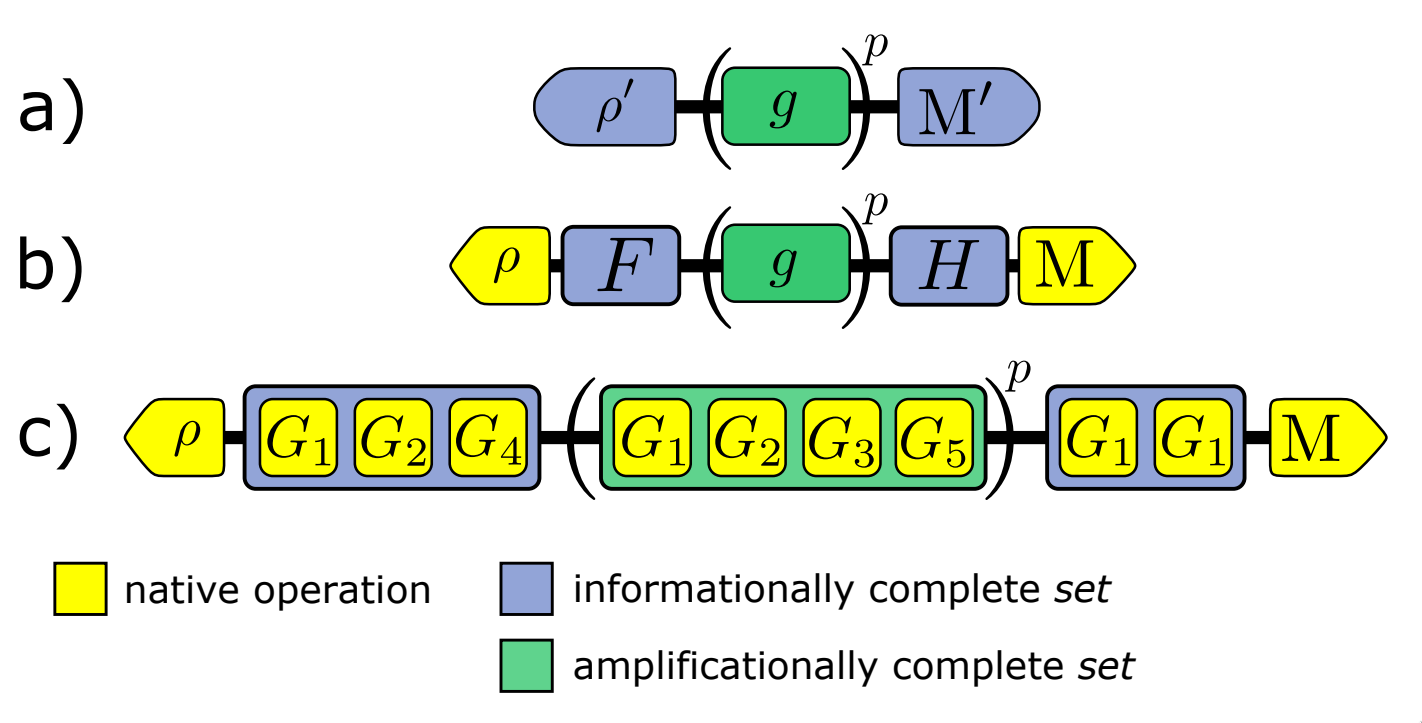}
    \caption{\textbf{Long Sequence Gate Set Tomography.}
    The structure of circuits in the standard LSGST experiment design, shown in increasing detail. 
    (a) Each GST circuit consists of an effective state preparation $\rho'$ followed by a \emph{germ} circuit $g$ repeated $p$ times, followed by an effective measurement $M' = \{E_i'\}$. 
    (b) Effective preparations are usually implemented by a native state preparation $\rho$ followed by a \emph{preparation fiducial circuit} $F$, and effective measurements are usually implemented by a \emph{measurement fiducial circuit} $H$ followed by a native measurement $M$. 
    (c) Writing the fiducials and germ in terms of the native gate operations reveals how the native operations of a gate set compose to form a GST circuit. 
    (Figure and caption reproduced with permission from \R\cite{nielsen2021gate}.)}
    \label{fig:LS-GST}
\end{figure}

\emph{\Ac{LSGST}} is an approach to self-consistent tomography of quantum gate sets that can beat the standard quantum limit \cite{Nielsen2021gatesettomography}. It achieves this by adding two innovations beyond LGST: i) longer quantum circuits that amplify gate errors, and ii) a new statistical estimation protocol to analyze long-circuit data.

LSGST circuits are formed in a similar fashion to LGST circuits. First, one selects informationally complete sets of state preparation and measurement fiducials, as in LGST. Where LGST uses these fiducials to probe each gate $G_k$ in the gate set, long sequence GST uses them to probe each of an \emph{amplificationally complete} list of ``germs''. A germ is a short sequence of native gates, chosen so as to amplify certain errors (see below). The circuits run by LSGST comprise a fiducial state preparation, an $p$-fold repeated germ, and a fiducial measurement (see Fig.~\ref{fig:LS-GST}). The ``germ powers'' $p$ are typically chosen to make the length of the $p$-fold repeated germ as close as possible to (but not greater than) a logarithmically-spaced integer, $1,2,4,8,\ldots, p_{\rm{max}}$. These many-fold repeated germs are what enable LSGST to achieve Heisenberg-limited scaling. 

To see how this works, consider repeating a single gate $G$ many ($p$) times. The resulting process can be computed from a spectral decomposition of the gate,
\begin{align}
    G &\doteq R \left(\begin{array}{ccc} e^{\phi_1} & & \\ & e^{\phi_2} & \\ & & \ddots \end{array}\right) R^{-1} ~, \\
    G^p &\doteq R \left(\begin{array}{ccc} e^{p\phi_1} & & \\ & e^{p\phi_2} & \\ & & \ddots \end{array}\right) R^{-1} ~,
\end{align}
where $\rm{diag}(e^{\phi_1}, e^{\phi_2}, \ldots)$ is a diagonal matrix of (generally complex) eigenvalues and $R$ is the change-of-basis matrix between the original and diagonalized bases. As we saw above, probing $G$ with an informationally complete set of fiducials allows us to estimate its eigenvalues, and thus the complex phases $\phi_i$, with uncertainty $\mathcal{O}(1/\sqrt{N})$. Similarly, probing $G^p$ allows us to estimate the \emph{amplified} phases $p \phi_i$ with uncertainty $\mathcal{O}(1/\sqrt{N})$, giving a $\mathcal{O}(1/p\sqrt{N})$ uncertainty in the estimate of $\phi_i$. This example also helps explain why GST does not exclusively use the longest sequences: knowledge of $e^{p\phi}$ for $p \gg 1$ is generally insufficient to deduce $\phi$, because the logarithm is multi-valued. Data from shorter sequences are necessary to determine the correct branch.

Repeating $G$ many times \emph{amplifies} its eigenvalues, allowing them to be estimated to high precision. But it does not increase sensitivity to \emph{axis errors}, so it does not enable high-precision estimation of the matrix of eigenvectors $R$. Amplifying the misalignment of the axis between the gates requires choosing and repeating composite germs that consist of products of native gates. Each germ defines a quantum operation, which can be described by a transfer matrix. Errors in a germ's component gates propagate to become errors in the germ. Some of them change the eigenvalues of the germ's transfer matrix. Those errors -- often linear combinations of errors on different gates -- are precisely the ones that the germ amplifies, and which can be measured to high precision by repeating that germ. A set of germs is called \emph{amplificationally complete} for a gate set $\mathcal{G}$ if knowledge of all the germs' eigenvalues is sufficient to reconstruct all of the non-gauge, non-SPAM degrees of freedom in the gate set. 

To analyze what errors a given germ amplifies, we assume that errors act as small perturbations to the target operation of a gate. In that limit, a perturbation to the germ's target operation is amplified by the germ if and only if it commutes with the target operation. It follows that each germ amplifies a subspace of perturbations to the gate set. A complete LSGST experiment design can be constructed by searching over possible germs (generally starting with the shortest) until the amplified directions span the non-gauge subspace; see \R\cite{nielsen2021gate} and the discussion of Sec.~\ref{sec:principles}. 

Fitting LSGST data is typically done using iterative maximum likelihood or least-squares optimization. This approach begins by fitting a model to the shortest ($p=1$) sequences, and then using that to seed the optimizer for the next round, which includes the $p=1$ and $p=2$ circuits. This procedure repeats until all circuits have been added and the optimizer has converged. This approach helps to avoid the wrong branch issue described above, and in practice is extremely robust. 

LSGST experiments designed this way are very overcomplete. For example, constructing a two-qubit LSGST experiment for a relatively simple gate set this way can easily prescribe more than 30,000 circuits! Many of these circuits provide redundant information, and can be removed from the experiment design with minimal loss of estimation accuracy. Ref.~\cite{Ostrove2023-xp} discusses techniques for constructing efficient, near-minimal GST experiments. In practice, these techniques have reduced the experimental complexity of 2-qubit GST by more than an order of magnitude, making it feasible and competive with randomized benchmarking.

%%%%%%%%%%%%%%%%%%%%%%% (In)validation of gate set models %%%%%%%%%%%%%%%%%%%%%%% 
\subsubsection{(In)validation of Gate Set Models}\label{sec:model_violation}

Many useful performance metrics can be extracted from a high-precision estimate of a gate set model, including both gauge-dependent and gauge-independent metrics discussed in Sec.~\ref{sec:gauge}. The \emph{error generator} framework discussed in Appendix \ref{sec:error_gen} can be used to identify observed errors with physical mechanisms. These analyses help quantify and classify the ``in model'' errors captured by the gate set model. 

But in practice, quantum computing systems often experience ``out of model'' errors that violate the assumptions of the gate set model. Such errors are often termed \emph{non-Markovian} because gate set models are designed to capture arbitrary Markovian errors \cite{Nielsen2021gatesettomography}, and thus (arguably) define what it means for errors to be Markovian, as discussed in Sec.~\ref{sec:nm_errors}. Examples of non-Markovian errors include low-frequency drift, leakage, and heating of auxiliary degrees of freedom (e.g., the trapped-ion motional mode used in Mølmer-Sørenson gates). 

When non-Markovian effects meaningfully impact circuit outcome statistics, it is very unlikely that any Markovian gate set model will be statistically consistent with the observations. This is an interesting difference between GST and QPT; because QPT experiments use the unknown process just once, and are usually not very overcomplete, they do not generally distinguish between Markovian and non-Markovian effects. In contrast, LSGST experiment designs use the unknown gate[s] many times (which amplifies non-Markovian effects) and are typically overcomplete. This overcompleteness can be leveraged to quantify, using statistical tests, how well or poorly an estimated gate set model fits the observed data. Model violation analysis provides insight into the presence, size, and sometimes nature of non-Markovian errors in GST experiments.

The primary tools for such a ``goodness-of-fit'' analysis are the \emph{log-likelihood ratio test} and \emph{Wilks' theorem} \cite{wilks1938large}. They make extensive use of the log-likelihood ratio statistic $\lambda \equiv 2(\log \mathcal{L}_\text{max} - \log \mathcal{L})$ that compares the likelihood $\mathcal{L}$ of an estimated GST model to the likelihood $\mathcal{L}_\text{max}$ of a much larger \emph{maximal} (a.k.a. \emph{saturated}) model that is guaranteed to fit the data well. The maximal model commonly used in GST treats each circuit as a totally independent experiment, assigning it an independent multinomial outcome distribution whose parameters are fit directly to that circuit's observed frequencies. Wilks' theorem \cite{wilks1938large} states that if the GST model is valid, then this log-likelihood ratio will be a $\chi^2_k$ random variable,
\begin{equation}
    2(\log \mathcal{L}_\text{max} - \log \mathcal{L}) \sim \chi^2_k  ~,
\end{equation}
where $k$ is the difference between the number of parameters in the maximal model, $N_\text{max}$, and the number of \emph{non-gauge} parameters in the estimate, $N_\text{nongauge}$: $k = N_\text{max} - N_\text{nongauge}$. A simple way to quantify the significance of the observed model violation is to compute and report the number of standard deviations by which the log-likelihood ratio exceeds its expected value under a $\chi_k^2$ hypothesis:
\begin{equation}\label{eq:model_violation}
    N_\sigma \equiv \frac{2(\log \mathcal{L}_\text{max} - \log \mathcal{L}) -k}{2 \sqrt{k}} ~. 
\end{equation}
If $N_\sigma \approx 1$, then the estimated gate set model fits the data as well as possible, suggesting that the device's errors are mostly Markovian. However $N_\sigma \gg 1$ indicates convincing statistical evidence for the presence of non-Markovian dynamics.

$N_\sigma$ is a statistical measure of model violation, and captures the strength of the observed evidence for non-Markovian dynamics. What it does \emph{not} do is quantify \emph{how much} non-Markovianity is present in any physically meaningful units. It is not a measure of \emph{effect size}. For example, simply doubling the number of shots performed for each circuit will (on average) double the log-likelihood ratio statistic and thus $N_\sigma$. The log-likelihood ratio statistic scales linearly with the amount of \emph{data} available. 

It is also useful to have a measure of the effect size of non-Markovian errors. In principle this could be extracted from a (much) larger model that is able to capture any expected non-Markovian effects. But such models do not yet exist (although methods do exist for low-frequency noise \cite{proctor2020detecting}), and doing so would require designing an experiment that is sensitive to all of the parameters and fitting it to data. \emph{Wildcard models} \cite{blume2020wildcard} provide an alternative. Wildcard models weaken the predictions of statistical error models just enough that they become consistent with observed data, and the amount of weakening required to do so constitutes a measure of model violation effect size . The parameters of a wildcard model can, with care, be interpreted as measuring how much non-Markovian error is present in the data. The wildcard error can then be compared to various error metrics, such as diamond distance (see Sec.~\ref{sec:ddist}), to determine whether or not the GST model is trustworthy. A full discussion of wildcard models is out of scope for this tutorial, but the interested reader is encouraged to consult \R\cite{blume2020wildcard}. Refs.~\cite{PRXQuantum.2.040338} and \cite{hashim2023benchmarking} provide examples of how this type of analysis can be used in practice.

\section{Randomized Benchmarks}\label{sec:randomized_benchmarks}

\emph{\Ac{RB}} protocols are a broad suite of methods that use varied-depth random circuits to quantify the rates of errors in a gate set (see Secs.~\ref{sec:gate_sets} and \ref{sec:quantum_gate_sets}). RB was initially developed in the mid- to late-2000s \cite{emerson2005scalable, dankert2009exact, knill2008randomized} to circumvent two of main the limitations of quantum process tomography (\ac{QPT}; see Sec.~\ref{sec:qpt}): QPT is corrupted by \ac{SPAM} errors, and it is inefficient in the number of qubits ($n$). There are now dozens of distinct RB protocols, each with their own purposes, strengths, and limitations. In this section, we review many of the most widely-used RB methods. In the first half of this section, we discuss the RB protocols that estimate a \emph{single} error rate for a set of gates:
\begin{itemize}
    \item \emph{Standard RB} (Sec.~\ref{sec:standardrb}). This is the \emph{de facto} standard RB protocol, which is typically used to benchmark gates that implement the one- or two-qubit Clifford group.
    \item \emph{Native Gate RB Protocols} (Sec.~\ref{sec:drb+}). These are a family of protocols that can directly benchmark a system's native gates, instead of using those gates to create all the Clifford group elements (as in standard RB). Protocols within this family include \emph{direct RB}, \emph{binary RB}, \emph{mirror RB}, and \emph{cross-entropy benchmarking}.
    \item \emph{RB for General Groups} (Sec.~\ref{sec:rb-general-groups}). This is a family of protocols for benchmarking sets of gates that form groups that are not unitary 2-designs. The most prominent such method is \emph{character RB}. 
\end{itemize}

There are a variety of RB protocols that measure quantities that are more complex or fine-grained than just a single error rate for a gate set (e.g., individual gate error rates). Many of these methods are adaptations of the foundational RB protocols presented in Sections~\ref{sec:standardrb}-\ref{sec:rb-general-groups}. We discuss:
\begin{itemize}
    \item \emph{Simultaneous RB} (Sec.~\ref{sec:simrb}). Simultaneous RB is a simple and widely-used technique for measuring the impact of simultaneous gate operations across multiple qubits. It can be used to quantify crosstalk errors between qubits.
    \item \emph{Interleaved RB} (Sec.~\ref{sec:irb}). Interleaved RB is a technique for estimating the infidelity of individual gates, but it has important limitations.
    \item \emph{Cycle Benchmarking} (Sec.~\ref{sec:cb}). Cycle benchmarking is a scalable method for estimating the infidelity of layers of gates. 
    \item \emph{Purity Benchmarking Protocols} (Sec.~\ref{sec:pb}). These are a family of protocols for estimating how much of a gate set's error is due to coherent and incoherent errors.
    \item \emph{RB Protocols for Non-Markovian Errors} (Sec.~\ref{sec:rb-nonmark}). These are a family of protocols for estimating the rates of various kinds of non-Markovian errors, such as leakage.
\end{itemize}

We begin this section with some mathematical background that is important for understanding and describing the various randomized benchmarks that we outline above.

%%%% Mathematical Preliminaries %%%%
\subsection{Mathematical Preliminaries}\label{sec:rb_math_twirling}

Despite being rather simple to implement, the mathematical theory of RB protocols is surprisingly deep and elegant. Describing it in full detail is well beyond the scope of this tutorial. However, several of the most important concepts from this theory are found commonly even in the experimental literature. In this subsection, we introduce those few mathematical concepts that are most helpful for reading and understanding articles on RB and related benchmarking protocols. These topics include:
\begin{itemize}
    \item twirling over a group,
    \item Schur's lemma, and
    \item unitary 2-designs.
\end{itemize}
The pragmatic reader can skip to Sec.~\ref{sec:standardrb}, where the RB protocol discussions begin. 

A number of \ac{QCVV} techniques, including RB and other randomized benchmarks, utilize averages over circuits that contain random gates. Each time the circuit is run, a new gate is sampled from some ensemble, and the circuit outcomes are typically averaged together (so they are treated as though they came from the same circuit). For example, standard RB (see Sec.~\ref{sec:standardrb}) uses sequences of random Clifford operations, whereas the randomized compiling \cite{wallman2016noise, hashim2021randomized} used in cycle benchmarking (Sec.~\ref{sec:cb}) and Pauli noise learning techniques (Sec.~\ref{sec:pnr}) inserts (and typically compiles in) random Pauli gates. At some point in the analysis of these techniques, one will encounter a superoperator $A$ (see Sec.~\ref{sec:superop}) that is averaged over all conjugations by elements of a group $\mathbb{G}$:
\begin{equation}\label{eq:twirl_continuous}
    T_\mathbb{G}(A) = \int d\mu(g) g A g^{-1} ~,
\end{equation}
where $d\mu$ is the Haar measure for the group $\mathbb{G}$. If $\mathbb{G}$ is a discrete group, then the Haar measure is just the counting measure, and the integral is often written as a sum:
\begin{equation}\label{eq:twirl_discrete}
    T_\mathbb{G}(A) = \frac{1}{\abs{\mathbb{G}}} \sum_{g\in \mathbb{G}} g A g^{-1} ~.
\end{equation}
Equations \ref{eq:twirl_continuous} and \ref{eq:twirl_discrete} define the \emph{twirl} of the superoperator $A$ over the group $\mathbb{G}$. In both equations above, $g$ is the superoperator (e.g., transfer matrix) representation of a group element $g$. So, even if we are twirling over a single-qubit unitary group, we will be using $4 \times 4$ transfer matrices, rather that the usual $2 \times 2$ unitary matrices. See Appendix \ref{sec:twirling} for a practical introduction to twirling and randomization.

\begin{figure}[t]
    \centering
    \includegraphics[keepaspectratio=true,width=\columnwidth]{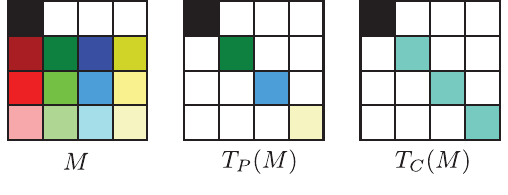}
    \caption{\textbf{Pauli and Clifford Twirling}. Twirling a PTM $M$ with the Pauli group, $T_\mathbb{P}(M)$ simply eliminates the off-diagonal entries. Twirling with the larger Clifford group, $T_\mathbb{C}(M)$, also averages all but one (the top left) of the diagonal entries.}
    \label{fig:PCtwirling}
\end{figure}

We can understand group twirls by taking a brief diversion into representation theory. Recall that quantum operations act on density matrices (see Sec.~\ref{sec:rep_quant_proc}). Transfer matrices are a representation of quantum operations that act on a vector space of \textit{vectorized} density matrices (see Sec.~\ref{sec:superop}). For a set of unitary superoperators that form a group $\mathbb{G}$, it turns out that we can divide this vector space into subspaces --- irreducible representation spaces --- in such a way that no element of $\mathbb{G}$ will mix distinct subspaces. This means that the entire set of superoperators in $\mathbb{G}$ can be simultaneously block diagonalized, with each block corresponding to an \textit{irreducible representation}, or \textit{irrep}. This decomposition into irreps is important, because Schur's lemma allows us to express the outcome of a twirl in terms of this decomposition. If the irreps are distinct (not related to each other by a similarity transform), then:
\begin{equation}
    T_\mathbb{G}(A) = \sum_\phi \frac{\Tr{A \mathbf{P}_\phi}}{\Tr{\mathbf{P}_\phi}} \mathbf{P}_\phi ~.
\end{equation}
where $\mathbf{P}_\phi$ is a projector onto the irreducible subspace of irrep $\phi$. 

The number and size of the irreps associated with the superoperator representation will depend on the group (representation) over which the twirl is being taken. The superoperator representation of the full unitary group $\mathsf{SU}(2^n)$ has just two irreps, a one-dimensional irrep that acts trivially on the trace of $\rho$, and a $(4^n-1)$-dimensional irrep that mixes all other components. This means that the twirl of any superoperator $A$ under the full unitary group will result in an $n$-qubit depolarizing channel --- a diagonal Pauli transfer matrix (\ac{PTM}; see Sec.~\ref{sec:ptm_rep}) with a single unit eigenvalue and a real number $p \in [0,1]$ repeated along the rest of the diagonal. Importantly, this $p$ is equal to the process polarization (see Sec.~\ref{sec:polarization}) of $A$, i.e., 
\begin{equation}
    p = f(A) ~.
\end{equation}
Equivalently, $A$ and $T_{\mathsf{SU}(2^n)}(A)$ have the same process (a.k.a.~entanglement) fidelity to the identity (\eq\ref{eq:process_fidelity_E}). So, an unknown error channel $\E$'s process fidelity can be learned by twirling it into a depolarizing channel and then learning that depolarizing channel's $p$, which is easy to do. This idea is foundational to RB.

\begin{figure}[t!]
    \centering
    \includegraphics[keepaspectratio=true,width=\columnwidth]{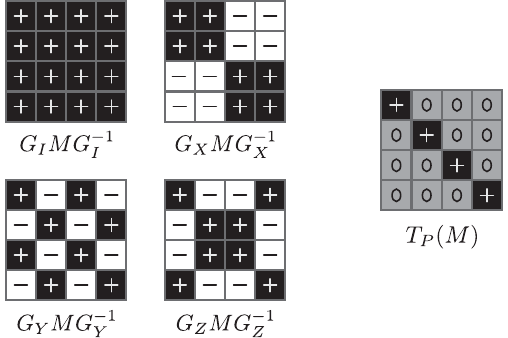}
    \caption{\textbf{Pauli Twirling}. 
    The PTMs for the Pauli operators are diagonal matrices of 1's and -1's. Conjugating a PTM $M$ by a Pauli operator's PTM just changes the signs of some of the entries, as illustrated here for a single-qubit $M$ and the four single-qubit Pauli operators. Averaging over all conjugations by Pauli operator PTMs, to get the twirled channel $T_P(M)$, preserves only the diagonal components.
    }
    \label{fig:Ptwirling}
\end{figure}

The unitary group is an infinite group, and twirling over it, even approximately, can be experimentally challenging. So, often we consider twirls over smaller, discrete subgroups of the unitaries, such as the Clifford group, the Pauli group, or one of the dihedral groups. One consequence of twirling over a subgroup of the full unitary group is that the superoperator representation of a subgroup of $\mathsf{SU}(2^n)$ could decompose into significantly more irreps. The superoperator representation of the Clifford group actually breaks into the exact same irreps as the unitary group. Groups whose superoperators have the same irrep structure as the full unitary group are known as unitary 2-designs (see Appendix \ref{sec:unitary_t_designs}), and are extremely important in QCVV. The superoperator representation of the $n$-qubit Pauli group, however, decomposes into $4^n$ one-dimensional irreps. Therefore, twirling a matrix $M$ over the Pauli group will remove the off-diagonal entries of the matrix but leave the $4^n$ diagonal elements unchanged, as illustrated in \fig\ref{fig:PCtwirling} and \fig\ref{fig:Ptwirling}. Twirling over the Clifford group will also project away the off-diagonal entries and will further replace all but one of the diagonal elements with their mean, as shown in \fig\ref{fig:PCtwirling}. 

%%%%%%%%%%%%%%%%%%%%%%% Standard Randomized Benchmarking %%%%%%%%%%%%%%%%%%%%%%%
\subsection{Standard Randomized Benchmarking (RB)}\label{sec:standardrb}

\begin{table*}[ht!]
\renewcommand{\arraystretch}{1.1}

\centering
% \resizebox{1.6\columnwidth}{!}{
\begin{tabular}{p{0.15\linewidth}<{\raggedright} | p{0.25\linewidth}<{\raggedright} | p{0.08\linewidth} | p{0.2\linewidth}<{\raggedright} | p{0.25\linewidth}<{\raggedright}}
    \hline
    \hline
    \textbf{Method} & \hfil \textbf{What it Measures} & \textbf{Scalability Score} & \hfil \textbf{Advantage(s)} & \hfil \textbf{Disadvantage(s)} \\
    \hline
    \hyperref[sec:standardrb]{Clifford RB} & Average error per Clifford gate ($r$) & \hfil 2  & \emph{de facto} standard protocol & Poor scaling; does not reliably measure error rate of native gates \\
    \hline
    \hyperref[sec:drb]{Direct RB} & Weighted-average error per gate in user-chosen gate set ($r_{\Omega}$)  & \hfil 3 & Can measure error per native gate  &  Scales better than standard RB but still limited, especially for non-Clifford gates \\
    \hline
    \hyperref[sec:birb]{Binary RB} & Weighted-average error per gate in user-chosen gate set ($r_{\Omega}$)  & \hfil 8 & Both reliable and scalable  & All gates must be Clifford \\
    \hline
    \hyperref[sec:mrb]{Mirror RB} & Weighted-average error per gate in user-chosen gate set ($r_{\Omega}$)   & \hfil 8  & Scalable, including for non-Clifford gates  & Small but systematic underestimate of gate error  \\
    \hline
    \hyperref[sec:xeb]{Cross-Entropy Benchmarking (XEB)} & Weighted-average error per gate in user-chosen gate set ($r_{\Omega}$) & \hfil 6 & Native gate does not need to be a Clifford & Need to classically simulate circuits, so very expensive beyond $n \sim 20$ qubits, and infeasible beyond $n \sim 60$ \\
    \hline
    \hyperref[sec:character_rb]{Character RB} & Average error per group-element gate  &  Depends on gate set used & Can benchmark gate sets that are not and do not generate unitary 2-designs & Sample inefficient (requires lots of data) for some groups  \\
    \hline
    \hyperref[sec:simrb]{Simultaneous RB} & Average error per simultaneous Clifford gate & \hfil 9  & Captures effects of crosstalk on qubits & Simple for one-qubit gates but complex scheduling for $n > 2$ qubit gates \\
    \hline
    \hyperref[sec:irb]{Interleaved RB} & Average error per dressed native Clifford gate & \hfil 1 & Simple and widely-used & Unreliable; native gate must be a Clifford \\
    \hline
    \hyperref[sec:cb]{Cycle Benchmarking} & Average error per dressed Clifford cycle/layer & \hfil 8  & Can estimate the error rate of dressed $n$-qubit cycle for large $n$ & Cycle must be Clifford (in most cases); need to measure multiple decays \\
    \hline
    \hyperref[sec:xrb]{eXtended RB} & Average unitarity and stochastic error per Clifford gate & \hfil 1 & Able to separate coherent from incoherent contributions to the \ac{AGSI} & Inefficient (does not amplify coherent errors); not scalable (uses state tomography) \\
    \hline
    \hyperref[sec:xeb_pb]{Speckle Purity Benchmarking} & Stochastic error per layer of native gates & \hfil 3 & Enables distinguishing coherent and stochastic contributions to XEB error rate & Similar disadvantages to XEB  \\
    \hline
    \hyperref[sec:itrb]{Iterative RB} & Coherent error per interleaved Clifford gate  & \hfil 1 & Conceptual simplicity, amplifies coherent errors & Only works for over/under-rotation errors  \\
    \hline
    \hyperref[sec:lrb]{Leakage RB} & Leakage rate per Clifford gate & \hfil 2 & Makes RB more reliable in the presence of leakage, measures leakage rate(s) & Data may not follow an exponential due to no randomization in the leakage space \\
    \hline
    \hline
\end{tabular}% }
\caption{
\textbf{Comparison of Randomized Benchmarks.} Here, we summarize the primary randomized benchmarks discussed in this tutorial (note that this tutorial does not cover every extant RB method). Randomized benchmarks do not all measure the same quantity, and so the first consideration when selecting an RB method is to consider what it measures (second column). The scalability of randomized benchmark is also often an important consideration when selecting an RB method, and here we give each method an \emph{informal} scalability score (between 0 and 10) that approximately summarizes how scalable it is. The scalability score is subjective, and we intend it only as very rough guide. For example, mirror RB, binary RB, and cycle benchmarking are all given a score of 8/10 for scaling because their classical computations required are simple and fast for any number of qubits ($n$), but they require both implementing an $n$-qubit layer/cycle of native gates and measuring all $n$ qubits with significantly non-zero fidelity. Key advantages and disadvantages of each method are also summarized.
}
\label{tab:rb_methods}

\end{table*}

There are many different RB protocols, and we cover many of them in this tutorial. But there is a \emph{de facto} standard version of RB \cite{magesan2011scalable} --- which we call \emph{standard RB} --- and we begin by explaining this protocol. This protocol is designed to benchmark any $n$-qubit gate set $\mathbb{G}_n$ (i.e., a set of $n$-qubit operations) that has the following two properties:
\begin{enumerate}
    \item $\mathbb{G}_n$ is a group  (see Appendix \ref{sec:groups_gatesets}), and 
    \item $\mathbb{G}_n$ is a unitary 2-design (see Appendix \ref{sec:unitary_t_designs}). 
\end{enumerate}
The $n$-qubit Clifford group $\mathbb{C}_n$ has these properties, and it is almost always the gate set that is benchmarked using standard RB. We refer to standard RB with the Clifford group as \emph{\ac{CRB}}. Most standard RB experiments are one- or two-qubit CRB. For three or more qubits, more scalable RB protocols are typically used (see Sec.~\ref{sec:drb+} for further discussion). Note that the random circuits of standard RB (defined below) can be constructed for any gate set that is a group, but when that group is not a unitary 2-design, the data from those circuits will not have the simple form of standard RB, and a different data analysis procedure and/or different circuits are required (see Sec.~\ref{sec:rb-general-groups}).

\begin{figure}[t!]
    \centering
    \includegraphics[width=0.8\columnwidth]{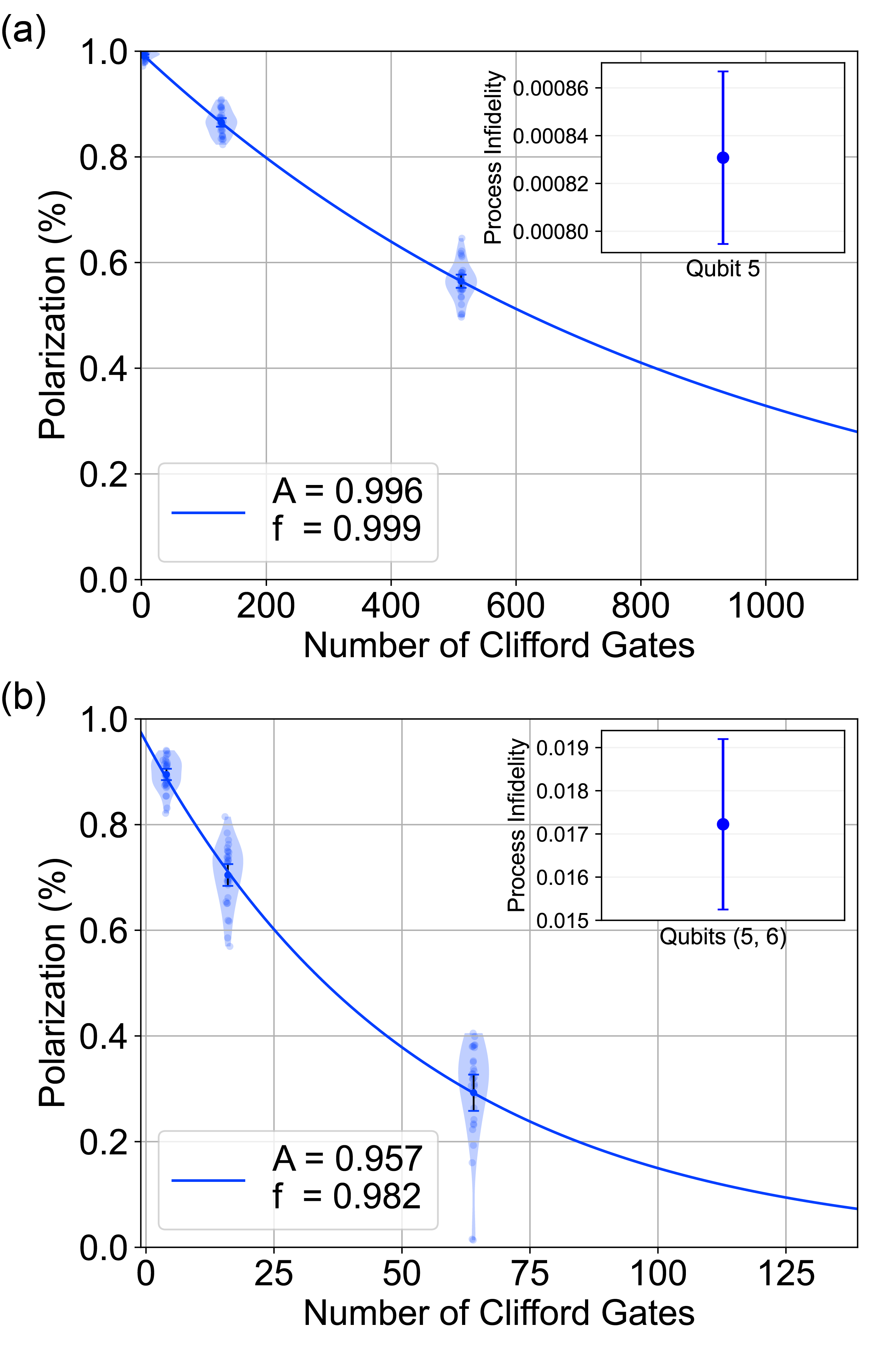}
    \caption{\textbf{One- and Two-Qubit Clifford-group RB.} 
    Exponential decays for CRB performed on \textbf{(a)} a single qubit (labeled 5) and \textbf{(b)} two qubits (labeled 5 and 6) on a superconducting quantum processor. The results are plotted in terms of the polarization (\eq\ref{eq:polarization}), a rescaling of the success probability. For both experiments, $K = 30$ random Clifford circuits were generated for each circuit depth $m$ at $L = 3$ different circuit depths. The SPAM parameters $A$ and exponential fit parameters $f$ are listed in the legends. At each circuit depth, circular data points plot the results of individual circuits and violin plots depict the distribution of results. Insets: the process infidelity for the (a) single-qubit CRB results [$e_F = 8.3(2) \times 10^{-4}$] and (b) two-qubit CRB results [$e_F = 1.7(1) \times 10^{-2}$]. The exponential decay curve for two-qubit CRB decays much faster than for single-qubit CRB, demonstrating that the error per Clifford is larger for two-qubit Cliffords than single-qubit Cliffords, as shown by their relative process infidelities.
    }
    \label{fig:rb_1q_2q}
\end{figure}

Standard RB measures a mean error rate (i.e., average gate set infidelity, or \ac{AGSI}; see Sec.~\ref{sec:ave_gate_fids}) for the gates in $\mathbb{G}_n$, and it is designed so that the AGSI is not corrupted by SPAM errors (as long as the SPAM errors are not too large). It is given by the following protocol:
\begin{enumerate}
    \item Run $K \gg 1$ random motion-reversal circuits for $L$ different circuit depths of depth $m \geq 0$ and record each circuit's success frequency \footnote{Here, we define a circuit of depth $m = 0$ to be the minimal benchmark depth, which contains only a single random gate (and its inverse). By defining it this way, the error in any gates in the $m = 0$ circuit all contributes to effective SPAM error. Therefore, any gates used for state-preparation or basis rotations for measurement can be compiled into the initial and final circuit layers, respectively.}. The circuit depths are typically linearly or logarithmically spaced, $K$ is typically between 20 and 1000, $L \ge 3$ in order to fit an exponential function to the observed data (see below), and it is considered best practice to have all $m > 1$. Each of the $K$ circuits at depth $m$ is sampled and run as follows:
    
    \begin{enumerate}
        \item Uniformly and independently sample $m + 1$ gates $C$ from $\mathbb{G}_n$, $\{ C_1, C_2, \dots, C_{m+1} \}$, and construct a sequence $C_{m+1} \circ C_{m} \circ \cdots \circ C_2 \circ C_1$. This sampling can be done efficiently (in $n$) if $\mathbb{G}_n$ is the $n$-qubit Clifford group \cite{gottesman1998heisenberg, Koenig2014-ih} (even though the size of $\mathbb{C}_n$ grows very quickly with $n$).
        
        \item Compute the inversion gate 
        \begin{equation}
            C_{m+2} = P(C_{m+1} \circ C_{m} \circ \cdots \circ C_2 \circ C_1)^{-1} ~,
        \end{equation}
        where $P$ is an $n$-qubit Pauli operator. The original description of the standard RB protocol does not include $P$ (i.e., it sets $P$ to the identity). However, it is now considered best-practice to sample a uniformly random $P$ \cite{fogarty2015nonexponential, muhonen2015quantifying, harper2019statistical}. Again, computing $C_{m+2}$ is efficient if $\mathbb{G}_n$ is the $n$-qubit Clifford group \cite{gottesman1998heisenberg}.
        
        \item Construct a circuit $\mathcal{C}_m$ composed of the $m+1$ randomly sampled gates and the inversion gate: 
        \begin{equation}
        \mathcal{C}_m = C_{m+2} \circ C_{m+1} \circ C_{m} \circ \dots \circ C_2 \circ C_1 ~.
        \end{equation}
        In the absence of errors, this circuit will always return the system to the original state, up to a final layer of Pauli gates determined by $P$. Thus, the ideal outcome is a particular bit string that is specified by $P$, which is the circuit's ``success'' outcome.
        
        \item Compile the circuit $\mathcal{C}_m$ into the native gates of the system being benchmarked, so that it can be measured experimentally. This compilation must simply replace each $n$-qubit Clifford in $\mathcal{C}_m$ with a sequence of those native gates that implements that particular unitary, i.e., ``compilation barriers'' must be placed between each layer in the circuit \footnote{This ensures that the entire circuit is not compiled down into a single gate layer, which would defeat the purpose of the benchmark.}.
        
        \item Execute the compiled circuit $N \geq 1$ times and compute its success frequency: 
        \begin{equation}
            p(\mathcal{C}_m) = N_{\textrm{success}}/N ~,
        \end{equation}
        where $N_{\textrm{success}}$ is the number of times the success outcome was observed. In experiments, typically $N$ is between 100 and 1000. For a fixed value of $K \times N$ (which is the total number of circuit executions in the RB experiment), $N=1$ is statistically optimal \cite{granade2015accelerated}, i.e., it results in the lowest uncertainties on the AGSI estimated by RB. However, due to the time required to compile circuits and upload waveforms in most experimental setups, sufficiently low uncertainty estimates of the AGSI can typically be achieved most quickly by setting $N \gg 1$. Each circuit execution is the following procedure:
            \begin{enumerate}
                \item Prepare each of the $n$ qubits in the $\ketbra{0}$ state.
                \item Apply the circuit $\mathcal{C}_m$.
                \item Measure all $n$ qubits in the computational basis, and check whether the ``success'' bit string was observed. 
            \end{enumerate}        
    \end{enumerate}
    
    \item Compute the average success probability $\bar{p}(m)$ for each depth $m$,
    \begin{equation}
        \Bar{p}(m) = \frac{1}{K} \sum_{\mathcal{C}_m} p(\mathcal{C}_m) ~.
    \end{equation}
    Then, fit this data to an exponential decay function:
    \begin{equation}\label{eq:rb-fit}
        \Bar{p}(m) = Af^m + B ~,
    \end{equation}
    where $A$, $f$, and $B$ are fit parameters. If the success bit string has been randomized (i.e., $P$ is uniformly random), fix $B = 1/2^n$ (which provides a higher-precision estimate of $f$ for the same amount of data \cite{fogarty2015nonexponential, muhonen2015quantifying, harper2019statistical}). $A$ is typically called the ``SPAM parameter'' because it includes information about the size of the SPAM errors \footnote{$Ap^m+B$ is the success probability of circuits containing $m+2$ Clifford gates and a SPAM operation, so $Ap^{-2}+B$, i.e., extrapolating the decay curve back to $m=-2$, is a heuristic for SPAM error's contribution to the success probabilities deviation from 1. Note that $A$ is not invariant under changes in the convention for `depth` in RB circuits --- e.g., defining $m$ to be the number of uniformly random Clifford gates in the circuit ---- but this heuristic for SPAM fidelity is.)}. RB data is typically analyzed with simple curve fitting routines (e.g., weighted least squares), although there are a variety of alternative fitting approaches. Note that meaningful results will only be obtained if $\bar{p}(m_{\min})$ is significantly greater than $\bar{p}(m_{\max})$, where $m_{\min}$ and $m_{\max}$ are the minimum and maximum depths used, which requires that the SPAM errors are not catastrophically large.
        
    \item RB theory shows that under certain circumstances (see below) the fit $f$ is an estimate of the mean process polarization (\eq\ref{eq:process_polarization}) of the gates in $\mathbb{G}_n$, but it is more common to report (in)fidelities than polarization. An estimate of the \emph{mean} of the gates' infidelities is 
    given by the average gate infidelity (\eq\ref{eq:ave_gate_infidelity}),
    \begin{equation}\label{eq:rb_r}
        r = \frac{d - 1}{d} (1 - f) ~,
    \end{equation}
    or the process (i.e., entanglement) infidelity (\eq\ref{eq:process_infidelity}),
    \begin{equation}\label{eq:rb_ef}
        e_F = \frac{d^2 - 1}{d^2} (1 - f) ~,
    \end{equation}
    where $d = 2^n$ is the dimension of the Hilbert space for $n$ qubits (see Tab.~\ref{tab:table_rel_fid_dep} for a summary of the linear relationships between these different quantities).
    $r$ or $e_F$ is an estimate of the mean of the infidelities of the gates in $\mathbb{G}_n$, so in the case of CRB this is often called the \emph{\ac{EPC}}. When comparing RB error rates, it is important to check whether the convention in \eq\ref{eq:rb_r} or~\ref{eq:rb_ef} is being used, as the process infidelity is stable under tensor products of parallel gates, while the average gate infidelity is not (see the discussion in Sec.~\ref{sec:ent_fid}).
\end{enumerate}

\begin{figure}[t!]
    \centering
    \includegraphics[width=\columnwidth]{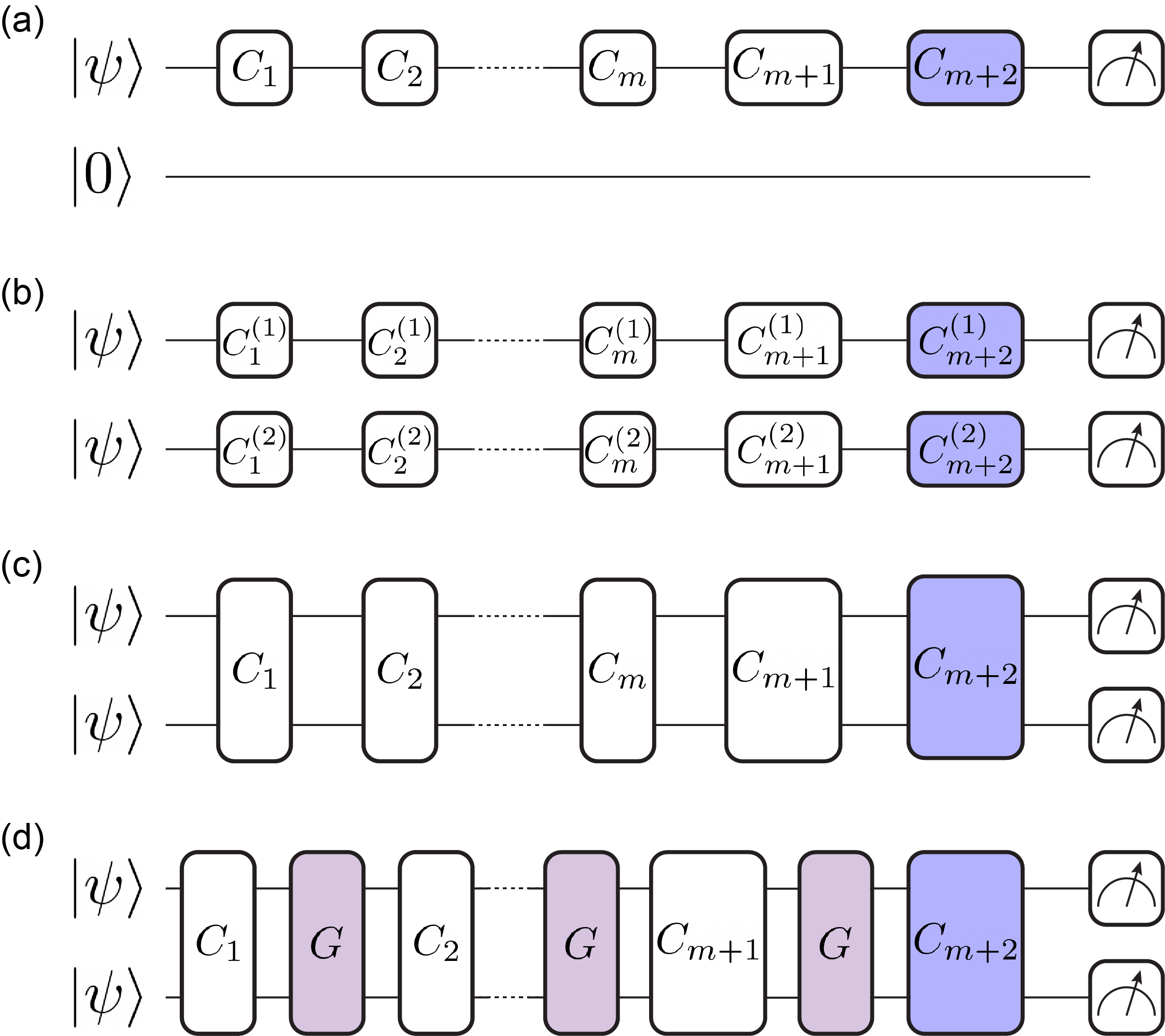}
    \caption[Randomized benchmarking sequences.]{\textbf{Standard, Simultaneous, and Interleaved RB.}
    \textbf{(a)} Circuit structure for standard single-qubit RB. Gates are only applied to the benchmarked qubit; all other qubits are assumed remain in their ground states.
    \textbf{(b)} Circuit structure for simultaneous RB. Here, gates are applied to two or more qubits simultaneously to benchmark the performance of simultaneous gate operations, where $C_m^{(i)}$ denotes the $m$th gate applied to the $i$th qubit.
    \textbf{(c)} Circuit structure for standard two-qubit RB. 
    \textbf{(d)} Circuit structure for interleaved RB. Here, $G$ (purple) is the interleaved gate whose infidelity can be estimated from analyzing (c) and (d) together (see Sec.~\ref{sec:irb}).
    For (a) -- (d), $C_{m+2}$ (blue) is the inversion gate for the entire sequence. 
    }
    \label{fig:rb_sequences}
\end{figure}

Examples of results for one-qubit and two-qubit CRB experiments are shown in \fig\ref{fig:rb_1q_2q}, and the forms of the circuits used in one- and two-qubit CRB are shown in \fig\ref{fig:rb_sequences}(a) and (c), respectively. In these figures, we plot \emph{polarization} rather than success probability. The polarization $S_{\textrm{pol}}$ is simply a rescaling of success probability ($p$), given by
\begin{equation}\label{eq:polarization}
    S_{\textrm{pol}} = \frac{p - 1/2^n}{1 - 1/2^n} ~.
\end{equation}
Polarization is sometimes more convenient because $S_{\textrm{pol}} = 0$ when all $n$ qubits are completely depolarized \cite{proctor2022measuring}. The measured one-qubit and two-qubit EPCs are $e_F = 8.3(2) \times 10^{-4}$ and $e_F = 1.7(1) \times 10^{-2}$, respectively. In most systems, it is expected that the EPC for two-qubit CRB will be higher than for single-qubit CRB, since two-qubit CRB requires two-qubit entangling gates, which are typically noisier than single-qubit gates.

To run $n$-qubit CRB experiments, each $n$-qubit Clifford operation must be decomposed into the system's native gates (step 1d above). For example, in a widely-used compilation \cite{barends2014superconducting} of the 24 single-qubit Clifford gates ($\mathbb{C}_1$) into rotations around $X$ and $Y$, the average number of single-qubit native gates per single-qubit Clifford gate is 1.875. CRB estimates the average error rate of the composite $n$-qubit Clifford gates (the EPC), \emph{not} the average error rate of the fundamental gates from which those gates are composed, and the EPC depends on the compilation used. However, it is common practice to rescale single-qubit CRB's EPC ($r_{\mathbb{C}_1}$) to a native gate error rate. For example, for the compilation of \R\cite{barends2014superconducting}, the EPC is typically related to the error per native single-qubit gate ($r_\text{SQ}$) with the simple heuristic:
\begin{equation}
    r_{\mathbb{C}_1} = 1.875 r_\text{SQ} ~.
\end{equation}
An alternate heuristic for estimating the error per native single-qubit gate is to use the common compilation strategy of decomposing all $U_3$ gates (i.e., arbitrary single-qubit rotations parametrized by three Euler angles) into a sequence consisting of three parametrized \emph{virtual} $Z$ gates and two \emph{physical} native $X_{\pi/2}$ gates \cite{mckay2017efficient}:
\begin{equation}\label{eq:zxzxz}
    U_3(\phi, \theta, \lambda) = Z_{\phi - \frac{\pi}{2}} X_\frac{\pi}{2} Z_{\pi - \theta} X_\frac{\pi}{2} Z_{\lambda - \frac{\pi}{2}} ~.
\end{equation}
Now, there are always two real gates (i.e., physical pulses) per single-qubit Clifford \footnote{Virtual $Z$ gates do not implement physical pulses; rather, they provide a frame update (i.e., a shift in phase) for the subsequent physical pulse.}, and thus the EPC is twice the error per native gate. One advantage of this approach is that it is straightforward to generalize to higher dimensions \cite{qutritrb}, and has been used to estimate native gate fidelities in single-qutrit ($d=3$) and single-ququart ($d=4$) CRB experiments \cite{nguyen2023empowering}, where 6 and 12 native gates are needed per single-qutrit and single-ququart Clifford gate, respectively; see Appendix \ref{sec:qrb} for an overview of randomized benchmarks for qudits.

Similarly, for two-qubit CRB and the widely-used compilation of \R\cite{barends2014superconducting}, a two-qubit Clifford gate contains 1.5 CNOT or CZ gates and 8.25 single-qubit gates, on average. For this compilation, two-qubit CRB's EPC ($r_{\mathbb{C}_2}$) is then often related to $r_\text{SQ}$ and the two-qubit gate error rate ($r_{CZ}$) using the simple heuristic
\begin{equation}
    r_{\mathbb{C}_2} = \frac{3}{2} r_\mathrm{CZ} + \frac{33}{4}r_\text{SQ ~}.
\end{equation}
These rescalings of the EPC are only heuristics though --- they are known to \emph{not} reliably estimate the native gate infidelities, in general. Importantly, the estimated error per native gate will typically change if different compilations are used. This is because CRB circuits prevent systematic addition or cancellation of coherent errors between different $n$-qubit Clifford gates, but not within the gate sequences used to create each $n$-qubit Clifford gate.

We now explain how to interpret RB results, and why RB works, by concisely summarizing the practical implications of the theory of standard RB. Standard RB works because the random gates twirl the errors in the gates, and because each gate is sampled from a unitary 2-design (such as the Clifford group) this twirl maps the gates' (potentially complicated) error maps into depolarization channels (as outlined in Sec.~\ref{sec:rb_math_twirling}). Turning this into a precise theory for RB is simple in the ``gate-independent noise'' idealization, where every gate in $\mathbb{G}_n$ is subject to the same CPTP error map $\E$. In this case, it is possible to show that standard RB's average success probability satisfies
\begin{equation}
    \bar{p}(m) =   A f(\E)^{m+1} + B ~.
\end{equation}
Here, $f(\E)$ is $\E$'s process polarization, and $A$ and $B$ absorb all SPAM error (and also have contributions from gate error). Straightforward derivations of this equation can be found throughout the literature on RB theory (e.g., see the ``zeroth-order model'' in \R\cite{magesan2012characterizing}). Therefore, for gate-independent noise, $r$ (\eq\ref{eq:rb_r}) or $e_F$ (\eq\ref{eq:rb_ef}) is a rigorous estimate of the average gate infidelity or process infidelity of each gate's error map $\E$, respectively. 

Understanding RB outside of the unrealistic setting of gate-independent noise is more complex. The modern theory of RB \cite{proctor2017randomized, wallman2018randomized, Merkel2021-ux, helsen2022framework} addresses the more realistic setting in which each gate has its own distinct error map. We will not delve into this theory here, but we highlight its main practical implications:
\begin{itemize}
    \item Standard RB's average success probability will decay exponentially as long as the gates experience only moderately small Markovian errors \cite{proctor2017randomized, wallman2018randomized, Merkel2021-ux, helsen2022framework} and the minimal circuit depth ($m$) used is sufficiently large. The theory shows that, under typical circumstances, $m \geq 1$ (corresponding to three Clifford gates in our depth convention) is sufficient. Therefore, standard RB data that is inconsistent with an exponential decay implies the presence of non-Markovian errors. For example, $1/f$ noise is well-known to cause non-exponential RB decays \cite{fogarty2015nonexponential}. 

    \item The simplest interpretation of standard RB's $f$ parameter is that it is equal to the mean of the gate's process polarizations, and therefore $r$ (or $e_F$) is equal to the mean of the gates' infidelities, i.e., 
    \begin{equation}
        \epsilon_{\textrm{uni}} = \frac{1}{|\mathbb{G}_n|} \sum_{G\in \mathbb{G}_n} \epsilon(G) ~,
    \end{equation}
    where $\epsilon(G)$ is the average gate or process infidelity of $G$. Unfortunately, although this interpretation contains the essence of what $r$ measures \cite{Carignan-Dugas2018-np}, it is subtly incorrect (in part because $\epsilon_{\textrm{uni}}$ is ill-defined, due to gauge ambiguities \cite{proctor2017randomized}). A mathematically precise understanding of $r$'s relationship to gate infidelity is not important for using RB. But it \emph{is} practically relevant when checking whether concurrent RB and tomography experiments have consistent results. Correctly predicting $r$ from measured transfer or process matrices requires either (a) using modern RB theory's predictions for how to compute $r$ from transfer or process matrices \cite{proctor2017randomized, wallman2018randomized, Merkel2021-ux, helsen2022framework}, or (b) simply simulating RB experiments using those transfer or process matrices.
\end{itemize}

\begin{figure}[t!]
    \centering
    \includegraphics[width=\columnwidth]{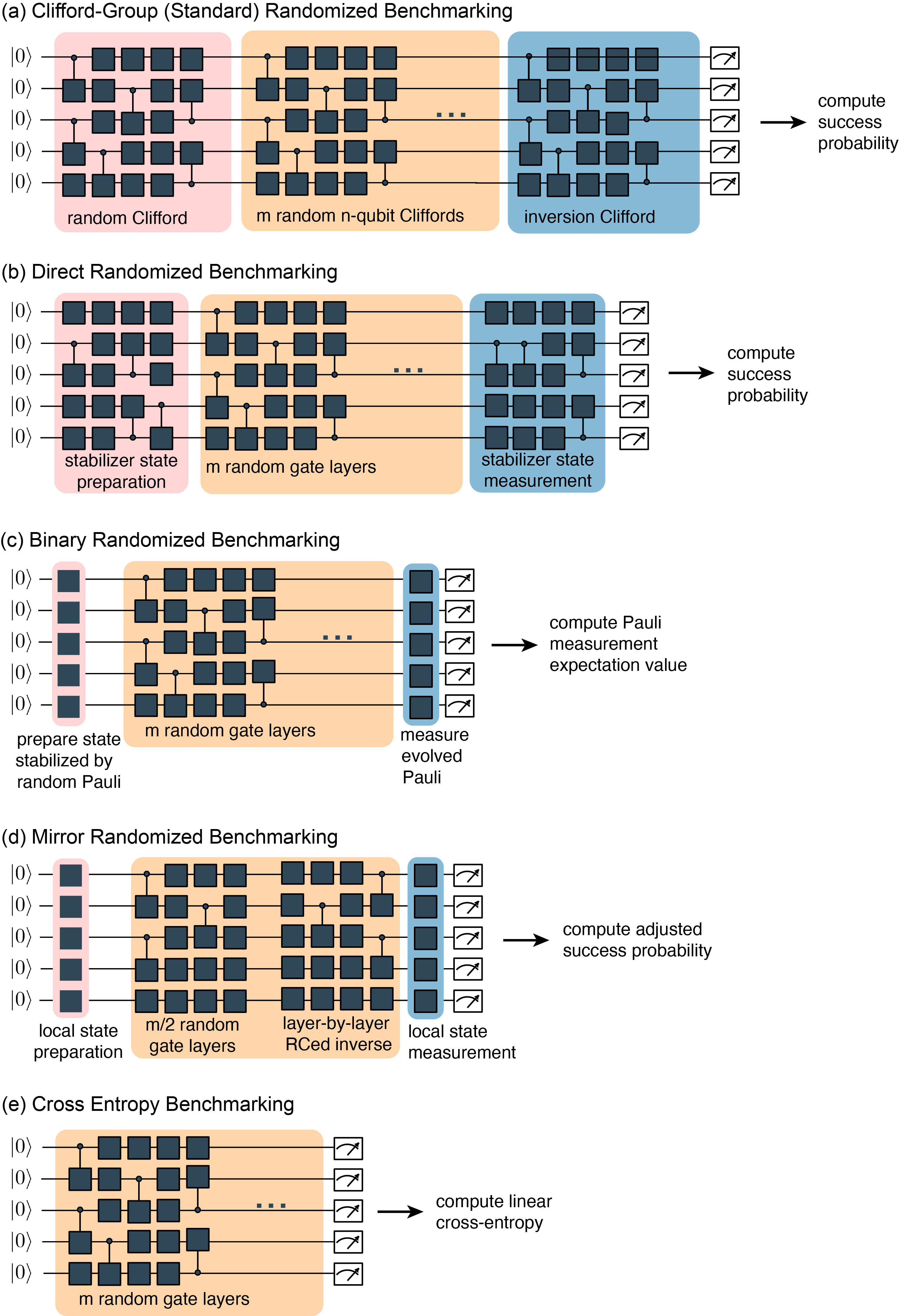}
    \caption{\textbf{Clifford-group and Native Gate RB Methods.}
    The structure of the random circuits and the success metrics used in \textbf{(a)} Clifford-group RB, \textbf{(b)} direct RB, \textbf{(c)} binary RB, \textbf{(d)} mirror RB, and \textbf{(e)} cross-entropy benchmarking. The red denotes state preparation layers (or sub-circuits), orange denotes the benchmarking sequence, and blue denotes the measurement basis rotations. CRB circuits consist of $m+1$ uniformly random $n$-qubit Clifford gates followed by the unique Clifford gate that inverts those $m+1$ gates, and so CRB measures the mean error rate of these $n$-qubit Clifford gates, often called the ``error per Clifford'' (EPC). The native gate RB protocols (b) -- (e) are all based on circuits containing $m$ layers of gates sampled from some distribution $\Omega$ over a layer set $\mathbb{G}_n$, and most of them surround that ``$\Omega$-distributed random circuit'' with additional circuits that implement different state preparations and measurements. The layer set used in these protocols is typically closely related to the set of all native layers for a system. These techniques all measure an error rate $r_{\Omega}$ that quantifies the $\Omega$-weighted average error rate of these layers. Each of the native gate RB protocols has its own strengths, limitations, and regimes of applicability, which are discussed in the main text.}
    \label{fig:native_rb_circuits}
\end{figure}

%%%%%%%%%%%%%%%%%%%%%%% Native Gate RB %%%%%%%%%%%%%%%%%%%%%%%
\subsection{Native Gate RB}\label{sec:drb+}

\emph{Native gate RB} is a family of methods that can directly benchmark a system's native $n$-qubit gates (which are often referred to as ``layers'' or ``cycles,'' but here we will follow RB convention and call them ``gates''). The main native gate RB techniques are:
\begin{itemize}
    \item direct RB (Sec.~\ref{sec:drb}), 
    \item binary RB (Sec.~\ref{sec:birb}),
    \item mirror RB (Sec.~\ref{sec:mrb}), and 
    \item linear cross-entropy benchmarking (Sec.~\ref{sec:xeb}).
\end{itemize}

\begin{figure*}[t]
    \centering
    \includegraphics[width=2\columnwidth]{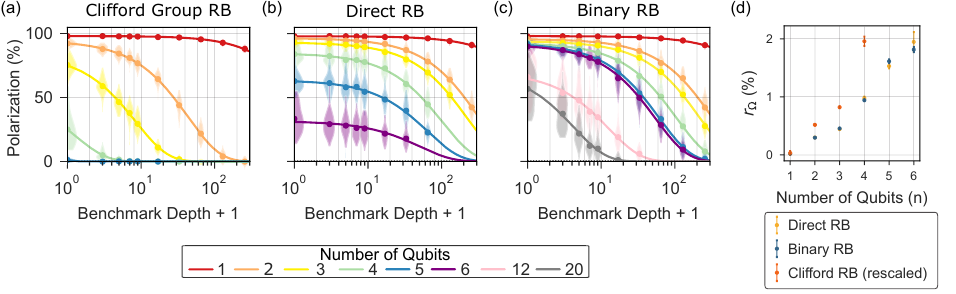}
    \caption{\textbf{Comparing Clifford-group RB, Direct RB, and Binary RB.} 
    The results of running \textbf{(a)} Clifford-group RB, \textbf{(b)} direct RB, and \textbf{(c)} binary RB on \texttt{ibm-hanoi}. For all protocols, the $m=0$ polarization drops as the number of benchmarked qubits increases, but this effect is smaller for protocols with shorter state preparation and measurement subroutines. 
    \textbf{(d)} The error rates extracted from each dataset. DRB and BiRB both measure the error rate of layers sampled from a distribution $\Omega$, and their results are similar. In contrast, CRB measures the error per $n$-qubit Clifford (EPC), which is often rescaled to estimate the error rate of native gates, but is not guaranteed do so accurately. (Figure adapted with permission from \R\cite{hines2023fully}).
    }
    \label{fig:native_rb_demo}
\end{figure*}

Native gate RB protocols address two practical limitations of standard CRB. Firstly, CRB is infeasible beyond a few qubits even with state-of-the-art gate error rates. This is because CRB runs circuits containing uniformly random elements of the $n$-qubit Clifford group, and the size of the circuits needed to create these $n$-qubit Clifford gates grows very rapidly with $n$ for typical native gate sets. In particular, a typical $n$-qubit Clifford gate requires $\mathcal{O}(n^2/\log n)$ two-qubit gates \cite{aaronson2004improved, Maslov2018-nl, Bravyi2020-hg, Proctor2023-ep, patel2008optimal}. The average success probability of even the shortest CRB circuits therefore quickly drops off to almost zero as $n$ increases \cite{polloreno2023theory}, as demonstrated in \fig\ref{fig:native_rb_demo}(a). This makes it impossible to estimate the EPC without impractical amounts of data when $n\gg 1$ (and the EPC rapidly converges to 1 as $n$ increases). Secondly, CRB measures the EPC, but most users of RB actually want to know the error per native gate. Although rescaling the EPC to estimate the error per native gate is common practice (see the discussion in the previous subsection), it has little theoretical justification \cite{epstein2014investigating}. Furthermore, beyond the one- and two-qubit setting, it is not even typically clear what would constitute a sensible and useful rescaling of the EPC.

Native gate RB protocols benchmark some user-specified set of $n$-qubit gates $\mathbb{G}_n$. In all the protocols described herein, this gate set is required to \emph{generate} a group that is a unitary 2-design, such as the Clifford group. A one-qubit example of such a gate set is
\begin{equation}\label{eq:GnEx}
    \mathbb{G}_1 = \{X_{\frac{\pi}{2}} , Y_{\frac{\pi}{2}} \} ~.
\end{equation}
In experimental uses of native gate RB methods to date, $\mathbb{G}_n$ has typically been chosen to be parallel applications of either (a) a system's native gates, or (b) one- and two-qubit gates that can easily be constructed from the native gates (e.g., all possible layers consisting of parallel applications of CNOT and single-qubit Clifford gates). Other choices for $\mathbb{G}_n$ are possible, however. 

Native gate RB protocols estimate an average error rate ($r_{\Omega}$) for the gates in $\mathbb{G}_n$ that is weighted by a user-specified probability distribution $\Omega$ over $\mathbb{G}_n$. This error rate is, in essence, the $\Omega$-weighted average infidelity of the gates, i.e., 
\begin{equation}
\epsilon_{\Omega} = \sum_{G \in \mathbb{G}_n} \Omega(G) e_F(G) ~,
\end{equation}
where $e_F(G)$ is the process infidelity of $G$ \cite{proctor2018direct, hines2022demonstrating, hines2023fully, polloreno2023theory} (although, as with CRB, there are some subtleties relating $r_{\Omega}$ to $\epsilon_{\Omega}$ because $\epsilon_{\Omega}$ is not gauge-invariant \cite{polloreno2023theory, hines2022demonstrating}). The distribution $\Omega$ can be chosen to measure the weighted error rate of most interest, and can even be varied to learn about which gates have higher error rates \cite{proctor2018direct, hines2022demonstrating}. For the gate set given in \eq\ref{eq:GnEx}, an example of such a distribution is 
\begin{equation}
   \Omega(X_{\frac{\pi}{2}})= 3/4 ~, \hspace{0.25cm} \Omega(Y_{\frac{\pi}{2}})= 1/4 ~.
\end{equation}

All native gate RB protocols follow a similar procedure to standard RB. Stated informally, they all have the following structure:
\begin{enumerate}
    \item Run random circuits of various depths $m$. The exact structure of the random circuits varies between different methods (see \fig\ref{fig:native_rb_circuits}), but in all cases the circuits consist of
    
    \begin{enumerate}
        \item $m$ random layers sampled from a user-specified distribution $\Omega$, called \emph{$\Omega$-distributed random circuits}, surrounded by 
        
        \item some additional, method-specific state-preparation and measurement layers (or sub-circuits).
    \end{enumerate}
    
    \item Estimate a success metric for each circuit, the details of which depend on the protocol \footnote{Here, a ``success metric'' does not necessarily refer to a ``metric'' in the strict mathematical sense; see the discussion in Sec.~\ref{sec:overview}.}.
    
    \item Estimate the average of this success metric at each depth, fit that average to the exponential decay function of \eq\ref{eq:rb-fit}, and from the fitted value for $f$ compute an error rate $r_{\Omega}$ using \eq\ref{eq:rb_r}.
\end{enumerate}

Native gate RB protocols work because random circuits randomize and spread errors (i.e., via ``scrambling'') \cite{polloreno2023theory} --- which must happen because a random sequence of elements from $\mathbb{G}_n$ converges to a random element generated from the group $\mathbb{G}_n$, which is a unitary 2-design. The implication of this is that the process fidelity of $\Omega$-distributed random circuits ($F_{\Omega,d}$) will decay exponentially with circuit depth at a rate given by $\epsilon_{\Omega}$ under broad conditions \cite{Carignan-Dugas2018-np, hines2022demonstrating, polloreno2023theory}. Each native gate RB protocol differs in (i) the state preparation and measurement structures used in its circuits, and (ii) its choice of success metric. These differences correspond to different ways to measure $F_{\Omega,d}$, each of which has its own strengths and weaknesses.

%%%%%%%%%%%%%%%%%%%%%%% Direct RB %%%%%%%%%%%%%%%%%%%%%%%
\subsubsection{Direct RB}\label{sec:drb}

\emph{\Ac{DRB}} \cite{proctor2018direct, polloreno2023theory} can benchmark any gate set that generates a group that is a unitary 2-design. It has been primarily used to benchmark gate sets that generate the $n$-qubit Clifford group \cite{McKay2023-bx, Chen2023-la, proctor2018direct}, and so we focus on that case. The circuits used in DRB (i) begin with a random sub-circuit that creates a uniformly random stabilizer state, (ii) have a depth $m$ $\Omega$-distributed circuit at their center, and (iii) end with a sub-circuit that maps the state that is (ideally) produced by the circuit so far to a random computational basis state. The structure of DRB circuits is shown in \fig\ref{fig:native_rb_circuits}(b).

Each DRB circuit always outputs a particular bit string $b$ when run without error, and the probability that this bit string is observed is DRB's success metric. The initial and final sub-circuits within a DRB circuit implement a (state) 2-design twirl on the error in the $\Omega$-distributed circuit. This guarantees that the mean success probability of DRB circuits decays exponentially and DRB's error rate ($r_{\Omega}$) approximately equals the weighted-average error rate $\epsilon_{\Omega}$ of the benchmarked gates \cite{proctor2018direct, polloreno2023theory}.

Figure~\ref{fig:native_rb_demo}(b) demonstrates DRB.
This figure shows the average polarization decay obtained when running $n$-qubit DRB experiments on an \texttt{IBM Q} system, for $n=1$ to $6$.
The key differences between DRB and CRB are illustrated by comparing \fig\ref{fig:native_rb_demo}(b) to the results of $n$-qubit CRB experiments run at the same time on the same system, shown in \fig\ref{fig:native_rb_demo}(a). First, the CRB polarization decays more quickly with depth than the DRB polarization does, i.e., they are measuring different error rates. Importantly, CRB measures the EPC, whereas DRB measures the error per layer of native gates. Second, the average polarization of the shallowest DRB circuits ($m=0$) is typically larger than that of the shallowest CRB circuits (also $m=0$). This is because generating a uniformly random stabilizer state requires only about $1/3$ as many two-qubit gates as generating a uniformly random Clifford gate \cite{Proctor2023-ep}. This means that DRB is feasible on more qubits than CRB. However, DRB is still not truly scalable, because DRB's state preparation and measurement subroutines require $\mathcal{O}(n^2/\log{n})$ two-qubit gates \cite{aaronson2004improved, Maslov2018-nl, Bravyi2020-hg, patel2008optimal, Proctor2023-ep}. These large subroutines mean that the polarization of $m=0$ DRB circuits still drops rapidly with increasing $n$ [see \fig\ref{fig:native_rb_circuits}(b)], even though it does not drop as quickly as CRB.

%%%%%%%%%%%%%%%%%%%%%%% Binary RB %%%%%%%%%%%%%%%%%%%%%%%
\subsubsection{Binary RB}\label{sec:birb}

\emph{\Ac{BiRB}} \cite{hines2023fully} is a native gate RB protocol that is designed to benchmark any gate set that generates the $n$-qubit Clifford group. BiRB's circuit structure is shown in \fig\ref{fig:native_rb_circuits}(c). Unlike most RB protocols, BiRB's circuits do not include an inversion gate or sub-circuit at their end --- i.e., they are not motion reversal circuits. BiRB circuits are therefore not definite outcome circuits, since they do not always return a particular ``success'' bit string if run without error. Instead, BiRB circuits consist of an $\Omega$-distribution random circuit with a layer of single-qubit gates at its start and at its end, and 50\% of all possible $n$-bit strings are designated as ``success'' bit strings and the other 50\% as ``fail'' bit strings.

The initial layer of single-qubit gates in BiRB circuits creates a tensor product eigenstate of a uniformly random $n$-qubit Pauli operator $P$. This simulates sending a uniformly random Pauli operator $P$ into an $\Omega$-distributed circuit --- a technique that enables scalable fidelity estimation, and which is also used in direct fidelity estimation (Sec.~\ref{sec:direct_fid_est}), cycle benchmarking (Sec.~\ref{sec:cb}), and Pauli noise learning (Sec.~\ref{sec:pnr}). In the absence of errors, the $\Omega$-distributed circuit transforms $P$ into another Pauli operator $P'$. The final layer of gates simply transforms $P'$ into a $Z$-type Pauli operator $P''$ (a tensor product of $Z$ and $I$ operators), enabling the measurement of whether $P$ ``survived'' the circuit (i.e., was correctly transformed by the circuit) using only a computational basis measurement. If the readout bit string is a $+1$ eigenstate of $P''$ we declare ``success,'' and otherwise we declare ``fail.'' We then (i) compute the success metric
\begin{equation}
    p = \nu_{\textrm{success}} - \nu_{\textrm{fail}} ~,
\end{equation}
where $\nu_{\textrm{success}}$ and $\nu_{\textrm{fail}}$ are the frequencies at which success and fail bit strings are observed, respectively; (ii) fit the mean of $p$ versus depth to the standard exponential decay function of \eq\ref{eq:rb-fit} with $B=0$; and (iii) estimate $r_{\Omega}$ using \eq\ref{eq:rb_r}.

An example of BiRB data is shown in \fig\ref{fig:native_rb_demo}(c). BiRB is more scalable than DRB (and CRB) and it is also arguably simpler to implement. This is because BiRB's circuits do not start and end with large sub-circuits [see \fig\ref{fig:native_rb_circuits}(c)]. The better scaling of BiRB can be seen by comparing \fig\ref{fig:native_rb_demo}(c) with (a) and (b). This shows that the polarization of the shallowest BiRB circuits decreases more slower than that of both DRB and CRB circuits as a function of the number of qubits. Note, however, that there is still a gradual decrease in the polarization of these shallowest circuits due to the increasing SPAM error with $n$.

%%%%%%%%%%%%%%%%%%%%%%% Mirror RB %%%%%%%%%%%%%%%%%%%%%%%
\subsubsection{Mirror RB}\label{sec:mrb}

\emph{\Ac{MRB}} \cite{hines2022demonstrating, proctor2021scalable, Mayer2021-vl, Amico2023-ze} is a native gate RB protocol that is scalable because it uses random ``mirror circuits'' \cite{proctor2022measuring} (see Secs.~\ref{sec:aces}, \ref{sec:mcfe}, and \ref{sec:mirror} for more on mirror circuits). MRB can efficiently benchmark both Clifford and universal gate sets. The structure of MRB circuits is shown in \fig\ref{fig:native_rb_circuits}(d). An MRB circuit of benchmark depth $m$ consists of (i) a layer of single-qubit gates each sampled independently from a 2-design, (ii) $m/2$ gates sampled from $\Omega$, and (iii) a depth-($1+m/2$) circuit consisting of each layer in the circuit so far, but in the reverse order, and each replaced with its inverse (i.e., a layer-by-layer inversion circuit), and (iv) randomized compiling applied to the entire circuit. Randomized compiling ensures that errors do not coherently add or cancel between a layer and its inverse in the second half of the circuit \cite{hines2022demonstrating, proctor2021scalable, proctor2022measuring}. Note that unlike in DRB, BiRB, and linear cross-entropy benchmarking (Sec.~\ref{sec:xeb}), the $m$ layers are \emph{not} all sampled independently from $\Omega$. Instead, $m/2$ layers are sampled independently from $\Omega$, and then the next $m/2$ layers are their inverses. MRB's use of a layer-by-layer inverse removes the large state-preparation and measurement subroutines used in DRB (and CRB) circuits.

\begin{figure}[t]
    \centering
    \includegraphics[width=\columnwidth]{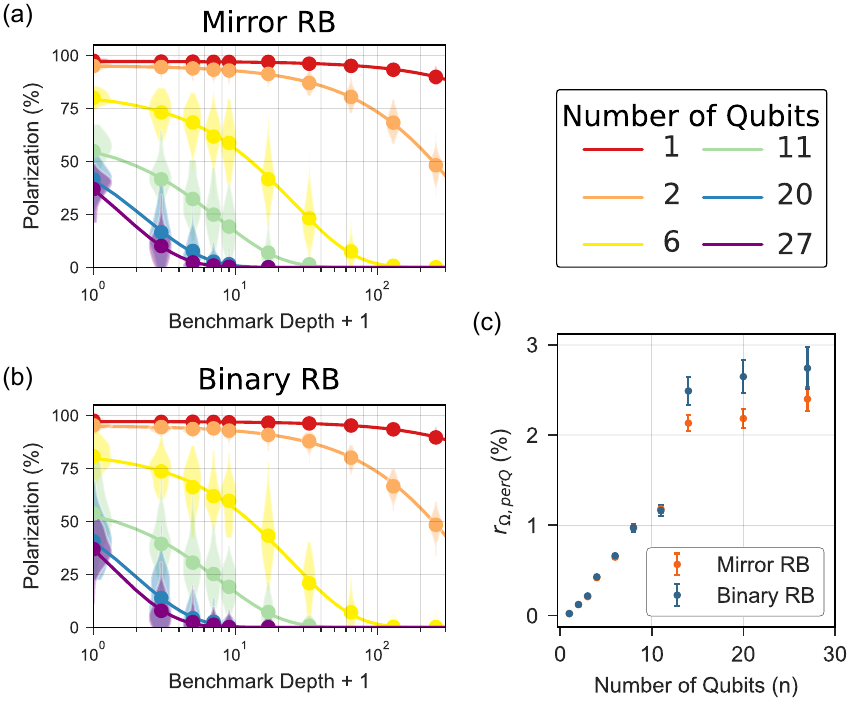}
    \caption{\textbf{Demonstrating Mirror RB and Binary RB.} 
    Results from running \textbf{(a)} mirror RB and \textbf{(b)} binary RB on \texttt{ibm-kolkata} to benchmark a Clifford gate set.
    \textbf{(c)} The RB error rate per qubit ($\sim r_{\Omega}/n$) extracted from each dataset. The error rate per qubit increases with $n$, which indicates the presence of crosstalk. MRB and BiRB are both designed to measure the error rate of layers sampled from a distribution $\Omega$, but MRB is known to typically slightly underestimate this error rate, which is consistent with the observations here. However, MRB can efficiently benchmark a universal gate set, whereas no other RB protocol can. (Figure adapted with permission from \R\cite{hines2023fully}.)
    }
    \label{fig:mrb_demonstration}
\end{figure}

MRB's state preparation and measurement layer of single-qubit gates are based on the insight that the infidelity of an error channel can be efficiently estimated using single-qubit 2-design twirling. However, this requires a more complex success metric than the frequency of observing the ``success'' bit string, used in DRB and CRB. In MRB, the success metric is based on the Hamming distance of the observed bit string from the success bit string. In particular, MRB's success metric --- called the \emph{adjusted success probability} --- is
\begin{equation}\label{eq:adjsp}
    p =  \frac{4^n}{4^n-1} \left[ \sum_{k=0}^{n} \left( -\frac{1}{2} \right)^k h_k \right] - \frac{1}{4^n -1} ~,
\end{equation}
where $h_k$ is the frequency that the circuit outputs a bit string with Hamming distance $k$ from its target bit string.
The theories in Refs.~\cite{proctor2022establishing,proctor2021scalable,emerson2005scalable} show that \eq\ref{eq:adjsp} is a reliable estimator of fidelity when using a local 2-design twirl.

Figure~\ref{fig:mrb_demonstration} demonstrates the use of MRB to benchmark a set of layers that generate the Clifford group, and compares MRB to BiRB of the same layer set. The correlations in MRB circuits enable creating motion reversal circuits without large ``overhead'' subroutines, as in DRB circuits, but they also have an unwanted side-effect. MRB theory \cite{hines2022demonstrating, proctor2021scalable} shows that if the error rates of a $\Omega$-distributed layer and its inverse are uncorrelated, then MRB accurately estimates $\epsilon_{\Omega}$, but that if these error rates are correlated then MRB \emph{slightly} underestimates $\epsilon_{\Omega}$. In real systems, these error rates typically are correlated, resulting in MRB slightly underestimating $\epsilon_{\Omega}$ \cite{hines2022demonstrating, proctor2021scalable}. We observed this effect in \fig\ref{fig:mrb_demonstration}, with BiRB's error rates slightly larger the MRB's error rates. BiRB is, therefore, expected to estimate $\epsilon_\Omega$ marginally more accurately than MRB, and BiRB is just as a scalable as MRB. However, MRB can efficiently benchmark universal gate sets (e.g., see the experiments in \R\cite{hines2022demonstrating}), where as BiRB (and DRB) cannot.

%%%%%%%%%%%%%%%%%%%%%%% XEB %%%%%%%%%%%%%%%%%%%%%%%
\subsubsection{Cross-Entropy Benchmarking}\label{sec:xeb}

\begin{figure*}[t]
    \centering
    \includegraphics[width=1.8\columnwidth]{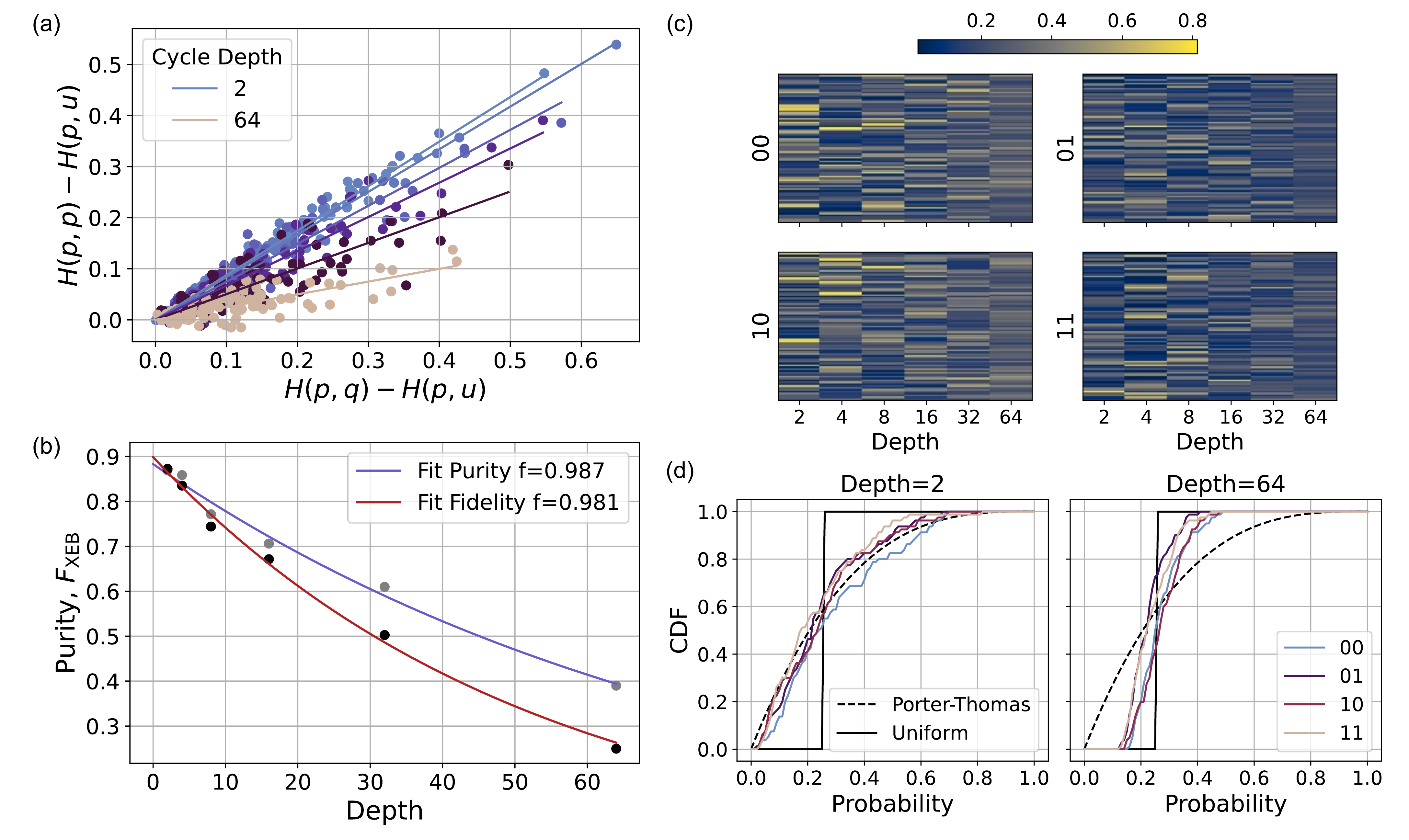}
    \caption{\textbf{Cross-Entropy Benchmarking of a Two-qubit CZ Gate.} 
    (a) Linear fits yielding $F_\text{XEB}(m)$ at each depth $m$. 
    (b) Exponential decay fits (solid lines) of measured (dots) $F_{\text{XEB}}(m)$ (black) and speckle purity (grey) as a function of depth $m$. The dressed CZ gate has a process infidelity of $e_F = 1.78\%$, of which 68\% can be attributed to stochastic errors based on the decay of the speckle purity.
    (c) The speckle pattern is plotted for each bit string across $N=30$ random circuits ($y$-axis) and at each depth $m$ ($x$-axis). We see that the speckled pattern at low depths (characteristic of the Porter-Thomas distribution) is smeared out at longer depths (as the uniform distribution is approached). 
    (d) Cumulative distribution function (CDF) of distributions of bit string probabilities (colored lines) at depths $m=2$ and $m=64$. At low depth, $m=2$, the probabilities of the various bit strings closely follow the Porter-Thomas distribution (dashed black line). At larger depths these distributions begin to converge to the uniform distribution (solid black line) where the probability of the given bit string is close to $1/d = 0.25$ for all of the $N=30$ random circuits. }
    \label{fig:xebdata}
\end{figure*}

\emph{\Ac{XEB}} is a collection of related protocols that run random circuits and quantify how well they performed by estimating the cross-entropy between the actual ($\mathbf{q}$) and ideal ($\mathbf{p}$) outcome distributions \cite{boixo2018characterizing, 2019GoogleSupremacy, neill2018blueprint, liu2021benchmarking, Heinrich2022-cs, Chen2022-hd}. In practice, these techniques typically use the \emph{linear} cross-entropy (see also \eq\ref{eq:lincrossentropy}):
\begin{equation}
    H_{\mathrm{lin}}(\mathbf{p}, \mathbf{q}) \equiv 2^n\sum_{x} {p_{x}{q}_x} - 1 ~, \label{eq:lincrossentropy2}
\end{equation}
where the sum is over all $n$-bit strings.

In the context of QCVV, the most important XEB methods are a family of protocols for measuring the average error rate ($\epsilon_{\Omega}$) of $n$-qubit circuit layers/gates --- i.e., they measure the same quantity as other native gate RB protocols discussed throughout Sec.~\ref{sec:drb+} --- and this is the type of protocol we detail below. But, first, we briefly discuss another meaning for ``XEB'' --- the protocol used for demonstrating ``quantum supremacy'' \cite{boixo2018characterizing, 2019GoogleSupremacy}. That XEB procedure is as follows: (i) run the $n$-qubit scrambling circuits described in the ``quantum supremacy'' literature \cite{boixo2018characterizing, 2019GoogleSupremacy}, (ii) run experiments to estimate $H_{\mathrm{lin}}(\mathbf{q}, \mathbf{p})$, where $\mathbf{q}$ is the actual and $\mathbf{p}$ the ideal outcome distributions for each sampled circuit (note that this estimation is challenging when $\mathbf{p}$ is infeasible to compute with classical simulations of the circuit), and (iii) use the value of this cross-entropy to quantify a quantum computer's performance. For sufficiently deep circuits on sufficiently many qubits, values of $H_{\mathrm{lin}}(\mathbf{p}, \mathbf{q})$ above some threshold $\delta$ are believed to be impossible to achieve in a reasonable amount of time using any existing classical computer \cite{2019GoogleSupremacy, 2021USTSupremacy, 2022USTSupremacy}. Obtaining such values for $H_{\mathrm{lin}}(\mathbf{p}, \mathbf{q})$ is sometimes referred to as demonstrating ``quantum supremacy,'' and this has now been achieved in multiple experiments \cite{2019GoogleSupremacy, 2021USTSupremacy, 2022USTSupremacy}.

We now turn to the XEB protocols that are designed to estimate the average infidelity of random $n$-qubit circuit layers ($\epsilon_{\Omega}$). These XEB protocols follow the same structure as all other native gate RB protocols (discussed throughout this subsection), using (i) plain $\Omega$-distributed random circuits [see \fig\ref{fig:native_rb_circuits}(e)] as its circuit family and (ii) a success metric related to the linear cross-entropy. The success metric is typically
\begin{equation}
    F_{\textrm{XEB}} = \frac{\hat{H}_{\textrm{lin}}(\mathbf{p}, \mathbf{q}) }{H_{\textrm{lin}}(\mathbf{p},\mathbf{p})} ~,\label{eq:lxeb-criteria}
\end{equation}
where $\hat{H}_{\textrm{lin}}(\mathbf{p}, \mathbf{q})$ is an estimate of $H_{\textrm{lin}}(\mathbf{p}, \mathbf{q})$. Typically, the estimate is computed using
\begin{equation} 
    {\hat{H}_{\textrm{lin}}(\mathbf{p},\mathbf{q})}  = \frac{2^n}{N} \sum_{x \in \mathbb{X}} p_x - 1 ~, \label{eq:lxeb-estimator}
\end{equation}
where $\mathbb{X}$ is the set of bit strings observed when running the circuit $N$ times. Calculating \eq\ref{eq:lxeb-estimator} requires classically computing $p_x$ for each $x \in \mathbb{X}$, which is generally exponentially expensive in the number of qubits. In the most well-known XEB experiments \cite{2019GoogleSupremacy}, the mean of $F_{\textrm{XEB}}$ over multiple random circuits of depth $m$ --- given ${\hat{H}_{\textrm{lin}}(\mathbf{p},\mathbf{q})}$ and $H_{\textrm{lin}}(\mathbf{p},\mathbf{p})$ for each circuit --- is estimated by plotting $\hat{H}_{\textrm{lin}}(\mathbf{p}, \mathbf{q})- \hat{H}_{\textrm{lin}}(\mathbf{p}, \mathbf{u})$ versus ${H_{\textrm{lin}}(\mathbf{p},\mathbf{p})}- \hat{H}_{\textrm{lin}}(\mathbf{p}, \mathbf{u})$ for every circuit of the same depth, where $\mathbf{u}$ is the uniform distribution in $d$ dimensions, and fitting that data to a line [as shown in the example of \fig\ref{fig:xebdata}(a)]. Note, however, that alternative success metrics or data analysis techniques can be used in XEB; for example, the success metric defined by \eq\ref{eq:lincrossentropy2} can be used instead of \eq\ref{eq:lxeb-criteria}, or the techniques of \emph{filtered RB} \cite{Heinrich2022-cs, helsen2022framework} can be applied to data from XEB circuits.

XEB's success metric is arguably less intuitive than the success probability used in most RB protocols, and so we now explain why $F_{\textrm{XEB}}$ enables estimating the average error rate of the benchmarked layers. Consider a depth-$m$ XEB circuit $\mathcal{C}$ and the observable $O_\mathcal{C} = \sum_x p_x \ketbra{x}$, where $\textbf{p} = \{p_x\}$ is $\mathcal{C}$'s ideal outcome distribution. Now, assume that $\mathcal{C}$'s imperfect implementation can be modelled by an $n$-qubit depolarizing error channel after each layer with process polarization $f$, i.e., the state output by $\mathcal{C}$ is 
\begin{equation}
    \rho_\mathcal{C} = f^m \ketbra{\psi_\mathcal{C}}{\psi_\mathcal{C}} + (1 - f^m) \frac{\mathbb{I}}{2^n} ~,
\end{equation}
where $m$ is the circuit's depth, and $\ketbra{\psi_\mathcal{C}}{\psi_\mathcal{C}}$ is the pure state that $\mathcal{C}$ would ideally create.
Then,
\begin{equation}
    \Tr[O_\mathcal{C} \rho_\mathcal{C}] = f^m \Tr[O_\mathcal{C} \psi_\mathcal{C}] + (1 - f^m) \frac{1}{2^n}\Tr[O_\mathcal{C}] ~.
\end{equation}
By substituting in $O_\mathcal{C}$, we find that
\begin{equation}
    \sum_x p_x q_x  = f^m\left(\sum_x p_x^2 - \frac{1}{2^n}\right) + \frac{1}{2^n} ~,
\end{equation}
where $q_x = \bra{x}\rho_\mathcal{C}\ket{x}$ is the actual probability of observing $x$. Rearranging, and substituting in the definition of the linear cross-entropy, we obtain:
\begin{equation}
    f^m = \frac{\hat{H}_{\textrm{lin}}(\mathbf{p}, \mathbf{q}) }{H_{\textrm{lin}}(\mathbf{p}, \mathbf{p})} ~.
\end{equation}
So, by estimating the RHS of this equation (for randomly sampled circuits) versus circuit depth, and fitting its mean versus $m$ to an exponential $F_{\textrm{XEB}} = Af^m$, we can extract the polarization $f$, from which we can calculate $r_{\Omega}$ using \eq\ref{eq:rb_r}.

The XEB protocol is defined for both Clifford \cite{Chen2022-hd} and non-Clifford \cite{boixo2018characterizing, 2019GoogleSupremacy, neill2018blueprint, liu2021benchmarking, Heinrich2022-cs}  circuits. However, we note that the canonical circuits for XEB are the same circuits as in the ``quantum supremacy'' demonstrations, and XEB with Clifford circuits requires an impractical number of samples for low-uncertainty estimates of $r_{\Omega}$ (BiRB is a reliable alternative to XEB with Clifford circuits that has better statistical properties). XEB reliably estimates the average (in)fidelity of $\Omega$-distributed layers under certain regularity conditions, including that the errors must be small \cite{Ware2023-zy}. As with other RB protocols, XEB is only a reliable, well-defined procedure if its success metric ($F_{\textrm{XEB}}$) decays exponentially. The theory of XEB shows that $F_{\textrm{XEB}}$ will be an exponential (assuming small Markovian errors), but only for XEB circuits that are deeper than some minimal depth $m_{\textrm{min}}$ \cite{helsen2022framework, liu2021benchmarking, Heinrich2022-cs, Chen2022-hd}. This minimal depth is related to the scrambling rate of the $\Omega$-distributed circuits that are chosen --- i.e., how many $\Omega$-random layers are needed to approximately transform any error map into an $n$-qubit depolarizing channel (note that the above theory simply assumes that each error map can be represented by such an $n$-qubit depolarizing channel). This minimum depth, therefore, depends on $\Omega$ and the layer set that $\Omega$ samples from (and therefore also on a device's connectivity) \cite{helsen2022framework, liu2021benchmarking, Heinrich2022-cs, Chen2022-hd}.

In addition to benchmarking the average infidelity of random $n$-qubit circuit layers, XEB can be structured to benchmark individual gates, layers of gates, or sub-circuits that are fully scrambling. Figure \ref{fig:xebdata} illustrates XEB performed on a two-qubit CZ gate. The benchmarked layers are composite layers consisting of (i) a layer of Haar-random single-qubit gates on each qubit, and then (ii) a two-qubit CZ gate. So, each random layer is a ``dressed'' CZ gate. From the results in Fig.~\ref{fig:xebdata}(b), one can extract a dressed process fidelity of $F_e = 98.2\%$. It should be noted, however, that unlike other methods for estimating individual gate (in)fidelities, such as interleaved RB (Sec.~\ref{sec:irb}) and cycle benchmarking (Sec.~\ref{sec:cb}), it is not as straightforward to separate the infidelity of the Haar-random twirling gates from the infidelity of the interleaved gate \footnote{Because XEB requires that an $n$-qubit circuit converges to an $n$-qubit Haar-random unitary, the infidelity of twirling layers consisting only of Haar-random single-qubit gates cannot be measured via an $n$-qubit XEB experiment; rather, it must be estimated from the combined infidelity of simultaneous XEB on all $n$ qubits. Or, instead, one could use $n$-qubit Haar-random unitaries for the twirl, in which case an $n$-qubit XEB experiment without the interleaved gate could be used to estimate the infidelity of the twirling layer. However, in this case, the decomposition of XEB circuits to native gates would scale poorly (similar to $n$-qubit CRB). Furthermore, note that the estimate of the interleaved gate's fidelity would be subject to systematic errors similar to those seen in IRB.}. Thus, by default, XEB always returns an estimate of the infidelity of the \emph{dressed} gate or layer. However, one of the utilities of XEB is that it does not require the interleaved gate or layer to be Clifford (unlike interleaved RB or cycle benchmarking), and has been used to benchmark the fidelity of multi-qubit non-Clifford gates, such as an $i$Toffoli \cite{kim2022high}, controlled-controlled-Z (CCZ) \cite{nguyen2022programmable}, and CCCZ gate \cite{nguyen2023empowering}. As outlined in Appendix \ref{sec:qudit_xeb}, XEB can be extended to benchmarking qudit gates as well.

%%%%%%%%%%%%%%%%%%%%%%% RB of General Groups %%%%%%%%%%%%%%%%%%%%%%%
\subsection{RB of General Groups}\label{sec:rb-general-groups}

Standard RB can benchmark any gate set that is both a group and a unitary 2-design (e.g., the 24 single-qubit Clifford gates), and modern native gate RB methods can directly benchmark a gate set that simply \emph{generates} a group that is a 2-design (e.g., $\{X_{\pi/2},Y_{\pi/2}\}$). However, some interesting gate sets either \emph{generate} groups that are not unitary 2-designs, or \emph{are} groups but \emph{are not} unitary 2-designs. For example, the CNOT, Hadamard, and $Z$ gates generate the ``real Clifford group,'' which is not a unitary 2-design \cite{Hashagen2018-dk}. Gate sets like this cannot be benchmarked either indirectly by standard RB or directly by (existing) native gate RB methods. Here, we discuss RB techniques that address this problem, and enable RB of gate sets that are groups but not unitary 2-designs \cite{Brown2018-dx, Hashagen2018-dk, helsen2022framework, carignandugas2015characterizing, claes2021character, Helsen2022matchgate}. 

The random circuits of standard RB can be constructed for any gate set that is a group, but when that group is not a unitary 2-design the average success probability $\bar{p}(m)$ of these circuits will not generally follow the simple exponential form $\bar{p}(m) = Af^m + B$, even approximately. Instead, the theory of twirling over general groups (see Sec.~\ref{sec:rb_math_twirling}) implies that $\bar{p}(m)$ will be a sum over multiple exponential decays, and those exponential decays can be \emph{matrix} exponentials. Specifically,
\begin{equation}\label{eq:RBmulti}
    \bar{p}(m) \approx \sum_{\lambda=1}^{k}\Tr(A_{\lambda}M_{\lambda}^m) ~,
\end{equation}
where the $M_{\lambda}$ matrices contain average gate error information (i.e., together they can be used to compute the mean infidelity of the gates), and the $A_{\lambda}$ matrices absorb all SPAM errors \cite{helsen2022framework}. The exact functional form is determined by how the superoperator representation of a gate $G$ decomposes into irreducible representations (see Sec.~\ref{sec:rb_math_twirling}) of $G$. Each term in \eq\ref{eq:RBmulti} corresponds to an irreducible representation in the decomposition of the superoperator representation, and the dimensions of $M_{\lambda}$ and $A_{\lambda}$ depend on the multiplicity of the corresponding irreducible representation. Reliably fitting data to multi-exponentials is challenging \cite{helsen2022framework}, and it contrasts with the conceptual and practical simplicity of RB. 

The literature on RB of groups that are not unitary 2-designs \cite{Brown2018-dx, Hashagen2018-dk, helsen2022framework, carignandugas2015characterizing, Helsen2022matchgate, claes2021character, Franca2018-ow} is about creating (1) techniques for reliably analyzing data of the form given in \eq\ref{eq:RBmulti} and/or (2) techniques for adapting the RB circuits and data analysis so that it is possible to separate out the multi-exponential decay of \eq\ref{eq:RBmulti} into individual exponential decays that can be separately analyzed. There are a variety of protocols for RB of particular groups that are not unitary 2-designs, including dihedral RB \cite{carignandugas2015characterizing}, real RB \cite{Brown2018-dx, Hashagen2018-dk}, and filtered RB \cite{Heinrich2022-cs, helsen2022framework}. Perhaps the most well-known RB protocol for general groups is \emph{character RB} \cite{helsen2022framework, helsen2019new, claes2021character}, and this is the only such protocol we discuss further.

%%%%%%%%%%%%%%%%%%%%%%% Character RB %%%%%%%%%%%%%%%%%%%%%%%
\subsubsection{Character RB}\label{sec:character_rb}

Character RB \cite{helsen2022framework, helsen2019new, claes2021character} is a conceptually elegant technique for RB of general groups that is also of practical importance (but which also has some important limitations). Character RB uses techniques from group representation theory to robustly isolate individual exponential decays in the multi-exponential of \eq\ref{eq:RBmulti}. The general and somewhat abstract ideas underpinning character RB enable many practical RB protocols, including an RB technique designed for biased-noise qubits \cite{claes2023estimating}. Many other RB or RB-adjacent protocols, such as simultaneous RB (Sec.~\ref{sec:simrb}) and cycle benchmarking (Sec.~\ref{sec:cb}), use the same technique. A character RB experiment is determined by:
\begin{itemize}
    \item a \emph{benchmarking group} $\mathcal{G}$, which is the set of gates one aims to benchmark, and 
    \item a \emph{character group} $\hat{\mathcal{G}} \subseteq \mathcal{G}$, which is used to extract individual exponential decays robustly. 
\end{itemize}
When run without errors, each character RB circuit implements a uniformly random element $\hat{G}$ of $\hat{\mathcal{G}}$. The results of running different circuits are added together with weights determined by $\hat{G}$ and a character function, which depends on the structure of $\hat{\mathcal{G}}$ and the decay the experiment aims to isolate. 

Character RB is not capable of isolating each exponential decay for every group. If the benchmarking group is not multiplicity-free (i.e., the superoperator representation of the group contains multiple copies of one or more irreducible representations) the results of character RB will still include multi-exponential decays \cite{claes2021character}. Furthermore, character RB requires an impractical number of samples for some groups --- a limitation that is addressed by filtered RB \cite{Heinrich2022-cs, helsen2022framework}. See \R\cite{helsen2022framework} for a detailed discussion of RB of general groups.

%%%%%%%%%%%%%%%%%%%%%%% simultaneous RB %%%%%%%%%%%%%%%%%%%%%%% 
\subsection{Simultaneous RB}\label{sec:simrb}

\emph{\Ac{sRB}} is a widely-used method for quantifying crosstalk errors \cite{Gambetta2012-zd}. It is perhaps the simplest of a variety of ``advanced'' RB techniques (discussed in Sec.~\ref{sec:simrb} -- \ref{sec:cb}) that build on or expand standard RB (Sec.~\ref{sec:standardrb}) and/or the native gate RB protocols (Sec.~\ref{sec:drb+}). These advanced RB methods are designed to measure gate set properties beyond the AGSI that those foundational RB techniques target.

The original sRB protocol consists of running single-qubit CRB on a qubit while either (i) idling neighboring qubits [\fig\ref{fig:rb_sequences}(a)], or (ii) driving those qubits by independently running CRB in parallel on those qubits [\fig\ref{fig:rb_sequences}(b)] \cite{Gambetta2012-zd}. These two \emph{isolated} and \emph{simultaneous} RB experiments result in two decay parameters ($f_{\textrm{iso}}$ and $f_{\textrm{sim}}$) and corresponding error rates ($r_{\textrm{iso}}$ and $r_{\textrm{sim}}$). Comparing these error rates quantifies the change in a qubit's gate error rate caused by driving neighboring qubits. Typically, $r_{\textrm{sim}} > r_{\textrm{iso}}$ due to crosstalk errors. The size of these crosstalk errors is sometimes quantified with the sRB number \cite{Gambetta2012-zd},
\begin{equation}
    r_{\textrm{sRB}} = \frac{d - 1}{d} (1- f_{\textrm{sim}}/f_{\textrm{iso}}) \approx r_{\textrm{sim}} - r_{\textrm{iso}} ~.
\end{equation}

\begin{figure}[t]
    \centering
    \includegraphics[width=\columnwidth]{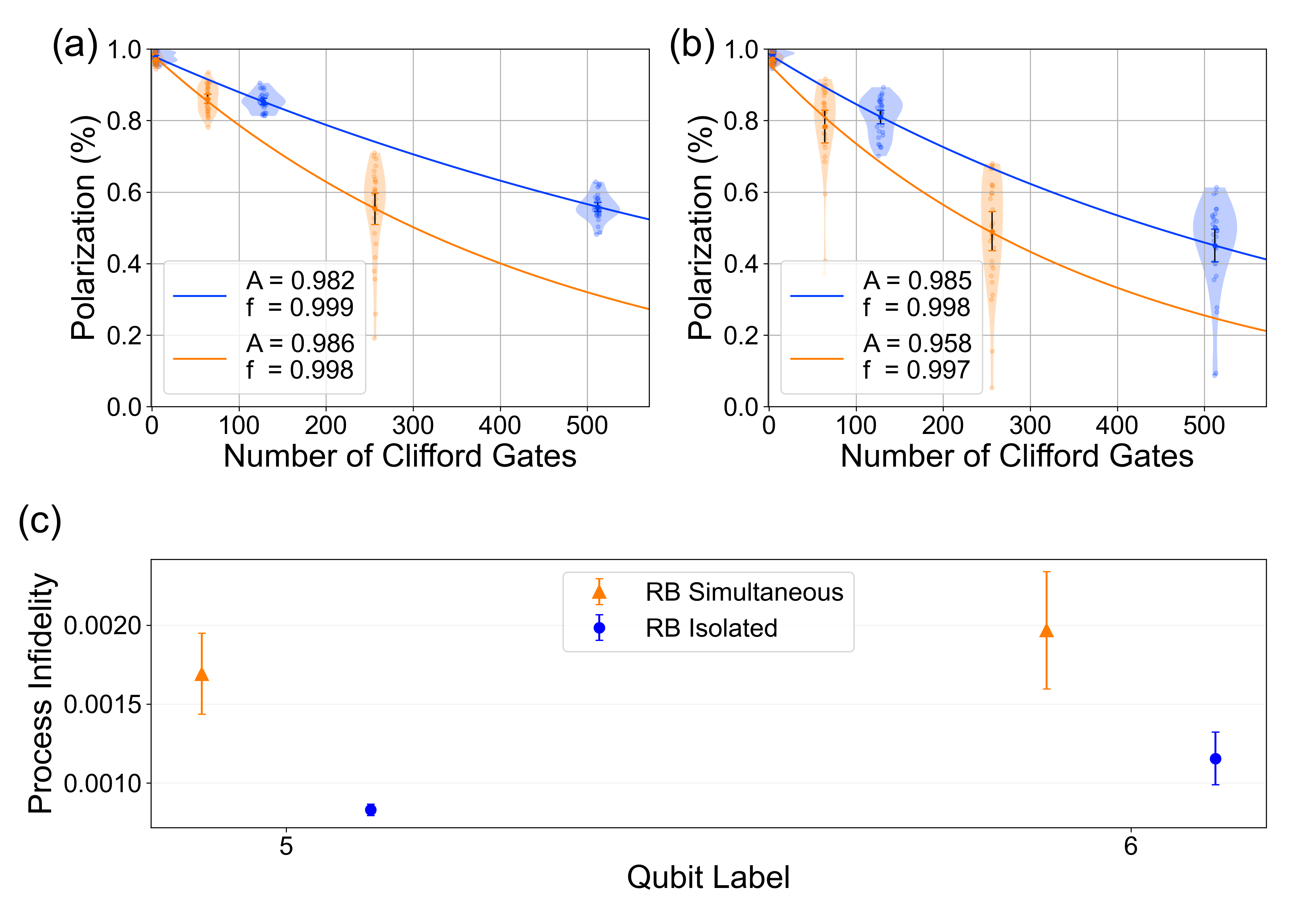}
    \caption[Simultaneous RB]{\textbf{Simultaneous RB.} A demonstration of sRB, whereby RB is run in isolation and in parallel on two superconducting qubits labeled 5 and 6. 
    \textbf{(a)} -- \textbf{(b)} Exponential decays for qubits 5 and 6, respectively, when running RB in isolation (blue) and simultaneously (orange). 
    \textbf{(c)} The process infidelity (i.e., EPC) for each qubit measured in isolation ($e_{F,\textrm{iso}}$, blue) and simultaneously ($e_{F,\textrm{sim}}$, orange). The difference between $e_{F,\textrm{iso}}$ and $e_{F,\textrm{sim}}$ can be used to quantify the crosstalk errors induced by driving the other qubit.}
    \label{fig:1q_rb}
\end{figure}

Figure~\ref{fig:1q_rb} shows data from running single-qubit CRB on two superconducting qubits while idling the other, as well as data from running CRB in parallel on the two qubits. We observe significant differences in the exponential decay rates between the isolated and parallel contexts, indicating that the single-qubit EPC is higher when gates are performed in parallel than in isolation. In this scenario --- and in other superconducting qubit systems in general --- the primary contributor to $r_{\textrm{sRB}}$ is likely crosstalk-induced coherent errors acting on both qubits when they are operated simultaneously.

Running sRB on all the qubits in an $n$-qubit system requires $n+1$ different RB experiments. So, it is now common to run only the simultaneous RB experiment (and to still refer to this as ``sRB''), measuring only $r_{\textrm{sim}}$ for each qubit. In a many-qubit processor, those error rates (one for each qubit) quantity the infidelity of each qubit's gates when running single-qubit gates in parallel on every qubit. Dividing this into contributions from local and crosstalk errors for every qubit requires $n$ more RB experiments, and is not necessary if the goal is to quantify the performance of many-qubit circuits. Therefore, these extra experiments are often skipped.

sRB can also be generalized to quantify crosstalk induced on or by multi-qubit gates by running $n \geq 2$ qubit RB on a set of qubits while either idling all other qubits or running RB on those other qubits \cite{mckay2019three, McKay2023-bx}. There are many ways to do this; for example, running single-qubit RB on all qubits quantifies simultaneous single-qubit crosstalk, running simultaneous two-qubit RB quantifies crosstalk between simultaneous two-qubit gates, or mixing single- and two-qubit RB captures crosstalk between simultaneous single- and two-qubit gates. Each choice for the parallel context will quantify a different aspect of device crosstalk. Notably, a version of simultaneous two-qubit RB is used to define the \emph{\ac{EPLG}} \cite{McKay2023-bx}, which quantifies the error per qubit when a random layer containing one- and two-qubit gates is applied to all qubits in a system simultaneously. As of 2025, IBM now reports the EPLG as one of its primary performance metrics for its cloud-access systems.

Implementing sRB requires addressing a scheduling problem that becomes worse as $n$ increases \cite{McKay2020-uh}. This is because CRB's random $n$-qubit Clifford gates get compiled into circuits of native gates of varying lengths (with typical depth increasing with $n$). This problem can be avoided if sRB does not use CRB, but instead uses a native gate RB protocol (Sec.~\ref{sec:drb+}), such as DRB, BiRB, or XEB \cite{McKay2023-bx}. Finally, we note that data from sRB experiments can also be used to learn more than just $r_{\textrm{sRB}}$, with the aid of a variety of more complex methods \cite{McKay2020-uh, harper2020efficient, harper2023learning} that enable estimating, for example, the rates of correlated errors between different pairs of qubits.

%%%%%%%%%%%%%%%%%%%%%%% Interleaved RB %%%%%%%%%%%%%%%%%%%%%%% 
\subsection{Interleaved RB}\label{sec:irb}

\begin{figure}[t!]
    \centering
    \includegraphics[width=\columnwidth]{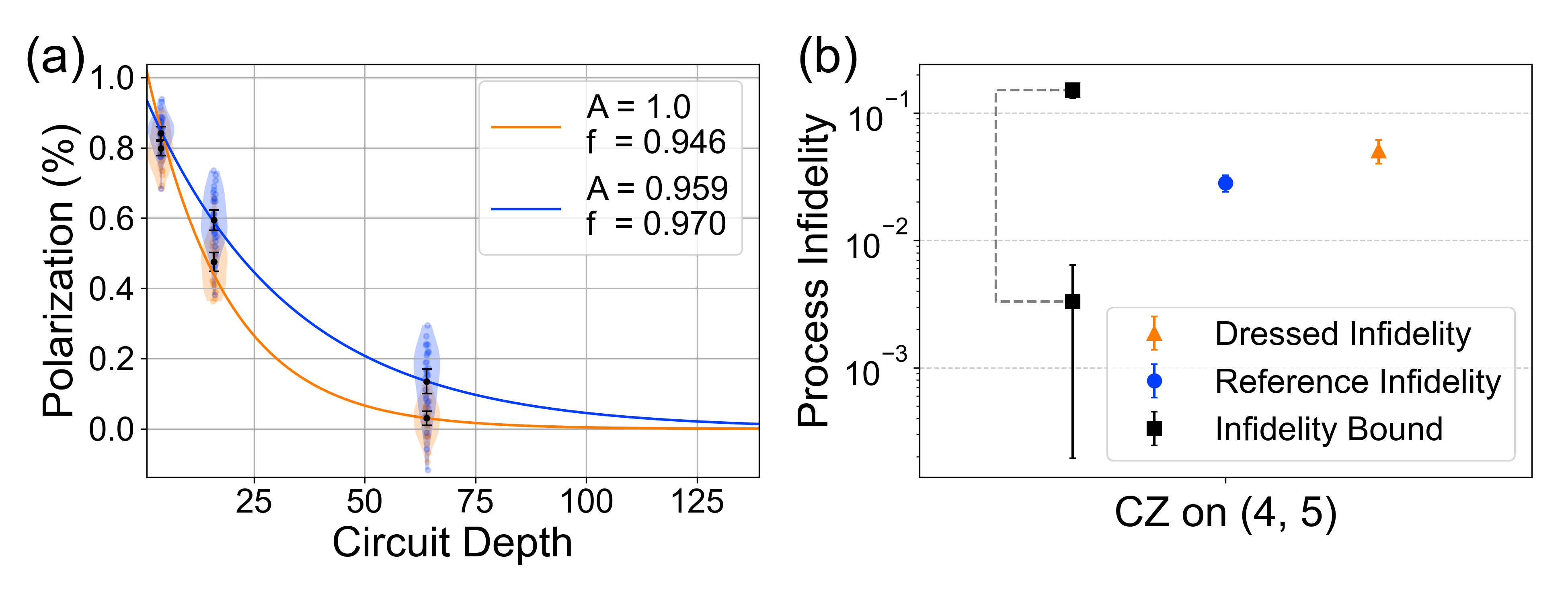}
    \caption[Interleaved RB.]{\textbf{Interleaved RB.} IRB of a CZ gate between two superconducting qubits. 
    \textbf{(a)} The \emph{reference} (blue) and \emph{interleaved} (orange) CRB decays. In the reference experiment, circuit depth ($m$) is the number of uniformly random two-qubit Clifford gates. In the interleaved experiment, each of the random Clifford gates is followed by a CZ gate, resulting in longer circuits and typically a lower average success probability at the same $m$ value. 
    \textbf{(b)} The estimated process infidelity extracted from the reference (blue, $e_F$) and interleaved (orange, $e_{F,D}$) experiments. We estimate the CZ gate's process infidelity [see \eq\ref{eq:ef-irb}] to be $e_{F,CZ} = 2.3(6)\%$. IRB is intended to estimate a gate $G$'s infidelity $\epsilon_G$, but the IRB error rate can be very different from $\epsilon_G$ due to systematic flaws in IRB. Upper and lower bounds [computed from \eq\ref{eq:irb_bounds}] on CZ's infidelity are shown in black. The error bars on this region are computed from the statistical uncertainties in the estimates of $e_F$ and $e_{F,D}$.}
    \label{fig:irb}
\end{figure}

\emph{\Ac{IRB}} \cite{magesan2012efficient} is a method for estimating the infidelity of an individual Clifford gate (extensions to some non-Clifford gates exist \cite{Garion2020-gi, Harper2017-xn}). IRB is typically used to estimate the infidelity of a one- or two-qubit gate, but in principle it can be applied to $n$-qubit gates for any $n$ (e.g., a many-qubit layer of parallel one- and two-qubit gates). IRB for an $n$-qubit gate $G$ is a simple extension of $n$-qubit CRB. It consists of two independent RB experiments. One experiment is often called the \emph{reference} RB experiment and it consists of simply running standard CRB to estimate the CRB decay parameter ($f$) and the corresponding EPC ($r$). The other experiment ---- the \emph{interleaved} RB experiment --- consists of again implementing CRB, but now each randomly sampled Clifford gate is followed by $G$ [see \fig\ref{fig:rb_sequences}(d)], i.e., a depth-$m$ interleaved circuit has the form
\begin{equation}
    \mathcal{C}_{m, G} = C_{m+2} \circ G \circ C_{m+1} \cdots \circ G \circ C_2 \circ G \circ C_1 ~,
\end{equation}
where $C_{1}$, $C_{2}$, $\dots$, $C_{m+1}$ are independent and uniformly sampled Clifford gates (as in CRB), and $C_{m+2}$ is the unique Clifford gate that inverts the entire preceding sequence. The interleaved RB experiment also produces an estimated decay parameter ($f_D$) and corresponding error rate ($r_D$). Figure~\ref{fig:irb}(a) shows reference and interleaved RB decay curves for IRB of a CZ gate between two superconducting qubits. The interleaved curve decays faster than the reference curve due to the additional gate $G$ inserted at each circuit depth.

\begin{figure*}[ht]
    \centering
    \includegraphics[width=2\columnwidth]{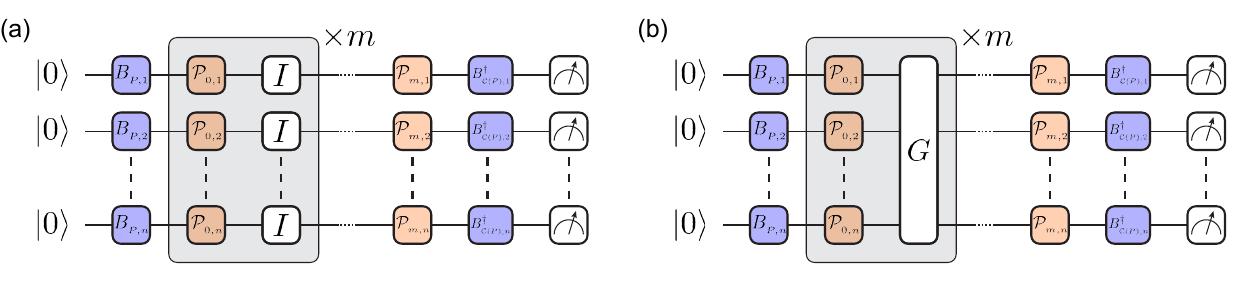}
    \caption{\textbf{Cycle Benchmarking Circuits.} 
        (a) CB of the all-identity ``reference'' cycle. 
        $B_{P,q}$ denotes the basis preparation gate on qubit $q$ for a random eigenstate of the Pauli $P$, $\mathcal{P}_{i, q}$ denotes the $i$th twirling operator acting on qubit $q$, and $B^\dagger_{\mathcal{C}(P),q}$ rotates qubit $q$ back to the initial eigenstate of $P$ at the end of the circuit.
        Explicit identity gates $I$ have been inserted for visual clarity, but this cycle is either skipped in compilation, or the identity gates can be implemented as true idles for the duration of a cycle of single-qubit or two-qubit gates; the choice is up to the experimenter. If the identity gates are skipped in compilation, then this measures the average fidelity of a cycle of random Pauli gates applied simultaneously to all $n$ qubits.
        (b) CB of an $n$-qubit gate cycle $G$. $G$ can be composed of any combination of single- and multi-qubit gates, as long as $G^{m} = \mathbb{I}$ for a sequence depth of $m$.}
    \label{fig:cb_circs}
\end{figure*}

The error rate $r_D$ is an estimate of the mean infidelity of $G$ composed with (i.e., ``dressed'' by) a uniformly random Clifford gate, \emph{not} an estimate of $G$'s infidelity. The standard IRB analysis attempts to subtract the contribution of the uniformly random Clifford gate's error to $r_D$, by comparing $r_D$ to $r$. Specifically, IRB's estimate of the gate $G$'s \emph{average gate infidelity} is defined by
\begin{equation}\label{eq:r-irb}
    r_G = \frac{d - 1}{d} \left(1 - \frac{f}{f_D}\right) \approx r_D - r ~.
\end{equation}
Alternatively, IRB's estimate of the gate $G$'s \emph{process infidelity} is given by
\begin{equation}\label{eq:ef-irb}
     e_{F, G} = \frac{d^2 - 1}{d^2} \left(1 - \frac{f}{f_D}\right) \approx e_{F,D} - e_F ~.
\end{equation}
Applying \eq\ref{eq:ef-irb} to our CZ gate data in \fig\ref{fig:irb}, we estimate CZ's process infidelity to be $e_{F,CZ} = 2.3(6)\%$ [see \fig\ref{fig:irb}(b)].

It is important to highlight that IRB is not generally a reliable method for estimating a gate's infidelity. This is primarily because unitary errors in $G$ can coherently add or cancel with errors in the random Clifford gates $C_i$, and this can even cause the interleaved curve to decay more slowly than the reference curve --- resulting in a \emph{negative} IRB error rate! --- even when $G$'s errors are large. This implies that there is a systematic and potentially large discrepancy between $r_G$ and $G$'s true infidelity, $\epsilon_G$ (we call these discrepancies \emph{systematic} as they are \emph{not} due to shot noise; i.e., they are not statistical in origin). For the average gate infidelity, IRB theory shows that $r_G$ and $\epsilon_G$ are related by the inequalities
\begin{equation}\label{eq:irb-eps}
    \epsilon_G - E <  r_G < \epsilon_G + E ~,
\end{equation}
where
\begin{equation}\label{eq:irb_bounds}
    E = \mathrm{min}  \left\{
    \begin{split}
    	&\frac{(d-1)}{d} \left[ \left| f - \frac{f_{D}}{f} \right| + (1-f) \right] \\
    	&\frac{2(d^2-1)(1-f)}{d^2 f}+\frac{4\sqrt{1-f}\sqrt{d^2-1}}{f}
    \end{split}\right\} ~,
\end{equation}  
and $d=2^n$. The upper- and lower-bounds in \eq\ref{eq:irb-eps} can span orders of magnitude, and a tighter relationship between $r_{G}$ and $\epsilon_G$ can only be guaranteed if more is known about the errors --- e.g., if it is known that coherent errors form a small contribution to infidelity \cite{carignan2019bounding}; see Sec.~\ref{sec:xrb}. In \fig\ref{fig:irb}(b), we plot these upper and lower bounds on the estimated process infidelity for the CZ gate. When this systematic error is combined with the statistically uncertainties in our estimates of $f$ and $f_D$, this range spans over two orders of magnitude, ranging from above $10^{-1}$ to below $10^{-3}$.

IRB has been widely-used, but its large systematic errors have caused it to become less popular in recent years. There are now a variety of alternatives to IRB, including many RB or RB-like techniques for measuring gate infidelities (as well as SPAM-error-robust tomographic techniques like gate set tomography; see Sec.~\ref{sec:gst}). Many of these techniques are also more scalable than IRB (IRB inherits the scaling problems of CRB, discussed in Sec.~\ref{sec:drb+}). One such technique is cycle benchmarking, which we discuss in detail in Sec.~\ref{sec:cb}. Other examples include fitting error models directly to RB data \cite{hines2022demonstrating, Hothem2023-bl}; running native gate RB protocols with different sampling distributions $\Omega$ and using simple linear algebra to estimate different gates' infidelities \cite{proctor2018direct, hines2022demonstrating}; interleaved versions of character RB (Sec.~\ref{sec:character_rb}), which is closely related to cycle benchmarking; and Pauli noise learning techniques (see Sec.~\ref{sec:pnr}).

%%%%%%% Cycle Benchmarking %%%%%%%%%%%%%%%%%%%%%%% 
\subsection{Cycle Benchmarking}\label{sec:cb}

\begin{figure*}[ht]
    \centering
    \includegraphics[width=2\columnwidth]{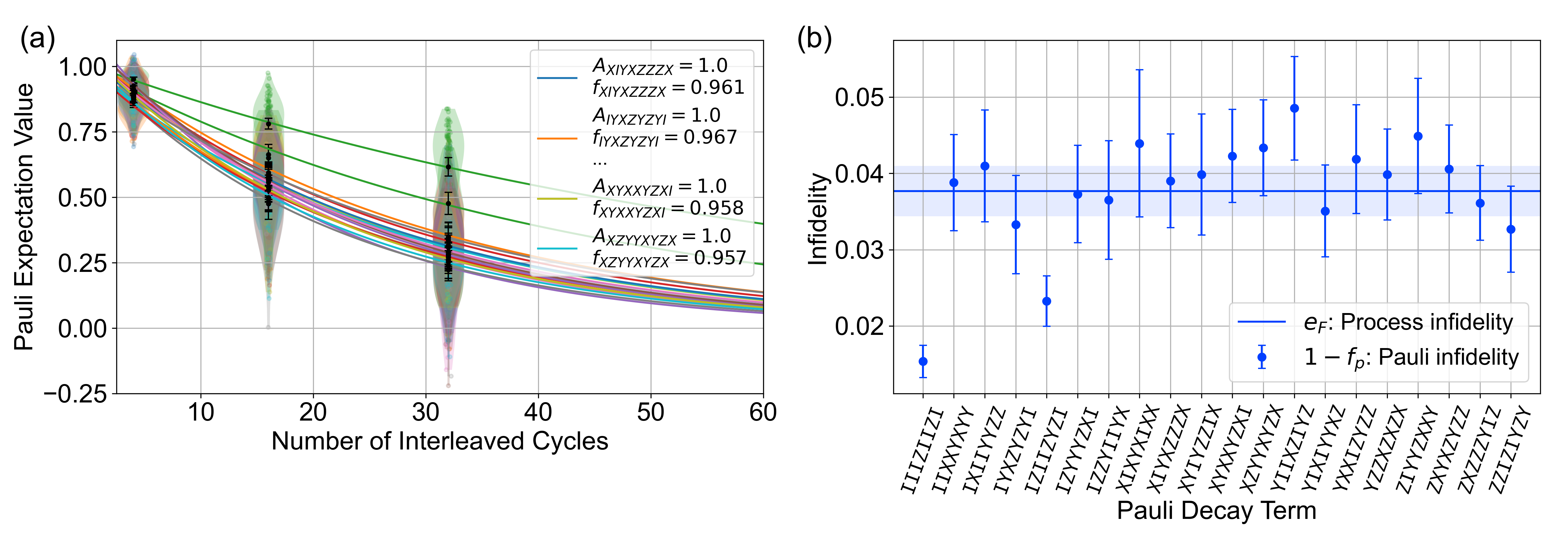}
    \caption{\textbf{Cycle Benchmarking of Simultaneous Single-qubit Paulis.} 
        (a) Pauli decays for each Pauli basis $P$, with the SPAM parameter $A_P$ and the exponential fit parameter $f_P$ listed in the legend for a subset of $P$. 
        (b) Pauli infidelities $e_P = 1 - f_P$ for each Pauli $P$, and the average process infidelity $e_F$ (horizontal blue line). The highlighted region denotes the 95\% confidence interval of $e_F$.}
    \label{fig:cb_all}
\end{figure*}

\emph{\Ac{CB}} \cite{erhard2019characterizing} is a protocol for estimating the process fidelity of an $n$-qubit gate, a.k.a.~a ``layer'' or ``cycle.'' Following CB convention, here we will use the ``cycle'' terminology, which is defined to be a set of gates acting on disjoint sets of qubits all occurring during the same moment in time, in analogy with a clock cycle on classical computers. CB is an alternative to IRB that is arguably more useful in practice. CB interleaves the cycle of interest in between layers of random Pauli gates (see \fig\ref{fig:cb_circs}), instead of the layers of random $n$-qubit Clifford gates used in IRB. The Pauli group implements a weaker twirl than the Clifford group --- it converts a general error map to a Pauli stochastic channel, rather than a global depolarizing channel (see Appendix \ref{sec:pauli_twirling}). But Pauli twirling requires only a single layer of parallel single-qubit Pauli gates, whereas $n$-qubit Clifford gates require many one- and two-qubit gates. This makes CB much more scalable than IRB, so CB enables benchmarking many-qubit cycles containing parallel one- and two-qubit gates.

CB estimates the process fidelity of a cycle of gates, and we describe CB for the case of an $n$-qubit cycle containing only Clifford gates:
\begin{enumerate}
    \item For $K$ different $n$-qubit Pauli operators $P$, that are uniformly sampled if $n\gg1$ but can consist of every possible Pauli operator if $n$ is small:
    
    \begin{enumerate}
        \item Use a layer of single-qubit gates to prepare the qubits in a random tensor-product eigenstate of $P$.
        
        \item Apply a circuit consisting of $m$ applications of the cycle of interest $G$ interleaved with cycles of randomly sampled $n$-qubit Pauli operators (applied via randomized compiling), for a range of values $m$ that all satisfy $G^m = \mathbb{I}$.
        
        \item Measure the Pauli operator $P$, whose estimated value we denote by $f_{P,m}$, which is typically achieved using a layer of single-qubit gates and a computational basis measurement.
        
        \item Fit $f_{P,m}$ to an exponential decay of the form 
        \begin{equation}
            f_{P,m} = Af_P^m ~,
        \end{equation}
        where $A$ absorbs all SPAM errors. The fit value for $f_P$ is an estimate of
        \begin{equation}\label{eq:cb-decay}
            f_P = \left(\prod_{k=1}^j \lambda_{G^kPG^{-k}}\right)^{1/j} ~,
        \end{equation}
        where $j$ is the smallest integer such that $G^j = \mathbb{I}$ and $\lambda_{P} = \Lambda_{PP}$, where $\Lambda$ is $G$'s error channel's PTM and $\Lambda_{PP}$ is the diagonal element indexed by Pauli operator $P$.
        
        \item Estimate the process fidelity of the cycle to be
        \begin{equation}
            F_e = \frac{1}{K} \sum_{i=1}^K f_{P_i} ~.
        \end{equation}
        
    \end{enumerate}
\end{enumerate}

\begin{figure*}[t]
    \centering
    \includegraphics[width=2\columnwidth]{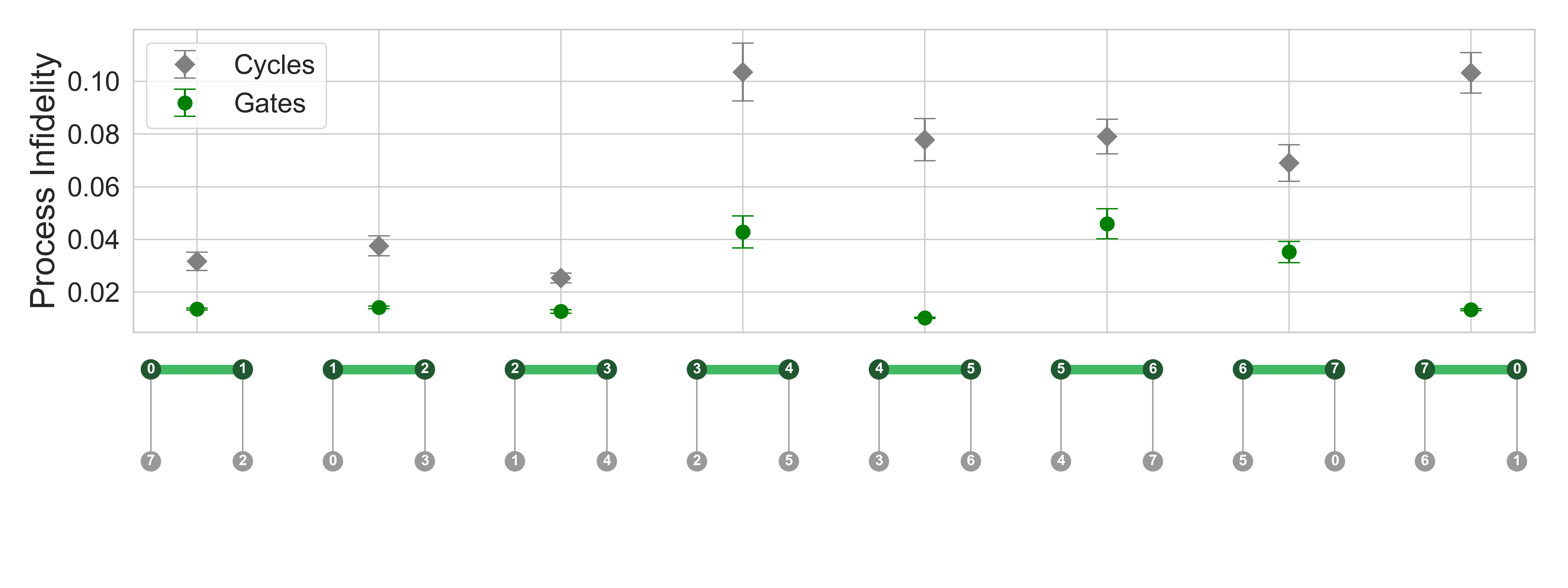}
    \caption{\textbf{CB of Gates vs.~Cycles.} 
    The process infidelities of eight different two-qubit CZ gates measured via CB are plotted in green. When the two nearest-neighbor idling qubits are benchmarked alongside each CZ gate (grey), the CB performance is worse in all cases. Notably, the good performance of an isolated CZ gate does not guarantee the good performance of the cycle which includes idling spectator qubits. For example, the CZ between qubits 4 and 5 has the lowest individual process infidelity, but has one of the worst process infidelities when the idles are included.}
    \label{fig:cb_gates_cycles}
\end{figure*}

Individual exponential decays in CB are often referred to as \emph{Pauli decays} and are labeled by the Pauli operator $P$ specifying the basis of the state preparation and measurement. If CB is applied to the idle (i.e., identity) cycle (see \fig\ref{fig:cb_all}), each Pauli decay curve measures the eigenvalues of the PTM of $G$, but for more general cycles some of the $f_P$ correspond to products of eigenvalues of $G$'s error channel (see Appendix \ref{sec:pauli_gauge}). This complication is encompassed by \eq\ref{eq:cb-decay}. As a result, $F_e$ is not an accurate estimate of the process fidelity in general. However, it is proven in \R\cite{erhard2019characterizing} that $F_e$ is a lower bound on the true process fidelity of the cycle (in the limit of infinite samples). The number of Pauli decays required to obtain a fixed estimation precision is independent of the number of qubits, and instead only depends on the infidelity of the cycle, which follows from standard statistical analysis of RB protocols \cite{erhard2019characterizing, harper2019statistical}. As a general guide, a minimum of $K = \min(20, 4^n - 1)$ Pauli operators should be sampled from $\mathbb{P}_n$ for low uncertainty estimates of the process fidelity \cite{erhard2019characterizing, trueq}.

CB measures the process (in)fidelity of a \emph{dressed} cycle. Therefore, the error rate measured by CB contains contributions from both errors in the interleaved cycle and the random Pauli gates. This is the relevant error rate for cycles that will be used in randomly compiled or Pauli frame randomized circuits \cite{hashim2021randomized}. But, it is also possible to approximately isolate the process infidelity of the ``bare'' interleaved cycle ($G$), by implementing CB with (i) the cycle of interest and (ii) a reference cycle containing no gates [see \fig\ref{fig:cb_circs}(a)], and then applying exactly the same analysis as in IRB (see \eq\ref{eq:ef-irb}). This has the same fundamental limitations as IRB (see the discussion in Sec.~\ref{sec:irb}), but in practice the systematic error in this estimation method is typically significantly smaller than in IRB. This is because the fidelity of a random Pauli gate (the randomizing gates in CB) is typically higher than the fidelity of a random Clifford gate (the randomizing gates in CRB). For example, \R\cite{mitchell2021hardware} used CB to estimate the fidelity of a CZ gate and (using \eq\ref{eq:irb_bounds}) found lower and upper bounds on its fidelity of 97.52(2)\% and 99.764(5)\%, respectively, whereas when using IRB these lower and upper bounds were 91.9(2)\% and 99.96(1)\%, respectively. 

CB is simplest and most efficient for benchmarking Clifford cycles, but it can also be used to benchmark non-Clifford gates. Doing so requires adding correction gates to the end of CB circuits to return the qubits to a Pauli eigenstate. This can require many multi-qubit gates at the end of each benchmarking circuit. Therefore, to reliably benchmark non-Clifford gates, the native gates required to implement the correction operations (which usually include the interleaved gate itself) must have high enough fidelity that the correction operations do not corrupt the measured fidelity of the dressed cycle. For example, \R\cite{hashim2022optimized} benchmarked non-Clifford $\text{CS} = \sqrt{\text{CZ}}$ and $\text{CS}^\dagger$ gates. CB was also used to benchmark a three-qubit non-Clifford $i$Toffoli gate \cite{kim2022high}, which would not have been feasible using three-qubit non-Clifford RB.

One utility of CB is that it can benchmark the process fidelity of an entire cycle of gates containing any combination of single- and multi-qubit gates (similar to methods like MRB), as long as the cycle composes to the identity operation at some circuit depth $m$ \footnote{By extension, it can also be used to measure the process fidelity of an entire sub-circuit, and can therefore be considered a form of SPAM-robust fidelity estimation; see Sec.~\ref{sec:fidelity_estimation}.}. Thus, CB can \emph{holistically} quantify the impact of crosstalk between gates in a parallel gate cycle. For example, it can be used to measure crosstalk experienced by idling spectator qubits during a two-qubit gate. To demonstrate this, in \fig\ref{fig:cb_gates_cycles} we plot the process infidelity of eight different CZ gates measured via CB on an eight-qubit superconducting quantum processor with a ring topology. Additionally, we measure the process infidelity of cycles containing each of the eight CZ gates, as well as idle gates on the spectator qubit on either side of the CZ gate (i.e., the interleaved gate cycle is $G = I \otimes \mathrm{CZ} \otimes I$). We observe that, in all cases, the cycle with the idle qubits has a larger process infidelity than the cycle containing just the CZ gates. This highlights two important concepts: (i) it should not be assumed that gates have no impact on idle qubits (and vice-versa \cite{krinner2020benchmarking}), and (ii) when understanding circuit performance, it is most informative to benchmark the constituent cycles as they appear in the circuit.

%%%%%%% Purity Benchmarking %%%%%%%%%%%%%%%%%%%%%%% 
\subsection{Purity Benchmarking}\label{sec:pb}

\begin{figure*}[t]
    \centering
    \includegraphics[width=2\columnwidth]{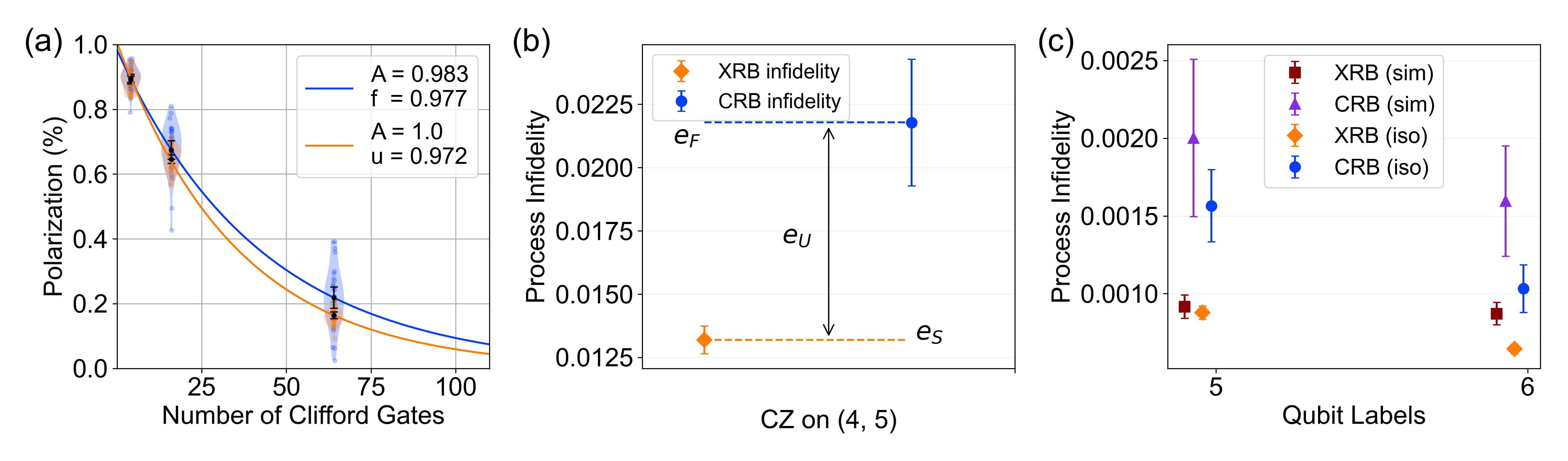}
    \caption[eXtended RB]{\textbf{eXtended RB.} 
        \textbf{(a)} Exponential decays for two-qubit CRB (blue) and XRB (orange). All two-qubit Cliffords are decomposed into native single-qubit gates and a native two-qubit CZ gate.
        \textbf{(b)} CRB (blue) and XRB (orange) process infidelities for the exponential decays in (a). The CRB process infidelity $e_F$ is the AGSI for two-qubit Cliffords, and the stochastic process infidelity $e_S$ is the approximate coherence limit for two-qubit Clifford gates. The difference between the two $e_U = e_F - e_S$ is the average error rate due to coherent errors.
        \textbf{(c)} Isolated vs.~simultaneous single-qubit CRB and XRB. The CRB process infidelity is larger for both qubits under simultaneous operation, whereas the stochastic process infidelities are approximately equal in both cases, indicating the presence of coherent crosstalk errors between simultaneous single-qubit gates.}
    \label{fig:xrb}
\end{figure*}

\emph{\Ac{PB}} \cite{wallman2015estimating, feng2016estimating, zhu2024purity} is a family of RB techniques for quantifying how coherent a gate set's errors are. Purity benchmarks provide complementary information to the foundational RB protocols (i.e., the group and native-gate RB protocols), which intentional mix together all kinds of errors into a single error rate. As discussed in Sec.~\ref{sec:errors}, Markovian errors can be broadly categorized as either coherent/unitary or incoherent/stochastic. PB methods can be used to quantify the relative contributions of coherent and stochastic errors, and they are based on the purity $\gamma$ of a quantum state $\rho$. The purity of a quantum state is
\begin{equation}\label{eq:purity}
    \gamma = \Tr(\rho^2) = \frac{1}{d} \left( 1 + ||\mathbf{r}(\rho)||^2 \right) ~,
\end{equation}
where 
\begin{equation}
    \rho = \frac{1}{d} [\Id + \mathbf{r}(\rho) \cdot \boldsymbol\sigma] ~,
\end{equation}
$\mathbf{r}(\rho)$ is the generalized $d$-dimensional Bloch vector, and $||\mathbf{r}(\rho)||^2$ is its Euclidean norm (i.e., average squared length). Here $\boldsymbol{\sigma}$ is the vector of Pauli matrices. 

In the context of randomized benchmarks, we prepare quantum states using sequences of random gates, and the final state $\rho' = \E(\rho)$ will have a purity $\gamma \le 1$, which depends on the nature of the error channel $\E$ (e.g., $\E$ can be some mixture of coherent and stochastic errors). One way to quantify how coherent the error channel $\E$ is in terms of \emph{unitarity} of $\E$,
\begin{equation}\label{eq:unitarity}
    u(\E) = \frac{1}{d - 1} \int d\psi \big|\big| \mathbf{r}\left[\E(\ketbra{\psi})\right] - \mathbf{r}\left[\E(\Id/d)\right] \big|\big|^2 ~,
\end{equation}
which is the Euclidean norm of the Bloch vector of the state $\E(\ketbra{\psi})$ (with the identity component subtracted off), averaged over all pure states. If $\E$ is a unitary channel, then $u(\E) = 1$, and $u(\E) < 1$ if $\E$ includes contributions from stochastic noise. While not all purity benchmarks utilize the unitarity, \eq\ref{eq:unitarity} demonstrates that it is possible to \emph{quantify} the relative contributions of stochastic and coherent errors to the AGSI of an RB experiment. In this subsection, we discuss several different randomized benchmarks which attempt to quantify the relative errors rates of coherent and stochastic errors in a gate set. While this goes beyond the scope of this tutorial, it should be noted that gate set tomography (Sec.~\ref{sec:gst}) can also be used to quantify the amount of coherent errors and stochastic noise in a gate \cite{mkadzik2021precision}.

%%%%%%% XRB %%%%%%%%%%%%%%%%%%%%%%% 
\subsubsection{eXtended RB}\label{sec:xrb}

\emph{\ac{XRB}} \cite{trueq, wallman2015estimating} is a PB protocol that is based on CRB. XRB estimates the average unitarity of a set of $n$-qubit Clifford gates. XRB requires only a small modification to the standard CRB protocol: XRB performs the standard CRB circuits introduced in Sec.~\ref{sec:standardrb}, but instead of performing an inverting operation at the end of the sequence, state tomography is performed on the resulting state in order to estimate the length of the Bloch vector. XRB characterizes the unitarity in terms of the decay rate of the average squared Bloch vector length with sequence depth. For a single qubit, the  average squared Bloch vector length,
\begin{equation}
    \Tilde\gamma = \vert\vert \boldsymbol{r}(\rho)\vert\vert^2 ~,
\end{equation} 
is equivalent to
\begin{equation}
    \Tilde{\gamma} = \langle X \rangle^2 + \langle Y \rangle^2 + \langle Z \rangle^2 ~.
\end{equation}
This is a shifted and rescaled version of \eq\ref{eq:purity}.

XRB consists of (i) running CRB circuits for various depths $m$ without the inversion gate, (ii) estimating $\Tilde{\gamma}$ for each circuit, and then (iii) fitting the mean of $\Tilde{\gamma}$ (which we denote by $\langle \Tilde{\gamma}(d) \rangle$) as a function of $m$, to
\begin{equation}
    \langle \Tilde{\gamma}(m) \rangle = A u^m ~. 
\end{equation}
The fit value for $u$ is an estimate of the mean unitarity of the benchmarked gates. This can then be used to estimate the \emph{stochastic process infidelity} $e_S(\E)$ defined by
\begin{equation}
    e_S = 1 - \sqrt{\frac{(d^2 - 1)u + 1}{d^2}} ~.
\end{equation}
If standard CRB is also performed in addition to XRB, then the process infidelity $e_F$ measured via CRB represents the total error. Together, one can estimate the \emph{coherent process infidelity} $e_U$ by $e_U = e_F - e_S$.

In \fig\ref{fig:xrb}(a), we plot exponential decays for CRB and XRB. In \fig\ref{fig:xrb}(b), we compare the process infidelity of the gates $e_F$ (measured via CRB) with the estimated stochastic process infidelity $e_S$ (measured via XRB); the difference between the two is the coherent process infidelity $e_U$. The stochastic process infidelity is an approximate measure for determining whether or not a gate set is \emph{coherence limited} (i.e., all gate errors are due to incoherent noise). If $e_F = e_S$, then $e_U = 0$ and, thus, the gates have no coherent errors. However, because infidelity is only sensitive to coherent errors at $\mathcal{O}(\theta^2)$, if an estimate of $e_S$ is equal to $e_F$ within error bars, it is possible that coherent errors still exist --- XRB does not amplify coherent errors, so it is an inefficient method for estimating the size of coherent errors.

One application of XRB is to quantify the magnitude of crosstalk errors. In \fig\ref{fig:xrb}(c), we show the CRB and XRB process infidelities for single-qubit gates performed in isolation and simultaneously for two qubits. We see that the CRB infidelity ($e_F$) for each qubit is larger for simultaneous CRB than for isolated CRB, but the XRB process infidelity ($e_S$) is approximately the same in both cases. This demonstrates that crosstalk-induced coherent errors make up a larger fraction of the total error rate under simultaneous operation.

%%%%%%%%%%%%%%%%%%%%%%% Speckle Purity Benchmarking %%%%%%%%%%%%%%%%%%%%%%% 
\subsubsection{Speckle Purity Benchmarking}\label{sec:xeb_pb}

\emph{\Ac{SPB}} \cite{2019GoogleSupremacy} is a protocol for estimating the decay of the purity of states produced by random circuits versus circuit depth. This method is based on the observation that, for long enough circuits, the probability $p(x)$ of observing a particular bit string $x$ will be a random variable with a Porter-Thomas distribution. For most circuits, $p(x)$ will be exponentially close to zero. But for some rare circuits $x$ will appear with  significantly higher probabilities. The observed data will then demonstrate a ``speckle pattern'' when presented visually. However, when depolarizing errors dominate, the speckle pattern will be smoothed out as the distribution approaches the uniform distribution [see \fig\ref{fig:xebdata}(c -- d)]. 

SPB is the following procedure: sample $N$ random scrambling circuits of depth $m$, i.e., circuits with properties similar to those typically used in XEB. Now, choose some bit string $x$, and let $\mathcal{P}$ be the probability of measuring bit string $x$ \textit{assuming the circuits are run without errors}. Because the circuit is random, $\mathcal{P}$ is a random variable. If the gates are unitary, then for sufficiently deep circuits, $\mathcal{P}$ will be distributed according to the Porter-Thomas (PT) distribution, whose probability density is:
\begin{equation}
    f_{\mathrm{PT}}(p) = (d-1)(1-p)^{d-2} ~,
\end{equation}
with variance
\begin{equation}
    \sigma_{\mathrm{PT}}^2 = \frac{d-1}{d^{2}(d+1)} ~.
\end{equation}
On the other hand, if the gates completely depolarize the state, then all output strings become equally probable. In this case, we can again describe the probability as a random variable, but with a trivial ($\mathcal{T}$) probability density function:
\begin{equation}
    f_{\mathcal{T}}(p) = \delta\left( p-\frac{1}{2^d} \right) ~,
\end{equation}
with $\delta$ the Dirac delta distribution, and the variance of $\mathcal{T}$ is zero
\begin{equation}
    \sigma_{\mathcal{T}}^2 = 0 ~.
\end{equation}

In real experiments, there are stochastic and coherent errors, and the stochastic errors push $\mathcal{P}$ towards the trivial distribution with increasing circuit depth, whereas coherent errors preserve the PT distribution. In particular, the probability $\mathcal{P}$ of measuring a given bit string in a depth $m$ circuit will be approximately given by a mixture of the PT distribution and the trivial distribution:
\begin{equation}
    f_{\mathrm{exp}}(p) \approx (1-\epsilon)^m f_{\mathrm{PT}}(p) + \epsilon^m f_{\mathcal{T}}(p) ~,
\end{equation}
with $\epsilon$ the rate of stochastic errors per circuit layer (more precisely, $1 - \epsilon$ is the process polarization corresponding to the stochastic portion of the error channels, and so $\epsilon$ is approximately the rate of stochastic errors except for very few qubits). SPB theory relates the variance of this distribution to the variance of the PT distribution as:
\begin{equation}
    \sigma^2_{\textrm{exp}} = \epsilon^{2m} \sigma_{\mathrm{PT}}^2 ~.
\end{equation}
Therefore, we can estimate the purity decay with cycle depth by simply observing the rate at which the variance of the distribution of bit string probabilities decays. For example, by comparing the XEB fidelity decay and purity decay for the data in \fig\ref{fig:xebdata}(b), we can investigate the relative size of coherent and incoherent errors in the system. Using SPB, we estimate that roughly 68\% of the dressed CZ's error can be attributed to stochastic errors; the remaining are attributed to coherent errors.

%%%%%%%%%%%%%%%%%%%%%%% Iterative RB %%%%%%%%%%%%%%%%%%%%%%% 
\subsubsection{Iterative RB}\label{sec:itrb}

\begin{figure*}[t]
    \centering
    \includegraphics[width=2\columnwidth]{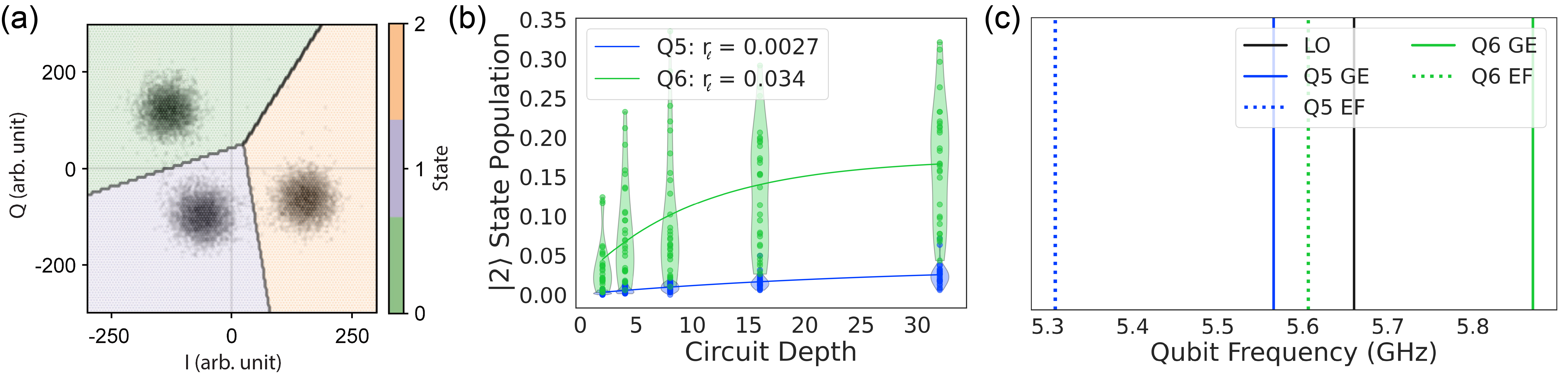}
    \caption[Leakage RB]{\textbf{Leakage RB.} There are a variety of RB protocols that can quantify rates of leakage errors. This figure demonstrates one such method, which we call \emph{leakage RB} (LRB), on transmon qubits. This method uses qutrit readout to discriminate between leaked and computational basis states. 
    \textbf{(a)} Example of readout and classification boundaries for classifying $\ket{0}$, $\ket{1}$, or $\ket{2}$ for a superconducting qubit. 
    \textbf{(b)} LRB data for two qubits (labeled Q5 and Q6) under simultaneous operation, showing the probability of observing the 2 outcome as a function of the RB circuit's depth. This data is fit to an exponential to estimate the leakage rate ($r_l$). Q6 has a much larger leakage rate ($r_l = 3.4\%$) and steady-state $\ket{2}$ state population than Q5 ($r_l = 0.27\%$). Leakage can be due to either coherent excitation to higher energy levels or thermal noise, but the frequency spectrum of the two qubits shown in \textbf{(c)} indicates that the $\ket{0} \rightarrow \ket{1}$ (i.e., ``GE'') transition frequency of Q5 is close to the $\ket{1} \rightarrow \ket{2}$ (i.e., ``EF'') transition frequency of Q6. Therefore, the leakage on Q6 is most likely due to crosstalk from Q5 when performing simultaneous single-qubit gates.}
    \label{fig:lrb}
\end{figure*}

\emph{Iterative RB} protocols are RB-like methods for amplifying gate errors so that it is possible to separate coherent errors from stochastic noise. These methods depend on the fact that constructively interfering coherent errors will grow quadratically with circuit depth. To see this, consider the simple example of applying many $R_x(2\pi)$ rotations to a qubit initially in the ground state, but each time the qubit over-rotates by a small angle $\theta$. The resulting state of the qubit after $M$ rotations is
\begin{equation}
    \ket{\psi} = \prod^M  e^{-i\theta\sigma_x} \ket{0} = \cos{(M\theta)}\ket{0} - i\sin{(M\theta)}\ket{1} ~.
\end{equation}
The fidelity of this state with respect to $\ket{0}$ is $F = \vert \braket{0 | \psi} \vert^2 = \cos^2{(M\theta)} \approx 1 - (M\theta)^2$, thus the infidelity is $\epsilon = 1 - F \approx (M\theta)^2$. Therefore, the infidelity scales quadratically in both the over-rotation angle $\theta$ and the number of rotations $M$. In contrast, stochastic errors typically only grow linearly with circuit depth: if we take $p$ to be the probability of a stochastic error per gate, then $1 - p$ is the probability of no error per gate, and $(1 - p)^M$ is the probability of no error after $M$ gates. For a circuit with $M$ total gates, $\epsilon = 1 - (1 - p)^M \approx Mp$ is the probability of an error after $M$ gates. Thus, stochastic errors accumulate linearly with circuit depth in the small error limit.

Iterative RB \cite{sheldon2016characterizing} interleaves $M$ repetitions of a target quantum gate within a standard CRB sequence (see Sec.~\ref{sec:irb}), with $M$ varied. Because the coherent errors in the gate will grow quadratically in $M$, one can fit the fidelity decay of the sequence to both quadratic and linear functions, with the quadratic component capturing the coherent contributions to the gate error, and the linear component capturing the incoherent contributions to the gate error. This method can be adapted to a variety of randomized benchmarks~\cite{2022MoskalenkoFluxonium, carignan2023estimating, debroy2023context}.

%%%%%%%%%%%%%%%%%%%%%%% RB with Non-Markovian Errors %%%%%%%%%%%%%%%%%%%%%%% 
\subsection{RB with Non-Markovian Errors}\label{sec:rb-nonmark}

All of the RB protocols discussed so far in this section are primarily based on theory that assumes Markovian errors. Those RB protocols are therefore not guaranteed to work correctly in the presence of non-Markovian errors. In general, standard RB data is not guaranteed to follow a simple exponential decay in the presence of non-Markovian errors \cite{wallman2018randomized, proctor2017randomized, epstein2014investigating}. For example, $1/f$ noise is well-known to cause non-exponential RB decays \cite{fogarty2015nonexponential}. Furthermore, the RB protocols discussed so far are not designed to learn anything about the rates of non-Markovian errors (although they will incorporate the rates of some kinds of non-Markovian errors into the measured RB error rates). There are, however, a variety of adaptations to RB protocols that enable learning about one or more kinds of non-Markovianity using RB. Examples include time-resolved RB \cite{proctor2020detecting} and loss RB \cite{Wallman2015-eg}, which measure drifting gate error rates versus time and qubit loss rates, respectively. Here, we discuss only the most-widely used RB protocols for non-Markovian errors: those designed for quantifying leakage.

%%%%%%%%%%%%%%%%%%%%%%% Leakage Benchmarking %%%%%%%%%%%%%%%%%%%%%%% 
\subsubsection{Leakage RB}\label{sec:lrb}

RB protocols that measure leakage rates are relatively simple to implement and are widely used. Leakage describes an error in which a qubit is excited out of the computational basis state to higher energy levels (see \fig\ref{fig:Leakage}). This is a common source of error in systems whose energy spacings are not sufficiently well-separated to isolate the $\ket{0} \longrightarrow \ket{1}$ transition from transitions to higher energy levels. Leakage cannot be captured by most RB protocols, and it can corrupt their results. However, many RB protocols can be modified to account for leakage \cite{chasseur2015complete, wallman2016robust, wood2018quantification}. There are a variety of ways to quantify leakage using RB methods, but the conceptually simplest methods consist of running standard RB experiments (or native gate RB experiments) while monitoring the $\ket{2}$ state population (or higher states). This method is often termed \emph{\ac{LRB}}, and we focus our discussion on this simple technique. 

LRB is the following simple adaptation to standard single-qubit CRB:
\begin{enumerate}
    \item Run standard CRB circuits and, at the end of each circuit, measure whether the final state of the qubit is $\ket{0}$, $\ket{1}$, or $\ket{2}$. 
    \item Fit the average $\ket{2}$ state population versus circuit depth to a simple exponential growth function, to estimate the leakage rate per Clifford gate ($r_l$).
    \end{enumerate} 
To perform LRB, it is therefore necessary to be able to readout the $\ket{2}$ state (although note that LRB is robust to errors in this readout). \fig\ref{fig:lrb} illustrates simultaneous LRB on two transmon qubits, and how it can be used to identify qubits with high leakage rates.

\section{Partial Tomography and Fidelity Estimation}\label{sec:partial_tomography}

\ac{QCVV} methods can be categorized by (i) the amount (and type) of information they provide, and (ii) the cost --- in both experimental and computational resources --- to run them. Gaining more information typically requires more resources, and so QCVV methods can often be placed on a sliding scale from (1) highly informative but costly, to (2) highly efficient but providing little information \cite{eisert2020quantum}. Tomography of quantum states, processes, measurements, or gate sets (see Sec.~\ref{sec:tomography}) is at one extreme of this continuum: these methods provide comprehensive information about the models and rates of all possible kinds of (Markovian) errors, but they require resources that are exponential in system size. This limits the application of these methods to the few-qubit setting. In contrast, most randomized benchmarks (see Sec.~\ref{sec:randomized_benchmarks}) are extremely efficient to run but they provide only one (or a handful) of numbers summarizing a gate set's performance (e.g., the average fidelity of a gate set). It is, however, both possible and often useful to obtain more information about a system (e.g., a gate set) than provided by randomized benchmarks, without resorting to exponentially expensive tomography. In this section, we discuss techniques that sit in the middle of this cost versus information sliding scale, which generally fit into one (or both) of two categories: (1) ``partial tomography'' methods, and (2) ``fidelity estimation'' methods.

There are many more resource efficient characterization and benchmarking techniques than can be covered here. Therefore, we list some important (and partially overlapping) categories of partial tomography and fidelity estimation, covering only some of them in detail below: 
\begin{itemize}
    \item \emph{Targeted Tomography}. An $n$-qubit state, gate, or gate set contains exponentially many (in $n$) independent parameters, so it is infeasible to learn all of them for $n \gg 1$. However, it possible to learn a subset of those parameters, or some function of those parameters. There are a variety of partial tomography techniques that target specific parameters in a state, process, or gate set \cite{Lopez2010-jc, Toth2010-xi, Bendersky2013-rv, Greganti2015-mg, Steffens2017-pm, Carmeli2017-fh, Helsen2019-fi, Helsen2021-zf}. For example, there are techniques for learning one or more of a transfer matrix's eigenvalues, e.g., spectral tomography \cite{Helsen2019-fi} and phase estimation (see Sec.~\ref{sec:phase_estimation}). One way to reduce the cost of process tomography is to instead learn its action as a classical, probabilistic gate. We discuss this method, sometimes called ``truth table tomography,'' in Sec.~\ref{sec:truthtable}.
    
    \item \emph{Randomized Measurement Methods}. A variety of partial tomography methods are built on the idea of measuring a quantum state or process in a small number of randomly chosen bases \cite{elben2023randomized}. Methods of this sort include shadow tomography \cite{aaronson2018shadow, huang2020predicting, kunjummen2023shadow}, direct fidelity estimation \cite{flammia2011direct, dasilva2011practical}, and cycle benchmarking (which we covered in Sec.~\ref{sec:cb}). We discuss direct fidelity estimation in Sec.~\ref{sec:direct_fid_est}.
    
    \item \emph{Pauli Noise Learning}. Stochastic Pauli channels (see Sec.~\ref{sec:stoch_pauli}) are a practically relevant class of error channels that contain only $4^n$ parameters, rather than the $16^n$ parameters of a general error map. There are a variety of techniques that (i) use twirling or randomized compiling \cite{wallman2016noise, hashim2021randomized} to enforce a stochastic Pauli noise model, and (ii) learn the parameters of those Pauli channels. Many of these methods have close connections to randomized benchmarks (Sec.~\ref{sec:randomized_benchmarks}), and some of them can efficiently learn \emph{sparse} Pauli channels that contain only a polynomial number of unknown parameters. We discuss some of these methods in Sec.~\ref{sec:pnr}. 
    
    \item \emph{Ansatz Tomography}. A range of tomographic methods exist that reconstruct states, processes, or gate sets more efficiently than the ``brute force'' methods discussed in Sec.~\ref{sec:tomography} by assuming or privileging some simplifying structure.  Some of these methods \emph{assume} a structure --- e.g., that the state is pure or local, or that the process is unitary or Pauli-stochastic --- and will give an incorrect estimate if the assumption is violated.  Better methods use a hierarchical \emph{ansatz} --- e.g., that the density matrix or process matrix has low rank --- and perform efficiently when the ansatz is satisfied, yet also recognize and characterize (less efficiently) objects that do not satisfy the ansatz. Examples include methods for tomography of pure states and unitaries \cite{Gutoski2014-wo, Ma2016-if}, states satisfying strong symmetries \cite{Lopez2010-jc, Toth2010-xi, moroder2012permutationally, Schwemmer2014-qz}, low-rank states and processes (including compressed sensing approaches) \cite{Guta2012-nx, Gross2010-ck, Riofrio2017-wk, PRXQuantum.4.010325}, matrix product and tensor network states \cite{Landon-Cardinal2012-nh, Baumgratz2013-js, Cramer2010-vo}, and techniques that assume a gate's errors can be described by few-parameter (i.e., sparse) Pauli channels \cite{flammia2021averaged} or Lindbladians \cite{van2023probabilistic, van2024techniques}. Methods also exist that use heuristics to improve efficiency, such as some machine learning approaches to tomography
    \cite{Kieferova2017-oi, Torlai2018-wt, Gao2018-xc, Carrasquilla2019-wh, Gebhart2023-dr}. With the exception of some efficient Pauli noise learning methods covered in Sec.~\ref{sec:pnr}, we do not discuss these techniques further.
\end{itemize}

%%%%%%%%%%%%%%%%%%%%%%% Truth Table %%%%%%%%%%%%%%%%%%%%%%% 
\subsection{Truth Table Tomography}\label{sec:truthtable}

\begin{figure}[t]
    \includegraphics[width=0.4\textwidth]{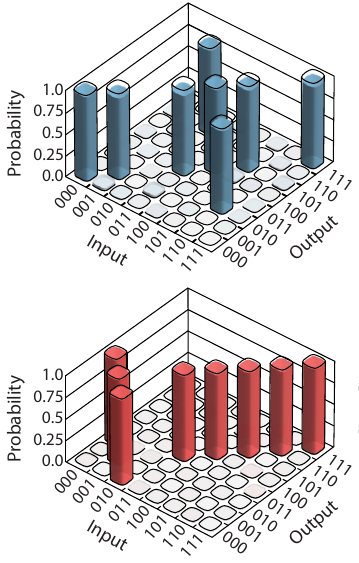}
    \caption{\textbf{Truth Table Tomography.}
    The stochastic matrices for experimental [top] Toffoli~\cite{nguyen2022programmable} and [bottom] $i$Toffoli gates~\cite{kim2022high} measured using truth table tomography on superconducting qubits. The Toffoli gate flips the first qubit if the second and third qubits are in the state $\ket{10}$. The $i$Toffoli gate flips the middle qubit if the first and third qubit are in the state $\ket{00}$. The corresponding gate fidelities are estimated to be $96.20(6)\%$ and $98.6(1)\%$, respectively.}
    \label{fig:truth_table}
\end{figure}

\emph{Truth table tomography} \cite{fedorov2011implementation, chu2023scalable} is perhaps the conceptually simplest form of partial tomography of an $n$-qubit process $\Lambda$. It consists of preparing the $n$ qubits in each of the $2^n$ different computational basis states, applying $\Lambda$, and then measuring in the computational basis. This directly estimates the $4^n$ different probabilities given by
\begin{equation}
    p_{y \mid x}(\Lambda) = \Tr[ \ketbra{y}{y} \Lambda(\ketbra{x}{x})] ~,
\end{equation}
where $x,y \in \{0,1\}^{n}$. The matrix of these probabilities $p_{y \mid x}(\Lambda)$ is a $2^n \times 2^n$ \emph{stochastic matrix} (each element of the matrix is a probability, and each row sums to one), which we denote by $S(\Lambda)$. This matrix is similar to the response (or confusion) matrix constructed when characterizing readout fidelities (see \eq\ref{eq:response_matrix}). 

As in quantum process tomography (\ac{QPT}, see Sec.~\ref{sec:qpt}), the measured stochastic matrix $S(\Lambda)$ is typically compared to the stochastic matrix $S(\mathcal{U})$ for the intended (ideal) superoperator $\mathcal{U}$ of a gate. For any gate that preserves the computational basis (i.e., each computational basis state is mapped to another computational basis state), $S(\mathcal{U})$ is a \emph{truth table}, i.e., it is the matrix for a deterministic (reversible) classical gate. For example, for the CNOT gate this matrix is
\begin{equation}
    S = \begin{pmatrix} 
        1 & 0 & 0 & 0 \\ 
        0 & 1 & 0 & 0 \\ 
        0 & 0 & 0 & 1 \\ 
        0 & 0 & 1 & 0 
        \end{pmatrix} ~.
\end{equation}
Note, however, that the stochastic matrix for a general unitary is instead a general (doubly \footnote{In a doubly stochastic matrix, both the rows and columns sum to 1.}) stochastic matrix, e.g., for a Hadamard gate
\begin{equation}
    S = \frac{1}{2} \begin{pmatrix} 1 & 1 \\ 1 & 1 \end{pmatrix} ~.
\end{equation}

To demonstrate truth table tomography, in \fig\ref{fig:truth_table} we plot measured and ideal stochastic matrices for experimental Toffoli~\cite{nguyen2022programmable} and $i$Toffoli~\cite{kim2022high} gates. Ideal Toffoli and $i$Toffoli gates preserve the computational basis. The Toffoli gate leaves the qubits unchanged unless the two control qubits (in this experiment, the second two qubits) are in the state $\ket{10}$, in which case the other qubit (in this experiment, the first qubit) is flipped, i.e., $\ket{010} \mapsto \ket{110}$ and $\ket{110} \mapsto \ket{010}$. The $i$Toffoli gate leaves the qubits unchanged unless unless the two control qubits (in this experiment, the first and last qubit) are in the state $|00\rangle$, in which case the other qubit (in this experiment, the middle qubit) is flipped, i.e., $\ket{000} \mapsto \ket{010}$ and $\ket{010} \mapsto \ket{000}$. 

For any gate that ideally maps computational basis states to computational basis states, the fidelity between an experimental and ideal gate's stochastic matrices, $S_{\textrm{exp}}$ and $S_{\textrm{ideal}}$ respectively, is given by~\cite{fedorov2011implementation, chu2023scalable}
\begin{equation}\label{eq:fidelity_tt}
     F_\mathrm{tt} = \frac{1}{2^n} \mathrm{Tr}(S_\mathrm{exp}^{\texttt{T}} S_\mathrm{ideal}) ~.
\end{equation}
For the experimental gates of \fig\ref{fig:truth_table}, we find fidelities of $96.20(6)\%$ and $98.6(1)\%$ for the Toffoli and $i$Toffoli gates, respectively.

Truth table tomography has a variety of limitations. Like full QPT, it requires a number of circuits that scales exponentially in the number of qubits. Furthermore, like QPT it is susceptible to \ac{SPAM} errors. However, unlike QPT, it is insensitive to phase errors in the gates. Therefore, fidelities measured using truth table tomography will typically disagree with those gate fidelities estimated using other techniques, such as interleaved RB (Sec.~\ref{sec:irb}) or cycle benchmarking (Sec.~\ref{sec:cb}) \cite{nguyen2022programmable}. To recover some information about the phases of a gate, the input states can be rotated to the $X$ basis. In combination with the $Z$-basis results, this data can be used to lower-bound the process fidelity~\cite{hofman2005complementary}, and this has been used to characterize the effects of three-qubit Toffoli gates in trapped-ions~\cite{figgatt2017complete} and neutral atoms~\cite{levine2019parallel}. Moreover, the method introduced in \R\cite{hofman2005complementary} can be generalized to upper- and lower-bound the fidelity of high-dimensional operations, such as $n$-qubit Toffoli gates~\cite{fang2023realization}. However, this is an inefficient approach to estimating gate fidelity compared to, e.g., direct fidelity estimation.

%%%%%%%%%%%%%%%%%%%%%%% Direct Fidelity Estimation %%%%%%%%%%%%%%%%%%%%%%% 
\subsection{Direct Fidelity Estimation}\label{sec:direct_fid_est}

\begin{figure}[t]
    \centering
    \includegraphics[width=\columnwidth]{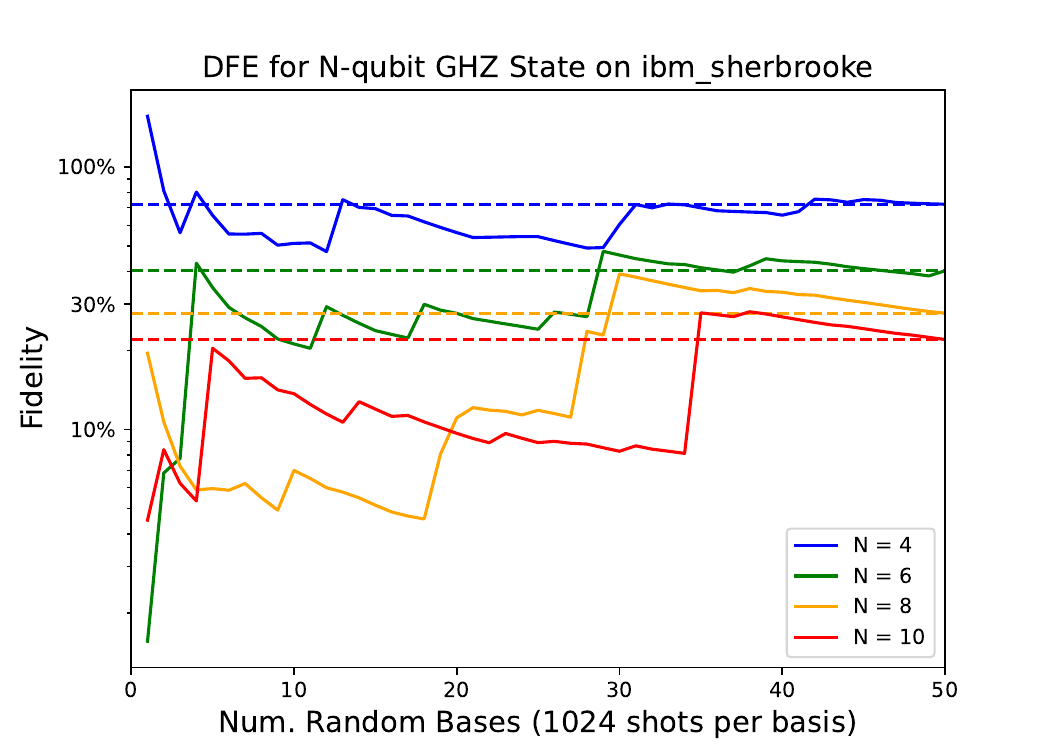}
    \caption{\textbf{Direct Fidelity Estimation.} 
    DFE performed on an $N=4,6,8,10$ qubit GHZ state on \texttt{ibm\_sherbrooke}. The running average of the GHZ state fidelity is plotted as a function of the number of random measurement bases, selected via importance sampling on the target state. The horizontal dashed lines is expected state fidelity calculated from the physical error rates listed for \texttt{ibm\_sherbrooke} at the time of the experiment.
    }
    \label{fig:dfe_experiment}
\end{figure}

Estimating the fidelity of a state or process requires computing the overlap between a system's state/process with the desired state/process (see Secs.~\ref{sec:state_fidelity} and \ref{sec:ent_fid}). While, in principle, a full tomographic reconstruction of the system could be used to accurately compute its overlap with the desired output, in practice full tomography becomes intractable beyond a few qubits. However, this overlap can be estimated by measuring only along axes of greater overlap with the desired state, while neglecting axes with little to no overlap. \emph{\Ac{DFE}} \cite{flammia2011direct, dasilva2011practical} is a protocol that uses this idea to measure state or process fidelities more efficiently than full tomography \cite{lu2020direct, zhang2021direct}. While DFE can in theory be used to compute the fidelity of entire circuits, more scalable methods have been developed specifically for estimating circuit fidelities (see Sec.~\ref{sec:fidelity_estimation}).

The purpose of DFE is to estimate the state fidelity (\eq\ref{eq:state_fidelity}) between an actual state $\rho$ and a desired pure state $\psi = \ketbra{\psi}$. This fidelity can be written as
\begin{equation}\label{eq:direct_fidelity_char_fn}
    F(\rho, \psi) = \Tr[\psi \rho] = \sum_{k=1}^{d^{2}} \chi_\psi(k) \chi_\rho(k) ~,
\end{equation}
where
\begin{equation}
    \chi_\rho(k) = \frac{\Tr[\rho P_k]}{\sqrt{d}}
\end{equation}
is known as the \emph{characteristic function} of $\rho$, $P_k$ are the $n$-qubit Pauli operators, and $d=2^n$ (for $n$ qubits). The quantity $\chi_\rho(k)$ is the $k^\text{th}$ expansion coefficient of $\rho$ in the normalized Pauli basis. While the exact expansion of \eq\ref{eq:direct_fidelity_char_fn} includes $d^2$ terms, $F(\rho, \psi)$ can be estimated by measuring only a subset of the most significant terms using importance sampling. To estimate the fidelity up to an additive error $\epsilon$ and failure probability $\delta$, one can take the following steps:
\begin{itemize}
    \item Choose a random value $k \in \{1, \dots, d^2\}$ with probability $p_k = \chi_\psi(k)^2$.
    \item Calculate $\chi_\rho(k)$ by measuring the expectation value of the Pauli operator $P_k$ for the unknown state $\rho$. Use this quantity to construct the estimator $X=\chi_\rho(k)/\chi_\psi(k)$.
    \item Repeat the steps above $l = 1/(\epsilon^2\delta)$ times and estimate $F(\rho,\psi)$ by the estimator 
    \begin{equation}
        \hat{F}(\rho, \psi) = \frac{1}{l} \sum_{i=1}^l X_i ~.
    \end{equation}
\end{itemize}
\R\cite{flammia2011direct} proves that if $\chi_{\rho}(k)$ is measured exactly, then
\begin{equation}
    \textrm{Pr}[ \vert \hat{F}(\rho,\psi)  - F(\rho, \psi) \rvert \geq \epsilon ] \leq \delta ~.
\end{equation}
However, there is always shot noise, i.e., each $\chi_{\rho}(k)$ is not measured perfectly. \R\cite{flammia2011direct} shows that $\rho$'s fidelity can be estimated using $m_k$ copies of $\rho$ for each $P_k$. While the exact number of copies varies with $k$, the average number of copies grows only linearly in $d$ (as opposed to quadratically, as for full state tomography). Specifically, if we have $m = \sum_{i=1}^l m_i$ total copies of $\rho$, then the expected number of copies required for a given $\delta$ and $\epsilon$ is
\begin{equation}
    \mathbb{E}(m) \leq 1 + \frac{1}{\epsilon^2\delta} + \frac{2d}{\epsilon^2}\textrm{log}(2/\delta) ~.
\end{equation}
Thus, we can generally reduce the cost from $d^2 \longrightarrow d$ by using DFE instead of state tomography.

The average number of copies can be significantly decreased for particular families of states. For example, let us consider the family of ``well-conditioned'' states, which includes all the states $\rho$ such that for every $k$, either $\Tr[\rho P_k] = 0$ or $|\Tr[\rho P_k] | \geq \alpha$ for $\alpha \leq 1$. For states in this family, we have
\begin{equation}
    m \leq \mathcal{O} \bigg( \frac{\log(1/\delta)}{\alpha^2\epsilon^2} \bigg) ~.
\end{equation}
For example, for stabilizer states $\alpha=1$, so the number of copies of $\rho$ needed is independent of the system's size, and for W states ($\alpha=1/n$), the average number of copies grows only as $n^2$.

DFE is cheaper than tomography, but its cost is still exponential in the number of qubits ($n$) for general states. Furthermore, like state tomography, DFE does not account for measurement errors, so DFE's estimates of state fidelity will also include errors from measurement. For this reason, fidelity estimation techniques which account for SPAM have been developed (see Sec.~\ref{sec:mcfe}).

To demonstrate how DFE depends on the number of random bases that are sampled, \fig\ref{fig:dfe_experiment} shows results from DFE of an $n$-qubit \ac{GHZ} state (for $n = \{4,6,8,10\}$) performed on \texttt{ibm\_sherbrooke}. Rather than measuring 1 shot for each sampled basis, as described above, the DFE prescription from Ref.~\cite{elben2023randomized} was followed. In particular, 1024 shots from each of 50 measurement bases randomly drawn via importance sampling from the ideal target GHZ state were measured. The running-average of the fidelity for each GHZ state is shown in \fig\ref{fig:dfe_experiment}. The estimated fidelities after 50 random measurement bases are in general agreement with the expected output fidelity based on a product-of-errors calculation for the physical error rates on \texttt{ibm\_sherbrooke}.

%%%%%%%%%%%%%%%%%%%%%%% Pauli Noise Learning %%%%%%%%%%%%%%%%%%%%%%% 
\subsection{Pauli Noise Learning}\label{sec:pnr}

Stochastic Pauli channels (see Sec.~\ref{sec:stoch_pauli}) are an important class of error channels that are particularly relevant for quantum error correction \cite{terhal2015quantum}. A general Pauli error map for $n$ qubits only contains $4^n - 1$ parameters --- the rates of each possible Pauli error --- which is fewer than the $\mathcal{O}(16^n)$ parameters of a general process matrix. While this is still exponential in the number of qubits, it can be further reduced using assumptions about the nature and locality of Pauli errors across an $n$-qubit device. Furthermore, while most quantum systems suffer from more complex error mechanisms than simply Pauli noise, one can experimentally \emph{design} stochastic channels \cite{graydon2022designing} using methods such as randomized compiling \cite{wallman2016noise, hashim2021randomized} and Pauli frame randomization \cite{kern2005quantum, ware2021experimental}, thus enforcing the same error model that can be efficiently characterized.

Several methods have been proposed for learning Pauli channels \cite{flammia2020efficient, harper2020efficient}. Pauli channels have diagonal Pauli transfer matrices (\ac{PTM}s, see Sec.~\ref{sec:ptm_rep}), denoted $\Lambda$, and these methods are designed to estimate the eigenvalues 
\begin{equation}
    \Lambda_{PP} = \frac{1}{d} \Tr[P\E(P)]~.
\end{equation}
The eigenvalue $\lambda_P = \Lambda_{PP}$ captures how much $\E$ attenuates the Pauli operator $P$. If $\Lambda_{PP} = 1$, then $\E$ preserves the Pauli operator $P$; if $\Lambda_{PP} < 1$, then $P$ is not preserved by $\E$.

The eigenvalues $\Lambda_{PP}$ can be related to the rates of each possible Pauli error, as we demonstrate using a one-qubit Pauli channel, which has the form
\begin{equation}
    \Lambda = 
    \begin{psmallmatrix}
        1 & 0 & 0 & 0 \\
        0 & 1 - 2(p_Y + p_Z) & 0 & 0 \\
        0 & 0 & 1 - 2(p_X + p_Z) & 0 \\
        0 & 0 & 0 & 1 - 2(p_X + p_Y)
   \end{psmallmatrix} ~,
\end{equation}
where $p_{Q}$ is the probability of the Pauli error $Q$. This example shows that a Pauli error $Q$ with probability $p_Q$ will attenuate the eigenvalue of any non-commuting Pauli operator $P$ by an amount $2p_Q$. To generalize the relationship between Pauli eigenvalues and Pauli error rates, we note that a stochastic Pauli channel's Kraus map is of the form $\E(\rho) = \sum_Q p_Q Q \rho Q^\dagger$, and this has a PTM given by
\begin{equation}
    \Lambda = \frac{1}{d} \sum_{P,Q \in \mathbb{P}_n} (-1)^{\langle P, Q \rangle} p_Q \ketbra{P \rangle}{\langle P} ~,
\end{equation}
where $\langle P, Q \rangle = 0$ if $[P, Q] = 0$, otherwise $\langle P, Q \rangle = 1$, and $\ket{\cdot\rangle}$ is the \emph{vectorization} defined in Sec.~\ref{sec:rep_quant_proc}. A single given eigenvalue $\Lambda_{PP}$ may therefore be computed as
\begin{equation}
    \Lambda_{PP} = \sum_{Q \in \mathbb{P}_n} (-1)^{\langle P, Q \rangle} p_Q ~.
\end{equation}
The inverse transformation --- computing a Pauli error rate $p_Q$ from Pauli eigenvalues --- is
\begin{equation}
    p_Q = \frac{1}{4^n} \sum_{P \in \mathbb{P}_n} (-1)^{\langle P, Q \rangle} \Lambda_{PP} ~.
\end{equation}
This transformation is the \emph{Walsh-Hadamard} transform, with the following matrix representation:
\begin{equation}\label{eq:walsh_hadamard}
    \mathbb{W}_{P,Q} = \frac{1}{d} \sum_{P,Q} (-1)^{\langle P, Q \rangle} \ketbra{P \rangle}{\langle Q} ~.
\end{equation}
This Walsh-Hadamard transform can be used to compute a vector of Pauli eigenvalues $\boldsymbol{\lambda}$ from a vector of Pauli error rates $\mathbf{p}$,
\begin{equation}
    \boldsymbol{\lambda} = \mathbb{W} \mathbf{p} ~,
\end{equation}
or, using the inverse transformation, to compute Pauli error rates from Pauli eigenvalues:
\begin{equation}\label{eq:inverse_walsh_hadamard_transform}
    \mathbf{p} = \mathbb{W}^{-1} \boldsymbol{\lambda} ~.
\end{equation}
Thus, the general strategy for learning Pauli channels is to measure a set of Pauli eigenvalues and then use \eq\ref{eq:inverse_walsh_hadamard_transform} to calculate the associated Pauli error rates. Measuring Pauli eigenvalues can be achieved using various methods, including cycle benchmarking (Sec.~\ref{sec:cb}), Pauli-twirled random Clifford circuits (see Sec.~\ref{sec:aces}), shadow tomography \cite{chen2023efficient}, etc. In what follows, we will describe two strategies for estimating Pauli eigenvalues.

%%%%%%%%%%%%%%%%%%%%%%% Cycle Error Reconstruction %%%%%%%%%%%%%%%%%%%%%%% 
\subsubsection{Cycle Error Reconstruction}\label{sec:cer}

\begin{figure*}[!htbp]
    \centering
    \includegraphics[width=1.9\columnwidth]{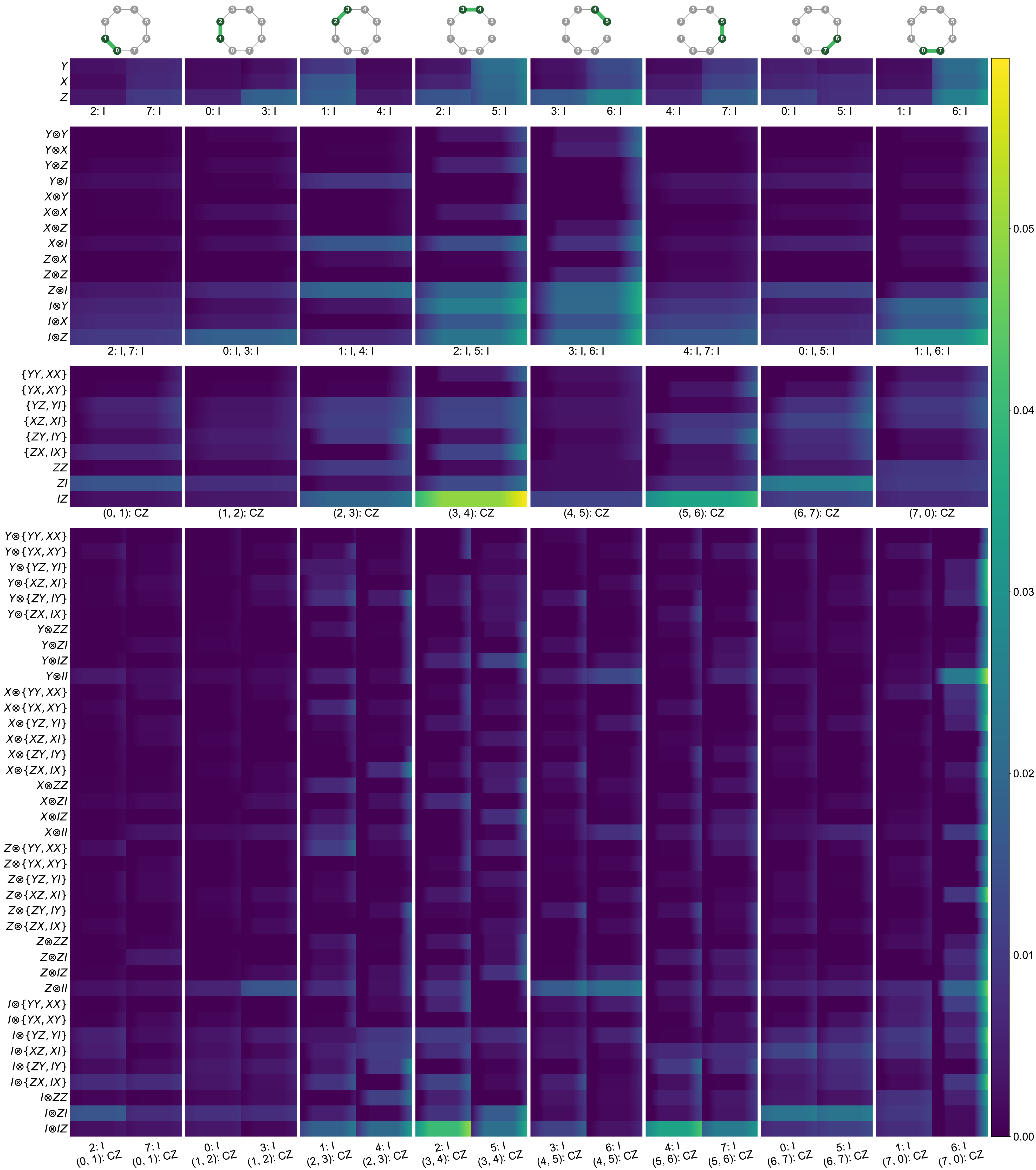}
    \caption{\textbf{Cycle Error Reconstruction.} Heatmap of Pauli errors measured via CER for an eight-qubit superconducting quantum processor. Each experiment consisted of benchmarking a two-qubit CZ gate (indicated by the graph pattern above each column) plus nearest-neighbor idling qubits. The $x$-axis labels the ideal gate operation on each subset of qubits in each benchmarked cycle. The $y$-axis labels the type of Pauli error, with the tensor notation $\otimes$ between Pauli errors indicating errors acting on product states, while the lack of tensors indicates errors on entangled qubits; curly brackets indicate gauge ambiguities (see Appendix \ref{sec:pauli_gauge}). The color of each cell indicates the marginalized error rate, and the gradient defines the 95\% confidence interval. The first row of subplots shows single-body errors acting on idling qubits; the second row of subplots shows correlated single- and two-body errors between idling qubits; the third row of subplots shows single-body errors acting on the CZ qubits; and the fourth row of subplots shows correlated single- and two-body errors between an idling spectator qubit and the CZ gate qubits. 
    }
    \label{fig:cer}
\end{figure*}

In Sec.~\ref{sec:cb}, we reviewed cycle benchmarking (\ac{CB}), a scalable protocol for measuring error rates for cycles containing parallel quantum gates. The goal of CB is to measure the eigenvalues of the PTM of a cycle. These eigenvalues can be estimated from a Pauli-twirled PTM; for this reason, CB uses randomized compiling. For example, CB performed on a two-qubit cycle would measure the following diagonal components of the PTM (up to any gauge ambiguities, see Appendix~\ref{sec:pauli_gauge}),
\begin{equation}\label{eq:ptm_cb}
    \Lambda = \begin{psmallmatrix}
                    1 &  &  &  &  &  &  &  &  &  &  &  &  &  &  &  \\
                      & \mkern-14mu f^m_{IX} &  &  &  &  &  &  &  &  &  &  &  &  &  &  \\
                      &  & \mkern-14mu f^m_{IY} &  &  &  &  &  &  &  &  &  &  &  &  &  \\
                      &  &  & \mkern-14mu f^m_{IZ} &  &  &  &  &  &  &  &  &  &  &  &  \\
                      &  &  &  & \mkern-14mu f^m_{XI} &  &  &  &  &  &  &  &  &  &  &  \\
                      &  &  &  &  & \mkern-14mu f^m_{XX} &  &  &  &  &  &  &  &  &  &  \\
                      &  &  &  &  &  & \mkern-14mu f^m_{XY} &  &  &  &  &  &  &  &  &  \\
                      &  &  &  &  &  &  & \mkern-14mu f^m_{XZ} &  &  &  &  &  &  &  &  \\
                      &  &  &  &  &  &  &  & \mkern-14mu f^m_{YI} &  &  &  &  &  &  &  \\
                      &  &  &  &  &  &  &  &  & \mkern-14mu f^m_{YX} &  &  &  &  &  &  \\
                      &  &  &  &  &  &  &  &  &  & \mkern-14mu f^m_{YY} &  &  &  &  &  \\
                      &  &  &  &  &  &  &  &  &  &  & \mkern-14mu f^m_{YZ} &  &  &  &  \\
                      &  &  &  &  &  &  &  &  &  &  &  & \mkern-14mu f^m_{ZI} &  &  &  \\
                      &  &  &  &  &  &  &  &  &  &  &  &  & \mkern-14mu f^m_{ZX} &  &  \\
                      &  &  &  &  &  &  &  &  &  &  &  &  &  & \mkern-14mu f^m_{ZY} &  \\
                      &  &  &  &  &  &  &  &  &  &  &  &  &  &  & \mkern-14mu f^m_{ZZ}
                \end{psmallmatrix} ~,
\end{equation}
where $f^m_P$ is the process polarization associated with the preparation and measurement basis of the Pauli $P$ at a circuit depth of $m$. Here, $f_P$ is distinct from the process polarization $f$ defined in Sec.~\ref{sec:polarization}, which is the \emph{average} process polarization in the unital block of a PTM. Under Clifford twirling, $f = \frac{1}{d^2 - 1} \sum_{P \ne I} f_P$, but under Pauli twirling each $f_P$ is unique (see Sec.~\ref{sec:rb_math_twirling} and Appendix \ref{sec:twirling}). In the limit that $m=0$, $f_P$ is an estimate of eigenvalue $\Lambda_{PP}$, representing how well $P$ is preserved by $\Lambda$. It should be noted that $f_P$ is often termed the \emph{Pauli fidelity} of $P$ in the literature when $P$ is unambiguous \cite{carignan2023error}, or the \emph{orbital fidelity} when individual Pauli fidelities cannot be learned due to gauge ambiguities; however, $f_P$ is not strictly a fidelity because it can be negative, since the elements of a PTM are bounded by [-1, 1].

\emph{\Ac{CER}} \cite{flammia2020efficient, carignan2023error} [also called \emph{\ac{KNR}} \cite{trueq}] is a protocol which leverages CB for efficiently estimating the eigenvalues of a cycle's PTM. CER results are based on targeted CB measurements in which specific Paulis are chosen to estimate the error rates afflicting subsets of the gates or idle qubits in the specific cycle of interest. Since the Pauli decays in CB are dual to the Pauli operators which cause errors, CER measures the error rate $p_Q$ of some fixed Pauli $Q$ by measuring a set of Pauli decays that commute and anti-commute with $Q$, and then using this information to reconstruct the probability $p_Q$ via linear inversion using \eq\ref{eq:inverse_walsh_hadamard_transform}. 

For the two-qubit PTM shown in \eq\ref{eq:ptm_cb}, it is feasible to measure all 15 operators, and thus reconstruct all weight-1 and weight-2 Pauli errors afflicting the cycle. However, for an arbitrary $n$-qubit cycle, while some high-weight errors can be estimated, it becomes exponentially expensive to measure all $4^n - 1$ Pauli errors. Instead, the usual strategy is to only reconstruct lower-weight Pauli errors, thus limiting the number of Pauli eigenvalues that must be measured. This strategy assumes that errors are relatively local to nearby qubits, and that long-range correlations are negligibly small.

In \fig\ref{fig:cer}, we plot a heatmap of the dominant Pauli errors on an 8-qubit superconducting quantum processor with a ring geometry. These errors are reconstructed by performing CB on the eight different cycles containing a single two-qubit CZ gate, as well as the idle qubits on either side of each CZ (i.e., the interleaved gate cycle is $G = I \otimes \text{CZ} \otimes I$, see \fig\ref{fig:cb_gates_cycles}). We see that the dominant Pauli errors on the quantum processor are weight-1 errors affecting the idle qubits or one of the entangled qubits. In fact, the largest error on the processor is a local $Z$ error on qubit 4 during the CZ gate between qubits 3 and 4. The source of this error is likely to a \emph{coherent} $Z$ error, not a \emph{stochastic} $Z$ error (by design, CER cannot distinguish between true stochastic Pauli errors and coherent errors which have been twirled into Pauli channels).

In \fig\ref{fig:cer}, we observe that some of the Pauli errors acting on entangled qubits appear grouped together in curly brackets. These groupings indicate error types that cannot be distinguished due to degeneracies, since some local errors acting on either qubit in the CZ gate will be transformed by the gate. This has to do with a fundamental gauge ambiguity in Pauli noise learning~\cite{chen2023learnability}, and is explained further in Appendix \ref{sec:pauli_gauge}. While only one- and two-body errors \footnote{Here, a $k$-\emph{body} error is any weight-$n$ Pauli error acting on $k$ gates. For example, a $Z$ error is a weight-1 error acting on a single qubit, but both $IZ$ and $ZZ$ are (weight-1 and weight-2, respectively) \emph{single}-body errors acting on two qubits involved in an entangling gate.} were measured in \fig\ref{fig:cer} (i.e., all $k \ge 2$-body errors were neglected), this is justified by the data, since we observe that two-body terms are largely suppressed compared to one-body terms. Moreover, the two-body error rates are the marginalized probabilities of \emph{all} $k$-body errors that act on the corresponding two bodies. Therefore, the fact that two-body errors are negligible proves that three- or more body errors are also negligible.

%%%%%%%%%%%%%%%%%%%%%%% ACES %%%%%%%%%%%%%%%%%%%%%%% 
\subsubsection{Averaged Circuit Eigenvalue Sampling}\label{sec:aces}

\emph{\Ac{ACES}} \cite{flammia2021averaged, pelaez2024average, hockings2024scalable} is an alternative scalable technique for learning the Pauli error rates of many layers of gates performed simultaneously. ACES uses \emph{random} Clifford circuits performed with randomized compiling to accomplish this, which allows many gates to be characterized in a single experiment. 

ACES estimates a Pauli channel for each gate in a set of Clifford gates. ACES can estimate arbitrary $n$-qubit Pauli error rates for each gate in principle, but, like CER, a reduced model is required for scalability. For example, a crosstalk-free error model \cite{PRXQuantum.2.040338, hashim2023benchmarking} can be used, in which each layer's error consists of tensor products of one- and two-qubit Pauli channels for each one- and two-qubit gate in the layer, respectively. Even for this highly restricted error model, the parameter space becomes large very quickly --- this model has $15N_{2Q}+3N_{1Q}$ parameters, where $N_{2Q}$ is the number of 2-qubit gates and $N_{1Q}$ is the number of single-qubit gates in the layer. For example, for a line of 100 qubits with bidirectional CNOT gates and 6 single-qubit Clifford gates per qubit, this amounts to 4,770 parameters. It is also possible to incorporate additional variables for state preparation and measurement error.

ACES uses measurements of Pauli observables of random Clifford circuits to learn many combinations of the eigenvalues of Pauli channels, which can then be used to estimate the individual eigenvalues themselves. Consider a circuit $\mathcal{C} = G_d G_{d-1} \cdots G_1$, where the $G_i$ are Clifford gates, and suppose each gate experiences a (gate-dependent) post-gate stochastic Pauli error $\Lambda_\E^{G_i}$, i.e., the noisy circuit is
\begin{equation}
    \tilde{\mathcal{C}} = \Lambda_\E^{G_d} \Lambda_d \cdots \Lambda_\E^{G_1} \Lambda_1 ~,
\end{equation} 
where $\Lambda_i$ denotes the PTM for $G_i$ and $\Lambda_\E^G$ denotes the PTM for $G$'s error channel. A Pauli measurement result is determined by the \emph{generalized eigenvalues} of the imperfect gates,
\begin{equation}\label{eq:aces}
    \Lambda_\E^{G} \Lambda[P] = \left( \Lambda_\E^{G} \right)_{GPG^{-1},GPG^{-1}} \left[ GPG^{-1} \right] ~.
\end{equation}
Stated differently, each Clifford operation transforms a Pauli $P$ into another Pauli $P'$ (see Appendix \ref{sec:clifford_group}), and $P'$ is always an eigenvector of the subsequent Pauli channel. By applying \eq\ref{eq:aces} to a sequence of gates, we see that the Pauli operators are generalized eigenvectors of any Clifford circuit that experiences only stochastic Pauli noise, and the circuit's generalized eigenvalues, denoted $\lambda^\mathcal{C}_{P}$, are
\begin{align}
    \Lambda_\E^{G_d} \Lambda_d \cdots \Lambda_\E^{G_1} \Lambda_1[P]
        & = \prod_{i = 1, \cdots, m} \left( \Lambda_\E^{G_i} \right)_{P_i, P_i} \left[ \mathcal{C}P\mathcal{C}^{-1} \right] ~, \label{eq:aces_gen_eigenval} \\
        & = \lambda^\mathcal{C}_{P} \left[ \mathcal{C}P\mathcal{C}^{-1} \right]
\end{align}
where $P_i$ denotes the Pauli $P$ evolved through the first $i$ gates, i.e., $P_i = \Lambda_i \cdots \Lambda_1[P]$. By (1) preparing properly-sampled eigenstates of $P$ (see Sec.~\ref{sec:direct_fid_est}), (2) performing $\mathcal{C}$, and then (3) measuring the final Pauli $\mathcal{C} P \mathcal{C}^{-1}$, the generalized eigenvalues $\lambda^\mathcal{C}_{P}$ can be determined experimentally.

\begin{figure}[t]
    \centering
    \includegraphics[width=\columnwidth]{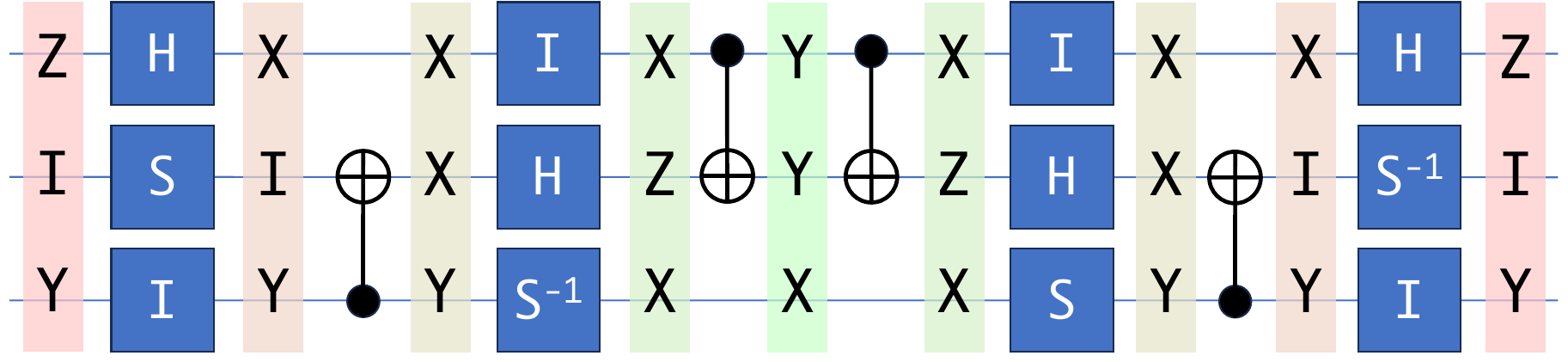}
    \caption{\textbf{Averaged Circuit Eigenvalue Sampling (ACES).}
    Example mirror circuit that can be used for the ACES protocol, consisting of single-qubit Clifford gates plus bi-directional CX gates. The three qubits are prepared in a +1 eigenstate of $ZIY$. Each layer of gates transforms the Pauli operator $ZIY$ into a different Pauli, and this Pauli describes which of the layer's generalized eigenvalues will contribute to the final measurement result. Under noise-free execution, the final state is the same as the input state, and therefore is a +1 eigenstate of $ZIY$. With execution under a stochastic Pauli noise model, the result of measuring $ZIY$ is attenuated by $\lambda^\mathcal{C}_{ZIY}$, which is the circuit eigenvalue.}
    \label{fig:ACES_example}
\end{figure}

To use the Pauli measurement results to estimate individual gate eigenvalues $\lambda^G_{P}$ (and hence the error model parameters), we construct a linear system of equations relating the circuit generalized eigenvalues to the $\lambda^G_{P}$. Taking the log of Eq.~\ref{eq:aces_gen_eigenval} (and assuming $\lambda_{P_i}^{G_i} > 0$ for all $i$, which is true as long as the gates have sufficiently low error rates), 
\begin{align}
    \ln{\lambda}^C_P & = \sum_{j=1}^d \ln\left[\left( \Lambda_\E^{G_j} \right)_{P_j,P_j}\right] ~.
\end{align}
Therefore, Pauli measurement results are related to the generalized eigenvalues of the gates by a system of linear equations $\mathbf{b} = A\mathbf{x}$, where $\mathbf{x}$ encodes the Pauli channel eigenvalues and $\mathbf{b}$ encodes the measurement results. $A$ is called the \emph{design matrix}, and each row of $A$ encodes how the gate eigenvalues relate to the result of a single Pauli observable measurement. It is determined by the choice of circuits and Pauli measurements, and can be efficiently computed since its computation only requires evolving Pauli operators through Clifford circuits. By running sufficiently many random Clifford circuits and performing sufficiently many independent Pauli measurements, a full-rank matrix $A$ can be generated. The Pauli eigenvalues of the gates can then be estimated by computing $\mathbf{x} = A^{+}\mathbf{b}$, where $A^{+}$ denotes the pseudoinverse of $A$. 

In principle, ACES can be run with any set of Clifford circuits, but because most processors are limited to computational basis measurements, a careful choice of circuit structure allows for more independent Pauli measurements to be performed with each circuit, rendering more information. ACES is often run with a form of mirror circuit (see \fig\ref{fig:ACES_example}), which enables measuring $\mathcal{O}(n)$ independent Pauli observables from each computational basis measurement. The version of mirror circuits in the original ACES protocol (Ref.~\cite{flammia2021averaged}) places a layer of random gates at the end of the circuit, so they are not identity circuits (or Pauli operators). The particular structure used means that any one- or two-qubit Pauli propagated through the full circuit has weight at most $6$ at the end of the circuit, which means that each measurement required to learn a crosstalk-free model with ACES requires at most $6$ qubits. 

\section{Estimating Circuit Fidelities}\label{sec:fidelity_estimation}

While all the individual components (states, gates, layers/cycles, and measurements) used in quantum circuits are becoming more accurate, they remain inherently noisy. When a few of them are combined to form a quantum circuit, their noise and errors accumulate. And when more than a few of them are combined (hundreds, thousands, or even --- eventually ---- millions), the accumulated noise becomes significant and can severely alter the outputs. So, given a quantum circuit of arbitrary size and nature, how can we determine whether its outputs are close to the ideal (noiseless) outputs?

Validating the outputs of quantum circuits turns out to be a puzzling problem. The simplest approach is to resort to classical simulations: when a quantum circuit is implemented, it is also simulated on a classical computer, and finally the outputs are compared. This approach is effective, but only for circuits that are feasible to simulate classically. Some experiments have already hit the boundary of what can be simulated classically in reasonable time \cite{2019GoogleSupremacy}. A different approach consists of individually characterizing the components used in the circuit of interest (e.g., benchmarking of gate layers, or tomography of individual gates), and then predicting the quality of its outputs from the characterization data. This approach can be scalable, but it is also often unreliable. Quantum circuits are more than the sum of their components, and the noise in a circuit may exhibit properties (such as drift, fluctuations, and temporal correlations) that may not be observed by inspecting individual components. This calls for protocols that can test the circuit as a whole, rather than its parts.

\begin{figure*}[t!]
    \includegraphics[width=2\columnwidth]{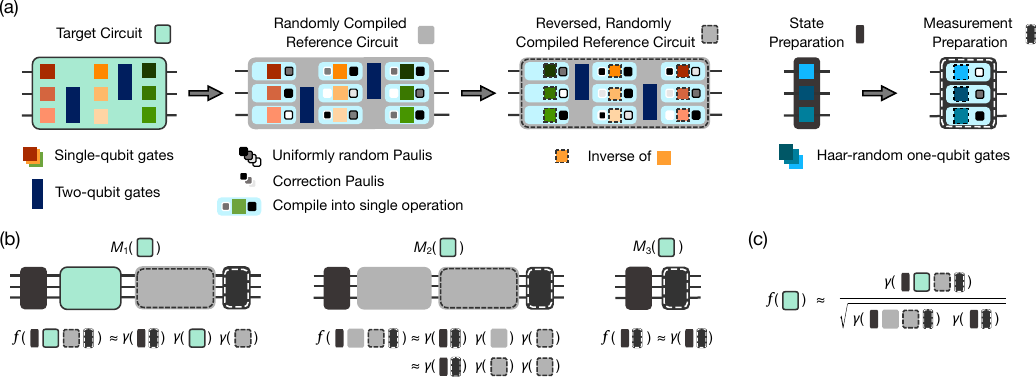} 
    \caption{\textbf{Mirror Circuit Fidelity Estimation.} 
    \textbf{(a)} MCFE estimates the process fidelity $F_e(\mathcal{C})$ with which a quantum computer can execute some target circuit $\mathcal{C}$ (green box), using motion reversal circuits built from $\mathcal{C}$ and the four reference (sub)circuits shown here. 
    \textbf{(b)} The three motion reversal circuits used in MCFE and \textbf{(c)} the simple data analysis that MCFE uses to estimate the $\mathcal{C}$'s process polarization $f(\mathcal{C})$, which can be re-scaled to estimate $F_e(C)$ (see Tab.~\ref{tab:table_rel_fid_dep}).
    (Figure adapted with permission from \R\cite{proctor2022establishing}.)
    }
    \label{fig:mcfe}
\end{figure*}

In this section, we provide an overview of some of the scalable methods to characterize the performance of quantum circuits. In particular, we describe the following methods:
\begin{itemize}
    \item \textit{Mirror Circuit Fidelity Estimation} (Sec.~\ref{sec:mcfe}). Mirror circuit fidelity estimation is a technique for estimating the process fidelity of any circuit $\mathcal{C}$ using mirror circuits of twice $\mathcal{C}$'s depth.
    \item \textit{Circuit Output Accreditation} (Sec.~\ref{sec:accreditation}). Circuit output accreditation is a technique for lower-bounding the process fidelity of a circuit $\mathcal{C}$ by running a set of ``trap'' circuits that are the same width and depth as $\mathcal{C}$, but contain only Clifford gates. 
\end{itemize}
Mirror circuit fidelity estimation and circuit output accreditation are complementary techniques with similar aims and slightly different properties and strengths (discussed later). These techniques have two important properties in common: they are both (i) efficient in the number of qubits, and (ii) robust to \ac{SPAM} errors. Both techniques run a number of circuits that is independent of the number of qubits, and require minimal classical computations. This contrasts with direct fidelity estimation (Sec.~\ref{sec:direct_fid_est}), which could be used to estimate a circuit's process fidelity, but is typically not used in practice because it is not robust to SPAM errors and it is exponentially expensive for general circuits. Finally, note that a variety of techniques exist for formal verification of the output of quantum algorithms or circuits, for example, interactive cryptographic protocols which allow a classical user to verify that a computation was carried out by a quantum device \cite{mahadev2018classical, brakerski2021cryptographic, kahanamoku2022classically, zhu2021interactive}. These methods are beyond the scope of this tutorial.

%%%%%%%%%%%%%%%%%%%%%%%%%%%% MCFE %%%%%%%%%%%%%%%%%%%%%%%%%%%%
\subsection{Mirror Circuit Fidelity Estimation}\label{sec:mcfe}

\emph{\Ac{MCFE}} \cite{proctor2022establishing} is a technique for efficiently measuring the process (i.e., entanglement) fidelity $F_e(\mathcal{C})$ with which a quantum computer can implement an $n$-qubit circuit $\mathcal{C}$. MCFE is robust in the presence of SPAM errors, and it is efficient in the number of qubits. MCFE consists of running circuits sampled from three ensembles of ``mirror circuits'' built from the circuit of interest $\mathcal{C}$, shown in \fig\ref{fig:mcfe}. The core idea is that by running circuits with three different structures, one of which contains $\mathcal{C}$, MCFE is able to approximately isolate the process fidelity of $\mathcal{C}$ from all other operations in those circuits using some simple algebra. Below we explain how MCFE works.

MCFE's first mirror circuit ensemble [$M_1(\mathcal{C})$] consists of (i) a layer $L$ of random single-qubit gates sampled from a unitary 2-design (e.g., Haar-random single-qubit gates), (ii) the circuit $\mathcal{C}$, (iii) a randomly compiled version of the inverse of $\mathcal{C}$ \cite{wallman2016noise, hashim2021randomized} (denoted $\mathcal{C}_{\textrm{rev}}$), and (iv) the inverse of $L$ compiled together with a random $n$-qubit Pauli gate. Each such $M_1(\mathcal{C})$ circuit embeds $\mathcal{C}$ within a larger (mirror) circuit that, if implemented without error, will always return an easy-to-compute ``success'' bit string. Therefore, we can easily assess how well each such circuit was run simply by looking at the frequency with which this success bit string is output from each $M_1(\mathcal{C})$ circuit --- suggesting that these circuits can be used to understand how well $\mathcal{C}$ can be executed. In particular, because of the randomization in the initial and final layer of gates, as well as the randomized compilation in $\mathcal{C}_{\textrm{rev}}$, it is possible to show that
\begin{equation}
    \mathbb{E}(\gamma[M_1(\mathcal{C})]) \approx f_0 f(\mathcal{C}_{\textrm{rev}}) f(\mathcal{C}) ~. \label{eq:mcfe-1}
\end{equation}
where $f(\cdot)$ is the process polarization (see \eq\ref{eq:process_polarization}), $\mathbb{E}(\cdot)$ denotes the expectation value over a circuit ensemble,
\begin{equation}
    \gamma(M) =  \frac{4^n}{4^n-1}\sum_{k=0}^n\left(-\frac{1}{2}\right)^{k}h_k(M) - \frac{1}{4^n -1} ~,
\end{equation}
where $h_k(M)$ is the frequency with which the output of mirror circuit $M$ is a Hamming distance of $k$ from its ``success'' bit string, and $f_0$ is a nuisance parameter called the ``effective SPAM polarization,'' which encompasses contributions from errors in the SPAM and in the layers of single-qubit gates [steps (i) and (iv)].

\begin{figure*}[t!]
    \centering
    \includegraphics[width=2\columnwidth]{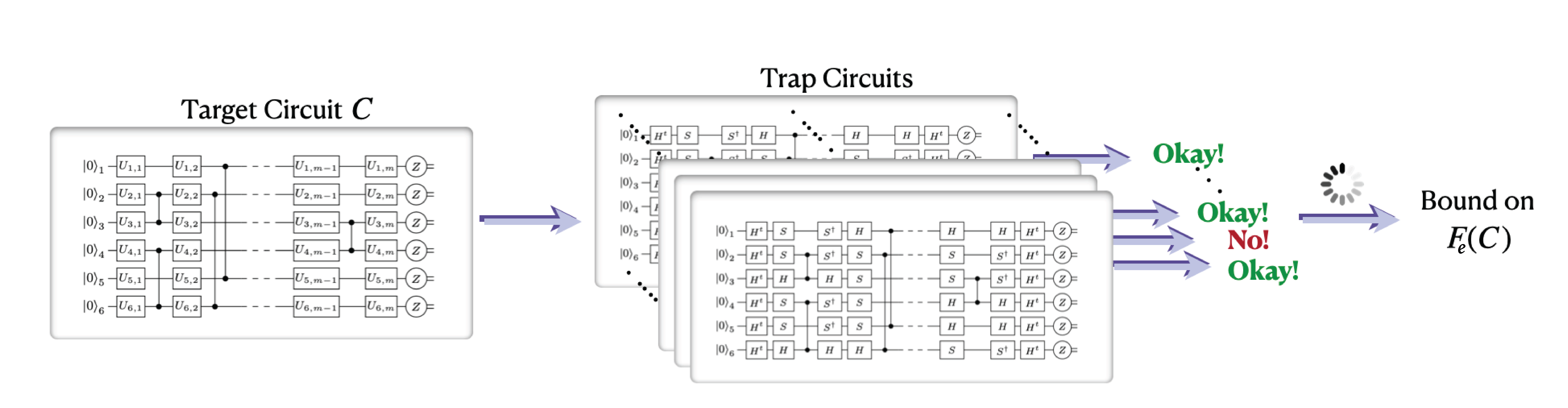} 
    \caption{\textbf{Circuit Output Accreditation.} Circuit output accreditation generates $N_c$ circuits based on the target quantum circuit $\mathcal{C}$. These circuits (called \textit{traps}) contain the same two-qubit gates as the target circuit, and after the neighboring layers of one-qubit gates are compiled into a single layer, every trap has the same width and depth as the target circuit. However, unlike the target circuit, in the absence of errors every trap returns a fixed, known output. By post-processing the number of incorrect trap outputs, accreditation protocols return a lower-bound on the process fidelity $F_e(\mathcal{C})$ of the target circuit.
    }
    \label{fig:cap}
\end{figure*}

The aim in MCFE is to measure $F_e(\mathcal{C})$ [a re-scaling of $f(C)$, see Tab.~\ref{tab:table_rel_fid_dep}], but if we only run circuits sampled from $M_1(\mathcal{C})$ to estimate $\mathbb{E}(\gamma[M_1(\mathcal{C})])$, we instead learn $f(\mathcal{C})$ multiplied by the unknowns $f_0$ and $f(\mathcal{C}_{\textrm{rev}})$. MCFE solves this problem by running circuits sampled from two additional ensembles [$M_2(\mathcal{C})$ and $M_3(\mathcal{C})$]. $M_3(\mathcal{C})$ is essentially a randomized SPAM experiment [steps (i) and (iv) above] that enables learning $f_0$:
\begin{equation}
    \mathbb{E}(\gamma[M_3(\mathcal{C})]) = f_0 ~. \label{eq:mcfe-3}
\end{equation}
Finally, $M_2(\mathcal{C})$ is a fully randomly compiled version of $M_1(\mathcal{C})$ (see \fig\ref{fig:mcfe}), which enables learning $f(\mathcal{C}_{\textrm{rev}})$: 
\begin{equation}
    \mathbb{E}(\gamma[M_2(\mathcal{C})]) \approx f_0 f(\mathcal{C}_{\textrm{rev}})^2 ~. \label{eq:mcfe-4}
\end{equation}
By applying simple algebra to Eqs.~\ref{eq:mcfe-1}, \ref{eq:mcfe-3}, and \ref{eq:mcfe-4}, we see that
\begin{equation}
    f(\mathcal{C}) \approx  \frac{
        \mathbb{E}(\gamma[M_1(\mathcal{C})]) 
    }{
        \sqrt{\mathbb{E}(\gamma[M_2(\mathcal{C})]) \mathbb{E}(\gamma[M_3(\mathcal{C})]) }
    } ~.
\end{equation}
This is the analysis used by MCFE to estimate $f(\mathcal{C})$, which can then be re-scaled to estimate $F_e(\mathcal{C})$ using \eq\ref{eq:process_polarization}.

%%%%%%%%%%%%%%%%%%%%%%%%%%%% Accreditation %%%%%%%%%%%%%%%%%%%%%%%%%%%%
\subsection{Circuit Output Accreditation}\label{sec:accreditation}

\emph{Circuit output accreditation} is an efficient strategy for lower-bounding the process fidelity in a ``target'' circuit of interest. It only requires implementing circuits with the same size and depth as the target circuit. Moreover, it is robust to SPAM errors, and it is scalable in the number of qubits and gates in the target circuit.

Different variants of circuit output accreditation have been proposed \cite{ferracin2018, ferracin2019, ferracin2020}, but they all rely on the idea of implementing the target circuit alongside a number $N_c$ of Clifford circuits, called ``traps'' (see \fig\ref{fig:cap}). These traps have the same width and depth as the target circuit, but they implement different computations. In particular, the traps are designed in such a way that, in the absence of noise, they return a fixed, known output. This allows us, in the presence of noise, to estimate the probability that a trap returns an incorrect output. This probability can then be used to bound the process fidelity of the target circuit. To describe circuit accreditation in more detail, we focus on the protocol in \R\cite{ferracin2020}, which provides the tightest bound on the fidelity of the target circuit.

The accreditation protocol takes as input a target circuit $\mathcal{C}$, alongside two numbers $\theta,\:\alpha\in(0, 1)$, which represent the desired statistical error on the bound and the confidence in the bound respectively. To bound the process fidelity $F_e(\mathcal{C})$ of $\mathcal{C}$, output accreditation makes the following assumptions:
\begin{enumerate}
    \item The circuit $\mathcal{C}$ (i) takes as input $n$ qubits in the state $\ket{0}$, (ii) implements the sequence of operations $U_{m+1}E_mU_m{\ldots}U_2E_1U_1$, where $U_j$ is a layer of single-qubit gates and $E_j$ is a layer of two-qubit (e.g., CZ) gates for every $j$, and (iii) ends with Pauli-$Z$ measurements on every qubit.
    \item The errors affecting the various layers in $\mathcal{C}$ are completely positive and trace-preserving (\ac{CPTP}).
    \item The errors affecting the layers of one-qubit gates $U_j$ are gate independent. That is, every layer of one-qubit gate suffers the same noise.
\end{enumerate}
The first assumption is made without loss of generality, since most quantum circuits can be recompiled in the required form. The second and third assumptions are standard in the literature and allow us to encompass a broad class of noise and error processes; notably, requiring that the noise is CPTP does not include errors such as leakage (see Sec.~\ref{sec:leakage}). Crucially, when combined together, these three assumptions enable us to use randomized compiling on the target circuit, that is, to transform arbitrary noise processes into Pauli noise. In the reminder of the subsection, we thus assume that every layer in $\mathcal{C}$ is subject to Pauli noise.

The trap circuits are generated by creating a copy of the target circuit, and by replacing every single-qubit gate in this copy with either $I$, $H$, or $S=\sqrt{Z}$. These extra gates are undone by compiling their inverses in the subsequent single-qubit gate layer. Note that by the third assumption, each trap generated in this way is affected by noise that is identical to that affecting the target circuit (i.e., gate-independent Pauli noise for single-qubit gates), as it is equal to the target circuit except for the individual single-qubit gate layers.

In the absence of errors, the traps always return a fixed outcome $(0, 0, \ldots, 0)$. However, if an error occurs, the trap returns an incorrect output with probability larger than $50\%$. The proof of this statement (which is provided in detail in \R\cite{ferracin2020}) requires commuting errors all the way to the end of the circuit, and showing that due to the effect of the randomly-chosen one-qubit gates, they have at least $50\%$ probability of flipping one or more bits in the output string. Building on this property of the traps, the accreditation protocol takes the following steps:
\begin{enumerate}
    \item Generate and run a trap circuit. If the trap circuit returns the bit-string $(0, 0, \ldots, 0)$, mark the run as ``successful.'' Otherwise, mark it as ``unsuccessful.''
    \item Repeat the step above a number $N_c = 2\ln(2 / (1-\alpha)) / \theta^2$ and calculate the total number $N_\textrm{uns} \in [0, N_c]$ of unsuccessful runs.
\end{enumerate}
After all the traps have been run, the process fidelity $F_e(\mathcal{C})$ of the target quantum circuit $\mathcal{C}$ is bounded above and below by $N_\textrm{uns}$ up to an error $\mathcal{O}(\theta)$:
\begin{equation}
    1 - \frac{N_\textrm{uns}}{N_c} \geq F_e(\mathcal{C}) \geq 1 - 2\frac{N_\textrm{uns}}{N_c} \:.
\end{equation}
Thus, circuit accreditation enables lower- and upper-bounding the circuit fidelity.

\section{Holistic Benchmarks}\label{sec:holistic}

Holistic benchmarks are methods for quantifying the overall performance of a quantum computer. These methods typically summarize important aspects of a quantum computer's performance in relatively few numbers or plots, such as the quantum volume \cite{cross2019validating} or capability regions \cite{proctor2022measuring}. Most holistic benchmarks quantify the impact of errors on overall performance, but they typically do not directly quantify gate (or layer) error rates, unlike \ac{RB} protocols (Sec.~\ref{sec:randomized_benchmarks}). Holistic benchmarks, therefore, complement and contrast with both detailed error characterization tools like tomography (Sec.~\ref{sec:tomography}) and RB protocols (Sec.~\ref{sec:randomized_benchmarks}). In this section, we discuss some of the most widely-used or important holistic benchmarking methods. We discuss the following areas within holistic benchmarking:
\begin{itemize}
    \item \textit{Volumetric Benchmarks} (Sec.~\ref{sec:volumetric_benchmarks}). Volumetric benchmarking \cite{Blume-Kohout2020-de} is a framework that encompasses many different benchmarks. We discuss this framework, and two of its specific benchmarks or benchmark families: the quantum volume benchmark \cite{cross2019validating} and mirror circuit benchmarks \cite{proctor2022measuring}.
    
    \item \textit{Application Benchmarks} (Sec.~\ref{sec:application_benchmarks}). Holistic benchmarks based on applications or algorithms are now widely used to benchmark and compare quantum computers. We overview some of these methods, using examples from two algorithmic benchmarking suites.
    
    \item \textit{Scalable Holistic Benchmarks} (Sec.~\ref{sec:scalable_holistic_benchmarks}). Many existing holistic benchmarks are not scalable, but there are now techniques for creating scalable benchmarks from any set of circuits or algorithms. We briefly discuss these methods.
\end{itemize}

%%%%%%%%%%%%%%%%%%%%%%% Volumetric Benchmarks %%%%%%%%%%%%%%%%%%%%%%% 
\subsection{Volumetric Benchmarks}\label{sec:volumetric_benchmarks}

\begin{figure}%[ht]
    \centering  
    \includegraphics[width=5.5cm]{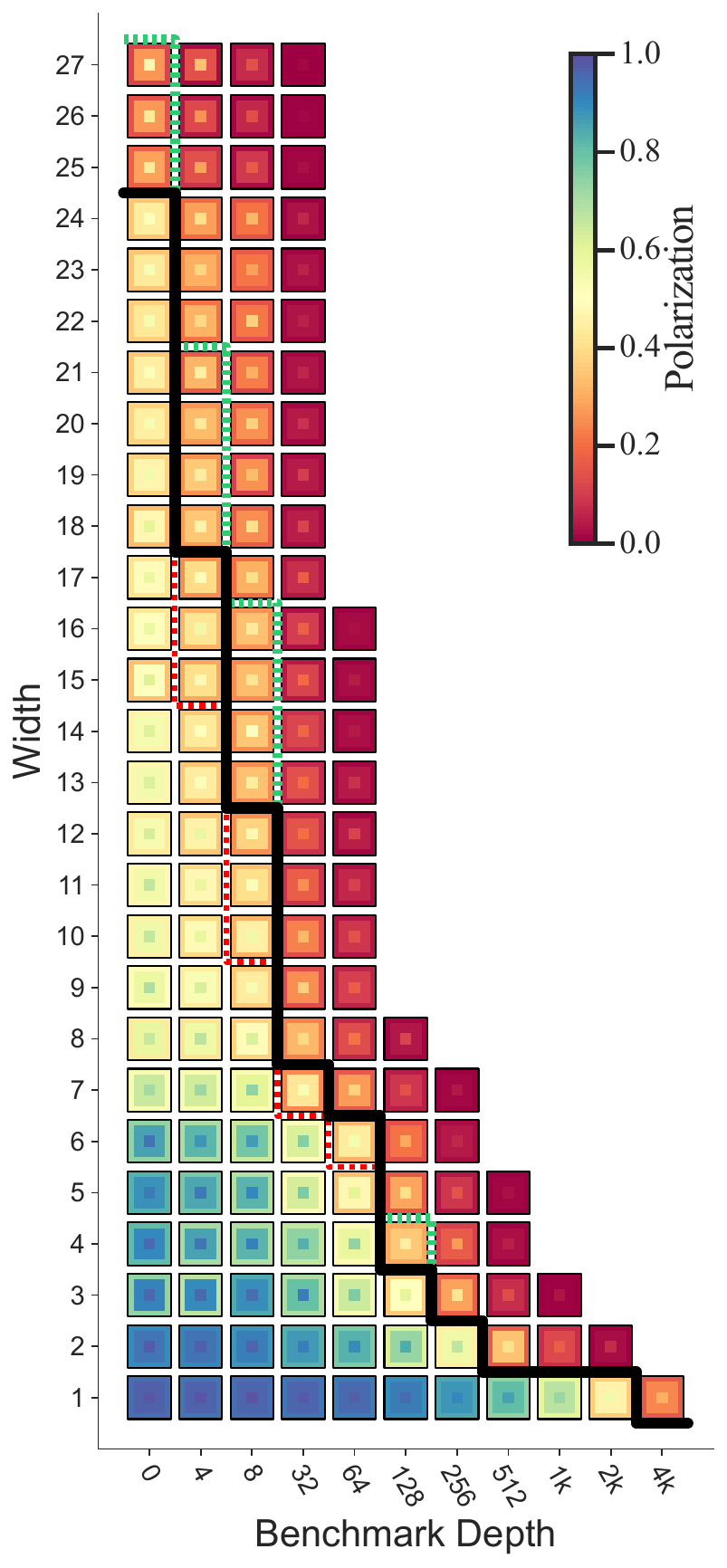}
    \caption{\textbf{Volumetric Benchmarking}. The results of a volumetric benchmark run on \texttt{ibmq\_montreal}. This benchmark consists of running randomized mirror circuits (Sec.~\ref{sec:mirror}) of various shapes. For each circuit width and benchmark depth (see main text), the concentric squares show the maximum (inner square), mean (middle square), and minimum (outer square) of the estimated polarizations $S_{\textrm{pol}}$ (\eq\ref{eq:polarization}) for all the circuits of that shape that were run. 
    % The polarization ($s_{\textrm{pol}}$) of the output distribution of a definite outcome $n$-qubit circuit $c$ is simply $s_{\textrm{pol}} = (s_{\textrm{sp}} -1/2^n)/(1-1/2^n)$ where $s_{\textrm{sp}}$ is $c$'s success probability \cite{proctor2022measuring}. 
    Frontiers (green, black, and red lines) show the circuit shapes at which these three statistics drop below the threshold value of $1/e$. (Figure reprinted with permission from \cite{Hothem2023-bl}.)
    }
    \label{fig:ibm_montreal_vb}
\end{figure}

Volumetric benchmarking \cite{Blume-Kohout2020-de} is a general methodology for benchmarking, rather than a specific benchmark. It generalizes ideas first introduced in the quantum volume benchmark (Sec.~\ref{sec:qv}). Volumetric benchmarks quantify a quantum computer's ability to run circuits with low error. In contrast to randomized benchmarks, volumetric benchmarks do not \emph{directly} quantify the error rates of a quantum computer's qubits or gates. Instead, they quantify a quantum computer's rate of errors when running circuits of various shapes. A specific volumetric benchmark is defined by: 
\begin{enumerate}
    \item A circuit family $\mathcal{C}_{W,D} = \{C\}$, that is indexed by circuit width ($W$, i.e., the number of qubits) and circuit depth ($D$). Note that ``circuit depth'' need not refer to the total number of layers of native gates in a low-level circuit; instead, for example, it could refer to the number of $n$-qubit Clifford gates in the circuit or the number of repetitions of an algorithmic subroutine (see discussion in Ref.~\cite{Blume-Kohout2020-de}). 

    \item A method for selecting circuits from $\mathcal{C}_{W,D}$ for each circuit shape $(W,D)$ (e.g., a probability distribution over $\mathcal{C}_{W,D}$ for each $W$ and $D$).
    
    \item An error metric or measure of success (e.g., total variation distance, classical fidelity, etc.; see Sec.~\ref{sec:overview}) with which to compute how well any circuit in $\mathcal{C}_{W,D}$ was performed on a quantum computer.
\end{enumerate}
Examples of circuit families for which a volumetric benchmark can be defined include the quantum volume circuits (see \fig\ref{fig:qv}), randomized mirror circuits (see Sec.~\ref{sec:mirror}), or the circuits from many algorithms.

\begin{figure}
    \centering  
    \includegraphics[width=5.5cm]{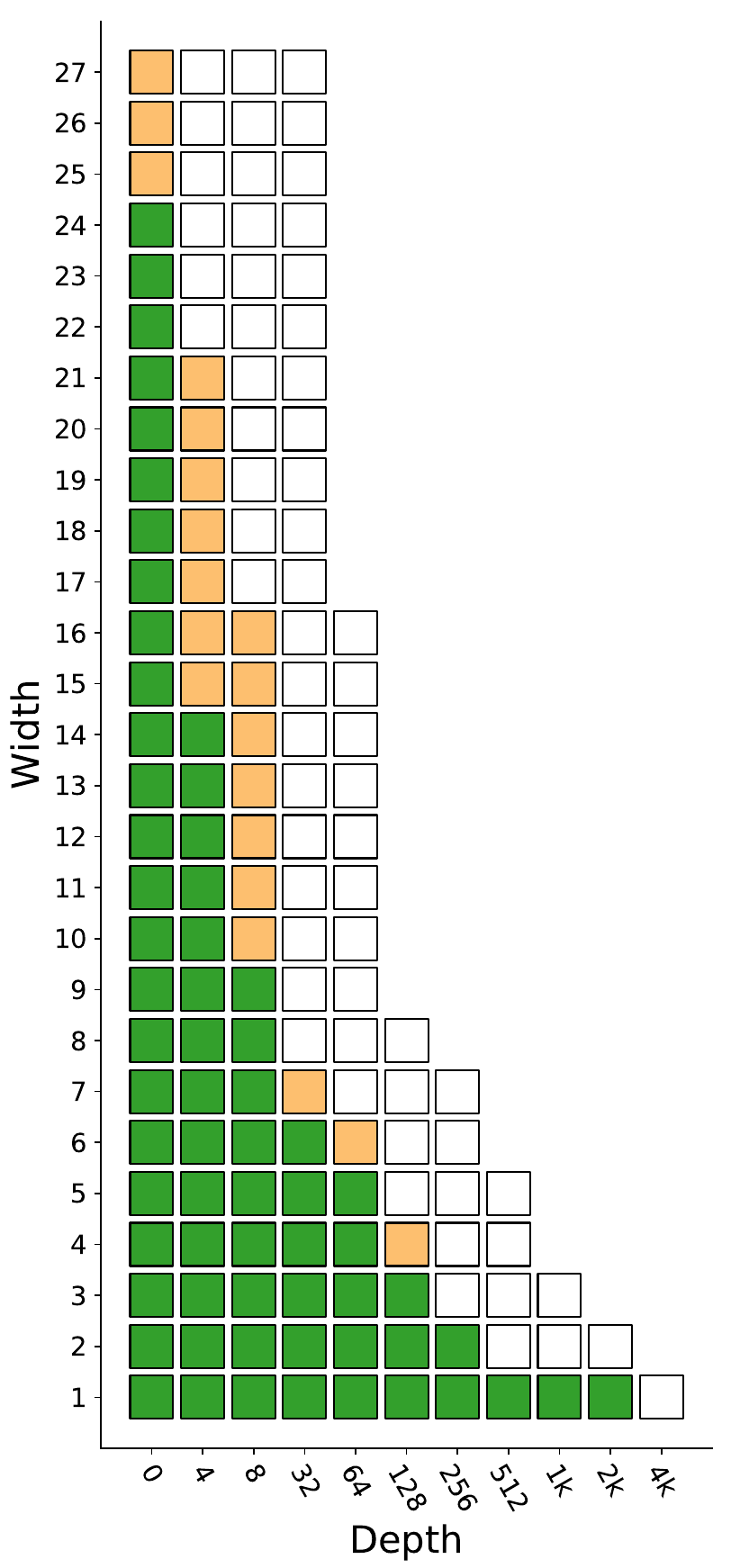}
    \caption{\textbf{Capability Regions}. 
    Capability regions \cite{proctor2022measuring} are high-level summaries of a quantum computer's ability to execute circuits with low error. This plot is a capability region created from the volumetric benchmarking data of \fig\ref{fig:ibm_montreal_vb}, and it summarizes the circuit shapes at which all randomized mirror circuits that were executed in this experiment ran successfully. Here, ``success'' means that a circuit's polarization (\eq\ref{eq:polarization}) is above the threshold of $1/e$. A square is green if all circuits ran successfully, orange if some succeeded and some failed, and white if no circuits succeeded. Capability regions will typically depend on the circuit family used to construct them, e.g., a circuit family with a higher two-qubit gate density will typically result in a smaller green region.}
    \label{fig:ibm_montreal_capreg}
\end{figure}

Applying a volumetric benchmark to a quantum computer consists of
\begin{enumerate}
    \item Picking a range of circuit shapes $(W,D)$ at which to run circuits.
    
    \item At each chosen $(W,D)$, selecting circuits from $\mathcal{C}_{W,D}$ and running them (the kinds of permissible compilation rules for each circuit depends on the benchmark).

    \item From each circuit's data, estimating how well the quantum computer ran that circuit using the benchmark's performance metric.
\end{enumerate}
This procedure generates data consisting of the quantum computer's performance on the benchmark's circuit family as a function of both width and depth. That data is then typically displayed on the width $\times$ depth plane --- sometimes called a ``volumetric benchmarking plot'' --- as demonstrated in \fig\ref{fig:ibm_montreal_vb}. Such a plot is a high-level overview of a quantum computer's performance on the circuits from that benchmark's circuit family. Volumetric plots can also be used to informally assess the kinds of errors occurring in the benchmarked system --- e.g., crosstalk errors cause circuit error rates to increase faster with increasing circuit width than would be predicted by gate error rates measured using isolated one- and two-qubit RB \cite{proctor2022measuring}.

The volumetric plot in \fig\ref{fig:ibm_montreal_vb} is a high-level performance summary, but it still contains a lot of detail. Therefore, it is sometimes useful to provide more concise and easily understood performance summaries. One way to do this is with ``capability regions'' \cite{proctor2022measuring}. Capability regions use volumetric benchmarking data to compute regions where a quantum computer can and cannot successfully run circuits, using some threshold for ``success.'' A capability region constructed from the data of \fig\ref{fig:ibm_montreal_vb} is shown in \fig\ref{fig:ibm_montreal_capreg}.

%%%%%%%%%%%%%%%%%%%%%%% Quantum Volume %%%%%%%%%%%%%%%%%%%%%%% 
\subsubsection{Quantum Volume}\label{sec:qv}

\begin{figure}
    \centering
    \includegraphics[width=\columnwidth]{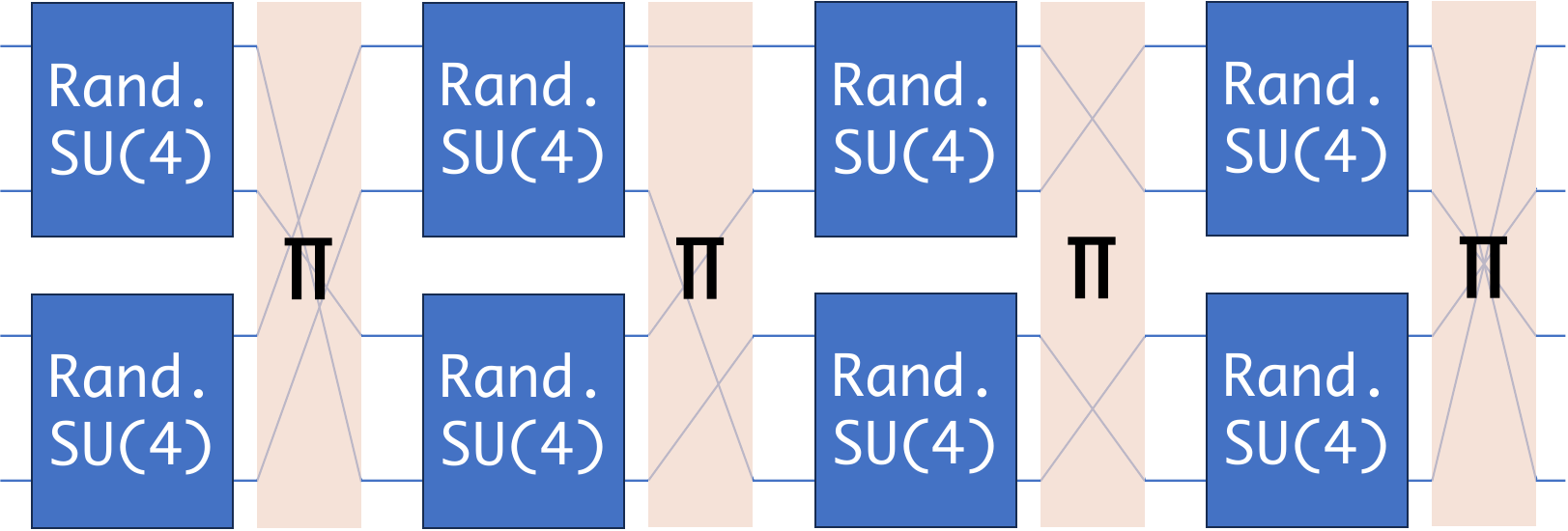}
    \caption{\textbf{Quantum Volume Circuits}. A depth $D$ (here $D=4$) quantum volume circuit on $W$ qubits (here $W=4$). Each layer consists of Haar-random two-qubit unitaries on a random pairing of the $W$ qubits (denoted here by random two-qubit gates between neighboring qubits and a random permutation of all the qubits). The quantum volume benchmark uses only square circuits, i.e., $D=W$. When $W$ is odd, one randomly-selected qubit idles during each layer.}
    \label{fig:qv}
\end{figure}

The \emph{\ac{QV}} benchmark \cite{cross2019validating} is a holistic benchmark that inspired volumetric benchmarking. The QV benchmark runs randomly sampled ``square'' circuits with a particular structure, and computes a single number --- the ``quantum volume'' --- summarizing a system's performance on those circuits. The QV benchmark explicitly permits compilation of its circuits, so it jointly tests a quantum computing system's compilers and gates, i.e., it is a ``full-stack'' benchmark \cite{cross2019validating, Amico2023-ze, Hines2023-be}. QV is currently one of the most widely-used metrics for comparing integrated quantum computing systems, and summarizing the field's overall progress (e.g., see Ref.~\cite{noauthor_undated-wz}). Note, however, that as of 2025 IBM (the developers of QV) are primarily using the error per layered gate (\ac{EPLG}) (see Sec.~\ref{sec:simrb}) to quantify their systems' overall performance.

The QV benchmark is based on the circuits shown in \fig\ref{fig:qv}, referred to as \emph{randomized model circuits} or simply \emph{quantum volume circuits}. The QV circuits are defined for any shape $(W,D)$, but the QV benchmark uses only circuits of this type that are ``square'' (i.e., have equal width and depth). Each layer in an $n$-qubit QV circuit comprises $n/2$ disjoint two-qubit gates, between random pairs of qubits (where $n/2$ is rounded down if $n$ is odd, and the single qubit that is not in any pair idles). Each gate is a uniformly random two-qubit unitary [i.e., it is drawn from the Haar measure on $\mathsf{SU(4)}$]. The QV analysis (described below) uses the concept of the \emph{heavy outputs} of a probability distribution (see Sec.~\ref{sec:linear_xe_heavy_output}). The \emph{heavy} outputs are the half of the outputs that are most likely to appear, i.e., those whose probability is above the median probability. For example, if the bit strings 00, 01, 10, and 11 have probabilities 0.1, 0.2, 0.3, and 0.4, respectively, the heavy outputs are 10 and 11.

The QV benchmark consists of applying the following ``quantum volume test'' for increasingly large $n$:
\begin{enumerate}
    \item Sample shape $(n, n)$ QV circuits.
    \item For each sampled circuit, compile it into a circuit that can be run on the specific system being tested. Approximate compilations are permissible (trading off fewer gates, and their associated errors, for intrinsic synthesis error in the circuit), but a faithful attempt to approximately implement each circuit's unitary is required.
    \item Run each compiled circuit many times, and for each circuit estimate the probability $h$ of a heavy output (this is estimated by simply computing the observed frequency of heavy outputs).
    \item Assess whether $\bar{h} > 2/3$ with 95\% confidence, where $\bar{h}$ is $h$ averaged over all sampled circuits of shape $(n, n)$. If $\bar{h} > 2/3$ with 95\% confidence, then the system passes the $n$-qubit QV test, and otherwise it fails.
\end{enumerate}
A system's QV is $2^{n}$, where $n+1$ is the smallest integer for which the system fails the QV test. For instance, if a 6-qubit QV test achieves $\bar{h} > 2/3$, but a 7-qubit QV test achieves $\bar{h} < 2/3$, the measured QV is $2^6 = 64$. 

\begin{figure}
    \centering
    \includegraphics[width=0.8\columnwidth]{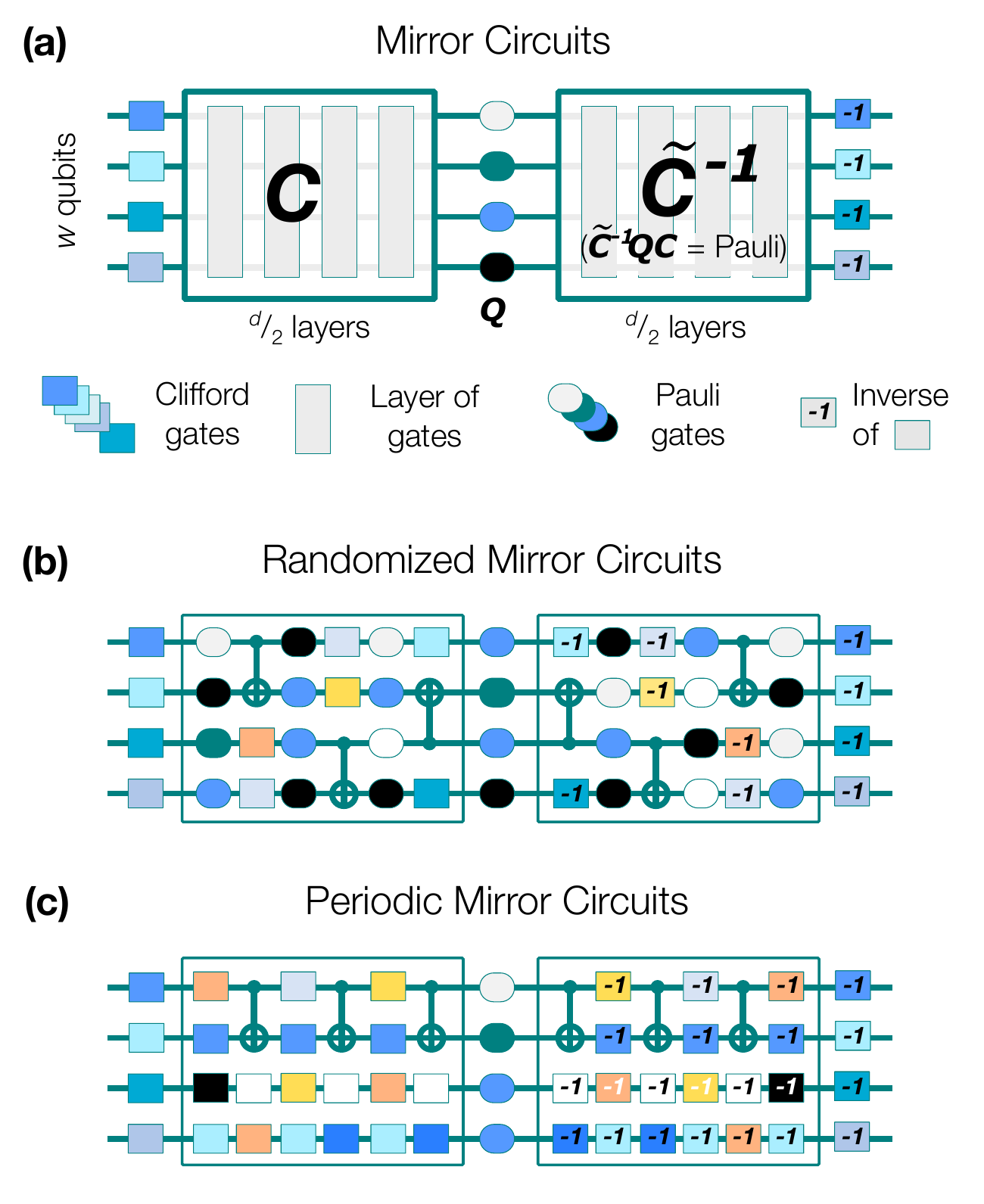}
    \caption{\textbf{Mirror circuit benchmarks}. 
    \textbf{(a)} Mirror circuits are a family of circuits with a motion reversal structure that are designed to enable scalable benchmarking. Two kinds of mirror circuit are \textbf{(b)} randomized mirror circuits and \textbf{(c)} periodic mirror circuits \cite{proctor2022measuring}.
    }
    \label{fig:mcbs}
\end{figure}

The QV threshold of $\bar{h} > 2/3$ is somewhat arbitrary, but it can be motivated as follows. In the absence of errors, deep and wide QV circuits have $h \approx (1 + \ln{2}) / 2 \approx 0.85$. In the presence of errors that completely depolarize all of the qubits by the end of a QV circuit --- i.e., the qubits are in the maximally mixed state by the end of the circuit --- then $h = 0.5$. The threshold value of $2/3$ is approximately half way between these two regimes, which corresponds to a probability of an error in the compiled QV circuits of approximately 50\%.

The QV benchmark favors quantum computers with high connectivity and gate set expressivity. For example, if one $n$-qubit system has linear connectivity and another has all-to-all connectivity, but they both have the same error rates on their one- and two-qubit gates, the all-to-all connectivity device will (typically) have a significantly higher QV. A benchmark that favors higher connectivity is reasonable, but it is not a universally good choice. Higher connectivity is likely broadly useful for NISQ algorithms, but is not useful under all circumstances (e.g., it is not needed to run quantum error correction using surface codes). Alternative versions of the QV benchmark with lower connectivity in the random circuits are possible \cite{Hines2023-be}.

\begin{figure}
    \centering
    \includegraphics[width=0.7\columnwidth]{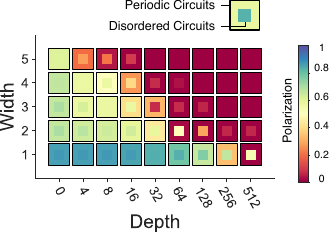}
    \caption{\textbf{Periodic and Random Circuit Benchmarks}. 
    A volumetric plot comparing a quantum computer's performance on disordered and highly structured circuits. This plot shows the polarization of the worst-performing circuit over 40 disordered and periodic circuits (randomized mirror circuits and periodic mirror circuits, respectively). Worse performance on periodic circuits (as seen here) is a signature of structured errors, such as coherent errors. Plot created using \texttt{ibmq\_london} data from Ref.~\cite{proctor2022measuring}.}
    \label{fig:mcb-volumetric-plot}
\end{figure}

The QV benchmark is not scalable, because estimating $h$ requires the simulation of the QV circuits to compute the heavy outputs, which is exponentially expensive. However, scalable adaptations of this benchmark have been proposed in Refs.~\cite{Amico2023-ze, Hines2023-be}.

%%%%%%%%%%%%%%%%%%%%%%% Mirror Circuit Benchmarks %%%%%%%%%%%%%%%%%%%%%%% 
\subsubsection{Mirror Circuit Benchmarks}\label{sec:mirror}

\emph{Mirror circuit benchmarks} \cite{proctor2022measuring} are a family of volumetric benchmarks based on mirror circuits. Mirror circuits (see \fig\ref{fig:mcbs}a) are a form of motion-reversal circuit that are constructed by (1) following a circuit ($\mathcal{C}$) with its layer-by-layer inverse ($\mathcal{C}^{-1}$), (2) adding in random Pauli gates between these two circuits, to prevent systematic error cancellation or addition between the two halves of the circuit, and (3) randomizing the state preparation and measurement basis of each qubit. Unlike the QV benchmark, mirror circuit benchmarks are \emph{not} full-stack benchmarks (although mirror circuit benchmarks can be adapted to full-stack benchmarking \cite{Hines2023-be}). This is because, like RB circuits, mirror circuits must not be arbitrarily compiled, as each mirror circuit's overall operation is simply bit flips on some of the qubits. Instead, mirror circuit benchmarks are designed to measure a system's ability to implement low-level circuits, complementing metrics that also incorporate compiler performance, like the QV. 

\begin{figure*}[!ht]
    \centering
    \begin{subfigure}[t]{0.47\textwidth}
        \includegraphics[width=0.9\textwidth]{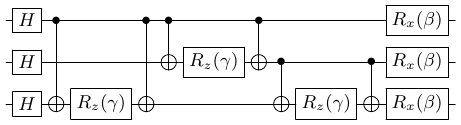}
        \caption{\texttt{Vanilla QAOA}}
    \end{subfigure}%
    \qquad
    \begin{subfigure}[t]{0.47\textwidth}
        \includegraphics[width=0.9\textwidth]{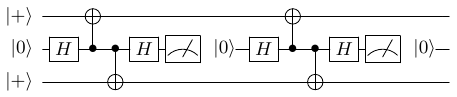}
        \caption{\texttt{Phase Code}}
    \end{subfigure}%
    \hfill
    \begin{subfigure}[t]{0.8\textwidth}
        \includegraphics[width=0.9\textwidth]{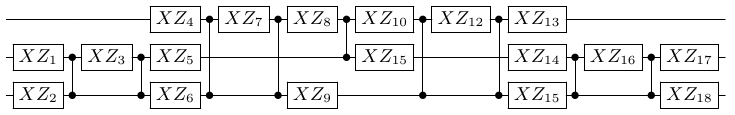}
        \caption{\texttt{QFT(1)}}
    \end{subfigure}
    \caption{\textbf{Example Application Benchmarks.} 
    \textbf{(a)} 3-qubit \texttt{Vanilla QAOA} benchmark from \texttt{SupermarQ}. In general, the $\gamma$ and $\beta$ parameters need to be optimized with respect to a specific objective function (e.g., a Hamiltonian ground state) in a variational quantum-classical loop. However, this benchmark can be designed to be measured for a single instance of $\gamma$ and $\beta$ to minimize errors due to drift (e.g., in cloud-based systems). 
    \textbf{(b)} 2-qubit \texttt{Phase Code} benchmark from \texttt{SupermarQ}. The phase code utilizes an ancilla qubit [middle] to detect phase flips on the data qubits [outer] using mid-circuit measurements. The ancilla qubit must be reset for each cycle of error detection.
    \textbf{(c)} 3-qubit \texttt{QFT(1)} benchmark from \texttt{QED-C} suite. The secret integer for this benchmark is $x=2$, and the $XZ_i$ gates are \texttt{PhasedXZ} gates where $XZ_i(a,x,z) = Z^z Z^a X^x Z^{-a}$.}
    \label{fig:app-benchmarks}
\end{figure*}

Mirror circuits can be used to construct scalable benchmarks from any set of circuits (see Sec.~\ref{sec:scalable_holistic_benchmarks}). Here, we discuss two specific mirror circuit benchmarks introduced in Ref.~\cite{proctor2022measuring}: \emph{randomized mirror circuit} and \emph{periodic mirror circuit} benchmarks, shown in \fig\ref{fig:mcbs}(b) and (c), respectively. Randomized mirror circuits are the same circuits that are used in \ac{MRB} (see Sec.~\ref{sec:mrb}) and \ac{ACES} (see Sec.~\ref{sec:aces}). Volumetric benchmarking with randomized mirror circuits is a scalable way to assess performance of a quantum computer on random, unstructured circuits. Like other random circuits (e.g., RB or \ac{XEB} circuits), these circuits scramble errors. Therefore, two different but equal-shape randomized mirror circuits typically have fairly similar performance, particularly as both circuit width and depth increases. In contrast, periodic mirror circuits are extremely ordered: they consist of repeating a short $n$-qubit ``germ'' circuit (that is randomly sampled from a distribution over possible short germ circuits). These circuits are not scrambling, but instead amplify particular errors --- with the particular errors that are amplified depending on the germ, as in long-sequence gate set tomography (see Sec.~\ref{sec:lsgst}). Therefore, in the presence of structured errors, such as coherent errors or biased stochastic Pauli errors, the variance in performance on periodic circuits will typically be much higher than with random circuits. Figure \ref{fig:mcb-volumetric-plot} shows how the performance of one system (\texttt{ibmq\_london}) differs on random and periodic mirror circuits, illustrating how these benchmarks can be used to reveal structured errors.

%%%%%%%%%%%%%%%%%%%%%%% Application Benchmarks %%%%%%%%%%%%%%%%%%%%%%% 
\subsection{Application Benchmarks}\label{sec:application_benchmarks}

Benchmarks based on algorithms and applications can be used to quantify the performance of quantum computing systems. Many different benchmarks fall under the category of ``application benchmarks,'' which encapsulates both high-level applications, such as solving a MaxCut problem or finding a ground state, variational algorithms such as the \ac{QAOA}, as well as key subroutines, such as the \ac{QFT} or quantum error correction. The primary purpose of application benchmarks is to measure the performance of a full-stack quantum computer for a specific use-case or algorithm. This contrasts with most characterization and other benchmarking protocols, which aim to measure specific properties (e.g., qubit coherence or gate fidelities) of low-level components.

Numerous application-oriented quantum benchmarking suites have been recently developed, including \texttt{SupermarQ} \cite{tomesh2022supermarq}, \texttt{QASMBench} \cite{li2022qasmbench}, and those developed by the \texttt{QED-C} \cite{lubinski2021application, lubinski2023optimization}. These various libraries are based on similar ideas, cover a range of application domains, and can be implemented on various quantum computing architectures (e.g., gate-based quantum computers, quantum annealers, etc.). In this subsection, we review several examples of application benchmarks from the \texttt{SupermarQ} and \texttt{QED-C} suites. However, because application benchmarking encompasses many diverse methodologies, the reader is encouraged to review others works for a broader perspective of the entire field of application benchmarks \cite{tomesh2022supermarq, li2022qasmbench, lubinski2021application, lubinski2023optimization, sawaya2023hamlib, miessen2024benchmarking}.

\begin{figure}[ht]
    \centering
    \includegraphics[width=\columnwidth]{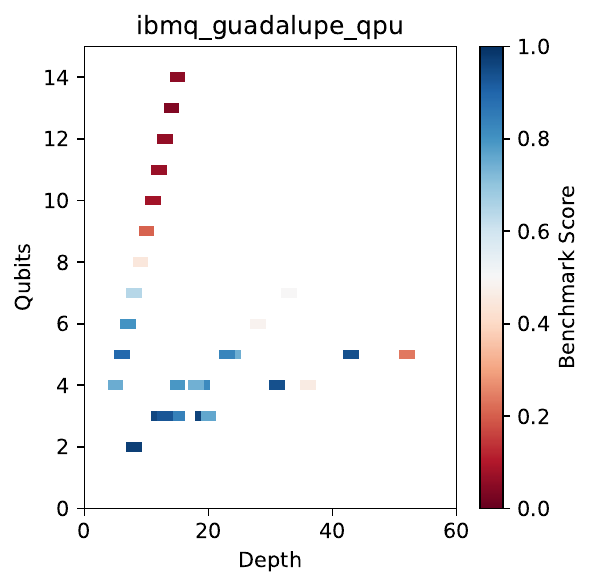}\label{fig:guadalupe_bmark}
    \caption{\textbf{Volumetric Benchmarking with Application Circuits.} 
    Evaluation of \texttt{SupermarQ} and \texttt{QED-C} benchmarks on a superconducting quantum computer (\texttt{ibmq\_quadalupe}) for a variety of different circuit widths and depths. 
    Each point represents the execution of a single application benchmark, where the color corresponds to the ``score'' achieved for that benchmark. Note that these benchmarks include the circuit compilation process in their evaluation (i.e., ``depth'' refers to the circuit depth of the uncompiled circuit). The exact expression used to obtain the score varies from benchmark to benchmark and can be found in the definitions of the benchmark suites \cite{tomesh2022supermarq, lubinski2021application}. Examples of a benchmark score include the measurement of an expectation value or classical fidelity. We observe that the performance falls off with circuit width; this could be because this system has limited connectivity, necessitating SWAP gates to implement non-local two-qubit gates.
    }
\label{fig:app_benchmark_performance}
\end{figure}

In \fig\ref{fig:app-benchmarks}, we show example circuits for three different benchmarks:  \texttt{SupermarQ}'s \texttt{Vanilla QAOA}, \texttt{SupermarQ}'s \texttt{Phase Code}, and the \texttt{QED-C} suite's \texttt{QFT(1)}. Both the \texttt{Vanilla QAOA} and \texttt{Phase Code} benchmarks are examples of \emph{proxy} applications which focus on a specific aspect of a larger, end-to-end application. For example, the \texttt{Vanilla QAOA} benchmark measures how well a quantum computer is able to execute a single instance of a variational circuit \cite{cerezo2021variational}, whereas full execution of the standard QAOA algorithm would involve iterating over a large number of circuit instances (and classical optimization). Similarly, the \texttt{Phase Code} benchmark tests a quantum computer's ability to execute circuits containing mid-circuit measurements --- which tests a component used in, e.g., syndrome extraction --- but does not use the results of these measurements to correct errors in the circuit. The \texttt{QFT(1)} benchmark, which contains both the QFT and its inverse, is an example of a key subroutine that appears in many quantum algorithms including Shor's algorithm \cite{shor1999polynomial} and the HHL algorithm \cite{harrow2009quantum}. Application benchmarks should specify the compiler optimizations that may be applied to the circuits prior to their execution.

Figure \ref{fig:app_benchmark_performance} shows the results of running application benchmarks from the \texttt{SupermarQ} and \texttt{QED-C} suites on \texttt{ibmq\_guadalupe}, represented as a volumetric plot. Each colored rectangle corresponds to an individual benchmark with the color indicating the score achieved. The benchmark score is a value ranging from 0 (poor performance) to 1 (best performance). The specific definition of the score function depends on the benchmark (see, e.g., the benchmark definitions given in \cite{tomesh2022supermarq} and \cite{lubinski2021application}), and typical examples include evaluating expectation values or computing the classical (Hellinger) fidelity (\eq\ref{eq:classical_fidelity}). Application benchmarks are now sometimes used to make high-level comparisons between different quantum computing systems \cite{murali2019full, li2022qasmbench, tomesh2022supermarq, lubinski2021application, lubinski2023optimization}.

\begin{figure*}[ht]
    \centering
    \subfloat[Quantum Program Profiles]{
    \includegraphics[width=0.3\textwidth]{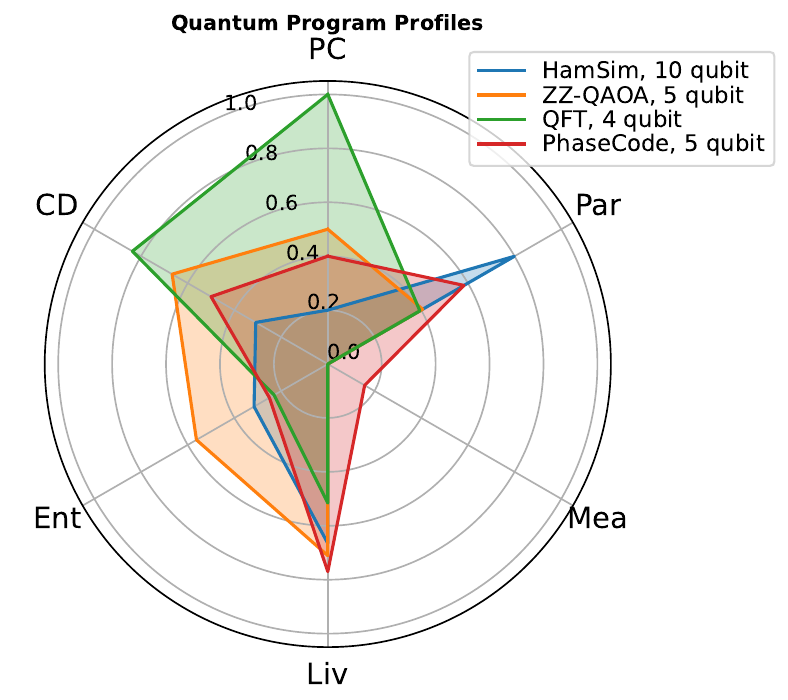}\label{fig:program_profiles}
        }
    % \qquad
    \subfloat[Correlation Heatmaps]{
    \includegraphics[width=0.7\textwidth]{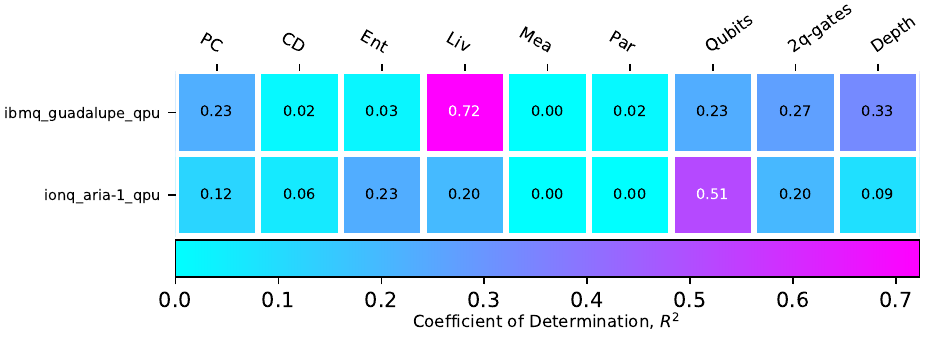}\label{fig:correlation_heatmap}
        }
    \caption{\textbf{Profiling Application Benchmarks}.
    \textbf{(a)} Quantum Program Profiles. Application benchmarks have many diverse properties that allow one to probe different properties of a quantum processor. We quantify the Program Connectivity (PC), Parallelism (Par), Measurement (Mea), Liveness (Liv), Entanglement-Ratio (Ent), and Critical Depth (CD) for the 10-qubit Hamiltonian Simulation (HamSim) benchmark, the 5-qubit ZZ-QAOA benchmark, the 4-qubit QFT circuit, and the 5-qubit Phase Code \cite{tomesh2022supermarq}
    \textbf{(b)} Correlation Heatmaps. After running the application benchmarks on a quantum computer, the observed performance can be correlated with the program features to produce the above heatmap. For example, the performance of the superconducting device, with relatively short coherence times, shows a high correlation (0.72) with the liveness feature of the benchmark circuits.}
    \label{fig:combined_profile_heatmap}
\end{figure*}

One useful aspect of application benchmarks is that their circuits typically have diverse properties, and so they may stress the quantum computer in diverse ways. This contrasts with benchmarks based on random circuits (e.g., RB protocols and the QV), which contain similar circuit structures. Initial efforts to profile quantum programs have underscored the distinct differences between applications originating from domains such as quantum chemistry and combinatorial optimization \cite{tomesh2022supermarq, li2022qasmbench}. Figure \ref{fig:combined_profile_heatmap} shows how the circuits of four different application benchmarks have significantly different properties \cite{tomesh2022supermarq}. It does so by plotting the values for 6 different features for each circuit: \emph{program connectivity} (PC), \emph{parallelism} (Par), \emph{measurement} (Mea), \emph{liveness} (Liv), \emph{entanglement-ratio} (Ent), and \emph{critical depth} (CD). 
The program connectivity of an $n$-qubit circuit is computed as $\sum_i^n \mathcal{D}(q_i) / (n^2 - n)$, where $\mathcal{D}(q_i)$ is the degree of qubit $q_i$ in the program's connectivity graph. This gives the program connectivity a range between zero, for programs without entangling gates, to one for a program with a complete connectivity graph.
The parallelism feature relates the total number of gates ($N_G$) and circuit depth ($D$) within the expression $(N_G - D) / (nD - D)$ to capture the amount of parallelism available within a quantum program. The parallel execution of gates often exhibit correlated crosstalk which degrades circuit performance.
The measurement feature is given by $L_{mcm} / D$ for a circuit with $L_{mcm}$ layers containing at least one mid-circuit measurement operation, during which the idling spectator qubits can dephase.
The liveness feature also considers the idling time of qubits. It is computed as $(\sum_{ij}A_{ij}) / (nD)$, where $A$ is a binary $(n \times D)$ matrix with entry $A_{ij} = 1$ if qubit $i$ is acted on by a gate at time step $j$, otherwise the entry is zero.
Entanglement-ratio is given by the fraction of entangling gates divided by the total gate count, and it can provide insights into program behavior if the specific hardware running the benchmark has large differences in one- and two-qubit gate error rates.
Finally, the critical-depth feature is computed by counting the number of entangling gates along the program's critical path and dividing by the total number of entangling gates in the circuit.

Each feature is meant to capture some salient aspect of a quantum program. These features, combined with the benchmark results in \fig\ref{fig:app_benchmark_performance}, can be used to correlate system performance with program profiles (see \fig\ref{fig:combined_profile_heatmap}). Each square in \fig\ref{fig:combined_profile_heatmap} corresponds to the coefficient of determination ($R^2$) for a particular pair of device and program feature. In other words, it shows the proportion of the variation in that device's performance, across all of the evaluated application benchmarks, which can be explained by that particular program feature. Each $R^2$ value is obtained by performing a linear regression over all of the device's scores on the benchmarks (dependent variable) and the specific values of the particular program feature for each benchmark (independent variable).

%%%%%%%%%%%%%%%%%%%%%%% Creating scalable volumetric benchmarks %%%%%%%%%%%%%%%%%%%%%%% 
\subsection{Scalable Holistic Benchmarks}\label{sec:scalable_holistic_benchmarks}

Benchmarking a quantum computer's performance on large circuits or applications is an inherently difficult task because it is not typically feasible to compute what the correct outcome should be using simulations on a classical computer. Therefore, many existing holistic benchmarks either do not scale beyond around 50 qubits (examples include the QV benchmark and XEB used for demonstrations of ``quantum supremacy'') or they only use circuits from some restricted circuit class that can be efficiently simulated classically (e.g., Clifford circuits, as in many RB methods). In the context of application benchmarks, the applications or circuits used are often designed to circumvent this ``verification'' problem, by ensuring that a quantum computer's performance on the benchmark can be quantified without inefficient circuit simulations \cite{tomesh2022supermarq}. Often this is achieved with algorithm-specific methods, e.g., by creating a benchmark that tests a quantum computer's ability to solve the one-dimensional transverse field Ising model \cite{pfeuty1970one}, or to prepare easy-to-verify states such as a GHZ state \cite{greenberger1989going}, or to measure operators with a known upper bound (as is done in the Mermin-Bell benchmark \cite{tomesh2022supermarq}). 

A complementary approach to creating scalable holistic benchmarks is to (1) choose a benchmark's circuits without addressing the efficiency problem, and then (2) measuring a quantum computer's performance on those circuits \emph{indirectly}. This can be achieved with any technique that can efficiently estimate a quantum computer's circuit execution fidelity (or some other interesting metric of circuit performance) for an arbitrary circuit. We discussed two such techniques --- mirror circuit fidelity estimation and circuit output accreditation --- in Sec.~\ref{sec:fidelity_estimation}. This can even enable scalable full-stack benchmarks, as discussed in Ref.~\cite{Hines2023-be}.
\section*{Acknowledgements}\label{sec:acknowledgements}
This work was supported by the U.S.~Department of Energy, Office of Science, Office of Advanced Scientific Computing Research Quantum Testbed Program under Contract No.~DE-AC02-05CH11231 and DE-SC0021526, as well as the Quantum Testbed Pathfinder Program. A.H.~acknowledges financial support from the Berkeley Initiative for Computational Transformation Fellows Program. T.P.~acknowledges support from an Office of Advanced Scientific Computing Research Early Career Award.

L.J.~and S.C.~acknowledge support from the ARO (W911NF-23-1-0077), ARO MURI (W911NF-21-1-0325), AFOSR MURI (FA9550-19-1-0399, FA9550-21-1-0209, FA9550-23-1-0338), DARPA (HR0011-24-9-0359, HR0011-24-9-0361), NSF (OMA-1936118, ERC-1941583, OMA-2137642, OSI-2326767, CCF-2312755), NTT Research, Packard Foundation (2020-71479).

A.H.~acknowledges fruitful discussions with Joel J.~Wallman, Joseph Emerson, Ian Hincks, and Arnaud Carignan-Dugas.

We acknowledge useful feedback from Juan Jesus Gonzalez De Mendoza Prada and Robin Harper.

Sandia National Laboratories is a multi-mission laboratory managed and operated by National Technology \& Engineering Solutions of Sandia, LLC (NTESS), a wholly owned subsidiary of Honeywell International Inc., for the U.S. Department of Energy’s National Nuclear Security Administration (DOE/NNSA) under contract DE-NA0003525. This written work is authored by an employee of NTESS. The employee, not NTESS, owns the right, title and interest in and to the written work and is responsible for its contents. Any subjective views or opinions that might be expressed in the written work do not necessarily represent the views of the U.S. Government. The publisher acknowledges that the U.S. Government retains a non-exclusive, paid-up, irrevocable, world-wide license to publish or reproduce the published form of this written work or allow others to do so, for U.S. Government purposes. The DOE will provide public access to results of federally sponsored research in accordance with the DOE Public Access Plan.

\section*{Author Contributions}\label{sec:author_contributions}
All authors contributed to the writing of the manuscript.

\section*{Competing Interests}\label{sec:competing_interests}
All authors declare no competing interests.

\section*{Data Availability}\label{sec:data_availability}
All data are available from the corresponding author upon reasonable request.

\bibliography{bibliography}

\begin{thebibliography}{460}%
\makeatletter
\providecommand \@ifxundefined [1]{%
 \@ifx{#1\undefined}
}%
\providecommand \@ifnum [1]{%
 \ifnum #1\expandafter \@firstoftwo
 \else \expandafter \@secondoftwo
 \fi
}%
\providecommand \@ifx [1]{%
 \ifx #1\expandafter \@firstoftwo
 \else \expandafter \@secondoftwo
 \fi
}%
\providecommand \natexlab [1]{#1}%
\providecommand \enquote  [1]{``#1''}%
\providecommand \bibnamefont  [1]{#1}%
\providecommand \bibfnamefont [1]{#1}%
\providecommand \citenamefont [1]{#1}%
\providecommand \href@noop [0]{\@secondoftwo}%
\providecommand \href [0]{\begingroup \@sanitize@url \@href}%
\providecommand \@href[1]{\@@startlink{#1}\@@href}%
\providecommand \@@href[1]{\endgroup#1\@@endlink}%
\providecommand \@sanitize@url [0]{\catcode `\\12\catcode `\$12\catcode
  `\&12\catcode `\#12\catcode `\^12\catcode `\_12\catcode `\%12\relax}%
\providecommand \@@startlink[1]{}%
\providecommand \@@endlink[0]{}%
\providecommand \url  [0]{\begingroup\@sanitize@url \@url }%
\providecommand \@url [1]{\endgroup\@href {#1}{\urlprefix }}%
\providecommand \urlprefix  [0]{URL }%
\providecommand \Eprint [0]{\href }%
\providecommand \doibase [0]{http://dx.doi.org/}%
\providecommand \selectlanguage [0]{\@gobble}%
\providecommand \bibinfo  [0]{\@secondoftwo}%
\providecommand \bibfield  [0]{\@secondoftwo}%
\providecommand \translation [1]{[#1]}%
\providecommand \BibitemOpen [0]{}%
\providecommand \bibitemStop [0]{}%
\providecommand \bibitemNoStop [0]{.\EOS\space}%
\providecommand \EOS [0]{\spacefactor3000\relax}%
\providecommand \BibitemShut  [1]{\csname bibitem#1\endcsname}%
\let\auto@bib@innerbib\@empty
%</preamble>
\bibitem [{\citenamefont {Shor}(1994)}]{shor1994algorithms}%
  \BibitemOpen
  \bibfield  {author} {\bibinfo {author} {\bibfnamefont {P.~W.}\ \bibnamefont
  {Shor}},\ }in\ \href@noop {} {\emph {\bibinfo {booktitle} {Proceedings 35th
  annual symposium on foundations of computer science}}}\ (\bibinfo
  {organization} {Ieee},\ \bibinfo {year} {1994})\ pp.\ \bibinfo {pages}
  {124--134}\BibitemShut {NoStop}%
\bibitem [{\citenamefont {Arute}\ \emph {et~al.}(2019)\citenamefont {Arute},
  \citenamefont {Arya}, \citenamefont {Babbush}, \citenamefont {Bacon},
  \citenamefont {Bardin}, \citenamefont {Barends}, \citenamefont {Biswas},
  \citenamefont {Boixo}, \citenamefont {Brandao}, \citenamefont {Buell},
  \citenamefont {Burkett}, \citenamefont {Chen}, \citenamefont {Chen},
  \citenamefont {Chiaro}, \citenamefont {Collins}, \citenamefont {Courtney},
  \citenamefont {Dunsworth}, \citenamefont {Farhi}, \citenamefont {Foxen},
  \citenamefont {Fowler}, \citenamefont {Gidney}, \citenamefont {Giustina},
  \citenamefont {Graff}, \citenamefont {Guerin}, \citenamefont {Habegger},
  \citenamefont {Harrigan}, \citenamefont {Hartmann}, \citenamefont {Ho},
  \citenamefont {Hoffmann}, \citenamefont {Huang}, \citenamefont {Humble},
  \citenamefont {Isakov}, \citenamefont {Jeffrey}, \citenamefont {Jiang},
  \citenamefont {Kafri}, \citenamefont {Kechedzhi}, \citenamefont {Kelly},
  \citenamefont {Klimov}, \citenamefont {Knysh}, \citenamefont {Korotkov},
  \citenamefont {Kostritsa}, \citenamefont {Landhuis}, \citenamefont
  {Lindmark}, \citenamefont {Lucero}, \citenamefont {Lyakh}, \citenamefont
  {Mandrà}, \citenamefont {McClean}, \citenamefont {McEwen}, \citenamefont
  {Megrant}, \citenamefont {Mi}, \citenamefont {Michielsen}, \citenamefont
  {Mohseni}, \citenamefont {Mutus}, \citenamefont {Naaman}, \citenamefont
  {Neeley}, \citenamefont {Neill}, \citenamefont {Niu}, \citenamefont {Ostby},
  \citenamefont {Petukhov}, \citenamefont {Platt}, \citenamefont {Quintana},
  \citenamefont {Rieffel}, \citenamefont {Roushan}, \citenamefont {Rubin},
  \citenamefont {Sank}, \citenamefont {Satzinger}, \citenamefont {Smelyanskiy},
  \citenamefont {Sung}, \citenamefont {Trevithick}, \citenamefont
  {Vainsencher}, \citenamefont {Villalonga}, \citenamefont {White},
  \citenamefont {Yao}, \citenamefont {Yeh}, \citenamefont {Zalcman},
  \citenamefont {Neven},\ and\ \citenamefont {Martinis}}]{2019GoogleSupremacy}%
  \BibitemOpen
  \bibfield  {author} {\bibinfo {author} {\bibfnamefont {F.}~\bibnamefont
  {Arute}}, \bibinfo {author} {\bibfnamefont {K.}~\bibnamefont {Arya}},
  \bibinfo {author} {\bibfnamefont {R.}~\bibnamefont {Babbush}}, \bibinfo
  {author} {\bibfnamefont {D.}~\bibnamefont {Bacon}}, \bibinfo {author}
  {\bibfnamefont {J.}~\bibnamefont {Bardin}}, \bibinfo {author} {\bibfnamefont
  {R.}~\bibnamefont {Barends}}, \bibinfo {author} {\bibfnamefont
  {R.}~\bibnamefont {Biswas}}, \bibinfo {author} {\bibfnamefont
  {S.}~\bibnamefont {Boixo}}, \bibinfo {author} {\bibfnamefont
  {F.}~\bibnamefont {Brandao}}, \bibinfo {author} {\bibfnamefont
  {D.}~\bibnamefont {Buell}}, \bibinfo {author} {\bibfnamefont
  {B.}~\bibnamefont {Burkett}}, \bibinfo {author} {\bibfnamefont
  {Y.}~\bibnamefont {Chen}}, \bibinfo {author} {\bibfnamefont {J.}~\bibnamefont
  {Chen}}, \bibinfo {author} {\bibfnamefont {B.}~\bibnamefont {Chiaro}},
  \bibinfo {author} {\bibfnamefont {R.}~\bibnamefont {Collins}}, \bibinfo
  {author} {\bibfnamefont {W.}~\bibnamefont {Courtney}}, \bibinfo {author}
  {\bibfnamefont {A.}~\bibnamefont {Dunsworth}}, \bibinfo {author}
  {\bibfnamefont {E.}~\bibnamefont {Farhi}}, \bibinfo {author} {\bibfnamefont
  {B.}~\bibnamefont {Foxen}}, \bibinfo {author} {\bibfnamefont
  {A.}~\bibnamefont {Fowler}}, \bibinfo {author} {\bibfnamefont {C.~M.}\
  \bibnamefont {Gidney}}, \bibinfo {author} {\bibfnamefont {M.}~\bibnamefont
  {Giustina}}, \bibinfo {author} {\bibfnamefont {R.}~\bibnamefont {Graff}},
  \bibinfo {author} {\bibfnamefont {K.}~\bibnamefont {Guerin}}, \bibinfo
  {author} {\bibfnamefont {S.}~\bibnamefont {Habegger}}, \bibinfo {author}
  {\bibfnamefont {M.}~\bibnamefont {Harrigan}}, \bibinfo {author}
  {\bibfnamefont {M.}~\bibnamefont {Hartmann}}, \bibinfo {author}
  {\bibfnamefont {A.}~\bibnamefont {Ho}}, \bibinfo {author} {\bibfnamefont
  {M.~R.}\ \bibnamefont {Hoffmann}}, \bibinfo {author} {\bibfnamefont
  {T.}~\bibnamefont {Huang}}, \bibinfo {author} {\bibfnamefont
  {T.}~\bibnamefont {Humble}}, \bibinfo {author} {\bibfnamefont
  {S.}~\bibnamefont {Isakov}}, \bibinfo {author} {\bibfnamefont
  {E.}~\bibnamefont {Jeffrey}}, \bibinfo {author} {\bibfnamefont
  {Z.}~\bibnamefont {Jiang}}, \bibinfo {author} {\bibfnamefont
  {D.}~\bibnamefont {Kafri}}, \bibinfo {author} {\bibfnamefont
  {K.}~\bibnamefont {Kechedzhi}}, \bibinfo {author} {\bibfnamefont
  {J.}~\bibnamefont {Kelly}}, \bibinfo {author} {\bibfnamefont
  {P.}~\bibnamefont {Klimov}}, \bibinfo {author} {\bibfnamefont
  {S.}~\bibnamefont {Knysh}}, \bibinfo {author} {\bibfnamefont
  {A.}~\bibnamefont {Korotkov}}, \bibinfo {author} {\bibfnamefont
  {F.}~\bibnamefont {Kostritsa}}, \bibinfo {author} {\bibfnamefont
  {D.}~\bibnamefont {Landhuis}}, \bibinfo {author} {\bibfnamefont
  {M.}~\bibnamefont {Lindmark}}, \bibinfo {author} {\bibfnamefont
  {E.}~\bibnamefont {Lucero}}, \bibinfo {author} {\bibfnamefont
  {D.}~\bibnamefont {Lyakh}}, \bibinfo {author} {\bibfnamefont
  {S.}~\bibnamefont {Mandrà}}, \bibinfo {author} {\bibfnamefont {J.~R.}\
  \bibnamefont {McClean}}, \bibinfo {author} {\bibfnamefont {M.}~\bibnamefont
  {McEwen}}, \bibinfo {author} {\bibfnamefont {A.}~\bibnamefont {Megrant}},
  \bibinfo {author} {\bibfnamefont {X.}~\bibnamefont {Mi}}, \bibinfo {author}
  {\bibfnamefont {K.}~\bibnamefont {Michielsen}}, \bibinfo {author}
  {\bibfnamefont {M.}~\bibnamefont {Mohseni}}, \bibinfo {author} {\bibfnamefont
  {J.}~\bibnamefont {Mutus}}, \bibinfo {author} {\bibfnamefont
  {O.}~\bibnamefont {Naaman}}, \bibinfo {author} {\bibfnamefont
  {M.}~\bibnamefont {Neeley}}, \bibinfo {author} {\bibfnamefont
  {C.}~\bibnamefont {Neill}}, \bibinfo {author} {\bibfnamefont {M.~Y.}\
  \bibnamefont {Niu}}, \bibinfo {author} {\bibfnamefont {E.}~\bibnamefont
  {Ostby}}, \bibinfo {author} {\bibfnamefont {A.}~\bibnamefont {Petukhov}},
  \bibinfo {author} {\bibfnamefont {J.}~\bibnamefont {Platt}}, \bibinfo
  {author} {\bibfnamefont {C.}~\bibnamefont {Quintana}}, \bibinfo {author}
  {\bibfnamefont {E.~G.}\ \bibnamefont {Rieffel}}, \bibinfo {author}
  {\bibfnamefont {P.}~\bibnamefont {Roushan}}, \bibinfo {author} {\bibfnamefont
  {N.}~\bibnamefont {Rubin}}, \bibinfo {author} {\bibfnamefont
  {D.}~\bibnamefont {Sank}}, \bibinfo {author} {\bibfnamefont {K.~J.}\
  \bibnamefont {Satzinger}}, \bibinfo {author} {\bibfnamefont {V.}~\bibnamefont
  {Smelyanskiy}}, \bibinfo {author} {\bibfnamefont {K.~J.}\ \bibnamefont
  {Sung}}, \bibinfo {author} {\bibfnamefont {M.}~\bibnamefont {Trevithick}},
  \bibinfo {author} {\bibfnamefont {A.}~\bibnamefont {Vainsencher}}, \bibinfo
  {author} {\bibfnamefont {B.}~\bibnamefont {Villalonga}}, \bibinfo {author}
  {\bibfnamefont {T.}~\bibnamefont {White}}, \bibinfo {author} {\bibfnamefont
  {Z.~J.}\ \bibnamefont {Yao}}, \bibinfo {author} {\bibfnamefont
  {P.}~\bibnamefont {Yeh}}, \bibinfo {author} {\bibfnamefont {A.}~\bibnamefont
  {Zalcman}}, \bibinfo {author} {\bibfnamefont {H.}~\bibnamefont {Neven}}, \
  and\ \bibinfo {author} {\bibfnamefont {J.}~\bibnamefont {Martinis}},\ }\href
  {https://www.nature.com/articles/s41586-019-1666-5} {\bibfield  {journal}
  {\bibinfo  {journal} {Nature}\ }\textbf {\bibinfo {volume} {574}},\ \bibinfo
  {pages} {505–510} (\bibinfo {year} {2019})}\BibitemShut {NoStop}%
\bibitem [{\citenamefont {Wu}\ \emph {et~al.}(2021)\citenamefont {Wu},
  \citenamefont {Bao}, \citenamefont {Cao}, \citenamefont {Chen}, \citenamefont
  {Chen}, \citenamefont {Chen}, \citenamefont {Chung}, \citenamefont {Deng},
  \citenamefont {Du}, \citenamefont {Fan}, \citenamefont {Gong}, \citenamefont
  {Guo}, \citenamefont {Guo}, \citenamefont {Guo}, \citenamefont {Han},
  \citenamefont {Hong}, \citenamefont {Huang}, \citenamefont {Huo},
  \citenamefont {Li}, \citenamefont {Li}, \citenamefont {Li}, \citenamefont
  {Li}, \citenamefont {Liang}, \citenamefont {Lin}, \citenamefont {Lin},
  \citenamefont {Qian}, \citenamefont {Qiao}, \citenamefont {Rong},
  \citenamefont {Su}, \citenamefont {Sun}, \citenamefont {Wang}, \citenamefont
  {Wang}, \citenamefont {Wu}, \citenamefont {Xu}, \citenamefont {Yan},
  \citenamefont {Yang}, \citenamefont {Yang}, \citenamefont {Ye}, \citenamefont
  {Yin}, \citenamefont {Ying}, \citenamefont {Yu}, \citenamefont {Zha},
  \citenamefont {Zhang}, \citenamefont {Zhang}, \citenamefont {Zhang},
  \citenamefont {Zhang}, \citenamefont {Zhao}, \citenamefont {Zhao},
  \citenamefont {Zhou}, \citenamefont {Zhu}, \citenamefont {Lu}, \citenamefont
  {Peng}, \citenamefont {Zhu},\ and\ \citenamefont {Pan}}]{2021USTSupremacy}%
  \BibitemOpen
  \bibfield  {author} {\bibinfo {author} {\bibfnamefont {Y.}~\bibnamefont
  {Wu}}, \bibinfo {author} {\bibfnamefont {W.-S.}\ \bibnamefont {Bao}},
  \bibinfo {author} {\bibfnamefont {S.}~\bibnamefont {Cao}}, \bibinfo {author}
  {\bibfnamefont {F.}~\bibnamefont {Chen}}, \bibinfo {author} {\bibfnamefont
  {M.-C.}\ \bibnamefont {Chen}}, \bibinfo {author} {\bibfnamefont
  {X.}~\bibnamefont {Chen}}, \bibinfo {author} {\bibfnamefont {T.-H.}\
  \bibnamefont {Chung}}, \bibinfo {author} {\bibfnamefont {H.}~\bibnamefont
  {Deng}}, \bibinfo {author} {\bibfnamefont {Y.}~\bibnamefont {Du}}, \bibinfo
  {author} {\bibfnamefont {D.}~\bibnamefont {Fan}}, \bibinfo {author}
  {\bibfnamefont {M.}~\bibnamefont {Gong}}, \bibinfo {author} {\bibfnamefont
  {C.}~\bibnamefont {Guo}}, \bibinfo {author} {\bibfnamefont {C.}~\bibnamefont
  {Guo}}, \bibinfo {author} {\bibfnamefont {S.}~\bibnamefont {Guo}}, \bibinfo
  {author} {\bibfnamefont {L.}~\bibnamefont {Han}}, \bibinfo {author}
  {\bibfnamefont {L.}~\bibnamefont {Hong}}, \bibinfo {author} {\bibfnamefont
  {H.-L.}\ \bibnamefont {Huang}}, \bibinfo {author} {\bibfnamefont {Y.-H.}\
  \bibnamefont {Huo}}, \bibinfo {author} {\bibfnamefont {L.}~\bibnamefont
  {Li}}, \bibinfo {author} {\bibfnamefont {N.}~\bibnamefont {Li}}, \bibinfo
  {author} {\bibfnamefont {S.}~\bibnamefont {Li}}, \bibinfo {author}
  {\bibfnamefont {Y.}~\bibnamefont {Li}}, \bibinfo {author} {\bibfnamefont
  {F.}~\bibnamefont {Liang}}, \bibinfo {author} {\bibfnamefont
  {C.}~\bibnamefont {Lin}}, \bibinfo {author} {\bibfnamefont {J.}~\bibnamefont
  {Lin}}, \bibinfo {author} {\bibfnamefont {H.}~\bibnamefont {Qian}}, \bibinfo
  {author} {\bibfnamefont {D.}~\bibnamefont {Qiao}}, \bibinfo {author}
  {\bibfnamefont {H.}~\bibnamefont {Rong}}, \bibinfo {author} {\bibfnamefont
  {H.}~\bibnamefont {Su}}, \bibinfo {author} {\bibfnamefont {L.}~\bibnamefont
  {Sun}}, \bibinfo {author} {\bibfnamefont {L.}~\bibnamefont {Wang}}, \bibinfo
  {author} {\bibfnamefont {S.}~\bibnamefont {Wang}}, \bibinfo {author}
  {\bibfnamefont {D.}~\bibnamefont {Wu}}, \bibinfo {author} {\bibfnamefont
  {Y.}~\bibnamefont {Xu}}, \bibinfo {author} {\bibfnamefont {K.}~\bibnamefont
  {Yan}}, \bibinfo {author} {\bibfnamefont {W.}~\bibnamefont {Yang}}, \bibinfo
  {author} {\bibfnamefont {Y.}~\bibnamefont {Yang}}, \bibinfo {author}
  {\bibfnamefont {Y.}~\bibnamefont {Ye}}, \bibinfo {author} {\bibfnamefont
  {J.}~\bibnamefont {Yin}}, \bibinfo {author} {\bibfnamefont {C.}~\bibnamefont
  {Ying}}, \bibinfo {author} {\bibfnamefont {J.}~\bibnamefont {Yu}}, \bibinfo
  {author} {\bibfnamefont {C.}~\bibnamefont {Zha}}, \bibinfo {author}
  {\bibfnamefont {C.}~\bibnamefont {Zhang}}, \bibinfo {author} {\bibfnamefont
  {H.}~\bibnamefont {Zhang}}, \bibinfo {author} {\bibfnamefont
  {K.}~\bibnamefont {Zhang}}, \bibinfo {author} {\bibfnamefont
  {Y.}~\bibnamefont {Zhang}}, \bibinfo {author} {\bibfnamefont
  {H.}~\bibnamefont {Zhao}}, \bibinfo {author} {\bibfnamefont {Y.}~\bibnamefont
  {Zhao}}, \bibinfo {author} {\bibfnamefont {L.}~\bibnamefont {Zhou}}, \bibinfo
  {author} {\bibfnamefont {Q.}~\bibnamefont {Zhu}}, \bibinfo {author}
  {\bibfnamefont {C.-Y.}\ \bibnamefont {Lu}}, \bibinfo {author} {\bibfnamefont
  {C.-Z.}\ \bibnamefont {Peng}}, \bibinfo {author} {\bibfnamefont
  {X.}~\bibnamefont {Zhu}}, \ and\ \bibinfo {author} {\bibfnamefont {J.-W.}\
  \bibnamefont {Pan}},\ }\href {\doibase 10.1103/PhysRevLett.127.180501}
  {\bibfield  {journal} {\bibinfo  {journal} {Phys. Rev. Lett.}\ }\textbf
  {\bibinfo {volume} {127}},\ \bibinfo {pages} {180501} (\bibinfo {year}
  {2021})}\BibitemShut {NoStop}%
\bibitem [{\citenamefont {Zhu}\ \emph {et~al.}(2022)\citenamefont {Zhu},
  \citenamefont {Cao}, \citenamefont {Chen}, \citenamefont {Chen},
  \citenamefont {Chen}, \citenamefont {Chung}, \citenamefont {Deng},
  \citenamefont {Du}, \citenamefont {Fan}, \citenamefont {Gong}, \citenamefont
  {Guo}, \citenamefont {Guo}, \citenamefont {Guo}, \citenamefont {Han},
  \citenamefont {Hong}, \citenamefont {Huang}, \citenamefont {Huo},
  \citenamefont {Li}, \citenamefont {Li}, \citenamefont {Li}, \citenamefont
  {Li}, \citenamefont {Liang}, \citenamefont {Lin}, \citenamefont {Lin},
  \citenamefont {Qian}, \citenamefont {Qiao}, \citenamefont {Rong},
  \citenamefont {Su}, \citenamefont {Sun}, \citenamefont {Wang}, \citenamefont
  {Wang}, \citenamefont {Wu}, \citenamefont {Wu}, \citenamefont {Xu},
  \citenamefont {Yan}, \citenamefont {Yang}, \citenamefont {Yang},
  \citenamefont {Ye}, \citenamefont {Yin}, \citenamefont {Ying}, \citenamefont
  {Yu}, \citenamefont {Zha}, \citenamefont {Zhang}, \citenamefont {Zhang},
  \citenamefont {Zhang}, \citenamefont {Zhang}, \citenamefont {Zhao},
  \citenamefont {Zhao}, \citenamefont {Zhou}, \citenamefont {Lu}, \citenamefont
  {Peng}, \citenamefont {Zhu},\ and\ \citenamefont {Pan}}]{2022USTSupremacy}%
  \BibitemOpen
  \bibfield  {author} {\bibinfo {author} {\bibfnamefont {Q.}~\bibnamefont
  {Zhu}}, \bibinfo {author} {\bibfnamefont {S.}~\bibnamefont {Cao}}, \bibinfo
  {author} {\bibfnamefont {F.}~\bibnamefont {Chen}}, \bibinfo {author}
  {\bibfnamefont {M.-C.}\ \bibnamefont {Chen}}, \bibinfo {author}
  {\bibfnamefont {X.}~\bibnamefont {Chen}}, \bibinfo {author} {\bibfnamefont
  {T.-H.}\ \bibnamefont {Chung}}, \bibinfo {author} {\bibfnamefont
  {H.}~\bibnamefont {Deng}}, \bibinfo {author} {\bibfnamefont {Y.}~\bibnamefont
  {Du}}, \bibinfo {author} {\bibfnamefont {D.}~\bibnamefont {Fan}}, \bibinfo
  {author} {\bibfnamefont {M.}~\bibnamefont {Gong}}, \bibinfo {author}
  {\bibfnamefont {C.}~\bibnamefont {Guo}}, \bibinfo {author} {\bibfnamefont
  {C.}~\bibnamefont {Guo}}, \bibinfo {author} {\bibfnamefont {S.}~\bibnamefont
  {Guo}}, \bibinfo {author} {\bibfnamefont {L.}~\bibnamefont {Han}}, \bibinfo
  {author} {\bibfnamefont {L.}~\bibnamefont {Hong}}, \bibinfo {author}
  {\bibfnamefont {H.-L.}\ \bibnamefont {Huang}}, \bibinfo {author}
  {\bibfnamefont {Y.-H.}\ \bibnamefont {Huo}}, \bibinfo {author} {\bibfnamefont
  {L.}~\bibnamefont {Li}}, \bibinfo {author} {\bibfnamefont {N.}~\bibnamefont
  {Li}}, \bibinfo {author} {\bibfnamefont {S.}~\bibnamefont {Li}}, \bibinfo
  {author} {\bibfnamefont {Y.}~\bibnamefont {Li}}, \bibinfo {author}
  {\bibfnamefont {F.}~\bibnamefont {Liang}}, \bibinfo {author} {\bibfnamefont
  {C.}~\bibnamefont {Lin}}, \bibinfo {author} {\bibfnamefont {J.}~\bibnamefont
  {Lin}}, \bibinfo {author} {\bibfnamefont {H.}~\bibnamefont {Qian}}, \bibinfo
  {author} {\bibfnamefont {D.}~\bibnamefont {Qiao}}, \bibinfo {author}
  {\bibfnamefont {H.}~\bibnamefont {Rong}}, \bibinfo {author} {\bibfnamefont
  {H.}~\bibnamefont {Su}}, \bibinfo {author} {\bibfnamefont {L.}~\bibnamefont
  {Sun}}, \bibinfo {author} {\bibfnamefont {L.}~\bibnamefont {Wang}}, \bibinfo
  {author} {\bibfnamefont {S.}~\bibnamefont {Wang}}, \bibinfo {author}
  {\bibfnamefont {D.}~\bibnamefont {Wu}}, \bibinfo {author} {\bibfnamefont
  {Y.}~\bibnamefont {Wu}}, \bibinfo {author} {\bibfnamefont {Y.}~\bibnamefont
  {Xu}}, \bibinfo {author} {\bibfnamefont {K.}~\bibnamefont {Yan}}, \bibinfo
  {author} {\bibfnamefont {W.}~\bibnamefont {Yang}}, \bibinfo {author}
  {\bibfnamefont {Y.}~\bibnamefont {Yang}}, \bibinfo {author} {\bibfnamefont
  {Y.}~\bibnamefont {Ye}}, \bibinfo {author} {\bibfnamefont {J.}~\bibnamefont
  {Yin}}, \bibinfo {author} {\bibfnamefont {C.}~\bibnamefont {Ying}}, \bibinfo
  {author} {\bibfnamefont {J.}~\bibnamefont {Yu}}, \bibinfo {author}
  {\bibfnamefont {C.}~\bibnamefont {Zha}}, \bibinfo {author} {\bibfnamefont
  {C.}~\bibnamefont {Zhang}}, \bibinfo {author} {\bibfnamefont
  {H.}~\bibnamefont {Zhang}}, \bibinfo {author} {\bibfnamefont
  {K.}~\bibnamefont {Zhang}}, \bibinfo {author} {\bibfnamefont
  {Y.}~\bibnamefont {Zhang}}, \bibinfo {author} {\bibfnamefont
  {H.}~\bibnamefont {Zhao}}, \bibinfo {author} {\bibfnamefont {Y.}~\bibnamefont
  {Zhao}}, \bibinfo {author} {\bibfnamefont {L.}~\bibnamefont {Zhou}}, \bibinfo
  {author} {\bibfnamefont {C.-Y.}\ \bibnamefont {Lu}}, \bibinfo {author}
  {\bibfnamefont {C.-Z.}\ \bibnamefont {Peng}}, \bibinfo {author}
  {\bibfnamefont {X.}~\bibnamefont {Zhu}}, \ and\ \bibinfo {author}
  {\bibfnamefont {J.-W.}\ \bibnamefont {Pan}},\ }\href {\doibase
  https://doi.org/10.1016/j.scib.2021.10.017} {\bibfield  {journal} {\bibinfo
  {journal} {Science Bulletin}\ }\textbf {\bibinfo {volume} {67}},\ \bibinfo
  {pages} {240} (\bibinfo {year} {2022})}\BibitemShut {NoStop}%
\bibitem [{\citenamefont {Madsen}\ \emph {et~al.}(2022)\citenamefont {Madsen},
  \citenamefont {Laudenbach}, \citenamefont {Askarani}, \citenamefont
  {Rortais}, \citenamefont {Vincent}, \citenamefont {Bulmer}, \citenamefont
  {Miatto}, \citenamefont {Neuhaus}, \citenamefont {Helt}, \citenamefont
  {Collins} \emph {et~al.}}]{madsen2022quantum}%
  \BibitemOpen
  \bibfield  {author} {\bibinfo {author} {\bibfnamefont {L.~S.}\ \bibnamefont
  {Madsen}}, \bibinfo {author} {\bibfnamefont {F.}~\bibnamefont {Laudenbach}},
  \bibinfo {author} {\bibfnamefont {M.~F.}\ \bibnamefont {Askarani}}, \bibinfo
  {author} {\bibfnamefont {F.}~\bibnamefont {Rortais}}, \bibinfo {author}
  {\bibfnamefont {T.}~\bibnamefont {Vincent}}, \bibinfo {author} {\bibfnamefont
  {J.~F.}\ \bibnamefont {Bulmer}}, \bibinfo {author} {\bibfnamefont {F.~M.}\
  \bibnamefont {Miatto}}, \bibinfo {author} {\bibfnamefont {L.}~\bibnamefont
  {Neuhaus}}, \bibinfo {author} {\bibfnamefont {L.~G.}\ \bibnamefont {Helt}},
  \bibinfo {author} {\bibfnamefont {M.~J.}\ \bibnamefont {Collins}},  \emph
  {et~al.},\ }\href@noop {} {\bibfield  {journal} {\bibinfo  {journal}
  {Nature}\ }\textbf {\bibinfo {volume} {606}},\ \bibinfo {pages} {75}
  (\bibinfo {year} {2022})}\BibitemShut {NoStop}%
\bibitem [{\citenamefont {Hacohen-Gourgy}\ \emph {et~al.}(2016)\citenamefont
  {Hacohen-Gourgy}, \citenamefont {Martin}, \citenamefont {Flurin},
  \citenamefont {Ramasesh}, \citenamefont {Whaley},\ and\ \citenamefont
  {Siddiqi}}]{hacohen2016quantum}%
  \BibitemOpen
  \bibfield  {author} {\bibinfo {author} {\bibfnamefont {S.}~\bibnamefont
  {Hacohen-Gourgy}}, \bibinfo {author} {\bibfnamefont {L.~S.}\ \bibnamefont
  {Martin}}, \bibinfo {author} {\bibfnamefont {E.}~\bibnamefont {Flurin}},
  \bibinfo {author} {\bibfnamefont {V.~V.}\ \bibnamefont {Ramasesh}}, \bibinfo
  {author} {\bibfnamefont {K.~B.}\ \bibnamefont {Whaley}}, \ and\ \bibinfo
  {author} {\bibfnamefont {I.}~\bibnamefont {Siddiqi}},\ }\href@noop {}
  {\bibfield  {journal} {\bibinfo  {journal} {Nature}\ }\textbf {\bibinfo
  {volume} {538}},\ \bibinfo {pages} {491} (\bibinfo {year}
  {2016})}\BibitemShut {NoStop}%
\bibitem [{\citenamefont {Colless}\ \emph {et~al.}(2018)\citenamefont
  {Colless}, \citenamefont {Ramasesh}, \citenamefont {Dahlen}, \citenamefont
  {Blok}, \citenamefont {Kimchi-Schwartz}, \citenamefont {McClean},
  \citenamefont {Carter}, \citenamefont {de~Jong},\ and\ \citenamefont
  {Siddiqi}}]{colless2018computation}%
  \BibitemOpen
  \bibfield  {author} {\bibinfo {author} {\bibfnamefont {J.~I.}\ \bibnamefont
  {Colless}}, \bibinfo {author} {\bibfnamefont {V.~V.}\ \bibnamefont
  {Ramasesh}}, \bibinfo {author} {\bibfnamefont {D.}~\bibnamefont {Dahlen}},
  \bibinfo {author} {\bibfnamefont {M.~S.}\ \bibnamefont {Blok}}, \bibinfo
  {author} {\bibfnamefont {M.~E.}\ \bibnamefont {Kimchi-Schwartz}}, \bibinfo
  {author} {\bibfnamefont {J.~R.}\ \bibnamefont {McClean}}, \bibinfo {author}
  {\bibfnamefont {J.}~\bibnamefont {Carter}}, \bibinfo {author} {\bibfnamefont
  {W.~A.}\ \bibnamefont {de~Jong}}, \ and\ \bibinfo {author} {\bibfnamefont
  {I.}~\bibnamefont {Siddiqi}},\ }\href@noop {} {\bibfield  {journal} {\bibinfo
   {journal} {Physical Review X}\ }\textbf {\bibinfo {volume} {8}},\ \bibinfo
  {pages} {011021} (\bibinfo {year} {2018})}\BibitemShut {NoStop}%
\bibitem [{\citenamefont {Blok}\ \emph {et~al.}(2021)\citenamefont {Blok},
  \citenamefont {Ramasesh}, \citenamefont {Schuster}, \citenamefont {O'Brien},
  \citenamefont {Kreikebaum}, \citenamefont {Dahlen}, \citenamefont {Morvan},
  \citenamefont {Yoshida}, \citenamefont {Yao},\ and\ \citenamefont
  {Siddiqi}}]{blok2021quantum}%
  \BibitemOpen
  \bibfield  {author} {\bibinfo {author} {\bibfnamefont {M.~S.}\ \bibnamefont
  {Blok}}, \bibinfo {author} {\bibfnamefont {V.~V.}\ \bibnamefont {Ramasesh}},
  \bibinfo {author} {\bibfnamefont {T.}~\bibnamefont {Schuster}}, \bibinfo
  {author} {\bibfnamefont {K.}~\bibnamefont {O'Brien}}, \bibinfo {author}
  {\bibfnamefont {J.~M.}\ \bibnamefont {Kreikebaum}}, \bibinfo {author}
  {\bibfnamefont {D.}~\bibnamefont {Dahlen}}, \bibinfo {author} {\bibfnamefont
  {A.}~\bibnamefont {Morvan}}, \bibinfo {author} {\bibfnamefont
  {B.}~\bibnamefont {Yoshida}}, \bibinfo {author} {\bibfnamefont {N.~Y.}\
  \bibnamefont {Yao}}, \ and\ \bibinfo {author} {\bibfnamefont
  {I.}~\bibnamefont {Siddiqi}},\ }\href {\doibase 10.1103/PhysRevX.11.021010}
  {\bibfield  {journal} {\bibinfo  {journal} {Phys. Rev. X}\ }\textbf {\bibinfo
  {volume} {11}},\ \bibinfo {pages} {021010} (\bibinfo {year}
  {2021})}\BibitemShut {NoStop}%
\bibitem [{\citenamefont {Mi}\ \emph {et~al.}(2022)\citenamefont {Mi},
  \citenamefont {Ippoliti}, \citenamefont {Quintana}, \citenamefont {Greene},
  \citenamefont {Chen}, \citenamefont {Gross}, \citenamefont {Arute},
  \citenamefont {Arya}, \citenamefont {Atalaya}, \citenamefont {Babbush} \emph
  {et~al.}}]{mi2022time}%
  \BibitemOpen
  \bibfield  {author} {\bibinfo {author} {\bibfnamefont {X.}~\bibnamefont
  {Mi}}, \bibinfo {author} {\bibfnamefont {M.}~\bibnamefont {Ippoliti}},
  \bibinfo {author} {\bibfnamefont {C.}~\bibnamefont {Quintana}}, \bibinfo
  {author} {\bibfnamefont {A.}~\bibnamefont {Greene}}, \bibinfo {author}
  {\bibfnamefont {Z.}~\bibnamefont {Chen}}, \bibinfo {author} {\bibfnamefont
  {J.}~\bibnamefont {Gross}}, \bibinfo {author} {\bibfnamefont
  {F.}~\bibnamefont {Arute}}, \bibinfo {author} {\bibfnamefont
  {K.}~\bibnamefont {Arya}}, \bibinfo {author} {\bibfnamefont {J.}~\bibnamefont
  {Atalaya}}, \bibinfo {author} {\bibfnamefont {R.}~\bibnamefont {Babbush}},
  \emph {et~al.},\ }\href@noop {} {\bibfield  {journal} {\bibinfo  {journal}
  {Nature}\ }\textbf {\bibinfo {volume} {601}},\ \bibinfo {pages} {531}
  (\bibinfo {year} {2022})}\BibitemShut {NoStop}%
\bibitem [{\citenamefont {Morvan}\ \emph {et~al.}(2022)\citenamefont {Morvan},
  \citenamefont {Andersen}, \citenamefont {Mi}, \citenamefont {Neill},
  \citenamefont {Petukhov}, \citenamefont {Kechedzhi}, \citenamefont {Abanin},
  \citenamefont {Michailidis}, \citenamefont {Acharya}, \citenamefont {Arute}
  \emph {et~al.}}]{morvan2022formation}%
  \BibitemOpen
  \bibfield  {author} {\bibinfo {author} {\bibfnamefont {A.}~\bibnamefont
  {Morvan}}, \bibinfo {author} {\bibfnamefont {T.}~\bibnamefont {Andersen}},
  \bibinfo {author} {\bibfnamefont {X.}~\bibnamefont {Mi}}, \bibinfo {author}
  {\bibfnamefont {C.}~\bibnamefont {Neill}}, \bibinfo {author} {\bibfnamefont
  {A.}~\bibnamefont {Petukhov}}, \bibinfo {author} {\bibfnamefont
  {K.}~\bibnamefont {Kechedzhi}}, \bibinfo {author} {\bibfnamefont
  {D.}~\bibnamefont {Abanin}}, \bibinfo {author} {\bibfnamefont
  {A.}~\bibnamefont {Michailidis}}, \bibinfo {author} {\bibfnamefont
  {R.}~\bibnamefont {Acharya}}, \bibinfo {author} {\bibfnamefont
  {F.}~\bibnamefont {Arute}},  \emph {et~al.},\ }\href@noop {} {\bibfield
  {journal} {\bibinfo  {journal} {Nature}\ }\textbf {\bibinfo {volume} {612}},\
  \bibinfo {pages} {240} (\bibinfo {year} {2022})}\BibitemShut {NoStop}%
\bibitem [{\citenamefont {Xiang}\ \emph {et~al.}(2024)\citenamefont {Xiang},
  \citenamefont {Jiang}, \citenamefont {Bao}, \citenamefont {Song},
  \citenamefont {Xu}, \citenamefont {Wang}, \citenamefont {Chen}, \citenamefont
  {Jin}, \citenamefont {Zhu}, \citenamefont {Zhu} \emph
  {et~al.}}]{xiang2024long}%
  \BibitemOpen
  \bibfield  {author} {\bibinfo {author} {\bibfnamefont {L.}~\bibnamefont
  {Xiang}}, \bibinfo {author} {\bibfnamefont {W.}~\bibnamefont {Jiang}},
  \bibinfo {author} {\bibfnamefont {Z.}~\bibnamefont {Bao}}, \bibinfo {author}
  {\bibfnamefont {Z.}~\bibnamefont {Song}}, \bibinfo {author} {\bibfnamefont
  {S.}~\bibnamefont {Xu}}, \bibinfo {author} {\bibfnamefont {K.}~\bibnamefont
  {Wang}}, \bibinfo {author} {\bibfnamefont {J.}~\bibnamefont {Chen}}, \bibinfo
  {author} {\bibfnamefont {F.}~\bibnamefont {Jin}}, \bibinfo {author}
  {\bibfnamefont {X.}~\bibnamefont {Zhu}}, \bibinfo {author} {\bibfnamefont
  {Z.}~\bibnamefont {Zhu}},  \emph {et~al.},\ }\href@noop {} {\bibfield
  {journal} {\bibinfo  {journal} {arXiv preprint arXiv:2401.04333}\ } (\bibinfo
  {year} {2024})}\BibitemShut {NoStop}%
\bibitem [{\citenamefont {Yamakawa}\ and\ \citenamefont
  {Zhandry}(2022)}]{yamakawa2022verifiable}%
  \BibitemOpen
  \bibfield  {author} {\bibinfo {author} {\bibfnamefont {T.}~\bibnamefont
  {Yamakawa}}\ and\ \bibinfo {author} {\bibfnamefont {M.}~\bibnamefont
  {Zhandry}},\ }in\ \href@noop {} {\emph {\bibinfo {booktitle} {2022 IEEE 63rd
  Annual Symposium on Foundations of Computer Science (FOCS)}}}\ (\bibinfo
  {organization} {IEEE},\ \bibinfo {year} {2022})\ pp.\ \bibinfo {pages}
  {69--74}\BibitemShut {NoStop}%
\bibitem [{\citenamefont {Aaronson}(2022)}]{aaronson2022much}%
  \BibitemOpen
  \bibfield  {author} {\bibinfo {author} {\bibfnamefont {S.}~\bibnamefont
  {Aaronson}},\ }\href@noop {} {\bibfield  {journal} {\bibinfo  {journal}
  {arXiv preprint arXiv:2209.06930}\ } (\bibinfo {year} {2022})}\BibitemShut
  {NoStop}%
\bibitem [{\citenamefont {Chen}\ \emph
  {et~al.}(2023{\natexlab{a}})\citenamefont {Chen}, \citenamefont {Cotler},
  \citenamefont {Huang},\ and\ \citenamefont {Li}}]{chen2023complexity}%
  \BibitemOpen
  \bibfield  {author} {\bibinfo {author} {\bibfnamefont {S.}~\bibnamefont
  {Chen}}, \bibinfo {author} {\bibfnamefont {J.}~\bibnamefont {Cotler}},
  \bibinfo {author} {\bibfnamefont {H.-Y.}\ \bibnamefont {Huang}}, \ and\
  \bibinfo {author} {\bibfnamefont {J.}~\bibnamefont {Li}},\ }\href@noop {}
  {\bibfield  {journal} {\bibinfo  {journal} {Nature Communications}\ }\textbf
  {\bibinfo {volume} {14}},\ \bibinfo {pages} {6001} (\bibinfo {year}
  {2023}{\natexlab{a}})}\BibitemShut {NoStop}%
\bibitem [{\citenamefont {Anshu}\ \emph {et~al.}(2023)\citenamefont {Anshu},
  \citenamefont {Breuckmann},\ and\ \citenamefont {Nirkhe}}]{anshu2023nlts}%
  \BibitemOpen
  \bibfield  {author} {\bibinfo {author} {\bibfnamefont {A.}~\bibnamefont
  {Anshu}}, \bibinfo {author} {\bibfnamefont {N.~P.}\ \bibnamefont
  {Breuckmann}}, \ and\ \bibinfo {author} {\bibfnamefont {C.}~\bibnamefont
  {Nirkhe}},\ }in\ \href@noop {} {\emph {\bibinfo {booktitle} {Proceedings of
  the 55th Annual ACM Symposium on Theory of Computing}}}\ (\bibinfo {year}
  {2023})\ pp.\ \bibinfo {pages} {1090--1096}\BibitemShut {NoStop}%
\bibitem [{\citenamefont {Aharonov}\ \emph {et~al.}(2023)\citenamefont
  {Aharonov}, \citenamefont {Gao}, \citenamefont {Landau}, \citenamefont
  {Liu},\ and\ \citenamefont {Vazirani}}]{aharonov2023polynomial}%
  \BibitemOpen
  \bibfield  {author} {\bibinfo {author} {\bibfnamefont {D.}~\bibnamefont
  {Aharonov}}, \bibinfo {author} {\bibfnamefont {X.}~\bibnamefont {Gao}},
  \bibinfo {author} {\bibfnamefont {Z.}~\bibnamefont {Landau}}, \bibinfo
  {author} {\bibfnamefont {Y.}~\bibnamefont {Liu}}, \ and\ \bibinfo {author}
  {\bibfnamefont {U.}~\bibnamefont {Vazirani}},\ }in\ \href@noop {} {\emph
  {\bibinfo {booktitle} {Proceedings of the 55th Annual ACM Symposium on Theory
  of Computing}}}\ (\bibinfo {year} {2023})\ pp.\ \bibinfo {pages}
  {945--957}\BibitemShut {NoStop}%
\bibitem [{\citenamefont {Cao}\ \emph {et~al.}(2018)\citenamefont {Cao},
  \citenamefont {Romero},\ and\ \citenamefont
  {Aspuru-Guzik}}]{cao2018potential}%
  \BibitemOpen
  \bibfield  {author} {\bibinfo {author} {\bibfnamefont {Y.}~\bibnamefont
  {Cao}}, \bibinfo {author} {\bibfnamefont {J.}~\bibnamefont {Romero}}, \ and\
  \bibinfo {author} {\bibfnamefont {A.}~\bibnamefont {Aspuru-Guzik}},\
  }\href@noop {} {\bibfield  {journal} {\bibinfo  {journal} {IBM Journal of
  Research and Development}\ }\textbf {\bibinfo {volume} {62}},\ \bibinfo
  {pages} {6} (\bibinfo {year} {2018})}\BibitemShut {NoStop}%
\bibitem [{\citenamefont {Siddiqi}(2021)}]{siddiqi2021engineering}%
  \BibitemOpen
  \bibfield  {author} {\bibinfo {author} {\bibfnamefont {I.}~\bibnamefont
  {Siddiqi}},\ }\href@noop {} {\bibfield  {journal} {\bibinfo  {journal}
  {Nature Reviews Materials}\ }\textbf {\bibinfo {volume} {6}},\ \bibinfo
  {pages} {875} (\bibinfo {year} {2021})}\BibitemShut {NoStop}%
\bibitem [{\citenamefont {Kitaev}(1995)}]{kitaev1995quantum}%
  \BibitemOpen
  \bibfield  {author} {\bibinfo {author} {\bibfnamefont {A.~Y.}\ \bibnamefont
  {Kitaev}},\ }\href@noop {} {\bibfield  {journal} {\bibinfo  {journal} {arXiv
  preprint quant-ph/9511026}\ } (\bibinfo {year} {1995})}\BibitemShut {NoStop}%
\bibitem [{\citenamefont {Grover}(1996)}]{grover1996fast}%
  \BibitemOpen
  \bibfield  {author} {\bibinfo {author} {\bibfnamefont {L.~K.}\ \bibnamefont
  {Grover}},\ }in\ \href@noop {} {\emph {\bibinfo {booktitle} {Proceedings of
  the twenty-eighth annual ACM symposium on Theory of computing}}}\ (\bibinfo
  {year} {1996})\ pp.\ \bibinfo {pages} {212--219}\BibitemShut {NoStop}%
\bibitem [{\citenamefont {Shor}(1999)}]{shor1999polynomial}%
  \BibitemOpen
  \bibfield  {author} {\bibinfo {author} {\bibfnamefont {P.~W.}\ \bibnamefont
  {Shor}},\ }\href@noop {} {\bibfield  {journal} {\bibinfo  {journal} {SIAM
  review}\ }\textbf {\bibinfo {volume} {41}},\ \bibinfo {pages} {303} (\bibinfo
  {year} {1999})}\BibitemShut {NoStop}%
\bibitem [{\citenamefont {Coppersmith}(2002)}]{coppersmith2002approximate}%
  \BibitemOpen
  \bibfield  {author} {\bibinfo {author} {\bibfnamefont {D.}~\bibnamefont
  {Coppersmith}},\ }\href@noop {} {\bibfield  {journal} {\bibinfo  {journal}
  {arXiv preprint quant-ph/0201067}\ } (\bibinfo {year} {2002})}\BibitemShut
  {NoStop}%
\bibitem [{\citenamefont {Harrow}\ \emph {et~al.}(2009)\citenamefont {Harrow},
  \citenamefont {Hassidim},\ and\ \citenamefont {Lloyd}}]{harrow2009quantum}%
  \BibitemOpen
  \bibfield  {author} {\bibinfo {author} {\bibfnamefont {A.~W.}\ \bibnamefont
  {Harrow}}, \bibinfo {author} {\bibfnamefont {A.}~\bibnamefont {Hassidim}}, \
  and\ \bibinfo {author} {\bibfnamefont {S.}~\bibnamefont {Lloyd}},\
  }\href@noop {} {\bibfield  {journal} {\bibinfo  {journal} {Physical review
  letters}\ }\textbf {\bibinfo {volume} {103}},\ \bibinfo {pages} {150502}
  (\bibinfo {year} {2009})}\BibitemShut {NoStop}%
\bibitem [{\citenamefont {Farhi}\ \emph {et~al.}(2014)\citenamefont {Farhi},
  \citenamefont {Goldstone},\ and\ \citenamefont {Gutmann}}]{farhi2014quantum}%
  \BibitemOpen
  \bibfield  {author} {\bibinfo {author} {\bibfnamefont {E.}~\bibnamefont
  {Farhi}}, \bibinfo {author} {\bibfnamefont {J.}~\bibnamefont {Goldstone}}, \
  and\ \bibinfo {author} {\bibfnamefont {S.}~\bibnamefont {Gutmann}},\
  }\href@noop {} {\bibfield  {journal} {\bibinfo  {journal} {arXiv preprint
  arXiv:1411.4028}\ } (\bibinfo {year} {2014})}\BibitemShut {NoStop}%
\bibitem [{\citenamefont {Liu}\ \emph {et~al.}(2021{\natexlab{a}})\citenamefont
  {Liu}, \citenamefont {Arunachalam},\ and\ \citenamefont
  {Temme}}]{liu2021rigorous}%
  \BibitemOpen
  \bibfield  {author} {\bibinfo {author} {\bibfnamefont {Y.}~\bibnamefont
  {Liu}}, \bibinfo {author} {\bibfnamefont {S.}~\bibnamefont {Arunachalam}}, \
  and\ \bibinfo {author} {\bibfnamefont {K.}~\bibnamefont {Temme}},\
  }\href@noop {} {\bibfield  {journal} {\bibinfo  {journal} {Nature Physics}\
  }\textbf {\bibinfo {volume} {17}},\ \bibinfo {pages} {1013} (\bibinfo {year}
  {2021}{\natexlab{a}})}\BibitemShut {NoStop}%
\bibitem [{\citenamefont {Daley}\ \emph {et~al.}(2022)\citenamefont {Daley},
  \citenamefont {Bloch}, \citenamefont {Kokail}, \citenamefont {Flannigan},
  \citenamefont {Pearson}, \citenamefont {Troyer},\ and\ \citenamefont
  {Zoller}}]{daley2022practical}%
  \BibitemOpen
  \bibfield  {author} {\bibinfo {author} {\bibfnamefont {A.~J.}\ \bibnamefont
  {Daley}}, \bibinfo {author} {\bibfnamefont {I.}~\bibnamefont {Bloch}},
  \bibinfo {author} {\bibfnamefont {C.}~\bibnamefont {Kokail}}, \bibinfo
  {author} {\bibfnamefont {S.}~\bibnamefont {Flannigan}}, \bibinfo {author}
  {\bibfnamefont {N.}~\bibnamefont {Pearson}}, \bibinfo {author} {\bibfnamefont
  {M.}~\bibnamefont {Troyer}}, \ and\ \bibinfo {author} {\bibfnamefont
  {P.}~\bibnamefont {Zoller}},\ }\href {\doibase 10.1038/s41586-022-04940-6}
  {\bibfield  {journal} {\bibinfo  {journal} {Nature}\ }\textbf {\bibinfo
  {volume} {607}},\ \bibinfo {pages} {667} (\bibinfo {year}
  {2022})}\BibitemShut {NoStop}%
\bibitem [{\citenamefont {Proctor}\ \emph {et~al.}(2025)\citenamefont
  {Proctor}, \citenamefont {Young}, \citenamefont {Baczewski},\ and\
  \citenamefont {Blume-Kohout}}]{proctor2024benchmarking}%
  \BibitemOpen
  \bibfield  {author} {\bibinfo {author} {\bibfnamefont {T.}~\bibnamefont
  {Proctor}}, \bibinfo {author} {\bibfnamefont {K.}~\bibnamefont {Young}},
  \bibinfo {author} {\bibfnamefont {A.~D.}\ \bibnamefont {Baczewski}}, \ and\
  \bibinfo {author} {\bibfnamefont {R.}~\bibnamefont {Blume-Kohout}},\
  }\href@noop {} {\bibfield  {journal} {\bibinfo  {journal} {Nature Reviews
  Physics}\ ,\ \bibinfo {pages} {1}} (\bibinfo {year} {2025})}\BibitemShut
  {NoStop}%
\bibitem [{Note1()}]{Note1}%
  \BibitemOpen
  \bibinfo {note} {The QCVV toolbox \protect \emph {also} includes other
  ``tools'' besides protocols. They include conceptual tools like \protect
  \emph {twirling} that are used by theorists to devise new protocols, and
  standardizing tools like \protect \emph {metrics} and \protect \emph {models}
  that enable clear communication between practitioners. But \protect \emph
  {protocols} are the heart of the field.}\BibitemShut {Stop}%
\bibitem [{\citenamefont {Eisert}\ \emph {et~al.}(2020)\citenamefont {Eisert},
  \citenamefont {Hangleiter}, \citenamefont {Walk}, \citenamefont {Roth},
  \citenamefont {Markham}, \citenamefont {Parekh}, \citenamefont {Chabaud},\
  and\ \citenamefont {Kashefi}}]{eisert2020quantum}%
  \BibitemOpen
  \bibfield  {author} {\bibinfo {author} {\bibfnamefont {J.}~\bibnamefont
  {Eisert}}, \bibinfo {author} {\bibfnamefont {D.}~\bibnamefont {Hangleiter}},
  \bibinfo {author} {\bibfnamefont {N.}~\bibnamefont {Walk}}, \bibinfo {author}
  {\bibfnamefont {I.}~\bibnamefont {Roth}}, \bibinfo {author} {\bibfnamefont
  {D.}~\bibnamefont {Markham}}, \bibinfo {author} {\bibfnamefont
  {R.}~\bibnamefont {Parekh}}, \bibinfo {author} {\bibfnamefont
  {U.}~\bibnamefont {Chabaud}}, \ and\ \bibinfo {author} {\bibfnamefont
  {E.}~\bibnamefont {Kashefi}},\ }\href@noop {} {\bibfield  {journal} {\bibinfo
   {journal} {Nature Reviews Physics}\ }\textbf {\bibinfo {volume} {2}},\
  \bibinfo {pages} {382} (\bibinfo {year} {2020})}\BibitemShut {NoStop}%
\bibitem [{\citenamefont {Kliesch}\ and\ \citenamefont
  {Roth}(2021)}]{kliesch2021theory}%
  \BibitemOpen
  \bibfield  {author} {\bibinfo {author} {\bibfnamefont {M.}~\bibnamefont
  {Kliesch}}\ and\ \bibinfo {author} {\bibfnamefont {I.}~\bibnamefont {Roth}},\
  }\href {\doibase 10.1103/PRXQuantum.2.010201} {\bibfield  {journal} {\bibinfo
   {journal} {PRX Quantum}\ }\textbf {\bibinfo {volume} {2}},\ \bibinfo {pages}
  {010201} (\bibinfo {year} {2021})}\BibitemShut {NoStop}%
\bibitem [{Note2()}]{Note2}%
  \BibitemOpen
  \bibinfo {note} {In mathematics, a vector space is a Hilbert space if and
  only if (iff) it is isomorphic to its dual space. But \protect \emph {all}
  finite-dimensional vector spaces are Hilbert spaces, and finite-dimensional
  spaces suffice to describe quantum data registers. So, the mathematical
  implications of ``Hilbert space'' are an unnecessary red herring for the
  purposes of this tutorial.}\BibitemShut {Stop}%
\bibitem [{\citenamefont {Sakurai}\ and\ \citenamefont
  {Napolitano}(2021)}]{sakurai2021modern}%
  \BibitemOpen
  \bibfield  {author} {\bibinfo {author} {\bibfnamefont {J.~J.}\ \bibnamefont
  {Sakurai}}\ and\ \bibinfo {author} {\bibfnamefont {J.}~\bibnamefont
  {Napolitano}},\ }\href@noop {} {\emph {\bibinfo {title} {Modern quantum
  mechanics}}}\ (\bibinfo  {publisher} {Cambridge University Press},\ \bibinfo
  {year} {2021})\BibitemShut {NoStop}%
\bibitem [{Note3()}]{Note3}%
  \BibitemOpen
  \bibinfo {note} {This subsection intentionally presents a simplified model of
  quantum mechanics consistent with most undergraduate textbooks. We neglect
  measurements of degenerate observables, which must be modeled by projectors
  of rank $>1$, for simplicity's sake. This important case is fully modeled by
  POVMs in the next subsection.}\BibitemShut {Stop}%
\bibitem [{Note4()}]{Note4}%
  \BibitemOpen
  \bibinfo {note} {This means that it is a variable in the \protect \emph
  {model} that has no physical reality, and can be varied without changing
  anything observable.}\BibitemShut {Stop}%
\bibitem [{Note5()}]{Note5}%
  \BibitemOpen
  \bibinfo {note} {It is ``projection-valued'' because it assigns a projection
  operator, rather than a non-negative real number, to each outcome. Born's
  Rule, with any state $\ketbra {\psi }$, defines a linear functional that maps
  a projection-valued measure to a standard probability measure, which is the
  probability distribution over that measurement's outcomes.}\BibitemShut
  {Stop}%
\bibitem [{Note6()}]{Note6}%
  \BibitemOpen
  \bibinfo {note} {$\protect \mathcal {B}(\protect \mathcal {H})$ means ``the
  space of bounded operators on $\protect \mathcal {H}$.'' Sometimes $\protect
  \mathcal {L}(\protect \mathcal {H})$, meaning ``the space of linear operators
  on $\protect \mathcal {H}$,'' is used instead. These coincide when $\protect
  \mathcal {H}$ is finite-dimensional.}\BibitemShut {Stop}%
\bibitem [{Note7()}]{Note7}%
  \BibitemOpen
  \bibinfo {note} {Technically, a POVM is a measure (like a probability
  distribution) over possible events, but which is ``operator-valued,'' meaning
  that instead of assigning a \protect \emph {probability} to each event, it
  assigns a \protect \emph {positive semi-definite operator} to each event,
  whose inner product with the system's state $\rho $ defines the event's
  probability.}\BibitemShut {Stop}%
\bibitem [{Note8()}]{Note8}%
  \BibitemOpen
  \bibinfo {note} {It is possible for initial correlation between the principal
  system and its environment to exist, and to be modeled. This scenario is
  advanced, conceptually tricky, and considered \protect \emph {non-Markovian}.
  It is not often considered in QCVV, and is outside the scope of this
  tutorial.}\BibitemShut {Stop}%
\bibitem [{\citenamefont {Stinespring}(1955)}]{stinespring1955positive}%
  \BibitemOpen
  \bibfield  {author} {\bibinfo {author} {\bibfnamefont {W.~F.}\ \bibnamefont
  {Stinespring}},\ }\href@noop {} {\bibfield  {journal} {\bibinfo  {journal}
  {Proceedings of the American Mathematical Society}\ }\textbf {\bibinfo
  {volume} {6}},\ \bibinfo {pages} {211} (\bibinfo {year} {1955})}\BibitemShut
  {NoStop}%
\bibitem [{\citenamefont {Wood}\ \emph {et~al.}(2015)\citenamefont {Wood},
  \citenamefont {Biamonte},\ and\ \citenamefont {Cory}}]{wood2011tensor}%
  \BibitemOpen
  \bibfield  {author} {\bibinfo {author} {\bibfnamefont {C.~J.}\ \bibnamefont
  {Wood}}, \bibinfo {author} {\bibfnamefont {J.~D.}\ \bibnamefont {Biamonte}},
  \ and\ \bibinfo {author} {\bibfnamefont {D.~G.}\ \bibnamefont {Cory}},\
  }\href@noop {} {\bibfield  {journal} {\bibinfo  {journal} {Quantum Info.
  Comput.}\ }\textbf {\bibinfo {volume} {15}},\ \bibinfo {pages} {759–811}
  (\bibinfo {year} {2015})}\BibitemShut {NoStop}%
\bibitem [{\citenamefont {Nielsen}\ and\ \citenamefont
  {Chuang}(2002)}]{nielsen2002quantum}%
  \BibitemOpen
  \bibfield  {author} {\bibinfo {author} {\bibfnamefont {M.~A.}\ \bibnamefont
  {Nielsen}}\ and\ \bibinfo {author} {\bibfnamefont {I.}~\bibnamefont
  {Chuang}},\ }\href@noop {} {\enquote {\bibinfo {title} {Quantum computation
  and quantum information},}\ } (\bibinfo {year} {2002})\BibitemShut {NoStop}%
\bibitem [{Note9()}]{Note9}%
  \BibitemOpen
  \bibinfo {note} {``Vectorization'' is used, in quantum information science
  and QCVV specifically, in at least two distinct ways. This one (Eq.~\ref
  {eq:vectorization}) maps a mathematical ``vector'' (an abstract element $\rho
  $ of Hilbert-Schmidt space) to a computer programming ``vector'' (a concrete
  list of numbers). But around Eq.~\ref {eq:vec}, we introduce another
  ``vectorization'' that maps mathematical vectors in $\protect \mathcal
  {B}(\protect \mathcal {H})$ to mathematical vectors in $\protect \mathcal
  {H}\otimes \protect \mathcal {H}$. Both are standard in the literature, and
  we apologize to the reader for our field's overloaded
  terminology.}\BibitemShut {Stop}%
\bibitem [{\citenamefont {Gilchrist}\ \emph {et~al.}(2009)\citenamefont
  {Gilchrist}, \citenamefont {Terno},\ and\ \citenamefont
  {Wood}}]{gilchrist2009vectorization}%
  \BibitemOpen
  \bibfield  {author} {\bibinfo {author} {\bibfnamefont {A.}~\bibnamefont
  {Gilchrist}}, \bibinfo {author} {\bibfnamefont {D.~R.}\ \bibnamefont
  {Terno}}, \ and\ \bibinfo {author} {\bibfnamefont {C.~J.}\ \bibnamefont
  {Wood}},\ }\href@noop {} {\bibfield  {journal} {\bibinfo  {journal} {arXiv
  preprint arXiv:0911.2539}\ } (\bibinfo {year} {2009})}\BibitemShut {NoStop}%
\bibitem [{Note10()}]{Note10}%
  \BibitemOpen
  \bibinfo {note} {It is possible to define quantum operations that map
  $\protect \mathcal {B}(\protect \mathcal {H})\DOTSB \mapstochar \rightarrow
  \protect \mathcal {B}(\protect \mathcal {H'})$, where $\protect \mathcal
  {H}\neq \protect \mathcal {H'}$, but these are used relatively rarely in QCVV
  and out of scope for this tutorial.}\BibitemShut {Stop}%
\bibitem [{\citenamefont {Blume-Kohout}\ \emph {et~al.}(2022)\citenamefont
  {Blume-Kohout}, \citenamefont {da~Silva}, \citenamefont {Nielsen},
  \citenamefont {Proctor}, \citenamefont {Rudinger}, \citenamefont {Sarovar},\
  and\ \citenamefont {Young}}]{blume2022taxonomy}%
  \BibitemOpen
  \bibfield  {author} {\bibinfo {author} {\bibfnamefont {R.}~\bibnamefont
  {Blume-Kohout}}, \bibinfo {author} {\bibfnamefont {M.~P.}\ \bibnamefont
  {da~Silva}}, \bibinfo {author} {\bibfnamefont {E.}~\bibnamefont {Nielsen}},
  \bibinfo {author} {\bibfnamefont {T.}~\bibnamefont {Proctor}}, \bibinfo
  {author} {\bibfnamefont {K.}~\bibnamefont {Rudinger}}, \bibinfo {author}
  {\bibfnamefont {M.}~\bibnamefont {Sarovar}}, \ and\ \bibinfo {author}
  {\bibfnamefont {K.}~\bibnamefont {Young}},\ }\href@noop {} {\bibfield
  {journal} {\bibinfo  {journal} {PRX Quantum}\ }\textbf {\bibinfo {volume}
  {3}},\ \bibinfo {pages} {020335} (\bibinfo {year} {2022})}\BibitemShut
  {NoStop}%
\bibitem [{\citenamefont {M{\k{a}}dzik}\ \emph {et~al.}(2022)\citenamefont
  {M{\k{a}}dzik}, \citenamefont {Asaad}, \citenamefont {Youssry}, \citenamefont
  {Joecker}, \citenamefont {Rudinger}, \citenamefont {Nielsen}, \citenamefont
  {Young}, \citenamefont {Proctor}, \citenamefont {Baczewski}, \citenamefont
  {Laucht} \emph {et~al.}}]{mkadzik2021precision}%
  \BibitemOpen
  \bibfield  {author} {\bibinfo {author} {\bibfnamefont {M.~T.}\ \bibnamefont
  {M{\k{a}}dzik}}, \bibinfo {author} {\bibfnamefont {S.}~\bibnamefont {Asaad}},
  \bibinfo {author} {\bibfnamefont {A.}~\bibnamefont {Youssry}}, \bibinfo
  {author} {\bibfnamefont {B.}~\bibnamefont {Joecker}}, \bibinfo {author}
  {\bibfnamefont {K.~M.}\ \bibnamefont {Rudinger}}, \bibinfo {author}
  {\bibfnamefont {E.}~\bibnamefont {Nielsen}}, \bibinfo {author} {\bibfnamefont
  {K.~C.}\ \bibnamefont {Young}}, \bibinfo {author} {\bibfnamefont {T.~J.}\
  \bibnamefont {Proctor}}, \bibinfo {author} {\bibfnamefont {A.~D.}\
  \bibnamefont {Baczewski}}, \bibinfo {author} {\bibfnamefont {A.}~\bibnamefont
  {Laucht}},  \emph {et~al.},\ }\href@noop {} {\bibfield  {journal} {\bibinfo
  {journal} {Nature}\ }\textbf {\bibinfo {volume} {601}},\ \bibinfo {pages}
  {348} (\bibinfo {year} {2022})}\BibitemShut {NoStop}%
\bibitem [{Note11()}]{Note11}%
  \BibitemOpen
  \bibinfo {note} {Note that some authors consider leakage, for example, to be
  a non-TP process, in which case the top row of the PTM captures
  state-dependent leakage. This is true if one only considers the qubit
  subspace within the full Hilbert space. However, strictly speaking, leakage
  is still TP, since the total probability of observing \protect \emph {some}
  outcome is preserved. For example, in some platforms leakage cannot be
  detected, and might instead be (erroneously) measured as 0 or 1, but the
  total number of shots will remain the same. In other platforms leakage can be
  more easily measured (see, for example, Fig.~\ref {fig:lrb}), in which case
  the total probability of observing 0, 1, or 2 is preserved. Therefore, when
  considering only the qubit subspace in the presence of leakage, it is
  sometimes common to relax the TP constraint, and instead simply require that
  the total probability must not increase (i.e., $\Tr [\protect \mathcal
  {E}(\rho )] \le \Tr [\rho ]$).}\BibitemShut {Stop}%
\bibitem [{\citenamefont {Choi}(1975)}]{choi1975completely}%
  \BibitemOpen
  \bibfield  {author} {\bibinfo {author} {\bibfnamefont {M.-D.}\ \bibnamefont
  {Choi}},\ }\href@noop {} {\bibfield  {journal} {\bibinfo  {journal} {Linear
  algebra and its applications}\ }\textbf {\bibinfo {volume} {10}},\ \bibinfo
  {pages} {285} (\bibinfo {year} {1975})}\BibitemShut {NoStop}%
\bibitem [{\citenamefont {Sudarshan}\ \emph {et~al.}(1961)\citenamefont
  {Sudarshan}, \citenamefont {Mathews},\ and\ \citenamefont
  {Rau}}]{sudarshan1961stochastic}%
  \BibitemOpen
  \bibfield  {author} {\bibinfo {author} {\bibfnamefont {E.}~\bibnamefont
  {Sudarshan}}, \bibinfo {author} {\bibfnamefont {P.}~\bibnamefont {Mathews}},
  \ and\ \bibinfo {author} {\bibfnamefont {J.}~\bibnamefont {Rau}},\
  }\href@noop {} {\bibfield  {journal} {\bibinfo  {journal} {Physical Review}\
  }\textbf {\bibinfo {volume} {121}},\ \bibinfo {pages} {920} (\bibinfo {year}
  {1961})}\BibitemShut {NoStop}%
\bibitem [{\citenamefont {Jamio{\l}kowski}(1972)}]{jamiolkowski1972linear}%
  \BibitemOpen
  \bibfield  {author} {\bibinfo {author} {\bibfnamefont {A.}~\bibnamefont
  {Jamio{\l}kowski}},\ }\href@noop {} {\bibfield  {journal} {\bibinfo
  {journal} {Reports on Mathematical Physics}\ }\textbf {\bibinfo {volume}
  {3}},\ \bibinfo {pages} {275} (\bibinfo {year} {1972})}\BibitemShut {NoStop}%
\bibitem [{\citenamefont {{\.Z}yczkowski}\ and\ \citenamefont
  {Bengtsson}(2004)}]{zyczkowski2004duality}%
  \BibitemOpen
  \bibfield  {author} {\bibinfo {author} {\bibfnamefont {K.}~\bibnamefont
  {{\.Z}yczkowski}}\ and\ \bibinfo {author} {\bibfnamefont {I.}~\bibnamefont
  {Bengtsson}},\ }\href@noop {} {\bibfield  {journal} {\bibinfo  {journal}
  {Open systems \& information dynamics}\ }\textbf {\bibinfo {volume} {11}},\
  \bibinfo {pages} {3} (\bibinfo {year} {2004})}\BibitemShut {NoStop}%
\bibitem [{\citenamefont {Hatridge}\ \emph {et~al.}(2013)\citenamefont
  {Hatridge}, \citenamefont {Shankar}, \citenamefont {Mirrahimi}, \citenamefont
  {Schackert}, \citenamefont {Geerlings}, \citenamefont {Brecht}, \citenamefont
  {Sliwa}, \citenamefont {Abdo}, \citenamefont {Frunzio}, \citenamefont
  {Girvin} \emph {et~al.}}]{hatridge2013quantum}%
  \BibitemOpen
  \bibfield  {author} {\bibinfo {author} {\bibfnamefont {M.}~\bibnamefont
  {Hatridge}}, \bibinfo {author} {\bibfnamefont {S.}~\bibnamefont {Shankar}},
  \bibinfo {author} {\bibfnamefont {M.}~\bibnamefont {Mirrahimi}}, \bibinfo
  {author} {\bibfnamefont {F.}~\bibnamefont {Schackert}}, \bibinfo {author}
  {\bibfnamefont {K.}~\bibnamefont {Geerlings}}, \bibinfo {author}
  {\bibfnamefont {T.}~\bibnamefont {Brecht}}, \bibinfo {author} {\bibfnamefont
  {K.}~\bibnamefont {Sliwa}}, \bibinfo {author} {\bibfnamefont
  {B.}~\bibnamefont {Abdo}}, \bibinfo {author} {\bibfnamefont {L.}~\bibnamefont
  {Frunzio}}, \bibinfo {author} {\bibfnamefont {S.~M.}\ \bibnamefont {Girvin}},
   \emph {et~al.},\ }\href@noop {} {\bibfield  {journal} {\bibinfo  {journal}
  {Science}\ }\textbf {\bibinfo {volume} {339}},\ \bibinfo {pages} {178}
  (\bibinfo {year} {2013})}\BibitemShut {NoStop}%
\bibitem [{\citenamefont {Cho}\ \emph {et~al.}(2019)\citenamefont {Cho},
  \citenamefont {Kim}, \citenamefont {Choi}, \citenamefont {Kim}, \citenamefont
  {Han}, \citenamefont {Lee}, \citenamefont {Moon},\ and\ \citenamefont
  {Kim}}]{cho2019emergence}%
  \BibitemOpen
  \bibfield  {author} {\bibinfo {author} {\bibfnamefont {Y.-W.}\ \bibnamefont
  {Cho}}, \bibinfo {author} {\bibfnamefont {Y.}~\bibnamefont {Kim}}, \bibinfo
  {author} {\bibfnamefont {Y.-H.}\ \bibnamefont {Choi}}, \bibinfo {author}
  {\bibfnamefont {Y.-S.}\ \bibnamefont {Kim}}, \bibinfo {author} {\bibfnamefont
  {S.-W.}\ \bibnamefont {Han}}, \bibinfo {author} {\bibfnamefont {S.-Y.}\
  \bibnamefont {Lee}}, \bibinfo {author} {\bibfnamefont {S.}~\bibnamefont
  {Moon}}, \ and\ \bibinfo {author} {\bibfnamefont {Y.-H.}\ \bibnamefont
  {Kim}},\ }\href@noop {} {\bibfield  {journal} {\bibinfo  {journal} {Nature
  Physics}\ }\textbf {\bibinfo {volume} {15}},\ \bibinfo {pages} {665}
  (\bibinfo {year} {2019})}\BibitemShut {NoStop}%
\bibitem [{\citenamefont {Davies}\ and\ \citenamefont
  {Lewis}(1970)}]{davies1970operational}%
  \BibitemOpen
  \bibfield  {author} {\bibinfo {author} {\bibfnamefont {E.~B.}\ \bibnamefont
  {Davies}}\ and\ \bibinfo {author} {\bibfnamefont {J.~T.}\ \bibnamefont
  {Lewis}},\ }\href@noop {} {\bibfield  {journal} {\bibinfo  {journal}
  {Communications in Mathematical Physics}\ }\textbf {\bibinfo {volume} {17}},\
  \bibinfo {pages} {239} (\bibinfo {year} {1970})}\BibitemShut {NoStop}%
\bibitem [{\citenamefont {Rudinger}\ \emph {et~al.}(2022)\citenamefont
  {Rudinger}, \citenamefont {Ribeill}, \citenamefont {Govia}, \citenamefont
  {Ware}, \citenamefont {Nielsen}, \citenamefont {Young}, \citenamefont {Ohki},
  \citenamefont {Blume-Kohout},\ and\ \citenamefont
  {Proctor}}]{rudinger2022characterizing}%
  \BibitemOpen
  \bibfield  {author} {\bibinfo {author} {\bibfnamefont {K.}~\bibnamefont
  {Rudinger}}, \bibinfo {author} {\bibfnamefont {G.~J.}\ \bibnamefont
  {Ribeill}}, \bibinfo {author} {\bibfnamefont {L.~C.}\ \bibnamefont {Govia}},
  \bibinfo {author} {\bibfnamefont {M.}~\bibnamefont {Ware}}, \bibinfo {author}
  {\bibfnamefont {E.}~\bibnamefont {Nielsen}}, \bibinfo {author} {\bibfnamefont
  {K.}~\bibnamefont {Young}}, \bibinfo {author} {\bibfnamefont {T.~A.}\
  \bibnamefont {Ohki}}, \bibinfo {author} {\bibfnamefont {R.}~\bibnamefont
  {Blume-Kohout}}, \ and\ \bibinfo {author} {\bibfnamefont {T.}~\bibnamefont
  {Proctor}},\ }\href@noop {} {\bibfield  {journal} {\bibinfo  {journal}
  {Physical Review Applied}\ }\textbf {\bibinfo {volume} {17}},\ \bibinfo
  {pages} {014014} (\bibinfo {year} {2022})}\BibitemShut {NoStop}%
\bibitem [{\citenamefont {von Neumann}(1932)}]{vonNeumann1932mathematische}%
  \BibitemOpen
  \bibfield  {author} {\bibinfo {author} {\bibfnamefont {J.}~\bibnamefont {von
  Neumann}},\ }\href@noop {} {\emph {\bibinfo {title} {Mathematische Grundlagen
  der Quantenmechanik}}}\ (\bibinfo  {publisher} {Springer},\ \bibinfo
  {address} {Berlin},\ \bibinfo {year} {1932})\BibitemShut {NoStop}%
\bibitem [{\citenamefont {Zurek}(1991)}]{zurek1991decoherence}%
  \BibitemOpen
  \bibfield  {author} {\bibinfo {author} {\bibfnamefont {W.~H.}\ \bibnamefont
  {Zurek}},\ }\href@noop {} {\bibfield  {journal} {\bibinfo  {journal} {Physics
  Today}\ }\textbf {\bibinfo {volume} {44}},\ \bibinfo {pages} {36} (\bibinfo
  {year} {1991})}\BibitemShut {NoStop}%
\bibitem [{\citenamefont {Braginsky}\ \emph {et~al.}(1980)\citenamefont
  {Braginsky}, \citenamefont {Vorontsov},\ and\ \citenamefont
  {Thorne}}]{braginsky1980quantum}%
  \BibitemOpen
  \bibfield  {author} {\bibinfo {author} {\bibfnamefont {V.~B.}\ \bibnamefont
  {Braginsky}}, \bibinfo {author} {\bibfnamefont {Y.~I.}\ \bibnamefont
  {Vorontsov}}, \ and\ \bibinfo {author} {\bibfnamefont {K.~S.}\ \bibnamefont
  {Thorne}},\ }\href@noop {} {\bibfield  {journal} {\bibinfo  {journal}
  {Science}\ }\textbf {\bibinfo {volume} {209}},\ \bibinfo {pages} {547}
  (\bibinfo {year} {1980})}\BibitemShut {NoStop}%
\bibitem [{\citenamefont {Braginsky}\ and\ \citenamefont
  {Khalili}(1996)}]{braginsky1996quantum}%
  \BibitemOpen
  \bibfield  {author} {\bibinfo {author} {\bibfnamefont {V.~B.}\ \bibnamefont
  {Braginsky}}\ and\ \bibinfo {author} {\bibfnamefont {F.~Y.}\ \bibnamefont
  {Khalili}},\ }\href@noop {} {\bibfield  {journal} {\bibinfo  {journal}
  {Reviews of Modern Physics}\ }\textbf {\bibinfo {volume} {68}},\ \bibinfo
  {pages} {1} (\bibinfo {year} {1996})}\BibitemShut {NoStop}%
\bibitem [{\citenamefont {Volz}\ \emph {et~al.}(2011)\citenamefont {Volz},
  \citenamefont {Gehr}, \citenamefont {Dubois}, \citenamefont {Est{\`e}ve},\
  and\ \citenamefont {Reichel}}]{volz2011measurement}%
  \BibitemOpen
  \bibfield  {author} {\bibinfo {author} {\bibfnamefont {J.}~\bibnamefont
  {Volz}}, \bibinfo {author} {\bibfnamefont {R.}~\bibnamefont {Gehr}}, \bibinfo
  {author} {\bibfnamefont {G.}~\bibnamefont {Dubois}}, \bibinfo {author}
  {\bibfnamefont {J.}~\bibnamefont {Est{\`e}ve}}, \ and\ \bibinfo {author}
  {\bibfnamefont {J.}~\bibnamefont {Reichel}},\ }\href@noop {} {\bibfield
  {journal} {\bibinfo  {journal} {Nature}\ }\textbf {\bibinfo {volume} {475}},\
  \bibinfo {pages} {210} (\bibinfo {year} {2011})}\BibitemShut {NoStop}%
\bibitem [{\citenamefont {Dassonneville}\ \emph {et~al.}(2020)\citenamefont
  {Dassonneville}, \citenamefont {Ramos}, \citenamefont {Milchakov},
  \citenamefont {Planat}, \citenamefont {Dumur}, \citenamefont {Foroughi},
  \citenamefont {Puertas}, \citenamefont {Leger}, \citenamefont {Bharadwaj},
  \citenamefont {Delaforce} \emph {et~al.}}]{dassonneville2020fast}%
  \BibitemOpen
  \bibfield  {author} {\bibinfo {author} {\bibfnamefont {R.}~\bibnamefont
  {Dassonneville}}, \bibinfo {author} {\bibfnamefont {T.}~\bibnamefont
  {Ramos}}, \bibinfo {author} {\bibfnamefont {V.}~\bibnamefont {Milchakov}},
  \bibinfo {author} {\bibfnamefont {L.}~\bibnamefont {Planat}}, \bibinfo
  {author} {\bibfnamefont {{\'E}.}~\bibnamefont {Dumur}}, \bibinfo {author}
  {\bibfnamefont {F.}~\bibnamefont {Foroughi}}, \bibinfo {author}
  {\bibfnamefont {J.}~\bibnamefont {Puertas}}, \bibinfo {author} {\bibfnamefont
  {S.}~\bibnamefont {Leger}}, \bibinfo {author} {\bibfnamefont
  {K.}~\bibnamefont {Bharadwaj}}, \bibinfo {author} {\bibfnamefont
  {J.}~\bibnamefont {Delaforce}},  \emph {et~al.},\ }\href@noop {} {\bibfield
  {journal} {\bibinfo  {journal} {Physical Review X}\ }\textbf {\bibinfo
  {volume} {10}},\ \bibinfo {pages} {011045} (\bibinfo {year}
  {2020})}\BibitemShut {NoStop}%
\bibitem [{\citenamefont {AI}(2023)}]{google2023suppressing}%
  \BibitemOpen
  \bibfield  {author} {\bibinfo {author} {\bibfnamefont {G.~Q.}\ \bibnamefont
  {AI}},\ }\href@noop {} {\bibfield  {journal} {\bibinfo  {journal} {Nature}\
  }\textbf {\bibinfo {volume} {614}},\ \bibinfo {pages} {676} (\bibinfo {year}
  {2023})}\BibitemShut {NoStop}%
\bibitem [{\citenamefont {Liu}\ \emph {et~al.}(2019)\citenamefont {Liu},
  \citenamefont {Zhang}, \citenamefont {Wan},\ and\ \citenamefont
  {Wang}}]{liu2019variational}%
  \BibitemOpen
  \bibfield  {author} {\bibinfo {author} {\bibfnamefont {J.-G.}\ \bibnamefont
  {Liu}}, \bibinfo {author} {\bibfnamefont {Y.-H.}\ \bibnamefont {Zhang}},
  \bibinfo {author} {\bibfnamefont {Y.}~\bibnamefont {Wan}}, \ and\ \bibinfo
  {author} {\bibfnamefont {L.}~\bibnamefont {Wang}},\ }\href@noop {} {\bibfield
   {journal} {\bibinfo  {journal} {Physical Review Research}\ }\textbf
  {\bibinfo {volume} {1}},\ \bibinfo {pages} {023025} (\bibinfo {year}
  {2019})}\BibitemShut {NoStop}%
\bibitem [{\citenamefont {Cong}\ \emph {et~al.}(2019)\citenamefont {Cong},
  \citenamefont {Choi},\ and\ \citenamefont {Lukin}}]{cong2019quantum}%
  \BibitemOpen
  \bibfield  {author} {\bibinfo {author} {\bibfnamefont {I.}~\bibnamefont
  {Cong}}, \bibinfo {author} {\bibfnamefont {S.}~\bibnamefont {Choi}}, \ and\
  \bibinfo {author} {\bibfnamefont {M.~D.}\ \bibnamefont {Lukin}},\ }\href@noop
  {} {\bibfield  {journal} {\bibinfo  {journal} {Nature Physics}\ }\textbf
  {\bibinfo {volume} {15}},\ \bibinfo {pages} {1273} (\bibinfo {year}
  {2019})}\BibitemShut {NoStop}%
\bibitem [{\citenamefont {Caves}\ \emph {et~al.}(1980)\citenamefont {Caves},
  \citenamefont {Thorne}, \citenamefont {Drever}, \citenamefont {Sandberg},\
  and\ \citenamefont {Zimmermann}}]{caves1980measurement}%
  \BibitemOpen
  \bibfield  {author} {\bibinfo {author} {\bibfnamefont {C.~M.}\ \bibnamefont
  {Caves}}, \bibinfo {author} {\bibfnamefont {K.~S.}\ \bibnamefont {Thorne}},
  \bibinfo {author} {\bibfnamefont {R.~W.}\ \bibnamefont {Drever}}, \bibinfo
  {author} {\bibfnamefont {V.~D.}\ \bibnamefont {Sandberg}}, \ and\ \bibinfo
  {author} {\bibfnamefont {M.}~\bibnamefont {Zimmermann}},\ }\href@noop {}
  {\bibfield  {journal} {\bibinfo  {journal} {Reviews of Modern Physics}\
  }\textbf {\bibinfo {volume} {52}},\ \bibinfo {pages} {341} (\bibinfo {year}
  {1980})}\BibitemShut {NoStop}%
\bibitem [{\citenamefont {Siddiqi}\ \emph {et~al.}(2006)\citenamefont
  {Siddiqi}, \citenamefont {Vijay}, \citenamefont {Metcalfe}, \citenamefont
  {Boaknin}, \citenamefont {Frunzio}, \citenamefont {Schoelkopf},\ and\
  \citenamefont {Devoret}}]{siddiqi2006dispersive}%
  \BibitemOpen
  \bibfield  {author} {\bibinfo {author} {\bibfnamefont {I.}~\bibnamefont
  {Siddiqi}}, \bibinfo {author} {\bibfnamefont {R.}~\bibnamefont {Vijay}},
  \bibinfo {author} {\bibfnamefont {M.}~\bibnamefont {Metcalfe}}, \bibinfo
  {author} {\bibfnamefont {E.}~\bibnamefont {Boaknin}}, \bibinfo {author}
  {\bibfnamefont {L.}~\bibnamefont {Frunzio}}, \bibinfo {author} {\bibfnamefont
  {R.}~\bibnamefont {Schoelkopf}}, \ and\ \bibinfo {author} {\bibfnamefont
  {M.}~\bibnamefont {Devoret}},\ }\href@noop {} {\bibfield  {journal} {\bibinfo
   {journal} {Physical Review B}\ }\textbf {\bibinfo {volume} {73}},\ \bibinfo
  {pages} {054510} (\bibinfo {year} {2006})}\BibitemShut {NoStop}%
\bibitem [{\citenamefont {Blais}\ \emph {et~al.}(2021)\citenamefont {Blais},
  \citenamefont {Grimsmo}, \citenamefont {Girvin},\ and\ \citenamefont
  {Wallraff}}]{blais2021circuit}%
  \BibitemOpen
  \bibfield  {author} {\bibinfo {author} {\bibfnamefont {A.}~\bibnamefont
  {Blais}}, \bibinfo {author} {\bibfnamefont {A.~L.}\ \bibnamefont {Grimsmo}},
  \bibinfo {author} {\bibfnamefont {S.~M.}\ \bibnamefont {Girvin}}, \ and\
  \bibinfo {author} {\bibfnamefont {A.}~\bibnamefont {Wallraff}},\ }\href@noop
  {} {\bibfield  {journal} {\bibinfo  {journal} {Reviews of Modern Physics}\
  }\textbf {\bibinfo {volume} {93}},\ \bibinfo {pages} {025005} (\bibinfo
  {year} {2021})}\BibitemShut {NoStop}%
\bibitem [{\citenamefont {Clerk}\ \emph {et~al.}(2010)\citenamefont {Clerk},
  \citenamefont {Devoret}, \citenamefont {Girvin}, \citenamefont {Marquardt},\
  and\ \citenamefont {Schoelkopf}}]{clerk2010introduction}%
  \BibitemOpen
  \bibfield  {author} {\bibinfo {author} {\bibfnamefont {A.~A.}\ \bibnamefont
  {Clerk}}, \bibinfo {author} {\bibfnamefont {M.~H.}\ \bibnamefont {Devoret}},
  \bibinfo {author} {\bibfnamefont {S.~M.}\ \bibnamefont {Girvin}}, \bibinfo
  {author} {\bibfnamefont {F.}~\bibnamefont {Marquardt}}, \ and\ \bibinfo
  {author} {\bibfnamefont {R.~J.}\ \bibnamefont {Schoelkopf}},\ }\href@noop {}
  {\bibfield  {journal} {\bibinfo  {journal} {Reviews of Modern Physics}\
  }\textbf {\bibinfo {volume} {82}},\ \bibinfo {pages} {1155} (\bibinfo {year}
  {2010})}\BibitemShut {NoStop}%
\bibitem [{\citenamefont {Korotkov}(2016)}]{korotkov2016quantum}%
  \BibitemOpen
  \bibfield  {author} {\bibinfo {author} {\bibfnamefont {A.~N.}\ \bibnamefont
  {Korotkov}},\ }\href@noop {} {\bibfield  {journal} {\bibinfo  {journal}
  {Physical Review A}\ }\textbf {\bibinfo {volume} {94}},\ \bibinfo {pages}
  {042326} (\bibinfo {year} {2016})}\BibitemShut {NoStop}%
\bibitem [{\citenamefont {Murch}\ \emph {et~al.}(2013)\citenamefont {Murch},
  \citenamefont {Weber}, \citenamefont {Macklin},\ and\ \citenamefont
  {Siddiqi}}]{murch2013observing}%
  \BibitemOpen
  \bibfield  {author} {\bibinfo {author} {\bibfnamefont {K.}~\bibnamefont
  {Murch}}, \bibinfo {author} {\bibfnamefont {S.}~\bibnamefont {Weber}},
  \bibinfo {author} {\bibfnamefont {C.}~\bibnamefont {Macklin}}, \ and\
  \bibinfo {author} {\bibfnamefont {I.}~\bibnamefont {Siddiqi}},\ }\href@noop
  {} {\bibfield  {journal} {\bibinfo  {journal} {Nature}\ }\textbf {\bibinfo
  {volume} {502}},\ \bibinfo {pages} {211} (\bibinfo {year}
  {2013})}\BibitemShut {NoStop}%
\bibitem [{\citenamefont {Weber}\ \emph {et~al.}(2014)\citenamefont {Weber},
  \citenamefont {Chantasri}, \citenamefont {Dressel}, \citenamefont {Jordan},
  \citenamefont {Murch},\ and\ \citenamefont {Siddiqi}}]{weber2014mapping}%
  \BibitemOpen
  \bibfield  {author} {\bibinfo {author} {\bibfnamefont {S.}~\bibnamefont
  {Weber}}, \bibinfo {author} {\bibfnamefont {A.}~\bibnamefont {Chantasri}},
  \bibinfo {author} {\bibfnamefont {J.}~\bibnamefont {Dressel}}, \bibinfo
  {author} {\bibfnamefont {A.~N.}\ \bibnamefont {Jordan}}, \bibinfo {author}
  {\bibfnamefont {K.}~\bibnamefont {Murch}}, \ and\ \bibinfo {author}
  {\bibfnamefont {I.}~\bibnamefont {Siddiqi}},\ }\href@noop {} {\bibfield
  {journal} {\bibinfo  {journal} {Nature}\ }\textbf {\bibinfo {volume} {511}},\
  \bibinfo {pages} {570} (\bibinfo {year} {2014})}\BibitemShut {NoStop}%
\bibitem [{\citenamefont {Koolstra}\ \emph {et~al.}(2022)\citenamefont
  {Koolstra}, \citenamefont {Stevenson}, \citenamefont {Barzili}, \citenamefont
  {Burns}, \citenamefont {Siva}, \citenamefont {Greenfield}, \citenamefont
  {Livingston}, \citenamefont {Hashim}, \citenamefont {Naik}, \citenamefont
  {Kreikebaum} \emph {et~al.}}]{koolstra2022monitoring}%
  \BibitemOpen
  \bibfield  {author} {\bibinfo {author} {\bibfnamefont {G.}~\bibnamefont
  {Koolstra}}, \bibinfo {author} {\bibfnamefont {N.}~\bibnamefont {Stevenson}},
  \bibinfo {author} {\bibfnamefont {S.}~\bibnamefont {Barzili}}, \bibinfo
  {author} {\bibfnamefont {L.}~\bibnamefont {Burns}}, \bibinfo {author}
  {\bibfnamefont {K.}~\bibnamefont {Siva}}, \bibinfo {author} {\bibfnamefont
  {S.}~\bibnamefont {Greenfield}}, \bibinfo {author} {\bibfnamefont
  {W.}~\bibnamefont {Livingston}}, \bibinfo {author} {\bibfnamefont
  {A.}~\bibnamefont {Hashim}}, \bibinfo {author} {\bibfnamefont
  {R.}~\bibnamefont {Naik}}, \bibinfo {author} {\bibfnamefont {J.}~\bibnamefont
  {Kreikebaum}},  \emph {et~al.},\ }\href@noop {} {\bibfield  {journal}
  {\bibinfo  {journal} {Physical Review X}\ }\textbf {\bibinfo {volume} {12}},\
  \bibinfo {pages} {031017} (\bibinfo {year} {2022})}\BibitemShut {NoStop}%
\bibitem [{\citenamefont {Kim}\ \emph {et~al.}(2018)\citenamefont {Kim},
  \citenamefont {Kim}, \citenamefont {Lee}, \citenamefont {Han}, \citenamefont
  {Moon}, \citenamefont {Kim},\ and\ \citenamefont {Cho}}]{kim2018direct}%
  \BibitemOpen
  \bibfield  {author} {\bibinfo {author} {\bibfnamefont {Y.}~\bibnamefont
  {Kim}}, \bibinfo {author} {\bibfnamefont {Y.-S.}\ \bibnamefont {Kim}},
  \bibinfo {author} {\bibfnamefont {S.-Y.}\ \bibnamefont {Lee}}, \bibinfo
  {author} {\bibfnamefont {S.-W.}\ \bibnamefont {Han}}, \bibinfo {author}
  {\bibfnamefont {S.}~\bibnamefont {Moon}}, \bibinfo {author} {\bibfnamefont
  {Y.-H.}\ \bibnamefont {Kim}}, \ and\ \bibinfo {author} {\bibfnamefont
  {Y.-W.}\ \bibnamefont {Cho}},\ }\href@noop {} {\bibfield  {journal} {\bibinfo
   {journal} {Nature communications}\ }\textbf {\bibinfo {volume} {9}},\
  \bibinfo {pages} {192} (\bibinfo {year} {2018})}\BibitemShut {NoStop}%
\bibitem [{\citenamefont {Siva}\ \emph {et~al.}(2023)\citenamefont {Siva},
  \citenamefont {Koolstra}, \citenamefont {Steinmetz}, \citenamefont
  {Livingston}, \citenamefont {Das}, \citenamefont {Chen}, \citenamefont
  {Kreikebaum}, \citenamefont {Stevenson}, \citenamefont {J{\"u}nger},
  \citenamefont {Santiago} \emph {et~al.}}]{siva2023time}%
  \BibitemOpen
  \bibfield  {author} {\bibinfo {author} {\bibfnamefont {K.}~\bibnamefont
  {Siva}}, \bibinfo {author} {\bibfnamefont {G.}~\bibnamefont {Koolstra}},
  \bibinfo {author} {\bibfnamefont {J.}~\bibnamefont {Steinmetz}}, \bibinfo
  {author} {\bibfnamefont {W.~P.}\ \bibnamefont {Livingston}}, \bibinfo
  {author} {\bibfnamefont {D.}~\bibnamefont {Das}}, \bibinfo {author}
  {\bibfnamefont {L.}~\bibnamefont {Chen}}, \bibinfo {author} {\bibfnamefont
  {J.~M.}\ \bibnamefont {Kreikebaum}}, \bibinfo {author} {\bibfnamefont
  {N.}~\bibnamefont {Stevenson}}, \bibinfo {author} {\bibfnamefont
  {C.}~\bibnamefont {J{\"u}nger}}, \bibinfo {author} {\bibfnamefont {D.~I.}\
  \bibnamefont {Santiago}},  \emph {et~al.},\ }\href@noop {} {\bibfield
  {journal} {\bibinfo  {journal} {PRX Quantum}\ }\textbf {\bibinfo {volume}
  {4}},\ \bibinfo {pages} {040324} (\bibinfo {year} {2023})}\BibitemShut
  {NoStop}%
\bibitem [{\citenamefont {Fuchs}\ and\ \citenamefont
  {Peres}(1996)}]{fuchs1996quantum}%
  \BibitemOpen
  \bibfield  {author} {\bibinfo {author} {\bibfnamefont {C.~A.}\ \bibnamefont
  {Fuchs}}\ and\ \bibinfo {author} {\bibfnamefont {A.}~\bibnamefont {Peres}},\
  }\href@noop {} {\bibfield  {journal} {\bibinfo  {journal} {Physical Review
  A}\ }\textbf {\bibinfo {volume} {53}},\ \bibinfo {pages} {2038} (\bibinfo
  {year} {1996})}\BibitemShut {NoStop}%
\bibitem [{\citenamefont {Hong}\ \emph {et~al.}(2022)\citenamefont {Hong},
  \citenamefont {Kim}, \citenamefont {Cho}, \citenamefont {Kim}, \citenamefont
  {Lee},\ and\ \citenamefont {Lim}}]{hong2022demonstration}%
  \BibitemOpen
  \bibfield  {author} {\bibinfo {author} {\bibfnamefont {S.}~\bibnamefont
  {Hong}}, \bibinfo {author} {\bibfnamefont {Y.-S.}\ \bibnamefont {Kim}},
  \bibinfo {author} {\bibfnamefont {Y.-W.}\ \bibnamefont {Cho}}, \bibinfo
  {author} {\bibfnamefont {J.}~\bibnamefont {Kim}}, \bibinfo {author}
  {\bibfnamefont {S.-W.}\ \bibnamefont {Lee}}, \ and\ \bibinfo {author}
  {\bibfnamefont {H.-T.}\ \bibnamefont {Lim}},\ }\href@noop {} {\bibfield
  {journal} {\bibinfo  {journal} {Physical review letters}\ }\textbf {\bibinfo
  {volume} {128}},\ \bibinfo {pages} {050401} (\bibinfo {year}
  {2022})}\BibitemShut {NoStop}%
\bibitem [{\citenamefont {Nielsen}\ \emph
  {et~al.}(2021{\natexlab{a}})\citenamefont {Nielsen}, \citenamefont
  {Rudinger}, \citenamefont {Proctor}, \citenamefont {Young},\ and\
  \citenamefont {Blume-Kohout}}]{Nielsen_2021}%
  \BibitemOpen
  \bibfield  {author} {\bibinfo {author} {\bibfnamefont {E.}~\bibnamefont
  {Nielsen}}, \bibinfo {author} {\bibfnamefont {K.}~\bibnamefont {Rudinger}},
  \bibinfo {author} {\bibfnamefont {T.}~\bibnamefont {Proctor}}, \bibinfo
  {author} {\bibfnamefont {K.}~\bibnamefont {Young}}, \ and\ \bibinfo {author}
  {\bibfnamefont {R.}~\bibnamefont {Blume-Kohout}},\ }\href {\doibase
  10.1088/1367-2630/ac20b9} {\bibfield  {journal} {\bibinfo  {journal} {New
  Journal of Physics}\ }\textbf {\bibinfo {volume} {23}},\ \bibinfo {pages}
  {093020} (\bibinfo {year} {2021}{\natexlab{a}})}\BibitemShut {NoStop}%
\bibitem [{\citenamefont {Rudinger}\ \emph {et~al.}(2021)\citenamefont
  {Rudinger}, \citenamefont {Hogle}, \citenamefont {Naik}, \citenamefont
  {Hashim}, \citenamefont {Lobser}, \citenamefont {Santiago}, \citenamefont
  {Grace}, \citenamefont {Nielsen}, \citenamefont {Proctor}, \citenamefont
  {Seritan}, \citenamefont {Clark}, \citenamefont {Blume-Kohout}, \citenamefont
  {Siddiqi},\ and\ \citenamefont {Young}}]{PRXQuantum.2.040338}%
  \BibitemOpen
  \bibfield  {author} {\bibinfo {author} {\bibfnamefont {K.}~\bibnamefont
  {Rudinger}}, \bibinfo {author} {\bibfnamefont {C.~W.}\ \bibnamefont {Hogle}},
  \bibinfo {author} {\bibfnamefont {R.~K.}\ \bibnamefont {Naik}}, \bibinfo
  {author} {\bibfnamefont {A.}~\bibnamefont {Hashim}}, \bibinfo {author}
  {\bibfnamefont {D.}~\bibnamefont {Lobser}}, \bibinfo {author} {\bibfnamefont
  {D.~I.}\ \bibnamefont {Santiago}}, \bibinfo {author} {\bibfnamefont {M.~D.}\
  \bibnamefont {Grace}}, \bibinfo {author} {\bibfnamefont {E.}~\bibnamefont
  {Nielsen}}, \bibinfo {author} {\bibfnamefont {T.}~\bibnamefont {Proctor}},
  \bibinfo {author} {\bibfnamefont {S.}~\bibnamefont {Seritan}}, \bibinfo
  {author} {\bibfnamefont {S.~M.}\ \bibnamefont {Clark}}, \bibinfo {author}
  {\bibfnamefont {R.}~\bibnamefont {Blume-Kohout}}, \bibinfo {author}
  {\bibfnamefont {I.}~\bibnamefont {Siddiqi}}, \ and\ \bibinfo {author}
  {\bibfnamefont {K.~C.}\ \bibnamefont {Young}},\ }\href {\doibase
  10.1103/PRXQuantum.2.040338} {\bibfield  {journal} {\bibinfo  {journal} {PRX
  Quantum}\ }\textbf {\bibinfo {volume} {2}},\ \bibinfo {pages} {040338}
  (\bibinfo {year} {2021})}\BibitemShut {NoStop}%
\bibitem [{\citenamefont {Hashim}\ \emph
  {et~al.}(2023{\natexlab{a}})\citenamefont {Hashim}, \citenamefont {Seritan},
  \citenamefont {Proctor}, \citenamefont {Rudinger}, \citenamefont {Goss},
  \citenamefont {Naik}, \citenamefont {Kreikebaum}, \citenamefont {Santiago},\
  and\ \citenamefont {Siddiqi}}]{hashim2023benchmarking}%
  \BibitemOpen
  \bibfield  {author} {\bibinfo {author} {\bibfnamefont {A.}~\bibnamefont
  {Hashim}}, \bibinfo {author} {\bibfnamefont {S.}~\bibnamefont {Seritan}},
  \bibinfo {author} {\bibfnamefont {T.}~\bibnamefont {Proctor}}, \bibinfo
  {author} {\bibfnamefont {K.}~\bibnamefont {Rudinger}}, \bibinfo {author}
  {\bibfnamefont {N.}~\bibnamefont {Goss}}, \bibinfo {author} {\bibfnamefont
  {R.}~\bibnamefont {Naik}}, \bibinfo {author} {\bibfnamefont {J.~M.}\
  \bibnamefont {Kreikebaum}}, \bibinfo {author} {\bibfnamefont
  {D.}~\bibnamefont {Santiago}}, \ and\ \bibinfo {author} {\bibfnamefont
  {I.}~\bibnamefont {Siddiqi}},\ }\href {\doibase 10.1038/s41534-023-00764-y}
  {\bibfield  {journal} {\bibinfo  {journal} {npj Quantum Inf}\ }\textbf
  {\bibinfo {volume} {9}} (\bibinfo {year} {2023}{\natexlab{a}}),\
  10.1038/s41534-023-00764-y}\BibitemShut {NoStop}%
\bibitem [{\citenamefont {Brieger}\ \emph {et~al.}(2023)\citenamefont
  {Brieger}, \citenamefont {Roth},\ and\ \citenamefont
  {Kliesch}}]{PRXQuantum.4.010325}%
  \BibitemOpen
  \bibfield  {author} {\bibinfo {author} {\bibfnamefont {R.}~\bibnamefont
  {Brieger}}, \bibinfo {author} {\bibfnamefont {I.}~\bibnamefont {Roth}}, \
  and\ \bibinfo {author} {\bibfnamefont {M.}~\bibnamefont {Kliesch}},\ }\href
  {\doibase 10.1103/PRXQuantum.4.010325} {\bibfield  {journal} {\bibinfo
  {journal} {PRX Quantum}\ }\textbf {\bibinfo {volume} {4}},\ \bibinfo {pages}
  {010325} (\bibinfo {year} {2023})}\BibitemShut {NoStop}%
\bibitem [{\citenamefont {Di~Matteo}\ \emph {et~al.}(2020)\citenamefont
  {Di~Matteo}, \citenamefont {Gamble}, \citenamefont {Granade}, \citenamefont
  {Rudinger},\ and\ \citenamefont {Wiebe}}]{DiMatteo2020operationalgauge}%
  \BibitemOpen
  \bibfield  {author} {\bibinfo {author} {\bibfnamefont {O.}~\bibnamefont
  {Di~Matteo}}, \bibinfo {author} {\bibfnamefont {J.}~\bibnamefont {Gamble}},
  \bibinfo {author} {\bibfnamefont {C.}~\bibnamefont {Granade}}, \bibinfo
  {author} {\bibfnamefont {K.}~\bibnamefont {Rudinger}}, \ and\ \bibinfo
  {author} {\bibfnamefont {N.}~\bibnamefont {Wiebe}},\ }\href {\doibase
  10.22331/q-2020-11-17-364} {\bibfield  {journal} {\bibinfo  {journal}
  {{Quantum}}\ }\textbf {\bibinfo {volume} {4}},\ \bibinfo {pages} {364}
  (\bibinfo {year} {2020})}\BibitemShut {NoStop}%
\bibitem [{\citenamefont {Marceaux}\ and\ \citenamefont
  {Young}(2023)}]{KalmanGST}%
  \BibitemOpen
  \bibfield  {author} {\bibinfo {author} {\bibfnamefont {J.~P.}\ \bibnamefont
  {Marceaux}}\ and\ \bibinfo {author} {\bibfnamefont {K.}~\bibnamefont
  {Young}},\ }in\ \href {\doibase 10.1109/QCE57702.2023.00159} {\emph {\bibinfo
  {booktitle} {2023 IEEE International Conference on Quantum Computing and
  Engineering (QCE)}}}\ (\bibinfo  {publisher} {IEEE Computer Society},\
  \bibinfo {address} {Los Alamitos, CA, USA},\ \bibinfo {year} {2023})\ pp.\
  \bibinfo {pages} {1401--1411}\BibitemShut {NoStop}%
\bibitem [{\citenamefont {Nielsen}\ \emph {et~al.}(2022)\citenamefont
  {Nielsen}, \citenamefont {Young},\ and\ \citenamefont
  {Blume-Kohout}}]{Nielsen2022-jx}%
  \BibitemOpen
  \bibfield  {author} {\bibinfo {author} {\bibfnamefont {E.}~\bibnamefont
  {Nielsen}}, \bibinfo {author} {\bibfnamefont {K.}~\bibnamefont {Young}}, \
  and\ \bibinfo {author} {\bibfnamefont {R.}~\bibnamefont {Blume-Kohout}}\
  }(\bibinfo {year} {2022})\ p.\ \bibinfo {pages} {M38.009}\BibitemShut
  {NoStop}%
\bibitem [{\citenamefont {Guillaud}\ and\ \citenamefont
  {Mirrahimi}(2019)}]{guillaud2019repetition}%
  \BibitemOpen
  \bibfield  {author} {\bibinfo {author} {\bibfnamefont {J.}~\bibnamefont
  {Guillaud}}\ and\ \bibinfo {author} {\bibfnamefont {M.}~\bibnamefont
  {Mirrahimi}},\ }\href {\doibase 10.1103/PhysRevX.9.041053} {\bibfield
  {journal} {\bibinfo  {journal} {Phys. Rev. X}\ }\textbf {\bibinfo {volume}
  {9}},\ \bibinfo {pages} {041053} (\bibinfo {year} {2019})}\BibitemShut
  {NoStop}%
\bibitem [{\citenamefont {Darmawan}\ \emph {et~al.}(2021)\citenamefont
  {Darmawan}, \citenamefont {Brown}, \citenamefont {Grimsmo}, \citenamefont
  {Tuckett},\ and\ \citenamefont {Puri}}]{darmawan2021practical}%
  \BibitemOpen
  \bibfield  {author} {\bibinfo {author} {\bibfnamefont {A.~S.}\ \bibnamefont
  {Darmawan}}, \bibinfo {author} {\bibfnamefont {B.~J.}\ \bibnamefont {Brown}},
  \bibinfo {author} {\bibfnamefont {A.~L.}\ \bibnamefont {Grimsmo}}, \bibinfo
  {author} {\bibfnamefont {D.~K.}\ \bibnamefont {Tuckett}}, \ and\ \bibinfo
  {author} {\bibfnamefont {S.}~\bibnamefont {Puri}},\ }\href {\doibase
  10.1103/PRXQuantum.2.030345} {\bibfield  {journal} {\bibinfo  {journal} {PRX
  Quantum}\ }\textbf {\bibinfo {volume} {2}},\ \bibinfo {pages} {030345}
  (\bibinfo {year} {2021})}\BibitemShut {NoStop}%
\bibitem [{\citenamefont {Nguyen}\ \emph {et~al.}(2022)\citenamefont {Nguyen},
  \citenamefont {Koolstra}, \citenamefont {Kim}, \citenamefont {Morvan},
  \citenamefont {Chistolini}, \citenamefont {Singh}, \citenamefont {Nesterov},
  \citenamefont {J\"unger}, \citenamefont {Chen}, \citenamefont {Pedramrazi},
  \citenamefont {Mitchell}, \citenamefont {Kreikebaum}, \citenamefont {Puri},
  \citenamefont {Santiago},\ and\ \citenamefont
  {Siddiqi}}]{nguyen2022blueprint}%
  \BibitemOpen
  \bibfield  {author} {\bibinfo {author} {\bibfnamefont {L.~B.}\ \bibnamefont
  {Nguyen}}, \bibinfo {author} {\bibfnamefont {G.}~\bibnamefont {Koolstra}},
  \bibinfo {author} {\bibfnamefont {Y.}~\bibnamefont {Kim}}, \bibinfo {author}
  {\bibfnamefont {A.}~\bibnamefont {Morvan}}, \bibinfo {author} {\bibfnamefont
  {T.}~\bibnamefont {Chistolini}}, \bibinfo {author} {\bibfnamefont
  {S.}~\bibnamefont {Singh}}, \bibinfo {author} {\bibfnamefont {K.~N.}\
  \bibnamefont {Nesterov}}, \bibinfo {author} {\bibfnamefont {C.}~\bibnamefont
  {J\"unger}}, \bibinfo {author} {\bibfnamefont {L.}~\bibnamefont {Chen}},
  \bibinfo {author} {\bibfnamefont {Z.}~\bibnamefont {Pedramrazi}}, \bibinfo
  {author} {\bibfnamefont {B.~K.}\ \bibnamefont {Mitchell}}, \bibinfo {author}
  {\bibfnamefont {J.~M.}\ \bibnamefont {Kreikebaum}}, \bibinfo {author}
  {\bibfnamefont {S.}~\bibnamefont {Puri}}, \bibinfo {author} {\bibfnamefont
  {D.~I.}\ \bibnamefont {Santiago}}, \ and\ \bibinfo {author} {\bibfnamefont
  {I.}~\bibnamefont {Siddiqi}},\ }\href {\doibase 10.1103/PRXQuantum.3.037001}
  {\bibfield  {journal} {\bibinfo  {journal} {PRX Quantum}\ }\textbf {\bibinfo
  {volume} {3}},\ \bibinfo {pages} {037001} (\bibinfo {year}
  {2022})}\BibitemShut {NoStop}%
\bibitem [{\citenamefont {Hahn}(1950)}]{hahn1950spin}%
  \BibitemOpen
  \bibfield  {author} {\bibinfo {author} {\bibfnamefont {E.~L.}\ \bibnamefont
  {Hahn}},\ }\href {\doibase 10.1103/PhysRev.80.580} {\bibfield  {journal}
  {\bibinfo  {journal} {Phys. Rev.}\ }\textbf {\bibinfo {volume} {80}},\
  \bibinfo {pages} {580} (\bibinfo {year} {1950})}\BibitemShut {NoStop}%
\bibitem [{\citenamefont {Carr}\ and\ \citenamefont
  {Purcell}(1954)}]{carr1954effects}%
  \BibitemOpen
  \bibfield  {author} {\bibinfo {author} {\bibfnamefont {H.~Y.}\ \bibnamefont
  {Carr}}\ and\ \bibinfo {author} {\bibfnamefont {E.~M.}\ \bibnamefont
  {Purcell}},\ }\href {\doibase 10.1103/PhysRev.94.630} {\bibfield  {journal}
  {\bibinfo  {journal} {Phys. Rev.}\ }\textbf {\bibinfo {volume} {94}},\
  \bibinfo {pages} {630} (\bibinfo {year} {1954})}\BibitemShut {NoStop}%
\bibitem [{\citenamefont {Meiboom}\ and\ \citenamefont
  {Gill}(1958)}]{meiboom1958modified}%
  \BibitemOpen
  \bibfield  {author} {\bibinfo {author} {\bibfnamefont {S.}~\bibnamefont
  {Meiboom}}\ and\ \bibinfo {author} {\bibfnamefont {D.}~\bibnamefont {Gill}},\
  }\href {\doibase 10.1063/1.1716296} {\bibfield  {journal} {\bibinfo
  {journal} {Review of scientific instruments}\ }\textbf {\bibinfo {volume}
  {29}},\ \bibinfo {pages} {688} (\bibinfo {year} {1958})}\BibitemShut
  {NoStop}%
\bibitem [{\citenamefont {Maudsley}(1986)}]{maudsley1986modified}%
  \BibitemOpen
  \bibfield  {author} {\bibinfo {author} {\bibfnamefont {A.}~\bibnamefont
  {Maudsley}},\ }\href@noop {} {\bibfield  {journal} {\bibinfo  {journal}
  {Journal of Magnetic Resonance (1969)}\ }\textbf {\bibinfo {volume} {69}},\
  \bibinfo {pages} {488} (\bibinfo {year} {1986})}\BibitemShut {NoStop}%
\bibitem [{\citenamefont {Ahmed}\ \emph {et~al.}(2013)\citenamefont {Ahmed},
  \citenamefont {Alvarez},\ and\ \citenamefont {Suter}}]{ahmed2013robustness}%
  \BibitemOpen
  \bibfield  {author} {\bibinfo {author} {\bibfnamefont {M.~A.~A.}\
  \bibnamefont {Ahmed}}, \bibinfo {author} {\bibfnamefont {G.~A.}\ \bibnamefont
  {Alvarez}}, \ and\ \bibinfo {author} {\bibfnamefont {D.}~\bibnamefont
  {Suter}},\ }\href@noop {} {\bibfield  {journal} {\bibinfo  {journal}
  {Physical Review A}\ }\textbf {\bibinfo {volume} {87}},\ \bibinfo {pages}
  {042309} (\bibinfo {year} {2013})}\BibitemShut {NoStop}%
\bibitem [{\citenamefont {Slichter}(2010)}]{Slichter2010-te}%
  \BibitemOpen
  \bibfield  {author} {\bibinfo {author} {\bibfnamefont {C.~P.}\ \bibnamefont
  {Slichter}},\ }\href {\doibase 10.1007/978-3-662-09441-9} {\emph {\bibinfo
  {title} {Principles of magnetic resonance}}},\ Springer Series in Solid-State
  Sciences\ (\bibinfo  {publisher} {Springer},\ \bibinfo {address} {Berlin,
  Germany},\ \bibinfo {year} {2010})\BibitemShut {NoStop}%
\bibitem [{\citenamefont {Kerman}(2010)}]{kerman2010metastable}%
  \BibitemOpen
  \bibfield  {author} {\bibinfo {author} {\bibfnamefont {A.~J.}\ \bibnamefont
  {Kerman}},\ }\href {\doibase 10.1103/PhysRevLett.104.027002} {\bibfield
  {journal} {\bibinfo  {journal} {Phys. Rev. Lett.}\ }\textbf {\bibinfo
  {volume} {104}},\ \bibinfo {pages} {027002} (\bibinfo {year}
  {2010})}\BibitemShut {NoStop}%
\bibitem [{\citenamefont {Lin}\ \emph {et~al.}(2018)\citenamefont {Lin},
  \citenamefont {Nguyen}, \citenamefont {Grabon}, \citenamefont {San~Miguel},
  \citenamefont {Pankratova},\ and\ \citenamefont
  {Manucharyan}}]{lin2018demonstration}%
  \BibitemOpen
  \bibfield  {author} {\bibinfo {author} {\bibfnamefont {Y.-H.}\ \bibnamefont
  {Lin}}, \bibinfo {author} {\bibfnamefont {L.~B.}\ \bibnamefont {Nguyen}},
  \bibinfo {author} {\bibfnamefont {N.}~\bibnamefont {Grabon}}, \bibinfo
  {author} {\bibfnamefont {J.}~\bibnamefont {San~Miguel}}, \bibinfo {author}
  {\bibfnamefont {N.}~\bibnamefont {Pankratova}}, \ and\ \bibinfo {author}
  {\bibfnamefont {V.~E.}\ \bibnamefont {Manucharyan}},\ }\href {\doibase
  10.1103/PhysRevLett.120.150503} {\bibfield  {journal} {\bibinfo  {journal}
  {Phys. Rev. Lett.}\ }\textbf {\bibinfo {volume} {120}},\ \bibinfo {pages}
  {150503} (\bibinfo {year} {2018})}\BibitemShut {NoStop}%
\bibitem [{\citenamefont {Earnest}\ \emph {et~al.}(2018)\citenamefont
  {Earnest}, \citenamefont {Chakram}, \citenamefont {Lu}, \citenamefont
  {Irons}, \citenamefont {Naik}, \citenamefont {Leung}, \citenamefont {Ocola},
  \citenamefont {Czaplewski}, \citenamefont {Baker}, \citenamefont {Lawrence},
  \citenamefont {Koch},\ and\ \citenamefont
  {Schuster}}]{earnest2018realization}%
  \BibitemOpen
  \bibfield  {author} {\bibinfo {author} {\bibfnamefont {N.}~\bibnamefont
  {Earnest}}, \bibinfo {author} {\bibfnamefont {S.}~\bibnamefont {Chakram}},
  \bibinfo {author} {\bibfnamefont {Y.}~\bibnamefont {Lu}}, \bibinfo {author}
  {\bibfnamefont {N.}~\bibnamefont {Irons}}, \bibinfo {author} {\bibfnamefont
  {R.~K.}\ \bibnamefont {Naik}}, \bibinfo {author} {\bibfnamefont
  {N.}~\bibnamefont {Leung}}, \bibinfo {author} {\bibfnamefont
  {L.}~\bibnamefont {Ocola}}, \bibinfo {author} {\bibfnamefont {D.~A.}\
  \bibnamefont {Czaplewski}}, \bibinfo {author} {\bibfnamefont
  {B.}~\bibnamefont {Baker}}, \bibinfo {author} {\bibfnamefont
  {J.}~\bibnamefont {Lawrence}}, \bibinfo {author} {\bibfnamefont
  {J.}~\bibnamefont {Koch}}, \ and\ \bibinfo {author} {\bibfnamefont {D.~I.}\
  \bibnamefont {Schuster}},\ }\href {\doibase 10.1103/PhysRevLett.120.150504}
  {\bibfield  {journal} {\bibinfo  {journal} {Phys. Rev. Lett.}\ }\textbf
  {\bibinfo {volume} {120}},\ \bibinfo {pages} {150504} (\bibinfo {year}
  {2018})}\BibitemShut {NoStop}%
\bibitem [{\citenamefont {Nguyen}\ \emph {et~al.}(2019)\citenamefont {Nguyen},
  \citenamefont {Lin}, \citenamefont {Somoroff}, \citenamefont {Mencia},
  \citenamefont {Grabon},\ and\ \citenamefont {Manucharyan}}]{nguyen2019high}%
  \BibitemOpen
  \bibfield  {author} {\bibinfo {author} {\bibfnamefont {L.~B.}\ \bibnamefont
  {Nguyen}}, \bibinfo {author} {\bibfnamefont {Y.-H.}\ \bibnamefont {Lin}},
  \bibinfo {author} {\bibfnamefont {A.}~\bibnamefont {Somoroff}}, \bibinfo
  {author} {\bibfnamefont {R.}~\bibnamefont {Mencia}}, \bibinfo {author}
  {\bibfnamefont {N.}~\bibnamefont {Grabon}}, \ and\ \bibinfo {author}
  {\bibfnamefont {V.~E.}\ \bibnamefont {Manucharyan}},\ }\href {\doibase
  10.1103/PhysRevX.9.041041} {\bibfield  {journal} {\bibinfo  {journal} {Phys.
  Rev. X}\ }\textbf {\bibinfo {volume} {9}},\ \bibinfo {pages} {041041}
  (\bibinfo {year} {2019})}\BibitemShut {NoStop}%
\bibitem [{\citenamefont {Wood}\ and\ \citenamefont
  {Gambetta}(2018)}]{wood2018quantification}%
  \BibitemOpen
  \bibfield  {author} {\bibinfo {author} {\bibfnamefont {C.~J.}\ \bibnamefont
  {Wood}}\ and\ \bibinfo {author} {\bibfnamefont {J.~M.}\ \bibnamefont
  {Gambetta}},\ }\href@noop {} {\bibfield  {journal} {\bibinfo  {journal}
  {Physical Review A}\ }\textbf {\bibinfo {volume} {97}},\ \bibinfo {pages}
  {032306} (\bibinfo {year} {2018})}\BibitemShut {NoStop}%
\bibitem [{\citenamefont {Proctor}\ \emph {et~al.}(2020)\citenamefont
  {Proctor}, \citenamefont {Revelle}, \citenamefont {Nielsen}, \citenamefont
  {Rudinger}, \citenamefont {Lobser}, \citenamefont {Maunz}, \citenamefont
  {Blume-Kohout},\ and\ \citenamefont {Young}}]{proctor2020detecting}%
  \BibitemOpen
  \bibfield  {author} {\bibinfo {author} {\bibfnamefont {T.}~\bibnamefont
  {Proctor}}, \bibinfo {author} {\bibfnamefont {M.}~\bibnamefont {Revelle}},
  \bibinfo {author} {\bibfnamefont {E.}~\bibnamefont {Nielsen}}, \bibinfo
  {author} {\bibfnamefont {K.}~\bibnamefont {Rudinger}}, \bibinfo {author}
  {\bibfnamefont {D.}~\bibnamefont {Lobser}}, \bibinfo {author} {\bibfnamefont
  {P.}~\bibnamefont {Maunz}}, \bibinfo {author} {\bibfnamefont
  {R.}~\bibnamefont {Blume-Kohout}}, \ and\ \bibinfo {author} {\bibfnamefont
  {K.}~\bibnamefont {Young}},\ }\href@noop {} {\bibfield  {journal} {\bibinfo
  {journal} {Nature communications}\ }\textbf {\bibinfo {volume} {11}},\
  \bibinfo {pages} {1} (\bibinfo {year} {2020})}\BibitemShut {NoStop}%
\bibitem [{\citenamefont {Ghosh}\ \emph {et~al.}(2013)\citenamefont {Ghosh},
  \citenamefont {Fowler}, \citenamefont {Martinis},\ and\ \citenamefont
  {Geller}}]{ghosh2013understanding}%
  \BibitemOpen
  \bibfield  {author} {\bibinfo {author} {\bibfnamefont {J.}~\bibnamefont
  {Ghosh}}, \bibinfo {author} {\bibfnamefont {A.~G.}\ \bibnamefont {Fowler}},
  \bibinfo {author} {\bibfnamefont {J.~M.}\ \bibnamefont {Martinis}}, \ and\
  \bibinfo {author} {\bibfnamefont {M.~R.}\ \bibnamefont {Geller}},\
  }\href@noop {} {\bibfield  {journal} {\bibinfo  {journal} {Physical Review
  A}\ }\textbf {\bibinfo {volume} {88}},\ \bibinfo {pages} {062329} (\bibinfo
  {year} {2013})}\BibitemShut {NoStop}%
\bibitem [{\citenamefont {Wallman}\ \emph {et~al.}(2016)\citenamefont
  {Wallman}, \citenamefont {Barnhill},\ and\ \citenamefont
  {Emerson}}]{wallman2016robust}%
  \BibitemOpen
  \bibfield  {author} {\bibinfo {author} {\bibfnamefont {J.~J.}\ \bibnamefont
  {Wallman}}, \bibinfo {author} {\bibfnamefont {M.}~\bibnamefont {Barnhill}}, \
  and\ \bibinfo {author} {\bibfnamefont {J.}~\bibnamefont {Emerson}},\
  }\href@noop {} {\bibfield  {journal} {\bibinfo  {journal} {New Journal of
  Physics}\ }\textbf {\bibinfo {volume} {18}},\ \bibinfo {pages} {043021}
  (\bibinfo {year} {2016})}\BibitemShut {NoStop}%
\bibitem [{\citenamefont {Chen}\ \emph {et~al.}(2016)\citenamefont {Chen},
  \citenamefont {Kelly}, \citenamefont {Quintana}, \citenamefont {Barends},
  \citenamefont {Campbell}, \citenamefont {Chen}, \citenamefont {Chiaro},
  \citenamefont {Dunsworth}, \citenamefont {Fowler}, \citenamefont {Lucero}
  \emph {et~al.}}]{chen2016measuring}%
  \BibitemOpen
  \bibfield  {author} {\bibinfo {author} {\bibfnamefont {Z.}~\bibnamefont
  {Chen}}, \bibinfo {author} {\bibfnamefont {J.}~\bibnamefont {Kelly}},
  \bibinfo {author} {\bibfnamefont {C.}~\bibnamefont {Quintana}}, \bibinfo
  {author} {\bibfnamefont {R.}~\bibnamefont {Barends}}, \bibinfo {author}
  {\bibfnamefont {B.}~\bibnamefont {Campbell}}, \bibinfo {author}
  {\bibfnamefont {Y.}~\bibnamefont {Chen}}, \bibinfo {author} {\bibfnamefont
  {B.}~\bibnamefont {Chiaro}}, \bibinfo {author} {\bibfnamefont
  {A.}~\bibnamefont {Dunsworth}}, \bibinfo {author} {\bibfnamefont
  {A.}~\bibnamefont {Fowler}}, \bibinfo {author} {\bibfnamefont
  {E.}~\bibnamefont {Lucero}},  \emph {et~al.},\ }\href@noop {} {\bibfield
  {journal} {\bibinfo  {journal} {Physical review letters}\ }\textbf {\bibinfo
  {volume} {116}},\ \bibinfo {pages} {020501} (\bibinfo {year}
  {2016})}\BibitemShut {NoStop}%
\bibitem [{\citenamefont {Hayes}\ \emph {et~al.}(2020)\citenamefont {Hayes},
  \citenamefont {Stack}, \citenamefont {Bjork}, \citenamefont {Potter},
  \citenamefont {Baldwin},\ and\ \citenamefont {Stutz}}]{hayes2020eliminating}%
  \BibitemOpen
  \bibfield  {author} {\bibinfo {author} {\bibfnamefont {D.}~\bibnamefont
  {Hayes}}, \bibinfo {author} {\bibfnamefont {D.}~\bibnamefont {Stack}},
  \bibinfo {author} {\bibfnamefont {B.}~\bibnamefont {Bjork}}, \bibinfo
  {author} {\bibfnamefont {A.}~\bibnamefont {Potter}}, \bibinfo {author}
  {\bibfnamefont {C.}~\bibnamefont {Baldwin}}, \ and\ \bibinfo {author}
  {\bibfnamefont {R.}~\bibnamefont {Stutz}},\ }\href@noop {} {\bibfield
  {journal} {\bibinfo  {journal} {Physical Review Letters}\ }\textbf {\bibinfo
  {volume} {124}},\ \bibinfo {pages} {170501} (\bibinfo {year}
  {2020})}\BibitemShut {NoStop}%
\bibitem [{\citenamefont {Babu}\ \emph {et~al.}(2021)\citenamefont {Babu},
  \citenamefont {Tuorila},\ and\ \citenamefont {Ala-Nissila}}]{babu2021state}%
  \BibitemOpen
  \bibfield  {author} {\bibinfo {author} {\bibfnamefont {A.~P.}\ \bibnamefont
  {Babu}}, \bibinfo {author} {\bibfnamefont {J.}~\bibnamefont {Tuorila}}, \
  and\ \bibinfo {author} {\bibfnamefont {T.}~\bibnamefont {Ala-Nissila}},\
  }\href@noop {} {\bibfield  {journal} {\bibinfo  {journal} {npj Quantum
  Information}\ }\textbf {\bibinfo {volume} {7}},\ \bibinfo {pages} {1}
  (\bibinfo {year} {2021})}\BibitemShut {NoStop}%
\bibitem [{\citenamefont {Li}\ \emph {et~al.}(2024)\citenamefont {Li},
  \citenamefont {Wang}, \citenamefont {Jiang}, \citenamefont {Xue},
  \citenamefont {Cai}, \citenamefont {Zhou}, \citenamefont {Gong},
  \citenamefont {Liu}, \citenamefont {Zheng}, \citenamefont {Ma} \emph
  {et~al.}}]{li2024direct}%
  \BibitemOpen
  \bibfield  {author} {\bibinfo {author} {\bibfnamefont {X.-G.}\ \bibnamefont
  {Li}}, \bibinfo {author} {\bibfnamefont {J.-H.}\ \bibnamefont {Wang}},
  \bibinfo {author} {\bibfnamefont {Y.-Y.}\ \bibnamefont {Jiang}}, \bibinfo
  {author} {\bibfnamefont {G.-M.}\ \bibnamefont {Xue}}, \bibinfo {author}
  {\bibfnamefont {X.-X.}\ \bibnamefont {Cai}}, \bibinfo {author} {\bibfnamefont
  {J.}~\bibnamefont {Zhou}}, \bibinfo {author} {\bibfnamefont {M.}~\bibnamefont
  {Gong}}, \bibinfo {author} {\bibfnamefont {Z.-F.}\ \bibnamefont {Liu}},
  \bibinfo {author} {\bibfnamefont {S.-Y.}\ \bibnamefont {Zheng}}, \bibinfo
  {author} {\bibfnamefont {D.-K.}\ \bibnamefont {Ma}},  \emph {et~al.},\
  }\href@noop {} {\bibfield  {journal} {\bibinfo  {journal} {arXiv preprint
  arXiv:2402.04245}\ } (\bibinfo {year} {2024})}\BibitemShut {NoStop}%
\bibitem [{\citenamefont {Harrington}\ \emph {et~al.}(2024)\citenamefont
  {Harrington}, \citenamefont {Li}, \citenamefont {Hays}, \citenamefont {Van
  De~Pontseele}, \citenamefont {Mayer}, \citenamefont {Pinckney}, \citenamefont
  {Contipelli}, \citenamefont {Gingras}, \citenamefont {Niedzielski},
  \citenamefont {Stickler} \emph {et~al.}}]{harrington2024synchronous}%
  \BibitemOpen
  \bibfield  {author} {\bibinfo {author} {\bibfnamefont {P.~M.}\ \bibnamefont
  {Harrington}}, \bibinfo {author} {\bibfnamefont {M.}~\bibnamefont {Li}},
  \bibinfo {author} {\bibfnamefont {M.}~\bibnamefont {Hays}}, \bibinfo {author}
  {\bibfnamefont {W.}~\bibnamefont {Van De~Pontseele}}, \bibinfo {author}
  {\bibfnamefont {D.}~\bibnamefont {Mayer}}, \bibinfo {author} {\bibfnamefont
  {H.~D.}\ \bibnamefont {Pinckney}}, \bibinfo {author} {\bibfnamefont
  {F.}~\bibnamefont {Contipelli}}, \bibinfo {author} {\bibfnamefont
  {M.}~\bibnamefont {Gingras}}, \bibinfo {author} {\bibfnamefont {B.~M.}\
  \bibnamefont {Niedzielski}}, \bibinfo {author} {\bibfnamefont
  {H.}~\bibnamefont {Stickler}},  \emph {et~al.},\ }\href@noop {} {\bibfield
  {journal} {\bibinfo  {journal} {arXiv preprint arXiv:2402.03208}\ } (\bibinfo
  {year} {2024})}\BibitemShut {NoStop}%
\bibitem [{\citenamefont {Mundada}\ \emph {et~al.}(2019)\citenamefont
  {Mundada}, \citenamefont {Zhang}, \citenamefont {Hazard},\ and\ \citenamefont
  {Houck}}]{mundada2019suppression}%
  \BibitemOpen
  \bibfield  {author} {\bibinfo {author} {\bibfnamefont {P.}~\bibnamefont
  {Mundada}}, \bibinfo {author} {\bibfnamefont {G.}~\bibnamefont {Zhang}},
  \bibinfo {author} {\bibfnamefont {T.}~\bibnamefont {Hazard}}, \ and\ \bibinfo
  {author} {\bibfnamefont {A.}~\bibnamefont {Houck}},\ }\href@noop {}
  {\bibfield  {journal} {\bibinfo  {journal} {Physical Review Applied}\
  }\textbf {\bibinfo {volume} {12}},\ \bibinfo {pages} {054023} (\bibinfo
  {year} {2019})}\BibitemShut {NoStop}%
\bibitem [{\citenamefont {Zhao}\ \emph {et~al.}(2020)\citenamefont {Zhao},
  \citenamefont {Xu}, \citenamefont {Lan}, \citenamefont {Chu}, \citenamefont
  {Tan}, \citenamefont {Yu},\ and\ \citenamefont {Yu}}]{zhao2020high}%
  \BibitemOpen
  \bibfield  {author} {\bibinfo {author} {\bibfnamefont {P.}~\bibnamefont
  {Zhao}}, \bibinfo {author} {\bibfnamefont {P.}~\bibnamefont {Xu}}, \bibinfo
  {author} {\bibfnamefont {D.}~\bibnamefont {Lan}}, \bibinfo {author}
  {\bibfnamefont {J.}~\bibnamefont {Chu}}, \bibinfo {author} {\bibfnamefont
  {X.}~\bibnamefont {Tan}}, \bibinfo {author} {\bibfnamefont {H.}~\bibnamefont
  {Yu}}, \ and\ \bibinfo {author} {\bibfnamefont {Y.}~\bibnamefont {Yu}},\
  }\href@noop {} {\bibfield  {journal} {\bibinfo  {journal} {Physical Review
  Letters}\ }\textbf {\bibinfo {volume} {125}},\ \bibinfo {pages} {200503}
  (\bibinfo {year} {2020})}\BibitemShut {NoStop}%
\bibitem [{\citenamefont {Ni}\ \emph {et~al.}(2021)\citenamefont {Ni},
  \citenamefont {Li}, \citenamefont {Zhang}, \citenamefont {Chu}, \citenamefont
  {Niu}, \citenamefont {Yan}, \citenamefont {Deng}, \citenamefont {Hu},
  \citenamefont {Li}, \citenamefont {Zhong} \emph {et~al.}}]{ni2021scalable}%
  \BibitemOpen
  \bibfield  {author} {\bibinfo {author} {\bibfnamefont {Z.}~\bibnamefont
  {Ni}}, \bibinfo {author} {\bibfnamefont {S.}~\bibnamefont {Li}}, \bibinfo
  {author} {\bibfnamefont {L.}~\bibnamefont {Zhang}}, \bibinfo {author}
  {\bibfnamefont {J.}~\bibnamefont {Chu}}, \bibinfo {author} {\bibfnamefont
  {J.}~\bibnamefont {Niu}}, \bibinfo {author} {\bibfnamefont {T.}~\bibnamefont
  {Yan}}, \bibinfo {author} {\bibfnamefont {X.}~\bibnamefont {Deng}}, \bibinfo
  {author} {\bibfnamefont {L.}~\bibnamefont {Hu}}, \bibinfo {author}
  {\bibfnamefont {J.}~\bibnamefont {Li}}, \bibinfo {author} {\bibfnamefont
  {Y.}~\bibnamefont {Zhong}},  \emph {et~al.},\ }\href@noop {} {\bibfield
  {journal} {\bibinfo  {journal} {arXiv preprint arXiv:2111.13292}\ } (\bibinfo
  {year} {2021})}\BibitemShut {NoStop}%
\bibitem [{\citenamefont {Serniak}\ \emph {et~al.}(2018)\citenamefont
  {Serniak}, \citenamefont {Hays}, \citenamefont {De~Lange}, \citenamefont
  {Diamond}, \citenamefont {Shankar}, \citenamefont {Burkhart}, \citenamefont
  {Frunzio}, \citenamefont {Houzet},\ and\ \citenamefont
  {Devoret}}]{serniak2018hot}%
  \BibitemOpen
  \bibfield  {author} {\bibinfo {author} {\bibfnamefont {K.}~\bibnamefont
  {Serniak}}, \bibinfo {author} {\bibfnamefont {M.}~\bibnamefont {Hays}},
  \bibinfo {author} {\bibfnamefont {G.}~\bibnamefont {De~Lange}}, \bibinfo
  {author} {\bibfnamefont {S.}~\bibnamefont {Diamond}}, \bibinfo {author}
  {\bibfnamefont {S.}~\bibnamefont {Shankar}}, \bibinfo {author} {\bibfnamefont
  {L.}~\bibnamefont {Burkhart}}, \bibinfo {author} {\bibfnamefont
  {L.}~\bibnamefont {Frunzio}}, \bibinfo {author} {\bibfnamefont
  {M.}~\bibnamefont {Houzet}}, \ and\ \bibinfo {author} {\bibfnamefont
  {M.}~\bibnamefont {Devoret}},\ }\href@noop {} {\bibfield  {journal} {\bibinfo
   {journal} {Physical review letters}\ }\textbf {\bibinfo {volume} {121}},\
  \bibinfo {pages} {157701} (\bibinfo {year} {2018})}\BibitemShut {NoStop}%
\bibitem [{\citenamefont {de~Graaf}\ \emph {et~al.}(2020)\citenamefont
  {de~Graaf}, \citenamefont {Faoro}, \citenamefont {Ioffe}, \citenamefont
  {Mahashabde}, \citenamefont {Burnett}, \citenamefont {Lindstr{\"o}m},
  \citenamefont {Kubatkin}, \citenamefont {Danilov},\ and\ \citenamefont
  {Tzalenchuk}}]{de2020two}%
  \BibitemOpen
  \bibfield  {author} {\bibinfo {author} {\bibfnamefont {S.}~\bibnamefont
  {de~Graaf}}, \bibinfo {author} {\bibfnamefont {L.}~\bibnamefont {Faoro}},
  \bibinfo {author} {\bibfnamefont {L.}~\bibnamefont {Ioffe}}, \bibinfo
  {author} {\bibfnamefont {S.}~\bibnamefont {Mahashabde}}, \bibinfo {author}
  {\bibfnamefont {J.}~\bibnamefont {Burnett}}, \bibinfo {author} {\bibfnamefont
  {T.}~\bibnamefont {Lindstr{\"o}m}}, \bibinfo {author} {\bibfnamefont
  {S.}~\bibnamefont {Kubatkin}}, \bibinfo {author} {\bibfnamefont
  {A.}~\bibnamefont {Danilov}}, \ and\ \bibinfo {author} {\bibfnamefont
  {A.~Y.}\ \bibnamefont {Tzalenchuk}},\ }\href@noop {} {\bibfield  {journal}
  {\bibinfo  {journal} {Science advances}\ }\textbf {\bibinfo {volume} {6}},\
  \bibinfo {pages} {eabc5055} (\bibinfo {year} {2020})}\BibitemShut {NoStop}%
\bibitem [{\citenamefont {Berlin-Udi}\ \emph {et~al.}(2021)\citenamefont
  {Berlin-Udi}, \citenamefont {Matthiesen}, \citenamefont {Lloyd},
  \citenamefont {Alonso}, \citenamefont {Noel}, \citenamefont {Orme},
  \citenamefont {Kim}, \citenamefont {Lordi},\ and\ \citenamefont
  {H{\"a}ffner}}]{berlin2021changes}%
  \BibitemOpen
  \bibfield  {author} {\bibinfo {author} {\bibfnamefont {M.}~\bibnamefont
  {Berlin-Udi}}, \bibinfo {author} {\bibfnamefont {C.}~\bibnamefont
  {Matthiesen}}, \bibinfo {author} {\bibfnamefont {P.}~\bibnamefont {Lloyd}},
  \bibinfo {author} {\bibfnamefont {A.}~\bibnamefont {Alonso}}, \bibinfo
  {author} {\bibfnamefont {C.}~\bibnamefont {Noel}}, \bibinfo {author}
  {\bibfnamefont {C.}~\bibnamefont {Orme}}, \bibinfo {author} {\bibfnamefont
  {C.-E.}\ \bibnamefont {Kim}}, \bibinfo {author} {\bibfnamefont
  {V.}~\bibnamefont {Lordi}}, \ and\ \bibinfo {author} {\bibfnamefont
  {H.}~\bibnamefont {H{\"a}ffner}},\ }\href@noop {} {\bibfield  {journal}
  {\bibinfo  {journal} {arXiv preprint arXiv:2103.04482}\ } (\bibinfo {year}
  {2021})}\BibitemShut {NoStop}%
\bibitem [{\citenamefont {Webb}\ \emph {et~al.}(2018)\citenamefont {Webb},
  \citenamefont {Webster}, \citenamefont {Collingbourne}, \citenamefont
  {Bretaud}, \citenamefont {Lawrence}, \citenamefont {Weidt}, \citenamefont
  {Mintert},\ and\ \citenamefont {Hensinger}}]{webb2018resilient}%
  \BibitemOpen
  \bibfield  {author} {\bibinfo {author} {\bibfnamefont {A.~E.}\ \bibnamefont
  {Webb}}, \bibinfo {author} {\bibfnamefont {S.~C.}\ \bibnamefont {Webster}},
  \bibinfo {author} {\bibfnamefont {S.}~\bibnamefont {Collingbourne}}, \bibinfo
  {author} {\bibfnamefont {D.}~\bibnamefont {Bretaud}}, \bibinfo {author}
  {\bibfnamefont {A.~M.}\ \bibnamefont {Lawrence}}, \bibinfo {author}
  {\bibfnamefont {S.}~\bibnamefont {Weidt}}, \bibinfo {author} {\bibfnamefont
  {F.}~\bibnamefont {Mintert}}, \ and\ \bibinfo {author} {\bibfnamefont
  {W.~K.}\ \bibnamefont {Hensinger}},\ }\href@noop {} {\bibfield  {journal}
  {\bibinfo  {journal} {Physical review letters}\ }\textbf {\bibinfo {volume}
  {121}},\ \bibinfo {pages} {180501} (\bibinfo {year} {2018})}\BibitemShut
  {NoStop}%
\bibitem [{\citenamefont {Burkard}(2009)}]{burkard2009non}%
  \BibitemOpen
  \bibfield  {author} {\bibinfo {author} {\bibfnamefont {G.}~\bibnamefont
  {Burkard}},\ }\href@noop {} {\bibfield  {journal} {\bibinfo  {journal}
  {Physical Review B}\ }\textbf {\bibinfo {volume} {79}},\ \bibinfo {pages}
  {125317} (\bibinfo {year} {2009})}\BibitemShut {NoStop}%
\bibitem [{\citenamefont {Groszkowski}\ \emph {et~al.}(2022)\citenamefont
  {Groszkowski}, \citenamefont {Seif}, \citenamefont {Koch},\ and\
  \citenamefont {Clerk}}]{groszkowski2022simple}%
  \BibitemOpen
  \bibfield  {author} {\bibinfo {author} {\bibfnamefont {P.}~\bibnamefont
  {Groszkowski}}, \bibinfo {author} {\bibfnamefont {A.}~\bibnamefont {Seif}},
  \bibinfo {author} {\bibfnamefont {J.}~\bibnamefont {Koch}}, \ and\ \bibinfo
  {author} {\bibfnamefont {A.}~\bibnamefont {Clerk}},\ }\href@noop {}
  {\bibfield  {journal} {\bibinfo  {journal} {arXiv preprint arXiv:2207.03980}\
  } (\bibinfo {year} {2022})}\BibitemShut {NoStop}%
\bibitem [{\citenamefont {Di{\'o}si}\ \emph {et~al.}(1998)\citenamefont
  {Di{\'o}si}, \citenamefont {Gisin},\ and\ \citenamefont
  {Strunz}}]{diosi1998non}%
  \BibitemOpen
  \bibfield  {author} {\bibinfo {author} {\bibfnamefont {L.}~\bibnamefont
  {Di{\'o}si}}, \bibinfo {author} {\bibfnamefont {N.}~\bibnamefont {Gisin}}, \
  and\ \bibinfo {author} {\bibfnamefont {W.~T.}\ \bibnamefont {Strunz}},\
  }\href@noop {} {\bibfield  {journal} {\bibinfo  {journal} {Physical Review
  A}\ }\textbf {\bibinfo {volume} {58}},\ \bibinfo {pages} {1699} (\bibinfo
  {year} {1998})}\BibitemShut {NoStop}%
\bibitem [{\citenamefont {Wolf}\ \emph {et~al.}(2008)\citenamefont {Wolf},
  \citenamefont {Eisert}, \citenamefont {Cubitt},\ and\ \citenamefont
  {Cirac}}]{wolf2008assessing}%
  \BibitemOpen
  \bibfield  {author} {\bibinfo {author} {\bibfnamefont {M.~M.}\ \bibnamefont
  {Wolf}}, \bibinfo {author} {\bibfnamefont {J.}~\bibnamefont {Eisert}},
  \bibinfo {author} {\bibfnamefont {T.~S.}\ \bibnamefont {Cubitt}}, \ and\
  \bibinfo {author} {\bibfnamefont {J.~I.}\ \bibnamefont {Cirac}},\ }\href@noop
  {} {\bibfield  {journal} {\bibinfo  {journal} {Physical review letters}\
  }\textbf {\bibinfo {volume} {101}},\ \bibinfo {pages} {150402} (\bibinfo
  {year} {2008})}\BibitemShut {NoStop}%
\bibitem [{\citenamefont {Piilo}\ \emph {et~al.}(2008)\citenamefont {Piilo},
  \citenamefont {Maniscalco}, \citenamefont {H{\"a}rk{\"o}nen},\ and\
  \citenamefont {Suominen}}]{piilo2008non}%
  \BibitemOpen
  \bibfield  {author} {\bibinfo {author} {\bibfnamefont {J.}~\bibnamefont
  {Piilo}}, \bibinfo {author} {\bibfnamefont {S.}~\bibnamefont {Maniscalco}},
  \bibinfo {author} {\bibfnamefont {K.}~\bibnamefont {H{\"a}rk{\"o}nen}}, \
  and\ \bibinfo {author} {\bibfnamefont {K.-A.}\ \bibnamefont {Suominen}},\
  }\href@noop {} {\bibfield  {journal} {\bibinfo  {journal} {Physical review
  letters}\ }\textbf {\bibinfo {volume} {100}},\ \bibinfo {pages} {180402}
  (\bibinfo {year} {2008})}\BibitemShut {NoStop}%
\bibitem [{\citenamefont {Breuer}\ \emph {et~al.}(2009)\citenamefont {Breuer},
  \citenamefont {Laine},\ and\ \citenamefont {Piilo}}]{breuer2009measure}%
  \BibitemOpen
  \bibfield  {author} {\bibinfo {author} {\bibfnamefont {H.-P.}\ \bibnamefont
  {Breuer}}, \bibinfo {author} {\bibfnamefont {E.-M.}\ \bibnamefont {Laine}}, \
  and\ \bibinfo {author} {\bibfnamefont {J.}~\bibnamefont {Piilo}},\
  }\href@noop {} {\bibfield  {journal} {\bibinfo  {journal} {Physical review
  letters}\ }\textbf {\bibinfo {volume} {103}},\ \bibinfo {pages} {210401}
  (\bibinfo {year} {2009})}\BibitemShut {NoStop}%
\bibitem [{\citenamefont {Liu}\ \emph {et~al.}(2011)\citenamefont {Liu},
  \citenamefont {Li}, \citenamefont {Huang}, \citenamefont {Li}, \citenamefont
  {Guo}, \citenamefont {Laine}, \citenamefont {Breuer},\ and\ \citenamefont
  {Piilo}}]{liu2011experimental}%
  \BibitemOpen
  \bibfield  {author} {\bibinfo {author} {\bibfnamefont {B.-H.}\ \bibnamefont
  {Liu}}, \bibinfo {author} {\bibfnamefont {L.}~\bibnamefont {Li}}, \bibinfo
  {author} {\bibfnamefont {Y.-F.}\ \bibnamefont {Huang}}, \bibinfo {author}
  {\bibfnamefont {C.-F.}\ \bibnamefont {Li}}, \bibinfo {author} {\bibfnamefont
  {G.-C.}\ \bibnamefont {Guo}}, \bibinfo {author} {\bibfnamefont {E.-M.}\
  \bibnamefont {Laine}}, \bibinfo {author} {\bibfnamefont {H.-P.}\ \bibnamefont
  {Breuer}}, \ and\ \bibinfo {author} {\bibfnamefont {J.}~\bibnamefont
  {Piilo}},\ }\href@noop {} {\bibfield  {journal} {\bibinfo  {journal} {Nature
  Physics}\ }\textbf {\bibinfo {volume} {7}},\ \bibinfo {pages} {931} (\bibinfo
  {year} {2011})}\BibitemShut {NoStop}%
\bibitem [{\citenamefont {De~Vega}\ and\ \citenamefont
  {Alonso}(2017)}]{de2017dynamics}%
  \BibitemOpen
  \bibfield  {author} {\bibinfo {author} {\bibfnamefont {I.}~\bibnamefont
  {De~Vega}}\ and\ \bibinfo {author} {\bibfnamefont {D.}~\bibnamefont
  {Alonso}},\ }\href@noop {} {\bibfield  {journal} {\bibinfo  {journal}
  {Reviews of Modern Physics}\ }\textbf {\bibinfo {volume} {89}},\ \bibinfo
  {pages} {015001} (\bibinfo {year} {2017})}\BibitemShut {NoStop}%
\bibitem [{\citenamefont {Glick}\ and\ \citenamefont
  {Adami}(2020)}]{glick2020markovian}%
  \BibitemOpen
  \bibfield  {author} {\bibinfo {author} {\bibfnamefont {J.~R.}\ \bibnamefont
  {Glick}}\ and\ \bibinfo {author} {\bibfnamefont {C.}~\bibnamefont {Adami}},\
  }\href@noop {} {\bibfield  {journal} {\bibinfo  {journal} {Foundations of
  Physics}\ }\textbf {\bibinfo {volume} {50}},\ \bibinfo {pages} {1008}
  (\bibinfo {year} {2020})}\BibitemShut {NoStop}%
\bibitem [{\citenamefont {Head-Marsden}\ \emph {et~al.}(2021)\citenamefont
  {Head-Marsden}, \citenamefont {Krastanov}, \citenamefont {Mazziotti},\ and\
  \citenamefont {Narang}}]{head2021capturing}%
  \BibitemOpen
  \bibfield  {author} {\bibinfo {author} {\bibfnamefont {K.}~\bibnamefont
  {Head-Marsden}}, \bibinfo {author} {\bibfnamefont {S.}~\bibnamefont
  {Krastanov}}, \bibinfo {author} {\bibfnamefont {D.~A.}\ \bibnamefont
  {Mazziotti}}, \ and\ \bibinfo {author} {\bibfnamefont {P.}~\bibnamefont
  {Narang}},\ }\href@noop {} {\bibfield  {journal} {\bibinfo  {journal}
  {Physical Review Research}\ }\textbf {\bibinfo {volume} {3}},\ \bibinfo
  {pages} {013182} (\bibinfo {year} {2021})}\BibitemShut {NoStop}%
\bibitem [{\citenamefont {Link}\ \emph {et~al.}(2022)\citenamefont {Link},
  \citenamefont {Strunz},\ and\ \citenamefont {Luoma}}]{link2022non}%
  \BibitemOpen
  \bibfield  {author} {\bibinfo {author} {\bibfnamefont {V.}~\bibnamefont
  {Link}}, \bibinfo {author} {\bibfnamefont {W.~T.}\ \bibnamefont {Strunz}}, \
  and\ \bibinfo {author} {\bibfnamefont {K.}~\bibnamefont {Luoma}},\
  }\href@noop {} {\bibfield  {journal} {\bibinfo  {journal} {Entropy}\ }\textbf
  {\bibinfo {volume} {24}},\ \bibinfo {pages} {352} (\bibinfo {year}
  {2022})}\BibitemShut {NoStop}%
\bibitem [{\citenamefont {Tserkis}\ \emph {et~al.}(2022)\citenamefont
  {Tserkis}, \citenamefont {Head-Marsden},\ and\ \citenamefont
  {Narang}}]{tserkis2022information}%
  \BibitemOpen
  \bibfield  {author} {\bibinfo {author} {\bibfnamefont {S.}~\bibnamefont
  {Tserkis}}, \bibinfo {author} {\bibfnamefont {K.}~\bibnamefont
  {Head-Marsden}}, \ and\ \bibinfo {author} {\bibfnamefont {P.}~\bibnamefont
  {Narang}},\ }\href@noop {} {\bibfield  {journal} {\bibinfo  {journal} {arXiv
  preprint arXiv:2203.00668}\ } (\bibinfo {year} {2022})}\BibitemShut {NoStop}%
\bibitem [{\citenamefont {Rivas}\ \emph {et~al.}(2014)\citenamefont {Rivas},
  \citenamefont {Huelga},\ and\ \citenamefont {Plenio}}]{rivas2014quantum}%
  \BibitemOpen
  \bibfield  {author} {\bibinfo {author} {\bibfnamefont {{\'A}.}~\bibnamefont
  {Rivas}}, \bibinfo {author} {\bibfnamefont {S.~F.}\ \bibnamefont {Huelga}}, \
  and\ \bibinfo {author} {\bibfnamefont {M.~B.}\ \bibnamefont {Plenio}},\
  }\href@noop {} {\bibfield  {journal} {\bibinfo  {journal} {Reports on
  Progress in Physics}\ }\textbf {\bibinfo {volume} {77}},\ \bibinfo {pages}
  {094001} (\bibinfo {year} {2014})}\BibitemShut {NoStop}%
\bibitem [{\citenamefont {Breuer}\ \emph {et~al.}(2016)\citenamefont {Breuer},
  \citenamefont {Laine}, \citenamefont {Piilo},\ and\ \citenamefont
  {Vacchini}}]{breuer2016colloquium}%
  \BibitemOpen
  \bibfield  {author} {\bibinfo {author} {\bibfnamefont {H.-P.}\ \bibnamefont
  {Breuer}}, \bibinfo {author} {\bibfnamefont {E.-M.}\ \bibnamefont {Laine}},
  \bibinfo {author} {\bibfnamefont {J.}~\bibnamefont {Piilo}}, \ and\ \bibinfo
  {author} {\bibfnamefont {B.}~\bibnamefont {Vacchini}},\ }\href@noop {}
  {\bibfield  {journal} {\bibinfo  {journal} {Reviews of Modern Physics}\
  }\textbf {\bibinfo {volume} {88}},\ \bibinfo {pages} {021002} (\bibinfo
  {year} {2016})}\BibitemShut {NoStop}%
\bibitem [{\citenamefont {Li}\ \emph {et~al.}(2018)\citenamefont {Li},
  \citenamefont {Hall},\ and\ \citenamefont {Wiseman}}]{li2018concepts}%
  \BibitemOpen
  \bibfield  {author} {\bibinfo {author} {\bibfnamefont {L.}~\bibnamefont
  {Li}}, \bibinfo {author} {\bibfnamefont {M.~J.}\ \bibnamefont {Hall}}, \ and\
  \bibinfo {author} {\bibfnamefont {H.~M.}\ \bibnamefont {Wiseman}},\
  }\href@noop {} {\bibfield  {journal} {\bibinfo  {journal} {Physics Reports}\
  }\textbf {\bibinfo {volume} {759}},\ \bibinfo {pages} {1} (\bibinfo {year}
  {2018})}\BibitemShut {NoStop}%
\bibitem [{\citenamefont {Li}\ \emph {et~al.}(2019)\citenamefont {Li},
  \citenamefont {Guo},\ and\ \citenamefont {Piilo}}]{li2019non}%
  \BibitemOpen
  \bibfield  {author} {\bibinfo {author} {\bibfnamefont {C.-F.}\ \bibnamefont
  {Li}}, \bibinfo {author} {\bibfnamefont {G.-C.}\ \bibnamefont {Guo}}, \ and\
  \bibinfo {author} {\bibfnamefont {J.}~\bibnamefont {Piilo}},\ }\href@noop {}
  {\bibfield  {journal} {\bibinfo  {journal} {EPL (Europhysics Letters)}\
  }\textbf {\bibinfo {volume} {127}},\ \bibinfo {pages} {50001} (\bibinfo
  {year} {2019})}\BibitemShut {NoStop}%
\bibitem [{\citenamefont {Milz}\ and\ \citenamefont
  {Modi}(2021)}]{milz2021quantum}%
  \BibitemOpen
  \bibfield  {author} {\bibinfo {author} {\bibfnamefont {S.}~\bibnamefont
  {Milz}}\ and\ \bibinfo {author} {\bibfnamefont {K.}~\bibnamefont {Modi}},\
  }\href@noop {} {\bibfield  {journal} {\bibinfo  {journal} {PRX Quantum}\
  }\textbf {\bibinfo {volume} {2}},\ \bibinfo {pages} {030201} (\bibinfo {year}
  {2021})}\BibitemShut {NoStop}%
\bibitem [{\citenamefont {White}\ \emph {et~al.}(2023)\citenamefont {White},
  \citenamefont {Jurcevic}, \citenamefont {Hill},\ and\ \citenamefont
  {Modi}}]{white2023unifying}%
  \BibitemOpen
  \bibfield  {author} {\bibinfo {author} {\bibfnamefont {G.~A.}\ \bibnamefont
  {White}}, \bibinfo {author} {\bibfnamefont {P.}~\bibnamefont {Jurcevic}},
  \bibinfo {author} {\bibfnamefont {C.~D.}\ \bibnamefont {Hill}}, \ and\
  \bibinfo {author} {\bibfnamefont {K.}~\bibnamefont {Modi}},\ }\href@noop {}
  {\bibfield  {journal} {\bibinfo  {journal} {arXiv preprint arXiv:2312.08454}\
  } (\bibinfo {year} {2023})}\BibitemShut {NoStop}%
\bibitem [{Note12()}]{Note12}%
  \BibitemOpen
  \bibinfo {note} {Here, we use the term ``metric'' loosely. For example, we
  discuss different types of fidelity in this section, but fidelity is strictly
  not a metric in a mathematical sense, as it does not obey the triangle
  inequality.}\BibitemShut {Stop}%
\bibitem [{\citenamefont {Bhattacharyya}(1943)}]{bhattacharyya1943measure}%
  \BibitemOpen
  \bibfield  {author} {\bibinfo {author} {\bibfnamefont {A.}~\bibnamefont
  {Bhattacharyya}},\ }\href@noop {} {\bibfield  {journal} {\bibinfo  {journal}
  {Bulletin of the Calcutta Mathematical Society}\ }\textbf {\bibinfo {volume}
  {35}},\ \bibinfo {pages} {99} (\bibinfo {year} {1943})}\BibitemShut {NoStop}%
\bibitem [{\citenamefont {Fuchs}(1996)}]{fuchs1996distinguishability}%
  \BibitemOpen
  \bibfield  {author} {\bibinfo {author} {\bibfnamefont {C.~A.}\ \bibnamefont
  {Fuchs}},\ }\href@noop {} {\bibfield  {journal} {\bibinfo  {journal} {arXiv
  preprint quant-ph/9601020}\ } (\bibinfo {year} {1996})}\BibitemShut {NoStop}%
\bibitem [{\citenamefont {Fuchs}\ and\ \citenamefont {Van
  De~Graaf}(1999)}]{fuchs1999cryptographic}%
  \BibitemOpen
  \bibfield  {author} {\bibinfo {author} {\bibfnamefont {C.~A.}\ \bibnamefont
  {Fuchs}}\ and\ \bibinfo {author} {\bibfnamefont {J.}~\bibnamefont {Van
  De~Graaf}},\ }\href@noop {} {\bibfield  {journal} {\bibinfo  {journal} {IEEE
  Transactions on Information Theory}\ }\textbf {\bibinfo {volume} {45}},\
  \bibinfo {pages} {1216} (\bibinfo {year} {1999})}\BibitemShut {NoStop}%
\bibitem [{Note13()}]{Note13}%
  \BibitemOpen
  \bibinfo {note} {The logarithm that appears in entropic quantities can be
  evaluated in any base; using $\log _2$ yields \protect \textit {bits} of
  entropy, while $\ln $ yields units called \protect \textit
  {nats}.}\BibitemShut {Stop}%
\bibitem [{Note14()}]{Note14}%
  \BibitemOpen
  \bibinfo {note} {Heavy output probability is not necessarily well-defined for
  highly degenerate distributions.}\BibitemShut {Stop}%
\bibitem [{\citenamefont {Boixo}\ \emph {et~al.}(2018)\citenamefont {Boixo},
  \citenamefont {Isakov}, \citenamefont {Smelyanskiy}, \citenamefont {Babbush},
  \citenamefont {Ding}, \citenamefont {Jiang}, \citenamefont {Bremner},
  \citenamefont {Martinis},\ and\ \citenamefont
  {Neven}}]{boixo2018characterizing}%
  \BibitemOpen
  \bibfield  {author} {\bibinfo {author} {\bibfnamefont {S.}~\bibnamefont
  {Boixo}}, \bibinfo {author} {\bibfnamefont {S.~V.}\ \bibnamefont {Isakov}},
  \bibinfo {author} {\bibfnamefont {V.~N.}\ \bibnamefont {Smelyanskiy}},
  \bibinfo {author} {\bibfnamefont {R.}~\bibnamefont {Babbush}}, \bibinfo
  {author} {\bibfnamefont {N.}~\bibnamefont {Ding}}, \bibinfo {author}
  {\bibfnamefont {Z.}~\bibnamefont {Jiang}}, \bibinfo {author} {\bibfnamefont
  {M.~J.}\ \bibnamefont {Bremner}}, \bibinfo {author} {\bibfnamefont {J.~M.}\
  \bibnamefont {Martinis}}, \ and\ \bibinfo {author} {\bibfnamefont
  {H.}~\bibnamefont {Neven}},\ }\href@noop {} {\bibfield  {journal} {\bibinfo
  {journal} {Nature Physics}\ }\textbf {\bibinfo {volume} {14}},\ \bibinfo
  {pages} {595} (\bibinfo {year} {2018})}\BibitemShut {NoStop}%
\bibitem [{\citenamefont {Cross}\ \emph {et~al.}(2019)\citenamefont {Cross},
  \citenamefont {Bishop}, \citenamefont {Sheldon}, \citenamefont {Nation},\
  and\ \citenamefont {Gambetta}}]{cross2019validating}%
  \BibitemOpen
  \bibfield  {author} {\bibinfo {author} {\bibfnamefont {A.~W.}\ \bibnamefont
  {Cross}}, \bibinfo {author} {\bibfnamefont {L.~S.}\ \bibnamefont {Bishop}},
  \bibinfo {author} {\bibfnamefont {S.}~\bibnamefont {Sheldon}}, \bibinfo
  {author} {\bibfnamefont {P.~D.}\ \bibnamefont {Nation}}, \ and\ \bibinfo
  {author} {\bibfnamefont {J.~M.}\ \bibnamefont {Gambetta}},\ }\href@noop {}
  {\bibfield  {journal} {\bibinfo  {journal} {Physical Review A}\ }\textbf
  {\bibinfo {volume} {100}},\ \bibinfo {pages} {032328} (\bibinfo {year}
  {2019})}\BibitemShut {NoStop}%
\bibitem [{\citenamefont {Helstrom}(1969)}]{helstrom1969quantum}%
  \BibitemOpen
  \bibfield  {author} {\bibinfo {author} {\bibfnamefont {C.~W.}\ \bibnamefont
  {Helstrom}},\ }\href@noop {} {\bibfield  {journal} {\bibinfo  {journal}
  {Journal of Statistical Physics}\ }\textbf {\bibinfo {volume} {1}},\ \bibinfo
  {pages} {231} (\bibinfo {year} {1969})}\BibitemShut {NoStop}%
\bibitem [{Note15()}]{Note15}%
  \BibitemOpen
  \bibinfo {note} {See, for example, the book \protect \textit {Quantum
  Computation and Quantum Information} \cite {nielsen2002quantum}.}\BibitemShut
  {Stop}%
\bibitem [{\citenamefont {Schumacher}(1995)}]{schumacher1995quantum}%
  \BibitemOpen
  \bibfield  {author} {\bibinfo {author} {\bibfnamefont {B.}~\bibnamefont
  {Schumacher}},\ }\href@noop {} {\bibfield  {journal} {\bibinfo  {journal}
  {Physical Review A}\ }\textbf {\bibinfo {volume} {51}},\ \bibinfo {pages}
  {2738} (\bibinfo {year} {1995})}\BibitemShut {NoStop}%
\bibitem [{\citenamefont {Uhlmann}(1976)}]{uhlmann1976transition}%
  \BibitemOpen
  \bibfield  {author} {\bibinfo {author} {\bibfnamefont {A.}~\bibnamefont
  {Uhlmann}},\ }\href@noop {} {\bibfield  {journal} {\bibinfo  {journal}
  {Reports on Mathematical Physics}\ }\textbf {\bibinfo {volume} {9}},\
  \bibinfo {pages} {273} (\bibinfo {year} {1976})}\BibitemShut {NoStop}%
\bibitem [{\citenamefont {Jozsa}(1994)}]{jozsa1994fidelity}%
  \BibitemOpen
  \bibfield  {author} {\bibinfo {author} {\bibfnamefont {R.}~\bibnamefont
  {Jozsa}},\ }\href@noop {} {\bibfield  {journal} {\bibinfo  {journal} {Journal
  of modern optics}\ }\textbf {\bibinfo {volume} {41}},\ \bibinfo {pages}
  {2315} (\bibinfo {year} {1994})}\BibitemShut {NoStop}%
\bibitem [{Note16()}]{Note16}%
  \BibitemOpen
  \bibinfo {note} {It is worth emphasizing that a \protect \emph {quantum
  process} is not analogous to a classical \protect \emph {stochastic process}.
  The classical analogue of a quantum process is a \protect \emph {stochastic
  matrix} (see Sec.~\ref {sec:truthtable}), which is related to stochastic
  processes, but quite distinct.}\BibitemShut {Stop}%
\bibitem [{\citenamefont {Kitaev}(1997)}]{kitaev1997quantum}%
  \BibitemOpen
  \bibfield  {author} {\bibinfo {author} {\bibfnamefont {A.~Y.}\ \bibnamefont
  {Kitaev}},\ }\href@noop {} {\bibfield  {journal} {\bibinfo  {journal}
  {Uspekhi Matematicheskikh Nauk}\ }\textbf {\bibinfo {volume} {52}},\ \bibinfo
  {pages} {53} (\bibinfo {year} {1997})}\BibitemShut {NoStop}%
\bibitem [{\citenamefont {Bennett}\ and\ \citenamefont
  {Wiesner}(1992)}]{BennettPRL92}%
  \BibitemOpen
  \bibfield  {author} {\bibinfo {author} {\bibfnamefont {C.~H.}\ \bibnamefont
  {Bennett}}\ and\ \bibinfo {author} {\bibfnamefont {S.~J.}\ \bibnamefont
  {Wiesner}},\ }\href@noop {} {\bibfield  {journal} {\bibinfo  {journal}
  {Physical review letters}\ }\textbf {\bibinfo {volume} {69}},\ \bibinfo
  {pages} {2881} (\bibinfo {year} {1992})}\BibitemShut {NoStop}%
\bibitem [{\citenamefont {Zurek}(2003)}]{ZurekPRL03}%
  \BibitemOpen
  \bibfield  {author} {\bibinfo {author} {\bibfnamefont {W.~H.}\ \bibnamefont
  {Zurek}},\ }\href@noop {} {\bibfield  {journal} {\bibinfo  {journal}
  {Physical review letters}\ }\textbf {\bibinfo {volume} {90}},\ \bibinfo
  {pages} {120404} (\bibinfo {year} {2003})}\BibitemShut {NoStop}%
\bibitem [{\citenamefont {Bennett}\ \emph {et~al.}(1993)\citenamefont
  {Bennett}, \citenamefont {Brassard}, \citenamefont {Cr{\'e}peau},
  \citenamefont {Jozsa}, \citenamefont {Peres},\ and\ \citenamefont
  {Wootters}}]{BennettPRL93}%
  \BibitemOpen
  \bibfield  {author} {\bibinfo {author} {\bibfnamefont {C.~H.}\ \bibnamefont
  {Bennett}}, \bibinfo {author} {\bibfnamefont {G.}~\bibnamefont {Brassard}},
  \bibinfo {author} {\bibfnamefont {C.}~\bibnamefont {Cr{\'e}peau}}, \bibinfo
  {author} {\bibfnamefont {R.}~\bibnamefont {Jozsa}}, \bibinfo {author}
  {\bibfnamefont {A.}~\bibnamefont {Peres}}, \ and\ \bibinfo {author}
  {\bibfnamefont {W.~K.}\ \bibnamefont {Wootters}},\ }\href@noop {} {\bibfield
  {journal} {\bibinfo  {journal} {Physical review letters}\ }\textbf {\bibinfo
  {volume} {70}},\ \bibinfo {pages} {1895} (\bibinfo {year}
  {1993})}\BibitemShut {NoStop}%
\bibitem [{\citenamefont {Aharonov}\ \emph {et~al.}(1998)\citenamefont
  {Aharonov}, \citenamefont {Kitaev},\ and\ \citenamefont
  {Nisan}}]{aharonov1998quantum}%
  \BibitemOpen
  \bibfield  {author} {\bibinfo {author} {\bibfnamefont {D.}~\bibnamefont
  {Aharonov}}, \bibinfo {author} {\bibfnamefont {A.}~\bibnamefont {Kitaev}}, \
  and\ \bibinfo {author} {\bibfnamefont {N.}~\bibnamefont {Nisan}},\ }in\
  \href@noop {} {\emph {\bibinfo {booktitle} {Proceedings of the thirtieth
  annual ACM symposium on Theory of computing}}}\ (\bibinfo {year} {1998})\
  pp.\ \bibinfo {pages} {20--30}\BibitemShut {NoStop}%
\bibitem [{\citenamefont {Viola}\ and\ \citenamefont
  {Lloyd}(1998)}]{viola1998dynamical}%
  \BibitemOpen
  \bibfield  {author} {\bibinfo {author} {\bibfnamefont {L.}~\bibnamefont
  {Viola}}\ and\ \bibinfo {author} {\bibfnamefont {S.}~\bibnamefont {Lloyd}},\
  }\href@noop {} {\bibfield  {journal} {\bibinfo  {journal} {Physical Review
  A}\ }\textbf {\bibinfo {volume} {58}},\ \bibinfo {pages} {2733} (\bibinfo
  {year} {1998})}\BibitemShut {NoStop}%
\bibitem [{\citenamefont {Viola}\ \emph {et~al.}(1999)\citenamefont {Viola},
  \citenamefont {Knill},\ and\ \citenamefont {Lloyd}}]{viola1999dynamical}%
  \BibitemOpen
  \bibfield  {author} {\bibinfo {author} {\bibfnamefont {L.}~\bibnamefont
  {Viola}}, \bibinfo {author} {\bibfnamefont {E.}~\bibnamefont {Knill}}, \ and\
  \bibinfo {author} {\bibfnamefont {S.}~\bibnamefont {Lloyd}},\ }\href@noop {}
  {\bibfield  {journal} {\bibinfo  {journal} {Physical Review Letters}\
  }\textbf {\bibinfo {volume} {82}},\ \bibinfo {pages} {2417} (\bibinfo {year}
  {1999})}\BibitemShut {NoStop}%
\bibitem [{\citenamefont {Knill}(2004)}]{knill2004fault}%
  \BibitemOpen
  \bibfield  {author} {\bibinfo {author} {\bibfnamefont {E.}~\bibnamefont
  {Knill}},\ }\href@noop {} {\bibfield  {journal} {\bibinfo  {journal} {arXiv
  preprint quant-ph/0404104}\ } (\bibinfo {year} {2004})}\BibitemShut {NoStop}%
\bibitem [{\citenamefont {Kern}\ \emph {et~al.}(2005)\citenamefont {Kern},
  \citenamefont {Alber},\ and\ \citenamefont {Shepelyansky}}]{kern2005quantum}%
  \BibitemOpen
  \bibfield  {author} {\bibinfo {author} {\bibfnamefont {O.}~\bibnamefont
  {Kern}}, \bibinfo {author} {\bibfnamefont {G.}~\bibnamefont {Alber}}, \ and\
  \bibinfo {author} {\bibfnamefont {D.~L.}\ \bibnamefont {Shepelyansky}},\
  }\href@noop {} {\bibfield  {journal} {\bibinfo  {journal} {The European
  Physical Journal D-Atomic, Molecular, Optical and Plasma Physics}\ }\textbf
  {\bibinfo {volume} {32}},\ \bibinfo {pages} {153} (\bibinfo {year}
  {2005})}\BibitemShut {NoStop}%
\bibitem [{\citenamefont {Ware}\ \emph {et~al.}(2021)\citenamefont {Ware},
  \citenamefont {Ribeill}, \citenamefont {Riste}, \citenamefont {Ryan},
  \citenamefont {Johnson},\ and\ \citenamefont
  {Da~Silva}}]{ware2021experimental}%
  \BibitemOpen
  \bibfield  {author} {\bibinfo {author} {\bibfnamefont {M.}~\bibnamefont
  {Ware}}, \bibinfo {author} {\bibfnamefont {G.}~\bibnamefont {Ribeill}},
  \bibinfo {author} {\bibfnamefont {D.}~\bibnamefont {Riste}}, \bibinfo
  {author} {\bibfnamefont {C.~A.}\ \bibnamefont {Ryan}}, \bibinfo {author}
  {\bibfnamefont {B.}~\bibnamefont {Johnson}}, \ and\ \bibinfo {author}
  {\bibfnamefont {M.~P.}\ \bibnamefont {Da~Silva}},\ }\href@noop {} {\bibfield
  {journal} {\bibinfo  {journal} {Physical Review A}\ }\textbf {\bibinfo
  {volume} {103}},\ \bibinfo {pages} {042604} (\bibinfo {year}
  {2021})}\BibitemShut {NoStop}%
\bibitem [{\citenamefont {Wallman}\ and\ \citenamefont
  {Emerson}(2016)}]{wallman2016noise}%
  \BibitemOpen
  \bibfield  {author} {\bibinfo {author} {\bibfnamefont {J.~J.}\ \bibnamefont
  {Wallman}}\ and\ \bibinfo {author} {\bibfnamefont {J.}~\bibnamefont
  {Emerson}},\ }\href {\doibase 10.1103/PhysRevA.94.052325} {\bibfield
  {journal} {\bibinfo  {journal} {Phys. Rev. A}\ }\textbf {\bibinfo {volume}
  {94}},\ \bibinfo {pages} {052325} (\bibinfo {year} {2016})}\BibitemShut
  {NoStop}%
\bibitem [{\citenamefont {Hashim}\ \emph {et~al.}(2021)\citenamefont {Hashim},
  \citenamefont {Naik}, \citenamefont {Morvan}, \citenamefont {Ville},
  \citenamefont {Mitchell}, \citenamefont {Kreikebaum}, \citenamefont {Davis},
  \citenamefont {Smith}, \citenamefont {Iancu}, \citenamefont {O'Brien},
  \citenamefont {Hincks}, \citenamefont {Wallman}, \citenamefont {Emerson},\
  and\ \citenamefont {Siddiqi}}]{hashim2021randomized}%
  \BibitemOpen
  \bibfield  {author} {\bibinfo {author} {\bibfnamefont {A.}~\bibnamefont
  {Hashim}}, \bibinfo {author} {\bibfnamefont {R.~K.}\ \bibnamefont {Naik}},
  \bibinfo {author} {\bibfnamefont {A.}~\bibnamefont {Morvan}}, \bibinfo
  {author} {\bibfnamefont {J.-L.}\ \bibnamefont {Ville}}, \bibinfo {author}
  {\bibfnamefont {B.}~\bibnamefont {Mitchell}}, \bibinfo {author}
  {\bibfnamefont {J.~M.}\ \bibnamefont {Kreikebaum}}, \bibinfo {author}
  {\bibfnamefont {M.}~\bibnamefont {Davis}}, \bibinfo {author} {\bibfnamefont
  {E.}~\bibnamefont {Smith}}, \bibinfo {author} {\bibfnamefont
  {C.}~\bibnamefont {Iancu}}, \bibinfo {author} {\bibfnamefont {K.~P.}\
  \bibnamefont {O'Brien}}, \bibinfo {author} {\bibfnamefont {I.}~\bibnamefont
  {Hincks}}, \bibinfo {author} {\bibfnamefont {J.~J.}\ \bibnamefont {Wallman}},
  \bibinfo {author} {\bibfnamefont {J.}~\bibnamefont {Emerson}}, \ and\
  \bibinfo {author} {\bibfnamefont {I.}~\bibnamefont {Siddiqi}},\ }\href
  {\doibase 10.1103/PhysRevX.11.041039} {\bibfield  {journal} {\bibinfo
  {journal} {Phys. Rev. X}\ }\textbf {\bibinfo {volume} {11}},\ \bibinfo
  {pages} {041039} (\bibinfo {year} {2021})}\BibitemShut {NoStop}%
\bibitem [{\citenamefont {Nielsen}\ \emph
  {et~al.}(2021{\natexlab{b}})\citenamefont {Nielsen}, \citenamefont {Gamble},
  \citenamefont {Rudinger}, \citenamefont {Scholten}, \citenamefont {Young},\
  and\ \citenamefont {Blume-Kohout}}]{nielsen2021gate}%
  \BibitemOpen
  \bibfield  {author} {\bibinfo {author} {\bibfnamefont {E.}~\bibnamefont
  {Nielsen}}, \bibinfo {author} {\bibfnamefont {J.~K.}\ \bibnamefont {Gamble}},
  \bibinfo {author} {\bibfnamefont {K.}~\bibnamefont {Rudinger}}, \bibinfo
  {author} {\bibfnamefont {T.}~\bibnamefont {Scholten}}, \bibinfo {author}
  {\bibfnamefont {K.}~\bibnamefont {Young}}, \ and\ \bibinfo {author}
  {\bibfnamefont {R.}~\bibnamefont {Blume-Kohout}},\ }\href@noop {} {\bibfield
  {journal} {\bibinfo  {journal} {Quantum}\ }\textbf {\bibinfo {volume} {5}},\
  \bibinfo {pages} {557} (\bibinfo {year} {2021}{\natexlab{b}})}\BibitemShut
  {NoStop}%
\bibitem [{\citenamefont {Aliferis}\ \emph {et~al.}(2006)\citenamefont
  {Aliferis}, \citenamefont {Gottesman},\ and\ \citenamefont
  {Preskill}}]{aliferis2006quantum}%
  \BibitemOpen
  \bibfield  {author} {\bibinfo {author} {\bibfnamefont {P.}~\bibnamefont
  {Aliferis}}, \bibinfo {author} {\bibfnamefont {D.}~\bibnamefont {Gottesman}},
  \ and\ \bibinfo {author} {\bibfnamefont {J.}~\bibnamefont {Preskill}},\
  }\href@noop {} {\bibfield  {journal} {\bibinfo  {journal} {Quantum
  Information \& Computation}\ }\textbf {\bibinfo {volume} {6}},\ \bibinfo
  {pages} {97} (\bibinfo {year} {2006})}\BibitemShut {NoStop}%
\bibitem [{\citenamefont {Sanders}\ \emph {et~al.}(2015)\citenamefont
  {Sanders}, \citenamefont {Wallman},\ and\ \citenamefont
  {Sanders}}]{sanders2015bounding}%
  \BibitemOpen
  \bibfield  {author} {\bibinfo {author} {\bibfnamefont {Y.~R.}\ \bibnamefont
  {Sanders}}, \bibinfo {author} {\bibfnamefont {J.~J.}\ \bibnamefont
  {Wallman}}, \ and\ \bibinfo {author} {\bibfnamefont {B.~C.}\ \bibnamefont
  {Sanders}},\ }\href@noop {} {\bibfield  {journal} {\bibinfo  {journal} {New
  Journal of Physics}\ }\textbf {\bibinfo {volume} {18}},\ \bibinfo {pages}
  {012002} (\bibinfo {year} {2015})}\BibitemShut {NoStop}%
\bibitem [{\citenamefont {Nielsen}(2002)}]{nielsen2002simple}%
  \BibitemOpen
  \bibfield  {author} {\bibinfo {author} {\bibfnamefont {M.~A.}\ \bibnamefont
  {Nielsen}},\ }\href@noop {} {\bibfield  {journal} {\bibinfo  {journal}
  {Physics Letters A}\ }\textbf {\bibinfo {volume} {303}},\ \bibinfo {pages}
  {249} (\bibinfo {year} {2002})}\BibitemShut {NoStop}%
\bibitem [{\citenamefont {Emerson}\ \emph {et~al.}(2005)\citenamefont
  {Emerson}, \citenamefont {Alicki},\ and\ \citenamefont
  {{\.Z}yczkowski}}]{emerson2005scalable}%
  \BibitemOpen
  \bibfield  {author} {\bibinfo {author} {\bibfnamefont {J.}~\bibnamefont
  {Emerson}}, \bibinfo {author} {\bibfnamefont {R.}~\bibnamefont {Alicki}}, \
  and\ \bibinfo {author} {\bibfnamefont {K.}~\bibnamefont {{\.Z}yczkowski}},\
  }\href@noop {} {\bibfield  {journal} {\bibinfo  {journal} {Journal of Optics
  B: Quantum and Semiclassical Optics}\ }\textbf {\bibinfo {volume} {7}},\
  \bibinfo {pages} {S347} (\bibinfo {year} {2005})}\BibitemShut {NoStop}%
\bibitem [{\citenamefont {Magesan}\ \emph
  {et~al.}(2011{\natexlab{a}})\citenamefont {Magesan}, \citenamefont
  {Blume-Kohout},\ and\ \citenamefont {Emerson}}]{magesan2011gate}%
  \BibitemOpen
  \bibfield  {author} {\bibinfo {author} {\bibfnamefont {E.}~\bibnamefont
  {Magesan}}, \bibinfo {author} {\bibfnamefont {R.}~\bibnamefont
  {Blume-Kohout}}, \ and\ \bibinfo {author} {\bibfnamefont {J.}~\bibnamefont
  {Emerson}},\ }\href@noop {} {\bibfield  {journal} {\bibinfo  {journal}
  {Physical Review A}\ }\textbf {\bibinfo {volume} {84}},\ \bibinfo {pages}
  {012309} (\bibinfo {year} {2011}{\natexlab{a}})}\BibitemShut {NoStop}%
\bibitem [{\citenamefont {Horodecki}\ \emph {et~al.}(1999)\citenamefont
  {Horodecki}, \citenamefont {Horodecki},\ and\ \citenamefont
  {Horodecki}}]{horodecki1999general}%
  \BibitemOpen
  \bibfield  {author} {\bibinfo {author} {\bibfnamefont {M.}~\bibnamefont
  {Horodecki}}, \bibinfo {author} {\bibfnamefont {P.}~\bibnamefont
  {Horodecki}}, \ and\ \bibinfo {author} {\bibfnamefont {R.}~\bibnamefont
  {Horodecki}},\ }\href@noop {} {\bibfield  {journal} {\bibinfo  {journal}
  {Physical Review A}\ }\textbf {\bibinfo {volume} {60}},\ \bibinfo {pages}
  {1888} (\bibinfo {year} {1999})}\BibitemShut {NoStop}%
\bibitem [{\citenamefont {Schumacher}(1996)}]{schumacher1996sending}%
  \BibitemOpen
  \bibfield  {author} {\bibinfo {author} {\bibfnamefont {B.}~\bibnamefont
  {Schumacher}},\ }\href@noop {} {\bibfield  {journal} {\bibinfo  {journal}
  {Physical Review A}\ }\textbf {\bibinfo {volume} {54}},\ \bibinfo {pages}
  {2614} (\bibinfo {year} {1996})}\BibitemShut {NoStop}%
\bibitem [{\citenamefont {Nielsen}(1996)}]{nielsen1996entanglement}%
  \BibitemOpen
  \bibfield  {author} {\bibinfo {author} {\bibfnamefont {M.~A.}\ \bibnamefont
  {Nielsen}},\ }\href@noop {} {\bibfield  {journal} {\bibinfo  {journal} {arXiv
  preprint quant-ph/9606012}\ } (\bibinfo {year} {1996})}\BibitemShut {NoStop}%
\bibitem [{\citenamefont {Gilchrist}\ \emph {et~al.}(2005)\citenamefont
  {Gilchrist}, \citenamefont {Langford},\ and\ \citenamefont
  {Nielsen}}]{gilchrist2005distance}%
  \BibitemOpen
  \bibfield  {author} {\bibinfo {author} {\bibfnamefont {A.}~\bibnamefont
  {Gilchrist}}, \bibinfo {author} {\bibfnamefont {N.~K.}\ \bibnamefont
  {Langford}}, \ and\ \bibinfo {author} {\bibfnamefont {M.~A.}\ \bibnamefont
  {Nielsen}},\ }\href@noop {} {\bibfield  {journal} {\bibinfo  {journal}
  {Physical Review A}\ }\textbf {\bibinfo {volume} {71}},\ \bibinfo {pages}
  {062310} (\bibinfo {year} {2005})}\BibitemShut {NoStop}%
\bibitem [{\citenamefont {Carignan-Dugas}(2019)}]{carignan2019walk}%
  \BibitemOpen
  \bibfield  {author} {\bibinfo {author} {\bibfnamefont {A.}~\bibnamefont
  {Carignan-Dugas}},\ }\emph {\bibinfo {title} {A walk through quantum noise: a
  study of error signatures and characterization methods}},\ \href@noop {}
  {Ph.D. thesis},\ \bibinfo  {school} {University of Waterloo} (\bibinfo {year}
  {2019})\BibitemShut {NoStop}%
\bibitem [{\citenamefont {Blume-Kohout}\ \emph {et~al.}(2017)\citenamefont
  {Blume-Kohout}, \citenamefont {Gamble}, \citenamefont {Nielsen},
  \citenamefont {Rudinger}, \citenamefont {Mizrahi}, \citenamefont {Fortier},\
  and\ \citenamefont {Maunz}}]{Blume_Kohout_2017}%
  \BibitemOpen
  \bibfield  {author} {\bibinfo {author} {\bibfnamefont {R.}~\bibnamefont
  {Blume-Kohout}}, \bibinfo {author} {\bibfnamefont {J.~K.}\ \bibnamefont
  {Gamble}}, \bibinfo {author} {\bibfnamefont {E.}~\bibnamefont {Nielsen}},
  \bibinfo {author} {\bibfnamefont {K.}~\bibnamefont {Rudinger}}, \bibinfo
  {author} {\bibfnamefont {J.}~\bibnamefont {Mizrahi}}, \bibinfo {author}
  {\bibfnamefont {K.}~\bibnamefont {Fortier}}, \ and\ \bibinfo {author}
  {\bibfnamefont {P.}~\bibnamefont {Maunz}},\ }\href {\doibase
  10.1038/ncomms14485} {\bibfield  {journal} {\bibinfo  {journal} {Nature
  Communications}\ }\textbf {\bibinfo {volume} {8}} (\bibinfo {year} {2017}),\
  10.1038/ncomms14485}\BibitemShut {NoStop}%
\bibitem [{\citenamefont {Watrous}(2009)}]{watrous2009semidefinite}%
  \BibitemOpen
  \bibfield  {author} {\bibinfo {author} {\bibfnamefont {J.}~\bibnamefont
  {Watrous}},\ }\href@noop {} {\bibfield  {journal} {\bibinfo  {journal}
  {Theory of Computing}\ }\textbf {\bibinfo {volume} {5}},\ \bibinfo {pages}
  {217} (\bibinfo {year} {2009})}\BibitemShut {NoStop}%
\bibitem [{\citenamefont {Watrous}(2012)}]{watrous2012simpler}%
  \BibitemOpen
  \bibfield  {author} {\bibinfo {author} {\bibfnamefont {J.}~\bibnamefont
  {Watrous}},\ }\href@noop {} {\bibfield  {journal} {\bibinfo  {journal} {arXiv
  preprint arXiv:1207.5726}\ } (\bibinfo {year} {2012})}\BibitemShut {NoStop}%
\bibitem [{\citenamefont {Wallman}\ and\ \citenamefont
  {Flammia}(2014)}]{wallman2014randomized}%
  \BibitemOpen
  \bibfield  {author} {\bibinfo {author} {\bibfnamefont {J.~J.}\ \bibnamefont
  {Wallman}}\ and\ \bibinfo {author} {\bibfnamefont {S.~T.}\ \bibnamefont
  {Flammia}},\ }\href@noop {} {\bibfield  {journal} {\bibinfo  {journal} {New
  Journal of Physics}\ }\textbf {\bibinfo {volume} {16}},\ \bibinfo {pages}
  {103032} (\bibinfo {year} {2014})}\BibitemShut {NoStop}%
\bibitem [{\citenamefont {Wallman}(2015)}]{wallman2015bounding}%
  \BibitemOpen
  \bibfield  {author} {\bibinfo {author} {\bibfnamefont {J.~J.}\ \bibnamefont
  {Wallman}},\ }\href@noop {} {\bibfield  {journal} {\bibinfo  {journal} {arXiv
  preprint arXiv:1511.00727}\ } (\bibinfo {year} {2015})}\BibitemShut {NoStop}%
\bibitem [{\citenamefont {Kueng}\ \emph {et~al.}(2016)\citenamefont {Kueng},
  \citenamefont {Long}, \citenamefont {Doherty},\ and\ \citenamefont
  {Flammia}}]{kueng2016comparing}%
  \BibitemOpen
  \bibfield  {author} {\bibinfo {author} {\bibfnamefont {R.}~\bibnamefont
  {Kueng}}, \bibinfo {author} {\bibfnamefont {D.~M.}\ \bibnamefont {Long}},
  \bibinfo {author} {\bibfnamefont {A.~C.}\ \bibnamefont {Doherty}}, \ and\
  \bibinfo {author} {\bibfnamefont {S.~T.}\ \bibnamefont {Flammia}},\
  }\href@noop {} {\bibfield  {journal} {\bibinfo  {journal} {Physical review
  letters}\ }\textbf {\bibinfo {volume} {117}},\ \bibinfo {pages} {170502}
  (\bibinfo {year} {2016})}\BibitemShut {NoStop}%
\bibitem [{\citenamefont {Luis}\ and\ \citenamefont
  {S{\'a}nchez-Soto}(1999)}]{Luis1999-ek}%
  \BibitemOpen
  \bibfield  {author} {\bibinfo {author} {\bibfnamefont {A.}~\bibnamefont
  {Luis}}\ and\ \bibinfo {author} {\bibfnamefont {L.~L.}\ \bibnamefont
  {S{\'a}nchez-Soto}},\ }\href@noop {} {\bibfield  {journal} {\bibinfo
  {journal} {Phys. Rev. Lett.}\ }\textbf {\bibinfo {volume} {83}},\ \bibinfo
  {pages} {3573} (\bibinfo {year} {1999})}\BibitemShut {NoStop}%
\bibitem [{\citenamefont {Ji}\ \emph {et~al.}(2006)\citenamefont {Ji},
  \citenamefont {Feng}, \citenamefont {Duan},\ and\ \citenamefont
  {Ying}}]{Ji2006-fp}%
  \BibitemOpen
  \bibfield  {author} {\bibinfo {author} {\bibfnamefont {Z.}~\bibnamefont
  {Ji}}, \bibinfo {author} {\bibfnamefont {Y.}~\bibnamefont {Feng}}, \bibinfo
  {author} {\bibfnamefont {R.}~\bibnamefont {Duan}}, \ and\ \bibinfo {author}
  {\bibfnamefont {M.}~\bibnamefont {Ying}},\ }\href {\doibase
  10.1103/PhysRevLett.96.200401} {\bibfield  {journal} {\bibinfo  {journal}
  {Phys. Rev. Lett.}\ }\textbf {\bibinfo {volume} {96}},\ \bibinfo {pages}
  {200401} (\bibinfo {year} {2006})}\BibitemShut {NoStop}%
\bibitem [{\citenamefont {Magesan}\ and\ \citenamefont
  {Cappellaro}(2013)}]{Magesan2013-vy}%
  \BibitemOpen
  \bibfield  {author} {\bibinfo {author} {\bibfnamefont {E.}~\bibnamefont
  {Magesan}}\ and\ \bibinfo {author} {\bibfnamefont {P.}~\bibnamefont
  {Cappellaro}},\ }\href {\doibase 10.1103/PhysRevA.88.022127} {\bibfield
  {journal} {\bibinfo  {journal} {Phys. Rev. A}\ }\textbf {\bibinfo {volume}
  {88}},\ \bibinfo {pages} {022127} (\bibinfo {year} {2013})},\ \Eprint
  {http://arxiv.org/abs/1301.5037} {arXiv:1301.5037 [quant-ph]} \BibitemShut
  {NoStop}%
\bibitem [{\citenamefont {Dressel}\ \emph {et~al.}(2014)\citenamefont
  {Dressel}, \citenamefont {Brun},\ and\ \citenamefont
  {Korotkov}}]{Dressel2014-rb}%
  \BibitemOpen
  \bibfield  {author} {\bibinfo {author} {\bibfnamefont {J.}~\bibnamefont
  {Dressel}}, \bibinfo {author} {\bibfnamefont {T.~A.}\ \bibnamefont {Brun}}, \
  and\ \bibinfo {author} {\bibfnamefont {A.~N.}\ \bibnamefont {Korotkov}},\
  }\href {\doibase 10.1103/PhysRevA.90.032302} {\bibfield  {journal} {\bibinfo
  {journal} {Phys. Rev. A}\ }\textbf {\bibinfo {volume} {90}},\ \bibinfo
  {pages} {032302} (\bibinfo {year} {2014})}\BibitemShut {NoStop}%
\bibitem [{\citenamefont {Blumoff}\ \emph {et~al.}(2016)\citenamefont
  {Blumoff}, \citenamefont {Chou}, \citenamefont {Shen}, \citenamefont
  {Reagor}, \citenamefont {Axline}, \citenamefont {Brierley}, \citenamefont
  {Silveri}, \citenamefont {Wang}, \citenamefont {Vlastakis}, \citenamefont
  {Nigg} \emph {et~al.}}]{blumoff2016implementing}%
  \BibitemOpen
  \bibfield  {author} {\bibinfo {author} {\bibfnamefont {J.~Z.}\ \bibnamefont
  {Blumoff}}, \bibinfo {author} {\bibfnamefont {K.}~\bibnamefont {Chou}},
  \bibinfo {author} {\bibfnamefont {C.}~\bibnamefont {Shen}}, \bibinfo {author}
  {\bibfnamefont {M.}~\bibnamefont {Reagor}}, \bibinfo {author} {\bibfnamefont
  {C.}~\bibnamefont {Axline}}, \bibinfo {author} {\bibfnamefont
  {R.}~\bibnamefont {Brierley}}, \bibinfo {author} {\bibfnamefont
  {M.}~\bibnamefont {Silveri}}, \bibinfo {author} {\bibfnamefont
  {C.}~\bibnamefont {Wang}}, \bibinfo {author} {\bibfnamefont {B.}~\bibnamefont
  {Vlastakis}}, \bibinfo {author} {\bibfnamefont {S.~E.}\ \bibnamefont {Nigg}},
   \emph {et~al.},\ }\href@noop {} {\bibfield  {journal} {\bibinfo  {journal}
  {Physical Review X}\ }\textbf {\bibinfo {volume} {6}},\ \bibinfo {pages}
  {031041} (\bibinfo {year} {2016})}\BibitemShut {NoStop}%
\bibitem [{\citenamefont {Mallet}\ \emph {et~al.}(2009)\citenamefont {Mallet},
  \citenamefont {Ong}, \citenamefont {Palacios-Laloy}, \citenamefont {Nguyen},
  \citenamefont {Bertet}, \citenamefont {Vion},\ and\ \citenamefont
  {Esteve}}]{mallet2009single}%
  \BibitemOpen
  \bibfield  {author} {\bibinfo {author} {\bibfnamefont {F.}~\bibnamefont
  {Mallet}}, \bibinfo {author} {\bibfnamefont {F.~R.}\ \bibnamefont {Ong}},
  \bibinfo {author} {\bibfnamefont {A.}~\bibnamefont {Palacios-Laloy}},
  \bibinfo {author} {\bibfnamefont {F.}~\bibnamefont {Nguyen}}, \bibinfo
  {author} {\bibfnamefont {P.}~\bibnamefont {Bertet}}, \bibinfo {author}
  {\bibfnamefont {D.}~\bibnamefont {Vion}}, \ and\ \bibinfo {author}
  {\bibfnamefont {D.}~\bibnamefont {Esteve}},\ }\href@noop {} {\bibfield
  {journal} {\bibinfo  {journal} {Nature Physics}\ }\textbf {\bibinfo {volume}
  {5}},\ \bibinfo {pages} {791} (\bibinfo {year} {2009})}\BibitemShut {NoStop}%
\bibitem [{\citenamefont {Johnson}\ \emph {et~al.}(2012)\citenamefont
  {Johnson}, \citenamefont {Macklin}, \citenamefont {Slichter}, \citenamefont
  {Vijay}, \citenamefont {Weingarten}, \citenamefont {Clarke},\ and\
  \citenamefont {Siddiqi}}]{johnson2012heralded}%
  \BibitemOpen
  \bibfield  {author} {\bibinfo {author} {\bibfnamefont {J.}~\bibnamefont
  {Johnson}}, \bibinfo {author} {\bibfnamefont {C.}~\bibnamefont {Macklin}},
  \bibinfo {author} {\bibfnamefont {D.}~\bibnamefont {Slichter}}, \bibinfo
  {author} {\bibfnamefont {R.}~\bibnamefont {Vijay}}, \bibinfo {author}
  {\bibfnamefont {E.}~\bibnamefont {Weingarten}}, \bibinfo {author}
  {\bibfnamefont {J.}~\bibnamefont {Clarke}}, \ and\ \bibinfo {author}
  {\bibfnamefont {I.}~\bibnamefont {Siddiqi}},\ }\href@noop {} {\bibfield
  {journal} {\bibinfo  {journal} {Physical review letters}\ }\textbf {\bibinfo
  {volume} {109}},\ \bibinfo {pages} {050506} (\bibinfo {year}
  {2012})}\BibitemShut {NoStop}%
\bibitem [{\citenamefont {Heinsoo}\ \emph {et~al.}(2018)\citenamefont
  {Heinsoo}, \citenamefont {Andersen}, \citenamefont {Remm}, \citenamefont
  {Krinner}, \citenamefont {Walter}, \citenamefont {Salath{\'e}}, \citenamefont
  {Gasparinetti}, \citenamefont {Besse}, \citenamefont {Poto{\v{c}}nik},
  \citenamefont {Wallraff} \emph {et~al.}}]{heinsoo2018rapid}%
  \BibitemOpen
  \bibfield  {author} {\bibinfo {author} {\bibfnamefont {J.}~\bibnamefont
  {Heinsoo}}, \bibinfo {author} {\bibfnamefont {C.~K.}\ \bibnamefont
  {Andersen}}, \bibinfo {author} {\bibfnamefont {A.}~\bibnamefont {Remm}},
  \bibinfo {author} {\bibfnamefont {S.}~\bibnamefont {Krinner}}, \bibinfo
  {author} {\bibfnamefont {T.}~\bibnamefont {Walter}}, \bibinfo {author}
  {\bibfnamefont {Y.}~\bibnamefont {Salath{\'e}}}, \bibinfo {author}
  {\bibfnamefont {S.}~\bibnamefont {Gasparinetti}}, \bibinfo {author}
  {\bibfnamefont {J.-C.}\ \bibnamefont {Besse}}, \bibinfo {author}
  {\bibfnamefont {A.}~\bibnamefont {Poto{\v{c}}nik}}, \bibinfo {author}
  {\bibfnamefont {A.}~\bibnamefont {Wallraff}},  \emph {et~al.},\ }\href@noop
  {} {\bibfield  {journal} {\bibinfo  {journal} {Physical Review Applied}\
  }\textbf {\bibinfo {volume} {10}},\ \bibinfo {pages} {034040} (\bibinfo
  {year} {2018})}\BibitemShut {NoStop}%
\bibitem [{\citenamefont {Elder}\ \emph {et~al.}(2020)\citenamefont {Elder},
  \citenamefont {Wang}, \citenamefont {Reinhold}, \citenamefont {Hann},
  \citenamefont {Chou}, \citenamefont {Lester}, \citenamefont {Rosenblum},
  \citenamefont {Frunzio}, \citenamefont {Jiang},\ and\ \citenamefont
  {Schoelkopf}}]{elder2020high}%
  \BibitemOpen
  \bibfield  {author} {\bibinfo {author} {\bibfnamefont {S.~S.}\ \bibnamefont
  {Elder}}, \bibinfo {author} {\bibfnamefont {C.~S.}\ \bibnamefont {Wang}},
  \bibinfo {author} {\bibfnamefont {P.}~\bibnamefont {Reinhold}}, \bibinfo
  {author} {\bibfnamefont {C.~T.}\ \bibnamefont {Hann}}, \bibinfo {author}
  {\bibfnamefont {K.~S.}\ \bibnamefont {Chou}}, \bibinfo {author}
  {\bibfnamefont {B.~J.}\ \bibnamefont {Lester}}, \bibinfo {author}
  {\bibfnamefont {S.}~\bibnamefont {Rosenblum}}, \bibinfo {author}
  {\bibfnamefont {L.}~\bibnamefont {Frunzio}}, \bibinfo {author} {\bibfnamefont
  {L.}~\bibnamefont {Jiang}}, \ and\ \bibinfo {author} {\bibfnamefont {R.~J.}\
  \bibnamefont {Schoelkopf}},\ }\href@noop {} {\bibfield  {journal} {\bibinfo
  {journal} {Physical Review X}\ }\textbf {\bibinfo {volume} {10}},\ \bibinfo
  {pages} {011001} (\bibinfo {year} {2020})}\BibitemShut {NoStop}%
\bibitem [{\citenamefont {Beale}\ and\ \citenamefont
  {Wallman}(2023)}]{beale2023randomized}%
  \BibitemOpen
  \bibfield  {author} {\bibinfo {author} {\bibfnamefont {S.~J.}\ \bibnamefont
  {Beale}}\ and\ \bibinfo {author} {\bibfnamefont {J.~J.}\ \bibnamefont
  {Wallman}},\ }\href@noop {} {\bibfield  {journal} {\bibinfo  {journal} {arXiv
  preprint arXiv:2304.06599}\ } (\bibinfo {year} {2023})}\BibitemShut {NoStop}%
\bibitem [{\citenamefont {Hashim}\ \emph
  {et~al.}(2023{\natexlab{b}})\citenamefont {Hashim}, \citenamefont
  {Carignan-Dugas}, \citenamefont {Chen}, \citenamefont {Juenger},
  \citenamefont {Fruitwala}, \citenamefont {Xu}, \citenamefont {Huang},
  \citenamefont {Wallman},\ and\ \citenamefont {Siddiqi}}]{hashim2023quasi}%
  \BibitemOpen
  \bibfield  {author} {\bibinfo {author} {\bibfnamefont {A.}~\bibnamefont
  {Hashim}}, \bibinfo {author} {\bibfnamefont {A.}~\bibnamefont
  {Carignan-Dugas}}, \bibinfo {author} {\bibfnamefont {L.}~\bibnamefont
  {Chen}}, \bibinfo {author} {\bibfnamefont {C.}~\bibnamefont {Juenger}},
  \bibinfo {author} {\bibfnamefont {N.}~\bibnamefont {Fruitwala}}, \bibinfo
  {author} {\bibfnamefont {Y.}~\bibnamefont {Xu}}, \bibinfo {author}
  {\bibfnamefont {G.}~\bibnamefont {Huang}}, \bibinfo {author} {\bibfnamefont
  {J.}~\bibnamefont {Wallman}}, \ and\ \bibinfo {author} {\bibfnamefont
  {I.}~\bibnamefont {Siddiqi}},\ }\href@noop {} {\bibfield  {journal} {\bibinfo
   {journal} {arXiv preprint arXiv:2312.14139}\ } (\bibinfo {year}
  {2023}{\natexlab{b}})}\BibitemShut {NoStop}%
\bibitem [{\citenamefont {Magesan}\ \emph
  {et~al.}(2012{\natexlab{a}})\citenamefont {Magesan}, \citenamefont
  {Gambetta}, \citenamefont {Johnson}, \citenamefont {Ryan}, \citenamefont
  {Chow}, \citenamefont {Merkel}, \citenamefont {Da~Silva}, \citenamefont
  {Keefe}, \citenamefont {Rothwell}, \citenamefont {Ohki} \emph
  {et~al.}}]{magesan2012efficient}%
  \BibitemOpen
  \bibfield  {author} {\bibinfo {author} {\bibfnamefont {E.}~\bibnamefont
  {Magesan}}, \bibinfo {author} {\bibfnamefont {J.~M.}\ \bibnamefont
  {Gambetta}}, \bibinfo {author} {\bibfnamefont {B.~R.}\ \bibnamefont
  {Johnson}}, \bibinfo {author} {\bibfnamefont {C.~A.}\ \bibnamefont {Ryan}},
  \bibinfo {author} {\bibfnamefont {J.~M.}\ \bibnamefont {Chow}}, \bibinfo
  {author} {\bibfnamefont {S.~T.}\ \bibnamefont {Merkel}}, \bibinfo {author}
  {\bibfnamefont {M.~P.}\ \bibnamefont {Da~Silva}}, \bibinfo {author}
  {\bibfnamefont {G.~A.}\ \bibnamefont {Keefe}}, \bibinfo {author}
  {\bibfnamefont {M.~B.}\ \bibnamefont {Rothwell}}, \bibinfo {author}
  {\bibfnamefont {T.~A.}\ \bibnamefont {Ohki}},  \emph {et~al.},\ }\href@noop
  {} {\bibfield  {journal} {\bibinfo  {journal} {Physical review letters}\
  }\textbf {\bibinfo {volume} {109}},\ \bibinfo {pages} {080505} (\bibinfo
  {year} {2012}{\natexlab{a}})}\BibitemShut {NoStop}%
\bibitem [{\citenamefont {Renes}\ \emph {et~al.}(2004)\citenamefont {Renes},
  \citenamefont {Blume-Kohout}, \citenamefont {Scott},\ and\ \citenamefont
  {Caves}}]{Renes2004-pc}%
  \BibitemOpen
  \bibfield  {author} {\bibinfo {author} {\bibfnamefont {J.~M.}\ \bibnamefont
  {Renes}}, \bibinfo {author} {\bibfnamefont {R.}~\bibnamefont {Blume-Kohout}},
  \bibinfo {author} {\bibfnamefont {A.~J.}\ \bibnamefont {Scott}}, \ and\
  \bibinfo {author} {\bibfnamefont {C.~M.}\ \bibnamefont {Caves}},\ }\href
  {\doibase 10.1063/1.1737053} {\bibfield  {journal} {\bibinfo  {journal} {J.
  Math. Phys.}\ }\textbf {\bibinfo {volume} {45}},\ \bibinfo {pages} {2171}
  (\bibinfo {year} {2004})}\BibitemShut {NoStop}%
\bibitem [{\citenamefont {Chow}\ \emph {et~al.}(2014)\citenamefont {Chow},
  \citenamefont {Gambetta}, \citenamefont {Magesan}, \citenamefont {Abraham},
  \citenamefont {Cross}, \citenamefont {Johnson}, \citenamefont {Masluk},
  \citenamefont {Ryan}, \citenamefont {Smolin}, \citenamefont {Srinivasan},\
  and\ \citenamefont {Steffen}}]{Chow2014-bc}%
  \BibitemOpen
  \bibfield  {author} {\bibinfo {author} {\bibfnamefont {J.~M.}\ \bibnamefont
  {Chow}}, \bibinfo {author} {\bibfnamefont {J.~M.}\ \bibnamefont {Gambetta}},
  \bibinfo {author} {\bibfnamefont {E.}~\bibnamefont {Magesan}}, \bibinfo
  {author} {\bibfnamefont {D.~W.}\ \bibnamefont {Abraham}}, \bibinfo {author}
  {\bibfnamefont {A.~W.}\ \bibnamefont {Cross}}, \bibinfo {author}
  {\bibfnamefont {B.~R.}\ \bibnamefont {Johnson}}, \bibinfo {author}
  {\bibfnamefont {N.~A.}\ \bibnamefont {Masluk}}, \bibinfo {author}
  {\bibfnamefont {C.~A.}\ \bibnamefont {Ryan}}, \bibinfo {author}
  {\bibfnamefont {J.~A.}\ \bibnamefont {Smolin}}, \bibinfo {author}
  {\bibfnamefont {S.~J.}\ \bibnamefont {Srinivasan}}, \ and\ \bibinfo {author}
  {\bibfnamefont {M.}~\bibnamefont {Steffen}},\ }\href {\doibase
  10.1038/ncomms5015} {\bibfield  {journal} {\bibinfo  {journal} {Nat.
  Commun.}\ }\textbf {\bibinfo {volume} {5}},\ \bibinfo {pages} {4015}
  (\bibinfo {year} {2014})}\BibitemShut {NoStop}%
\bibitem [{\citenamefont {Holevo}(1998)}]{Holevo98}%
  \BibitemOpen
  \bibfield  {author} {\bibinfo {author} {\bibfnamefont {A.~S.}\ \bibnamefont
  {Holevo}},\ }\href@noop {} {\bibfield  {journal} {\bibinfo  {journal}
  {Russian Mathematical Surveys}\ }\textbf {\bibinfo {volume} {53}},\ \bibinfo
  {pages} {1295} (\bibinfo {year} {1998})}\BibitemShut {NoStop}%
\bibitem [{\citenamefont {McLaren}\ \emph {et~al.}(2023)\citenamefont
  {McLaren}, \citenamefont {Graydon},\ and\ \citenamefont
  {Wallman}}]{McLaren2023-gf}%
  \BibitemOpen
  \bibfield  {author} {\bibinfo {author} {\bibfnamefont {D.}~\bibnamefont
  {McLaren}}, \bibinfo {author} {\bibfnamefont {M.~A.}\ \bibnamefont
  {Graydon}}, \ and\ \bibinfo {author} {\bibfnamefont {J.~J.}\ \bibnamefont
  {Wallman}},\ }\href@noop {} {\bibfield  {journal} {\bibinfo  {journal} {arXiv
  preprint arXiv:2306.07418}\ } (\bibinfo {year} {2023})}\BibitemShut {NoStop}%
\bibitem [{\citenamefont {Pereira}\ \emph {et~al.}(2022)\citenamefont
  {Pereira}, \citenamefont {Garc{\'\i}a-Ripoll},\ and\ \citenamefont
  {Ramos}}]{pereira2022complete}%
  \BibitemOpen
  \bibfield  {author} {\bibinfo {author} {\bibfnamefont {L.}~\bibnamefont
  {Pereira}}, \bibinfo {author} {\bibfnamefont {J.~J.}\ \bibnamefont
  {Garc{\'\i}a-Ripoll}}, \ and\ \bibinfo {author} {\bibfnamefont
  {T.}~\bibnamefont {Ramos}},\ }\href@noop {} {\bibfield  {journal} {\bibinfo
  {journal} {Physical Review Letters}\ }\textbf {\bibinfo {volume} {129}},\
  \bibinfo {pages} {010402} (\bibinfo {year} {2022})}\BibitemShut {NoStop}%
\bibitem [{\citenamefont {Pereira}\ \emph {et~al.}(2023)\citenamefont
  {Pereira}, \citenamefont {Garc{\'\i}a-Ripoll},\ and\ \citenamefont
  {Ramos}}]{pereira2023parallel}%
  \BibitemOpen
  \bibfield  {author} {\bibinfo {author} {\bibfnamefont {L.}~\bibnamefont
  {Pereira}}, \bibinfo {author} {\bibfnamefont {J.~J.}\ \bibnamefont
  {Garc{\'\i}a-Ripoll}}, \ and\ \bibinfo {author} {\bibfnamefont
  {T.}~\bibnamefont {Ramos}},\ }\href@noop {} {\bibfield  {journal} {\bibinfo
  {journal} {npj Quantum Information}\ }\textbf {\bibinfo {volume} {9}},\
  \bibinfo {pages} {22} (\bibinfo {year} {2023})}\BibitemShut {NoStop}%
\bibitem [{\citenamefont {Nakamura}\ \emph {et~al.}(1999)\citenamefont
  {Nakamura}, \citenamefont {Pashkin},\ and\ \citenamefont
  {Tsai}}]{nakamura1999coherent}%
  \BibitemOpen
  \bibfield  {author} {\bibinfo {author} {\bibfnamefont {Y.}~\bibnamefont
  {Nakamura}}, \bibinfo {author} {\bibfnamefont {Y.~A.}\ \bibnamefont
  {Pashkin}}, \ and\ \bibinfo {author} {\bibfnamefont {J.~S.}\ \bibnamefont
  {Tsai}},\ }\href {\doibase 10.1038/19718} {\bibfield  {journal} {\bibinfo
  {journal} {Nature}\ }\textbf {\bibinfo {volume} {398}},\ \bibinfo {pages}
  {786–788} (\bibinfo {year} {1999})}\BibitemShut {NoStop}%
\bibitem [{\citenamefont {Astafiev}\ \emph {et~al.}(2010)\citenamefont
  {Astafiev}, \citenamefont {Zagoskin}, \citenamefont {Abdumalikov},
  \citenamefont {Pashkin}, \citenamefont {Yamamoto}, \citenamefont {Inomata},
  \citenamefont {Nakamura},\ and\ \citenamefont
  {Tsai}}]{astafiev2010resonance}%
  \BibitemOpen
  \bibfield  {author} {\bibinfo {author} {\bibfnamefont {O.}~\bibnamefont
  {Astafiev}}, \bibinfo {author} {\bibfnamefont {A.~M.}\ \bibnamefont
  {Zagoskin}}, \bibinfo {author} {\bibfnamefont {A.~A.}\ \bibnamefont
  {Abdumalikov}}, \bibinfo {author} {\bibfnamefont {Y.~A.}\ \bibnamefont
  {Pashkin}}, \bibinfo {author} {\bibfnamefont {T.}~\bibnamefont {Yamamoto}},
  \bibinfo {author} {\bibfnamefont {K.}~\bibnamefont {Inomata}}, \bibinfo
  {author} {\bibfnamefont {Y.}~\bibnamefont {Nakamura}}, \ and\ \bibinfo
  {author} {\bibfnamefont {J.~S.}\ \bibnamefont {Tsai}},\ }\href {\doibase
  10.1126/science.1181918} {\bibfield  {journal} {\bibinfo  {journal}
  {Science}\ }\textbf {\bibinfo {volume} {327}},\ \bibinfo {pages} {840–843}
  (\bibinfo {year} {2010})}\BibitemShut {NoStop}%
\bibitem [{\citenamefont {Cottet}\ \emph {et~al.}(2021)\citenamefont {Cottet},
  \citenamefont {Xiong}, \citenamefont {Nguyen}, \citenamefont {Lin},\ and\
  \citenamefont {Manucharyan}}]{cottet2021electron}%
  \BibitemOpen
  \bibfield  {author} {\bibinfo {author} {\bibfnamefont {N.}~\bibnamefont
  {Cottet}}, \bibinfo {author} {\bibfnamefont {H.}~\bibnamefont {Xiong}},
  \bibinfo {author} {\bibfnamefont {L.~B.}\ \bibnamefont {Nguyen}}, \bibinfo
  {author} {\bibfnamefont {Y.-H.}\ \bibnamefont {Lin}}, \ and\ \bibinfo
  {author} {\bibfnamefont {V.~E.}\ \bibnamefont {Manucharyan}},\ }\href
  {\doibase 10.1038/s41467-021-26686-x} {\bibfield  {journal} {\bibinfo
  {journal} {Nature Communications}\ }\textbf {\bibinfo {volume} {12}},\
  \bibinfo {pages} {1} (\bibinfo {year} {2021})}\BibitemShut {NoStop}%
\bibitem [{\citenamefont {Merkel}\ \emph
  {et~al.}(2013{\natexlab{a}})\citenamefont {Merkel}, \citenamefont {Gambetta},
  \citenamefont {Smolin}, \citenamefont {Poletto}, \citenamefont {C\'orcoles},
  \citenamefont {Johnson}, \citenamefont {Ryan},\ and\ \citenamefont
  {Steffen}}]{merkel2013selfconsistent}%
  \BibitemOpen
  \bibfield  {author} {\bibinfo {author} {\bibfnamefont {S.~T.}\ \bibnamefont
  {Merkel}}, \bibinfo {author} {\bibfnamefont {J.~M.}\ \bibnamefont
  {Gambetta}}, \bibinfo {author} {\bibfnamefont {J.~A.}\ \bibnamefont
  {Smolin}}, \bibinfo {author} {\bibfnamefont {S.}~\bibnamefont {Poletto}},
  \bibinfo {author} {\bibfnamefont {A.~D.}\ \bibnamefont {C\'orcoles}},
  \bibinfo {author} {\bibfnamefont {B.~R.}\ \bibnamefont {Johnson}}, \bibinfo
  {author} {\bibfnamefont {C.~A.}\ \bibnamefont {Ryan}}, \ and\ \bibinfo
  {author} {\bibfnamefont {M.}~\bibnamefont {Steffen}},\ }\href {\doibase
  10.1103/PhysRevA.87.062119} {\bibfield  {journal} {\bibinfo  {journal} {Phys.
  Rev. A}\ }\textbf {\bibinfo {volume} {87}},\ \bibinfo {pages} {062119}
  (\bibinfo {year} {2013}{\natexlab{a}})}\BibitemShut {NoStop}%
\bibitem [{\citenamefont {Blume-Kohout}\ \emph {et~al.}(2013)\citenamefont
  {Blume-Kohout}, \citenamefont {Gamble}, \citenamefont {Nielsen},
  \citenamefont {Mizrahi}, \citenamefont {Sterk},\ and\ \citenamefont
  {Maunz}}]{Blume-Kohout2013gst}%
  \BibitemOpen
  \bibfield  {author} {\bibinfo {author} {\bibfnamefont {R.}~\bibnamefont
  {Blume-Kohout}}, \bibinfo {author} {\bibfnamefont {J.~K.}\ \bibnamefont
  {Gamble}}, \bibinfo {author} {\bibfnamefont {E.}~\bibnamefont {Nielsen}},
  \bibinfo {author} {\bibfnamefont {J.}~\bibnamefont {Mizrahi}}, \bibinfo
  {author} {\bibfnamefont {J.~D.}\ \bibnamefont {Sterk}}, \ and\ \bibinfo
  {author} {\bibfnamefont {P.}~\bibnamefont {Maunz}},\ }\href@noop {}
  {\bibfield  {journal} {\bibinfo  {journal} {arXiv preprint arXiv:1310.4492}\
  } (\bibinfo {year} {2013})}\BibitemShut {NoStop}%
\bibitem [{\citenamefont {Proctor}\ \emph {et~al.}(2017)\citenamefont
  {Proctor}, \citenamefont {Rudinger}, \citenamefont {Young}, \citenamefont
  {Sarovar},\ and\ \citenamefont {Blume-Kohout}}]{proctor2017randomized}%
  \BibitemOpen
  \bibfield  {author} {\bibinfo {author} {\bibfnamefont {T.}~\bibnamefont
  {Proctor}}, \bibinfo {author} {\bibfnamefont {K.}~\bibnamefont {Rudinger}},
  \bibinfo {author} {\bibfnamefont {K.}~\bibnamefont {Young}}, \bibinfo
  {author} {\bibfnamefont {M.}~\bibnamefont {Sarovar}}, \ and\ \bibinfo
  {author} {\bibfnamefont {R.}~\bibnamefont {Blume-Kohout}},\ }\href@noop {}
  {\bibfield  {journal} {\bibinfo  {journal} {Physical review letters}\
  }\textbf {\bibinfo {volume} {119}},\ \bibinfo {pages} {130502} (\bibinfo
  {year} {2017})}\BibitemShut {NoStop}%
\bibitem [{\citenamefont {Nielsen}\ \emph
  {et~al.}(2021{\natexlab{c}})\citenamefont {Nielsen}, \citenamefont {Gamble},
  \citenamefont {Rudinger}, \citenamefont {Scholten}, \citenamefont {Young},\
  and\ \citenamefont {Blume-Kohout}}]{Nielsen2021gatesettomography}%
  \BibitemOpen
  \bibfield  {author} {\bibinfo {author} {\bibfnamefont {E.}~\bibnamefont
  {Nielsen}}, \bibinfo {author} {\bibfnamefont {J.~K.}\ \bibnamefont {Gamble}},
  \bibinfo {author} {\bibfnamefont {K.}~\bibnamefont {Rudinger}}, \bibinfo
  {author} {\bibfnamefont {T.}~\bibnamefont {Scholten}}, \bibinfo {author}
  {\bibfnamefont {K.}~\bibnamefont {Young}}, \ and\ \bibinfo {author}
  {\bibfnamefont {R.}~\bibnamefont {Blume-Kohout}},\ }\href {\doibase
  10.22331/q-2021-10-05-557} {\bibfield  {journal} {\bibinfo  {journal}
  {{Quantum}}\ }\textbf {\bibinfo {volume} {5}},\ \bibinfo {pages} {557}
  (\bibinfo {year} {2021}{\natexlab{c}})}\BibitemShut {NoStop}%
\bibitem [{\citenamefont {Wallman}(2018)}]{wallman2018randomized}%
  \BibitemOpen
  \bibfield  {author} {\bibinfo {author} {\bibfnamefont {J.~J.}\ \bibnamefont
  {Wallman}},\ }\href@noop {} {\bibfield  {journal} {\bibinfo  {journal}
  {Quantum}\ }\textbf {\bibinfo {volume} {2}},\ \bibinfo {pages} {47} (\bibinfo
  {year} {2018})}\BibitemShut {NoStop}%
\bibitem [{\citenamefont {Zhong}\ \emph {et~al.}(2020)\citenamefont {Zhong},
  \citenamefont {Wang}, \citenamefont {Deng}, \citenamefont {Chen},
  \citenamefont {Peng}, \citenamefont {Luo}, \citenamefont {Qin}, \citenamefont
  {Wu}, \citenamefont {Ding}, \citenamefont {Hu} \emph
  {et~al.}}]{zhong2020quantum}%
  \BibitemOpen
  \bibfield  {author} {\bibinfo {author} {\bibfnamefont {H.-S.}\ \bibnamefont
  {Zhong}}, \bibinfo {author} {\bibfnamefont {H.}~\bibnamefont {Wang}},
  \bibinfo {author} {\bibfnamefont {Y.-H.}\ \bibnamefont {Deng}}, \bibinfo
  {author} {\bibfnamefont {M.-C.}\ \bibnamefont {Chen}}, \bibinfo {author}
  {\bibfnamefont {L.-C.}\ \bibnamefont {Peng}}, \bibinfo {author}
  {\bibfnamefont {Y.-H.}\ \bibnamefont {Luo}}, \bibinfo {author} {\bibfnamefont
  {J.}~\bibnamefont {Qin}}, \bibinfo {author} {\bibfnamefont {D.}~\bibnamefont
  {Wu}}, \bibinfo {author} {\bibfnamefont {X.}~\bibnamefont {Ding}}, \bibinfo
  {author} {\bibfnamefont {Y.}~\bibnamefont {Hu}},  \emph {et~al.},\
  }\href@noop {} {\bibfield  {journal} {\bibinfo  {journal} {Science}\ }\textbf
  {\bibinfo {volume} {370}},\ \bibinfo {pages} {1460} (\bibinfo {year}
  {2020})}\BibitemShut {NoStop}%
\bibitem [{\citenamefont {Dasgupta}\ and\ \citenamefont
  {Humble}(2022)}]{Dasgupta2022-nv}%
  \BibitemOpen
  \bibfield  {author} {\bibinfo {author} {\bibfnamefont {S.}~\bibnamefont
  {Dasgupta}}\ and\ \bibinfo {author} {\bibfnamefont {T.~S.}\ \bibnamefont
  {Humble}},\ }\href {\doibase 10.3390/e24020244} {\bibfield  {journal}
  {\bibinfo  {journal} {Entropy}\ }\textbf {\bibinfo {volume} {24}} (\bibinfo
  {year} {2022}),\ 10.3390/e24020244}\BibitemShut {NoStop}%
\bibitem [{\citenamefont {Gross}\ \emph
  {et~al.}(2010{\natexlab{a}})\citenamefont {Gross}, \citenamefont {Liu},
  \citenamefont {Flammia}, \citenamefont {Becker},\ and\ \citenamefont
  {Eisert}}]{Gross2010-lb}%
  \BibitemOpen
  \bibfield  {author} {\bibinfo {author} {\bibfnamefont {D.}~\bibnamefont
  {Gross}}, \bibinfo {author} {\bibfnamefont {Y.-K.}\ \bibnamefont {Liu}},
  \bibinfo {author} {\bibfnamefont {S.~T.}\ \bibnamefont {Flammia}}, \bibinfo
  {author} {\bibfnamefont {S.}~\bibnamefont {Becker}}, \ and\ \bibinfo {author}
  {\bibfnamefont {J.}~\bibnamefont {Eisert}},\ }\href {\doibase
  10.1103/PhysRevLett.105.150401} {\bibfield  {journal} {\bibinfo  {journal}
  {Phys. Rev. Lett.}\ }\textbf {\bibinfo {volume} {105}},\ \bibinfo {pages}
  {150401} (\bibinfo {year} {2010}{\natexlab{a}})}\BibitemShut {NoStop}%
\bibitem [{\citenamefont {Riofr{\'\i}o}\ \emph
  {et~al.}(2017{\natexlab{a}})\citenamefont {Riofr{\'\i}o}, \citenamefont
  {Gross}, \citenamefont {Flammia}, \citenamefont {Monz}, \citenamefont {Nigg},
  \citenamefont {Blatt},\ and\ \citenamefont {Eisert}}]{Riofrio2017-yq}%
  \BibitemOpen
  \bibfield  {author} {\bibinfo {author} {\bibfnamefont {C.~A.}\ \bibnamefont
  {Riofr{\'\i}o}}, \bibinfo {author} {\bibfnamefont {D.}~\bibnamefont {Gross}},
  \bibinfo {author} {\bibfnamefont {S.~T.}\ \bibnamefont {Flammia}}, \bibinfo
  {author} {\bibfnamefont {T.}~\bibnamefont {Monz}}, \bibinfo {author}
  {\bibfnamefont {D.}~\bibnamefont {Nigg}}, \bibinfo {author} {\bibfnamefont
  {R.}~\bibnamefont {Blatt}}, \ and\ \bibinfo {author} {\bibfnamefont
  {J.}~\bibnamefont {Eisert}},\ }\href {\doibase 10.1038/ncomms15305}
  {\bibfield  {journal} {\bibinfo  {journal} {Nat. Commun.}\ }\textbf {\bibinfo
  {volume} {8}},\ \bibinfo {pages} {15305} (\bibinfo {year}
  {2017}{\natexlab{a}})}\BibitemShut {NoStop}%
\bibitem [{\citenamefont {Meyer}(2021)}]{Meyer2021-lp}%
  \BibitemOpen
  \bibfield  {author} {\bibinfo {author} {\bibfnamefont {J.~J.}\ \bibnamefont
  {Meyer}},\ }\href {\doibase 10.22331/q-2021-09-09-539} {\bibfield  {journal}
  {\bibinfo  {journal} {Quantum}\ }\textbf {\bibinfo {volume} {5}},\ \bibinfo
  {pages} {539} (\bibinfo {year} {2021})}\BibitemShut {NoStop}%
\bibitem [{\citenamefont {Ostrove}\ \emph {et~al.}(2023)\citenamefont
  {Ostrove}, \citenamefont {Rudinger}, \citenamefont {Seritan}, \citenamefont
  {Young},\ and\ \citenamefont {Blume-Kohout}}]{Ostrove2023-xp}%
  \BibitemOpen
  \bibfield  {author} {\bibinfo {author} {\bibfnamefont {C.}~\bibnamefont
  {Ostrove}}, \bibinfo {author} {\bibfnamefont {K.}~\bibnamefont {Rudinger}},
  \bibinfo {author} {\bibfnamefont {S.}~\bibnamefont {Seritan}}, \bibinfo
  {author} {\bibfnamefont {K.}~\bibnamefont {Young}}, \ and\ \bibinfo {author}
  {\bibfnamefont {R.}~\bibnamefont {Blume-Kohout}},\ }in\ \href@noop {} {\emph
  {\bibinfo {booktitle} {2023 IEEE International Conference on Quantum
  Computing and Engineering (QCE)}}},\ Vol.~\bibinfo {volume} {1}\ (\bibinfo
  {organization} {IEEE},\ \bibinfo {year} {2023})\ pp.\ \bibinfo {pages}
  {1422--1432}\BibitemShut {NoStop}%
\bibitem [{\citenamefont {Tong}\ \emph {et~al.}(2024)\citenamefont {Tong},
  \citenamefont {Zhang},\ and\ \citenamefont {Pokharel}}]{Tong2024-el}%
  \BibitemOpen
  \bibfield  {author} {\bibinfo {author} {\bibfnamefont {C.}~\bibnamefont
  {Tong}}, \bibinfo {author} {\bibfnamefont {H.}~\bibnamefont {Zhang}}, \ and\
  \bibinfo {author} {\bibfnamefont {B.}~\bibnamefont {Pokharel}},\ }\href@noop
  {} {\bibfield  {journal} {\bibinfo  {journal} {arXiv [quant-ph]}\ } (\bibinfo
  {year} {2024})},\ \Eprint {http://arxiv.org/abs/2403.02294} {arXiv:2403.02294
  [quant-ph]} \BibitemShut {NoStop}%
\bibitem [{\citenamefont {Proctor}\ \emph
  {et~al.}(2022{\natexlab{a}})\citenamefont {Proctor}, \citenamefont {Seritan},
  \citenamefont {Rudinger}, \citenamefont {Nielsen}, \citenamefont
  {Blume-Kohout},\ and\ \citenamefont {Young}}]{proctor2021scalable}%
  \BibitemOpen
  \bibfield  {author} {\bibinfo {author} {\bibfnamefont {T.}~\bibnamefont
  {Proctor}}, \bibinfo {author} {\bibfnamefont {S.}~\bibnamefont {Seritan}},
  \bibinfo {author} {\bibfnamefont {K.}~\bibnamefont {Rudinger}}, \bibinfo
  {author} {\bibfnamefont {E.}~\bibnamefont {Nielsen}}, \bibinfo {author}
  {\bibfnamefont {R.}~\bibnamefont {Blume-Kohout}}, \ and\ \bibinfo {author}
  {\bibfnamefont {K.}~\bibnamefont {Young}},\ }\href {\doibase
  10.1103/PhysRevLett.129.150502} {\bibfield  {journal} {\bibinfo  {journal}
  {Phys. Rev. Lett.}\ }\textbf {\bibinfo {volume} {129}},\ \bibinfo {pages}
  {150502} (\bibinfo {year} {2022}{\natexlab{a}})}\BibitemShut {NoStop}%
\bibitem [{\citenamefont {Torlai}\ \emph {et~al.}(2023)\citenamefont {Torlai},
  \citenamefont {Wood}, \citenamefont {Acharya}, \citenamefont {Carleo},
  \citenamefont {Carrasquilla},\ and\ \citenamefont {Aolita}}]{Torlai2023-ca}%
  \BibitemOpen
  \bibfield  {author} {\bibinfo {author} {\bibfnamefont {G.}~\bibnamefont
  {Torlai}}, \bibinfo {author} {\bibfnamefont {C.~J.}\ \bibnamefont {Wood}},
  \bibinfo {author} {\bibfnamefont {A.}~\bibnamefont {Acharya}}, \bibinfo
  {author} {\bibfnamefont {G.}~\bibnamefont {Carleo}}, \bibinfo {author}
  {\bibfnamefont {J.}~\bibnamefont {Carrasquilla}}, \ and\ \bibinfo {author}
  {\bibfnamefont {L.}~\bibnamefont {Aolita}},\ }\href {\doibase
  10.1038/s41467-023-38332-9} {\bibfield  {journal} {\bibinfo  {journal} {Nat.
  Commun.}\ }\textbf {\bibinfo {volume} {14}},\ \bibinfo {pages} {2858}
  (\bibinfo {year} {2023})}\BibitemShut {NoStop}%
\bibitem [{\citenamefont {Evans}\ \emph {et~al.}(2019)\citenamefont {Evans},
  \citenamefont {Harper},\ and\ \citenamefont {Flammia}}]{Evans2019-kp}%
  \BibitemOpen
  \bibfield  {author} {\bibinfo {author} {\bibfnamefont {T.~J.}\ \bibnamefont
  {Evans}}, \bibinfo {author} {\bibfnamefont {R.}~\bibnamefont {Harper}}, \
  and\ \bibinfo {author} {\bibfnamefont {S.~T.}\ \bibnamefont {Flammia}},\
  }\href@noop {} {\bibfield  {journal} {\bibinfo  {journal} {arXiv preprint
  arXiv:1912.07636}\ } (\bibinfo {year} {2019})}\BibitemShut {NoStop}%
\bibitem [{\citenamefont {Huang}\ \emph {et~al.}(2021)\citenamefont {Huang},
  \citenamefont {Kueng},\ and\ \citenamefont {Preskill}}]{Huang2021-bn}%
  \BibitemOpen
  \bibfield  {author} {\bibinfo {author} {\bibfnamefont {H.-Y.}\ \bibnamefont
  {Huang}}, \bibinfo {author} {\bibfnamefont {R.}~\bibnamefont {Kueng}}, \ and\
  \bibinfo {author} {\bibfnamefont {J.}~\bibnamefont {Preskill}},\ }\href
  {\doibase 10.1103/PhysRevLett.127.030503} {\bibfield  {journal} {\bibinfo
  {journal} {Phys. Rev. Lett.}\ }\textbf {\bibinfo {volume} {127}},\ \bibinfo
  {pages} {030503} (\bibinfo {year} {2021})}\BibitemShut {NoStop}%
\bibitem [{\citenamefont {Huang}\ \emph {et~al.}(2020)\citenamefont {Huang},
  \citenamefont {Kueng},\ and\ \citenamefont {Preskill}}]{huang2020predicting}%
  \BibitemOpen
  \bibfield  {author} {\bibinfo {author} {\bibfnamefont {H.-Y.}\ \bibnamefont
  {Huang}}, \bibinfo {author} {\bibfnamefont {R.}~\bibnamefont {Kueng}}, \ and\
  \bibinfo {author} {\bibfnamefont {J.}~\bibnamefont {Preskill}},\ }\href@noop
  {} {\bibfield  {journal} {\bibinfo  {journal} {Nature Physics}\ }\textbf
  {\bibinfo {volume} {16}},\ \bibinfo {pages} {1050} (\bibinfo {year}
  {2020})}\BibitemShut {NoStop}%
\bibitem [{\citenamefont {Andrews}\ \emph {et~al.}(2019)\citenamefont
  {Andrews}, \citenamefont {Jones}, \citenamefont {Reed}, \citenamefont
  {Jones}, \citenamefont {Ha}, \citenamefont {Jura}, \citenamefont {Kerckhoff},
  \citenamefont {Levendorf}, \citenamefont {Meenehan}, \citenamefont {Merkel}
  \emph {et~al.}}]{andrews2019quantifying}%
  \BibitemOpen
  \bibfield  {author} {\bibinfo {author} {\bibfnamefont {R.~W.}\ \bibnamefont
  {Andrews}}, \bibinfo {author} {\bibfnamefont {C.}~\bibnamefont {Jones}},
  \bibinfo {author} {\bibfnamefont {M.~D.}\ \bibnamefont {Reed}}, \bibinfo
  {author} {\bibfnamefont {A.~M.}\ \bibnamefont {Jones}}, \bibinfo {author}
  {\bibfnamefont {S.~D.}\ \bibnamefont {Ha}}, \bibinfo {author} {\bibfnamefont
  {M.~P.}\ \bibnamefont {Jura}}, \bibinfo {author} {\bibfnamefont
  {J.}~\bibnamefont {Kerckhoff}}, \bibinfo {author} {\bibfnamefont
  {M.}~\bibnamefont {Levendorf}}, \bibinfo {author} {\bibfnamefont
  {S.}~\bibnamefont {Meenehan}}, \bibinfo {author} {\bibfnamefont {S.~T.}\
  \bibnamefont {Merkel}},  \emph {et~al.},\ }\href@noop {} {\bibfield
  {journal} {\bibinfo  {journal} {Nature nanotechnology}\ }\textbf {\bibinfo
  {volume} {14}},\ \bibinfo {pages} {747} (\bibinfo {year} {2019})}\BibitemShut
  {NoStop}%
\bibitem [{\citenamefont {Helsen}\ \emph
  {et~al.}(2019{\natexlab{a}})\citenamefont {Helsen}, \citenamefont {Xue},
  \citenamefont {Vandersypen},\ and\ \citenamefont {Wehner}}]{helsen2019new}%
  \BibitemOpen
  \bibfield  {author} {\bibinfo {author} {\bibfnamefont {J.}~\bibnamefont
  {Helsen}}, \bibinfo {author} {\bibfnamefont {X.}~\bibnamefont {Xue}},
  \bibinfo {author} {\bibfnamefont {L.~M.}\ \bibnamefont {Vandersypen}}, \ and\
  \bibinfo {author} {\bibfnamefont {S.}~\bibnamefont {Wehner}},\ }\href@noop {}
  {\bibfield  {journal} {\bibinfo  {journal} {npj Quantum Information}\
  }\textbf {\bibinfo {volume} {5}},\ \bibinfo {pages} {71} (\bibinfo {year}
  {2019}{\natexlab{a}})}\BibitemShut {NoStop}%
\bibitem [{\citenamefont {Gupta}\ \emph {et~al.}(2020)\citenamefont {Gupta},
  \citenamefont {Govia},\ and\ \citenamefont {Biercuk}}]{Gupta2020-kl}%
  \BibitemOpen
  \bibfield  {author} {\bibinfo {author} {\bibfnamefont {R.~S.}\ \bibnamefont
  {Gupta}}, \bibinfo {author} {\bibfnamefont {L.~C.~G.}\ \bibnamefont {Govia}},
  \ and\ \bibinfo {author} {\bibfnamefont {M.~J.}\ \bibnamefont {Biercuk}},\
  }\href {\doibase 10.1103/PhysRevA.102.042611} {\bibfield  {journal} {\bibinfo
   {journal} {Phys. Rev. A}\ }\textbf {\bibinfo {volume} {102}},\ \bibinfo
  {pages} {042611} (\bibinfo {year} {2020})}\BibitemShut {NoStop}%
\bibitem [{\citenamefont {Majumder}\ \emph {et~al.}(2020)\citenamefont
  {Majumder}, \citenamefont {de~Castro},\ and\ \citenamefont
  {Brown}}]{Majumder2020-ut}%
  \BibitemOpen
  \bibfield  {author} {\bibinfo {author} {\bibfnamefont {S.}~\bibnamefont
  {Majumder}}, \bibinfo {author} {\bibfnamefont {L.~A.}\ \bibnamefont
  {de~Castro}}, \ and\ \bibinfo {author} {\bibfnamefont {K.~R.}\ \bibnamefont
  {Brown}},\ }\href {\doibase 10.1038/s41534-020-0251-y} {\bibfield  {journal}
  {\bibinfo  {journal} {npj Quantum Information}\ }\textbf {\bibinfo {volume}
  {6}},\ \bibinfo {pages} {1} (\bibinfo {year} {2020})}\BibitemShut {NoStop}%
\bibitem [{\citenamefont {Fruitwala}\ \emph {et~al.}(2024)\citenamefont
  {Fruitwala}, \citenamefont {Hashim}, \citenamefont {Rajagopala},
  \citenamefont {Xu}, \citenamefont {Hines}, \citenamefont {Naik},
  \citenamefont {Siddiqi}, \citenamefont {Klymko}, \citenamefont {Huang},\ and\
  \citenamefont {Nowrouzi}}]{fruitwala2024hardware}%
  \BibitemOpen
  \bibfield  {author} {\bibinfo {author} {\bibfnamefont {N.}~\bibnamefont
  {Fruitwala}}, \bibinfo {author} {\bibfnamefont {A.}~\bibnamefont {Hashim}},
  \bibinfo {author} {\bibfnamefont {A.~D.}\ \bibnamefont {Rajagopala}},
  \bibinfo {author} {\bibfnamefont {Y.}~\bibnamefont {Xu}}, \bibinfo {author}
  {\bibfnamefont {J.}~\bibnamefont {Hines}}, \bibinfo {author} {\bibfnamefont
  {R.~K.}\ \bibnamefont {Naik}}, \bibinfo {author} {\bibfnamefont
  {I.}~\bibnamefont {Siddiqi}}, \bibinfo {author} {\bibfnamefont
  {K.}~\bibnamefont {Klymko}}, \bibinfo {author} {\bibfnamefont
  {G.}~\bibnamefont {Huang}}, \ and\ \bibinfo {author} {\bibfnamefont
  {K.}~\bibnamefont {Nowrouzi}},\ }\href@noop {} {\bibfield  {journal}
  {\bibinfo  {journal} {arXiv preprint arXiv:2406.13967}\ } (\bibinfo {year}
  {2024})}\BibitemShut {NoStop}%
\bibitem [{\citenamefont {Schervish}(1996)}]{Schervish1996-in}%
  \BibitemOpen
  \bibfield  {author} {\bibinfo {author} {\bibfnamefont {M.~J.}\ \bibnamefont
  {Schervish}},\ }\href {\doibase 10.1080/00031305.1996.10474380} {\bibfield
  {journal} {\bibinfo  {journal} {Am. Stat.}\ }\textbf {\bibinfo {volume}
  {50}},\ \bibinfo {pages} {203} (\bibinfo {year} {1996})}\BibitemShut
  {NoStop}%
\bibitem [{\citenamefont {Blume-Kohout}\ \emph {et~al.}(2020)\citenamefont
  {Blume-Kohout}, \citenamefont {Rudinger}, \citenamefont {Nielsen},
  \citenamefont {Proctor},\ and\ \citenamefont {Young}}]{blume2020wildcard}%
  \BibitemOpen
  \bibfield  {author} {\bibinfo {author} {\bibfnamefont {R.}~\bibnamefont
  {Blume-Kohout}}, \bibinfo {author} {\bibfnamefont {K.}~\bibnamefont
  {Rudinger}}, \bibinfo {author} {\bibfnamefont {E.}~\bibnamefont {Nielsen}},
  \bibinfo {author} {\bibfnamefont {T.}~\bibnamefont {Proctor}}, \ and\
  \bibinfo {author} {\bibfnamefont {K.}~\bibnamefont {Young}},\ }\href@noop {}
  {\bibfield  {journal} {\bibinfo  {journal} {Preprint at
  https://arxiv.org/abs/2012.12231}\ } (\bibinfo {year} {2020})}\BibitemShut
  {NoStop}%
\bibitem [{\citenamefont {Nguyen}\ \emph
  {et~al.}(2024{\natexlab{a}})\citenamefont {Nguyen}, \citenamefont {Kim},
  \citenamefont {Hashim}, \citenamefont {Goss}, \citenamefont {Marinelli},
  \citenamefont {Bhandari}, \citenamefont {Das}, \citenamefont {Naik},
  \citenamefont {Kreikebaum}, \citenamefont {Jordan} \emph
  {et~al.}}]{nguyen2022programmable}%
  \BibitemOpen
  \bibfield  {author} {\bibinfo {author} {\bibfnamefont {L.~B.}\ \bibnamefont
  {Nguyen}}, \bibinfo {author} {\bibfnamefont {Y.}~\bibnamefont {Kim}},
  \bibinfo {author} {\bibfnamefont {A.}~\bibnamefont {Hashim}}, \bibinfo
  {author} {\bibfnamefont {N.}~\bibnamefont {Goss}}, \bibinfo {author}
  {\bibfnamefont {B.}~\bibnamefont {Marinelli}}, \bibinfo {author}
  {\bibfnamefont {B.}~\bibnamefont {Bhandari}}, \bibinfo {author}
  {\bibfnamefont {D.}~\bibnamefont {Das}}, \bibinfo {author} {\bibfnamefont
  {R.~K.}\ \bibnamefont {Naik}}, \bibinfo {author} {\bibfnamefont {J.~M.}\
  \bibnamefont {Kreikebaum}}, \bibinfo {author} {\bibfnamefont {A.~N.}\
  \bibnamefont {Jordan}},  \emph {et~al.},\ }\href {\doibase
  10.1038/s41567-023-02326-7} {\bibfield  {journal} {\bibinfo  {journal}
  {Nature Physics}\ }\textbf {\bibinfo {volume} {20}},\ \bibinfo {pages} {240}
  (\bibinfo {year} {2024}{\natexlab{a}})}\BibitemShut {NoStop}%
\bibitem [{\citenamefont {Demtr{\"o}der}(1973)}]{demtroder1973laser}%
  \BibitemOpen
  \bibfield  {author} {\bibinfo {author} {\bibfnamefont {W.}~\bibnamefont
  {Demtr{\"o}der}},\ }\href@noop {} {\emph {\bibinfo {title} {Laser
  spectroscopy}}},\ Vol.~\bibinfo {volume} {2}\ (\bibinfo  {publisher}
  {Springer},\ \bibinfo {year} {1973})\BibitemShut {NoStop}%
\bibitem [{\citenamefont {Ramsey}(1950)}]{ramsey1950molecular}%
  \BibitemOpen
  \bibfield  {author} {\bibinfo {author} {\bibfnamefont {N.~F.}\ \bibnamefont
  {Ramsey}},\ }\href@noop {} {\bibfield  {journal} {\bibinfo  {journal}
  {Physical Review}\ }\textbf {\bibinfo {volume} {78}},\ \bibinfo {pages} {695}
  (\bibinfo {year} {1950})}\BibitemShut {NoStop}%
\bibitem [{\citenamefont {Somoroff}\ \emph {et~al.}(2023)\citenamefont
  {Somoroff}, \citenamefont {Ficheux}, \citenamefont {Mencia}, \citenamefont
  {Xiong}, \citenamefont {Kuzmin},\ and\ \citenamefont
  {Manucharyan}}]{somoroff2023millisecond}%
  \BibitemOpen
  \bibfield  {author} {\bibinfo {author} {\bibfnamefont {A.}~\bibnamefont
  {Somoroff}}, \bibinfo {author} {\bibfnamefont {Q.}~\bibnamefont {Ficheux}},
  \bibinfo {author} {\bibfnamefont {R.~A.}\ \bibnamefont {Mencia}}, \bibinfo
  {author} {\bibfnamefont {H.}~\bibnamefont {Xiong}}, \bibinfo {author}
  {\bibfnamefont {R.}~\bibnamefont {Kuzmin}}, \ and\ \bibinfo {author}
  {\bibfnamefont {V.~E.}\ \bibnamefont {Manucharyan}},\ }\href@noop {}
  {\bibfield  {journal} {\bibinfo  {journal} {Physical Review Letters}\
  }\textbf {\bibinfo {volume} {130}},\ \bibinfo {pages} {267001} (\bibinfo
  {year} {2023})}\BibitemShut {NoStop}%
\bibitem [{\citenamefont {Wang}\ \emph {et~al.}(2021)\citenamefont {Wang},
  \citenamefont {Luan}, \citenamefont {Qiao}, \citenamefont {Um}, \citenamefont
  {Zhang}, \citenamefont {Wang}, \citenamefont {Yuan}, \citenamefont {Gu},
  \citenamefont {Zhang},\ and\ \citenamefont {Kim}}]{wang2021single}%
  \BibitemOpen
  \bibfield  {author} {\bibinfo {author} {\bibfnamefont {P.}~\bibnamefont
  {Wang}}, \bibinfo {author} {\bibfnamefont {C.-Y.}\ \bibnamefont {Luan}},
  \bibinfo {author} {\bibfnamefont {M.}~\bibnamefont {Qiao}}, \bibinfo {author}
  {\bibfnamefont {M.}~\bibnamefont {Um}}, \bibinfo {author} {\bibfnamefont
  {J.}~\bibnamefont {Zhang}}, \bibinfo {author} {\bibfnamefont
  {Y.}~\bibnamefont {Wang}}, \bibinfo {author} {\bibfnamefont {X.}~\bibnamefont
  {Yuan}}, \bibinfo {author} {\bibfnamefont {M.}~\bibnamefont {Gu}}, \bibinfo
  {author} {\bibfnamefont {J.}~\bibnamefont {Zhang}}, \ and\ \bibinfo {author}
  {\bibfnamefont {K.}~\bibnamefont {Kim}},\ }\href@noop {} {\bibfield
  {journal} {\bibinfo  {journal} {Nature communications}\ }\textbf {\bibinfo
  {volume} {12}},\ \bibinfo {pages} {233} (\bibinfo {year} {2021})}\BibitemShut
  {NoStop}%
\bibitem [{\citenamefont {Redfield}(1957)}]{redfield1957theory}%
  \BibitemOpen
  \bibfield  {author} {\bibinfo {author} {\bibfnamefont {A.~G.}\ \bibnamefont
  {Redfield}},\ }\href@noop {} {\bibfield  {journal} {\bibinfo  {journal} {IBM
  Journal of Research and Development}\ }\textbf {\bibinfo {volume} {1}},\
  \bibinfo {pages} {19} (\bibinfo {year} {1957})}\BibitemShut {NoStop}%
\bibitem [{Note17()}]{Note17}%
  \BibitemOpen
  \bibinfo {note} {The term ``phase estimation'' has multiple and closely
  related uses, as it also refers to the similar but distinct task of
  estimating a Hamiltonian's eigenvalues \cite
  {kitaev1995quantum}.}\BibitemShut {Stop}%
\bibitem [{\citenamefont {Kimmel}\ \emph {et~al.}(2015)\citenamefont {Kimmel},
  \citenamefont {Low},\ and\ \citenamefont {Yoder}}]{kimmel2015robust}%
  \BibitemOpen
  \bibfield  {author} {\bibinfo {author} {\bibfnamefont {S.}~\bibnamefont
  {Kimmel}}, \bibinfo {author} {\bibfnamefont {G.~H.}\ \bibnamefont {Low}}, \
  and\ \bibinfo {author} {\bibfnamefont {T.~J.}\ \bibnamefont {Yoder}},\
  }\href@noop {} {\bibfield  {journal} {\bibinfo  {journal} {Physical Review
  A}\ }\textbf {\bibinfo {volume} {92}},\ \bibinfo {pages} {062315} (\bibinfo
  {year} {2015})}\BibitemShut {NoStop}%
\bibitem [{Note18()}]{Note18}%
  \BibitemOpen
  \bibinfo {note} {A unit-normalized linear combination of Pauli matrices also
  works, i.e., $\protect \bf {\protect \hat {n}} \cdot \protect \bm {\sigma }$,
  but makes state preparation and measurement mildly more
  complicated.}\BibitemShut {Stop}%
\bibitem [{\citenamefont {Russo}\ \emph
  {et~al.}(2021{\natexlab{a}})\citenamefont {Russo}, \citenamefont {Kirby},
  \citenamefont {Rudinger}, \citenamefont {Baczewski},\ and\ \citenamefont
  {Kimmel}}]{russo2021consistency}%
  \BibitemOpen
  \bibfield  {author} {\bibinfo {author} {\bibfnamefont {A.~E.}\ \bibnamefont
  {Russo}}, \bibinfo {author} {\bibfnamefont {W.~M.}\ \bibnamefont {Kirby}},
  \bibinfo {author} {\bibfnamefont {K.~M.}\ \bibnamefont {Rudinger}}, \bibinfo
  {author} {\bibfnamefont {A.~D.}\ \bibnamefont {Baczewski}}, \ and\ \bibinfo
  {author} {\bibfnamefont {S.}~\bibnamefont {Kimmel}},\ }\href@noop {}
  {\bibfield  {journal} {\bibinfo  {journal} {Physical Review A}\ }\textbf
  {\bibinfo {volume} {103}},\ \bibinfo {pages} {042609} (\bibinfo {year}
  {2021}{\natexlab{a}})}\BibitemShut {NoStop}%
\bibitem [{\citenamefont {Rudinger}\ \emph {et~al.}(2017)\citenamefont
  {Rudinger}, \citenamefont {Kimmel}, \citenamefont {Lobser},\ and\
  \citenamefont {Maunz}}]{rudinger2017experimental}%
  \BibitemOpen
  \bibfield  {author} {\bibinfo {author} {\bibfnamefont {K.}~\bibnamefont
  {Rudinger}}, \bibinfo {author} {\bibfnamefont {S.}~\bibnamefont {Kimmel}},
  \bibinfo {author} {\bibfnamefont {D.}~\bibnamefont {Lobser}}, \ and\ \bibinfo
  {author} {\bibfnamefont {P.}~\bibnamefont {Maunz}},\ }\href@noop {}
  {\bibfield  {journal} {\bibinfo  {journal} {Physical review letters}\
  }\textbf {\bibinfo {volume} {118}},\ \bibinfo {pages} {190502} (\bibinfo
  {year} {2017})}\BibitemShut {NoStop}%
\bibitem [{pyr(2021)}]{pyrpe2021repo}%
  \BibitemOpen
  \href@noop {} {\enquote {\bibinfo {title} {pyrpe},}\ }\bibinfo {howpublished}
  {\url{https://https://gitlab.com/quapack/pyrpe}} (\bibinfo {year}
  {2021})\BibitemShut {NoStop}%
\bibitem [{\citenamefont {Russo}\ \emph
  {et~al.}(2021{\natexlab{b}})\citenamefont {Russo}, \citenamefont {Rudinger},
  \citenamefont {Morrison},\ and\ \citenamefont
  {Baczewski}}]{russo2021evaluating}%
  \BibitemOpen
  \bibfield  {author} {\bibinfo {author} {\bibfnamefont {A.~E.}\ \bibnamefont
  {Russo}}, \bibinfo {author} {\bibfnamefont {K.~M.}\ \bibnamefont {Rudinger}},
  \bibinfo {author} {\bibfnamefont {B.~C.~A.}\ \bibnamefont {Morrison}}, \ and\
  \bibinfo {author} {\bibfnamefont {A.~D.}\ \bibnamefont {Baczewski}},\ }\href
  {\doibase 10.1103/PhysRevLett.126.210501} {\bibfield  {journal} {\bibinfo
  {journal} {Phys. Rev. Lett.}\ }\textbf {\bibinfo {volume} {126}},\ \bibinfo
  {pages} {210501} (\bibinfo {year} {2021}{\natexlab{b}})}\BibitemShut
  {NoStop}%
\bibitem [{\citenamefont {Rudinger}\ \emph {et~al.}(2025)\citenamefont
  {Rudinger}, \citenamefont {Marceaux}, \citenamefont {Hashim}, \citenamefont
  {Santiago}, \citenamefont {Siddiqi},\ and\ \citenamefont
  {Young}}]{rudinger2025heisenberg}%
  \BibitemOpen
  \bibfield  {author} {\bibinfo {author} {\bibfnamefont {K.}~\bibnamefont
  {Rudinger}}, \bibinfo {author} {\bibfnamefont {J.~P.}\ \bibnamefont
  {Marceaux}}, \bibinfo {author} {\bibfnamefont {A.}~\bibnamefont {Hashim}},
  \bibinfo {author} {\bibfnamefont {D.~I.}\ \bibnamefont {Santiago}}, \bibinfo
  {author} {\bibfnamefont {I.}~\bibnamefont {Siddiqi}}, \ and\ \bibinfo
  {author} {\bibfnamefont {K.~C.}\ \bibnamefont {Young}},\ }\href@noop {}
  {\bibfield  {journal} {\bibinfo  {journal} {arXiv preprint arXiv:2502.06698}\
  } (\bibinfo {year} {2025})}\BibitemShut {NoStop}%
\bibitem [{\citenamefont {Gale}\ \emph {et~al.}(1968)\citenamefont {Gale},
  \citenamefont {Guth},\ and\ \citenamefont
  {Trammell}}]{gale1968determination}%
  \BibitemOpen
  \bibfield  {author} {\bibinfo {author} {\bibfnamefont {W.}~\bibnamefont
  {Gale}}, \bibinfo {author} {\bibfnamefont {E.}~\bibnamefont {Guth}}, \ and\
  \bibinfo {author} {\bibfnamefont {G.}~\bibnamefont {Trammell}},\ }\href@noop
  {} {\bibfield  {journal} {\bibinfo  {journal} {Physical Review}\ }\textbf
  {\bibinfo {volume} {165}},\ \bibinfo {pages} {1434} (\bibinfo {year}
  {1968})}\BibitemShut {NoStop}%
\bibitem [{\citenamefont {Chuang}\ and\ \citenamefont
  {Nielsen}(1997)}]{chuang1997prescription}%
  \BibitemOpen
  \bibfield  {author} {\bibinfo {author} {\bibfnamefont {I.~L.}\ \bibnamefont
  {Chuang}}\ and\ \bibinfo {author} {\bibfnamefont {M.~A.}\ \bibnamefont
  {Nielsen}},\ }\href@noop {} {\bibfield  {journal} {\bibinfo  {journal}
  {Journal of Modern Optics}\ }\textbf {\bibinfo {volume} {44}},\ \bibinfo
  {pages} {2455} (\bibinfo {year} {1997})}\BibitemShut {NoStop}%
\bibitem [{\citenamefont {Wootters}\ and\ \citenamefont
  {Fields}(1989)}]{wootters1989optimal}%
  \BibitemOpen
  \bibfield  {author} {\bibinfo {author} {\bibfnamefont {W.~K.}\ \bibnamefont
  {Wootters}}\ and\ \bibinfo {author} {\bibfnamefont {B.~D.}\ \bibnamefont
  {Fields}},\ }\href@noop {} {\bibfield  {journal} {\bibinfo  {journal} {Annals
  of Physics}\ }\textbf {\bibinfo {volume} {191}},\ \bibinfo {pages} {363}
  (\bibinfo {year} {1989})}\BibitemShut {NoStop}%
\bibitem [{\citenamefont {Adamson}\ and\ \citenamefont
  {Steinberg}(2010)}]{adamson2010improving}%
  \BibitemOpen
  \bibfield  {author} {\bibinfo {author} {\bibfnamefont {R.}~\bibnamefont
  {Adamson}}\ and\ \bibinfo {author} {\bibfnamefont {A.~M.}\ \bibnamefont
  {Steinberg}},\ }\href@noop {} {\bibfield  {journal} {\bibinfo  {journal}
  {Physical review letters}\ }\textbf {\bibinfo {volume} {105}},\ \bibinfo
  {pages} {030406} (\bibinfo {year} {2010})}\BibitemShut {NoStop}%
\bibitem [{\citenamefont {Stricker}\ \emph {et~al.}(2022)\citenamefont
  {Stricker}, \citenamefont {Meth}, \citenamefont {Postler}, \citenamefont
  {Edmunds}, \citenamefont {Ferrie}, \citenamefont {Blatt}, \citenamefont
  {Schindler}, \citenamefont {Monz}, \citenamefont {Kueng},\ and\ \citenamefont
  {Ringbauer}}]{stricker2022experimental}%
  \BibitemOpen
  \bibfield  {author} {\bibinfo {author} {\bibfnamefont {R.}~\bibnamefont
  {Stricker}}, \bibinfo {author} {\bibfnamefont {M.}~\bibnamefont {Meth}},
  \bibinfo {author} {\bibfnamefont {L.}~\bibnamefont {Postler}}, \bibinfo
  {author} {\bibfnamefont {C.}~\bibnamefont {Edmunds}}, \bibinfo {author}
  {\bibfnamefont {C.}~\bibnamefont {Ferrie}}, \bibinfo {author} {\bibfnamefont
  {R.}~\bibnamefont {Blatt}}, \bibinfo {author} {\bibfnamefont
  {P.}~\bibnamefont {Schindler}}, \bibinfo {author} {\bibfnamefont
  {T.}~\bibnamefont {Monz}}, \bibinfo {author} {\bibfnamefont {R.}~\bibnamefont
  {Kueng}}, \ and\ \bibinfo {author} {\bibfnamefont {M.}~\bibnamefont
  {Ringbauer}},\ }\href@noop {} {\bibfield  {journal} {\bibinfo  {journal} {PRX
  Quantum}\ }\textbf {\bibinfo {volume} {3}},\ \bibinfo {pages} {040310}
  (\bibinfo {year} {2022})}\BibitemShut {NoStop}%
\bibitem [{\citenamefont {Flammia}\ \emph {et~al.}(2005)\citenamefont
  {Flammia}, \citenamefont {Silberfarb},\ and\ \citenamefont
  {Caves}}]{flammia2005minimal}%
  \BibitemOpen
  \bibfield  {author} {\bibinfo {author} {\bibfnamefont {S.~T.}\ \bibnamefont
  {Flammia}}, \bibinfo {author} {\bibfnamefont {A.}~\bibnamefont {Silberfarb}},
  \ and\ \bibinfo {author} {\bibfnamefont {C.~M.}\ \bibnamefont {Caves}},\
  }\href@noop {} {\bibfield  {journal} {\bibinfo  {journal} {Foundations of
  Physics}\ }\textbf {\bibinfo {volume} {35}},\ \bibinfo {pages} {1985}
  (\bibinfo {year} {2005})}\BibitemShut {NoStop}%
\bibitem [{\citenamefont {Filippov}\ and\ \citenamefont
  {Man'Ko}(2011)}]{filippov2011mutually}%
  \BibitemOpen
  \bibfield  {author} {\bibinfo {author} {\bibfnamefont {S.}~\bibnamefont
  {Filippov}}\ and\ \bibinfo {author} {\bibfnamefont {V.}~\bibnamefont
  {Man'Ko}},\ }\href@noop {} {\bibfield  {journal} {\bibinfo  {journal}
  {Physica Scripta}\ }\textbf {\bibinfo {volume} {2011}},\ \bibinfo {pages}
  {014010} (\bibinfo {year} {2011})}\BibitemShut {NoStop}%
\bibitem [{\citenamefont {Gottesman}(1997)}]{gottesman1997stabilizer}%
  \BibitemOpen
  \bibfield  {author} {\bibinfo {author} {\bibfnamefont {D.~E.}\ \bibnamefont
  {Gottesman}},\ }\emph {\bibinfo {title} {Stabilizer Codes and Quantum Error
  Correction}},\ \href@noop {} {Ph.D. thesis},\ \bibinfo  {school} {California
  Institute of Technology} (\bibinfo {year} {1997})\BibitemShut {NoStop}%
\bibitem [{Note19()}]{Note19}%
  \BibitemOpen
  \bibinfo {note} {Shot noise is not the only source of fluctuations and
  errors. Laboratory measurements are also subject to, for example, imperfect
  signal amplification, electronic noise, poor quantum efficiency, imperfect
  signal conversion, errors in digitization and classification, and a
  terrifying range of \protect \emph {systematic} errors like drift over the
  duration of a tomography experiment. However, there is no systematic
  theoretical treatment of these noise sources. In practice, the techniques
  used to deal with shot noise (which \protect \emph {does} have a solid
  theory) can deal with these noise sources too, although not
  optimally.}\BibitemShut {Stop}%
\bibitem [{\citenamefont {Smolin}\ \emph {et~al.}(2012)\citenamefont {Smolin},
  \citenamefont {Gambetta},\ and\ \citenamefont {Smith}}]{smolin2012efficient}%
  \BibitemOpen
  \bibfield  {author} {\bibinfo {author} {\bibfnamefont {J.~A.}\ \bibnamefont
  {Smolin}}, \bibinfo {author} {\bibfnamefont {J.~M.}\ \bibnamefont
  {Gambetta}}, \ and\ \bibinfo {author} {\bibfnamefont {G.}~\bibnamefont
  {Smith}},\ }\href@noop {} {\bibfield  {journal} {\bibinfo  {journal}
  {Physical review letters}\ }\textbf {\bibinfo {volume} {108}},\ \bibinfo
  {pages} {070502} (\bibinfo {year} {2012})}\BibitemShut {NoStop}%
\bibitem [{\citenamefont {Gu{\c{t}}{\u{a}}}\ \emph {et~al.}(2020)\citenamefont
  {Gu{\c{t}}{\u{a}}}, \citenamefont {Kahn}, \citenamefont {Kueng},\ and\
  \citenamefont {Tropp}}]{guta2020fast}%
  \BibitemOpen
  \bibfield  {author} {\bibinfo {author} {\bibfnamefont {M.}~\bibnamefont
  {Gu{\c{t}}{\u{a}}}}, \bibinfo {author} {\bibfnamefont {J.}~\bibnamefont
  {Kahn}}, \bibinfo {author} {\bibfnamefont {R.}~\bibnamefont {Kueng}}, \ and\
  \bibinfo {author} {\bibfnamefont {J.~A.}\ \bibnamefont {Tropp}},\ }\href@noop
  {} {\bibfield  {journal} {\bibinfo  {journal} {Journal of Physics A:
  Mathematical and Theoretical}\ }\textbf {\bibinfo {volume} {53}},\ \bibinfo
  {pages} {204001} (\bibinfo {year} {2020})}\BibitemShut {NoStop}%
\bibitem [{\citenamefont {Surawy-Stepney}\ \emph {et~al.}(2022)\citenamefont
  {Surawy-Stepney}, \citenamefont {Kahn}, \citenamefont {Kueng},\ and\
  \citenamefont {Guta}}]{surawy2022projected}%
  \BibitemOpen
  \bibfield  {author} {\bibinfo {author} {\bibfnamefont {T.}~\bibnamefont
  {Surawy-Stepney}}, \bibinfo {author} {\bibfnamefont {J.}~\bibnamefont
  {Kahn}}, \bibinfo {author} {\bibfnamefont {R.}~\bibnamefont {Kueng}}, \ and\
  \bibinfo {author} {\bibfnamefont {M.}~\bibnamefont {Guta}},\ }\href@noop {}
  {\bibfield  {journal} {\bibinfo  {journal} {Quantum}\ }\textbf {\bibinfo
  {volume} {6}},\ \bibinfo {pages} {844} (\bibinfo {year} {2022})}\BibitemShut
  {NoStop}%
\bibitem [{\citenamefont {Hradil}(1997)}]{hradil1997quantum}%
  \BibitemOpen
  \bibfield  {author} {\bibinfo {author} {\bibfnamefont {Z.}~\bibnamefont
  {Hradil}},\ }\href {\doibase 10.1103/PhysRevA.55.R1561} {\bibfield  {journal}
  {\bibinfo  {journal} {Phys. Rev. A}\ }\textbf {\bibinfo {volume} {55}},\
  \bibinfo {pages} {R1561} (\bibinfo {year} {1997})}\BibitemShut {NoStop}%
\bibitem [{\citenamefont {Banaszek}\ \emph {et~al.}(1999)\citenamefont
  {Banaszek}, \citenamefont {D'Ariano}, \citenamefont {Paris},\ and\
  \citenamefont {Sacchi}}]{banaszek1999maximum}%
  \BibitemOpen
  \bibfield  {author} {\bibinfo {author} {\bibfnamefont {K.}~\bibnamefont
  {Banaszek}}, \bibinfo {author} {\bibfnamefont {G.~M.}\ \bibnamefont
  {D'Ariano}}, \bibinfo {author} {\bibfnamefont {M.~G.~A.}\ \bibnamefont
  {Paris}}, \ and\ \bibinfo {author} {\bibfnamefont {M.~F.}\ \bibnamefont
  {Sacchi}},\ }\href {\doibase 10.1103/PhysRevA.61.010304} {\bibfield
  {journal} {\bibinfo  {journal} {Phys. Rev. A}\ }\textbf {\bibinfo {volume}
  {61}},\ \bibinfo {pages} {010304} (\bibinfo {year} {1999})}\BibitemShut
  {NoStop}%
\bibitem [{\citenamefont
  {Blume-Kohout}(2010{\natexlab{a}})}]{blume2010optimal}%
  \BibitemOpen
  \bibfield  {author} {\bibinfo {author} {\bibfnamefont {R.}~\bibnamefont
  {Blume-Kohout}},\ }\href@noop {} {\bibfield  {journal} {\bibinfo  {journal}
  {New Journal of Physics}\ }\textbf {\bibinfo {volume} {12}},\ \bibinfo
  {pages} {043034} (\bibinfo {year} {2010}{\natexlab{a}})}\BibitemShut
  {NoStop}%
\bibitem [{\citenamefont {Bu{\v{z}}ek}\ \emph {et~al.}(1998)\citenamefont
  {Bu{\v{z}}ek}, \citenamefont {Derka}, \citenamefont {Adam},\ and\
  \citenamefont {Knight}}]{buvzek1998reconstruction}%
  \BibitemOpen
  \bibfield  {author} {\bibinfo {author} {\bibfnamefont {V.}~\bibnamefont
  {Bu{\v{z}}ek}}, \bibinfo {author} {\bibfnamefont {R.}~\bibnamefont {Derka}},
  \bibinfo {author} {\bibfnamefont {G.}~\bibnamefont {Adam}}, \ and\ \bibinfo
  {author} {\bibfnamefont {P.~L.}\ \bibnamefont {Knight}},\ }\href@noop {}
  {\bibfield  {journal} {\bibinfo  {journal} {Annals of Physics}\ }\textbf
  {\bibinfo {volume} {266}},\ \bibinfo {pages} {454} (\bibinfo {year}
  {1998})}\BibitemShut {NoStop}%
\bibitem [{\citenamefont {Granade}\ \emph {et~al.}(2016)\citenamefont
  {Granade}, \citenamefont {Combes},\ and\ \citenamefont
  {Cory}}]{granade2016practical}%
  \BibitemOpen
  \bibfield  {author} {\bibinfo {author} {\bibfnamefont {C.}~\bibnamefont
  {Granade}}, \bibinfo {author} {\bibfnamefont {J.}~\bibnamefont {Combes}}, \
  and\ \bibinfo {author} {\bibfnamefont {D.}~\bibnamefont {Cory}},\ }\href@noop
  {} {\bibfield  {journal} {\bibinfo  {journal} {New Journal of Physics}\
  }\textbf {\bibinfo {volume} {18}},\ \bibinfo {pages} {033024} (\bibinfo
  {year} {2016})}\BibitemShut {NoStop}%
\bibitem [{\citenamefont {Caves}\ \emph {et~al.}(2002)\citenamefont {Caves},
  \citenamefont {Fuchs},\ and\ \citenamefont {Schack}}]{caves2002quantum}%
  \BibitemOpen
  \bibfield  {author} {\bibinfo {author} {\bibfnamefont {C.~M.}\ \bibnamefont
  {Caves}}, \bibinfo {author} {\bibfnamefont {C.~A.}\ \bibnamefont {Fuchs}}, \
  and\ \bibinfo {author} {\bibfnamefont {R.}~\bibnamefont {Schack}},\
  }\href@noop {} {\bibfield  {journal} {\bibinfo  {journal} {Physical review
  A}\ }\textbf {\bibinfo {volume} {65}},\ \bibinfo {pages} {022305} (\bibinfo
  {year} {2002})}\BibitemShut {NoStop}%
\bibitem [{\citenamefont {Von~Toussaint}(2011)}]{von2011bayesian}%
  \BibitemOpen
  \bibfield  {author} {\bibinfo {author} {\bibfnamefont {U.}~\bibnamefont
  {Von~Toussaint}},\ }\href@noop {} {\bibfield  {journal} {\bibinfo  {journal}
  {Reviews of Modern Physics}\ }\textbf {\bibinfo {volume} {83}},\ \bibinfo
  {pages} {943} (\bibinfo {year} {2011})}\BibitemShut {NoStop}%
\bibitem [{\citenamefont {Barnett}\ and\ \citenamefont
  {Croke}(2009)}]{barnett2009quantum}%
  \BibitemOpen
  \bibfield  {author} {\bibinfo {author} {\bibfnamefont {S.~M.}\ \bibnamefont
  {Barnett}}\ and\ \bibinfo {author} {\bibfnamefont {S.}~\bibnamefont
  {Croke}},\ }\href@noop {} {\bibfield  {journal} {\bibinfo  {journal}
  {Advances in Optics and Photonics}\ }\textbf {\bibinfo {volume} {1}},\
  \bibinfo {pages} {238} (\bibinfo {year} {2009})}\BibitemShut {NoStop}%
\bibitem [{\citenamefont {Bae}\ and\ \citenamefont
  {Kwek}(2015)}]{bae2015quantum}%
  \BibitemOpen
  \bibfield  {author} {\bibinfo {author} {\bibfnamefont {J.}~\bibnamefont
  {Bae}}\ and\ \bibinfo {author} {\bibfnamefont {L.-C.}\ \bibnamefont {Kwek}},\
  }\href@noop {} {\bibfield  {journal} {\bibinfo  {journal} {Journal of Physics
  A: Mathematical and Theoretical}\ }\textbf {\bibinfo {volume} {48}},\
  \bibinfo {pages} {083001} (\bibinfo {year} {2015})}\BibitemShut {NoStop}%
\bibitem [{\citenamefont {Siddhu}(2019)}]{siddhu2019maximum}%
  \BibitemOpen
  \bibfield  {author} {\bibinfo {author} {\bibfnamefont {V.}~\bibnamefont
  {Siddhu}},\ }\href@noop {} {\bibfield  {journal} {\bibinfo  {journal}
  {Physical Review A}\ }\textbf {\bibinfo {volume} {99}},\ \bibinfo {pages}
  {012342} (\bibinfo {year} {2019})}\BibitemShut {NoStop}%
\bibitem [{\citenamefont {Blume-Kohout}(2010{\natexlab{b}})}]{blume2010hedged}%
  \BibitemOpen
  \bibfield  {author} {\bibinfo {author} {\bibfnamefont {R.}~\bibnamefont
  {Blume-Kohout}},\ }\href@noop {} {\bibfield  {journal} {\bibinfo  {journal}
  {Physical review letters}\ }\textbf {\bibinfo {volume} {105}},\ \bibinfo
  {pages} {200504} (\bibinfo {year} {2010}{\natexlab{b}})}\BibitemShut
  {NoStop}%
\bibitem [{\citenamefont {Husz{\'a}r}\ and\ \citenamefont
  {Houlsby}(2012)}]{huszar2012adaptive}%
  \BibitemOpen
  \bibfield  {author} {\bibinfo {author} {\bibfnamefont {F.}~\bibnamefont
  {Husz{\'a}r}}\ and\ \bibinfo {author} {\bibfnamefont {N.~M.}\ \bibnamefont
  {Houlsby}},\ }\href@noop {} {\bibfield  {journal} {\bibinfo  {journal}
  {Physical Review A—Atomic, Molecular, and Optical Physics}\ }\textbf
  {\bibinfo {volume} {85}},\ \bibinfo {pages} {052120} (\bibinfo {year}
  {2012})}\BibitemShut {NoStop}%
\bibitem [{\citenamefont {Kravtsov}\ \emph {et~al.}(2013)\citenamefont
  {Kravtsov}, \citenamefont {Straupe}, \citenamefont {Radchenko}, \citenamefont
  {Houlsby}, \citenamefont {Husz{\'a}r},\ and\ \citenamefont
  {Kulik}}]{kravtsov2013experimental}%
  \BibitemOpen
  \bibfield  {author} {\bibinfo {author} {\bibfnamefont {K.~S.}\ \bibnamefont
  {Kravtsov}}, \bibinfo {author} {\bibfnamefont {S.~S.}\ \bibnamefont
  {Straupe}}, \bibinfo {author} {\bibfnamefont {I.~V.}\ \bibnamefont
  {Radchenko}}, \bibinfo {author} {\bibfnamefont {N.~M.}\ \bibnamefont
  {Houlsby}}, \bibinfo {author} {\bibfnamefont {F.}~\bibnamefont {Husz{\'a}r}},
  \ and\ \bibinfo {author} {\bibfnamefont {S.~P.}\ \bibnamefont {Kulik}},\
  }\href@noop {} {\bibfield  {journal} {\bibinfo  {journal} {Physical Review
  A—Atomic, Molecular, and Optical Physics}\ }\textbf {\bibinfo {volume}
  {87}},\ \bibinfo {pages} {062122} (\bibinfo {year} {2013})}\BibitemShut
  {NoStop}%
\bibitem [{\citenamefont {Granade}\ \emph {et~al.}(2017)\citenamefont
  {Granade}, \citenamefont {Ferrie},\ and\ \citenamefont
  {Flammia}}]{granade2017practical}%
  \BibitemOpen
  \bibfield  {author} {\bibinfo {author} {\bibfnamefont {C.}~\bibnamefont
  {Granade}}, \bibinfo {author} {\bibfnamefont {C.}~\bibnamefont {Ferrie}}, \
  and\ \bibinfo {author} {\bibfnamefont {S.~T.}\ \bibnamefont {Flammia}},\
  }\href@noop {} {\bibfield  {journal} {\bibinfo  {journal} {New Journal of
  Physics}\ }\textbf {\bibinfo {volume} {19}},\ \bibinfo {pages} {113017}
  (\bibinfo {year} {2017})}\BibitemShut {NoStop}%
\bibitem [{\citenamefont {Christandl}\ and\ \citenamefont
  {Renner}(2012)}]{christandl2012reliable}%
  \BibitemOpen
  \bibfield  {author} {\bibinfo {author} {\bibfnamefont {M.}~\bibnamefont
  {Christandl}}\ and\ \bibinfo {author} {\bibfnamefont {R.}~\bibnamefont
  {Renner}},\ }\href@noop {} {\bibfield  {journal} {\bibinfo  {journal}
  {Physical Review Letters}\ }\textbf {\bibinfo {volume} {109}},\ \bibinfo
  {pages} {120403} (\bibinfo {year} {2012})}\BibitemShut {NoStop}%
\bibitem [{\citenamefont {Oh}\ \emph {et~al.}(2019)\citenamefont {Oh},
  \citenamefont {Teo},\ and\ \citenamefont {Jeong}}]{oh2019efficient}%
  \BibitemOpen
  \bibfield  {author} {\bibinfo {author} {\bibfnamefont {C.}~\bibnamefont
  {Oh}}, \bibinfo {author} {\bibfnamefont {Y.~S.}\ \bibnamefont {Teo}}, \ and\
  \bibinfo {author} {\bibfnamefont {H.}~\bibnamefont {Jeong}},\ }\href@noop {}
  {\bibfield  {journal} {\bibinfo  {journal} {Physical Review A}\ }\textbf
  {\bibinfo {volume} {100}},\ \bibinfo {pages} {012345} (\bibinfo {year}
  {2019})}\BibitemShut {NoStop}%
\bibitem [{\citenamefont {Lukens}\ \emph {et~al.}(2020)\citenamefont {Lukens},
  \citenamefont {Law}, \citenamefont {Jasra},\ and\ \citenamefont
  {Lougovski}}]{lukens2020practical}%
  \BibitemOpen
  \bibfield  {author} {\bibinfo {author} {\bibfnamefont {J.~M.}\ \bibnamefont
  {Lukens}}, \bibinfo {author} {\bibfnamefont {K.~J.}\ \bibnamefont {Law}},
  \bibinfo {author} {\bibfnamefont {A.}~\bibnamefont {Jasra}}, \ and\ \bibinfo
  {author} {\bibfnamefont {P.}~\bibnamefont {Lougovski}},\ }\href@noop {}
  {\bibfield  {journal} {\bibinfo  {journal} {New Journal of Physics}\ }\textbf
  {\bibinfo {volume} {22}},\ \bibinfo {pages} {063038} (\bibinfo {year}
  {2020})}\BibitemShut {NoStop}%
\bibitem [{\citenamefont {Evans}\ \emph {et~al.}(2022)\citenamefont {Evans},
  \citenamefont {Huang}, \citenamefont {Yoneda}, \citenamefont {Harper},
  \citenamefont {Tanttu}, \citenamefont {Chan}, \citenamefont {Hudson},
  \citenamefont {Itoh}, \citenamefont {Saraiva}, \citenamefont {Yang} \emph
  {et~al.}}]{evans2022fast}%
  \BibitemOpen
  \bibfield  {author} {\bibinfo {author} {\bibfnamefont {T.}~\bibnamefont
  {Evans}}, \bibinfo {author} {\bibfnamefont {W.}~\bibnamefont {Huang}},
  \bibinfo {author} {\bibfnamefont {J.}~\bibnamefont {Yoneda}}, \bibinfo
  {author} {\bibfnamefont {R.}~\bibnamefont {Harper}}, \bibinfo {author}
  {\bibfnamefont {T.}~\bibnamefont {Tanttu}}, \bibinfo {author} {\bibfnamefont
  {K.}~\bibnamefont {Chan}}, \bibinfo {author} {\bibfnamefont {F.}~\bibnamefont
  {Hudson}}, \bibinfo {author} {\bibfnamefont {K.}~\bibnamefont {Itoh}},
  \bibinfo {author} {\bibfnamefont {A.}~\bibnamefont {Saraiva}}, \bibinfo
  {author} {\bibfnamefont {C.}~\bibnamefont {Yang}},  \emph {et~al.},\
  }\href@noop {} {\bibfield  {journal} {\bibinfo  {journal} {Physical Review
  Applied}\ }\textbf {\bibinfo {volume} {17}},\ \bibinfo {pages} {024068}
  (\bibinfo {year} {2022})}\BibitemShut {NoStop}%
\bibitem [{\citenamefont {Poyatos}\ \emph {et~al.}(1997)\citenamefont
  {Poyatos}, \citenamefont {Cirac},\ and\ \citenamefont
  {Zoller}}]{poyatos1997complete}%
  \BibitemOpen
  \bibfield  {author} {\bibinfo {author} {\bibfnamefont {J.}~\bibnamefont
  {Poyatos}}, \bibinfo {author} {\bibfnamefont {J.~I.}\ \bibnamefont {Cirac}},
  \ and\ \bibinfo {author} {\bibfnamefont {P.}~\bibnamefont {Zoller}},\
  }\href@noop {} {\bibfield  {journal} {\bibinfo  {journal} {Physical Review
  Letters}\ }\textbf {\bibinfo {volume} {78}},\ \bibinfo {pages} {390}
  (\bibinfo {year} {1997})}\BibitemShut {NoStop}%
\bibitem [{\citenamefont {Kim}\ \emph {et~al.}(2022)\citenamefont {Kim},
  \citenamefont {Morvan}, \citenamefont {Nguyen}, \citenamefont {Naik},
  \citenamefont {J{\"u}nger}, \citenamefont {Chen}, \citenamefont {Kreikebaum},
  \citenamefont {Santiago},\ and\ \citenamefont {Siddiqi}}]{kim2022high}%
  \BibitemOpen
  \bibfield  {author} {\bibinfo {author} {\bibfnamefont {Y.}~\bibnamefont
  {Kim}}, \bibinfo {author} {\bibfnamefont {A.}~\bibnamefont {Morvan}},
  \bibinfo {author} {\bibfnamefont {L.~B.}\ \bibnamefont {Nguyen}}, \bibinfo
  {author} {\bibfnamefont {R.~K.}\ \bibnamefont {Naik}}, \bibinfo {author}
  {\bibfnamefont {C.}~\bibnamefont {J{\"u}nger}}, \bibinfo {author}
  {\bibfnamefont {L.}~\bibnamefont {Chen}}, \bibinfo {author} {\bibfnamefont
  {J.~M.}\ \bibnamefont {Kreikebaum}}, \bibinfo {author} {\bibfnamefont
  {D.~I.}\ \bibnamefont {Santiago}}, \ and\ \bibinfo {author} {\bibfnamefont
  {I.}~\bibnamefont {Siddiqi}},\ }\href {\doibase 10.1038/s41567-022-01590-3}
  {\bibfield  {journal} {\bibinfo  {journal} {Nat. Phys.}\ }\textbf {\bibinfo
  {volume} {18}},\ \bibinfo {pages} {783} (\bibinfo {year} {2022})}\BibitemShut
  {NoStop}%
\bibitem [{\citenamefont {Chow}\ \emph {et~al.}(2012)\citenamefont {Chow},
  \citenamefont {Gambetta}, \citenamefont {C\'orcoles}, \citenamefont {Merkel},
  \citenamefont {Smolin}, \citenamefont {Rigetti}, \citenamefont {Poletto},
  \citenamefont {Keefe}, \citenamefont {Rothwell}, \citenamefont {Rozen},
  \citenamefont {Ketchen},\ and\ \citenamefont {Steffen}}]{chow2012universal}%
  \BibitemOpen
  \bibfield  {author} {\bibinfo {author} {\bibfnamefont {J.~M.}\ \bibnamefont
  {Chow}}, \bibinfo {author} {\bibfnamefont {J.~M.}\ \bibnamefont {Gambetta}},
  \bibinfo {author} {\bibfnamefont {A.~D.}\ \bibnamefont {C\'orcoles}},
  \bibinfo {author} {\bibfnamefont {S.~T.}\ \bibnamefont {Merkel}}, \bibinfo
  {author} {\bibfnamefont {J.~A.}\ \bibnamefont {Smolin}}, \bibinfo {author}
  {\bibfnamefont {C.}~\bibnamefont {Rigetti}}, \bibinfo {author} {\bibfnamefont
  {S.}~\bibnamefont {Poletto}}, \bibinfo {author} {\bibfnamefont {G.~A.}\
  \bibnamefont {Keefe}}, \bibinfo {author} {\bibfnamefont {M.~B.}\ \bibnamefont
  {Rothwell}}, \bibinfo {author} {\bibfnamefont {J.~R.}\ \bibnamefont {Rozen}},
  \bibinfo {author} {\bibfnamefont {M.~B.}\ \bibnamefont {Ketchen}}, \ and\
  \bibinfo {author} {\bibfnamefont {M.}~\bibnamefont {Steffen}},\ }\href
  {\doibase 10.1103/PhysRevLett.109.060501} {\bibfield  {journal} {\bibinfo
  {journal} {Phys. Rev. Lett.}\ }\textbf {\bibinfo {volume} {109}},\ \bibinfo
  {pages} {060501} (\bibinfo {year} {2012})}\BibitemShut {NoStop}%
\bibitem [{\citenamefont {C\'orcoles}\ \emph {et~al.}(2013)\citenamefont
  {C\'orcoles}, \citenamefont {Gambetta}, \citenamefont {Chow}, \citenamefont
  {Smolin}, \citenamefont {Ware}, \citenamefont {Strand}, \citenamefont
  {Plourde},\ and\ \citenamefont {Steffen}}]{corcoles2013process}%
  \BibitemOpen
  \bibfield  {author} {\bibinfo {author} {\bibfnamefont {A.~D.}\ \bibnamefont
  {C\'orcoles}}, \bibinfo {author} {\bibfnamefont {J.~M.}\ \bibnamefont
  {Gambetta}}, \bibinfo {author} {\bibfnamefont {J.~M.}\ \bibnamefont {Chow}},
  \bibinfo {author} {\bibfnamefont {J.~A.}\ \bibnamefont {Smolin}}, \bibinfo
  {author} {\bibfnamefont {M.}~\bibnamefont {Ware}}, \bibinfo {author}
  {\bibfnamefont {J.}~\bibnamefont {Strand}}, \bibinfo {author} {\bibfnamefont
  {B.~L.~T.}\ \bibnamefont {Plourde}}, \ and\ \bibinfo {author} {\bibfnamefont
  {M.}~\bibnamefont {Steffen}},\ }\href {\doibase 10.1103/PhysRevA.87.030301}
  {\bibfield  {journal} {\bibinfo  {journal} {Phys. Rev. A}\ }\textbf {\bibinfo
  {volume} {87}},\ \bibinfo {pages} {030301} (\bibinfo {year}
  {2013})}\BibitemShut {NoStop}%
\bibitem [{\citenamefont {Blume-Kohout}\ and\ \citenamefont
  {Proctor}(2024)}]{blume2024easy}%
  \BibitemOpen
  \bibfield  {author} {\bibinfo {author} {\bibfnamefont {R.}~\bibnamefont
  {Blume-Kohout}}\ and\ \bibinfo {author} {\bibfnamefont {T.}~\bibnamefont
  {Proctor}},\ }\href@noop {} {\bibfield  {journal} {\bibinfo  {journal} {arXiv
  preprint arXiv:2412.16293}\ } (\bibinfo {year} {2024})}\BibitemShut {NoStop}%
\bibitem [{\citenamefont {Mitchell}\ \emph {et~al.}(2003)\citenamefont
  {Mitchell}, \citenamefont {Ellenor}, \citenamefont {Schneider},\ and\
  \citenamefont {Steinberg}}]{mitchell2003diagnosis}%
  \BibitemOpen
  \bibfield  {author} {\bibinfo {author} {\bibfnamefont {M.}~\bibnamefont
  {Mitchell}}, \bibinfo {author} {\bibfnamefont {C.}~\bibnamefont {Ellenor}},
  \bibinfo {author} {\bibfnamefont {S.}~\bibnamefont {Schneider}}, \ and\
  \bibinfo {author} {\bibfnamefont {A.}~\bibnamefont {Steinberg}},\ }\href@noop
  {} {\bibfield  {journal} {\bibinfo  {journal} {Physical review letters}\
  }\textbf {\bibinfo {volume} {91}},\ \bibinfo {pages} {120402} (\bibinfo
  {year} {2003})}\BibitemShut {NoStop}%
\bibitem [{\citenamefont {O'Brien}\ \emph {et~al.}(2004)\citenamefont
  {O'Brien}, \citenamefont {Pryde}, \citenamefont {Gilchrist}, \citenamefont
  {James}, \citenamefont {Langford}, \citenamefont {Ralph},\ and\ \citenamefont
  {White}}]{o2004quantum}%
  \BibitemOpen
  \bibfield  {author} {\bibinfo {author} {\bibfnamefont {J.~L.}\ \bibnamefont
  {O'Brien}}, \bibinfo {author} {\bibfnamefont {G.~J.}\ \bibnamefont {Pryde}},
  \bibinfo {author} {\bibfnamefont {A.}~\bibnamefont {Gilchrist}}, \bibinfo
  {author} {\bibfnamefont {D.~F.}\ \bibnamefont {James}}, \bibinfo {author}
  {\bibfnamefont {N.~K.}\ \bibnamefont {Langford}}, \bibinfo {author}
  {\bibfnamefont {T.~C.}\ \bibnamefont {Ralph}}, \ and\ \bibinfo {author}
  {\bibfnamefont {A.~G.}\ \bibnamefont {White}},\ }\href@noop {} {\bibfield
  {journal} {\bibinfo  {journal} {Physical review letters}\ }\textbf {\bibinfo
  {volume} {93}},\ \bibinfo {pages} {080502} (\bibinfo {year}
  {2004})}\BibitemShut {NoStop}%
\bibitem [{\citenamefont {Knee}\ \emph {et~al.}(2018)\citenamefont {Knee},
  \citenamefont {Bolduc}, \citenamefont {Leach},\ and\ \citenamefont
  {Gauger}}]{knee2018quantum}%
  \BibitemOpen
  \bibfield  {author} {\bibinfo {author} {\bibfnamefont {G.~C.}\ \bibnamefont
  {Knee}}, \bibinfo {author} {\bibfnamefont {E.}~\bibnamefont {Bolduc}},
  \bibinfo {author} {\bibfnamefont {J.}~\bibnamefont {Leach}}, \ and\ \bibinfo
  {author} {\bibfnamefont {E.~M.}\ \bibnamefont {Gauger}},\ }\href@noop {}
  {\bibfield  {journal} {\bibinfo  {journal} {Physical Review A}\ }\textbf
  {\bibinfo {volume} {98}},\ \bibinfo {pages} {062336} (\bibinfo {year}
  {2018})}\BibitemShut {NoStop}%
\bibitem [{\citenamefont {Fiur{\'a}{\v{s}}ek}(2001)}]{fiuravsek2001maximum}%
  \BibitemOpen
  \bibfield  {author} {\bibinfo {author} {\bibfnamefont {J.}~\bibnamefont
  {Fiur{\'a}{\v{s}}ek}},\ }\href@noop {} {\bibfield  {journal} {\bibinfo
  {journal} {Physical Review A}\ }\textbf {\bibinfo {volume} {64}},\ \bibinfo
  {pages} {024102} (\bibinfo {year} {2001})}\BibitemShut {NoStop}%
\bibitem [{\citenamefont {Lundeen}\ \emph {et~al.}(2009)\citenamefont
  {Lundeen}, \citenamefont {Feito}, \citenamefont {Coldenstrodt-Ronge},
  \citenamefont {Pregnell}, \citenamefont {Silberhorn}, \citenamefont {Ralph},
  \citenamefont {Eisert}, \citenamefont {Plenio},\ and\ \citenamefont
  {Walmsley}}]{lundeen2009tomography}%
  \BibitemOpen
  \bibfield  {author} {\bibinfo {author} {\bibfnamefont {J.~S.}\ \bibnamefont
  {Lundeen}}, \bibinfo {author} {\bibfnamefont {A.}~\bibnamefont {Feito}},
  \bibinfo {author} {\bibfnamefont {H.}~\bibnamefont {Coldenstrodt-Ronge}},
  \bibinfo {author} {\bibfnamefont {K.~L.}\ \bibnamefont {Pregnell}}, \bibinfo
  {author} {\bibfnamefont {C.}~\bibnamefont {Silberhorn}}, \bibinfo {author}
  {\bibfnamefont {T.~C.}\ \bibnamefont {Ralph}}, \bibinfo {author}
  {\bibfnamefont {J.}~\bibnamefont {Eisert}}, \bibinfo {author} {\bibfnamefont
  {M.~B.}\ \bibnamefont {Plenio}}, \ and\ \bibinfo {author} {\bibfnamefont
  {I.~A.}\ \bibnamefont {Walmsley}},\ }\href@noop {} {\bibfield  {journal}
  {\bibinfo  {journal} {Nature Physics}\ }\textbf {\bibinfo {volume} {5}},\
  \bibinfo {pages} {27} (\bibinfo {year} {2009})}\BibitemShut {NoStop}%
\bibitem [{\citenamefont {Bravyi}\ \emph {et~al.}(2021)\citenamefont {Bravyi},
  \citenamefont {Sheldon}, \citenamefont {Kandala}, \citenamefont {Mckay},\
  and\ \citenamefont {Gambetta}}]{bravyi2021mitigating}%
  \BibitemOpen
  \bibfield  {author} {\bibinfo {author} {\bibfnamefont {S.}~\bibnamefont
  {Bravyi}}, \bibinfo {author} {\bibfnamefont {S.}~\bibnamefont {Sheldon}},
  \bibinfo {author} {\bibfnamefont {A.}~\bibnamefont {Kandala}}, \bibinfo
  {author} {\bibfnamefont {D.~C.}\ \bibnamefont {Mckay}}, \ and\ \bibinfo
  {author} {\bibfnamefont {J.~M.}\ \bibnamefont {Gambetta}},\ }\href@noop {}
  {\bibfield  {journal} {\bibinfo  {journal} {Physical Review A}\ }\textbf
  {\bibinfo {volume} {103}},\ \bibinfo {pages} {042605} (\bibinfo {year}
  {2021})}\BibitemShut {NoStop}%
\bibitem [{\citenamefont {Chen}\ \emph {et~al.}(2019)\citenamefont {Chen},
  \citenamefont {Farahzad}, \citenamefont {Yoo},\ and\ \citenamefont
  {Wei}}]{chen2019detector}%
  \BibitemOpen
  \bibfield  {author} {\bibinfo {author} {\bibfnamefont {Y.}~\bibnamefont
  {Chen}}, \bibinfo {author} {\bibfnamefont {M.}~\bibnamefont {Farahzad}},
  \bibinfo {author} {\bibfnamefont {S.}~\bibnamefont {Yoo}}, \ and\ \bibinfo
  {author} {\bibfnamefont {T.-C.}\ \bibnamefont {Wei}},\ }\href@noop {}
  {\bibfield  {journal} {\bibinfo  {journal} {Physical Review A}\ }\textbf
  {\bibinfo {volume} {100}},\ \bibinfo {pages} {052315} (\bibinfo {year}
  {2019})}\BibitemShut {NoStop}%
\bibitem [{\citenamefont {Nguyen}\ \emph
  {et~al.}(2024{\natexlab{b}})\citenamefont {Nguyen}, \citenamefont {Goss},
  \citenamefont {Siva}, \citenamefont {Kim}, \citenamefont {Younis},
  \citenamefont {Qing}, \citenamefont {Hashim}, \citenamefont {Santiago},\ and\
  \citenamefont {Siddiqi}}]{nguyen2023empowering}%
  \BibitemOpen
  \bibfield  {author} {\bibinfo {author} {\bibfnamefont {L.~B.}\ \bibnamefont
  {Nguyen}}, \bibinfo {author} {\bibfnamefont {N.}~\bibnamefont {Goss}},
  \bibinfo {author} {\bibfnamefont {K.}~\bibnamefont {Siva}}, \bibinfo {author}
  {\bibfnamefont {Y.}~\bibnamefont {Kim}}, \bibinfo {author} {\bibfnamefont
  {E.}~\bibnamefont {Younis}}, \bibinfo {author} {\bibfnamefont
  {B.}~\bibnamefont {Qing}}, \bibinfo {author} {\bibfnamefont {A.}~\bibnamefont
  {Hashim}}, \bibinfo {author} {\bibfnamefont {D.~I.}\ \bibnamefont
  {Santiago}}, \ and\ \bibinfo {author} {\bibfnamefont {I.}~\bibnamefont
  {Siddiqi}},\ }\href {\doibase 10.1038/s41467-024-51434-2} {\bibfield
  {journal} {\bibinfo  {journal} {Nature Communications}\ ,\ \bibinfo {pages}
  {7117}} (\bibinfo {year} {2024}{\natexlab{b}})}\BibitemShut {NoStop}%
\bibitem [{\citenamefont {Greenbaum}(2015)}]{greenbaum2015introduction}%
  \BibitemOpen
  \bibfield  {author} {\bibinfo {author} {\bibfnamefont {D.}~\bibnamefont
  {Greenbaum}},\ }\href@noop {} {\bibfield  {journal} {\bibinfo  {journal}
  {arXiv preprint arXiv:1509.02921}\ } (\bibinfo {year} {2015})}\BibitemShut
  {NoStop}%
\bibitem [{\citenamefont {Merkel}\ \emph
  {et~al.}(2013{\natexlab{b}})\citenamefont {Merkel}, \citenamefont {Gambetta},
  \citenamefont {Smolin}, \citenamefont {Poletto}, \citenamefont
  {C{\'o}rcoles}, \citenamefont {Johnson}, \citenamefont {Ryan},\ and\
  \citenamefont {Steffen}}]{Merkel2013gst}%
  \BibitemOpen
  \bibfield  {author} {\bibinfo {author} {\bibfnamefont {S.~T.}\ \bibnamefont
  {Merkel}}, \bibinfo {author} {\bibfnamefont {J.~M.}\ \bibnamefont
  {Gambetta}}, \bibinfo {author} {\bibfnamefont {J.~A.}\ \bibnamefont
  {Smolin}}, \bibinfo {author} {\bibfnamefont {S.}~\bibnamefont {Poletto}},
  \bibinfo {author} {\bibfnamefont {A.~D.}\ \bibnamefont {C{\'o}rcoles}},
  \bibinfo {author} {\bibfnamefont {B.~R.}\ \bibnamefont {Johnson}}, \bibinfo
  {author} {\bibfnamefont {C.~A.}\ \bibnamefont {Ryan}}, \ and\ \bibinfo
  {author} {\bibfnamefont {M.}~\bibnamefont {Steffen}},\ }\href {\doibase
  10.1103/PhysRevA.87.062119} {\bibfield  {journal} {\bibinfo  {journal} {Phys.
  Rev. A}\ }\textbf {\bibinfo {volume} {87}},\ \bibinfo {pages} {062119}
  (\bibinfo {year} {2013}{\natexlab{b}})}\BibitemShut {NoStop}%
\bibitem [{\citenamefont {Gu}\ \emph {et~al.}(2021)\citenamefont {Gu},
  \citenamefont {Mishra}, \citenamefont {Englert},\ and\ \citenamefont
  {Ng}}]{PRXQuantum.2.030328}%
  \BibitemOpen
  \bibfield  {author} {\bibinfo {author} {\bibfnamefont {Y.}~\bibnamefont
  {Gu}}, \bibinfo {author} {\bibfnamefont {R.}~\bibnamefont {Mishra}}, \bibinfo
  {author} {\bibfnamefont {B.-G.}\ \bibnamefont {Englert}}, \ and\ \bibinfo
  {author} {\bibfnamefont {H.~K.}\ \bibnamefont {Ng}},\ }\href {\doibase
  10.1103/PRXQuantum.2.030328} {\bibfield  {journal} {\bibinfo  {journal} {PRX
  Quantum}\ }\textbf {\bibinfo {volume} {2}},\ \bibinfo {pages} {030328}
  (\bibinfo {year} {2021})}\BibitemShut {NoStop}%
\bibitem [{\citenamefont {Dehollain}\ \emph {et~al.}(2016)\citenamefont
  {Dehollain}, \citenamefont {Muhonen}, \citenamefont {Blume-Kohout},
  \citenamefont {Rudinger}, \citenamefont {Gamble}, \citenamefont {Nielsen},
  \citenamefont {Laucht}, \citenamefont {Simmons}, \citenamefont {Kalra},
  \citenamefont {Dzurak} \emph {et~al.}}]{dehollain2016optimization}%
  \BibitemOpen
  \bibfield  {author} {\bibinfo {author} {\bibfnamefont {J.~P.}\ \bibnamefont
  {Dehollain}}, \bibinfo {author} {\bibfnamefont {J.~T.}\ \bibnamefont
  {Muhonen}}, \bibinfo {author} {\bibfnamefont {R.}~\bibnamefont
  {Blume-Kohout}}, \bibinfo {author} {\bibfnamefont {K.~M.}\ \bibnamefont
  {Rudinger}}, \bibinfo {author} {\bibfnamefont {J.~K.}\ \bibnamefont
  {Gamble}}, \bibinfo {author} {\bibfnamefont {E.}~\bibnamefont {Nielsen}},
  \bibinfo {author} {\bibfnamefont {A.}~\bibnamefont {Laucht}}, \bibinfo
  {author} {\bibfnamefont {S.}~\bibnamefont {Simmons}}, \bibinfo {author}
  {\bibfnamefont {R.}~\bibnamefont {Kalra}}, \bibinfo {author} {\bibfnamefont
  {A.~S.}\ \bibnamefont {Dzurak}},  \emph {et~al.},\ }\href@noop {} {\bibfield
  {journal} {\bibinfo  {journal} {New Journal of Physics}\ }\textbf {\bibinfo
  {volume} {18}},\ \bibinfo {pages} {103018} (\bibinfo {year}
  {2016})}\BibitemShut {NoStop}%
\bibitem [{\citenamefont {Xue}\ \emph {et~al.}(2022)\citenamefont {Xue},
  \citenamefont {Russ}, \citenamefont {Samkharadze}, \citenamefont {Undseth},
  \citenamefont {Sammak}, \citenamefont {Scappucci},\ and\ \citenamefont
  {Vandersypen}}]{xue2022quantum}%
  \BibitemOpen
  \bibfield  {author} {\bibinfo {author} {\bibfnamefont {X.}~\bibnamefont
  {Xue}}, \bibinfo {author} {\bibfnamefont {M.}~\bibnamefont {Russ}}, \bibinfo
  {author} {\bibfnamefont {N.}~\bibnamefont {Samkharadze}}, \bibinfo {author}
  {\bibfnamefont {B.}~\bibnamefont {Undseth}}, \bibinfo {author} {\bibfnamefont
  {A.}~\bibnamefont {Sammak}}, \bibinfo {author} {\bibfnamefont
  {G.}~\bibnamefont {Scappucci}}, \ and\ \bibinfo {author} {\bibfnamefont
  {L.~M.~K.}\ \bibnamefont {Vandersypen}},\ }\href {\doibase
  https://doi.org/10.1038/s41586-021-04273-w} {\bibfield  {journal} {\bibinfo
  {journal} {Nature}\ }\textbf {\bibinfo {volume} {601}},\ \bibinfo {pages}
  {343} (\bibinfo {year} {2022})}\BibitemShut {NoStop}%
\bibitem [{\citenamefont {Nielsen}\ \emph {et~al.}(2020)\citenamefont
  {Nielsen}, \citenamefont {Rudinger}, \citenamefont {Proctor}, \citenamefont
  {Russo}, \citenamefont {Young},\ and\ \citenamefont
  {Blume-Kohout}}]{Nielsen2020pygsti}%
  \BibitemOpen
  \bibfield  {author} {\bibinfo {author} {\bibfnamefont {E.}~\bibnamefont
  {Nielsen}}, \bibinfo {author} {\bibfnamefont {K.}~\bibnamefont {Rudinger}},
  \bibinfo {author} {\bibfnamefont {T.}~\bibnamefont {Proctor}}, \bibinfo
  {author} {\bibfnamefont {A.}~\bibnamefont {Russo}}, \bibinfo {author}
  {\bibfnamefont {K.}~\bibnamefont {Young}}, \ and\ \bibinfo {author}
  {\bibfnamefont {R.}~\bibnamefont {Blume-Kohout}},\ }\href {\doibase
  10.1088/2058-9565/ab8aa4} {\bibfield  {journal} {\bibinfo  {journal} {Quantum
  Sci. Technol.}\ }\textbf {\bibinfo {volume} {5}},\ \bibinfo {pages} {044002}
  (\bibinfo {year} {2020})}\BibitemShut {NoStop}%
\bibitem [{\citenamefont {Giovannetti}\ \emph {et~al.}(2004)\citenamefont
  {Giovannetti}, \citenamefont {Lloyd},\ and\ \citenamefont
  {Maccone}}]{Giovannetti2004quantumlimit}%
  \BibitemOpen
  \bibfield  {author} {\bibinfo {author} {\bibfnamefont {V.}~\bibnamefont
  {Giovannetti}}, \bibinfo {author} {\bibfnamefont {S.}~\bibnamefont {Lloyd}},
  \ and\ \bibinfo {author} {\bibfnamefont {L.}~\bibnamefont {Maccone}},\ }\href
  {\doibase 10.1126/science.1104149} {\bibfield  {journal} {\bibinfo  {journal}
  {Science}\ }\textbf {\bibinfo {volume} {306}},\ \bibinfo {pages} {1330}
  (\bibinfo {year} {2004})}\BibitemShut {NoStop}%
\bibitem [{\citenamefont {Wilks}(1938)}]{wilks1938large}%
  \BibitemOpen
  \bibfield  {author} {\bibinfo {author} {\bibfnamefont {S.~S.}\ \bibnamefont
  {Wilks}},\ }\href@noop {} {\bibfield  {journal} {\bibinfo  {journal} {The
  annals of mathematical statistics}\ }\textbf {\bibinfo {volume} {9}},\
  \bibinfo {pages} {60} (\bibinfo {year} {1938})}\BibitemShut {NoStop}%
\bibitem [{\citenamefont {Dankert}\ \emph {et~al.}(2009)\citenamefont
  {Dankert}, \citenamefont {Cleve}, \citenamefont {Emerson},\ and\
  \citenamefont {Livine}}]{dankert2009exact}%
  \BibitemOpen
  \bibfield  {author} {\bibinfo {author} {\bibfnamefont {C.}~\bibnamefont
  {Dankert}}, \bibinfo {author} {\bibfnamefont {R.}~\bibnamefont {Cleve}},
  \bibinfo {author} {\bibfnamefont {J.}~\bibnamefont {Emerson}}, \ and\
  \bibinfo {author} {\bibfnamefont {E.}~\bibnamefont {Livine}},\ }\href@noop {}
  {\bibfield  {journal} {\bibinfo  {journal} {Physical Review A}\ }\textbf
  {\bibinfo {volume} {80}},\ \bibinfo {pages} {012304} (\bibinfo {year}
  {2009})}\BibitemShut {NoStop}%
\bibitem [{\citenamefont {Knill}\ \emph {et~al.}(2008)\citenamefont {Knill},
  \citenamefont {Leibfried}, \citenamefont {Reichle}, \citenamefont {Britton},
  \citenamefont {Blakestad}, \citenamefont {Jost}, \citenamefont {Langer},
  \citenamefont {Ozeri}, \citenamefont {Seidelin},\ and\ \citenamefont
  {Wineland}}]{knill2008randomized}%
  \BibitemOpen
  \bibfield  {author} {\bibinfo {author} {\bibfnamefont {E.}~\bibnamefont
  {Knill}}, \bibinfo {author} {\bibfnamefont {D.}~\bibnamefont {Leibfried}},
  \bibinfo {author} {\bibfnamefont {R.}~\bibnamefont {Reichle}}, \bibinfo
  {author} {\bibfnamefont {J.}~\bibnamefont {Britton}}, \bibinfo {author}
  {\bibfnamefont {R.~B.}\ \bibnamefont {Blakestad}}, \bibinfo {author}
  {\bibfnamefont {J.~D.}\ \bibnamefont {Jost}}, \bibinfo {author}
  {\bibfnamefont {C.}~\bibnamefont {Langer}}, \bibinfo {author} {\bibfnamefont
  {R.}~\bibnamefont {Ozeri}}, \bibinfo {author} {\bibfnamefont
  {S.}~\bibnamefont {Seidelin}}, \ and\ \bibinfo {author} {\bibfnamefont
  {D.~J.}\ \bibnamefont {Wineland}},\ }\href@noop {} {\bibfield  {journal}
  {\bibinfo  {journal} {Physical Review A}\ }\textbf {\bibinfo {volume} {77}},\
  \bibinfo {pages} {012307} (\bibinfo {year} {2008})}\BibitemShut {NoStop}%
\bibitem [{\citenamefont {Magesan}\ \emph
  {et~al.}(2011{\natexlab{b}})\citenamefont {Magesan}, \citenamefont
  {Gambetta},\ and\ \citenamefont {Emerson}}]{magesan2011scalable}%
  \BibitemOpen
  \bibfield  {author} {\bibinfo {author} {\bibfnamefont {E.}~\bibnamefont
  {Magesan}}, \bibinfo {author} {\bibfnamefont {J.~M.}\ \bibnamefont
  {Gambetta}}, \ and\ \bibinfo {author} {\bibfnamefont {J.}~\bibnamefont
  {Emerson}},\ }\href@noop {} {\bibfield  {journal} {\bibinfo  {journal}
  {Physical Review Letters}\ }\textbf {\bibinfo {volume} {106}},\ \bibinfo
  {pages} {180504} (\bibinfo {year} {2011}{\natexlab{b}})}\BibitemShut
  {NoStop}%
\bibitem [{Note20()}]{Note20}%
  \BibitemOpen
  \bibinfo {note} {Here, we define a circuit of depth $m = 0$ to be the minimal
  benchmark depth, which contains only a single random gate (and its inverse).
  By defining it this way, the error in any gates in the $m = 0$ circuit all
  contributes to effective SPAM error. Therefore, any gates used for
  state-preparation or basis rotations for measurement can be compiled into the
  initial and final circuit layers, respectively.}\BibitemShut {Stop}%
\bibitem [{\citenamefont
  {Gottesman}(1998{\natexlab{a}})}]{gottesman1998heisenberg}%
  \BibitemOpen
  \bibfield  {author} {\bibinfo {author} {\bibfnamefont {D.}~\bibnamefont
  {Gottesman}},\ }\href@noop {} {\bibfield  {journal} {\bibinfo  {journal}
  {arXiv preprint quant-ph/9807006}\ } (\bibinfo {year}
  {1998}{\natexlab{a}})}\BibitemShut {NoStop}%
\bibitem [{\citenamefont {Koenig}\ and\ \citenamefont
  {Smolin}(2014)}]{Koenig2014-ih}%
  \BibitemOpen
  \bibfield  {author} {\bibinfo {author} {\bibfnamefont {R.}~\bibnamefont
  {Koenig}}\ and\ \bibinfo {author} {\bibfnamefont {J.~A.}\ \bibnamefont
  {Smolin}},\ }\href {\doibase 10.1063/1.4903507} {\bibfield  {journal}
  {\bibinfo  {journal} {J. Math. Phys.}\ }\textbf {\bibinfo {volume} {55}},\
  \bibinfo {pages} {122202} (\bibinfo {year} {2014})}\BibitemShut {NoStop}%
\bibitem [{\citenamefont {Fogarty}\ \emph {et~al.}(2015)\citenamefont
  {Fogarty}, \citenamefont {Veldhorst}, \citenamefont {Harper}, \citenamefont
  {Yang}, \citenamefont {Bartlett}, \citenamefont {Flammia},\ and\
  \citenamefont {Dzurak}}]{fogarty2015nonexponential}%
  \BibitemOpen
  \bibfield  {author} {\bibinfo {author} {\bibfnamefont {M.~A.}\ \bibnamefont
  {Fogarty}}, \bibinfo {author} {\bibfnamefont {M.}~\bibnamefont {Veldhorst}},
  \bibinfo {author} {\bibfnamefont {R.}~\bibnamefont {Harper}}, \bibinfo
  {author} {\bibfnamefont {C.}~\bibnamefont {Yang}}, \bibinfo {author}
  {\bibfnamefont {S.}~\bibnamefont {Bartlett}}, \bibinfo {author}
  {\bibfnamefont {S.~T.}\ \bibnamefont {Flammia}}, \ and\ \bibinfo {author}
  {\bibfnamefont {A.}~\bibnamefont {Dzurak}},\ }\href@noop {} {\bibfield
  {journal} {\bibinfo  {journal} {Physical Review A}\ }\textbf {\bibinfo
  {volume} {92}},\ \bibinfo {pages} {022326} (\bibinfo {year}
  {2015})}\BibitemShut {NoStop}%
\bibitem [{\citenamefont {Muhonen}\ \emph {et~al.}(2015)\citenamefont
  {Muhonen}, \citenamefont {Laucht}, \citenamefont {Simmons}, \citenamefont
  {Dehollain}, \citenamefont {Kalra}, \citenamefont {Hudson}, \citenamefont
  {Freer}, \citenamefont {Itoh}, \citenamefont {Jamieson}, \citenamefont
  {McCallum} \emph {et~al.}}]{muhonen2015quantifying}%
  \BibitemOpen
  \bibfield  {author} {\bibinfo {author} {\bibfnamefont {J.~T.}\ \bibnamefont
  {Muhonen}}, \bibinfo {author} {\bibfnamefont {A.}~\bibnamefont {Laucht}},
  \bibinfo {author} {\bibfnamefont {S.}~\bibnamefont {Simmons}}, \bibinfo
  {author} {\bibfnamefont {J.~P.}\ \bibnamefont {Dehollain}}, \bibinfo {author}
  {\bibfnamefont {R.}~\bibnamefont {Kalra}}, \bibinfo {author} {\bibfnamefont
  {F.~E.}\ \bibnamefont {Hudson}}, \bibinfo {author} {\bibfnamefont
  {S.}~\bibnamefont {Freer}}, \bibinfo {author} {\bibfnamefont {K.~M.}\
  \bibnamefont {Itoh}}, \bibinfo {author} {\bibfnamefont {D.~N.}\ \bibnamefont
  {Jamieson}}, \bibinfo {author} {\bibfnamefont {J.~C.}\ \bibnamefont
  {McCallum}},  \emph {et~al.},\ }\href@noop {} {\bibfield  {journal} {\bibinfo
   {journal} {Journal of Physics: Condensed Matter}\ }\textbf {\bibinfo
  {volume} {27}},\ \bibinfo {pages} {154205} (\bibinfo {year}
  {2015})}\BibitemShut {NoStop}%
\bibitem [{\citenamefont {Harper}\ \emph {et~al.}(2019)\citenamefont {Harper},
  \citenamefont {Hincks}, \citenamefont {Ferrie}, \citenamefont {Flammia},\
  and\ \citenamefont {Wallman}}]{harper2019statistical}%
  \BibitemOpen
  \bibfield  {author} {\bibinfo {author} {\bibfnamefont {R.}~\bibnamefont
  {Harper}}, \bibinfo {author} {\bibfnamefont {I.}~\bibnamefont {Hincks}},
  \bibinfo {author} {\bibfnamefont {C.}~\bibnamefont {Ferrie}}, \bibinfo
  {author} {\bibfnamefont {S.~T.}\ \bibnamefont {Flammia}}, \ and\ \bibinfo
  {author} {\bibfnamefont {J.~J.}\ \bibnamefont {Wallman}},\ }\href@noop {}
  {\bibfield  {journal} {\bibinfo  {journal} {Physical Review A}\ }\textbf
  {\bibinfo {volume} {99}},\ \bibinfo {pages} {052350} (\bibinfo {year}
  {2019})}\BibitemShut {NoStop}%
\bibitem [{Note21()}]{Note21}%
  \BibitemOpen
  \bibinfo {note} {This ensures that the entire circuit is not compiled down
  into a single gate layer, which would defeat the purpose of the
  benchmark.}\BibitemShut {Stop}%
\bibitem [{\citenamefont {Granade}\ \emph {et~al.}(2015)\citenamefont
  {Granade}, \citenamefont {Ferrie},\ and\ \citenamefont
  {Cory}}]{granade2015accelerated}%
  \BibitemOpen
  \bibfield  {author} {\bibinfo {author} {\bibfnamefont {C.}~\bibnamefont
  {Granade}}, \bibinfo {author} {\bibfnamefont {C.}~\bibnamefont {Ferrie}}, \
  and\ \bibinfo {author} {\bibfnamefont {D.~G.}\ \bibnamefont {Cory}},\
  }\href@noop {} {\bibfield  {journal} {\bibinfo  {journal} {New Journal of
  Physics}\ }\textbf {\bibinfo {volume} {17}},\ \bibinfo {pages} {013042}
  (\bibinfo {year} {2015})}\BibitemShut {NoStop}%
\bibitem [{Note22()}]{Note22}%
  \BibitemOpen
  \bibinfo {note} {$Ap^m+B$ is the success probability of circuits containing
  $m+2$ Clifford gates and a SPAM operation, so $Ap^{-2}+B$, i.e.,
  extrapolating the decay curve back to $m=-2$, is a heuristic for SPAM error's
  contribution to the success probabilities deviation from 1. Note that $A$ is
  not invariant under changes in the convention for `depth` in RB circuits ---
  e.g., defining $m$ to be the number of uniformly random Clifford gates in the
  circuit ---- but this heuristic for SPAM fidelity is.)}\BibitemShut {NoStop}%
\bibitem [{\citenamefont {Proctor}\ \emph
  {et~al.}(2022{\natexlab{b}})\citenamefont {Proctor}, \citenamefont
  {Rudinger}, \citenamefont {Young}, \citenamefont {Nielsen},\ and\
  \citenamefont {Blume-Kohout}}]{proctor2022measuring}%
  \BibitemOpen
  \bibfield  {author} {\bibinfo {author} {\bibfnamefont {T.}~\bibnamefont
  {Proctor}}, \bibinfo {author} {\bibfnamefont {K.}~\bibnamefont {Rudinger}},
  \bibinfo {author} {\bibfnamefont {K.}~\bibnamefont {Young}}, \bibinfo
  {author} {\bibfnamefont {E.}~\bibnamefont {Nielsen}}, \ and\ \bibinfo
  {author} {\bibfnamefont {R.}~\bibnamefont {Blume-Kohout}},\ }\href@noop {}
  {\bibfield  {journal} {\bibinfo  {journal} {Nature Physics}\ }\textbf
  {\bibinfo {volume} {18}},\ \bibinfo {pages} {75} (\bibinfo {year}
  {2022}{\natexlab{b}})}\BibitemShut {NoStop}%
\bibitem [{\citenamefont {Barends}\ \emph {et~al.}(2014)\citenamefont
  {Barends}, \citenamefont {Kelly}, \citenamefont {Megrant}, \citenamefont
  {Veitia}, \citenamefont {Sank}, \citenamefont {Jeffrey}, \citenamefont
  {White}, \citenamefont {Mutus}, \citenamefont {Fowler}, \citenamefont
  {Campbell} \emph {et~al.}}]{barends2014superconducting}%
  \BibitemOpen
  \bibfield  {author} {\bibinfo {author} {\bibfnamefont {R.}~\bibnamefont
  {Barends}}, \bibinfo {author} {\bibfnamefont {J.}~\bibnamefont {Kelly}},
  \bibinfo {author} {\bibfnamefont {A.}~\bibnamefont {Megrant}}, \bibinfo
  {author} {\bibfnamefont {A.}~\bibnamefont {Veitia}}, \bibinfo {author}
  {\bibfnamefont {D.}~\bibnamefont {Sank}}, \bibinfo {author} {\bibfnamefont
  {E.}~\bibnamefont {Jeffrey}}, \bibinfo {author} {\bibfnamefont {T.~C.}\
  \bibnamefont {White}}, \bibinfo {author} {\bibfnamefont {J.}~\bibnamefont
  {Mutus}}, \bibinfo {author} {\bibfnamefont {A.~G.}\ \bibnamefont {Fowler}},
  \bibinfo {author} {\bibfnamefont {B.}~\bibnamefont {Campbell}},  \emph
  {et~al.},\ }\href@noop {} {\bibfield  {journal} {\bibinfo  {journal}
  {Nature}\ }\textbf {\bibinfo {volume} {508}},\ \bibinfo {pages} {500}
  (\bibinfo {year} {2014})}\BibitemShut {NoStop}%
\bibitem [{\citenamefont {McKay}\ \emph {et~al.}(2017)\citenamefont {McKay},
  \citenamefont {Wood}, \citenamefont {Sheldon}, \citenamefont {Chow},\ and\
  \citenamefont {Gambetta}}]{mckay2017efficient}%
  \BibitemOpen
  \bibfield  {author} {\bibinfo {author} {\bibfnamefont {D.~C.}\ \bibnamefont
  {McKay}}, \bibinfo {author} {\bibfnamefont {C.~J.}\ \bibnamefont {Wood}},
  \bibinfo {author} {\bibfnamefont {S.}~\bibnamefont {Sheldon}}, \bibinfo
  {author} {\bibfnamefont {J.~M.}\ \bibnamefont {Chow}}, \ and\ \bibinfo
  {author} {\bibfnamefont {J.~M.}\ \bibnamefont {Gambetta}},\ }\href@noop {}
  {\bibfield  {journal} {\bibinfo  {journal} {Physical Review A}\ }\textbf
  {\bibinfo {volume} {96}},\ \bibinfo {pages} {022330} (\bibinfo {year}
  {2017})}\BibitemShut {NoStop}%
\bibitem [{Note23()}]{Note23}%
  \BibitemOpen
  \bibinfo {note} {Virtual $Z$ gates do not implement physical pulses; rather,
  they provide a frame update (i.e., a shift in phase) for the subsequent
  physical pulse.}\BibitemShut {Stop}%
\bibitem [{\citenamefont {Morvan}\ \emph {et~al.}(2021)\citenamefont {Morvan},
  \citenamefont {Ramasesh}, \citenamefont {Blok}, \citenamefont {Kreikebaum},
  \citenamefont {O’Brien}, \citenamefont {Chen}, \citenamefont {Mitchell},
  \citenamefont {Naik}, \citenamefont {Santiago},\ and\ \citenamefont
  {Siddiqi}}]{qutritrb}%
  \BibitemOpen
  \bibfield  {author} {\bibinfo {author} {\bibfnamefont {A.}~\bibnamefont
  {Morvan}}, \bibinfo {author} {\bibfnamefont {V.}~\bibnamefont {Ramasesh}},
  \bibinfo {author} {\bibfnamefont {M.}~\bibnamefont {Blok}}, \bibinfo {author}
  {\bibfnamefont {J.}~\bibnamefont {Kreikebaum}}, \bibinfo {author}
  {\bibfnamefont {K.}~\bibnamefont {O’Brien}}, \bibinfo {author}
  {\bibfnamefont {L.}~\bibnamefont {Chen}}, \bibinfo {author} {\bibfnamefont
  {B.}~\bibnamefont {Mitchell}}, \bibinfo {author} {\bibfnamefont
  {R.}~\bibnamefont {Naik}}, \bibinfo {author} {\bibfnamefont {D.}~\bibnamefont
  {Santiago}}, \ and\ \bibinfo {author} {\bibfnamefont {I.}~\bibnamefont
  {Siddiqi}},\ }\href@noop {} {\bibfield  {journal} {\bibinfo  {journal}
  {Physical review letters}\ }\textbf {\bibinfo {volume} {126}},\ \bibinfo
  {pages} {210504} (\bibinfo {year} {2021})}\BibitemShut {NoStop}%
\bibitem [{\citenamefont {Magesan}\ \emph
  {et~al.}(2012{\natexlab{b}})\citenamefont {Magesan}, \citenamefont
  {Gambetta},\ and\ \citenamefont {Emerson}}]{magesan2012characterizing}%
  \BibitemOpen
  \bibfield  {author} {\bibinfo {author} {\bibfnamefont {E.}~\bibnamefont
  {Magesan}}, \bibinfo {author} {\bibfnamefont {J.~M.}\ \bibnamefont
  {Gambetta}}, \ and\ \bibinfo {author} {\bibfnamefont {J.}~\bibnamefont
  {Emerson}},\ }\href@noop {} {\bibfield  {journal} {\bibinfo  {journal}
  {Physical Review A}\ }\textbf {\bibinfo {volume} {85}},\ \bibinfo {pages}
  {042311} (\bibinfo {year} {2012}{\natexlab{b}})}\BibitemShut {NoStop}%
\bibitem [{\citenamefont {Merkel}\ \emph {et~al.}(2021)\citenamefont {Merkel},
  \citenamefont {Pritchett},\ and\ \citenamefont {Fong}}]{Merkel2021-ux}%
  \BibitemOpen
  \bibfield  {author} {\bibinfo {author} {\bibfnamefont {S.~T.}\ \bibnamefont
  {Merkel}}, \bibinfo {author} {\bibfnamefont {E.~J.}\ \bibnamefont
  {Pritchett}}, \ and\ \bibinfo {author} {\bibfnamefont {B.~H.}\ \bibnamefont
  {Fong}},\ }\href {\doibase 10.22331/q-2021-11-16-581} {\bibfield  {journal}
  {\bibinfo  {journal} {Quantum}\ }\textbf {\bibinfo {volume} {5}},\ \bibinfo
  {pages} {581} (\bibinfo {year} {2021})}\BibitemShut {NoStop}%
\bibitem [{\citenamefont {Helsen}\ \emph
  {et~al.}(2022{\natexlab{a}})\citenamefont {Helsen}, \citenamefont {Roth},
  \citenamefont {Onorati}, \citenamefont {Werner},\ and\ \citenamefont
  {Eisert}}]{helsen2022framework}%
  \BibitemOpen
  \bibfield  {author} {\bibinfo {author} {\bibfnamefont {J.}~\bibnamefont
  {Helsen}}, \bibinfo {author} {\bibfnamefont {I.}~\bibnamefont {Roth}},
  \bibinfo {author} {\bibfnamefont {E.}~\bibnamefont {Onorati}}, \bibinfo
  {author} {\bibfnamefont {A.}~\bibnamefont {Werner}}, \ and\ \bibinfo {author}
  {\bibfnamefont {J.}~\bibnamefont {Eisert}},\ }\href {\doibase
  10.1103/PRXQuantum.3.020357} {\bibfield  {journal} {\bibinfo  {journal} {PRX
  Quantum}\ }\textbf {\bibinfo {volume} {3}},\ \bibinfo {pages} {020357}
  (\bibinfo {year} {2022}{\natexlab{a}})}\BibitemShut {NoStop}%
\bibitem [{\citenamefont {Carignan-Dugas}\ \emph {et~al.}(2018)\citenamefont
  {Carignan-Dugas}, \citenamefont {Boone}, \citenamefont {Wallman},\ and\
  \citenamefont {Emerson}}]{Carignan-Dugas2018-np}%
  \BibitemOpen
  \bibfield  {author} {\bibinfo {author} {\bibfnamefont {A.}~\bibnamefont
  {Carignan-Dugas}}, \bibinfo {author} {\bibfnamefont {K.}~\bibnamefont
  {Boone}}, \bibinfo {author} {\bibfnamefont {J.~J.}\ \bibnamefont {Wallman}},
  \ and\ \bibinfo {author} {\bibfnamefont {J.}~\bibnamefont {Emerson}},\ }\href
  {\doibase 10.1088/1367-2630/aadcc7} {\bibfield  {journal} {\bibinfo
  {journal} {New J. Phys.}\ }\textbf {\bibinfo {volume} {20}},\ \bibinfo
  {pages} {092001} (\bibinfo {year} {2018})}\BibitemShut {NoStop}%
\bibitem [{\citenamefont {Aaronson}\ and\ \citenamefont
  {Gottesman}(2004)}]{aaronson2004improved}%
  \BibitemOpen
  \bibfield  {author} {\bibinfo {author} {\bibfnamefont {S.}~\bibnamefont
  {Aaronson}}\ and\ \bibinfo {author} {\bibfnamefont {D.}~\bibnamefont
  {Gottesman}},\ }\href {\doibase 10.1103/PhysRevA.70.052328} {\bibfield
  {journal} {\bibinfo  {journal} {Phys. Rev. A}\ }\textbf {\bibinfo {volume}
  {70}},\ \bibinfo {pages} {052328} (\bibinfo {year} {2004})}\BibitemShut
  {NoStop}%
\bibitem [{\citenamefont {Maslov}\ and\ \citenamefont
  {Roetteler}(2018)}]{Maslov2018-nl}%
  \BibitemOpen
  \bibfield  {author} {\bibinfo {author} {\bibfnamefont {D.}~\bibnamefont
  {Maslov}}\ and\ \bibinfo {author} {\bibfnamefont {M.}~\bibnamefont
  {Roetteler}},\ }\href {\doibase 10.1109/TIT.2018.2825602} {\bibfield
  {journal} {\bibinfo  {journal} {IEEE Trans. Inf. Theory}\ }\textbf {\bibinfo
  {volume} {64}},\ \bibinfo {pages} {4729} (\bibinfo {year}
  {2018})}\BibitemShut {NoStop}%
\bibitem [{\citenamefont {Bravyi}\ and\ \citenamefont
  {Maslov}(2021)}]{Bravyi2020-hg}%
  \BibitemOpen
  \bibfield  {author} {\bibinfo {author} {\bibfnamefont {S.}~\bibnamefont
  {Bravyi}}\ and\ \bibinfo {author} {\bibfnamefont {D.}~\bibnamefont
  {Maslov}},\ }\href@noop {} {\bibfield  {journal} {\bibinfo  {journal} {IEEE
  Transactions on Information Theory}\ }\textbf {\bibinfo {volume} {67}},\
  \bibinfo {pages} {4546} (\bibinfo {year} {2021})}\BibitemShut {NoStop}%
\bibitem [{\citenamefont {Proctor}\ and\ \citenamefont
  {Young}(2023)}]{Proctor2023-ep}%
  \BibitemOpen
  \bibfield  {author} {\bibinfo {author} {\bibfnamefont {T.}~\bibnamefont
  {Proctor}}\ and\ \bibinfo {author} {\bibfnamefont {K.}~\bibnamefont
  {Young}},\ }\href@noop {} {\bibfield  {journal} {\bibinfo  {journal} {arXiv
  preprint arXiv:2310.10882}\ } (\bibinfo {year} {2023})}\BibitemShut {NoStop}%
\bibitem [{\citenamefont {Patel}\ \emph {et~al.}(2008)\citenamefont {Patel},
  \citenamefont {Markov},\ and\ \citenamefont {Hayes}}]{patel2008optimal}%
  \BibitemOpen
  \bibfield  {author} {\bibinfo {author} {\bibfnamefont {K.~N.}\ \bibnamefont
  {Patel}}, \bibinfo {author} {\bibfnamefont {I.~L.}\ \bibnamefont {Markov}}, \
  and\ \bibinfo {author} {\bibfnamefont {J.~P.}\ \bibnamefont {Hayes}},\
  }\href@noop {} {\bibfield  {journal} {\bibinfo  {journal} {Quantum Info.
  Comput.}\ }\textbf {\bibinfo {volume} {8}},\ \bibinfo {pages} {282–294}
  (\bibinfo {year} {2008})}\BibitemShut {NoStop}%
\bibitem [{\citenamefont {Polloreno}\ \emph {et~al.}(2023)\citenamefont
  {Polloreno}, \citenamefont {Carignan-Dugas}, \citenamefont {Hines},
  \citenamefont {Blume-Kohout}, \citenamefont {Young},\ and\ \citenamefont
  {Proctor}}]{polloreno2023theory}%
  \BibitemOpen
  \bibfield  {author} {\bibinfo {author} {\bibfnamefont {A.~M.}\ \bibnamefont
  {Polloreno}}, \bibinfo {author} {\bibfnamefont {A.}~\bibnamefont
  {Carignan-Dugas}}, \bibinfo {author} {\bibfnamefont {J.}~\bibnamefont
  {Hines}}, \bibinfo {author} {\bibfnamefont {R.}~\bibnamefont {Blume-Kohout}},
  \bibinfo {author} {\bibfnamefont {K.}~\bibnamefont {Young}}, \ and\ \bibinfo
  {author} {\bibfnamefont {T.}~\bibnamefont {Proctor}},\ }\href@noop {}
  {\enquote {\bibinfo {title} {A theory of direct randomized benchmarking},}\ }
  (\bibinfo {year} {2023}),\ \Eprint {http://arxiv.org/abs/2302.13853}
  {arXiv:2302.13853 [quant-ph]} \BibitemShut {NoStop}%
\bibitem [{\citenamefont {Epstein}\ \emph {et~al.}(2014)\citenamefont
  {Epstein}, \citenamefont {Cross}, \citenamefont {Magesan},\ and\
  \citenamefont {Gambetta}}]{epstein2014investigating}%
  \BibitemOpen
  \bibfield  {author} {\bibinfo {author} {\bibfnamefont {J.~M.}\ \bibnamefont
  {Epstein}}, \bibinfo {author} {\bibfnamefont {A.~W.}\ \bibnamefont {Cross}},
  \bibinfo {author} {\bibfnamefont {E.}~\bibnamefont {Magesan}}, \ and\
  \bibinfo {author} {\bibfnamefont {J.~M.}\ \bibnamefont {Gambetta}},\ }\href
  {\doibase 10.1103/PhysRevA.89.062321} {\bibfield  {journal} {\bibinfo
  {journal} {Phys. Rev. A}\ }\textbf {\bibinfo {volume} {89}},\ \bibinfo
  {pages} {062321} (\bibinfo {year} {2014})}\BibitemShut {NoStop}%
\bibitem [{\citenamefont {Proctor}\ \emph {et~al.}(2019)\citenamefont
  {Proctor}, \citenamefont {Carignan-Dugas}, \citenamefont {Rudinger},
  \citenamefont {Nielsen}, \citenamefont {Blume-Kohout},\ and\ \citenamefont
  {Young}}]{proctor2018direct}%
  \BibitemOpen
  \bibfield  {author} {\bibinfo {author} {\bibfnamefont {T.~J.}\ \bibnamefont
  {Proctor}}, \bibinfo {author} {\bibfnamefont {A.}~\bibnamefont
  {Carignan-Dugas}}, \bibinfo {author} {\bibfnamefont {K.}~\bibnamefont
  {Rudinger}}, \bibinfo {author} {\bibfnamefont {E.}~\bibnamefont {Nielsen}},
  \bibinfo {author} {\bibfnamefont {R.}~\bibnamefont {Blume-Kohout}}, \ and\
  \bibinfo {author} {\bibfnamefont {K.}~\bibnamefont {Young}},\ }\href
  {https://journals.aps.org/prl/abstract/10.1103/PhysRevLett.123.030503}
  {\bibfield  {journal} {\bibinfo  {journal} {Phys. Rev. Lett.}\ }\textbf
  {\bibinfo {volume} {123}} (\bibinfo {year} {2019})}\BibitemShut {NoStop}%
\bibitem [{\citenamefont {Hines}\ \emph {et~al.}(2023)\citenamefont {Hines},
  \citenamefont {Lu}, \citenamefont {Naik}, \citenamefont {Hashim},
  \citenamefont {Ville}, \citenamefont {Mitchell}, \citenamefont {Kriekebaum},
  \citenamefont {Santiago}, \citenamefont {Seritan}, \citenamefont {Nielsen},
  \citenamefont {Blume-Kohout}, \citenamefont {Young}, \citenamefont {Siddiqi},
  \citenamefont {Whaley},\ and\ \citenamefont
  {Proctor}}]{hines2022demonstrating}%
  \BibitemOpen
  \bibfield  {author} {\bibinfo {author} {\bibfnamefont {J.}~\bibnamefont
  {Hines}}, \bibinfo {author} {\bibfnamefont {M.}~\bibnamefont {Lu}}, \bibinfo
  {author} {\bibfnamefont {R.~K.}\ \bibnamefont {Naik}}, \bibinfo {author}
  {\bibfnamefont {A.}~\bibnamefont {Hashim}}, \bibinfo {author} {\bibfnamefont
  {J.-L.}\ \bibnamefont {Ville}}, \bibinfo {author} {\bibfnamefont
  {B.}~\bibnamefont {Mitchell}}, \bibinfo {author} {\bibfnamefont {J.~M.}\
  \bibnamefont {Kriekebaum}}, \bibinfo {author} {\bibfnamefont {D.~I.}\
  \bibnamefont {Santiago}}, \bibinfo {author} {\bibfnamefont {S.}~\bibnamefont
  {Seritan}}, \bibinfo {author} {\bibfnamefont {E.}~\bibnamefont {Nielsen}},
  \bibinfo {author} {\bibfnamefont {R.}~\bibnamefont {Blume-Kohout}}, \bibinfo
  {author} {\bibfnamefont {K.}~\bibnamefont {Young}}, \bibinfo {author}
  {\bibfnamefont {I.}~\bibnamefont {Siddiqi}}, \bibinfo {author} {\bibfnamefont
  {B.}~\bibnamefont {Whaley}}, \ and\ \bibinfo {author} {\bibfnamefont
  {T.}~\bibnamefont {Proctor}},\ }\href {\doibase 10.1103/PhysRevX.13.041030}
  {\bibfield  {journal} {\bibinfo  {journal} {Phys. Rev. X}\ }\textbf {\bibinfo
  {volume} {13}},\ \bibinfo {pages} {041030} (\bibinfo {year}
  {2023})}\BibitemShut {NoStop}%
\bibitem [{\citenamefont {Hines}\ \emph {et~al.}(2024)\citenamefont {Hines},
  \citenamefont {Hothem}, \citenamefont {Blume-Kohout}, \citenamefont
  {Whaley},\ and\ \citenamefont {Proctor}}]{hines2023fully}%
  \BibitemOpen
  \bibfield  {author} {\bibinfo {author} {\bibfnamefont {J.}~\bibnamefont
  {Hines}}, \bibinfo {author} {\bibfnamefont {D.}~\bibnamefont {Hothem}},
  \bibinfo {author} {\bibfnamefont {R.}~\bibnamefont {Blume-Kohout}}, \bibinfo
  {author} {\bibfnamefont {B.}~\bibnamefont {Whaley}}, \ and\ \bibinfo {author}
  {\bibfnamefont {T.}~\bibnamefont {Proctor}},\ }\href {\doibase
  10.1103/PRXQuantum.5.030334} {\bibfield  {journal} {\bibinfo  {journal} {PRX
  Quantum}\ }\textbf {\bibinfo {volume} {5}},\ \bibinfo {pages} {030334}
  (\bibinfo {year} {2024})}\BibitemShut {NoStop}%
\bibitem [{Note24()}]{Note24}%
  \BibitemOpen
  \bibinfo {note} {Here, a ``success metric'' does not necessarily refer to a
  ``metric'' in the strict mathematical sense; see the discussion in Sec.~\ref
  {sec:overview}.}\BibitemShut {Stop}%
\bibitem [{\citenamefont {McKay}\ \emph {et~al.}(2023)\citenamefont {McKay},
  \citenamefont {Hincks}, \citenamefont {Pritchett}, \citenamefont {Carroll},
  \citenamefont {Govia},\ and\ \citenamefont {Merkel}}]{McKay2023-bx}%
  \BibitemOpen
  \bibfield  {author} {\bibinfo {author} {\bibfnamefont {D.~C.}\ \bibnamefont
  {McKay}}, \bibinfo {author} {\bibfnamefont {I.}~\bibnamefont {Hincks}},
  \bibinfo {author} {\bibfnamefont {E.~J.}\ \bibnamefont {Pritchett}}, \bibinfo
  {author} {\bibfnamefont {M.}~\bibnamefont {Carroll}}, \bibinfo {author}
  {\bibfnamefont {L.~C.}\ \bibnamefont {Govia}}, \ and\ \bibinfo {author}
  {\bibfnamefont {S.~T.}\ \bibnamefont {Merkel}},\ }\href@noop {} {\bibfield
  {journal} {\bibinfo  {journal} {arXiv preprint arXiv:2311.05933}\ } (\bibinfo
  {year} {2023})}\BibitemShut {NoStop}%
\bibitem [{\citenamefont {Chen}\ \emph
  {et~al.}(2023{\natexlab{b}})\citenamefont {Chen}, \citenamefont {Nielsen},
  \citenamefont {Ebert}, \citenamefont {Inlek}, \citenamefont {Wright},
  \citenamefont {Chaplin}, \citenamefont {Maksymov}, \citenamefont {P{\'a}ez},
  \citenamefont {Poudel}, \citenamefont {Maunz} \emph {et~al.}}]{Chen2023-la}%
  \BibitemOpen
  \bibfield  {author} {\bibinfo {author} {\bibfnamefont {J.-S.}\ \bibnamefont
  {Chen}}, \bibinfo {author} {\bibfnamefont {E.}~\bibnamefont {Nielsen}},
  \bibinfo {author} {\bibfnamefont {M.}~\bibnamefont {Ebert}}, \bibinfo
  {author} {\bibfnamefont {V.}~\bibnamefont {Inlek}}, \bibinfo {author}
  {\bibfnamefont {K.}~\bibnamefont {Wright}}, \bibinfo {author} {\bibfnamefont
  {V.}~\bibnamefont {Chaplin}}, \bibinfo {author} {\bibfnamefont
  {A.}~\bibnamefont {Maksymov}}, \bibinfo {author} {\bibfnamefont
  {E.}~\bibnamefont {P{\'a}ez}}, \bibinfo {author} {\bibfnamefont
  {A.}~\bibnamefont {Poudel}}, \bibinfo {author} {\bibfnamefont
  {P.}~\bibnamefont {Maunz}},  \emph {et~al.},\ }\href@noop {} {\bibfield
  {journal} {\bibinfo  {journal} {arXiv preprint arXiv:2308.05071}\ } (\bibinfo
  {year} {2023}{\natexlab{b}})}\BibitemShut {NoStop}%
\bibitem [{\citenamefont {Mayer}\ \emph {et~al.}(2021)\citenamefont {Mayer},
  \citenamefont {Hall}, \citenamefont {Gatterman}, \citenamefont {Halit},
  \citenamefont {Lee}, \citenamefont {Bohnet}, \citenamefont {Gresh},
  \citenamefont {Hankin}, \citenamefont {Gilmore}, \citenamefont {Gerber} \emph
  {et~al.}}]{Mayer2021-vl}%
  \BibitemOpen
  \bibfield  {author} {\bibinfo {author} {\bibfnamefont {K.}~\bibnamefont
  {Mayer}}, \bibinfo {author} {\bibfnamefont {A.}~\bibnamefont {Hall}},
  \bibinfo {author} {\bibfnamefont {T.}~\bibnamefont {Gatterman}}, \bibinfo
  {author} {\bibfnamefont {S.~K.}\ \bibnamefont {Halit}}, \bibinfo {author}
  {\bibfnamefont {K.}~\bibnamefont {Lee}}, \bibinfo {author} {\bibfnamefont
  {J.}~\bibnamefont {Bohnet}}, \bibinfo {author} {\bibfnamefont
  {D.}~\bibnamefont {Gresh}}, \bibinfo {author} {\bibfnamefont
  {A.}~\bibnamefont {Hankin}}, \bibinfo {author} {\bibfnamefont
  {K.}~\bibnamefont {Gilmore}}, \bibinfo {author} {\bibfnamefont
  {J.}~\bibnamefont {Gerber}},  \emph {et~al.},\ }\href@noop {} {\bibfield
  {journal} {\bibinfo  {journal} {arXiv preprint arXiv:2108.10431}\ } (\bibinfo
  {year} {2021})}\BibitemShut {NoStop}%
\bibitem [{\citenamefont {Amico}\ \emph {et~al.}(2023)\citenamefont {Amico},
  \citenamefont {Zhang}, \citenamefont {Jurcevic}, \citenamefont {Bishop},
  \citenamefont {Nation}, \citenamefont {Wack},\ and\ \citenamefont
  {McKay}}]{Amico2023-ze}%
  \BibitemOpen
  \bibfield  {author} {\bibinfo {author} {\bibfnamefont {M.}~\bibnamefont
  {Amico}}, \bibinfo {author} {\bibfnamefont {H.}~\bibnamefont {Zhang}},
  \bibinfo {author} {\bibfnamefont {P.}~\bibnamefont {Jurcevic}}, \bibinfo
  {author} {\bibfnamefont {L.~S.}\ \bibnamefont {Bishop}}, \bibinfo {author}
  {\bibfnamefont {P.}~\bibnamefont {Nation}}, \bibinfo {author} {\bibfnamefont
  {A.}~\bibnamefont {Wack}}, \ and\ \bibinfo {author} {\bibfnamefont {D.~C.}\
  \bibnamefont {McKay}},\ }\href@noop {} {\bibfield  {journal} {\bibinfo
  {journal} {arXiv preprint arXiv:2303.02108}\ } (\bibinfo {year}
  {2023})}\BibitemShut {NoStop}%
\bibitem [{\citenamefont {Proctor}\ \emph
  {et~al.}(2022{\natexlab{c}})\citenamefont {Proctor}, \citenamefont {Seritan},
  \citenamefont {Nielsen}, \citenamefont {Rudinger}, \citenamefont {Young},
  \citenamefont {Blume-Kohout},\ and\ \citenamefont
  {Sarovar}}]{proctor2022establishing}%
  \BibitemOpen
  \bibfield  {author} {\bibinfo {author} {\bibfnamefont {T.}~\bibnamefont
  {Proctor}}, \bibinfo {author} {\bibfnamefont {S.}~\bibnamefont {Seritan}},
  \bibinfo {author} {\bibfnamefont {E.}~\bibnamefont {Nielsen}}, \bibinfo
  {author} {\bibfnamefont {K.}~\bibnamefont {Rudinger}}, \bibinfo {author}
  {\bibfnamefont {K.}~\bibnamefont {Young}}, \bibinfo {author} {\bibfnamefont
  {R.}~\bibnamefont {Blume-Kohout}}, \ and\ \bibinfo {author} {\bibfnamefont
  {M.}~\bibnamefont {Sarovar}},\ }\href@noop {} {\bibfield  {journal} {\bibinfo
   {journal} {arXiv preprint arXiv:2204.07568}\ } (\bibinfo {year}
  {2022}{\natexlab{c}})}\BibitemShut {NoStop}%
\bibitem [{\citenamefont {Neill}\ \emph {et~al.}(2018)\citenamefont {Neill},
  \citenamefont {Roushan}, \citenamefont {Kechedzhi}, \citenamefont {Boixo},
  \citenamefont {Isakov}, \citenamefont {Smelyanskiy}, \citenamefont {Megrant},
  \citenamefont {Chiaro}, \citenamefont {Dunsworth}, \citenamefont {Arya},
  \citenamefont {Barends}, \citenamefont {Burkett}, \citenamefont {Chen},
  \citenamefont {Chen}, \citenamefont {Fowler}, \citenamefont {Foxen},
  \citenamefont {Giustina}, \citenamefont {Graff}, \citenamefont {Jeffrey},
  \citenamefont {Huang}, \citenamefont {Kelly}, \citenamefont {Klimov},
  \citenamefont {Lucero}, \citenamefont {Mutus}, \citenamefont {Neeley},
  \citenamefont {Quintana}, \citenamefont {Sank}, \citenamefont {Vainsencher},
  \citenamefont {Wenner}, \citenamefont {White}, \citenamefont {Neven},\ and\
  \citenamefont {Martinis}}]{neill2018blueprint}%
  \BibitemOpen
  \bibfield  {author} {\bibinfo {author} {\bibfnamefont {C.}~\bibnamefont
  {Neill}}, \bibinfo {author} {\bibfnamefont {P.}~\bibnamefont {Roushan}},
  \bibinfo {author} {\bibfnamefont {K.}~\bibnamefont {Kechedzhi}}, \bibinfo
  {author} {\bibfnamefont {S.}~\bibnamefont {Boixo}}, \bibinfo {author}
  {\bibfnamefont {S.~V.}\ \bibnamefont {Isakov}}, \bibinfo {author}
  {\bibfnamefont {V.}~\bibnamefont {Smelyanskiy}}, \bibinfo {author}
  {\bibfnamefont {A.}~\bibnamefont {Megrant}}, \bibinfo {author} {\bibfnamefont
  {B.}~\bibnamefont {Chiaro}}, \bibinfo {author} {\bibfnamefont
  {A.}~\bibnamefont {Dunsworth}}, \bibinfo {author} {\bibfnamefont
  {K.}~\bibnamefont {Arya}}, \bibinfo {author} {\bibfnamefont {R.}~\bibnamefont
  {Barends}}, \bibinfo {author} {\bibfnamefont {B.}~\bibnamefont {Burkett}},
  \bibinfo {author} {\bibfnamefont {Y.}~\bibnamefont {Chen}}, \bibinfo {author}
  {\bibfnamefont {Z.}~\bibnamefont {Chen}}, \bibinfo {author} {\bibfnamefont
  {A.}~\bibnamefont {Fowler}}, \bibinfo {author} {\bibfnamefont
  {B.}~\bibnamefont {Foxen}}, \bibinfo {author} {\bibfnamefont
  {M.}~\bibnamefont {Giustina}}, \bibinfo {author} {\bibfnamefont
  {R.}~\bibnamefont {Graff}}, \bibinfo {author} {\bibfnamefont
  {E.}~\bibnamefont {Jeffrey}}, \bibinfo {author} {\bibfnamefont
  {T.}~\bibnamefont {Huang}}, \bibinfo {author} {\bibfnamefont
  {J.}~\bibnamefont {Kelly}}, \bibinfo {author} {\bibfnamefont
  {P.}~\bibnamefont {Klimov}}, \bibinfo {author} {\bibfnamefont
  {E.}~\bibnamefont {Lucero}}, \bibinfo {author} {\bibfnamefont
  {J.}~\bibnamefont {Mutus}}, \bibinfo {author} {\bibfnamefont
  {M.}~\bibnamefont {Neeley}}, \bibinfo {author} {\bibfnamefont
  {C.}~\bibnamefont {Quintana}}, \bibinfo {author} {\bibfnamefont
  {D.}~\bibnamefont {Sank}}, \bibinfo {author} {\bibfnamefont {A.}~\bibnamefont
  {Vainsencher}}, \bibinfo {author} {\bibfnamefont {J.}~\bibnamefont {Wenner}},
  \bibinfo {author} {\bibfnamefont {T.~C.}\ \bibnamefont {White}}, \bibinfo
  {author} {\bibfnamefont {H.}~\bibnamefont {Neven}}, \ and\ \bibinfo {author}
  {\bibfnamefont {J.~M.}\ \bibnamefont {Martinis}},\ }\href {\doibase
  10.1126/science.aao4309} {\bibfield  {journal} {\bibinfo  {journal}
  {Science}\ }\textbf {\bibinfo {volume} {360}},\ \bibinfo {pages} {195}
  (\bibinfo {year} {2018})},\ \Eprint
  {http://arxiv.org/abs/https://www.science.org/doi/pdf/10.1126/science.aao4309}
  {https://www.science.org/doi/pdf/10.1126/science.aao4309} \BibitemShut
  {NoStop}%
\bibitem [{\citenamefont {Liu}\ \emph {et~al.}(2021{\natexlab{b}})\citenamefont
  {Liu}, \citenamefont {Otten}, \citenamefont {Bassirianjahromi}, \citenamefont
  {Jiang},\ and\ \citenamefont {Fefferman}}]{liu2021benchmarking}%
  \BibitemOpen
  \bibfield  {author} {\bibinfo {author} {\bibfnamefont {Y.}~\bibnamefont
  {Liu}}, \bibinfo {author} {\bibfnamefont {M.}~\bibnamefont {Otten}}, \bibinfo
  {author} {\bibfnamefont {R.}~\bibnamefont {Bassirianjahromi}}, \bibinfo
  {author} {\bibfnamefont {L.}~\bibnamefont {Jiang}}, \ and\ \bibinfo {author}
  {\bibfnamefont {B.}~\bibnamefont {Fefferman}},\ }\href@noop {} {\bibfield
  {journal} {\bibinfo  {journal} {arXiv preprint arXiv:2105.05232}\ } (\bibinfo
  {year} {2021}{\natexlab{b}})}\BibitemShut {NoStop}%
\bibitem [{\citenamefont {Heinrich}\ \emph {et~al.}(2022)\citenamefont
  {Heinrich}, \citenamefont {Kliesch},\ and\ \citenamefont
  {Roth}}]{Heinrich2022-cs}%
  \BibitemOpen
  \bibfield  {author} {\bibinfo {author} {\bibfnamefont {M.}~\bibnamefont
  {Heinrich}}, \bibinfo {author} {\bibfnamefont {M.}~\bibnamefont {Kliesch}}, \
  and\ \bibinfo {author} {\bibfnamefont {I.}~\bibnamefont {Roth}},\ }\href@noop
  {} {\bibfield  {journal} {\bibinfo  {journal} {arXiv preprint
  arXiv:2212.06181}\ } (\bibinfo {year} {2022})}\BibitemShut {NoStop}%
\bibitem [{\citenamefont {Chen}\ \emph
  {et~al.}(2023{\natexlab{c}})\citenamefont {Chen}, \citenamefont {Ding},
  \citenamefont {Huang},\ and\ \citenamefont {Kong}}]{Chen2022-hd}%
  \BibitemOpen
  \bibfield  {author} {\bibinfo {author} {\bibfnamefont {J.}~\bibnamefont
  {Chen}}, \bibinfo {author} {\bibfnamefont {D.}~\bibnamefont {Ding}}, \bibinfo
  {author} {\bibfnamefont {C.}~\bibnamefont {Huang}}, \ and\ \bibinfo {author}
  {\bibfnamefont {L.}~\bibnamefont {Kong}},\ }\href@noop {} {\bibfield
  {journal} {\bibinfo  {journal} {Physical Review A}\ }\textbf {\bibinfo
  {volume} {108}},\ \bibinfo {pages} {052613} (\bibinfo {year}
  {2023}{\natexlab{c}})}\BibitemShut {NoStop}%
\bibitem [{\citenamefont {Ware}\ \emph {et~al.}(2023)\citenamefont {Ware},
  \citenamefont {Deshpande}, \citenamefont {Hangleiter}, \citenamefont
  {Niroula}, \citenamefont {Fefferman}, \citenamefont {Gorshkov},\ and\
  \citenamefont {Gullans}}]{Ware2023-zy}%
  \BibitemOpen
  \bibfield  {author} {\bibinfo {author} {\bibfnamefont {B.}~\bibnamefont
  {Ware}}, \bibinfo {author} {\bibfnamefont {A.}~\bibnamefont {Deshpande}},
  \bibinfo {author} {\bibfnamefont {D.}~\bibnamefont {Hangleiter}}, \bibinfo
  {author} {\bibfnamefont {P.}~\bibnamefont {Niroula}}, \bibinfo {author}
  {\bibfnamefont {B.}~\bibnamefont {Fefferman}}, \bibinfo {author}
  {\bibfnamefont {A.~V.}\ \bibnamefont {Gorshkov}}, \ and\ \bibinfo {author}
  {\bibfnamefont {M.~J.}\ \bibnamefont {Gullans}},\ }\href@noop {} {\bibfield
  {journal} {\bibinfo  {journal} {arXiv preprint arXiv:2305.04954}\ } (\bibinfo
  {year} {2023})}\BibitemShut {NoStop}%
\bibitem [{Note25()}]{Note25}%
  \BibitemOpen
  \bibinfo {note} {Because XEB requires that an $n$-qubit circuit converges to
  an $n$-qubit Haar-random unitary, the infidelity of twirling layers
  consisting only of Haar-random single-qubit gates cannot be measured via an
  $n$-qubit XEB experiment; rather, it must be estimated from the combined
  infidelity of simultaneous XEB on all $n$ qubits. Or, instead, one could use
  $n$-qubit Haar-random unitaries for the twirl, in which case an $n$-qubit XEB
  experiment without the interleaved gate could be used to estimate the
  infidelity of the twirling layer. However, in this case, the decomposition of
  XEB circuits to native gates would scale poorly (similar to $n$-qubit CRB).
  Furthermore, note that the estimate of the interleaved gate's fidelity would
  be subject to systematic errors similar to those seen in IRB.}\BibitemShut
  {Stop}%
\bibitem [{\citenamefont {Hashagen}\ \emph {et~al.}(2018)\citenamefont
  {Hashagen}, \citenamefont {Flammia}, \citenamefont {Gross},\ and\
  \citenamefont {Wallman}}]{Hashagen2018-dk}%
  \BibitemOpen
  \bibfield  {author} {\bibinfo {author} {\bibfnamefont {A.~K.}\ \bibnamefont
  {Hashagen}}, \bibinfo {author} {\bibfnamefont {S.~T.}\ \bibnamefont
  {Flammia}}, \bibinfo {author} {\bibfnamefont {D.}~\bibnamefont {Gross}}, \
  and\ \bibinfo {author} {\bibfnamefont {J.~J.}\ \bibnamefont {Wallman}},\
  }\href {\doibase 10.22331/q-2018-08-22-85} {\bibfield  {journal} {\bibinfo
  {journal} {Quantum}\ }\textbf {\bibinfo {volume} {2}},\ \bibinfo {pages} {85}
  (\bibinfo {year} {2018})}\BibitemShut {NoStop}%
\bibitem [{\citenamefont {Brown}\ and\ \citenamefont
  {Eastin}(2018)}]{Brown2018-dx}%
  \BibitemOpen
  \bibfield  {author} {\bibinfo {author} {\bibfnamefont {W.~G.}\ \bibnamefont
  {Brown}}\ and\ \bibinfo {author} {\bibfnamefont {B.}~\bibnamefont {Eastin}},\
  }\href@noop {} {\bibfield  {journal} {\bibinfo  {journal} {Physical Review
  A}\ }\textbf {\bibinfo {volume} {97}},\ \bibinfo {pages} {062323} (\bibinfo
  {year} {2018})}\BibitemShut {NoStop}%
\bibitem [{\citenamefont {Carignan-Dugas}\ \emph {et~al.}(2015)\citenamefont
  {Carignan-Dugas}, \citenamefont {Wallman},\ and\ \citenamefont
  {Emerson}}]{carignandugas2015characterizing}%
  \BibitemOpen
  \bibfield  {author} {\bibinfo {author} {\bibfnamefont {A.}~\bibnamefont
  {Carignan-Dugas}}, \bibinfo {author} {\bibfnamefont {J.~J.}\ \bibnamefont
  {Wallman}}, \ and\ \bibinfo {author} {\bibfnamefont {J.}~\bibnamefont
  {Emerson}},\ }\href {\doibase 10.1103/PhysRevA.92.060302} {\bibfield
  {journal} {\bibinfo  {journal} {Phys. Rev. A}\ }\textbf {\bibinfo {volume}
  {92}},\ \bibinfo {pages} {060302} (\bibinfo {year} {2015})}\BibitemShut
  {NoStop}%
\bibitem [{\citenamefont {Claes}\ \emph {et~al.}(2021)\citenamefont {Claes},
  \citenamefont {Rieffel},\ and\ \citenamefont {Wang}}]{claes2021character}%
  \BibitemOpen
  \bibfield  {author} {\bibinfo {author} {\bibfnamefont {J.}~\bibnamefont
  {Claes}}, \bibinfo {author} {\bibfnamefont {E.}~\bibnamefont {Rieffel}}, \
  and\ \bibinfo {author} {\bibfnamefont {Z.}~\bibnamefont {Wang}},\ }\href
  {\doibase 10.1103/PRXQuantum.2.010351} {\bibfield  {journal} {\bibinfo
  {journal} {PRX Quantum}\ }\textbf {\bibinfo {volume} {2}},\ \bibinfo {pages}
  {010351} (\bibinfo {year} {2021})}\BibitemShut {NoStop}%
\bibitem [{\citenamefont {Helsen}\ \emph
  {et~al.}(2022{\natexlab{b}})\citenamefont {Helsen}, \citenamefont {Nezami},
  \citenamefont {Reagor},\ and\ \citenamefont {Walter}}]{Helsen2022matchgate}%
  \BibitemOpen
  \bibfield  {author} {\bibinfo {author} {\bibfnamefont {J.}~\bibnamefont
  {Helsen}}, \bibinfo {author} {\bibfnamefont {S.}~\bibnamefont {Nezami}},
  \bibinfo {author} {\bibfnamefont {M.}~\bibnamefont {Reagor}}, \ and\ \bibinfo
  {author} {\bibfnamefont {M.}~\bibnamefont {Walter}},\ }\href {\doibase
  10.22331/q-2022-02-21-657} {\bibfield  {journal} {\bibinfo  {journal}
  {{Quantum}}\ }\textbf {\bibinfo {volume} {6}},\ \bibinfo {pages} {657}
  (\bibinfo {year} {2022}{\natexlab{b}})}\BibitemShut {NoStop}%
\bibitem [{\citenamefont {França}\ and\ \citenamefont
  {Hashagen}(2018)}]{Franca2018-ow}%
  \BibitemOpen
  \bibfield  {author} {\bibinfo {author} {\bibfnamefont {D.~S.}\ \bibnamefont
  {França}}\ and\ \bibinfo {author} {\bibfnamefont {A.~K.}\ \bibnamefont
  {Hashagen}},\ }\href {\doibase 10.1088/1751-8121/aad6fa} {\bibfield
  {journal} {\bibinfo  {journal} {J. Phys. A: Math. Theor.}\ }\textbf {\bibinfo
  {volume} {51}},\ \bibinfo {pages} {395302} (\bibinfo {year}
  {2018})}\BibitemShut {NoStop}%
\bibitem [{\citenamefont {Claes}\ and\ \citenamefont
  {Puri}(2023)}]{claes2023estimating}%
  \BibitemOpen
  \bibfield  {author} {\bibinfo {author} {\bibfnamefont {J.}~\bibnamefont
  {Claes}}\ and\ \bibinfo {author} {\bibfnamefont {S.}~\bibnamefont {Puri}},\
  }\href {\doibase 10.1103/PRXQuantum.4.010307} {\bibfield  {journal} {\bibinfo
   {journal} {PRX Quantum}\ }\textbf {\bibinfo {volume} {4}},\ \bibinfo {pages}
  {010307} (\bibinfo {year} {2023})}\BibitemShut {NoStop}%
\bibitem [{\citenamefont {Gambetta}\ \emph {et~al.}(2012)\citenamefont
  {Gambetta}, \citenamefont {C{\'o}rcoles}, \citenamefont {Merkel},
  \citenamefont {Johnson}, \citenamefont {Smolin}, \citenamefont {Chow},
  \citenamefont {Ryan}, \citenamefont {Rigetti}, \citenamefont {Poletto},
  \citenamefont {Ohki}, \citenamefont {Ketchen},\ and\ \citenamefont
  {Steffen}}]{Gambetta2012-zd}%
  \BibitemOpen
  \bibfield  {author} {\bibinfo {author} {\bibfnamefont {J.~M.}\ \bibnamefont
  {Gambetta}}, \bibinfo {author} {\bibfnamefont {A.~D.}\ \bibnamefont
  {C{\'o}rcoles}}, \bibinfo {author} {\bibfnamefont {S.~T.}\ \bibnamefont
  {Merkel}}, \bibinfo {author} {\bibfnamefont {B.~R.}\ \bibnamefont {Johnson}},
  \bibinfo {author} {\bibfnamefont {J.~A.}\ \bibnamefont {Smolin}}, \bibinfo
  {author} {\bibfnamefont {J.~M.}\ \bibnamefont {Chow}}, \bibinfo {author}
  {\bibfnamefont {C.~A.}\ \bibnamefont {Ryan}}, \bibinfo {author}
  {\bibfnamefont {C.}~\bibnamefont {Rigetti}}, \bibinfo {author} {\bibfnamefont
  {S.}~\bibnamefont {Poletto}}, \bibinfo {author} {\bibfnamefont {T.~A.}\
  \bibnamefont {Ohki}}, \bibinfo {author} {\bibfnamefont {M.~B.}\ \bibnamefont
  {Ketchen}}, \ and\ \bibinfo {author} {\bibfnamefont {M.}~\bibnamefont
  {Steffen}},\ }\href {\doibase 10.1103/PhysRevLett.109.240504} {\bibfield
  {journal} {\bibinfo  {journal} {Phys. Rev. Lett.}\ }\textbf {\bibinfo
  {volume} {109}},\ \bibinfo {pages} {240504} (\bibinfo {year}
  {2012})}\BibitemShut {NoStop}%
\bibitem [{\citenamefont {McKay}\ \emph {et~al.}(2019)\citenamefont {McKay},
  \citenamefont {Sheldon}, \citenamefont {Smolin}, \citenamefont {Chow},\ and\
  \citenamefont {Gambetta}}]{mckay2019three}%
  \BibitemOpen
  \bibfield  {author} {\bibinfo {author} {\bibfnamefont {D.~C.}\ \bibnamefont
  {McKay}}, \bibinfo {author} {\bibfnamefont {S.}~\bibnamefont {Sheldon}},
  \bibinfo {author} {\bibfnamefont {J.~A.}\ \bibnamefont {Smolin}}, \bibinfo
  {author} {\bibfnamefont {J.~M.}\ \bibnamefont {Chow}}, \ and\ \bibinfo
  {author} {\bibfnamefont {J.~M.}\ \bibnamefont {Gambetta}},\ }\href@noop {}
  {\bibfield  {journal} {\bibinfo  {journal} {Physical review letters}\
  }\textbf {\bibinfo {volume} {122}},\ \bibinfo {pages} {200502} (\bibinfo
  {year} {2019})}\BibitemShut {NoStop}%
\bibitem [{\citenamefont {McKay}\ \emph {et~al.}(2020)\citenamefont {McKay},
  \citenamefont {Cross}, \citenamefont {Wood},\ and\ \citenamefont
  {Gambetta}}]{McKay2020-uh}%
  \BibitemOpen
  \bibfield  {author} {\bibinfo {author} {\bibfnamefont {D.~C.}\ \bibnamefont
  {McKay}}, \bibinfo {author} {\bibfnamefont {A.~W.}\ \bibnamefont {Cross}},
  \bibinfo {author} {\bibfnamefont {C.~J.}\ \bibnamefont {Wood}}, \ and\
  \bibinfo {author} {\bibfnamefont {J.~M.}\ \bibnamefont {Gambetta}},\
  }\href@noop {} {\bibfield  {journal} {\bibinfo  {journal} {arXiv preprint
  arXiv:2003.02354}\ } (\bibinfo {year} {2020})}\BibitemShut {NoStop}%
\bibitem [{\citenamefont {Harper}\ \emph {et~al.}(2020)\citenamefont {Harper},
  \citenamefont {Flammia},\ and\ \citenamefont
  {Wallman}}]{harper2020efficient}%
  \BibitemOpen
  \bibfield  {author} {\bibinfo {author} {\bibfnamefont {R.}~\bibnamefont
  {Harper}}, \bibinfo {author} {\bibfnamefont {S.~T.}\ \bibnamefont {Flammia}},
  \ and\ \bibinfo {author} {\bibfnamefont {J.~J.}\ \bibnamefont {Wallman}},\
  }\href@noop {} {\bibfield  {journal} {\bibinfo  {journal} {Nature Physics}\
  }\textbf {\bibinfo {volume} {16}},\ \bibinfo {pages} {1184} (\bibinfo {year}
  {2020})}\BibitemShut {NoStop}%
\bibitem [{\citenamefont {Harper}\ and\ \citenamefont
  {Flammia}(2023)}]{harper2023learning}%
  \BibitemOpen
  \bibfield  {author} {\bibinfo {author} {\bibfnamefont {R.}~\bibnamefont
  {Harper}}\ and\ \bibinfo {author} {\bibfnamefont {S.~T.}\ \bibnamefont
  {Flammia}},\ }\href {\doibase 10.1103/PRXQuantum.4.040311} {\bibfield
  {journal} {\bibinfo  {journal} {PRX Quantum}\ }\textbf {\bibinfo {volume}
  {4}},\ \bibinfo {pages} {040311} (\bibinfo {year} {2023})}\BibitemShut
  {NoStop}%
\bibitem [{\citenamefont {Garion}\ \emph {et~al.}(2021)\citenamefont {Garion},
  \citenamefont {Kanazawa}, \citenamefont {Landa}, \citenamefont {McKay},
  \citenamefont {Sheldon}, \citenamefont {Cross},\ and\ \citenamefont
  {Wood}}]{Garion2020-gi}%
  \BibitemOpen
  \bibfield  {author} {\bibinfo {author} {\bibfnamefont {S.}~\bibnamefont
  {Garion}}, \bibinfo {author} {\bibfnamefont {N.}~\bibnamefont {Kanazawa}},
  \bibinfo {author} {\bibfnamefont {H.}~\bibnamefont {Landa}}, \bibinfo
  {author} {\bibfnamefont {D.~C.}\ \bibnamefont {McKay}}, \bibinfo {author}
  {\bibfnamefont {S.}~\bibnamefont {Sheldon}}, \bibinfo {author} {\bibfnamefont
  {A.~W.}\ \bibnamefont {Cross}}, \ and\ \bibinfo {author} {\bibfnamefont
  {C.~J.}\ \bibnamefont {Wood}},\ }\href@noop {} {\bibfield  {journal}
  {\bibinfo  {journal} {Physical Review Research}\ }\textbf {\bibinfo {volume}
  {3}},\ \bibinfo {pages} {013204} (\bibinfo {year} {2021})}\BibitemShut
  {NoStop}%
\bibitem [{\citenamefont {Harper}\ and\ \citenamefont
  {Flammia}(2017)}]{Harper2017-xn}%
  \BibitemOpen
  \bibfield  {author} {\bibinfo {author} {\bibfnamefont {R.}~\bibnamefont
  {Harper}}\ and\ \bibinfo {author} {\bibfnamefont {S.~T.}\ \bibnamefont
  {Flammia}},\ }\href@noop {} {\bibfield  {journal} {\bibinfo  {journal}
  {Quantum Science and Technology}\ }\textbf {\bibinfo {volume} {2}},\ \bibinfo
  {pages} {015008} (\bibinfo {year} {2017})}\BibitemShut {NoStop}%
\bibitem [{\citenamefont {Carignan-Dugas}\ \emph {et~al.}(2019)\citenamefont
  {Carignan-Dugas}, \citenamefont {Wallman},\ and\ \citenamefont
  {Emerson}}]{carignan2019bounding}%
  \BibitemOpen
  \bibfield  {author} {\bibinfo {author} {\bibfnamefont {A.}~\bibnamefont
  {Carignan-Dugas}}, \bibinfo {author} {\bibfnamefont {J.~J.}\ \bibnamefont
  {Wallman}}, \ and\ \bibinfo {author} {\bibfnamefont {J.}~\bibnamefont
  {Emerson}},\ }\href@noop {} {\bibfield  {journal} {\bibinfo  {journal} {New
  Journal of Physics}\ }\textbf {\bibinfo {volume} {21}},\ \bibinfo {pages}
  {053016} (\bibinfo {year} {2019})}\BibitemShut {NoStop}%
\bibitem [{\citenamefont {Hothem}\ \emph {et~al.}(2023)\citenamefont {Hothem},
  \citenamefont {Hines}, \citenamefont {Nataraj}, \citenamefont
  {Blume-Kohout},\ and\ \citenamefont {Proctor}}]{Hothem2023-bl}%
  \BibitemOpen
  \bibfield  {author} {\bibinfo {author} {\bibfnamefont {D.}~\bibnamefont
  {Hothem}}, \bibinfo {author} {\bibfnamefont {J.}~\bibnamefont {Hines}},
  \bibinfo {author} {\bibfnamefont {K.}~\bibnamefont {Nataraj}}, \bibinfo
  {author} {\bibfnamefont {R.}~\bibnamefont {Blume-Kohout}}, \ and\ \bibinfo
  {author} {\bibfnamefont {T.}~\bibnamefont {Proctor}},\ }in\ \href@noop {}
  {\emph {\bibinfo {booktitle} {2023 IEEE International Conference on Quantum
  Computing and Engineering (QCE)}}},\ Vol.~\bibinfo {volume} {1}\ (\bibinfo
  {organization} {IEEE},\ \bibinfo {year} {2023})\ pp.\ \bibinfo {pages}
  {709--714}\BibitemShut {NoStop}%
\bibitem [{\citenamefont {Erhard}\ \emph {et~al.}(2019)\citenamefont {Erhard},
  \citenamefont {Wallman}, \citenamefont {Postler}, \citenamefont {Meth},
  \citenamefont {Stricker}, \citenamefont {Martinez}, \citenamefont
  {Schindler}, \citenamefont {Monz}, \citenamefont {Emerson},\ and\
  \citenamefont {Blatt}}]{erhard2019characterizing}%
  \BibitemOpen
  \bibfield  {author} {\bibinfo {author} {\bibfnamefont {A.}~\bibnamefont
  {Erhard}}, \bibinfo {author} {\bibfnamefont {J.~J.}\ \bibnamefont {Wallman}},
  \bibinfo {author} {\bibfnamefont {L.}~\bibnamefont {Postler}}, \bibinfo
  {author} {\bibfnamefont {M.}~\bibnamefont {Meth}}, \bibinfo {author}
  {\bibfnamefont {R.}~\bibnamefont {Stricker}}, \bibinfo {author}
  {\bibfnamefont {E.~A.}\ \bibnamefont {Martinez}}, \bibinfo {author}
  {\bibfnamefont {P.}~\bibnamefont {Schindler}}, \bibinfo {author}
  {\bibfnamefont {T.}~\bibnamefont {Monz}}, \bibinfo {author} {\bibfnamefont
  {J.}~\bibnamefont {Emerson}}, \ and\ \bibinfo {author} {\bibfnamefont
  {R.}~\bibnamefont {Blatt}},\ }\href@noop {} {\bibfield  {journal} {\bibinfo
  {journal} {Nature communications}\ }\textbf {\bibinfo {volume} {10}},\
  \bibinfo {pages} {1} (\bibinfo {year} {2019})}\BibitemShut {NoStop}%
\bibitem [{\citenamefont {Beale}\ \emph {et~al.}(2020)\citenamefont {Beale},
  \citenamefont {Carignan-Dugas}, \citenamefont {Dahlen}, \citenamefont
  {Emerson}, \citenamefont {Hincks}, \citenamefont {Iyer}, \citenamefont
  {Jain}, \citenamefont {Hufnagel}, \citenamefont {Ospadov}, \citenamefont
  {Saunders}, \citenamefont {Stasiuk}, \citenamefont {Wallman},\ and\
  \citenamefont {Winick}}]{trueq}%
  \BibitemOpen
  \bibfield  {author} {\bibinfo {author} {\bibfnamefont {S.~J.}\ \bibnamefont
  {Beale}}, \bibinfo {author} {\bibfnamefont {A.}~\bibnamefont
  {Carignan-Dugas}}, \bibinfo {author} {\bibfnamefont {D.}~\bibnamefont
  {Dahlen}}, \bibinfo {author} {\bibfnamefont {J.}~\bibnamefont {Emerson}},
  \bibinfo {author} {\bibfnamefont {I.}~\bibnamefont {Hincks}}, \bibinfo
  {author} {\bibfnamefont {P.}~\bibnamefont {Iyer}}, \bibinfo {author}
  {\bibfnamefont {A.}~\bibnamefont {Jain}}, \bibinfo {author} {\bibfnamefont
  {D.}~\bibnamefont {Hufnagel}}, \bibinfo {author} {\bibfnamefont
  {E.}~\bibnamefont {Ospadov}}, \bibinfo {author} {\bibfnamefont
  {J.}~\bibnamefont {Saunders}}, \bibinfo {author} {\bibfnamefont
  {A.}~\bibnamefont {Stasiuk}}, \bibinfo {author} {\bibfnamefont {J.~J.}\
  \bibnamefont {Wallman}}, \ and\ \bibinfo {author} {\bibfnamefont
  {A.}~\bibnamefont {Winick}},\ }\href {\doibase 10.5281/zenodo.3945250}
  {\enquote {\bibinfo {title} {True-q},}\ } (\bibinfo {year}
  {2020})\BibitemShut {NoStop}%
\bibitem [{\citenamefont {Mitchell}\ \emph {et~al.}(2021)\citenamefont
  {Mitchell}, \citenamefont {Naik}, \citenamefont {Morvan}, \citenamefont
  {Hashim}, \citenamefont {Kreikebaum}, \citenamefont {Marinelli},
  \citenamefont {Lavrijsen}, \citenamefont {Nowrouzi}, \citenamefont
  {Santiago},\ and\ \citenamefont {Siddiqi}}]{mitchell2021hardware}%
  \BibitemOpen
  \bibfield  {author} {\bibinfo {author} {\bibfnamefont {B.~K.}\ \bibnamefont
  {Mitchell}}, \bibinfo {author} {\bibfnamefont {R.~K.}\ \bibnamefont {Naik}},
  \bibinfo {author} {\bibfnamefont {A.}~\bibnamefont {Morvan}}, \bibinfo
  {author} {\bibfnamefont {A.}~\bibnamefont {Hashim}}, \bibinfo {author}
  {\bibfnamefont {J.~M.}\ \bibnamefont {Kreikebaum}}, \bibinfo {author}
  {\bibfnamefont {B.}~\bibnamefont {Marinelli}}, \bibinfo {author}
  {\bibfnamefont {W.}~\bibnamefont {Lavrijsen}}, \bibinfo {author}
  {\bibfnamefont {K.}~\bibnamefont {Nowrouzi}}, \bibinfo {author}
  {\bibfnamefont {D.~I.}\ \bibnamefont {Santiago}}, \ and\ \bibinfo {author}
  {\bibfnamefont {I.}~\bibnamefont {Siddiqi}},\ }\href@noop {} {\bibfield
  {journal} {\bibinfo  {journal} {Physical review letters}\ }\textbf {\bibinfo
  {volume} {127}},\ \bibinfo {pages} {200502} (\bibinfo {year}
  {2021})}\BibitemShut {NoStop}%
\bibitem [{\citenamefont {Hashim}\ \emph {et~al.}(2022)\citenamefont {Hashim},
  \citenamefont {Rines}, \citenamefont {Omole}, \citenamefont {Naik},
  \citenamefont {Kreikebaum}, \citenamefont {Santiago}, \citenamefont {Chong},
  \citenamefont {Siddiqi},\ and\ \citenamefont
  {Gokhale}}]{hashim2022optimized}%
  \BibitemOpen
  \bibfield  {author} {\bibinfo {author} {\bibfnamefont {A.}~\bibnamefont
  {Hashim}}, \bibinfo {author} {\bibfnamefont {R.}~\bibnamefont {Rines}},
  \bibinfo {author} {\bibfnamefont {V.}~\bibnamefont {Omole}}, \bibinfo
  {author} {\bibfnamefont {R.~K.}\ \bibnamefont {Naik}}, \bibinfo {author}
  {\bibfnamefont {J.~M.}\ \bibnamefont {Kreikebaum}}, \bibinfo {author}
  {\bibfnamefont {D.~I.}\ \bibnamefont {Santiago}}, \bibinfo {author}
  {\bibfnamefont {F.~T.}\ \bibnamefont {Chong}}, \bibinfo {author}
  {\bibfnamefont {I.}~\bibnamefont {Siddiqi}}, \ and\ \bibinfo {author}
  {\bibfnamefont {P.}~\bibnamefont {Gokhale}},\ }\href@noop {} {\bibfield
  {journal} {\bibinfo  {journal} {Physical Review Research}\ }\textbf {\bibinfo
  {volume} {4}},\ \bibinfo {pages} {033028} (\bibinfo {year}
  {2022})}\BibitemShut {NoStop}%
\bibitem [{Note26()}]{Note26}%
  \BibitemOpen
  \bibinfo {note} {By extension, it can also be used to measure the process
  fidelity of an entire sub-circuit, and can therefore be considered a form of
  SPAM-robust fidelity estimation; see Sec.~\ref
  {sec:fidelity_estimation}.}\BibitemShut {Stop}%
\bibitem [{\citenamefont {Krinner}\ \emph {et~al.}(2020)\citenamefont
  {Krinner}, \citenamefont {Lazar}, \citenamefont {Remm}, \citenamefont
  {Andersen}, \citenamefont {Lacroix}, \citenamefont {Norris}, \citenamefont
  {Hellings}, \citenamefont {Gabureac}, \citenamefont {Eichler},\ and\
  \citenamefont {Wallraff}}]{krinner2020benchmarking}%
  \BibitemOpen
  \bibfield  {author} {\bibinfo {author} {\bibfnamefont {S.}~\bibnamefont
  {Krinner}}, \bibinfo {author} {\bibfnamefont {S.}~\bibnamefont {Lazar}},
  \bibinfo {author} {\bibfnamefont {A.}~\bibnamefont {Remm}}, \bibinfo {author}
  {\bibfnamefont {C.}~\bibnamefont {Andersen}}, \bibinfo {author}
  {\bibfnamefont {N.}~\bibnamefont {Lacroix}}, \bibinfo {author} {\bibfnamefont
  {G.}~\bibnamefont {Norris}}, \bibinfo {author} {\bibfnamefont
  {C.}~\bibnamefont {Hellings}}, \bibinfo {author} {\bibfnamefont
  {M.}~\bibnamefont {Gabureac}}, \bibinfo {author} {\bibfnamefont
  {C.}~\bibnamefont {Eichler}}, \ and\ \bibinfo {author} {\bibfnamefont
  {A.}~\bibnamefont {Wallraff}},\ }\href@noop {} {\bibfield  {journal}
  {\bibinfo  {journal} {Physical Review Applied}\ }\textbf {\bibinfo {volume}
  {14}},\ \bibinfo {pages} {024042} (\bibinfo {year} {2020})}\BibitemShut
  {NoStop}%
\bibitem [{\citenamefont {Wallman}\ \emph
  {et~al.}(2015{\natexlab{a}})\citenamefont {Wallman}, \citenamefont {Granade},
  \citenamefont {Harper},\ and\ \citenamefont
  {Flammia}}]{wallman2015estimating}%
  \BibitemOpen
  \bibfield  {author} {\bibinfo {author} {\bibfnamefont {J.}~\bibnamefont
  {Wallman}}, \bibinfo {author} {\bibfnamefont {C.}~\bibnamefont {Granade}},
  \bibinfo {author} {\bibfnamefont {R.}~\bibnamefont {Harper}}, \ and\ \bibinfo
  {author} {\bibfnamefont {S.~T.}\ \bibnamefont {Flammia}},\ }\href@noop {}
  {\bibfield  {journal} {\bibinfo  {journal} {New Journal of Physics}\ }\textbf
  {\bibinfo {volume} {17}},\ \bibinfo {pages} {113020} (\bibinfo {year}
  {2015}{\natexlab{a}})}\BibitemShut {NoStop}%
\bibitem [{\citenamefont {Feng}\ \emph {et~al.}(2016)\citenamefont {Feng},
  \citenamefont {Wallman}, \citenamefont {Buonacorsi}, \citenamefont {Cho},
  \citenamefont {Park}, \citenamefont {Xin}, \citenamefont {Lu}, \citenamefont
  {Baugh},\ and\ \citenamefont {Laflamme}}]{feng2016estimating}%
  \BibitemOpen
  \bibfield  {author} {\bibinfo {author} {\bibfnamefont {G.}~\bibnamefont
  {Feng}}, \bibinfo {author} {\bibfnamefont {J.~J.}\ \bibnamefont {Wallman}},
  \bibinfo {author} {\bibfnamefont {B.}~\bibnamefont {Buonacorsi}}, \bibinfo
  {author} {\bibfnamefont {F.~H.}\ \bibnamefont {Cho}}, \bibinfo {author}
  {\bibfnamefont {D.~K.}\ \bibnamefont {Park}}, \bibinfo {author}
  {\bibfnamefont {T.}~\bibnamefont {Xin}}, \bibinfo {author} {\bibfnamefont
  {D.}~\bibnamefont {Lu}}, \bibinfo {author} {\bibfnamefont {J.}~\bibnamefont
  {Baugh}}, \ and\ \bibinfo {author} {\bibfnamefont {R.}~\bibnamefont
  {Laflamme}},\ }\href@noop {} {\bibfield  {journal} {\bibinfo  {journal}
  {Physical review letters}\ }\textbf {\bibinfo {volume} {117}},\ \bibinfo
  {pages} {260501} (\bibinfo {year} {2016})}\BibitemShut {NoStop}%
\bibitem [{\citenamefont {Zhu}\ \emph {et~al.}(2024)\citenamefont {Zhu},
  \citenamefont {B{\'e}janin}, \citenamefont {Xu},\ and\ \citenamefont
  {Mariantoni}}]{zhu2024purity}%
  \BibitemOpen
  \bibfield  {author} {\bibinfo {author} {\bibfnamefont {A.}~\bibnamefont
  {Zhu}}, \bibinfo {author} {\bibfnamefont {J.~H.}\ \bibnamefont
  {B{\'e}janin}}, \bibinfo {author} {\bibfnamefont {X.}~\bibnamefont {Xu}}, \
  and\ \bibinfo {author} {\bibfnamefont {M.}~\bibnamefont {Mariantoni}},\
  }\href@noop {} {\bibfield  {journal} {\bibinfo  {journal} {arXiv preprint
  arXiv:2407.07960}\ } (\bibinfo {year} {2024})}\BibitemShut {NoStop}%
\bibitem [{\citenamefont {Sheldon}\ \emph {et~al.}(2016)\citenamefont
  {Sheldon}, \citenamefont {Bishop}, \citenamefont {Magesan}, \citenamefont
  {Filipp}, \citenamefont {Chow},\ and\ \citenamefont
  {Gambetta}}]{sheldon2016characterizing}%
  \BibitemOpen
  \bibfield  {author} {\bibinfo {author} {\bibfnamefont {S.}~\bibnamefont
  {Sheldon}}, \bibinfo {author} {\bibfnamefont {L.~S.}\ \bibnamefont {Bishop}},
  \bibinfo {author} {\bibfnamefont {E.}~\bibnamefont {Magesan}}, \bibinfo
  {author} {\bibfnamefont {S.}~\bibnamefont {Filipp}}, \bibinfo {author}
  {\bibfnamefont {J.~M.}\ \bibnamefont {Chow}}, \ and\ \bibinfo {author}
  {\bibfnamefont {J.~M.}\ \bibnamefont {Gambetta}},\ }\href@noop {} {\bibfield
  {journal} {\bibinfo  {journal} {Physical Review A}\ }\textbf {\bibinfo
  {volume} {93}},\ \bibinfo {pages} {012301} (\bibinfo {year}
  {2016})}\BibitemShut {NoStop}%
\bibitem [{\citenamefont {Moskalenko}\ \emph {et~al.}(2022)\citenamefont
  {Moskalenko}, \citenamefont {Simakov}, \citenamefont {Abramov}, \citenamefont
  {Grigorev}, \citenamefont {Moskalev}, \citenamefont {Pishchimova},
  \citenamefont {Smirnov}, \citenamefont {Zikiy}, \citenamefont {Rodionov},\
  and\ \citenamefont {Besedin}}]{2022MoskalenkoFluxonium}%
  \BibitemOpen
  \bibfield  {author} {\bibinfo {author} {\bibfnamefont {I.~N.}\ \bibnamefont
  {Moskalenko}}, \bibinfo {author} {\bibfnamefont {I.~A.}\ \bibnamefont
  {Simakov}}, \bibinfo {author} {\bibfnamefont {N.~N.}\ \bibnamefont
  {Abramov}}, \bibinfo {author} {\bibfnamefont {A.~A.}\ \bibnamefont
  {Grigorev}}, \bibinfo {author} {\bibfnamefont {D.~O.}\ \bibnamefont
  {Moskalev}}, \bibinfo {author} {\bibfnamefont {A.~A.}\ \bibnamefont
  {Pishchimova}}, \bibinfo {author} {\bibfnamefont {N.~S.}\ \bibnamefont
  {Smirnov}}, \bibinfo {author} {\bibfnamefont {E.~V.}\ \bibnamefont {Zikiy}},
  \bibinfo {author} {\bibfnamefont {I.~A.}\ \bibnamefont {Rodionov}}, \ and\
  \bibinfo {author} {\bibfnamefont {I.~S.}\ \bibnamefont {Besedin}},\ }\href
  {\doibase 10.48550/ARXIV.2203.16302} {\enquote {\bibinfo {title} {High
  fidelity two-qubit gates on fluxoniums using a tunable coupler},}\ }
  (\bibinfo {year} {2022})\BibitemShut {NoStop}%
\bibitem [{\citenamefont {Carignan-Dugas}\ \emph {et~al.}(2024)\citenamefont
  {Carignan-Dugas}, \citenamefont {Ranu},\ and\ \citenamefont
  {Dreher}}]{carignan2023estimating}%
  \BibitemOpen
  \bibfield  {author} {\bibinfo {author} {\bibfnamefont {A.}~\bibnamefont
  {Carignan-Dugas}}, \bibinfo {author} {\bibfnamefont {S.~K.}\ \bibnamefont
  {Ranu}}, \ and\ \bibinfo {author} {\bibfnamefont {P.}~\bibnamefont
  {Dreher}},\ }\href {\doibase 10.22331/q-2024-06-13-1367} {\bibfield
  {journal} {\bibinfo  {journal} {{Quantum}}\ }\textbf {\bibinfo {volume}
  {8}},\ \bibinfo {pages} {1367} (\bibinfo {year} {2024})}\BibitemShut
  {NoStop}%
\bibitem [{\citenamefont {Debroy}\ \emph {et~al.}(2023)\citenamefont {Debroy},
  \citenamefont {Genois}, \citenamefont {Gross}, \citenamefont {Mruczkiewicz},
  \citenamefont {Lee}, \citenamefont {Hong}, \citenamefont {Chen},
  \citenamefont {Smelyanskiy},\ and\ \citenamefont
  {Jiang}}]{debroy2023context}%
  \BibitemOpen
  \bibfield  {author} {\bibinfo {author} {\bibfnamefont {D.~M.}\ \bibnamefont
  {Debroy}}, \bibinfo {author} {\bibfnamefont {E.}~\bibnamefont {Genois}},
  \bibinfo {author} {\bibfnamefont {J.~A.}\ \bibnamefont {Gross}}, \bibinfo
  {author} {\bibfnamefont {W.}~\bibnamefont {Mruczkiewicz}}, \bibinfo {author}
  {\bibfnamefont {K.}~\bibnamefont {Lee}}, \bibinfo {author} {\bibfnamefont
  {S.}~\bibnamefont {Hong}}, \bibinfo {author} {\bibfnamefont {Z.}~\bibnamefont
  {Chen}}, \bibinfo {author} {\bibfnamefont {V.}~\bibnamefont {Smelyanskiy}}, \
  and\ \bibinfo {author} {\bibfnamefont {Z.}~\bibnamefont {Jiang}},\
  }\href@noop {} {\bibfield  {journal} {\bibinfo  {journal} {arXiv preprint
  arXiv:2303.17565}\ } (\bibinfo {year} {2023})}\BibitemShut {NoStop}%
\bibitem [{\citenamefont {Wallman}\ \emph
  {et~al.}(2015{\natexlab{b}})\citenamefont {Wallman}, \citenamefont
  {Barnhill},\ and\ \citenamefont {Emerson}}]{Wallman2015-eg}%
  \BibitemOpen
  \bibfield  {author} {\bibinfo {author} {\bibfnamefont {J.~J.}\ \bibnamefont
  {Wallman}}, \bibinfo {author} {\bibfnamefont {M.}~\bibnamefont {Barnhill}}, \
  and\ \bibinfo {author} {\bibfnamefont {J.}~\bibnamefont {Emerson}},\ }\href
  {\doibase 10.1103/PhysRevLett.115.060501} {\bibfield  {journal} {\bibinfo
  {journal} {Phys. Rev. Lett.}\ }\textbf {\bibinfo {volume} {115}},\ \bibinfo
  {pages} {060501} (\bibinfo {year} {2015}{\natexlab{b}})}\BibitemShut
  {NoStop}%
\bibitem [{\citenamefont {Chasseur}\ and\ \citenamefont
  {Wilhelm}(2015)}]{chasseur2015complete}%
  \BibitemOpen
  \bibfield  {author} {\bibinfo {author} {\bibfnamefont {T.}~\bibnamefont
  {Chasseur}}\ and\ \bibinfo {author} {\bibfnamefont {F.~K.}\ \bibnamefont
  {Wilhelm}},\ }\href@noop {} {\bibfield  {journal} {\bibinfo  {journal}
  {Physical Review A}\ }\textbf {\bibinfo {volume} {92}},\ \bibinfo {pages}
  {042333} (\bibinfo {year} {2015})}\BibitemShut {NoStop}%
\bibitem [{\citenamefont {L{\'o}pez}\ \emph {et~al.}(2010)\citenamefont
  {L{\'o}pez}, \citenamefont {Bendersky}, \citenamefont {Paz},\ and\
  \citenamefont {Cory}}]{Lopez2010-jc}%
  \BibitemOpen
  \bibfield  {author} {\bibinfo {author} {\bibfnamefont {C.~C.}\ \bibnamefont
  {L{\'o}pez}}, \bibinfo {author} {\bibfnamefont {A.}~\bibnamefont
  {Bendersky}}, \bibinfo {author} {\bibfnamefont {J.~P.}\ \bibnamefont {Paz}},
  \ and\ \bibinfo {author} {\bibfnamefont {D.~G.}\ \bibnamefont {Cory}},\
  }\href {\doibase 10.1103/PhysRevA.81.062113} {\bibfield  {journal} {\bibinfo
  {journal} {Phys. Rev. A}\ }\textbf {\bibinfo {volume} {81}},\ \bibinfo
  {pages} {062113} (\bibinfo {year} {2010})}\BibitemShut {NoStop}%
\bibitem [{\citenamefont {T{\'o}th}\ \emph {et~al.}(2010)\citenamefont
  {T{\'o}th}, \citenamefont {Wieczorek}, \citenamefont {Gross}, \citenamefont
  {Krischek}, \citenamefont {Schwemmer},\ and\ \citenamefont
  {Weinfurter}}]{Toth2010-xi}%
  \BibitemOpen
  \bibfield  {author} {\bibinfo {author} {\bibfnamefont {G.}~\bibnamefont
  {T{\'o}th}}, \bibinfo {author} {\bibfnamefont {W.}~\bibnamefont {Wieczorek}},
  \bibinfo {author} {\bibfnamefont {D.}~\bibnamefont {Gross}}, \bibinfo
  {author} {\bibfnamefont {R.}~\bibnamefont {Krischek}}, \bibinfo {author}
  {\bibfnamefont {C.}~\bibnamefont {Schwemmer}}, \ and\ \bibinfo {author}
  {\bibfnamefont {H.}~\bibnamefont {Weinfurter}},\ }\href {\doibase
  10.1103/PhysRevLett.105.250403} {\bibfield  {journal} {\bibinfo  {journal}
  {Phys. Rev. Lett.}\ }\textbf {\bibinfo {volume} {105}},\ \bibinfo {pages}
  {250403} (\bibinfo {year} {2010})}\BibitemShut {NoStop}%
\bibitem [{\citenamefont {Bendersky}\ and\ \citenamefont
  {Paz}(2013)}]{Bendersky2013-rv}%
  \BibitemOpen
  \bibfield  {author} {\bibinfo {author} {\bibfnamefont {A.}~\bibnamefont
  {Bendersky}}\ and\ \bibinfo {author} {\bibfnamefont {J.~P.}\ \bibnamefont
  {Paz}},\ }\href {\doibase 10.1103/PhysRevA.87.012122} {\bibfield  {journal}
  {\bibinfo  {journal} {Phys. Rev. A}\ }\textbf {\bibinfo {volume} {87}},\
  \bibinfo {pages} {012122} (\bibinfo {year} {2013})}\BibitemShut {NoStop}%
\bibitem [{\citenamefont {Greganti}\ \emph {et~al.}(2015)\citenamefont
  {Greganti}, \citenamefont {Roehsner}, \citenamefont {Barz}, \citenamefont
  {Waegell},\ and\ \citenamefont {Walther}}]{Greganti2015-mg}%
  \BibitemOpen
  \bibfield  {author} {\bibinfo {author} {\bibfnamefont {C.}~\bibnamefont
  {Greganti}}, \bibinfo {author} {\bibfnamefont {M.-C.}\ \bibnamefont
  {Roehsner}}, \bibinfo {author} {\bibfnamefont {S.}~\bibnamefont {Barz}},
  \bibinfo {author} {\bibfnamefont {M.}~\bibnamefont {Waegell}}, \ and\
  \bibinfo {author} {\bibfnamefont {P.}~\bibnamefont {Walther}},\ }\href
  {\doibase 10.1103/PhysRevA.91.022325} {\bibfield  {journal} {\bibinfo
  {journal} {Phys. Rev. A}\ }\textbf {\bibinfo {volume} {91}},\ \bibinfo
  {pages} {022325} (\bibinfo {year} {2015})}\BibitemShut {NoStop}%
\bibitem [{\citenamefont {Steffens}\ \emph {et~al.}(2017)\citenamefont
  {Steffens}, \citenamefont {Rebentrost}, \citenamefont {Marvian},
  \citenamefont {Eisert},\ and\ \citenamefont {Lloyd}}]{Steffens2017-pm}%
  \BibitemOpen
  \bibfield  {author} {\bibinfo {author} {\bibfnamefont {A.}~\bibnamefont
  {Steffens}}, \bibinfo {author} {\bibfnamefont {P.}~\bibnamefont
  {Rebentrost}}, \bibinfo {author} {\bibfnamefont {I.}~\bibnamefont {Marvian}},
  \bibinfo {author} {\bibfnamefont {J.}~\bibnamefont {Eisert}}, \ and\ \bibinfo
  {author} {\bibfnamefont {S.}~\bibnamefont {Lloyd}},\ }\href {\doibase
  10.1088/1367-2630/aa5e48} {\bibfield  {journal} {\bibinfo  {journal} {New J.
  Phys.}\ }\textbf {\bibinfo {volume} {19}},\ \bibinfo {pages} {033005}
  (\bibinfo {year} {2017})}\BibitemShut {NoStop}%
\bibitem [{\citenamefont {Carmeli}\ \emph {et~al.}(2017)\citenamefont
  {Carmeli}, \citenamefont {Heinosaari}, \citenamefont {Schultz},\ and\
  \citenamefont {Toigo}}]{Carmeli2017-fh}%
  \BibitemOpen
  \bibfield  {author} {\bibinfo {author} {\bibfnamefont {C.}~\bibnamefont
  {Carmeli}}, \bibinfo {author} {\bibfnamefont {T.}~\bibnamefont {Heinosaari}},
  \bibinfo {author} {\bibfnamefont {J.}~\bibnamefont {Schultz}}, \ and\
  \bibinfo {author} {\bibfnamefont {A.}~\bibnamefont {Toigo}},\ }\href
  {\doibase 10.1098/rspa.2016.0866} {\bibfield  {journal} {\bibinfo  {journal}
  {Proc. Math. Phys. Eng. Sci.}\ }\textbf {\bibinfo {volume} {473}},\ \bibinfo
  {pages} {20160866} (\bibinfo {year} {2017})}\BibitemShut {NoStop}%
\bibitem [{\citenamefont {Helsen}\ \emph
  {et~al.}(2019{\natexlab{b}})\citenamefont {Helsen}, \citenamefont
  {Battistel},\ and\ \citenamefont {Terhal}}]{Helsen2019-fi}%
  \BibitemOpen
  \bibfield  {author} {\bibinfo {author} {\bibfnamefont {J.}~\bibnamefont
  {Helsen}}, \bibinfo {author} {\bibfnamefont {F.}~\bibnamefont {Battistel}}, \
  and\ \bibinfo {author} {\bibfnamefont {B.~M.}\ \bibnamefont {Terhal}},\
  }\href {\doibase 10.1038/s41534-019-0189-0} {\bibfield  {journal} {\bibinfo
  {journal} {npj Quantum Information}\ }\textbf {\bibinfo {volume} {5}},\
  \bibinfo {pages} {74} (\bibinfo {year} {2019}{\natexlab{b}})}\BibitemShut
  {NoStop}%
\bibitem [{\citenamefont {Helsen}\ \emph {et~al.}(2023)\citenamefont {Helsen},
  \citenamefont {Ioannou}, \citenamefont {Kitzinger}, \citenamefont {Onorati},
  \citenamefont {Werner}, \citenamefont {Eisert},\ and\ \citenamefont
  {Roth}}]{Helsen2021-zf}%
  \BibitemOpen
  \bibfield  {author} {\bibinfo {author} {\bibfnamefont {J.}~\bibnamefont
  {Helsen}}, \bibinfo {author} {\bibfnamefont {M.}~\bibnamefont {Ioannou}},
  \bibinfo {author} {\bibfnamefont {J.}~\bibnamefont {Kitzinger}}, \bibinfo
  {author} {\bibfnamefont {E.}~\bibnamefont {Onorati}}, \bibinfo {author}
  {\bibfnamefont {A.}~\bibnamefont {Werner}}, \bibinfo {author} {\bibfnamefont
  {J.}~\bibnamefont {Eisert}}, \ and\ \bibinfo {author} {\bibfnamefont
  {I.}~\bibnamefont {Roth}},\ }\href@noop {} {\bibfield  {journal} {\bibinfo
  {journal} {Nature Communications}\ }\textbf {\bibinfo {volume} {14}},\
  \bibinfo {pages} {5039} (\bibinfo {year} {2023})}\BibitemShut {NoStop}%
\bibitem [{\citenamefont {Elben}\ \emph {et~al.}(2023)\citenamefont {Elben},
  \citenamefont {Flammia}, \citenamefont {Huang}, \citenamefont {Kueng},
  \citenamefont {Preskill}, \citenamefont {Vermersch},\ and\ \citenamefont
  {Zoller}}]{elben2023randomized}%
  \BibitemOpen
  \bibfield  {author} {\bibinfo {author} {\bibfnamefont {A.}~\bibnamefont
  {Elben}}, \bibinfo {author} {\bibfnamefont {S.~T.}\ \bibnamefont {Flammia}},
  \bibinfo {author} {\bibfnamefont {H.-Y.}\ \bibnamefont {Huang}}, \bibinfo
  {author} {\bibfnamefont {R.}~\bibnamefont {Kueng}}, \bibinfo {author}
  {\bibfnamefont {J.}~\bibnamefont {Preskill}}, \bibinfo {author}
  {\bibfnamefont {B.}~\bibnamefont {Vermersch}}, \ and\ \bibinfo {author}
  {\bibfnamefont {P.}~\bibnamefont {Zoller}},\ }\href@noop {} {\bibfield
  {journal} {\bibinfo  {journal} {Nature Reviews Physics}\ }\textbf {\bibinfo
  {volume} {5}},\ \bibinfo {pages} {9} (\bibinfo {year} {2023})}\BibitemShut
  {NoStop}%
\bibitem [{\citenamefont {Aaronson}(2018)}]{aaronson2018shadow}%
  \BibitemOpen
  \bibfield  {author} {\bibinfo {author} {\bibfnamefont {S.}~\bibnamefont
  {Aaronson}},\ }in\ \href@noop {} {\emph {\bibinfo {booktitle} {Proceedings of
  the 50th annual ACM SIGACT symposium on theory of computing}}}\ (\bibinfo
  {year} {2018})\ pp.\ \bibinfo {pages} {325--338}\BibitemShut {NoStop}%
\bibitem [{\citenamefont {Kunjummen}\ \emph {et~al.}(2023)\citenamefont
  {Kunjummen}, \citenamefont {Tran}, \citenamefont {Carney},\ and\
  \citenamefont {Taylor}}]{kunjummen2023shadow}%
  \BibitemOpen
  \bibfield  {author} {\bibinfo {author} {\bibfnamefont {J.}~\bibnamefont
  {Kunjummen}}, \bibinfo {author} {\bibfnamefont {M.~C.}\ \bibnamefont {Tran}},
  \bibinfo {author} {\bibfnamefont {D.}~\bibnamefont {Carney}}, \ and\ \bibinfo
  {author} {\bibfnamefont {J.~M.}\ \bibnamefont {Taylor}},\ }\href@noop {}
  {\bibfield  {journal} {\bibinfo  {journal} {Physical Review A}\ }\textbf
  {\bibinfo {volume} {107}},\ \bibinfo {pages} {042403} (\bibinfo {year}
  {2023})}\BibitemShut {NoStop}%
\bibitem [{\citenamefont {Flammia}\ and\ \citenamefont
  {Liu}(2011)}]{flammia2011direct}%
  \BibitemOpen
  \bibfield  {author} {\bibinfo {author} {\bibfnamefont {S.~T.}\ \bibnamefont
  {Flammia}}\ and\ \bibinfo {author} {\bibfnamefont {Y.-K.}\ \bibnamefont
  {Liu}},\ }\href@noop {} {\bibfield  {journal} {\bibinfo  {journal} {Physical
  Review Letters}\ }\textbf {\bibinfo {volume} {106}} (\bibinfo {year}
  {2011})}\BibitemShut {NoStop}%
\bibitem [{\citenamefont {da~Silva}\ \emph {et~al.}(2011)\citenamefont
  {da~Silva}, \citenamefont {Landon-Cardinal},\ and\ \citenamefont
  {Poulin}}]{dasilva2011practical}%
  \BibitemOpen
  \bibfield  {author} {\bibinfo {author} {\bibfnamefont {M.~P.}\ \bibnamefont
  {da~Silva}}, \bibinfo {author} {\bibfnamefont {O.}~\bibnamefont
  {Landon-Cardinal}}, \ and\ \bibinfo {author} {\bibfnamefont {D.}~\bibnamefont
  {Poulin}},\ }\href@noop {} {\bibfield  {journal} {\bibinfo  {journal}
  {Physical Review Letters}\ }\textbf {\bibinfo {volume} {107}} (\bibinfo
  {year} {2011})}\BibitemShut {NoStop}%
\bibitem [{\citenamefont {Gutoski}\ and\ \citenamefont
  {Johnston}(2014)}]{Gutoski2014-wo}%
  \BibitemOpen
  \bibfield  {author} {\bibinfo {author} {\bibfnamefont {G.}~\bibnamefont
  {Gutoski}}\ and\ \bibinfo {author} {\bibfnamefont {N.}~\bibnamefont
  {Johnston}},\ }\href {\doibase 10.1063/1.4867625} {\bibfield  {journal}
  {\bibinfo  {journal} {J. Math. Phys.}\ }\textbf {\bibinfo {volume} {55}},\
  \bibinfo {pages} {032201} (\bibinfo {year} {2014})}\BibitemShut {NoStop}%
\bibitem [{\citenamefont {Ma}\ \emph {et~al.}(2016)\citenamefont {Ma},
  \citenamefont {Jackson}, \citenamefont {Zhou}, \citenamefont {Chen},
  \citenamefont {Lu}, \citenamefont {Mazurek}, \citenamefont {Fisher},
  \citenamefont {Peng}, \citenamefont {Kribs}, \citenamefont {Resch},\ and\
  \citenamefont {{Others}}}]{Ma2016-if}%
  \BibitemOpen
  \bibfield  {author} {\bibinfo {author} {\bibfnamefont {X.}~\bibnamefont
  {Ma}}, \bibinfo {author} {\bibfnamefont {T.}~\bibnamefont {Jackson}},
  \bibinfo {author} {\bibfnamefont {H.}~\bibnamefont {Zhou}}, \bibinfo {author}
  {\bibfnamefont {J.}~\bibnamefont {Chen}}, \bibinfo {author} {\bibfnamefont
  {D.}~\bibnamefont {Lu}}, \bibinfo {author} {\bibfnamefont {M.~D.}\
  \bibnamefont {Mazurek}}, \bibinfo {author} {\bibfnamefont {K.~A.~G.}\
  \bibnamefont {Fisher}}, \bibinfo {author} {\bibfnamefont {X.}~\bibnamefont
  {Peng}}, \bibinfo {author} {\bibfnamefont {D.}~\bibnamefont {Kribs}},
  \bibinfo {author} {\bibfnamefont {K.~J.}\ \bibnamefont {Resch}}, \ and\
  \bibinfo {author} {\bibnamefont {{Others}}},\ }\href@noop {} {\bibfield
  {journal} {\bibinfo  {journal} {Phys. Rev. A}\ }\textbf {\bibinfo {volume}
  {93}},\ \bibinfo {pages} {032140} (\bibinfo {year} {2016})}\BibitemShut
  {NoStop}%
\bibitem [{\citenamefont {Moroder}\ \emph {et~al.}(2012)\citenamefont
  {Moroder}, \citenamefont {Hyllus}, \citenamefont {T{\'o}th}, \citenamefont
  {Schwemmer}, \citenamefont {Niggebaum}, \citenamefont {Gaile}, \citenamefont
  {G{\"u}hne},\ and\ \citenamefont {Weinfurter}}]{moroder2012permutationally}%
  \BibitemOpen
  \bibfield  {author} {\bibinfo {author} {\bibfnamefont {T.}~\bibnamefont
  {Moroder}}, \bibinfo {author} {\bibfnamefont {P.}~\bibnamefont {Hyllus}},
  \bibinfo {author} {\bibfnamefont {G.}~\bibnamefont {T{\'o}th}}, \bibinfo
  {author} {\bibfnamefont {C.}~\bibnamefont {Schwemmer}}, \bibinfo {author}
  {\bibfnamefont {A.}~\bibnamefont {Niggebaum}}, \bibinfo {author}
  {\bibfnamefont {S.}~\bibnamefont {Gaile}}, \bibinfo {author} {\bibfnamefont
  {O.}~\bibnamefont {G{\"u}hne}}, \ and\ \bibinfo {author} {\bibfnamefont
  {H.}~\bibnamefont {Weinfurter}},\ }\href@noop {} {\bibfield  {journal}
  {\bibinfo  {journal} {New Journal of Physics}\ }\textbf {\bibinfo {volume}
  {14}},\ \bibinfo {pages} {105001} (\bibinfo {year} {2012})}\BibitemShut
  {NoStop}%
\bibitem [{\citenamefont {Schwemmer}\ \emph {et~al.}(2014)\citenamefont
  {Schwemmer}, \citenamefont {T{\'o}th}, \citenamefont {Niggebaum},
  \citenamefont {Moroder}, \citenamefont {Gross}, \citenamefont {G{\"u}hne},\
  and\ \citenamefont {Weinfurter}}]{Schwemmer2014-qz}%
  \BibitemOpen
  \bibfield  {author} {\bibinfo {author} {\bibfnamefont {C.}~\bibnamefont
  {Schwemmer}}, \bibinfo {author} {\bibfnamefont {G.}~\bibnamefont {T{\'o}th}},
  \bibinfo {author} {\bibfnamefont {A.}~\bibnamefont {Niggebaum}}, \bibinfo
  {author} {\bibfnamefont {T.}~\bibnamefont {Moroder}}, \bibinfo {author}
  {\bibfnamefont {D.}~\bibnamefont {Gross}}, \bibinfo {author} {\bibfnamefont
  {O.}~\bibnamefont {G{\"u}hne}}, \ and\ \bibinfo {author} {\bibfnamefont
  {H.}~\bibnamefont {Weinfurter}},\ }\href {\doibase
  10.1103/PhysRevLett.113.040503} {\bibfield  {journal} {\bibinfo  {journal}
  {Phys. Rev. Lett.}\ }\textbf {\bibinfo {volume} {113}},\ \bibinfo {pages}
  {040503} (\bibinfo {year} {2014})}\BibitemShut {NoStop}%
\bibitem [{\citenamefont {Gu{\c t}{\u a}}\ \emph {et~al.}(2012)\citenamefont
  {Gu{\c t}{\u a}}, \citenamefont {Kypraios},\ and\ \citenamefont
  {Dryden}}]{Guta2012-nx}%
  \BibitemOpen
  \bibfield  {author} {\bibinfo {author} {\bibfnamefont {M.}~\bibnamefont
  {Gu{\c t}{\u a}}}, \bibinfo {author} {\bibfnamefont {T.}~\bibnamefont
  {Kypraios}}, \ and\ \bibinfo {author} {\bibfnamefont {I.}~\bibnamefont
  {Dryden}},\ }\href@noop {} {\bibfield  {journal} {\bibinfo  {journal} {New J.
  Phys.}\ } (\bibinfo {year} {2012})}\BibitemShut {NoStop}%
\bibitem [{\citenamefont {Gross}\ \emph
  {et~al.}(2010{\natexlab{b}})\citenamefont {Gross}, \citenamefont {Liu},
  \citenamefont {Flammia}, \citenamefont {Becker},\ and\ \citenamefont
  {Eisert}}]{Gross2010-ck}%
  \BibitemOpen
  \bibfield  {author} {\bibinfo {author} {\bibfnamefont {D.}~\bibnamefont
  {Gross}}, \bibinfo {author} {\bibfnamefont {Y.-K.}\ \bibnamefont {Liu}},
  \bibinfo {author} {\bibfnamefont {S.~T.}\ \bibnamefont {Flammia}}, \bibinfo
  {author} {\bibfnamefont {S.}~\bibnamefont {Becker}}, \ and\ \bibinfo {author}
  {\bibfnamefont {J.}~\bibnamefont {Eisert}},\ }\href {\doibase
  10.1103/PhysRevLett.105.150401} {\bibfield  {journal} {\bibinfo  {journal}
  {Phys. Rev. Lett.}\ }\textbf {\bibinfo {volume} {105}},\ \bibinfo {pages}
  {150401} (\bibinfo {year} {2010}{\natexlab{b}})}\BibitemShut {NoStop}%
\bibitem [{\citenamefont {Riofr{\'\i}o}\ \emph
  {et~al.}(2017{\natexlab{b}})\citenamefont {Riofr{\'\i}o}, \citenamefont
  {Gross}, \citenamefont {Flammia}, \citenamefont {Monz}, \citenamefont {Nigg},
  \citenamefont {Blatt},\ and\ \citenamefont {Eisert}}]{Riofrio2017-wk}%
  \BibitemOpen
  \bibfield  {author} {\bibinfo {author} {\bibfnamefont {C.~A.}\ \bibnamefont
  {Riofr{\'\i}o}}, \bibinfo {author} {\bibfnamefont {D.}~\bibnamefont {Gross}},
  \bibinfo {author} {\bibfnamefont {S.~T.}\ \bibnamefont {Flammia}}, \bibinfo
  {author} {\bibfnamefont {T.}~\bibnamefont {Monz}}, \bibinfo {author}
  {\bibfnamefont {D.}~\bibnamefont {Nigg}}, \bibinfo {author} {\bibfnamefont
  {R.}~\bibnamefont {Blatt}}, \ and\ \bibinfo {author} {\bibfnamefont
  {J.}~\bibnamefont {Eisert}},\ }\href {\doibase 10.1038/ncomms15305}
  {\bibfield  {journal} {\bibinfo  {journal} {Nat. Commun.}\ }\textbf {\bibinfo
  {volume} {8}},\ \bibinfo {pages} {15305} (\bibinfo {year}
  {2017}{\natexlab{b}})}\BibitemShut {NoStop}%
\bibitem [{\citenamefont {Landon-Cardinal}\ and\ \citenamefont
  {Poulin}(2012)}]{Landon-Cardinal2012-nh}%
  \BibitemOpen
  \bibfield  {author} {\bibinfo {author} {\bibfnamefont {O.}~\bibnamefont
  {Landon-Cardinal}}\ and\ \bibinfo {author} {\bibfnamefont {D.}~\bibnamefont
  {Poulin}},\ }\href {\doibase 10.1088/1367-2630/14/8/085004} {\bibfield
  {journal} {\bibinfo  {journal} {New J. Phys.}\ }\textbf {\bibinfo {volume}
  {14}},\ \bibinfo {pages} {085004} (\bibinfo {year} {2012})}\BibitemShut
  {NoStop}%
\bibitem [{\citenamefont {Baumgratz}\ \emph {et~al.}(2013)\citenamefont
  {Baumgratz}, \citenamefont {Gross}, \citenamefont {Cramer},\ and\
  \citenamefont {Plenio}}]{Baumgratz2013-js}%
  \BibitemOpen
  \bibfield  {author} {\bibinfo {author} {\bibfnamefont {T.}~\bibnamefont
  {Baumgratz}}, \bibinfo {author} {\bibfnamefont {D.}~\bibnamefont {Gross}},
  \bibinfo {author} {\bibfnamefont {M.}~\bibnamefont {Cramer}}, \ and\ \bibinfo
  {author} {\bibfnamefont {M.~B.}\ \bibnamefont {Plenio}},\ }\href {\doibase
  10.1103/PhysRevLett.111.020401} {\bibfield  {journal} {\bibinfo  {journal}
  {Phys. Rev. Lett.}\ }\textbf {\bibinfo {volume} {111}},\ \bibinfo {pages}
  {020401} (\bibinfo {year} {2013})}\BibitemShut {NoStop}%
\bibitem [{\citenamefont {Cramer}\ \emph {et~al.}(2010)\citenamefont {Cramer},
  \citenamefont {Plenio}, \citenamefont {Flammia}, \citenamefont {Somma},
  \citenamefont {Gross}, \citenamefont {Bartlett}, \citenamefont
  {Landon-Cardinal}, \citenamefont {Poulin},\ and\ \citenamefont
  {Liu}}]{Cramer2010-vo}%
  \BibitemOpen
  \bibfield  {author} {\bibinfo {author} {\bibfnamefont {M.}~\bibnamefont
  {Cramer}}, \bibinfo {author} {\bibfnamefont {M.~B.}\ \bibnamefont {Plenio}},
  \bibinfo {author} {\bibfnamefont {S.~T.}\ \bibnamefont {Flammia}}, \bibinfo
  {author} {\bibfnamefont {R.}~\bibnamefont {Somma}}, \bibinfo {author}
  {\bibfnamefont {D.}~\bibnamefont {Gross}}, \bibinfo {author} {\bibfnamefont
  {S.~D.}\ \bibnamefont {Bartlett}}, \bibinfo {author} {\bibfnamefont
  {O.}~\bibnamefont {Landon-Cardinal}}, \bibinfo {author} {\bibfnamefont
  {D.}~\bibnamefont {Poulin}}, \ and\ \bibinfo {author} {\bibfnamefont {Y.-K.}\
  \bibnamefont {Liu}},\ }\href {\doibase 10.1038/ncomms1147} {\bibfield
  {journal} {\bibinfo  {journal} {Nat. Commun.}\ }\textbf {\bibinfo {volume}
  {1}},\ \bibinfo {pages} {149} (\bibinfo {year} {2010})}\BibitemShut {NoStop}%
\bibitem [{\citenamefont {Flammia}(2022)}]{flammia2021averaged}%
  \BibitemOpen
  \bibfield  {author} {\bibinfo {author} {\bibfnamefont {S.~T.}\ \bibnamefont
  {Flammia}},\ }in\ \href {\doibase 10.4230/LIPIcs.TQC.2022.4} {\emph {\bibinfo
  {booktitle} {17th Conference on the Theory of Quantum Computation,
  Communication and Cryptography (TQC 2022)}}},\ \bibinfo {series} {Leibniz
  International Proceedings in Informatics (LIPIcs)}, Vol.\ \bibinfo {volume}
  {232},\ \bibinfo {editor} {edited by\ \bibinfo {editor} {\bibfnamefont
  {F.}~\bibnamefont {Le~Gall}}\ and\ \bibinfo {editor} {\bibfnamefont
  {T.}~\bibnamefont {Morimae}}}\ (\bibinfo  {publisher} {Schloss Dagstuhl --
  Leibniz-Zentrum f{\"u}r Informatik},\ \bibinfo {address} {Dagstuhl,
  Germany},\ \bibinfo {year} {2022})\ pp.\ \bibinfo {pages}
  {4:1--4:10}\BibitemShut {NoStop}%
\bibitem [{\citenamefont {Van Den~Berg}\ \emph {et~al.}(2023)\citenamefont {Van
  Den~Berg}, \citenamefont {Minev}, \citenamefont {Kandala},\ and\
  \citenamefont {Temme}}]{van2023probabilistic}%
  \BibitemOpen
  \bibfield  {author} {\bibinfo {author} {\bibfnamefont {E.}~\bibnamefont {Van
  Den~Berg}}, \bibinfo {author} {\bibfnamefont {Z.~K.}\ \bibnamefont {Minev}},
  \bibinfo {author} {\bibfnamefont {A.}~\bibnamefont {Kandala}}, \ and\
  \bibinfo {author} {\bibfnamefont {K.}~\bibnamefont {Temme}},\ }\href@noop {}
  {\bibfield  {journal} {\bibinfo  {journal} {Nature Physics}\ } (\bibinfo
  {year} {2023})}\BibitemShut {NoStop}%
\bibitem [{\citenamefont {van~den Berg}\ and\ \citenamefont
  {Wocjan}(2024)}]{van2024techniques}%
  \BibitemOpen
  \bibfield  {author} {\bibinfo {author} {\bibfnamefont {E.}~\bibnamefont
  {van~den Berg}}\ and\ \bibinfo {author} {\bibfnamefont {P.}~\bibnamefont
  {Wocjan}},\ }\href@noop {} {\bibfield  {journal} {\bibinfo  {journal}
  {Quantum}\ }\textbf {\bibinfo {volume} {8}},\ \bibinfo {pages} {1556}
  (\bibinfo {year} {2024})}\BibitemShut {NoStop}%
\bibitem [{\citenamefont {Kieferov{\'a}}\ and\ \citenamefont
  {Wiebe}(2017)}]{Kieferova2017-oi}%
  \BibitemOpen
  \bibfield  {author} {\bibinfo {author} {\bibfnamefont {M.}~\bibnamefont
  {Kieferov{\'a}}}\ and\ \bibinfo {author} {\bibfnamefont {N.}~\bibnamefont
  {Wiebe}},\ }\href {\doibase 10.1103/PhysRevA.96.062327} {\bibfield  {journal}
  {\bibinfo  {journal} {Phys. Rev. A}\ }\textbf {\bibinfo {volume} {96}},\
  \bibinfo {pages} {062327} (\bibinfo {year} {2017})}\BibitemShut {NoStop}%
\bibitem [{\citenamefont {Torlai}\ \emph {et~al.}(2018)\citenamefont {Torlai},
  \citenamefont {Mazzola}, \citenamefont {Carrasquilla}, \citenamefont
  {Troyer}, \citenamefont {Melko},\ and\ \citenamefont
  {Carleo}}]{Torlai2018-wt}%
  \BibitemOpen
  \bibfield  {author} {\bibinfo {author} {\bibfnamefont {G.}~\bibnamefont
  {Torlai}}, \bibinfo {author} {\bibfnamefont {G.}~\bibnamefont {Mazzola}},
  \bibinfo {author} {\bibfnamefont {J.}~\bibnamefont {Carrasquilla}}, \bibinfo
  {author} {\bibfnamefont {M.}~\bibnamefont {Troyer}}, \bibinfo {author}
  {\bibfnamefont {R.}~\bibnamefont {Melko}}, \ and\ \bibinfo {author}
  {\bibfnamefont {G.}~\bibnamefont {Carleo}},\ }\href {\doibase
  10.1038/s41567-018-0048-5} {\bibfield  {journal} {\bibinfo  {journal} {Nat.
  Phys.}\ }\textbf {\bibinfo {volume} {14}},\ \bibinfo {pages} {447} (\bibinfo
  {year} {2018})}\BibitemShut {NoStop}%
\bibitem [{\citenamefont {Gao}\ \emph {et~al.}(2018)\citenamefont {Gao},
  \citenamefont {Qiao}, \citenamefont {Jiao}, \citenamefont {Ma}, \citenamefont
  {Hu}, \citenamefont {Ren}, \citenamefont {Yang}, \citenamefont {Tang},
  \citenamefont {Yung},\ and\ \citenamefont {Jin}}]{Gao2018-xc}%
  \BibitemOpen
  \bibfield  {author} {\bibinfo {author} {\bibfnamefont {J.}~\bibnamefont
  {Gao}}, \bibinfo {author} {\bibfnamefont {L.-F.}\ \bibnamefont {Qiao}},
  \bibinfo {author} {\bibfnamefont {Z.-Q.}\ \bibnamefont {Jiao}}, \bibinfo
  {author} {\bibfnamefont {Y.-C.}\ \bibnamefont {Ma}}, \bibinfo {author}
  {\bibfnamefont {C.-Q.}\ \bibnamefont {Hu}}, \bibinfo {author} {\bibfnamefont
  {R.-J.}\ \bibnamefont {Ren}}, \bibinfo {author} {\bibfnamefont {A.-L.}\
  \bibnamefont {Yang}}, \bibinfo {author} {\bibfnamefont {H.}~\bibnamefont
  {Tang}}, \bibinfo {author} {\bibfnamefont {M.-H.}\ \bibnamefont {Yung}}, \
  and\ \bibinfo {author} {\bibfnamefont {X.-M.}\ \bibnamefont {Jin}},\ }\href
  {\doibase 10.1103/PhysRevLett.120.240501} {\bibfield  {journal} {\bibinfo
  {journal} {Phys. Rev. Lett.}\ }\textbf {\bibinfo {volume} {120}},\ \bibinfo
  {pages} {240501} (\bibinfo {year} {2018})}\BibitemShut {NoStop}%
\bibitem [{\citenamefont {Carrasquilla}\ \emph {et~al.}(2019)\citenamefont
  {Carrasquilla}, \citenamefont {Torlai}, \citenamefont {Melko},\ and\
  \citenamefont {Aolita}}]{Carrasquilla2019-wh}%
  \BibitemOpen
  \bibfield  {author} {\bibinfo {author} {\bibfnamefont {J.}~\bibnamefont
  {Carrasquilla}}, \bibinfo {author} {\bibfnamefont {G.}~\bibnamefont
  {Torlai}}, \bibinfo {author} {\bibfnamefont {R.~G.}\ \bibnamefont {Melko}}, \
  and\ \bibinfo {author} {\bibfnamefont {L.}~\bibnamefont {Aolita}},\ }\href
  {\doibase 10.1038/s42256-019-0028-1} {\bibfield  {journal} {\bibinfo
  {journal} {Nature Machine Intelligence}\ }\textbf {\bibinfo {volume} {1}},\
  \bibinfo {pages} {155} (\bibinfo {year} {2019})}\BibitemShut {NoStop}%
\bibitem [{\citenamefont {Gebhart}\ \emph {et~al.}(2023)\citenamefont
  {Gebhart}, \citenamefont {Santagati}, \citenamefont {Gentile}, \citenamefont
  {Gauger}, \citenamefont {Craig}, \citenamefont {Ares}, \citenamefont
  {Banchi}, \citenamefont {Marquardt}, \citenamefont {Pezz{\`e}},\ and\
  \citenamefont {Bonato}}]{Gebhart2023-dr}%
  \BibitemOpen
  \bibfield  {author} {\bibinfo {author} {\bibfnamefont {V.}~\bibnamefont
  {Gebhart}}, \bibinfo {author} {\bibfnamefont {R.}~\bibnamefont {Santagati}},
  \bibinfo {author} {\bibfnamefont {A.~A.}\ \bibnamefont {Gentile}}, \bibinfo
  {author} {\bibfnamefont {E.~M.}\ \bibnamefont {Gauger}}, \bibinfo {author}
  {\bibfnamefont {D.}~\bibnamefont {Craig}}, \bibinfo {author} {\bibfnamefont
  {N.}~\bibnamefont {Ares}}, \bibinfo {author} {\bibfnamefont {L.}~\bibnamefont
  {Banchi}}, \bibinfo {author} {\bibfnamefont {F.}~\bibnamefont {Marquardt}},
  \bibinfo {author} {\bibfnamefont {L.}~\bibnamefont {Pezz{\`e}}}, \ and\
  \bibinfo {author} {\bibfnamefont {C.}~\bibnamefont {Bonato}},\ }\href
  {\doibase 10.1038/s42254-022-00552-1} {\bibfield  {journal} {\bibinfo
  {journal} {Nature Reviews Physics}\ }\textbf {\bibinfo {volume} {5}},\
  \bibinfo {pages} {141} (\bibinfo {year} {2023})}\BibitemShut {NoStop}%
\bibitem [{\citenamefont {Fedorov}\ \emph {et~al.}(2011)\citenamefont
  {Fedorov}, \citenamefont {Steffen}, \citenamefont {Baur}, \citenamefont
  {da~Silva},\ and\ \citenamefont {Wallraff}}]{fedorov2011implementation}%
  \BibitemOpen
  \bibfield  {author} {\bibinfo {author} {\bibfnamefont {A.}~\bibnamefont
  {Fedorov}}, \bibinfo {author} {\bibfnamefont {L.}~\bibnamefont {Steffen}},
  \bibinfo {author} {\bibfnamefont {M.}~\bibnamefont {Baur}}, \bibinfo {author}
  {\bibfnamefont {M.~P.}\ \bibnamefont {da~Silva}}, \ and\ \bibinfo {author}
  {\bibfnamefont {A.}~\bibnamefont {Wallraff}},\ }\href {\doibase
  10.1038/nature10713} {\bibfield  {journal} {\bibinfo  {journal} {Nature}\
  }\textbf {\bibinfo {volume} {481}},\ \bibinfo {pages} {170} (\bibinfo {year}
  {2011})}\BibitemShut {NoStop}%
\bibitem [{\citenamefont {Chu}\ \emph {et~al.}(2023)\citenamefont {Chu},
  \citenamefont {He}, \citenamefont {Zhou}, \citenamefont {Yuan}, \citenamefont
  {Zhang}, \citenamefont {Guo}, \citenamefont {Hai}, \citenamefont {Han},
  \citenamefont {Hu}, \citenamefont {Huang} \emph {et~al.}}]{chu2023scalable}%
  \BibitemOpen
  \bibfield  {author} {\bibinfo {author} {\bibfnamefont {J.}~\bibnamefont
  {Chu}}, \bibinfo {author} {\bibfnamefont {X.}~\bibnamefont {He}}, \bibinfo
  {author} {\bibfnamefont {Y.}~\bibnamefont {Zhou}}, \bibinfo {author}
  {\bibfnamefont {J.}~\bibnamefont {Yuan}}, \bibinfo {author} {\bibfnamefont
  {L.}~\bibnamefont {Zhang}}, \bibinfo {author} {\bibfnamefont
  {Q.}~\bibnamefont {Guo}}, \bibinfo {author} {\bibfnamefont {Y.}~\bibnamefont
  {Hai}}, \bibinfo {author} {\bibfnamefont {Z.}~\bibnamefont {Han}}, \bibinfo
  {author} {\bibfnamefont {C.-K.}\ \bibnamefont {Hu}}, \bibinfo {author}
  {\bibfnamefont {W.}~\bibnamefont {Huang}},  \emph {et~al.},\ }\href {\doibase
  10.1038/s41567-022-01813-7} {\bibfield  {journal} {\bibinfo  {journal}
  {Nature Physics}\ }\textbf {\bibinfo {volume} {19}},\ \bibinfo {pages} {126}
  (\bibinfo {year} {2023})}\BibitemShut {NoStop}%
\bibitem [{Note27()}]{Note27}%
  \BibitemOpen
  \bibinfo {note} {In a doubly stochastic matrix, both the rows and columns sum
  to 1.}\BibitemShut {Stop}%
\bibitem [{\citenamefont {Hofmann}(2005)}]{hofman2005complementary}%
  \BibitemOpen
  \bibfield  {author} {\bibinfo {author} {\bibfnamefont {H.~F.}\ \bibnamefont
  {Hofmann}},\ }\href {\doibase 10.1103/PhysRevLett.94.160504} {\bibfield
  {journal} {\bibinfo  {journal} {Phys. Rev. Lett.}\ }\textbf {\bibinfo
  {volume} {94}},\ \bibinfo {pages} {160504} (\bibinfo {year}
  {2005})}\BibitemShut {NoStop}%
\bibitem [{\citenamefont {Figgatt}\ \emph {et~al.}(2017)\citenamefont
  {Figgatt}, \citenamefont {Maslov}, \citenamefont {Landsman}, \citenamefont
  {Linke}, \citenamefont {Debnath},\ and\ \citenamefont
  {Monroe}}]{figgatt2017complete}%
  \BibitemOpen
  \bibfield  {author} {\bibinfo {author} {\bibfnamefont {C.}~\bibnamefont
  {Figgatt}}, \bibinfo {author} {\bibfnamefont {D.}~\bibnamefont {Maslov}},
  \bibinfo {author} {\bibfnamefont {K.~A.}\ \bibnamefont {Landsman}}, \bibinfo
  {author} {\bibfnamefont {N.~M.}\ \bibnamefont {Linke}}, \bibinfo {author}
  {\bibfnamefont {S.}~\bibnamefont {Debnath}}, \ and\ \bibinfo {author}
  {\bibfnamefont {C.}~\bibnamefont {Monroe}},\ }\href {\doibase
  10.1038/s41467-017-01904-7} {\bibfield  {journal} {\bibinfo  {journal}
  {Nature communications}\ }\textbf {\bibinfo {volume} {8}},\ \bibinfo {pages}
  {1918} (\bibinfo {year} {2017})}\BibitemShut {NoStop}%
\bibitem [{\citenamefont {Levine}\ \emph {et~al.}(2019)\citenamefont {Levine},
  \citenamefont {Keesling}, \citenamefont {Semeghini}, \citenamefont {Omran},
  \citenamefont {Wang}, \citenamefont {Ebadi}, \citenamefont {Bernien},
  \citenamefont {Greiner}, \citenamefont {Vuleti\ifmmode~\acute{c}\else
  \'{c}\fi{}}, \citenamefont {Pichler},\ and\ \citenamefont
  {Lukin}}]{levine2019parallel}%
  \BibitemOpen
  \bibfield  {author} {\bibinfo {author} {\bibfnamefont {H.}~\bibnamefont
  {Levine}}, \bibinfo {author} {\bibfnamefont {A.}~\bibnamefont {Keesling}},
  \bibinfo {author} {\bibfnamefont {G.}~\bibnamefont {Semeghini}}, \bibinfo
  {author} {\bibfnamefont {A.}~\bibnamefont {Omran}}, \bibinfo {author}
  {\bibfnamefont {T.~T.}\ \bibnamefont {Wang}}, \bibinfo {author}
  {\bibfnamefont {S.}~\bibnamefont {Ebadi}}, \bibinfo {author} {\bibfnamefont
  {H.}~\bibnamefont {Bernien}}, \bibinfo {author} {\bibfnamefont
  {M.}~\bibnamefont {Greiner}}, \bibinfo {author} {\bibfnamefont
  {V.}~\bibnamefont {Vuleti\ifmmode~\acute{c}\else \'{c}\fi{}}}, \bibinfo
  {author} {\bibfnamefont {H.}~\bibnamefont {Pichler}}, \ and\ \bibinfo
  {author} {\bibfnamefont {M.~D.}\ \bibnamefont {Lukin}},\ }\href {\doibase
  10.1103/PhysRevLett.123.170503} {\bibfield  {journal} {\bibinfo  {journal}
  {Phys. Rev. Lett.}\ }\textbf {\bibinfo {volume} {123}},\ \bibinfo {pages}
  {170503} (\bibinfo {year} {2019})}\BibitemShut {NoStop}%
\bibitem [{\citenamefont {Fang}\ \emph {et~al.}(2023)\citenamefont {Fang},
  \citenamefont {Wang}, \citenamefont {Sun},\ and\ \citenamefont
  {Kim}}]{fang2023realization}%
  \BibitemOpen
  \bibfield  {author} {\bibinfo {author} {\bibfnamefont {C.}~\bibnamefont
  {Fang}}, \bibinfo {author} {\bibfnamefont {Y.}~\bibnamefont {Wang}}, \bibinfo
  {author} {\bibfnamefont {K.}~\bibnamefont {Sun}}, \ and\ \bibinfo {author}
  {\bibfnamefont {J.}~\bibnamefont {Kim}},\ }\href@noop {} {\enquote {\bibinfo
  {title} {Realization of scalable cirac-zoller multi-qubit gates},}\ }
  (\bibinfo {year} {2023}),\ \Eprint {http://arxiv.org/abs/2301.07564}
  {arXiv:2301.07564 [quant-ph]} \BibitemShut {NoStop}%
\bibitem [{\citenamefont {Lu}\ \emph {et~al.}(2020)\citenamefont {Lu},
  \citenamefont {Sim}, \citenamefont {Suzuki}, \citenamefont {Englert},\ and\
  \citenamefont {Ng}}]{lu2020direct}%
  \BibitemOpen
  \bibfield  {author} {\bibinfo {author} {\bibfnamefont {Y.}~\bibnamefont
  {Lu}}, \bibinfo {author} {\bibfnamefont {J.~Y.}\ \bibnamefont {Sim}},
  \bibinfo {author} {\bibfnamefont {J.}~\bibnamefont {Suzuki}}, \bibinfo
  {author} {\bibfnamefont {B.-G.}\ \bibnamefont {Englert}}, \ and\ \bibinfo
  {author} {\bibfnamefont {H.~K.}\ \bibnamefont {Ng}},\ }\href@noop {}
  {\bibfield  {journal} {\bibinfo  {journal} {Physical Review A}\ }\textbf
  {\bibinfo {volume} {102}} (\bibinfo {year} {2020})}\BibitemShut {NoStop}%
\bibitem [{\citenamefont {Zhang}\ \emph {et~al.}(2021)\citenamefont {Zhang},
  \citenamefont {Luo}, \citenamefont {Wen}, \citenamefont {Feng}, \citenamefont
  {Pang}, \citenamefont {Luo},\ and\ \citenamefont {Zhou}}]{zhang2021direct}%
  \BibitemOpen
  \bibfield  {author} {\bibinfo {author} {\bibfnamefont {X.}~\bibnamefont
  {Zhang}}, \bibinfo {author} {\bibfnamefont {M.}~\bibnamefont {Luo}}, \bibinfo
  {author} {\bibfnamefont {Z.}~\bibnamefont {Wen}}, \bibinfo {author}
  {\bibfnamefont {Q.}~\bibnamefont {Feng}}, \bibinfo {author} {\bibfnamefont
  {S.}~\bibnamefont {Pang}}, \bibinfo {author} {\bibfnamefont {W.}~\bibnamefont
  {Luo}}, \ and\ \bibinfo {author} {\bibfnamefont {X.}~\bibnamefont {Zhou}},\
  }\href@noop {} {\bibfield  {journal} {\bibinfo  {journal} {Physical Review
  Letters}\ }\textbf {\bibinfo {volume} {127}} (\bibinfo {year}
  {2021})}\BibitemShut {NoStop}%
\bibitem [{\citenamefont {Terhal}(2015)}]{terhal2015quantum}%
  \BibitemOpen
  \bibfield  {author} {\bibinfo {author} {\bibfnamefont {B.~M.}\ \bibnamefont
  {Terhal}},\ }\href@noop {} {\bibfield  {journal} {\bibinfo  {journal}
  {Reviews of Modern Physics}\ }\textbf {\bibinfo {volume} {87}},\ \bibinfo
  {pages} {307} (\bibinfo {year} {2015})}\BibitemShut {NoStop}%
\bibitem [{\citenamefont {Graydon}\ \emph {et~al.}(2022)\citenamefont
  {Graydon}, \citenamefont {Skanes-Norman},\ and\ \citenamefont
  {Wallman}}]{graydon2022designing}%
  \BibitemOpen
  \bibfield  {author} {\bibinfo {author} {\bibfnamefont {M.~A.}\ \bibnamefont
  {Graydon}}, \bibinfo {author} {\bibfnamefont {J.}~\bibnamefont
  {Skanes-Norman}}, \ and\ \bibinfo {author} {\bibfnamefont {J.~J.}\
  \bibnamefont {Wallman}},\ }\href@noop {} {\bibfield  {journal} {\bibinfo
  {journal} {arXiv preprint arXiv:2201.07156}\ } (\bibinfo {year}
  {2022})}\BibitemShut {NoStop}%
\bibitem [{\citenamefont {Flammia}\ and\ \citenamefont
  {Wallman}(2020)}]{flammia2020efficient}%
  \BibitemOpen
  \bibfield  {author} {\bibinfo {author} {\bibfnamefont {S.~T.}\ \bibnamefont
  {Flammia}}\ and\ \bibinfo {author} {\bibfnamefont {J.~J.}\ \bibnamefont
  {Wallman}},\ }\href@noop {} {\bibfield  {journal} {\bibinfo  {journal} {ACM
  Transactions on Quantum Computing}\ }\textbf {\bibinfo {volume} {1}},\
  \bibinfo {pages} {1} (\bibinfo {year} {2020})}\BibitemShut {NoStop}%
\bibitem [{\citenamefont {Chen}\ \emph
  {et~al.}(2023{\natexlab{d}})\citenamefont {Chen}, \citenamefont {Yu},
  \citenamefont {Zhu},\ and\ \citenamefont {Wang}}]{chen2023efficient}%
  \BibitemOpen
  \bibfield  {author} {\bibinfo {author} {\bibfnamefont {Y.}~\bibnamefont
  {Chen}}, \bibinfo {author} {\bibfnamefont {Z.}~\bibnamefont {Yu}}, \bibinfo
  {author} {\bibfnamefont {C.}~\bibnamefont {Zhu}}, \ and\ \bibinfo {author}
  {\bibfnamefont {X.}~\bibnamefont {Wang}},\ }\href@noop {} {\bibfield
  {journal} {\bibinfo  {journal} {arXiv preprint arXiv:2305.04148}\ } (\bibinfo
  {year} {2023}{\natexlab{d}})}\BibitemShut {NoStop}%
\bibitem [{\citenamefont {Carignan-Dugas}\ \emph {et~al.}(2023)\citenamefont
  {Carignan-Dugas}, \citenamefont {Dahlen}, \citenamefont {Hincks},
  \citenamefont {Ospadov}, \citenamefont {Beale}, \citenamefont {Ferracin},
  \citenamefont {Skanes-Norman}, \citenamefont {Emerson},\ and\ \citenamefont
  {Wallman}}]{carignan2023error}%
  \BibitemOpen
  \bibfield  {author} {\bibinfo {author} {\bibfnamefont {A.}~\bibnamefont
  {Carignan-Dugas}}, \bibinfo {author} {\bibfnamefont {D.}~\bibnamefont
  {Dahlen}}, \bibinfo {author} {\bibfnamefont {I.}~\bibnamefont {Hincks}},
  \bibinfo {author} {\bibfnamefont {E.}~\bibnamefont {Ospadov}}, \bibinfo
  {author} {\bibfnamefont {S.~J.}\ \bibnamefont {Beale}}, \bibinfo {author}
  {\bibfnamefont {S.}~\bibnamefont {Ferracin}}, \bibinfo {author}
  {\bibfnamefont {J.}~\bibnamefont {Skanes-Norman}}, \bibinfo {author}
  {\bibfnamefont {J.}~\bibnamefont {Emerson}}, \ and\ \bibinfo {author}
  {\bibfnamefont {J.~J.}\ \bibnamefont {Wallman}},\ }\href@noop {} {\bibfield
  {journal} {\bibinfo  {journal} {arXiv preprint arXiv:2303.17714}\ } (\bibinfo
  {year} {2023})}\BibitemShut {NoStop}%
\bibitem [{\citenamefont {Chen}\ \emph
  {et~al.}(2023{\natexlab{e}})\citenamefont {Chen}, \citenamefont {Liu},
  \citenamefont {Otten}, \citenamefont {Seif}, \citenamefont {Fefferman},\ and\
  \citenamefont {Jiang}}]{chen2023learnability}%
  \BibitemOpen
  \bibfield  {author} {\bibinfo {author} {\bibfnamefont {S.}~\bibnamefont
  {Chen}}, \bibinfo {author} {\bibfnamefont {Y.}~\bibnamefont {Liu}}, \bibinfo
  {author} {\bibfnamefont {M.}~\bibnamefont {Otten}}, \bibinfo {author}
  {\bibfnamefont {A.}~\bibnamefont {Seif}}, \bibinfo {author} {\bibfnamefont
  {B.}~\bibnamefont {Fefferman}}, \ and\ \bibinfo {author} {\bibfnamefont
  {L.}~\bibnamefont {Jiang}},\ }\href@noop {} {\bibfield  {journal} {\bibinfo
  {journal} {Nature Communications}\ }\textbf {\bibinfo {volume} {14}},\
  \bibinfo {pages} {52} (\bibinfo {year} {2023}{\natexlab{e}})}\BibitemShut
  {NoStop}%
\bibitem [{Note28()}]{Note28}%
  \BibitemOpen
  \bibinfo {note} {Here, a $k$-\protect \emph {body} error is any weight-$n$
  Pauli error acting on $k$ gates. For example, a $Z$ error is a weight-1 error
  acting on a single qubit, but both $IZ$ and $ZZ$ are (weight-1 and weight-2,
  respectively) \protect \emph {single}-body errors acting on two qubits
  involved in an entangling gate.}\BibitemShut {Stop}%
\bibitem [{\citenamefont {Pelaez~Cisneros}\ \emph {et~al.}(2024)\citenamefont
  {Pelaez~Cisneros}, \citenamefont {Omole}, \citenamefont {Gokhale},
  \citenamefont {Rines}, \citenamefont {Smith}, \citenamefont {Perlin},\ and\
  \citenamefont {Hashim}}]{pelaez2024average}%
  \BibitemOpen
  \bibfield  {author} {\bibinfo {author} {\bibfnamefont {E.}~\bibnamefont
  {Pelaez~Cisneros}}, \bibinfo {author} {\bibfnamefont {V.}~\bibnamefont
  {Omole}}, \bibinfo {author} {\bibfnamefont {P.}~\bibnamefont {Gokhale}},
  \bibinfo {author} {\bibfnamefont {R.}~\bibnamefont {Rines}}, \bibinfo
  {author} {\bibfnamefont {K.~N.}\ \bibnamefont {Smith}}, \bibinfo {author}
  {\bibfnamefont {M.~A.}\ \bibnamefont {Perlin}}, \ and\ \bibinfo {author}
  {\bibfnamefont {A.}~\bibnamefont {Hashim}},\ }\href@noop {} {\bibfield
  {journal} {\bibinfo  {journal} {arXiv e-prints}\ ,\ \bibinfo {pages} {arXiv}}
  (\bibinfo {year} {2024})}\BibitemShut {NoStop}%
\bibitem [{\citenamefont {Hockings}\ \emph {et~al.}(2024)\citenamefont
  {Hockings}, \citenamefont {Doherty},\ and\ \citenamefont
  {Harper}}]{hockings2024scalable}%
  \BibitemOpen
  \bibfield  {author} {\bibinfo {author} {\bibfnamefont {E.~T.}\ \bibnamefont
  {Hockings}}, \bibinfo {author} {\bibfnamefont {A.~C.}\ \bibnamefont
  {Doherty}}, \ and\ \bibinfo {author} {\bibfnamefont {R.}~\bibnamefont
  {Harper}},\ }\href {https://arxiv.org/abs/2404.06545} {\enquote {\bibinfo
  {title} {Scalable noise characterisation of syndrome extraction circuits with
  averaged circuit eigenvalue sampling},}\ } (\bibinfo {year} {2024}),\ \Eprint
  {http://arxiv.org/abs/2404.06545} {arXiv:2404.06545 [quant-ph]} \BibitemShut
  {NoStop}%
\bibitem [{\citenamefont {Mahadev}(2018)}]{mahadev2018classical}%
  \BibitemOpen
  \bibfield  {author} {\bibinfo {author} {\bibfnamefont {U.}~\bibnamefont
  {Mahadev}},\ }in\ \href@noop {} {\emph {\bibinfo {booktitle} {2018 IEEE 59th
  Annual Symposium on Foundations of Computer Science (FOCS)}}}\ (\bibinfo
  {organization} {IEEE},\ \bibinfo {year} {2018})\ pp.\ \bibinfo {pages}
  {259--267}\BibitemShut {NoStop}%
\bibitem [{\citenamefont {Brakerski}\ \emph {et~al.}(2021)\citenamefont
  {Brakerski}, \citenamefont {Christiano}, \citenamefont {Mahadev},
  \citenamefont {Vazirani},\ and\ \citenamefont
  {Vidick}}]{brakerski2021cryptographic}%
  \BibitemOpen
  \bibfield  {author} {\bibinfo {author} {\bibfnamefont {Z.}~\bibnamefont
  {Brakerski}}, \bibinfo {author} {\bibfnamefont {P.}~\bibnamefont
  {Christiano}}, \bibinfo {author} {\bibfnamefont {U.}~\bibnamefont {Mahadev}},
  \bibinfo {author} {\bibfnamefont {U.}~\bibnamefont {Vazirani}}, \ and\
  \bibinfo {author} {\bibfnamefont {T.}~\bibnamefont {Vidick}},\ }\href@noop {}
  {\bibfield  {journal} {\bibinfo  {journal} {Journal of the ACM (JACM)}\
  }\textbf {\bibinfo {volume} {68}},\ \bibinfo {pages} {1} (\bibinfo {year}
  {2021})}\BibitemShut {NoStop}%
\bibitem [{\citenamefont {Kahanamoku-Meyer}\ \emph {et~al.}(2022)\citenamefont
  {Kahanamoku-Meyer}, \citenamefont {Choi}, \citenamefont {Vazirani},\ and\
  \citenamefont {Yao}}]{kahanamoku2022classically}%
  \BibitemOpen
  \bibfield  {author} {\bibinfo {author} {\bibfnamefont {G.~D.}\ \bibnamefont
  {Kahanamoku-Meyer}}, \bibinfo {author} {\bibfnamefont {S.}~\bibnamefont
  {Choi}}, \bibinfo {author} {\bibfnamefont {U.~V.}\ \bibnamefont {Vazirani}},
  \ and\ \bibinfo {author} {\bibfnamefont {N.~Y.}\ \bibnamefont {Yao}},\
  }\href@noop {} {\bibfield  {journal} {\bibinfo  {journal} {Nature Physics}\
  }\textbf {\bibinfo {volume} {18}},\ \bibinfo {pages} {918} (\bibinfo {year}
  {2022})}\BibitemShut {NoStop}%
\bibitem [{\citenamefont {Zhu}\ \emph {et~al.}(2021)\citenamefont {Zhu},
  \citenamefont {Kahanamoku-Meyer}, \citenamefont {Lewis}, \citenamefont
  {Noel}, \citenamefont {Katz}, \citenamefont {Harraz}, \citenamefont {Wang},
  \citenamefont {Risinger}, \citenamefont {Feng}, \citenamefont {Biswas} \emph
  {et~al.}}]{zhu2021interactive}%
  \BibitemOpen
  \bibfield  {author} {\bibinfo {author} {\bibfnamefont {D.}~\bibnamefont
  {Zhu}}, \bibinfo {author} {\bibfnamefont {G.~D.}\ \bibnamefont
  {Kahanamoku-Meyer}}, \bibinfo {author} {\bibfnamefont {L.}~\bibnamefont
  {Lewis}}, \bibinfo {author} {\bibfnamefont {C.}~\bibnamefont {Noel}},
  \bibinfo {author} {\bibfnamefont {O.}~\bibnamefont {Katz}}, \bibinfo {author}
  {\bibfnamefont {B.}~\bibnamefont {Harraz}}, \bibinfo {author} {\bibfnamefont
  {Q.}~\bibnamefont {Wang}}, \bibinfo {author} {\bibfnamefont {A.}~\bibnamefont
  {Risinger}}, \bibinfo {author} {\bibfnamefont {L.}~\bibnamefont {Feng}},
  \bibinfo {author} {\bibfnamefont {D.}~\bibnamefont {Biswas}},  \emph
  {et~al.},\ }\href@noop {} {\bibfield  {journal} {\bibinfo  {journal} {arXiv
  preprint arXiv:2112.05156}\ } (\bibinfo {year} {2021})}\BibitemShut {NoStop}%
\bibitem [{\citenamefont {Ferracin}\ \emph {et~al.}(2018)\citenamefont
  {Ferracin}, \citenamefont {Kapourniotis},\ and\ \citenamefont
  {Datta}}]{ferracin2018}%
  \BibitemOpen
  \bibfield  {author} {\bibinfo {author} {\bibfnamefont {S.}~\bibnamefont
  {Ferracin}}, \bibinfo {author} {\bibfnamefont {T.}~\bibnamefont
  {Kapourniotis}}, \ and\ \bibinfo {author} {\bibfnamefont {A.}~\bibnamefont
  {Datta}},\ }\href@noop {} {\bibfield  {journal} {\bibinfo  {journal}
  {Physical Review A 98, 022323}\ } (\bibinfo {year} {2018})}\BibitemShut
  {NoStop}%
\bibitem [{\citenamefont {Ferracin}\ \emph {et~al.}(2019)\citenamefont
  {Ferracin}, \citenamefont {Kapourniotis},\ and\ \citenamefont
  {Datta}}]{ferracin2019}%
  \BibitemOpen
  \bibfield  {author} {\bibinfo {author} {\bibfnamefont {S.}~\bibnamefont
  {Ferracin}}, \bibinfo {author} {\bibfnamefont {T.}~\bibnamefont
  {Kapourniotis}}, \ and\ \bibinfo {author} {\bibfnamefont {A.}~\bibnamefont
  {Datta}},\ }\href@noop {} {\bibfield  {journal} {\bibinfo  {journal} {New J.
  Phys. 21 113038}\ } (\bibinfo {year} {2019})}\BibitemShut {NoStop}%
\bibitem [{\citenamefont {Ferracin}\ \emph {et~al.}(2020)\citenamefont
  {Ferracin}, \citenamefont {Merkel}, \citenamefont {McKay},\ and\
  \citenamefont {Datta}}]{ferracin2020}%
  \BibitemOpen
  \bibfield  {author} {\bibinfo {author} {\bibfnamefont {S.}~\bibnamefont
  {Ferracin}}, \bibinfo {author} {\bibfnamefont {S.}~\bibnamefont {Merkel}},
  \bibinfo {author} {\bibfnamefont {D.}~\bibnamefont {McKay}}, \ and\ \bibinfo
  {author} {\bibfnamefont {A.}~\bibnamefont {Datta}},\ }\href@noop {}
  {\bibfield  {journal} {\bibinfo  {journal} {Physical Review A 104, 042603}\ }
  (\bibinfo {year} {2020})}\BibitemShut {NoStop}%
\bibitem [{\citenamefont {Blume-Kohout}\ and\ \citenamefont
  {Young}(2020)}]{Blume-Kohout2020-de}%
  \BibitemOpen
  \bibfield  {author} {\bibinfo {author} {\bibfnamefont {R.}~\bibnamefont
  {Blume-Kohout}}\ and\ \bibinfo {author} {\bibfnamefont {K.~C.}\ \bibnamefont
  {Young}},\ }\href {\doibase 10.22331/q-2020-11-15-362} {\bibfield  {journal}
  {\bibinfo  {journal} {Quantum}\ }\textbf {\bibinfo {volume} {4}},\ \bibinfo
  {pages} {362} (\bibinfo {year} {2020})}\BibitemShut {NoStop}%
\bibitem [{\citenamefont {Hines}\ and\ \citenamefont
  {Proctor}(2024)}]{Hines2023-be}%
  \BibitemOpen
  \bibfield  {author} {\bibinfo {author} {\bibfnamefont {J.}~\bibnamefont
  {Hines}}\ and\ \bibinfo {author} {\bibfnamefont {T.}~\bibnamefont
  {Proctor}},\ }\href@noop {} {\bibfield  {journal} {\bibinfo  {journal} {IEEE
  Transactions on Quantum Engineering}\ } (\bibinfo {year} {2024})}\BibitemShut
  {NoStop}%
\bibitem [{noa(2024)}]{noauthor_undated-wz}%
  \BibitemOpen
  \href {https://metriq.info/} {\enquote {\bibinfo {title} {Metriq -
  community-driven quantum benchmarks},}\ }\bibinfo {howpublished}
  {\url{https://metriq.info/}} (\bibinfo {year} {2024}),\ \bibinfo {note}
  {accessed: 2024-1-30}\BibitemShut {NoStop}%
\bibitem [{\citenamefont {Tomesh}\ \emph {et~al.}(2022)\citenamefont {Tomesh},
  \citenamefont {Gokhale}, \citenamefont {Omole}, \citenamefont {Ravi},
  \citenamefont {Smith}, \citenamefont {Viszlai}, \citenamefont {Wu},
  \citenamefont {Hardavellas}, \citenamefont {Martonosi},\ and\ \citenamefont
  {Chong}}]{tomesh2022supermarq}%
  \BibitemOpen
  \bibfield  {author} {\bibinfo {author} {\bibfnamefont {T.}~\bibnamefont
  {Tomesh}}, \bibinfo {author} {\bibfnamefont {P.}~\bibnamefont {Gokhale}},
  \bibinfo {author} {\bibfnamefont {V.}~\bibnamefont {Omole}}, \bibinfo
  {author} {\bibfnamefont {G.~S.}\ \bibnamefont {Ravi}}, \bibinfo {author}
  {\bibfnamefont {K.~N.}\ \bibnamefont {Smith}}, \bibinfo {author}
  {\bibfnamefont {J.}~\bibnamefont {Viszlai}}, \bibinfo {author} {\bibfnamefont
  {X.-C.}\ \bibnamefont {Wu}}, \bibinfo {author} {\bibfnamefont
  {N.}~\bibnamefont {Hardavellas}}, \bibinfo {author} {\bibfnamefont {M.~R.}\
  \bibnamefont {Martonosi}}, \ and\ \bibinfo {author} {\bibfnamefont {F.~T.}\
  \bibnamefont {Chong}},\ }in\ \href@noop {} {\emph {\bibinfo {booktitle} {2022
  IEEE International Symposium on High-Performance Computer Architecture
  (HPCA)}}}\ (\bibinfo {organization} {IEEE},\ \bibinfo {year} {2022})\ pp.\
  \bibinfo {pages} {587--603}\BibitemShut {NoStop}%
\bibitem [{\citenamefont {Li}\ \emph {et~al.}(2022)\citenamefont {Li},
  \citenamefont {Stein}, \citenamefont {Krishnamoorthy},\ and\ \citenamefont
  {Ang}}]{li2022qasmbench}%
  \BibitemOpen
  \bibfield  {author} {\bibinfo {author} {\bibfnamefont {A.}~\bibnamefont
  {Li}}, \bibinfo {author} {\bibfnamefont {S.}~\bibnamefont {Stein}}, \bibinfo
  {author} {\bibfnamefont {S.}~\bibnamefont {Krishnamoorthy}}, \ and\ \bibinfo
  {author} {\bibfnamefont {J.}~\bibnamefont {Ang}},\ }\href@noop {} {\bibfield
  {journal} {\bibinfo  {journal} {ACM Transactions on Quantum Computing}\ }
  (\bibinfo {year} {2022})}\BibitemShut {NoStop}%
\bibitem [{\citenamefont {Lubinski}\ \emph {et~al.}(2021)\citenamefont
  {Lubinski}, \citenamefont {Johri}, \citenamefont {Varosy}, \citenamefont
  {Coleman}, \citenamefont {Zhao}, \citenamefont {Necaise}, \citenamefont
  {Baldwin}, \citenamefont {Mayer},\ and\ \citenamefont
  {Proctor}}]{lubinski2021application}%
  \BibitemOpen
  \bibfield  {author} {\bibinfo {author} {\bibfnamefont {T.}~\bibnamefont
  {Lubinski}}, \bibinfo {author} {\bibfnamefont {S.}~\bibnamefont {Johri}},
  \bibinfo {author} {\bibfnamefont {P.}~\bibnamefont {Varosy}}, \bibinfo
  {author} {\bibfnamefont {J.}~\bibnamefont {Coleman}}, \bibinfo {author}
  {\bibfnamefont {L.}~\bibnamefont {Zhao}}, \bibinfo {author} {\bibfnamefont
  {J.}~\bibnamefont {Necaise}}, \bibinfo {author} {\bibfnamefont {C.~H.}\
  \bibnamefont {Baldwin}}, \bibinfo {author} {\bibfnamefont {K.}~\bibnamefont
  {Mayer}}, \ and\ \bibinfo {author} {\bibfnamefont {T.}~\bibnamefont
  {Proctor}},\ }\href@noop {} {\bibfield  {journal} {\bibinfo  {journal} {arXiv
  preprint arXiv:2110.03137}\ } (\bibinfo {year} {2021})}\BibitemShut {NoStop}%
\bibitem [{\citenamefont {Lubinski}\ \emph {et~al.}(2023)\citenamefont
  {Lubinski}, \citenamefont {Coffrin}, \citenamefont {McGeoch}, \citenamefont
  {Sathe}, \citenamefont {Apanavicius},\ and\ \citenamefont
  {Neira}}]{lubinski2023optimization}%
  \BibitemOpen
  \bibfield  {author} {\bibinfo {author} {\bibfnamefont {T.}~\bibnamefont
  {Lubinski}}, \bibinfo {author} {\bibfnamefont {C.}~\bibnamefont {Coffrin}},
  \bibinfo {author} {\bibfnamefont {C.}~\bibnamefont {McGeoch}}, \bibinfo
  {author} {\bibfnamefont {P.}~\bibnamefont {Sathe}}, \bibinfo {author}
  {\bibfnamefont {J.}~\bibnamefont {Apanavicius}}, \ and\ \bibinfo {author}
  {\bibfnamefont {D.~E.~B.}\ \bibnamefont {Neira}},\ }\href@noop {} {\bibfield
  {journal} {\bibinfo  {journal} {arXiv preprint arXiv:2302.02278}\ } (\bibinfo
  {year} {2023})}\BibitemShut {NoStop}%
\bibitem [{\citenamefont {Sawaya}\ \emph {et~al.}(2023)\citenamefont {Sawaya},
  \citenamefont {Marti-Dafcik}, \citenamefont {Ho}, \citenamefont {Tabor},
  \citenamefont {Bernal}, \citenamefont {Magann}, \citenamefont {Premaratne},
  \citenamefont {Dubey}, \citenamefont {Matsuura}, \citenamefont {Bishop} \emph
  {et~al.}}]{sawaya2023hamlib}%
  \BibitemOpen
  \bibfield  {author} {\bibinfo {author} {\bibfnamefont {N.~P.}\ \bibnamefont
  {Sawaya}}, \bibinfo {author} {\bibfnamefont {D.}~\bibnamefont
  {Marti-Dafcik}}, \bibinfo {author} {\bibfnamefont {Y.}~\bibnamefont {Ho}},
  \bibinfo {author} {\bibfnamefont {D.~P.}\ \bibnamefont {Tabor}}, \bibinfo
  {author} {\bibfnamefont {D.}~\bibnamefont {Bernal}}, \bibinfo {author}
  {\bibfnamefont {A.~B.}\ \bibnamefont {Magann}}, \bibinfo {author}
  {\bibfnamefont {S.}~\bibnamefont {Premaratne}}, \bibinfo {author}
  {\bibfnamefont {P.}~\bibnamefont {Dubey}}, \bibinfo {author} {\bibfnamefont
  {A.}~\bibnamefont {Matsuura}}, \bibinfo {author} {\bibfnamefont
  {N.}~\bibnamefont {Bishop}},  \emph {et~al.},\ }\href@noop {} {\bibfield
  {journal} {\bibinfo  {journal} {arXiv preprint arXiv:2306.13126}\ } (\bibinfo
  {year} {2023})}\BibitemShut {NoStop}%
\bibitem [{\citenamefont {Miessen}\ \emph {et~al.}(2024)\citenamefont
  {Miessen}, \citenamefont {Egger}, \citenamefont {Tavernelli},\ and\
  \citenamefont {Mazzola}}]{miessen2024benchmarking}%
  \BibitemOpen
  \bibfield  {author} {\bibinfo {author} {\bibfnamefont {A.}~\bibnamefont
  {Miessen}}, \bibinfo {author} {\bibfnamefont {D.~J.}\ \bibnamefont {Egger}},
  \bibinfo {author} {\bibfnamefont {I.}~\bibnamefont {Tavernelli}}, \ and\
  \bibinfo {author} {\bibfnamefont {G.}~\bibnamefont {Mazzola}},\ }\href@noop
  {} {\bibfield  {journal} {\bibinfo  {journal} {arXiv preprint
  arXiv:2404.08053}\ } (\bibinfo {year} {2024})}\BibitemShut {NoStop}%
\bibitem [{\citenamefont {Cerezo}\ \emph {et~al.}(2021)\citenamefont {Cerezo},
  \citenamefont {Arrasmith}, \citenamefont {Babbush}, \citenamefont {Benjamin},
  \citenamefont {Endo}, \citenamefont {Fujii}, \citenamefont {McClean},
  \citenamefont {Mitarai}, \citenamefont {Yuan}, \citenamefont {Cincio} \emph
  {et~al.}}]{cerezo2021variational}%
  \BibitemOpen
  \bibfield  {author} {\bibinfo {author} {\bibfnamefont {M.}~\bibnamefont
  {Cerezo}}, \bibinfo {author} {\bibfnamefont {A.}~\bibnamefont {Arrasmith}},
  \bibinfo {author} {\bibfnamefont {R.}~\bibnamefont {Babbush}}, \bibinfo
  {author} {\bibfnamefont {S.~C.}\ \bibnamefont {Benjamin}}, \bibinfo {author}
  {\bibfnamefont {S.}~\bibnamefont {Endo}}, \bibinfo {author} {\bibfnamefont
  {K.}~\bibnamefont {Fujii}}, \bibinfo {author} {\bibfnamefont {J.~R.}\
  \bibnamefont {McClean}}, \bibinfo {author} {\bibfnamefont {K.}~\bibnamefont
  {Mitarai}}, \bibinfo {author} {\bibfnamefont {X.}~\bibnamefont {Yuan}},
  \bibinfo {author} {\bibfnamefont {L.}~\bibnamefont {Cincio}},  \emph
  {et~al.},\ }\href@noop {} {\bibfield  {journal} {\bibinfo  {journal} {Nature
  Reviews Physics}\ }\textbf {\bibinfo {volume} {3}},\ \bibinfo {pages} {625}
  (\bibinfo {year} {2021})}\BibitemShut {NoStop}%
\bibitem [{\citenamefont {Murali}\ \emph {et~al.}(2019)\citenamefont {Murali},
  \citenamefont {Linke}, \citenamefont {Martonosi}, \citenamefont {Abhari},
  \citenamefont {Nguyen},\ and\ \citenamefont {Alderete}}]{murali2019full}%
  \BibitemOpen
  \bibfield  {author} {\bibinfo {author} {\bibfnamefont {P.}~\bibnamefont
  {Murali}}, \bibinfo {author} {\bibfnamefont {N.~M.}\ \bibnamefont {Linke}},
  \bibinfo {author} {\bibfnamefont {M.}~\bibnamefont {Martonosi}}, \bibinfo
  {author} {\bibfnamefont {A.~J.}\ \bibnamefont {Abhari}}, \bibinfo {author}
  {\bibfnamefont {N.~H.}\ \bibnamefont {Nguyen}}, \ and\ \bibinfo {author}
  {\bibfnamefont {C.~H.}\ \bibnamefont {Alderete}},\ }in\ \href@noop {} {\emph
  {\bibinfo {booktitle} {Proceedings of the 46th International Symposium on
  Computer Architecture}}}\ (\bibinfo {year} {2019})\ pp.\ \bibinfo {pages}
  {527--540}\BibitemShut {NoStop}%
\bibitem [{\citenamefont {Pfeuty}(1970)}]{pfeuty1970one}%
  \BibitemOpen
  \bibfield  {author} {\bibinfo {author} {\bibfnamefont {P.}~\bibnamefont
  {Pfeuty}},\ }\href@noop {} {\bibfield  {journal} {\bibinfo  {journal} {ANNALS
  of Physics}\ }\textbf {\bibinfo {volume} {57}},\ \bibinfo {pages} {79}
  (\bibinfo {year} {1970})}\BibitemShut {NoStop}%
\bibitem [{\citenamefont {Greenberger}\ \emph {et~al.}(1989)\citenamefont
  {Greenberger}, \citenamefont {Horne},\ and\ \citenamefont
  {Zeilinger}}]{greenberger1989going}%
  \BibitemOpen
  \bibfield  {author} {\bibinfo {author} {\bibfnamefont {D.~M.}\ \bibnamefont
  {Greenberger}}, \bibinfo {author} {\bibfnamefont {M.~A.}\ \bibnamefont
  {Horne}}, \ and\ \bibinfo {author} {\bibfnamefont {A.}~\bibnamefont
  {Zeilinger}},\ }\href@noop {} {\bibfield  {journal} {\bibinfo  {journal}
  {Bell’s theorem, quantum theory and conceptions of the universe}\ ,\
  \bibinfo {pages} {69}} (\bibinfo {year} {1989})}\BibitemShut {NoStop}%
\bibitem [{\citenamefont {Webb}(2015)}]{webb2015clifford}%
  \BibitemOpen
  \bibfield  {author} {\bibinfo {author} {\bibfnamefont {Z.}~\bibnamefont
  {Webb}},\ }\href@noop {} {\bibfield  {journal} {\bibinfo  {journal} {arXiv
  preprint arXiv:1510.02769}\ } (\bibinfo {year} {2015})}\BibitemShut {NoStop}%
\bibitem [{\citenamefont {Zhu}(2017)}]{zhu2017multiqubit}%
  \BibitemOpen
  \bibfield  {author} {\bibinfo {author} {\bibfnamefont {H.}~\bibnamefont
  {Zhu}},\ }\href@noop {} {\bibfield  {journal} {\bibinfo  {journal} {Physical
  Review A}\ }\textbf {\bibinfo {volume} {96}},\ \bibinfo {pages} {062336}
  (\bibinfo {year} {2017})}\BibitemShut {NoStop}%
\bibitem [{\citenamefont {Graydon}\ \emph {et~al.}(2021)\citenamefont
  {Graydon}, \citenamefont {Skanes-Norman},\ and\ \citenamefont
  {Wallman}}]{graydon2021clifford}%
  \BibitemOpen
  \bibfield  {author} {\bibinfo {author} {\bibfnamefont {M.~A.}\ \bibnamefont
  {Graydon}}, \bibinfo {author} {\bibfnamefont {J.}~\bibnamefont
  {Skanes-Norman}}, \ and\ \bibinfo {author} {\bibfnamefont {J.~J.}\
  \bibnamefont {Wallman}},\ }\href@noop {} {\bibfield  {journal} {\bibinfo
  {journal} {arXiv preprint arXiv:2108.04200}\ } (\bibinfo {year}
  {2021})}\BibitemShut {NoStop}%
\bibitem [{Note29()}]{Note29}%
  \BibitemOpen
  \bibinfo {note} {While we model our errors using a \protect \emph {post-gate}
  error matrix, it is equally valid to model errors using a \protect \emph
  {pre-gate} error matrix, or in some cases one which occurs concurrently with
  the gate \cite {wallman2018randomized}.}\BibitemShut {Stop}%
\bibitem [{\citenamefont {Lindblad}(1976)}]{lindblad1976generators}%
  \BibitemOpen
  \bibfield  {author} {\bibinfo {author} {\bibfnamefont {G.}~\bibnamefont
  {Lindblad}},\ }\href@noop {} {\bibfield  {journal} {\bibinfo  {journal}
  {Communications in Mathematical Physics}\ }\textbf {\bibinfo {volume} {48}},\
  \bibinfo {pages} {119} (\bibinfo {year} {1976})}\BibitemShut {NoStop}%
\bibitem [{\citenamefont
  {Gottesman}(1998{\natexlab{b}})}]{gottesman1998theory}%
  \BibitemOpen
  \bibfield  {author} {\bibinfo {author} {\bibfnamefont {D.}~\bibnamefont
  {Gottesman}},\ }\href@noop {} {\bibfield  {journal} {\bibinfo  {journal}
  {Physical Review A}\ }\textbf {\bibinfo {volume} {57}},\ \bibinfo {pages}
  {127} (\bibinfo {year} {1998}{\natexlab{b}})}\BibitemShut {NoStop}%
\bibitem [{\citenamefont {Crooks}(2020)}]{crooks2020gates}%
  \BibitemOpen
  \bibfield  {author} {\bibinfo {author} {\bibfnamefont {G.~E.}\ \bibnamefont
  {Crooks}},\ }\href@noop {} {\bibfield  {journal} {\bibinfo  {journal} {Gates
  states and circuits}\ } (\bibinfo {year} {2020})}\BibitemShut {NoStop}%
\bibitem [{\citenamefont {Pozniak}\ \emph {et~al.}(1998)\citenamefont
  {Pozniak}, \citenamefont {Zyczkowski},\ and\ \citenamefont
  {Kus}}]{pozniak1998composed}%
  \BibitemOpen
  \bibfield  {author} {\bibinfo {author} {\bibfnamefont {M.}~\bibnamefont
  {Pozniak}}, \bibinfo {author} {\bibfnamefont {K.}~\bibnamefont {Zyczkowski}},
  \ and\ \bibinfo {author} {\bibfnamefont {M.}~\bibnamefont {Kus}},\
  }\href@noop {} {\bibfield  {journal} {\bibinfo  {journal} {Journal of Physics
  A: Mathematical and General}\ }\textbf {\bibinfo {volume} {31}},\ \bibinfo
  {pages} {1059} (\bibinfo {year} {1998})}\BibitemShut {NoStop}%
\bibitem [{\citenamefont {Bannai}\ and\ \citenamefont
  {Bannai}(2009)}]{bannai2009survey}%
  \BibitemOpen
  \bibfield  {author} {\bibinfo {author} {\bibfnamefont {E.}~\bibnamefont
  {Bannai}}\ and\ \bibinfo {author} {\bibfnamefont {E.}~\bibnamefont
  {Bannai}},\ }\href@noop {} {\bibfield  {journal} {\bibinfo  {journal}
  {European Journal of Combinatorics}\ }\textbf {\bibinfo {volume} {30}},\
  \bibinfo {pages} {1392} (\bibinfo {year} {2009})}\BibitemShut {NoStop}%
\bibitem [{Note30()}]{Note30}%
  \BibitemOpen
  \bibinfo {note} {The Clifford group also forms a unitary 3-design only when
  the qudit dimension $D=2$ \cite {webb2015clifford, zhu2017multiqubit,
  graydon2021clifford}.}\BibitemShut {Stop}%
\bibitem [{\citenamefont {Gross}\ \emph {et~al.}(2007)\citenamefont {Gross},
  \citenamefont {Audenaert},\ and\ \citenamefont {Eisert}}]{gross2007evenly}%
  \BibitemOpen
  \bibfield  {author} {\bibinfo {author} {\bibfnamefont {D.}~\bibnamefont
  {Gross}}, \bibinfo {author} {\bibfnamefont {K.}~\bibnamefont {Audenaert}}, \
  and\ \bibinfo {author} {\bibfnamefont {J.}~\bibnamefont {Eisert}},\
  }\href@noop {} {\bibfield  {journal} {\bibinfo  {journal} {Journal of
  mathematical physics}\ }\textbf {\bibinfo {volume} {48}} (\bibinfo {year}
  {2007})}\BibitemShut {NoStop}%
\bibitem [{\citenamefont {Goss}\ \emph {et~al.}(2024)\citenamefont {Goss},
  \citenamefont {Ferracin}, \citenamefont {Hashim}, \citenamefont
  {Carignan-Dugas}, \citenamefont {Kreikebaum}, \citenamefont {Naik},
  \citenamefont {Santiago},\ and\ \citenamefont {Siddiqi}}]{goss2023extending}%
  \BibitemOpen
  \bibfield  {author} {\bibinfo {author} {\bibfnamefont {N.}~\bibnamefont
  {Goss}}, \bibinfo {author} {\bibfnamefont {S.}~\bibnamefont {Ferracin}},
  \bibinfo {author} {\bibfnamefont {A.}~\bibnamefont {Hashim}}, \bibinfo
  {author} {\bibfnamefont {A.}~\bibnamefont {Carignan-Dugas}}, \bibinfo
  {author} {\bibfnamefont {J.~M.}\ \bibnamefont {Kreikebaum}}, \bibinfo
  {author} {\bibfnamefont {R.~K.}\ \bibnamefont {Naik}}, \bibinfo {author}
  {\bibfnamefont {D.~I.}\ \bibnamefont {Santiago}}, \ and\ \bibinfo {author}
  {\bibfnamefont {I.}~\bibnamefont {Siddiqi}},\ }\href@noop {} {\bibfield
  {journal} {\bibinfo  {journal} {npj Quantum Information}\ }\textbf {\bibinfo
  {volume} {10}},\ \bibinfo {pages} {101} (\bibinfo {year} {2024})}\BibitemShut
  {NoStop}%
\bibitem [{\citenamefont {Hoeffding}(1963)}]{H63}%
  \BibitemOpen
  \bibfield  {author} {\bibinfo {author} {\bibfnamefont {W.}~\bibnamefont
  {Hoeffding}},\ }\href
  {https://www.tandfonline.com/doi/abs/10.1080/01621459.1963.10500830}
  {\bibfield  {journal} {\bibinfo  {journal} {Journal of the American
  Statistical Association, 58 (301), pp 13–30}\ } (\bibinfo {year}
  {1963})}\BibitemShut {NoStop}%
\bibitem [{\citenamefont {Goss}\ \emph {et~al.}(2022)\citenamefont {Goss},
  \citenamefont {Morvan}, \citenamefont {Marinelli}, \citenamefont {Mitchell},
  \citenamefont {Nguyen}, \citenamefont {Naik}, \citenamefont {Chen},
  \citenamefont {J{\"u}nger}, \citenamefont {Kreikebaum}, \citenamefont
  {Santiago} \emph {et~al.}}]{goss2022high}%
  \BibitemOpen
  \bibfield  {author} {\bibinfo {author} {\bibfnamefont {N.}~\bibnamefont
  {Goss}}, \bibinfo {author} {\bibfnamefont {A.}~\bibnamefont {Morvan}},
  \bibinfo {author} {\bibfnamefont {B.}~\bibnamefont {Marinelli}}, \bibinfo
  {author} {\bibfnamefont {B.~K.}\ \bibnamefont {Mitchell}}, \bibinfo {author}
  {\bibfnamefont {L.~B.}\ \bibnamefont {Nguyen}}, \bibinfo {author}
  {\bibfnamefont {R.~K.}\ \bibnamefont {Naik}}, \bibinfo {author}
  {\bibfnamefont {L.}~\bibnamefont {Chen}}, \bibinfo {author} {\bibfnamefont
  {C.}~\bibnamefont {J{\"u}nger}}, \bibinfo {author} {\bibfnamefont {J.~M.}\
  \bibnamefont {Kreikebaum}}, \bibinfo {author} {\bibfnamefont {D.~I.}\
  \bibnamefont {Santiago}},  \emph {et~al.},\ }\href@noop {} {\bibfield
  {journal} {\bibinfo  {journal} {Nature Communications}\ }\textbf {\bibinfo
  {volume} {13}},\ \bibinfo {pages} {7481} (\bibinfo {year}
  {2022})}\BibitemShut {NoStop}%
\bibitem [{\citenamefont {Liu}\ \emph {et~al.}(2023)\citenamefont {Liu},
  \citenamefont {Wang}, \citenamefont {Zhang}, \citenamefont {Zhang},
  \citenamefont {Cai}, \citenamefont {Xu}, \citenamefont {Li}, \citenamefont
  {Han}, \citenamefont {Li}, \citenamefont {Xue}, \citenamefont {Liu},
  \citenamefont {You}, \citenamefont {Jin},\ and\ \citenamefont
  {Yu}}]{PhysRevX.13.021028}%
  \BibitemOpen
  \bibfield  {author} {\bibinfo {author} {\bibfnamefont {P.}~\bibnamefont
  {Liu}}, \bibinfo {author} {\bibfnamefont {R.}~\bibnamefont {Wang}}, \bibinfo
  {author} {\bibfnamefont {J.-N.}\ \bibnamefont {Zhang}}, \bibinfo {author}
  {\bibfnamefont {Y.}~\bibnamefont {Zhang}}, \bibinfo {author} {\bibfnamefont
  {X.}~\bibnamefont {Cai}}, \bibinfo {author} {\bibfnamefont {H.}~\bibnamefont
  {Xu}}, \bibinfo {author} {\bibfnamefont {Z.}~\bibnamefont {Li}}, \bibinfo
  {author} {\bibfnamefont {J.}~\bibnamefont {Han}}, \bibinfo {author}
  {\bibfnamefont {X.}~\bibnamefont {Li}}, \bibinfo {author} {\bibfnamefont
  {G.}~\bibnamefont {Xue}}, \bibinfo {author} {\bibfnamefont {W.}~\bibnamefont
  {Liu}}, \bibinfo {author} {\bibfnamefont {L.}~\bibnamefont {You}}, \bibinfo
  {author} {\bibfnamefont {Y.}~\bibnamefont {Jin}}, \ and\ \bibinfo {author}
  {\bibfnamefont {H.}~\bibnamefont {Yu}},\ }\href {\doibase
  10.1103/PhysRevX.13.021028} {\bibfield  {journal} {\bibinfo  {journal} {Phys.
  Rev. X}\ }\textbf {\bibinfo {volume} {13}},\ \bibinfo {pages} {021028}
  (\bibinfo {year} {2023})}\BibitemShut {NoStop}%
\bibitem [{\citenamefont {Cao}\ \emph {et~al.}(2023)\citenamefont {Cao},
  \citenamefont {Bakr}, \citenamefont {Campanaro}, \citenamefont {Fasciati},
  \citenamefont {Wills}, \citenamefont {Lall}, \citenamefont {Shteynas},
  \citenamefont {Chidambaram}, \citenamefont {Rungger},\ and\ \citenamefont
  {Leek}}]{cao2023emulating}%
  \BibitemOpen
  \bibfield  {author} {\bibinfo {author} {\bibfnamefont {S.}~\bibnamefont
  {Cao}}, \bibinfo {author} {\bibfnamefont {M.}~\bibnamefont {Bakr}}, \bibinfo
  {author} {\bibfnamefont {G.}~\bibnamefont {Campanaro}}, \bibinfo {author}
  {\bibfnamefont {S.~D.}\ \bibnamefont {Fasciati}}, \bibinfo {author}
  {\bibfnamefont {J.}~\bibnamefont {Wills}}, \bibinfo {author} {\bibfnamefont
  {D.}~\bibnamefont {Lall}}, \bibinfo {author} {\bibfnamefont {B.}~\bibnamefont
  {Shteynas}}, \bibinfo {author} {\bibfnamefont {V.}~\bibnamefont
  {Chidambaram}}, \bibinfo {author} {\bibfnamefont {I.}~\bibnamefont
  {Rungger}}, \ and\ \bibinfo {author} {\bibfnamefont {P.}~\bibnamefont
  {Leek}},\ }\href@noop {} {\enquote {\bibinfo {title} {Emulating two qubits
  with a four-level transmon qudit for variational quantum algorithms},}\ }
  (\bibinfo {year} {2023}),\ \Eprint {http://arxiv.org/abs/2303.04796}
  {arXiv:2303.04796 [quant-ph]} \BibitemShut {NoStop}%
\bibitem [{\citenamefont {Ringbauer}\ \emph {et~al.}(2022)\citenamefont
  {Ringbauer}, \citenamefont {Meth}, \citenamefont {Postler}, \citenamefont
  {Stricker}, \citenamefont {Blatt}, \citenamefont {Schindler},\ and\
  \citenamefont {Monz}}]{ringbauer2021universal}%
  \BibitemOpen
  \bibfield  {author} {\bibinfo {author} {\bibfnamefont {M.}~\bibnamefont
  {Ringbauer}}, \bibinfo {author} {\bibfnamefont {M.}~\bibnamefont {Meth}},
  \bibinfo {author} {\bibfnamefont {L.}~\bibnamefont {Postler}}, \bibinfo
  {author} {\bibfnamefont {R.}~\bibnamefont {Stricker}}, \bibinfo {author}
  {\bibfnamefont {R.}~\bibnamefont {Blatt}}, \bibinfo {author} {\bibfnamefont
  {P.}~\bibnamefont {Schindler}}, \ and\ \bibinfo {author} {\bibfnamefont
  {T.}~\bibnamefont {Monz}},\ }\href {\doibase 10.1038/s41567-022-01658-0}
  {\bibfield  {journal} {\bibinfo  {journal} {Nature Physics}\ }\textbf
  {\bibinfo {volume} {18}},\ \bibinfo {pages} {1053} (\bibinfo {year}
  {2022})}\BibitemShut {NoStop}%
\bibitem [{\citenamefont {Hrmo}\ \emph {et~al.}(2023)\citenamefont {Hrmo},
  \citenamefont {Wilhelm}, \citenamefont {Gerster}, \citenamefont {van Mourik},
  \citenamefont {Huber}, \citenamefont {Blatt}, \citenamefont {Schindler},
  \citenamefont {Monz},\ and\ \citenamefont {Ringbauer}}]{native-qudit-ion}%
  \BibitemOpen
  \bibfield  {author} {\bibinfo {author} {\bibfnamefont {P.}~\bibnamefont
  {Hrmo}}, \bibinfo {author} {\bibfnamefont {B.}~\bibnamefont {Wilhelm}},
  \bibinfo {author} {\bibfnamefont {L.}~\bibnamefont {Gerster}}, \bibinfo
  {author} {\bibfnamefont {M.~W.}\ \bibnamefont {van Mourik}}, \bibinfo
  {author} {\bibfnamefont {M.}~\bibnamefont {Huber}}, \bibinfo {author}
  {\bibfnamefont {R.}~\bibnamefont {Blatt}}, \bibinfo {author} {\bibfnamefont
  {P.}~\bibnamefont {Schindler}}, \bibinfo {author} {\bibfnamefont
  {T.}~\bibnamefont {Monz}}, \ and\ \bibinfo {author} {\bibfnamefont
  {M.}~\bibnamefont {Ringbauer}},\ }\href {\doibase 10.1038/s41467-023-37375-2}
  {\bibfield  {journal} {\bibinfo  {journal} {Nature Communications}\ }\textbf
  {\bibinfo {volume} {14}},\ \bibinfo {pages} {2242} (\bibinfo {year}
  {2023})}\BibitemShut {NoStop}%
\bibitem [{\citenamefont {Lanyon}\ \emph {et~al.}(2008)\citenamefont {Lanyon},
  \citenamefont {Weinhold}, \citenamefont {Langford}, \citenamefont {O'Brien},
  \citenamefont {Resch}, \citenamefont {Gilchrist},\ and\ \citenamefont
  {White}}]{lanyon2008manipulating}%
  \BibitemOpen
  \bibfield  {author} {\bibinfo {author} {\bibfnamefont {B.~P.}\ \bibnamefont
  {Lanyon}}, \bibinfo {author} {\bibfnamefont {T.~J.}\ \bibnamefont
  {Weinhold}}, \bibinfo {author} {\bibfnamefont {N.~K.}\ \bibnamefont
  {Langford}}, \bibinfo {author} {\bibfnamefont {J.~L.}\ \bibnamefont
  {O'Brien}}, \bibinfo {author} {\bibfnamefont {K.~J.}\ \bibnamefont {Resch}},
  \bibinfo {author} {\bibfnamefont {A.}~\bibnamefont {Gilchrist}}, \ and\
  \bibinfo {author} {\bibfnamefont {A.~G.}\ \bibnamefont {White}},\ }\href
  {\doibase 10.1103/PhysRevLett.100.060504} {\bibfield  {journal} {\bibinfo
  {journal} {Phys. Rev. Lett.}\ }\textbf {\bibinfo {volume} {100}},\ \bibinfo
  {pages} {060504} (\bibinfo {year} {2008})}\BibitemShut {NoStop}%
\bibitem [{\citenamefont {Chi}\ \emph {et~al.}(2022)\citenamefont {Chi},
  \citenamefont {Huang}, \citenamefont {Zhang}, \citenamefont {Mao},
  \citenamefont {Zhou}, \citenamefont {Chen}, \citenamefont {Zhai},
  \citenamefont {Bao}, \citenamefont {Dai}, \citenamefont {Yuan}, \citenamefont
  {Zhang}, \citenamefont {Dai}, \citenamefont {Tang}, \citenamefont {Yang},
  \citenamefont {Li}, \citenamefont {Ding}, \citenamefont {Oxenl{\o}we},
  \citenamefont {Thompson}, \citenamefont {O'Brien}, \citenamefont {Li},
  \citenamefont {Gong},\ and\ \citenamefont {Wang}}]{photonic-qudit}%
  \BibitemOpen
  \bibfield  {author} {\bibinfo {author} {\bibfnamefont {Y.}~\bibnamefont
  {Chi}}, \bibinfo {author} {\bibfnamefont {J.}~\bibnamefont {Huang}}, \bibinfo
  {author} {\bibfnamefont {Z.}~\bibnamefont {Zhang}}, \bibinfo {author}
  {\bibfnamefont {J.}~\bibnamefont {Mao}}, \bibinfo {author} {\bibfnamefont
  {Z.}~\bibnamefont {Zhou}}, \bibinfo {author} {\bibfnamefont {X.}~\bibnamefont
  {Chen}}, \bibinfo {author} {\bibfnamefont {C.}~\bibnamefont {Zhai}}, \bibinfo
  {author} {\bibfnamefont {J.}~\bibnamefont {Bao}}, \bibinfo {author}
  {\bibfnamefont {T.}~\bibnamefont {Dai}}, \bibinfo {author} {\bibfnamefont
  {H.}~\bibnamefont {Yuan}}, \bibinfo {author} {\bibfnamefont {M.}~\bibnamefont
  {Zhang}}, \bibinfo {author} {\bibfnamefont {D.}~\bibnamefont {Dai}}, \bibinfo
  {author} {\bibfnamefont {B.}~\bibnamefont {Tang}}, \bibinfo {author}
  {\bibfnamefont {Y.}~\bibnamefont {Yang}}, \bibinfo {author} {\bibfnamefont
  {Z.}~\bibnamefont {Li}}, \bibinfo {author} {\bibfnamefont {Y.}~\bibnamefont
  {Ding}}, \bibinfo {author} {\bibfnamefont {L.~K.}\ \bibnamefont
  {Oxenl{\o}we}}, \bibinfo {author} {\bibfnamefont {M.~G.}\ \bibnamefont
  {Thompson}}, \bibinfo {author} {\bibfnamefont {J.~L.}\ \bibnamefont
  {O'Brien}}, \bibinfo {author} {\bibfnamefont {Y.}~\bibnamefont {Li}},
  \bibinfo {author} {\bibfnamefont {Q.}~\bibnamefont {Gong}}, \ and\ \bibinfo
  {author} {\bibfnamefont {J.}~\bibnamefont {Wang}},\ }\href {\doibase
  10.1038/s41467-022-28767-x} {\bibfield  {journal} {\bibinfo  {journal}
  {Nature Communications}\ }\textbf {\bibinfo {volume} {13}},\ \bibinfo {pages}
  {1166} (\bibinfo {year} {2022})}\BibitemShut {NoStop}%
\bibitem [{\citenamefont {Duclos-Cianci}\ and\ \citenamefont
  {Poulin}(2013)}]{PhysRevA.87.062338}%
  \BibitemOpen
  \bibfield  {author} {\bibinfo {author} {\bibfnamefont {G.}~\bibnamefont
  {Duclos-Cianci}}\ and\ \bibinfo {author} {\bibfnamefont {D.}~\bibnamefont
  {Poulin}},\ }\href {\doibase 10.1103/PhysRevA.87.062338} {\bibfield
  {journal} {\bibinfo  {journal} {Phys. Rev. A}\ }\textbf {\bibinfo {volume}
  {87}},\ \bibinfo {pages} {062338} (\bibinfo {year} {2013})}\BibitemShut
  {NoStop}%
\bibitem [{\citenamefont {Anwar}\ \emph {et~al.}(2014)\citenamefont {Anwar},
  \citenamefont {Brown}, \citenamefont {Campbell},\ and\ \citenamefont
  {Browne}}]{qudit-toric-codes-threshold}%
  \BibitemOpen
  \bibfield  {author} {\bibinfo {author} {\bibfnamefont {H.}~\bibnamefont
  {Anwar}}, \bibinfo {author} {\bibfnamefont {B.~J.}\ \bibnamefont {Brown}},
  \bibinfo {author} {\bibfnamefont {E.~T.}\ \bibnamefont {Campbell}}, \ and\
  \bibinfo {author} {\bibfnamefont {D.~E.}\ \bibnamefont {Browne}},\
  }\href@noop {} {\bibfield  {journal} {\bibinfo  {journal} {New Journal of
  Physics}\ }\textbf {\bibinfo {volume} {16}},\ \bibinfo {pages} {063038}
  (\bibinfo {year} {2014})}\BibitemShut {NoStop}%
\bibitem [{\citenamefont {Muralidharan}\ \emph {et~al.}(2017)\citenamefont
  {Muralidharan}, \citenamefont {Zou}, \citenamefont {Li}, \citenamefont
  {Wen},\ and\ \citenamefont {Jiang}}]{muralidharan_zou_li_wen_jiang_2017}%
  \BibitemOpen
  \bibfield  {author} {\bibinfo {author} {\bibfnamefont {S.}~\bibnamefont
  {Muralidharan}}, \bibinfo {author} {\bibfnamefont {C.-L.}\ \bibnamefont
  {Zou}}, \bibinfo {author} {\bibfnamefont {L.}~\bibnamefont {Li}}, \bibinfo
  {author} {\bibfnamefont {J.}~\bibnamefont {Wen}}, \ and\ \bibinfo {author}
  {\bibfnamefont {L.}~\bibnamefont {Jiang}},\ }\href {\doibase
  10.1088/1367-2630/aa573a} {\bibfield  {journal} {\bibinfo  {journal} {New
  Journal of Physics}\ }\textbf {\bibinfo {volume} {19}},\ \bibinfo {pages}
  {013026} (\bibinfo {year} {2017})}\BibitemShut {NoStop}%
\bibitem [{\citenamefont {Campbell}\ \emph {et~al.}(2012)\citenamefont
  {Campbell}, \citenamefont {Anwar},\ and\ \citenamefont
  {Browne}}]{PhysRevX.2.041021}%
  \BibitemOpen
  \bibfield  {author} {\bibinfo {author} {\bibfnamefont {E.~T.}\ \bibnamefont
  {Campbell}}, \bibinfo {author} {\bibfnamefont {H.}~\bibnamefont {Anwar}}, \
  and\ \bibinfo {author} {\bibfnamefont {D.~E.}\ \bibnamefont {Browne}},\
  }\href {\doibase 10.1103/PhysRevX.2.041021} {\bibfield  {journal} {\bibinfo
  {journal} {Phys. Rev. X}\ }\textbf {\bibinfo {volume} {2}},\ \bibinfo {pages}
  {041021} (\bibinfo {year} {2012})}\BibitemShut {NoStop}%
\bibitem [{\citenamefont {Gokhale}\ \emph {et~al.}(2019)\citenamefont
  {Gokhale}, \citenamefont {Baker}, \citenamefont {Duckering}, \citenamefont
  {Brown}, \citenamefont {Brown},\ and\ \citenamefont
  {Chong}}]{10.1145/3307650.3322253}%
  \BibitemOpen
  \bibfield  {author} {\bibinfo {author} {\bibfnamefont {P.}~\bibnamefont
  {Gokhale}}, \bibinfo {author} {\bibfnamefont {J.~M.}\ \bibnamefont {Baker}},
  \bibinfo {author} {\bibfnamefont {C.}~\bibnamefont {Duckering}}, \bibinfo
  {author} {\bibfnamefont {N.~C.}\ \bibnamefont {Brown}}, \bibinfo {author}
  {\bibfnamefont {K.~R.}\ \bibnamefont {Brown}}, \ and\ \bibinfo {author}
  {\bibfnamefont {F.~T.}\ \bibnamefont {Chong}},\ }in\ \href {\doibase
  10.1145/3307650.3322253} {\emph {\bibinfo {booktitle} {Proceedings of the
  46th International Symposium on Computer Architecture}}},\ \bibinfo {series
  and number} {ISCA '19}\ (\bibinfo  {publisher} {Association for Computing
  Machinery},\ \bibinfo {address} {New York, NY, USA},\ \bibinfo {year}
  {2019})\ p.\ \bibinfo {pages} {554–566}\BibitemShut {NoStop}%
\bibitem [{\citenamefont {Gustafson}(2022)}]{sqed-simulation}%
  \BibitemOpen
  \bibfield  {author} {\bibinfo {author} {\bibfnamefont {E.}~\bibnamefont
  {Gustafson}},\ }\href {\doibase 10.48550/ARXIV.2201.04546} {\enquote
  {\bibinfo {title} {Noise improvements in quantum simulations of sqed using
  qutrits},}\ } (\bibinfo {year} {2022})\BibitemShut {NoStop}%
\bibitem [{\citenamefont {Truflandier}\ \emph {et~al.}(2016)\citenamefont
  {Truflandier}, \citenamefont {Dianzinga},\ and\ \citenamefont
  {Bowler}}]{purification}%
  \BibitemOpen
  \bibfield  {author} {\bibinfo {author} {\bibfnamefont {L.~A.}\ \bibnamefont
  {Truflandier}}, \bibinfo {author} {\bibfnamefont {R.~M.}\ \bibnamefont
  {Dianzinga}}, \ and\ \bibinfo {author} {\bibfnamefont {D.~R.}\ \bibnamefont
  {Bowler}},\ }\href {\doibase 10.1063/1.4943213} {\bibfield  {journal}
  {\bibinfo  {journal} {The Journal of Chemical Physics}\ }\textbf {\bibinfo
  {volume} {144}} (\bibinfo {year} {2016}),\ 10.1063/1.4943213},\ \bibinfo
  {note} {091102},\ \Eprint
  {http://arxiv.org/abs/https://pubs.aip.org/aip/jcp/article-pdf/doi/10.1063/1.4943213/13330452/091102\_1\_online.pdf}
  {https://pubs.aip.org/aip/jcp/article-pdf/doi/10.1063/1.4943213/13330452/091102\_1\_online.pdf}
  \BibitemShut {NoStop}%
\bibitem [{\citenamefont {Cao}\ \emph {et~al.}(2022)\citenamefont {Cao},
  \citenamefont {Lall}, \citenamefont {Bakr}, \citenamefont {Campanaro},
  \citenamefont {Fasciati}, \citenamefont {Wills}, \citenamefont {Chidambaram},
  \citenamefont {Shteynas}, \citenamefont {Rungger},\ and\ \citenamefont
  {Leek}}]{cao2022efficient}%
  \BibitemOpen
  \bibfield  {author} {\bibinfo {author} {\bibfnamefont {S.}~\bibnamefont
  {Cao}}, \bibinfo {author} {\bibfnamefont {D.}~\bibnamefont {Lall}}, \bibinfo
  {author} {\bibfnamefont {M.}~\bibnamefont {Bakr}}, \bibinfo {author}
  {\bibfnamefont {G.}~\bibnamefont {Campanaro}}, \bibinfo {author}
  {\bibfnamefont {S.}~\bibnamefont {Fasciati}}, \bibinfo {author}
  {\bibfnamefont {J.}~\bibnamefont {Wills}}, \bibinfo {author} {\bibfnamefont
  {V.}~\bibnamefont {Chidambaram}}, \bibinfo {author} {\bibfnamefont
  {B.}~\bibnamefont {Shteynas}}, \bibinfo {author} {\bibfnamefont
  {I.}~\bibnamefont {Rungger}}, \ and\ \bibinfo {author} {\bibfnamefont
  {P.}~\bibnamefont {Leek}},\ }\href@noop {} {\enquote {\bibinfo {title}
  {Efficient qutrit gate-set tomography on a transmon},}\ } (\bibinfo {year}
  {2022}),\ \Eprint {http://arxiv.org/abs/2210.04857} {arXiv:2210.04857
  [quant-ph]} \BibitemShut {NoStop}%
\bibitem [{\citenamefont {Seifert}\ \emph {et~al.}(2023)\citenamefont
  {Seifert}, \citenamefont {Li}, \citenamefont {Roy}, \citenamefont {Schuster},
  \citenamefont {Chong},\ and\ \citenamefont {Baker}}]{seifert2023exploring}%
  \BibitemOpen
  \bibfield  {author} {\bibinfo {author} {\bibfnamefont {L.~M.}\ \bibnamefont
  {Seifert}}, \bibinfo {author} {\bibfnamefont {Z.}~\bibnamefont {Li}},
  \bibinfo {author} {\bibfnamefont {T.}~\bibnamefont {Roy}}, \bibinfo {author}
  {\bibfnamefont {D.~I.}\ \bibnamefont {Schuster}}, \bibinfo {author}
  {\bibfnamefont {F.~T.}\ \bibnamefont {Chong}}, \ and\ \bibinfo {author}
  {\bibfnamefont {J.~M.}\ \bibnamefont {Baker}},\ }\href@noop {} {\enquote
  {\bibinfo {title} {Exploring ququart computation on a transmon using optimal
  control},}\ } (\bibinfo {year} {2023}),\ \Eprint
  {http://arxiv.org/abs/2304.11159} {arXiv:2304.11159 [quant-ph]} \BibitemShut
  {NoStop}%
\bibitem [{\citenamefont {Bollob{\'a}s}(1998)}]{bollobas1998modern}%
  \BibitemOpen
  \bibfield  {author} {\bibinfo {author} {\bibfnamefont {B.}~\bibnamefont
  {Bollob{\'a}s}},\ }\href@noop {} {\emph {\bibinfo {title} {Modern graph
  theory}}},\ Vol.\ \bibinfo {volume} {184}\ (\bibinfo  {publisher} {Springer
  Science \& Business Media},\ \bibinfo {year} {1998})\BibitemShut {NoStop}%
\bibitem [{\citenamefont {Chen}\ \emph {et~al.}(2024)\citenamefont {Chen},
  \citenamefont {Zhang}, \citenamefont {Jiang},\ and\ \citenamefont
  {Flammia}}]{chen2024efficient}%
  \BibitemOpen
  \bibfield  {author} {\bibinfo {author} {\bibfnamefont {S.}~\bibnamefont
  {Chen}}, \bibinfo {author} {\bibfnamefont {Z.}~\bibnamefont {Zhang}},
  \bibinfo {author} {\bibfnamefont {L.}~\bibnamefont {Jiang}}, \ and\ \bibinfo
  {author} {\bibfnamefont {S.~T.}\ \bibnamefont {Flammia}},\ }\href@noop {}
  {\bibfield  {journal} {\bibinfo  {journal} {arXiv preprint arXiv:2410.03906}\
  } (\bibinfo {year} {2024})}\BibitemShut {NoStop}%
\bibitem [{\citenamefont {Zhang}\ \emph {et~al.}(2024)\citenamefont {Zhang},
  \citenamefont {Chen}, \citenamefont {Liu},\ and\ \citenamefont
  {Jiang}}]{zhang2024generalized}%
  \BibitemOpen
  \bibfield  {author} {\bibinfo {author} {\bibfnamefont {Z.}~\bibnamefont
  {Zhang}}, \bibinfo {author} {\bibfnamefont {S.}~\bibnamefont {Chen}},
  \bibinfo {author} {\bibfnamefont {Y.}~\bibnamefont {Liu}}, \ and\ \bibinfo
  {author} {\bibfnamefont {L.}~\bibnamefont {Jiang}},\ }\href@noop {}
  {\bibfield  {journal} {\bibinfo  {journal} {arXiv preprint arXiv:2406.02669}\
  } (\bibinfo {year} {2024})}\BibitemShut {NoStop}%
\bibitem [{\citenamefont {Endo}\ \emph {et~al.}(2018)\citenamefont {Endo},
  \citenamefont {Benjamin},\ and\ \citenamefont {Li}}]{endo2018practical}%
  \BibitemOpen
  \bibfield  {author} {\bibinfo {author} {\bibfnamefont {S.}~\bibnamefont
  {Endo}}, \bibinfo {author} {\bibfnamefont {S.~C.}\ \bibnamefont {Benjamin}},
  \ and\ \bibinfo {author} {\bibfnamefont {Y.}~\bibnamefont {Li}},\ }\href@noop
  {} {\bibfield  {journal} {\bibinfo  {journal} {Physical Review X}\ }\textbf
  {\bibinfo {volume} {8}},\ \bibinfo {pages} {031027} (\bibinfo {year}
  {2018})}\BibitemShut {NoStop}%
\bibitem [{\citenamefont {Ferracin}\ \emph {et~al.}(2024)\citenamefont
  {Ferracin}, \citenamefont {Hashim}, \citenamefont {Ville}, \citenamefont
  {Naik}, \citenamefont {Carignan-Dugas}, \citenamefont {Qassim}, \citenamefont
  {Morvan}, \citenamefont {Santiago}, \citenamefont {Siddiqi},\ and\
  \citenamefont {Wallman}}]{ferracin2022efficiently}%
  \BibitemOpen
  \bibfield  {author} {\bibinfo {author} {\bibfnamefont {S.}~\bibnamefont
  {Ferracin}}, \bibinfo {author} {\bibfnamefont {A.}~\bibnamefont {Hashim}},
  \bibinfo {author} {\bibfnamefont {J.-L.}\ \bibnamefont {Ville}}, \bibinfo
  {author} {\bibfnamefont {R.}~\bibnamefont {Naik}}, \bibinfo {author}
  {\bibfnamefont {A.}~\bibnamefont {Carignan-Dugas}}, \bibinfo {author}
  {\bibfnamefont {H.}~\bibnamefont {Qassim}}, \bibinfo {author} {\bibfnamefont
  {A.}~\bibnamefont {Morvan}}, \bibinfo {author} {\bibfnamefont {D.~I.}\
  \bibnamefont {Santiago}}, \bibinfo {author} {\bibfnamefont {I.}~\bibnamefont
  {Siddiqi}}, \ and\ \bibinfo {author} {\bibfnamefont {J.~J.}\ \bibnamefont
  {Wallman}},\ }\href {\doibase 10.22331/q-2024-07-15-1410} {\bibfield
  {journal} {\bibinfo  {journal} {{Quantum}}\ }\textbf {\bibinfo {volume}
  {8}},\ \bibinfo {pages} {1410} (\bibinfo {year} {2024})}\BibitemShut
  {NoStop}%
\end{thebibliography}%

\clearpage

\appendix
\setcounter{table}{0}
\renewcommand{\thetable}{A\arabic{table}}

\setcounter{figure}{0}
\renewcommand{\thefigure}{A\arabic{figure}}

%%%%%%%%%%%%%%%%%%%%%% Error Generators %%%%%%%%%%%%%%%%%%%%%%% 
\section{Error Generators}\label{sec:error_gen}

In Sec.~\ref{sec:rep_quant_proc}, we introduced various representations for modeling errors acting on quantum processes. These representations, including Kraus operators, transfer matrices, process matrices, and Choi matrices, are able to capture arbitrary CPTP gate errors. Unfortunately, it can be difficult to tease apart an arbitrary \ac{CPTP} map and relate components of observed error matrices to known error sources, such as qubit fluctuations or systematic calibration errors in gates. A more immediate connection can be made using \emph{error generators}. Error generators are designed to capture and categorize those \textit{small} Markovian errors in quantum gates that appear in reasonably well behaved quantum computers. In what follows, we denote the ideal transfer matrix of a gate $G$ to be $\Lambda_G$. Now, using the composition property of transfer matrices (see Sec.~\ref{sec:superop}), we may write the transfer matrix of the noisy quantum gate $\tilde{\Lambda}_G$ as
\begin{equation}\label{eq:post_gate_error}
    \tilde{\Lambda}_G = \Lambda_\E \Lambda_G ~,
\end{equation}
where $\Lambda_\E$ is the transfer matrix (e.g., \ac{PTM}) that captures the noise and errors impacting $\Lambda_G$ \footnote{While we model our errors using a \emph{post-gate} error matrix, it is equally valid to model errors using a \emph{pre-gate} error matrix, or in some cases one which occurs concurrently with the gate \cite{wallman2018randomized}.}. If the error is small, then the noisy gate is close to the target unitary (i.e., $\tilde{\Lambda}_G \approx \Lambda_G$) and $\vert\vert \Lambda_\E - \mathbb{I} \vert\vert \ll 1$ . By taking the $\log(\Lambda_\E)$, we can learn how much $\Lambda_\E$ deviates from $\mathbb{I}$, or rather how much $\tilde{\Lambda}_G$ deviates from $\Lambda_G$. Thus, we define the error generator \cite{blume2022taxonomy} of $\Lambda_\E$ to be
\begin{equation}\label{eq:error_gen}
    \mathcal{L} = \log(\Lambda_\E) \simeq \Lambda_\E - \mathbb{I} ~,
\end{equation}
such that we may write \eq\ref{eq:post_gate_error} as
\begin{equation}\label{eq:post_gate_error_gen}
    \tilde{\Lambda}_G = e^{\mathcal{L}} \Lambda_G ~.
\end{equation}
Here, $\mathcal{L}$ is the generator of $\Lambda_\E$ in analogy with how Hamiltonians are the generators of unitary transformations. 

In order to ensure that $\mathcal{L}$ generates a CPTP map, it must be expressible as a Lindblad superoperator \cite{lindblad1976generators}. Expanding the Lindblad equation in a basis of Pauli operators $\{P_j\}$ we have:
\begin{align}\label{eq:lindblad}
    \mathcal{L}(\rho) = \sum_j \epsilon_j [P_j,\rho] 
        + \sum_{j,k} h_{j,k} \left(
            P_j \rho P_k \rho P_j - \frac{1}{2} \left\{ P_j^\dagger P_k,\rho\right\} 
        \right) ~.
\end{align} 
Here, $\epsilon_j \in \mathbb{R}$ characterizes the size of unitary errors, and $h_{j,k}$ is positive semi-definite and quantifies the type and rate of the dissipative dynamics. 

While the (H)amiltonian error rates ($\epsilon_j$) that describe the unitary dynamics in Eq.~\ref{eq:lindblad} are often relatively easy to understand (e.g., a Pauli-$X$ error on an $X$ gate corresponds to an over- or under-rotation), the dissipative part can be more challenging. So, the error generator framework splits those errors into symmetric (or \textit{stochastic}) and antisymmetric (or \textit{active}) components. The (A)ctive components can be derived from couplings to quantum degrees of freedom and are responsible for, e.g., non-unital effects such as amplitude damping. The stochastic components are precisely those that might arise from fluctuating Hamiltonian terms. Because stochastic \textit{Pauli} errors appear so frequently (e.g., in Pauli frame randomization, randomized compiling, and models for quantum error correction), the symmetric sector is further decomposed into a Pauli (S)tochastic and stochastic (C)orrelation sectors corresponding to the diagonal and off-diagonal symmetric terms. Thus, we define the following elementary generators $\{ H, S, C, A \}$:
\begin{align}\label{eq:elem_generators}
    H_P(\rho) &= -i [P,\rho] ~, \\
    S_P(\rho) &= P\rho P - \rho ~, \\
    C_{P,Q}(\rho) &= P \rho Q + Q \rho P - \frac{1}{2} \{ \{P,Q\}, \rho\} ~, \\
    A_{P,Q}(\rho) &= i \left( P \rho Q - Q \rho P + \frac{1}{2} \{ [P,Q], \rho\} \right) ~.
\end{align}
They form a complete basis for superoperators. An error generator $\mathcal{L}$ is a Lindbladian superoperator that generates coherent, stochastic, and/or nonunital gate errors. Therefore, we may write $\mathcal{L}$ as a linear combination of elementary error generators,
\begin{align}
    \mathcal{L} &= \mathcal{L}_\mathbb{H} + \mathcal{L}_\mathbb{S} + \mathcal{L}_\mathbb{C} + \mathcal{L}_\mathbb{A} ~, \\
                &= \sum_P h_P H_P \, + \, \sum_P s_P S_P ~, \\
                &\,\,\,\,\, + \sum_{P, Q>P} c_{P, Q} C_{P,Q} \, + \, \sum_{P, Q>P} a_{P, Q} A_{P,Q} ~,
\end{align}
where the coefficients $\{h, s, c, a\}$ denote the \emph{error rates} of each error process, and where all errors and their corresponding rates are indexed by one (or two) distinct Pauli operators $P$ (and $Q$). Thus, any arbitrary error generator $\mathcal{L}$ can be written as a linear combination of all of the elementary error generators. This framework enables one to quantify the rates of different errors afflicting quantum gates \cite{mkadzik2021precision}.

%%%%%%%%%%%%%%%%%%%%%%% Groups and Gate sets %%%%%%%%%%%%%%%%%%%%%%%
\section{Groups and Gate Sets}\label{sec:groups_gatesets}

\subsection{Groups}

A \emph{group} $\mathbb{G}$ is a mathematical set of operational elements $\{ G_i \}$ that satisfy the following basic properties:
\begin{enumerate}
    \item Closure: $\forall ~ G_i, G_j \in \mathbb{G}$, $G_i \cdot G_j = G_k \in \mathbb{G}$.
    \item Associativity: $(G_i \cdot G_j) \cdot G_k = G_i \cdot (G_j \cdot G_k)$.
    \item Identity element: $\exists ~ \mathbb{I} \in \mathbb{G}$ s.t.~$\mathbb{I} \cdot G_i = G_i \cdot \mathbb{I} = G_i$, $\forall ~ G_i \in \mathbb{G}$.
    \item Inverse element: $\forall ~ G_i \in \mathbb{G}$, $\exists ~ G_i^{-1} \in \mathbb{G}$ s.t.~$G_i \cdot G_i^{-1} = G_i^{-1} \cdot G_i = \mathbb{I}$.
\end{enumerate}
Below, we introduce some important groups and gate sets in quantum computing, and highlight the properties that distinguish each from the others.

%%%%%%%%%%%%%%%%%%%%%%% The Pauli Group %%%%%%%%%%%%%%%%%%%%%%%
\subsection{The Pauli Group}\label{sec:pauli_group}

The $n$-qubit \emph{Pauli group}, denoted $\mathbb{P}_n$, is the set of Pauli operators formed by the tensor product of all combinations of single-qubit Paulis for $n$ qubits, multiplied by factors of $\pm 1$ or $\pm i$:
\begin{equation}
    \mathbb{P}_n = \{\pm 1, \pm i\} \times \{I, X, Y, Z\}^{\otimes n} ~,
\end{equation}
where
\begin{align}
     I &= \sigma_0 = \begin{pmatrix} 1 & 0 \\ 0 & 1 \end{pmatrix} ~, \\
     X &= \sigma_x = \begin{pmatrix} 0 & 1 \\ 1 & 0 \end{pmatrix} ~, \\
     Y &= \sigma_y = \begin{pmatrix} 0 & -i \\ i & 0 \end{pmatrix} ~, \\
     Z &= \sigma_z = \begin{pmatrix} 1 & 0 \\ 0 & -1 \end{pmatrix}
\end{align}
are the single-qubit Pauli operators. The $n$-qubit Pauli operators have a number of helpful properties, namely:
\begin{enumerate}
    \item they form a projective group under matrix multiplication, 
    \item they are unitary and Hermitian,
    \item they are a trace-orthogonal basis for the space of operators,
    \item they correspond to natural Hamiltonians, and
    \item they form a unitary 1-design in $d = 2^n$-dimensional Hilbert space (see Sec.~\ref{sec:unitary_t_designs}).
\end{enumerate}

%%%%%%%%%%%%%%%%%%%%%%% The Clifford Group %%%%%%%%%%%%%%%%%%%%%%%
\subsection{The Clifford Group}\label{sec:clifford_group}

The $n$-qubit \emph{Clifford group} \cite{gottesman1998theory}, denoted $\mathbb{C}_n$, is the set of operations that \emph{normalize} the $n$-qubit Pauli group. This means that any element from the Clifford group $C \in \mathbb{C}_n$ maps each Pauli operator to another Pauli operator under conjugation:
\begin{equation}
    \forall ~ C \in \mathbb{C}_n: C P C^\dagger \mapsto P' \in \mathbb{P}_n, ~ \forall ~ P \in \mathbb{P}_n ~.
\end{equation}
Typical examples of Clifford gates which are not in the Pauli group are the Hadamard $H$, $S = \sqrt{Z}$, CNOT, SWAP, and iSWAP gates. In fact, the subgroup of Clifford gates $\{H, S, \text{CNOT}\}$ is sufficient to generate the full Clifford group between any pair of qubits. In Tables \ref{tab:H_conjugation} and \ref{tab:S_conjugation}, we show the action of all single-qubit Pauli operators under conjugation by the Hadamard $H$ and $S$ gates, respectively. In Tables \ref{tab:CNOT_conjugation} and \ref{tab:iSWAP_conjugation}, we show the action of all two-qubit Pauli operators under conjugation by the CNOT and iSWAP gates, respectively. In all cases, we find that the resulting gate is a Pauli which belongs to $\mathbb{P}_n$.

\begin{table}
    \centering
    \begin{minipage}[t]{.40\textwidth}
        \centering
        \begin{tabular}{ | c | c |}
            \hline
            $P$ & $HPH^\dagger$ \\ \hline \hline
            $I$ & $I$ \\
            $X$ & $Z$ \\
            $Y$ & -$Y$ \\
            $Z$ & $X$ \\
            \hline
        \end{tabular}
        \caption[$\mathbb{P}_1$ under $H$ conjugation.]{\textbf{$\mathbb{P}_1$ under $H$ conjugation.}}
        \label{tab:H_conjugation}
    \end{minipage}\qquad
    \vskip.5\baselineskip
    \begin{minipage}[t]{.40\textwidth}
        \centering
        \begin{tabular}{ | c | c |}
            \hline
            $P$ & $SPS^\dagger$ \\ \hline \hline
            $I$ & $I$ \\
            $X$ & $Y$ \\
            $Y$ & -$X$ \\
            $Z$ & $Z$ \\
            \hline
        \end{tabular}
        \caption[$\mathbb{P}_1$ under $S$ conjugation.]{\textbf{$\mathbb{P}_1$ under $S$ conjugation.}}
        \label{tab:S_conjugation}
    \end{minipage}
\end{table}

\begin{table}
    \centering
    \begin{minipage}[h]{.40\textwidth}
        \centering
        \begin{tabular}{ | c | c |}
            \hline
            $P$ & $\text{CNOT}(P)\text{CNOT}^\dagger$ \\ \hline \hline
            $I \otimes I$ & $I \otimes I$ \\
            $I \otimes X$ & $I \otimes X$ \\
            $I \otimes Y$ & $Z \otimes Y$ \\
            $I \otimes Z$ & $Z \otimes Z$ \\
            $X \otimes I$ & $X \otimes X$ \\
            $X \otimes X$ & $X \otimes I$ \\
            $X \otimes Y$ & $Y \otimes Z$ \\
            $X \otimes Z$ & -$Y \otimes Y$ \\
            $Y \otimes I$ & $Y \otimes X$ \\
            $Y \otimes X$ & $Y \otimes I$ \\
            $Y \otimes Y$ & -$X \otimes Z$ \\
            $Y \otimes Z$ & $X \otimes Y$ \\
            $Z \otimes I$ & $Z \otimes I$ \\
            $Z \otimes X$ & $Z \otimes X$ \\
            $Z \otimes Y$ & $I \otimes Y$ \\
            $Z \otimes Z$ & $I \otimes Z$ \\
            \hline
        \end{tabular}
        \caption[$\mathbb{P}_2$ under CNOT conjugation.]{\textbf{$\mathbb{P}_2$ under CNOT conjugation.}}
        \label{tab:CNOT_conjugation}
    \end{minipage}\qquad
    \vskip.5\baselineskip
    \begin{minipage}[h]{.40\textwidth}
        \centering
        \begin{tabular}{ | c | c |}
            \hline
            $P$ & $\text{iSWAP}(P)\text{iSWAP}^\dagger$ \\ \hline \hline
            $I \otimes I$ & $I \otimes I$ \\
            $I \otimes X$ & $Y \otimes Z$ \\
            $I \otimes Y$ & -$X \otimes Z$ \\
            $I \otimes Z$ & $Z \otimes I$ \\
            $X \otimes I$ & $Z \otimes Y$ \\
            $X \otimes X$ & $X \otimes X$ \\
            $X \otimes Y$ & $Y \otimes X$ \\
            $X \otimes Z$ & $I \otimes Y$ \\
            $Y \otimes I$ & -$Z \otimes X$ \\
            $Y \otimes X$ & $X \otimes Y$ \\
            $Y \otimes Y$ & $Y \otimes Y$ \\
            $Y \otimes Z$ & -$I \otimes X$ \\
            $Z \otimes I$ & $I \otimes Z$ \\
            $Z \otimes X$ & $Y \otimes I$ \\
            $Z \otimes Y$ & -$X \otimes I$ \\
            $Z \otimes Z$ & $Z \otimes Z$ \\
            \hline
        \end{tabular}
        \caption[$\mathbb{P}_2$ under iSWAP conjugation.]{\textbf{$\mathbb{P}_2$ under iSWAP conjugation.}}
        \label{tab:iSWAP_conjugation}
    \end{minipage}
\end{table}

The single-qubit Clifford group $\mathbb{C}_1$ contains 24 single-qubit gates; these include any integer number of $\pi/2$ rotations about any of the six cardinal axes of the Bloch sphere ($\pm \hat{x}$, $\pm \hat{y}$, and $\pm \hat{z}$), which includes all single-qubit Pauli gates ($\mathbb{P}_n \subset \mathbb{C}_n$). The size of the $n$-qubit Clifford group is given by \cite{crooks2020gates}:
\begin{equation}
    |\mathbb{C}_n| = 2^{n^2 + 2n} \prod_{j = 1, n} 4^j - 1 ~.
\end{equation}
For example, the two-qubit Clifford group contains 11,520 elements, the three-qubit Clifford group contains 92,897,280 elements, the four-qubit Clifford group contains 12,128,668,876,800 elements, etc.

The Clifford group holds a special place in quantum computing. According to the \emph{Gottesman-Knill theorem} \cite{gottesman1998heisenberg}, quantum circuits containing only Clifford gates and Pauli basis measurements can be efficiently simulated in polynomial time on a classical computer. Therefore, Clifford circuits are insufficient to realize the full potential of quantum computers over classical computers. In fact, in order to perform universal quantum computation, one requires a gate set which also contains a non-Clifford gate, such as the $T = \sqrt{S}$ gate (sometimes called the ``$\pi/8$'' gate for historical reasons). Nonetheless, Clifford gates are ubiquitous in quantum computations and are essential to a number of important applications. For example, stabilizer codes in quantum error correction use Clifford gates for encoding and decoding. Additionally, benchmarking procedures for measuring average error rates of quantum gate sets, such as randomized benchmarking (Sec.~\ref{sec:standardrb}), are constructed entirely of Clifford gates. Importantly, Clifford gates are used in these protocols because they form a unitary 2-design (and sometimes a unitary 3-design; see Sec.~\ref{sec:unitary_t_designs}).

%%%%%%%%%%%%%%%%%%%%%%% The Weyl and Gell-Mann Bases %%%%%%%%%%%%%%%%%%%%%%%
\subsection{Qudit Groups and Bases}\label{sec:weyl_gellman}

The qubit Pauli group serves as a natural basis for analyzing qubit systems. However, no set of operators with the same properties exists for higher dimensional systems. Nevertheless, we can define two natural generalizations of the Pauli operators to higher dimensions, the \emph{Weyl} and \emph{Gell-Mann} operators that, taken together, satisfy all of the properties of the single-qubit Pauli group. In general, the Weyl operators allow one to more naturally generalize the machinery of qubit-based benchmarking routines. In contrast, the Gell-Mann matrices correspond more immediately to the underlying physical operations performed on a qudit based quantum processor, such as Rabi oscillations and Z gates in a two-level subspace of the qudit. In what follows, we refer to $D$ as the dimension of the qudit, and $d = D^n$ as the dimension of the Hilbert space for $n$ qudits.

%%%%%%%%%%%%%%%%%%%%%%% The Gell-Mann basis %%%%%%%%%%%%%%%%%%%%%%%
\subsubsection{The Gell-Mann Basis}

To construct the Gell-Mann operators, we can begin by embedding the single-qubit Pauli operators into two-dimensional subspaces of the higher-dimensional qudit space. Specifically, for a $D$-dimensional qudit, we can define these operators as
\begin{align}
    X^{jk} = \ketbra{j}{k} + \ketbra{k}{j} ~,\\
    Y^{jk} = i\ketbra{j}{k} - i \ketbra{k}{j} ~,\\
    Z^{jk} = \ketbra{j} - \ketbra{k} ~,
\end{align}
where $0 \leq j \leq k \leq D$. We note that while the $Z^{jk}$ are Hermitian operators, they are not linearly independent and thus cannot serve as a sufficient basis for qudit tomography. We can therefore extend the set $\{ X^{jk}, Y^{jk}: 0 \leq j <k <D \}$ to a trace-orthogonal basis with the inclusion of additional diagonal operators:
\begin{equation}
    W^j = -j\ketbra{j}{j} + \sum_{0\leq k < j}\ketbra{k}{k}
\end{equation}
for $1 \leq j < D$. The combined set $\mathbb{G}_{D} = \{I, X^{jk}, Y^{jk}: 0 \leq j <k <D\} \cup \{ W^j: 1 \leq j < D \}$ forms the \emph{Gell-Mann basis} for qudit dimension $D$. Like the qubit Pauli matrices, every element of the Gell-Mann group is both traceless and Hermitian, and thus serves as a suitable choice for constructing qudit transfer matrices.

%%%%%%%%%%%%%%%%%%%%%%% The Weyl Group %%%%%%%%%%%%%%%%%%%%%%%
\subsubsection{The Weyl Group}

Having constructed the qudit Gell-Mann basis, we now turn our attention to generalizing the qubit Pauli operators over the entire qudit space rather than embedding them in two-level subspaces of the qudit. This is known as the \emph{Weyl group}, from which we can generalize qubit-based quantum algorithms and codes to qudits, as well as generalize the Clifford group to higher dimensions. We begin by defining $\mathbb{Z}_D = \{0, 1, \dots, D-1 \}$, the additive group of the integers modulo $D$. Using $\mathbb{Z}_D$, we can generalize the qubit $X$ and $Z$ operators to the qudit space as
\begin{align}
    X &= \sum_{j \in \mathbb{Z}_D} \ketbra{j \oplus_D 1}{j} ~, \\
    Z &= \sum_{j \in \mathbb{Z}_D} \exp(\frac{2\pi i}{D}j)\ketbra{j}{j} ~,
\end{align}
where $\oplus_D$ denotes addition modulo $D$. We note here that both $X$ and $Z$ compose to the identity under $D$ rounds of self-multiplication, which is the natural generalization of the qubit $X$ and $Z$ operators squaring to the identity. The Weyl basis follows as a trace-orthogonal basis over $\mathbb{C}^{D\times D}$ defined as $\mathbb{W}_D = \{ W_{xz} = X^x Z^z: x,z \in \mathbb{Z}_D \}$. Often in the literature, the qudit $X$ operator is referred to as the ``shift'' operator as it increments the qudit state modulo $D$, and the qudit $Z$ operator as the ``clock'' operator as it applies phases corresponding to multiples of the $D$-th root of unity. Finally, we can define the $n$-qudit Weyl group by taking the $n$-fold tensor product of all single-qudit Weyl matrices: $\mathbb{W}_{D,n} = \mathbb{W}_D^{\otimes n}$. It is worth noting that one can define multi-qudit gates directly from the definitions of the Weyl matrices, similar to how the Hamiltonians for two-qubit gates are often defined in terms of $n$-qubit Paulis.

%%%%%%%%%%%%%%%%%%%%%%% The $n$-Qudit Clifford Group %%%%%%%%%%%%%%%%%%%%%%%
\subsubsection{The $n$-Qudit Clifford Group}

One important caveat about the Weyl basis is that, owing to the fact that it is not closed under multiplication, it is not a proper group. However, every element of the closure of the Weyl basis is related to an element of the Weyl basis up to an overall phase. We can therefore get rid of this overall phase by considering solely the adjoint action of the Weyl operators, and therefore defining a proper group we refer to as the ``extended Weyl group," $\mathbb{E}\mathbb{W}_D = \mathsf{U}(1)\mathbb{W}_D$ and the ``extended $n$-qudit Weyl group" as $\mathbb{E}\mathbb{W}_{D,n} = \mathsf{U}(1)\mathbb{W}_{D,n}$, where $\mathsf{U}(1)$ is the 1-dimension unitary group, allowing for arbitrary phases. Finally, with all this formalism defined, we can naturally define the $n$-qudit Clifford group to be the set of operators that normalize the extended Weyl group, or stated explicitly, the set $\mathbb{C}_{D,n}$ where:
\begin{equation}
    \mathbb{C}_{D,n} = \{ U \in \mathsf{U}(D^n) : U \mathbb{E}\mathbb{W}_{D,n}U^\dag = \mathbb{E}\mathbb{W}_{D,n} \} ~.
\end{equation}

%%%%%%%%%%%%%%%%%%%%%%% Randomization and Twirling %%%%%%%%%%%%%%%%%%%%%%% 
\section{Randomization and Twirling} \label{sec:twirling}

The notion of \emph{twirling} a quantum channel is a central component of many benchmarking methods based on randomized gate sampling. The basic concept of twirling is to average a quantum channel $\E$ over some unitary group. This allows one to measure the average performance of a quantum operation (e.g., a gate) for different combinations of input and output states, while reducing the resources needed for measuring the process fidelity of a gate compared to full quantum process tomography. As we will see in this section, twirling maps a dense \ac{CPTP} matrix (modeled as a $d^2 \otimes d^2$ superoperator, where $d = 2^n$ for $n$ qubits) into a block diagonal matrix, effectively condensing information about the physical process into the eigenvalues of the matrix. Below, we give a precise definition of twirling, and discuss the difference between twirling over the Pauli and Clifford groups.

%%%%%%%%%%%%%%%%%%%%%%% The Haar Measure %%%%%%%%%%%%%%%%%%%%%%% 
\subsection{The Haar Measure}\label{sec:haar}

When twirling a quantum channel $\E$ over a unitary group in $d$ dimensions, $\mathsf{U}(d)$, it is necessary to uniformly sample at random unitaries from $\mathsf{U}(d)$. The \emph{uniform Haar measure}, denoted $\mu(U)$, is mathematical measure that is unique to each locally compact topological group which assigns equal weights to all elements of the group. $\mu(U)$ defines an integral over $\mathsf{U}(d)$ that is invariant under group transformations, and is normalized such that the total measure of the group is $\int d\mu = 1$. The uniform Haar measure defines how different elements of $\mathsf{U}(d)$ are weighted over unitary space, and therefore can be used to integrate functions over all of $\mathsf{U}(d)$. 

To better understand the role that the uniform Haar measure plays in twirling, consider the simple example of the integral of some function $f(r, \theta, \phi)$ in spherical coordinates over all of $\mathbb{R}^3$,
\begin{equation}
    V = \iiint_{\mathbb{R}^3} f(r, \theta, \phi) r^2 \sin(\theta) dr d\theta d\phi ~.
\end{equation}
Here, $r^2 \sin(\theta) dr d\theta d\phi$ is the Lebesgue \emph{measure} over $\mathbb{R}^3$, which ensures that the integral is taken uniformly over all of $\mathbb{R}^3$. For the special unitary group in 2 dimensions, $\mathsf{SU}(2)$ --- the relevant group for single-qubit gates --- the Haar measure is given as
\begin{equation}
    d\mu = \sin(\theta) d\theta d\phi d\gamma ~,
\end{equation}
which is nearly identical to the Haar measure for a sphere, except it contains no radial component and instead includes an additional phase term $\gamma$ which comes from the U$_3$ parametrization of single-qubit rotations (see \eq\ref{eq:zxzxz}). Although the exact Haar measure in $\mathsf{SU}(d)$ is needed for integration over the entire unitary space that is relevant to qudits, some knowledge of the Haar measure is sufficient for the task of twirling, which only requires that we sample enough points uniformly at random that \emph{approximate} $\mathsf{SU}(d)$.

%%%%%%%%%%%%%%%%%%%%%%% Unitary $t$-Designs %%%%%%%%%%%%%%%%%%%%%%% 
\subsection{Unitary $t$-Designs}\label{sec:unitary_t_designs}

Twirling involves averaging a channel $\E$ over a unitary group $\mathsf{U}(d)$. However, any continuous group has an infinite number of elements. For example, in the case of $\mathsf{SU}(2)$, there are an infinite number of points on the surface of the Bloch sphere. Therefore, it is impossible to average $\E$ over \emph{all} of $\mathsf{U}(d)$. Instead, one typically samples unitaries from some subgroup $\mathbb{G} \subset \mathsf{U}(d)$ which \emph{approximates} $\mathsf{U}(d)$. This forms the basis of what is called a \emph{unitary $t$-design}, which describes a unitary group that simulates the statistical properties of uniformly distributed Haar-random matrices \cite{pozniak1998composed} up to the $t$'th moment.

Classically, the notion of spherical $t$-designs defines a finite collection of points on the surface of a unit sphere which provide a ``good'' approximation to the integral over the entire unit sphere \cite{bannai2009survey}. \emph{Unitary} $t$-designs are the extension of spherical $t$-designs to the quantum domain, for which we desire to reproduce the basic properties of an entire unitary group $\mathsf{U}(d)$. Formally, a unitary $t$-design in $d$ dimensions is a set of unitary operators $\{U_1, ..., U_K\}$ such that the sum over every polynomial $P_{t,t}(U_k) = U_k^{\otimes t} \otimes (U_k^*)^{\otimes t}$ of degree no larger than $t$ in the matrix elements of $U$ and their complex conjugates is equal to the integral of $P_{(t,t)}(U)$ over $\mathsf{U}(d)$,
\begin{equation}
    \frac{1}{K} \sum_{k=1}^K P_{(t,t)}(U_k) = \int_{\mathsf{U}(d)} d\mu(U) P_{(t,t)}(U) ~.
\end{equation}
Very heuristically, unitaries in a $t$-design are evenly spaced around the $d$-dimensional unit sphere defining $\mathsf{U}(d)$, with larger values of $t$ defining more densely spaced points. For example, the $d$-dimensional Pauli group forms a unitary 1-design, and the $d$-dimensional Clifford group ($\mathbb{C}_{D,n}$) forms a unitary 2-design when the qudit dimension $D$ is prime \footnote{The Clifford group also forms a unitary 3-design only when the qudit dimension $D=2$ \cite{webb2015clifford, zhu2017multiqubit, graydon2021clifford}.}. According to \R\cite{gross2007evenly}, a set of unitaries $\{ U_k \}_{k=1}^K$ forms a unitary 2-design iff
\begin{equation}
    \frac{1}{K^2} \sum_{k, k^\prime = 1}^K \left| \Tr(U_{k^\prime}^\dagger U_k) \right|^4 = 2 ~.
\end{equation}

%%%%%%%%%%%%%%%%%%%%%%% Twirling Quantum Channels %%%%%%%%%%%%%%%%%%%%%%% 
\subsection{Twirling Quantum Channels}\label{sec:twirling_channels}

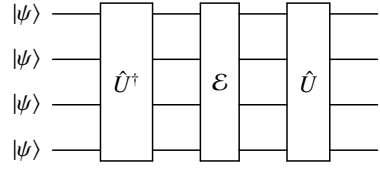
\begin{figure}[t]
    \begin{equation*}
        {\Qcircuit @C=2em @R=1em {
            \lstick{\ket{\psi}} & \multigate{3}{\hat{U}^\dagger} & \multigate{3}{\E} & \multigate{3}{\hat{U}} & \qw \\
            \lstick{\ket{\psi}} & \ghost{\hat{U}^\dagger} & \ghost{\E} & \ghost{\hat{U}} & \qw \\
            \lstick{\ket{\psi}} & \ghost{\hat{U}^\dagger} & \ghost{\E} & \ghost{\hat{U}} & \qw \\
            \lstick{\ket{\psi}} & \ghost{\hat{U}^\dagger} & \ghost{\E} & \ghost{\hat{U}} & \qw
                                }}
    \end{equation*}
    \caption[Twirling.]{\textbf{Twirling.} Twirling a quantum channel results in the map $\E \mapsto \hat{U} \circ \E \circ \hat{U}^\dagger$.}
    \label{fig:twirling}
\end{figure}

To understand how one constructs an average quantum channel by twirling, first consider a quantum channel $\E$ that represents (the error in) some quantum gate or process. Next, consider a unitary operator $\hat{U}$ that belongs to $\mathsf{U}(d)$. Suppose that $\E$ is conjugated by $\hat{U}$, mapping $\E \mapsto \hat{U} \circ \E \circ \hat{U}^\dagger$ (see \fig\ref{fig:twirling}). Using this notation, a twirled channel $\Bar{\E}$ is given by
\begin{equation}
    \Bar{\E} = \int_{\mathsf{U}(d)} d\mu(\hat{U}) \hat{U} \circ \E \circ \hat{U}^\dagger ~.
\end{equation}
Thus, $\Bar{\E}$ can be thought of as the expected value of $\E$ conjugated with all possible unitaries $\hat{U} \in \mathsf{U}(d)$. Because $\E = \E(\rho)$ is a linear map on a quantum state $\rho$, $\hat{U}$ also acts on $\rho$ by conjugation:
\begin{equation}
    \hat{U}(\rho) = U \rho U^\dagger~, ~ \hat{U}^\dagger(\rho) = U^\dagger \rho U ~.
\end{equation}
Therefore, the twirled channel of a density operator $\Bar{\E}(\rho)$ can be written
\begin{equation}\label{eq:twirled_channel}
    \Bar{\E}(\rho) = \int_{\mathsf{U}(d)} d\mu(U) U^\dagger \E (U \rho U^\dagger) U ~.
\end{equation}

As discussed in the previous section, it is not practical to twirl over all of a unitary group $\mathsf{U}(d)$. Rather, it is much more common to twirl a channel over a discrete set of unitaries that approximates some properties of $\mathsf{U}(d)$. For example, consider the channel $\E(\rho) = A \rho B$, where $\{A, B\}$ are arbitrary linear operators. Next, consider some group $\{ U_k \}_{k=1}^K$ consisting of $K$ unitary operators. In the discrete case, the twirled channel $\Bar{\E}(\rho)$ can be written as the weighted average over all $K$ operators \cite{dankert2009exact}:
\begin{equation}\label{eq:twirl_finite_group}
    \Bar{\E}(\rho) = \frac{1}{K} \sum_{k=1}^K U_k^\dagger A U_k \rho U_k^\dagger B U_k ~.
\end{equation}

\begin{figure}[t]
    \centering
    \includegraphics[width=\columnwidth]{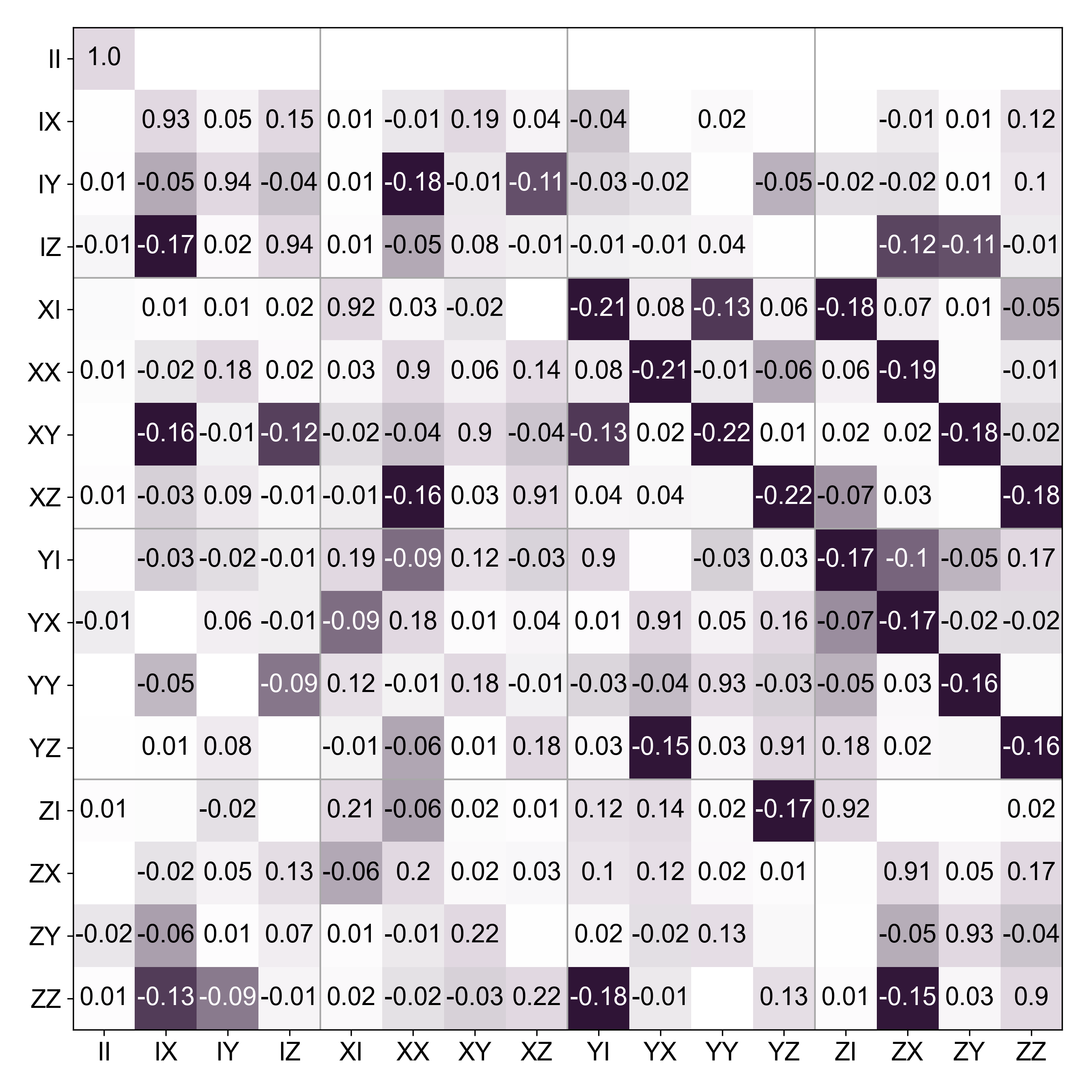}
    \caption{\textbf{PTM of a Random Two-Qubit CPTP Channel.} The color (transparency) of each cell is determined by the sign (magnitude) of each entry.}
    \label{fig:rand_ptm}
\end{figure}

Note that twirling a channel does not change the average gate fidelity or process fidelity of the channel. To see this, we replace $\E(\rho)$ in \eq\ref{eq:ave_gate_fidelity} with the twirled channel $\Bar{\E}(\rho)$ in \eq\ref{eq:twirled_channel}, and find that
\begin{align}
    F_{\mathrm{avg}}(\Bar{\E}) &= \int d\psi \int d\mu(U) \bra{\psi} U^\dagger \E (U \rho U^\dagger) U \ket{\psi} ~, \\
                          &= \int d\mu(U) \int d\psi \bra{\psi} U^\dagger \E (U \rho U^\dagger) U \ket{\psi} ~, \\
                          &= \int d\mu(U) F_{\mathrm{avg}}(\E) ~, \\
                          &= F_{\mathrm{avg}}(\E) ~,
\end{align}
where, in the second to last step we made a change of variables $\ket{\psi'} \equiv U \ket{\psi}$, and in the final step utilized the fact that $\int d\mu(U) = 1$.

\begin{figure*}
    \centering
    \begin{subfigure}[b]{0.23\textwidth}
        \centering 
        \includegraphics[width=\textwidth]{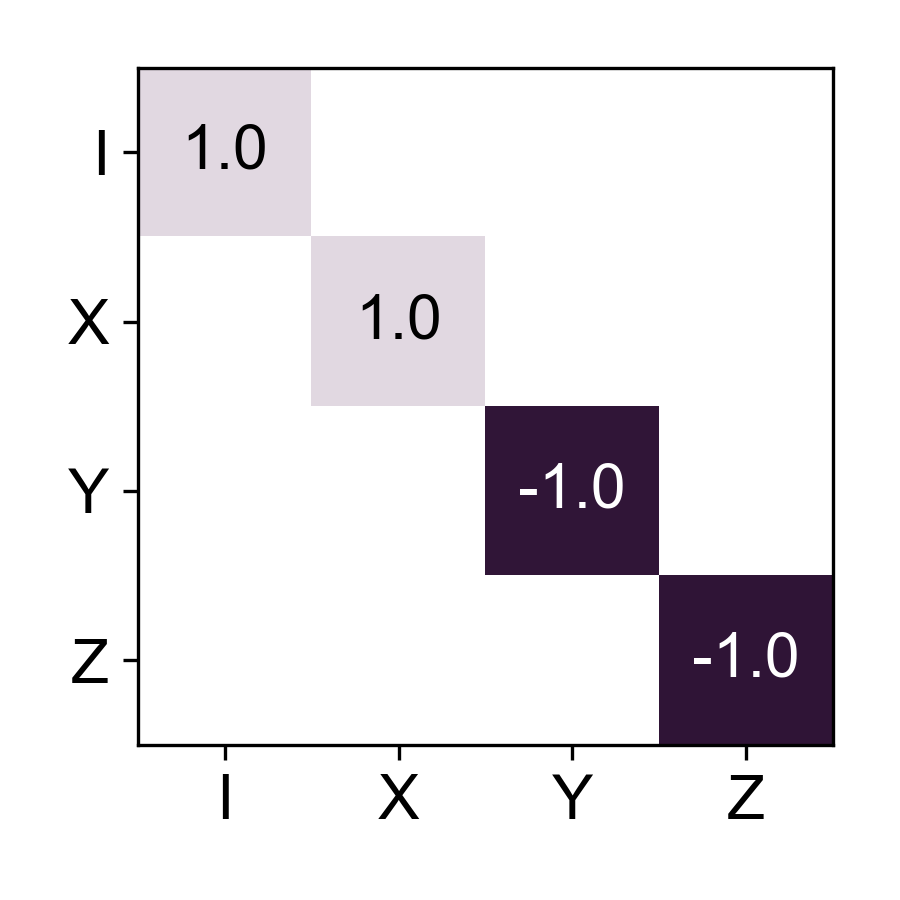}
        \caption{PTM for Pauli $X$}
    \end{subfigure}
    \hspace{5pt}
    \begin{subfigure}[b]{0.23\textwidth}
        \centering
        \includegraphics[width=\textwidth]{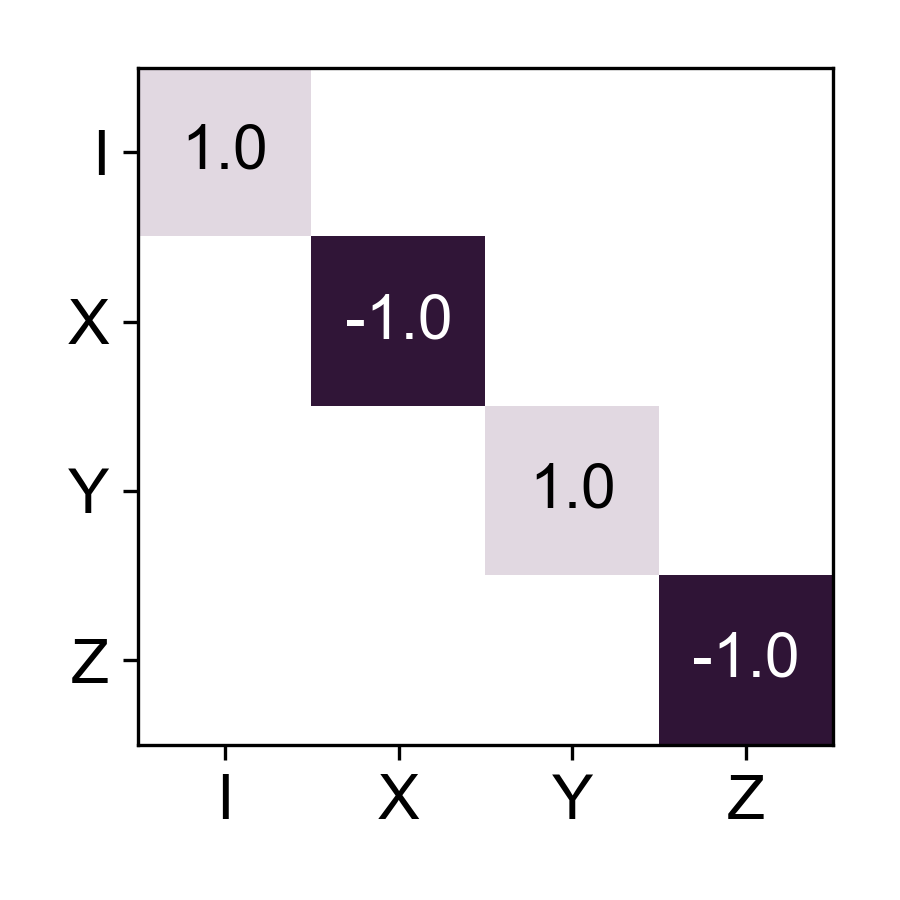}
        \caption{PTM for Pauli $Y$}
    \end{subfigure}
    \hspace{5pt}
    \begin{subfigure}[b]{0.23\textwidth}
        \centering
        \includegraphics[width=\textwidth]{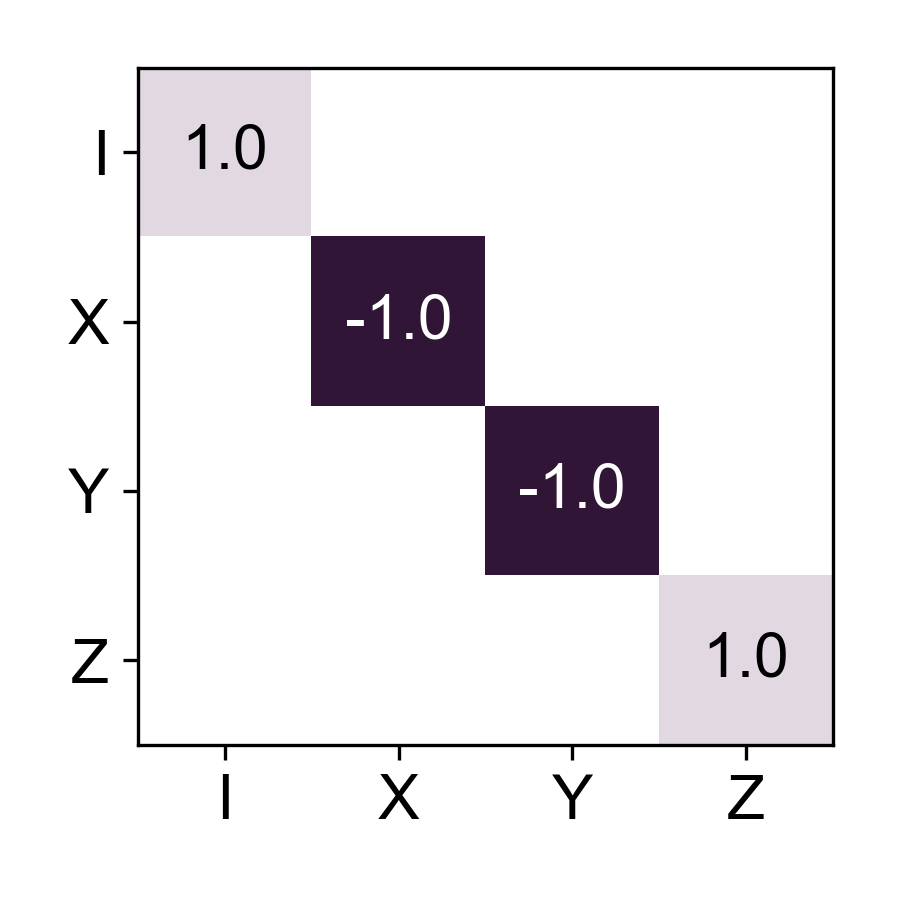}
        \caption{PTM for Pauli $Z$}
    \end{subfigure}
    \hspace{5pt}
    \begin{subfigure}[b]{0.23\textwidth}
        \centering
        \includegraphics[width=\textwidth]{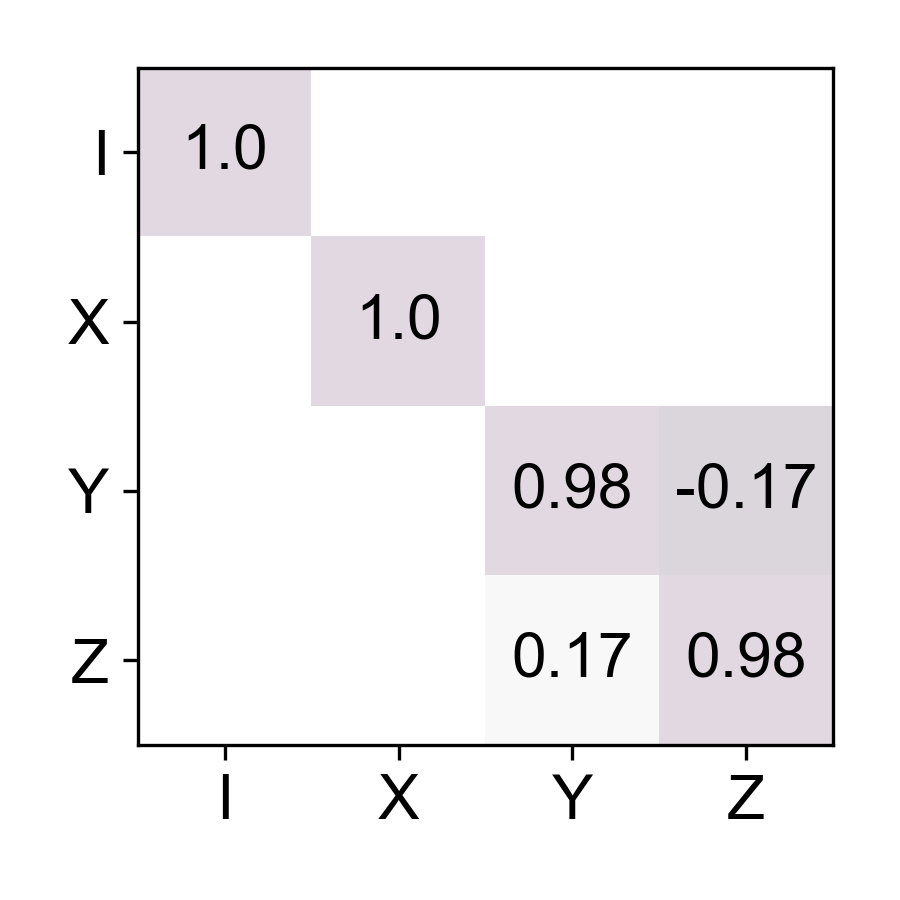}
        \caption{PTM for $R_x(\pi / 18)$}
        \label{fig:ptm_rx}
    \end{subfigure}
    \\
    \vspace{5pt}
    \begin{subfigure}[b]{0.23\textwidth}
        \centering
        \includegraphics[width=\textwidth]{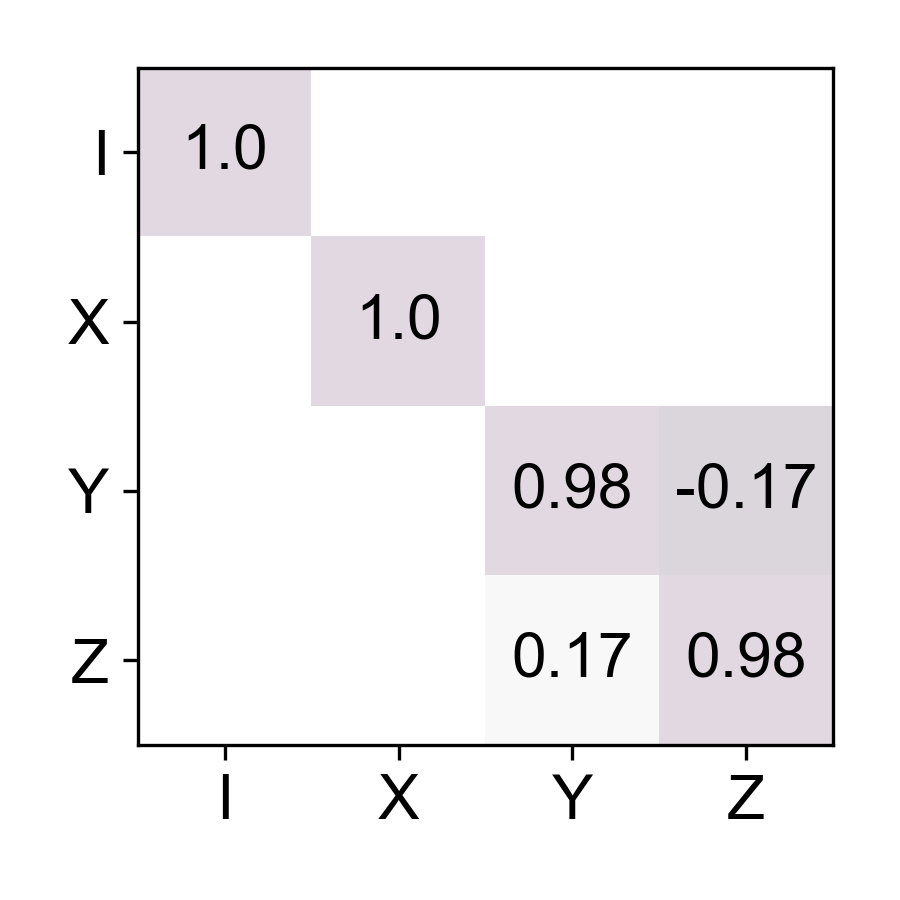}
        \caption{$X \circ R_x \circ X$}
    \end{subfigure}
    \hspace{5pt}
    \begin{subfigure}[b]{0.23\textwidth}
        \centering
        \includegraphics[width=\textwidth]{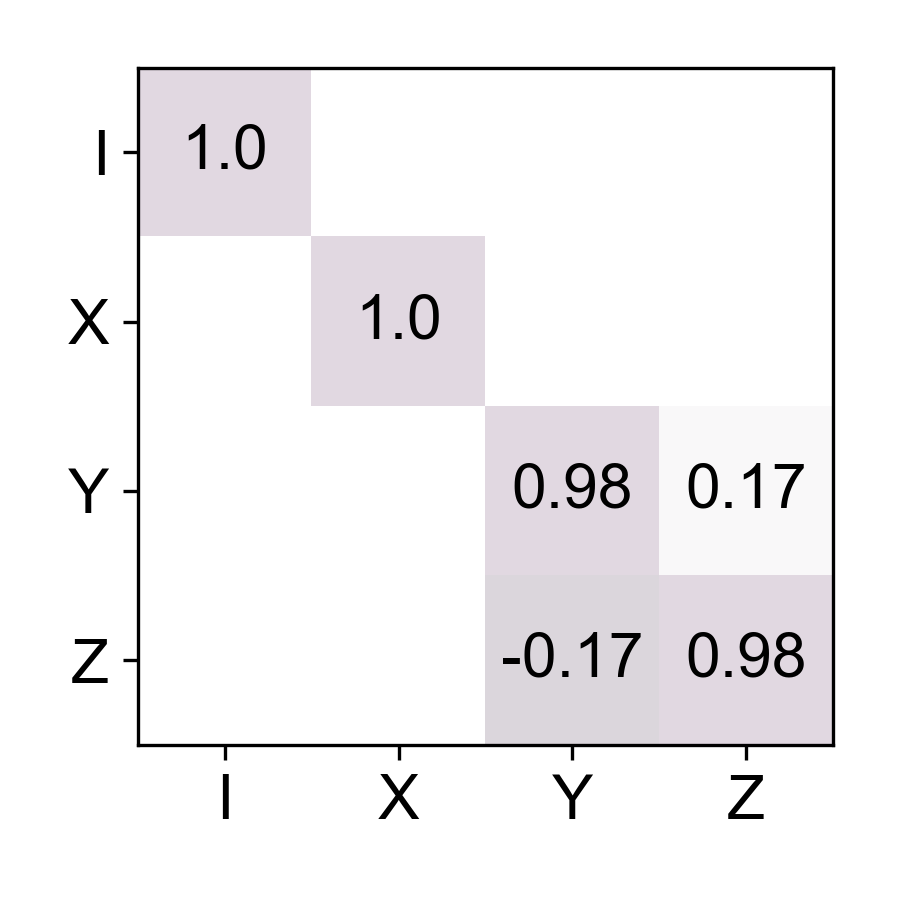}
        \caption{$Y \circ R_x \circ Y$}
    \end{subfigure}
    \hspace{5pt}
    \begin{subfigure}[b]{0.23\textwidth}
        \centering
        \includegraphics[width=\textwidth]{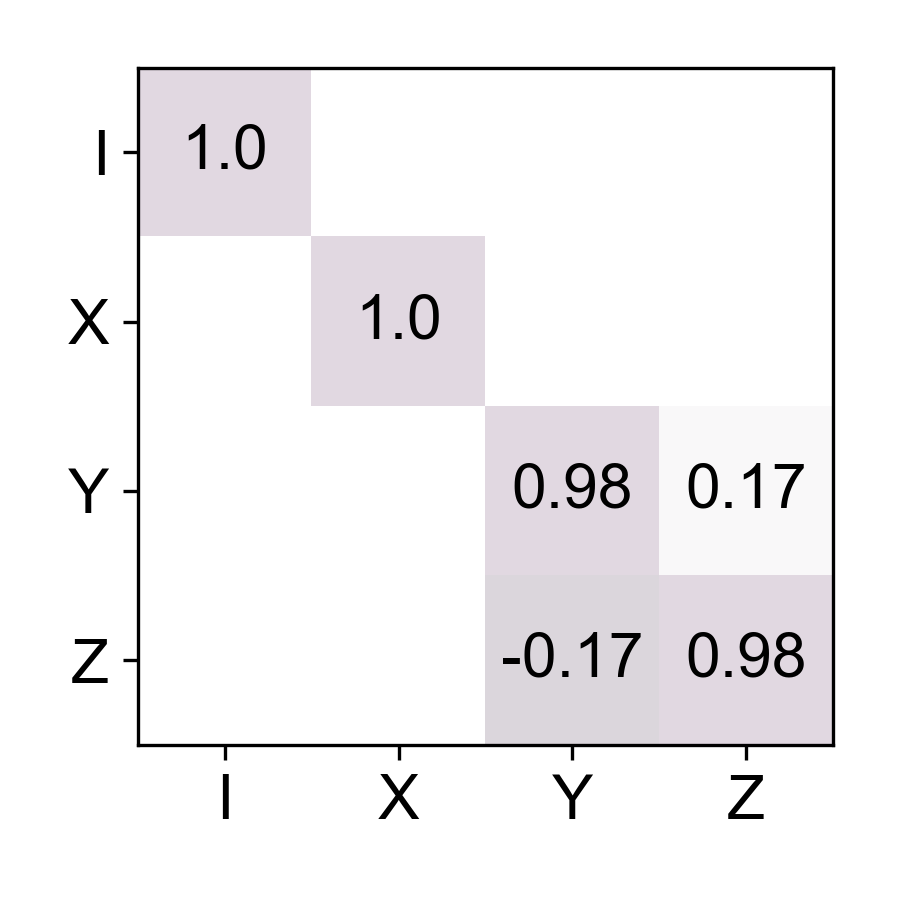}
        \caption{$Z \circ R_x \circ Z$}
    \end{subfigure}
    \hspace{5pt}
    \begin{subfigure}[b]{0.23\textwidth}
        \centering
        \includegraphics[width=\textwidth]{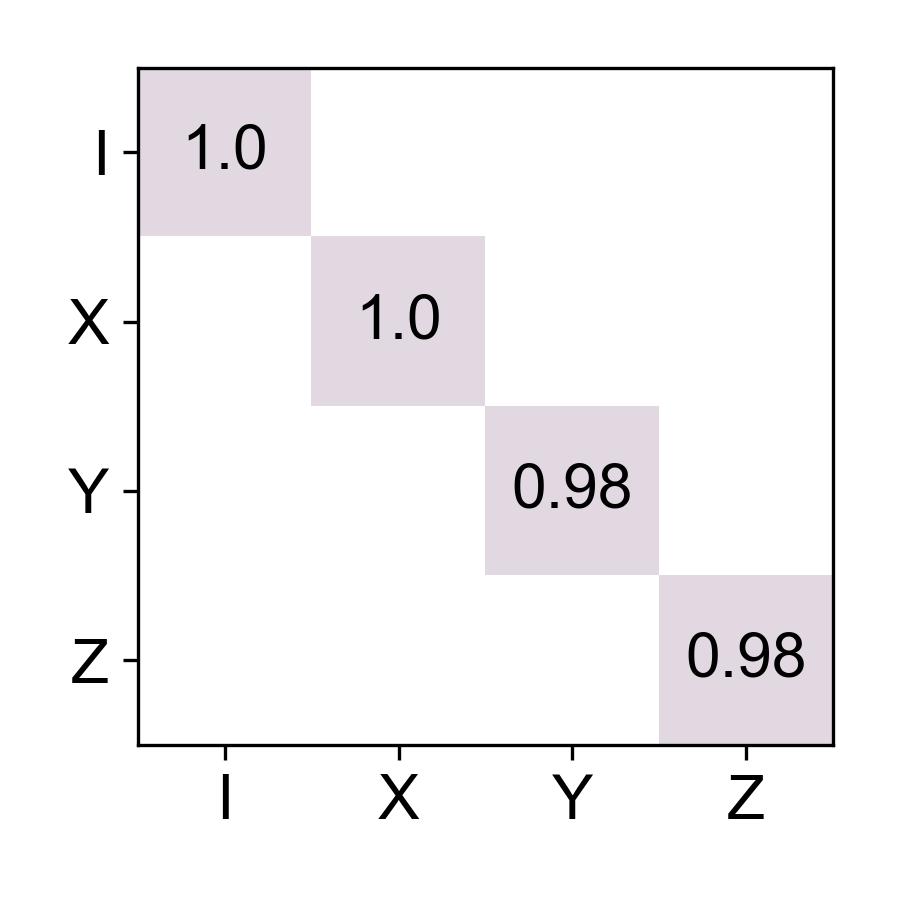}
        \caption{$\frac{1}{2} \Big( X \circ R_x \circ X + Z \circ R_x \circ Z \Big)$}
        \label{fig:pauli_twirling_lucky}
    \end{subfigure}
    \caption[Basics of Pauli Twirling.]{\textbf{Basics of Pauli Twirling.} 
    (a) PTM for the Pauli-$X$ gate. 
    (b) PTM for the Pauli-$Y$ gate. 
    (c) PTM for the Pauli-$Z$ gate. 
    (d) PTM for an $R_x(\pi / 18)$ rotation. 
    (e) $R_x(\pi / 18)$ conjugated with $X$ gates; because $R_x(\pi / 18)$ commutes with $X$, the PTM is left unchanged. 
    (f) $R_x(\pi / 18)$ conjugated with $Y$ gates; because $R_x(\pi / 18)$ does not commutes with $Y$, the signs of the off-diagonal terms have been flipped relative to $R_x(\pi / 18)$. 
    (g) $R_x(\pi / 18)$ conjugated with $Z$ gates; because $R_x(\pi / 18)$ does not commutes with $Z$, the signs of the off-diagonal terms have been flipped relative to $R_x(\pi / 18)$. 
    (h) Average of $R_x(\pi / 18)$ twirled with the Pauli-$X$ and and Pauli-$Z$ gates, resulting in a PTM in which the off-diagonal terms have been exactly averaged to zero. For all plots, the color (transparency) of each cell is determined by the sign (magnitude) of each entry.}
    \label{fig:pauli_twirling_basics}
\end{figure*}

%%%%%%%%%%%%%%%%%%%%%%% Pauli Twirling %%%%%%%%%%%%%%%%%%%%%%% 
\subsection{Pauli Twirling}\label{sec:pauli_twirling}

One can twirl a channel $\E$ over any group. However, it is often convenient to choose a particular group, such as the Pauli or Clifford group (see Appendix \ref{sec:groups_gatesets}). Because we often represent our channels in the Pauli basis (e.g., in the \ac{PTM} representation; see Sec.~\ref{sec:ptm_rep}), it is educational to first understand the basics of Pauli twirling, before considering twirling over any other unitary group. Pauli twirling an arbitrary channel $\E$ can be understood with the following example (shown in \fig\ref{fig:pauli_twirling_basics}): consider the PTM of a $R_x(\pi / 18)$ rotation (\fig\ref{fig:ptm_rx}), which contains off-diagonal terms only in the lower right-hand block of the PTM. When conjugating $R_x(\pi / 18)$ with a Pauli from the Pauli group $P \in \{I, X, Y, Z \}$, the off-diagonal elements of $R_x(\pi / 18)$ remain unchanged for $I \circ R_x(\pi / 18) \circ I$ and $X \circ R_x(\pi / 18) \circ X$, but have their signs flipped for $Y \circ R_x(\pi / 18) \circ Y$ and $Z \circ R_x(\pi / 18) \circ Z$. More generally, for any arbitrary channel $\E$, the signs of the off-diagonal terms remain the same for the elements of $\E$ with which $P$ commutes, and are reversed for the elements of $\E$ with which $P$ anti-commutes.

\begin{figure*}
    \centering
    \begin{subfigure}[b]{0.31\textwidth}
        \centering 
        \includegraphics[width=\textwidth]{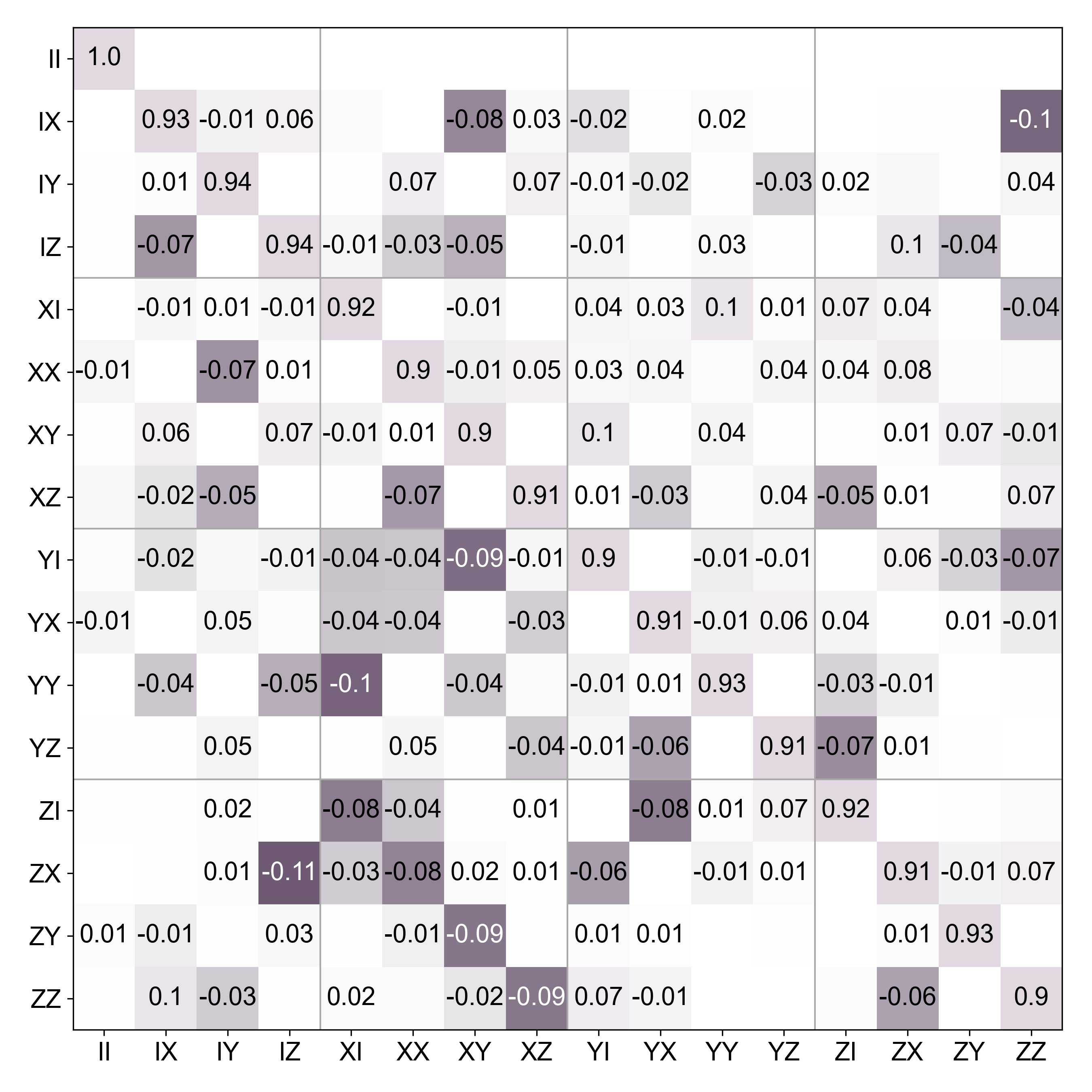}
        \caption{Pauli twirling, $N=10$}
    \end{subfigure}
    \hspace{5pt}
    \begin{subfigure}[b]{0.31\textwidth}
        \centering
        \includegraphics[width=\textwidth]{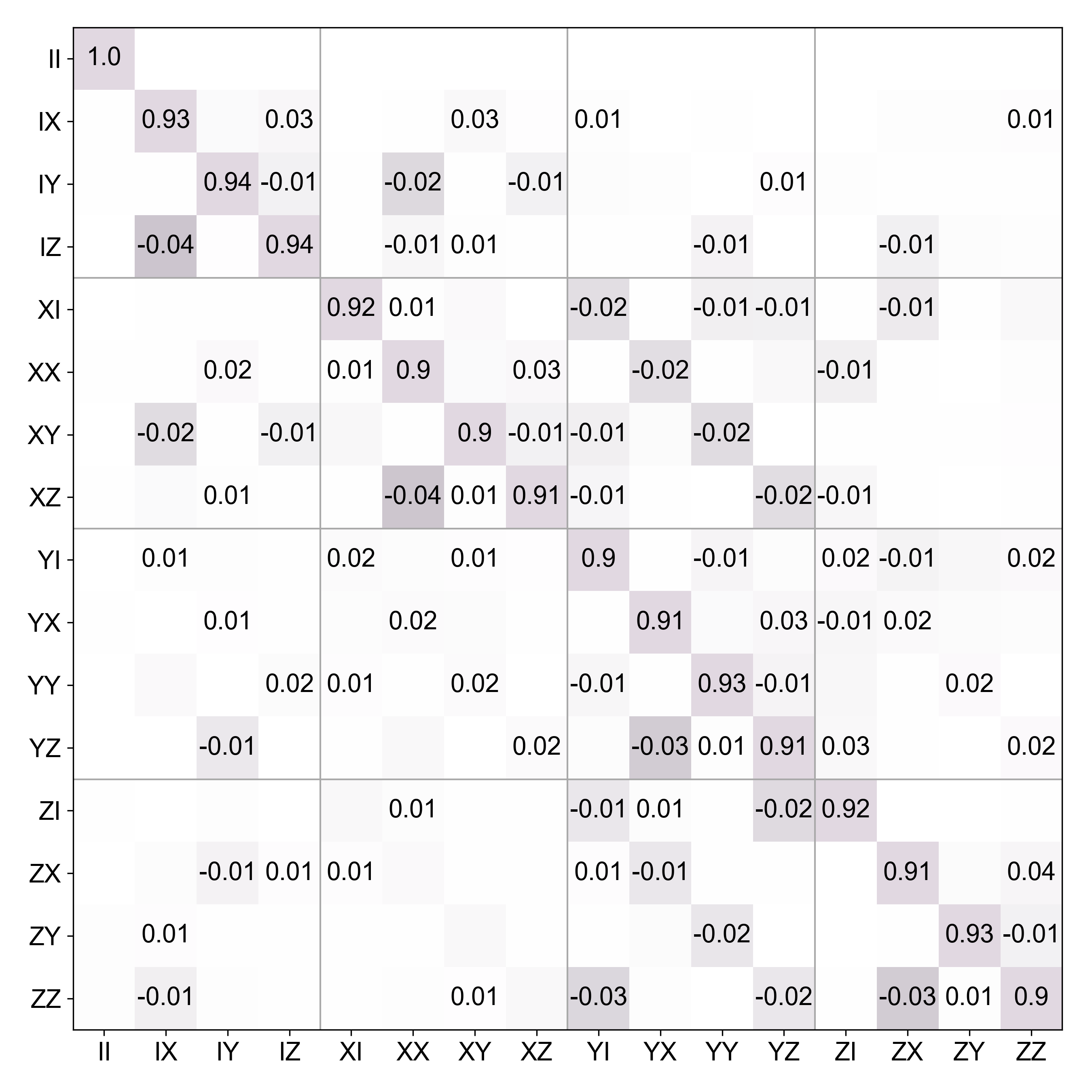}
        \caption{Pauli twirling, $N=100$}
    \end{subfigure}
    \hspace{5pt}
    \begin{subfigure}[b]{0.31\textwidth}
        \centering
        \includegraphics[width=\textwidth]{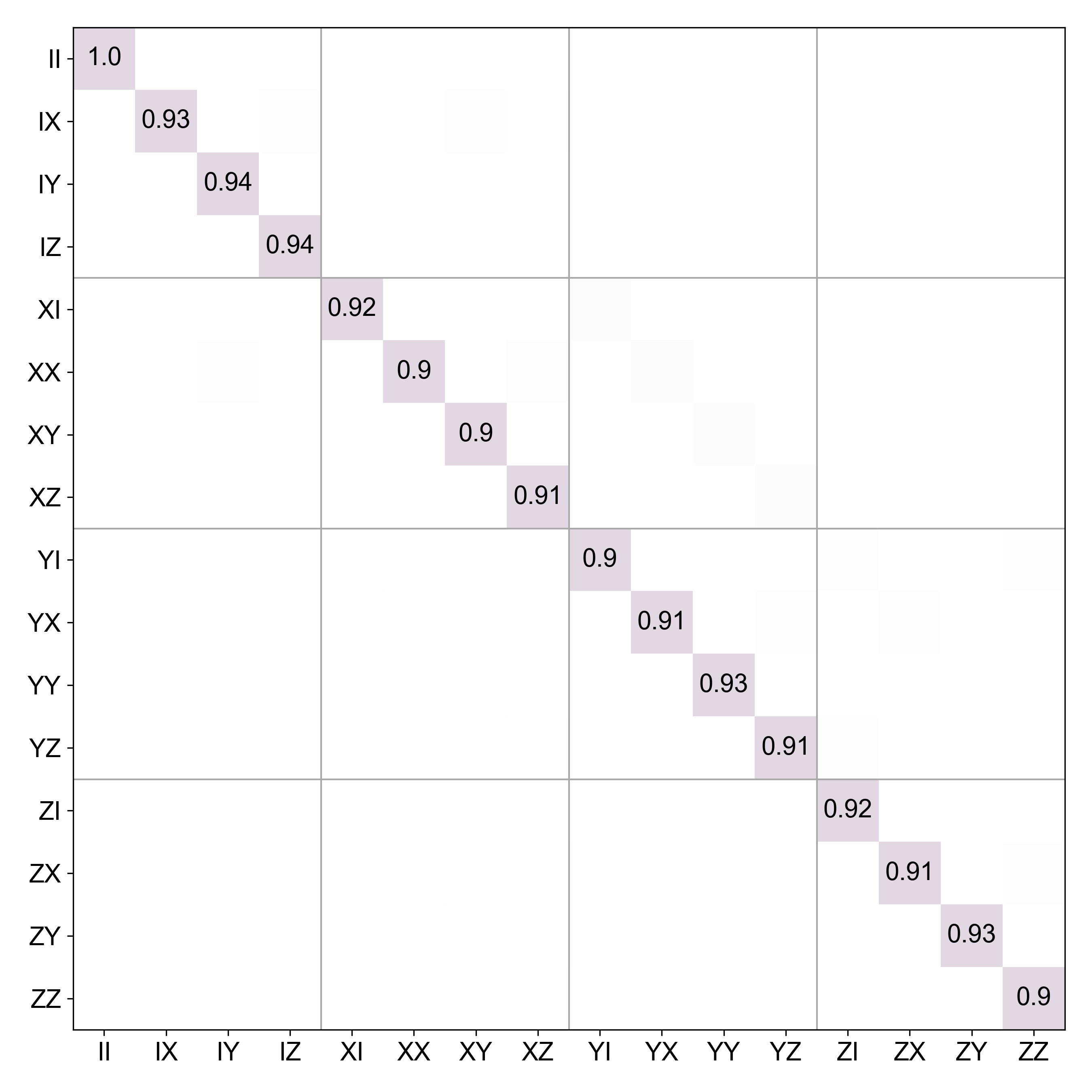}
        \caption{Pauli twirling, $N=10$k}
    \end{subfigure}
    \\
    \vspace{5pt}
    \begin{subfigure}[b]{0.31\textwidth}
        \centering
        \includegraphics[width=\textwidth]{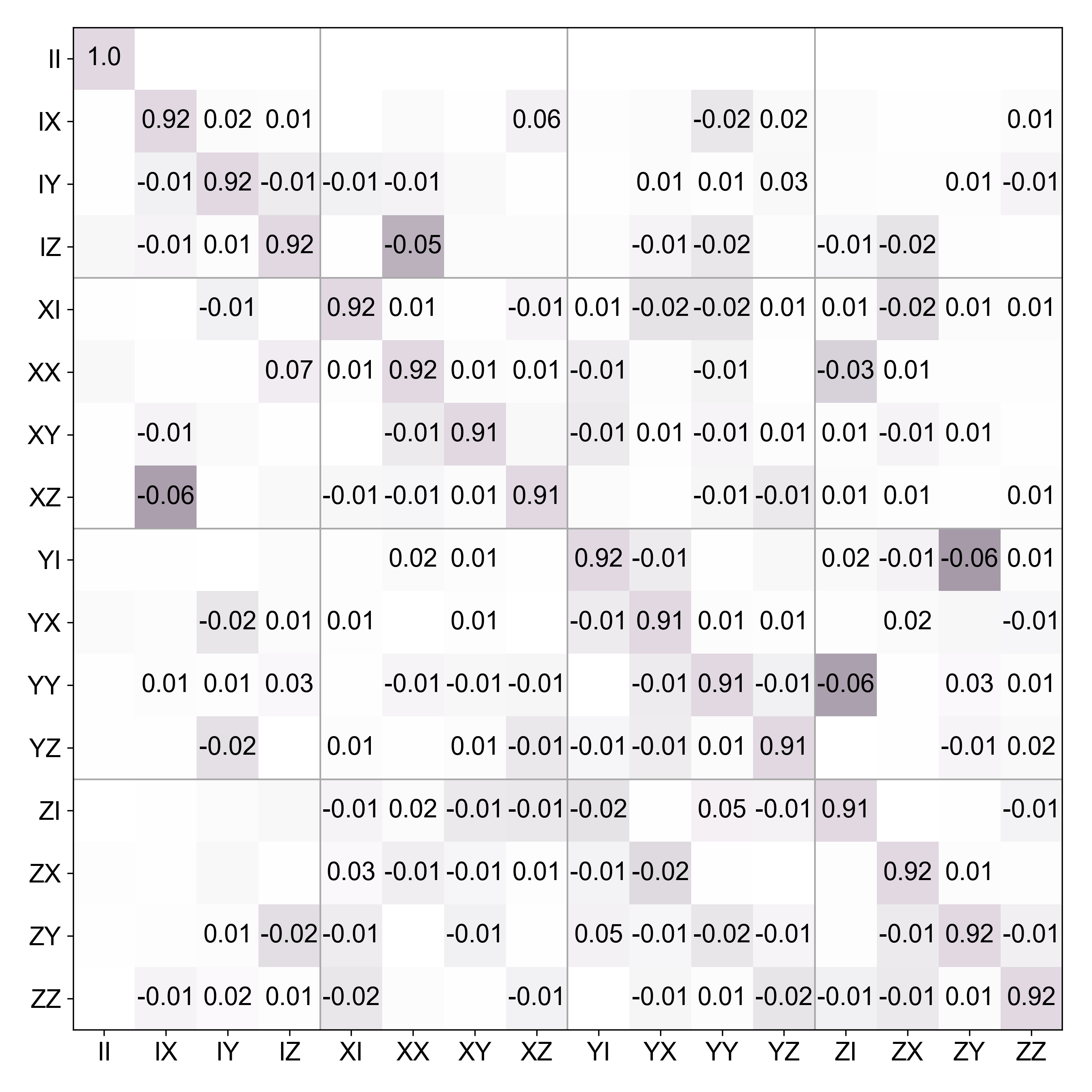}
        \caption{Clifford twirling, $N=10$}
    \end{subfigure}
    \hspace{5pt}
    \begin{subfigure}[b]{0.31\textwidth}
        \centering
        \includegraphics[width=\textwidth]{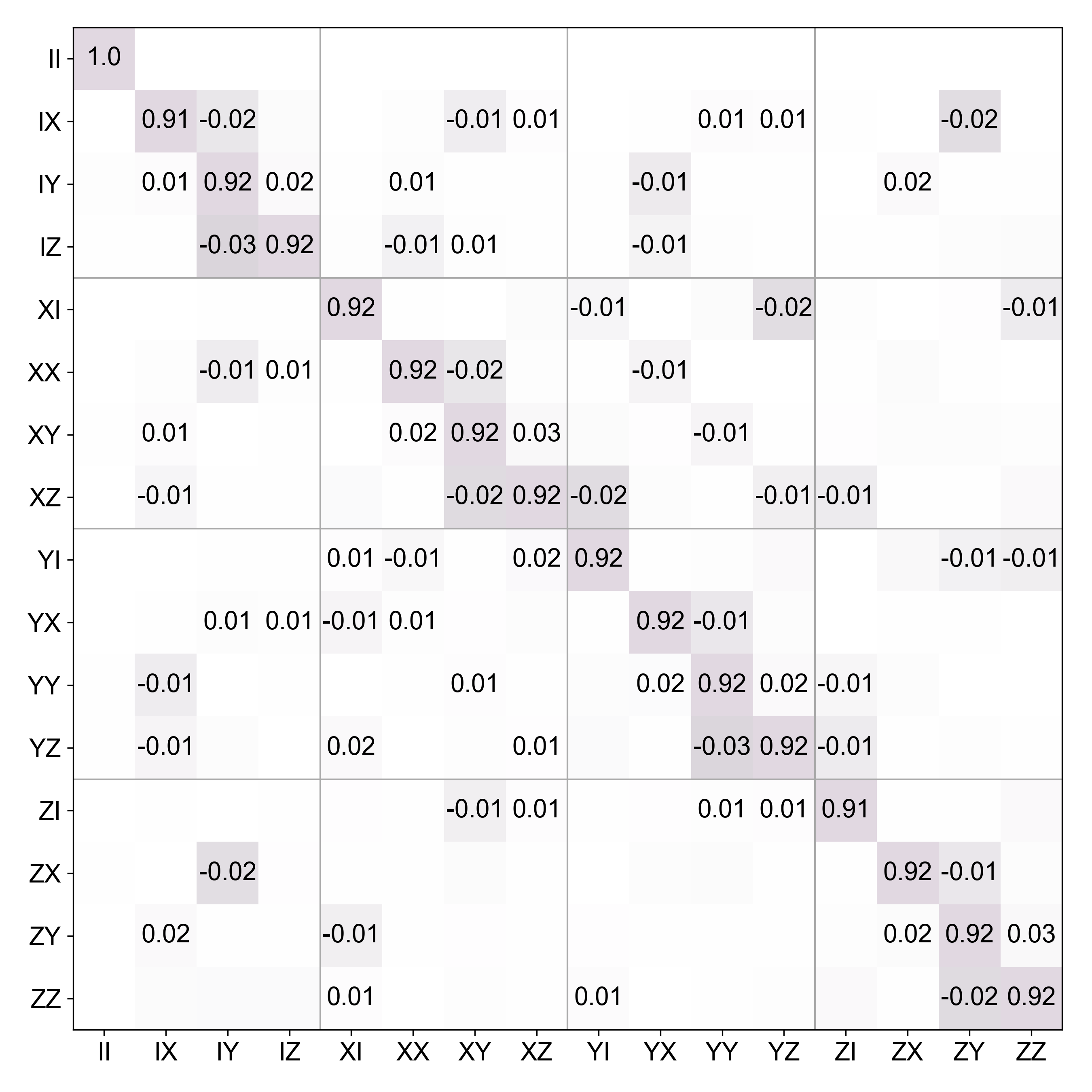}
        \caption{Clifford twirling, $N=100$}
    \end{subfigure}
    \hspace{5pt}
    \begin{subfigure}[b]{0.31\textwidth}
        \centering
        \includegraphics[width=\textwidth]{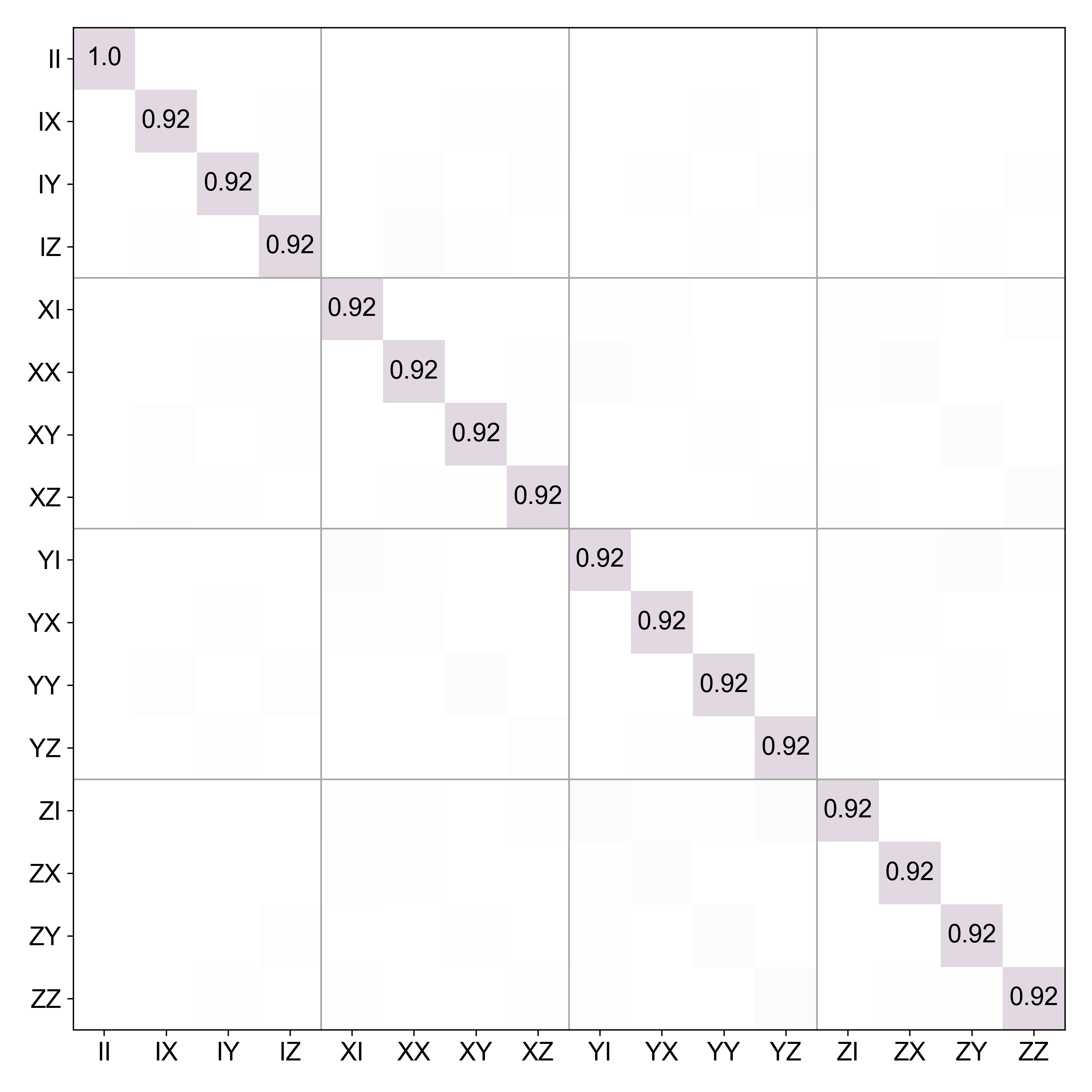}
        \caption{Clifford twirling, $N=10$k}
    \end{subfigure}
    \caption{\textbf{Pauli Twirling vs.~Clifford Twirling.} Twirling of the random CPTP channel in \fig\ref{fig:rand_ptm} as a function of the number $N$ of randomly-selected Pauli and Clifford gates.
    Pauli twirling preserves the eigenvalues along the diagonal of the PTM, whereas Clifford twirling averages them together into a global depolarizing channel.}
    \label{fig:pauli_vs_clifford_twirling}
\end{figure*}

When twirling with respect to the uniform distribution over the Pauli group, using \eq\ref{eq:twirl_finite_group} we may write the twirled channel as
\begin{equation}\label{eq:pauli_twirl}
    \Bar{\E}(\rho) = \frac{1}{d^2} \sum_{P \in \mathbb{P}_n} P^\dagger \E (P \rho P^\dagger) P ~.
\end{equation}
However, it is often unnecessary (and inefficient) to twirl over the entire Pauli group, depending on the size of the system we are considering. In general, Pauli twirling is implemented by averaging over $N$ randomly sampled Paulis,
\begin{equation}
    \Bar{\E}(\rho) = \frac{1}{N} \sum_{P \in_R \mathbb{P}_n}^N P^\dagger \E (P \rho P^\dagger) P ~,
\end{equation}
where $R$ denotes that $P$ is chosen at random from the $n$-qubit Pauli group $\mathbb{P}_n$ each time. 
In this case, the off-diagonal terms of $\E(\rho)$ change sign with a 50\% probability upon conjugation with a randomly-selected Pauli. When averaging a channel over $N$ Paulis, the magnitudes of the off-diagonal terms scale as $\theta/\sqrt{N}$, reminiscent  of a random walk, and thus vanish as $N \longrightarrow \infty$ or if by luck the correct Paulis were sampled which average to zero (see \fig\ref{fig:pauli_twirling_lucky}). This feature of twirling is often referred to as ``noise tailoring,'' as the noise profile of a channel is modified as the number of averages increases. In fact, in the limit of $N \longrightarrow \infty$, any arbitrary Markovian error channel is mapped into a stochastic Pauli channel (see Sec.~\ref{sec:stoch_pauli}) by Pauli twirling. As a concrete example, in \fig\ref{fig:rand_ptm} we plot a random CPTP two-qubit PTM and show how the off-diagonal terms are averaged to zero under Pauli twirling as $N$ is increased from 10, to 100, to $10^4$ (see \fig\ref{fig:pauli_vs_clifford_twirling}). Note that the diagonal terms in the PTM remain unchanged for all $N$. More specifically, Pauli twirling tailors all noise into Pauli channels, in which the diagonal entries of the PTM remain unchanged (Eq.~\ref{eq:ptm_stoch_pauli}).

%%%%%%%%%%%%%%%%%%%%%%% Clifford Twirling %%%%%%%%%%%%%%%%%%%%%%% 
\subsection{Clifford Twirling}\label{sec:clifford_twirling}

The $n$-qubit Clifford group $\mathbb{C}_n$ normalizes the $n$-qubit Pauli group. Functionally, this means that Clifford gates $C \in \mathbb{C}_n$ map Paulis to Paulis under conjugation: $ C P C^\dagger \mapsto P' \in \mathbb{P}_n, \forall P \in \mathbb{P}_n$; see Sec.~\ref{sec:clifford_group} for several examples. Twirling over the entire $n$-qubit Clifford group can be done discretely for any channel $\E$:
\begin{equation}\label{eq:clifford_twirl}
    \Bar{\E}(\rho) = \frac{1}{| \mathbb{C}_n |} \sum_{C \in \mathbb{C}_n} C^\dagger \E (C \rho C^\dagger) C ~.
\end{equation}
Because the Clifford group forms a unitary 2-design, Clifford twirling replicates the properties of twirling over all of $\mathsf{U}(d)$ up to the second moment. In practice, however, one does not twirl over the entire $n$-qubit Clifford group due to the large number of Clifford elements (see Appendix \ref{sec:clifford_group}). Similar to Pauli twirling, sampling $N$ Clifford gates at random is usually sufficient for suitably large $N$:
\begin{equation}
    \Bar{\E}(\rho) = \frac{1}{N} \sum_{C \in_R \mathbb{C}_n}^N C^\dagger \E (C \rho C^\dagger) C ~,
\end{equation}
where $R$ denotes that each $C$ is selected uniformly at random from $\mathbb{C}_n$.

Clifford twirling a quantum channel $\E$ has the same effect as Pauli twirling on the off-diagonal matrix elements of $\E$. Namely, in the limit of large $N$, all off-diagonal elements are averaged to zero. This is demonstrated in \fig\ref{fig:pauli_vs_clifford_twirling}, where we show the impact of Clifford twirling on the random PTM shown in \fig\ref{fig:rand_ptm} for $N = 10, 100, 10^4$. However, we observe that in contrast to Pauli twirling, Clifford twirling does not preserve the eigenvalues of the PTM. Rather, Clifford twirling averages all diagonal elements of the PTM to the same value (except for the first element), effectively tailoring noise into a global depolarizing channel (see Secs.~\ref{sec:dep_noise} and \ref{sec:polarization}). This is due to the fact that conjugating a Pauli operator by Clifford gates can map the Pauli into a different Pauli. Thus, in the limit of large $N$, Clifford twirling averages the eigenvalues of the PTM. Note that, similar to Pauli twirling, Clifford twirling does not change the process fidelity of a PTM.

%%%%%%%%%%%%%%%%%%%%%%% Weyl Twirling %%%%%%%%%%%%%%%%%%%%%%% 
\subsection{Weyl Twirling}\label{sec:twirling_qudits}

\begin{figure*}
    \centering
    \includegraphics[width = 0.9\textwidth]{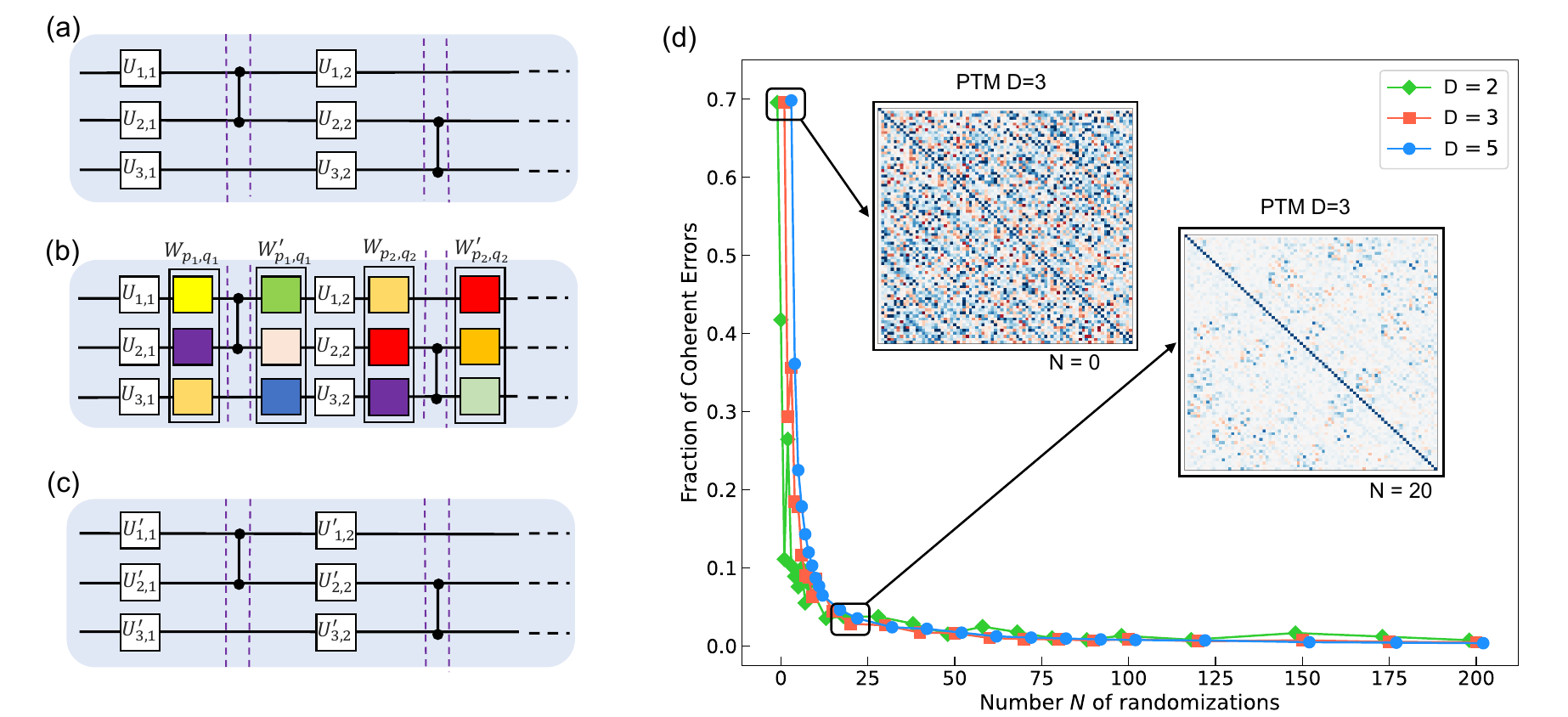}
    \caption{\textbf{Weyl Twirling.} 
    \textbf{(a)} A hypothetical input circuit, which alternates cycles of one- and multi-qudit gates. 
    \textbf{(b)} Randomized compiling of the circuit in (a). Random Weyl gates ($W_{\bar{p}_j, \bar{q}_j}$) and their inverses ($W_{\bar{p}_j, \bar{q}_j}^\prime = W_{\bar{p}_j, \bar{q}_j}^\dagger H_j W_{\bar{p}_j, \bar{q}_j}$) are added before and after every cycle of multi-qudit gates $H_j$, respectively, to generate logically-equivalent circuits. 
    \textbf{(c)} Before executing the circuit, the twirling gates are recompiled into the existing native one-qudit gates. In this way, the returned circuit has the same depth as the input one. 
    \textbf{(d)} Numerical study of the fraction of coherent errors under Weyl twirling in randomly generated two-qudit (Weyl) transfer matrices for $D \in \{2,\:3,\:5\}$. All transfer matrices are generated with a coherent fraction of 70\%. The numerics demonstrate that twirling has the same overhead regardless of qudit dimension: suppression of off-diagonal terms in the transfer matrices for all dimensions $D$ is approximately equal as a function of the number of random of randomizations $N$. The inset transfer matrices visualize the suppression of off-diagonal terms as a function of $N$ for $D = 3$.
    (Figure reproduced with permission from \cite{goss2023extending}.)}
    \label{fig:qudit-twirling}
\end{figure*}

The Weyl-Heisenberg group forms a unitary 1-design. Therefore, it is possible to use Weyl operators to twirl in higher dimensions, tailoring noise into a stochastic Weyl channel,
\begin{equation}
    \E(\rho) = \sum_{\Bar{p},\Bar{q}}^{D^{2n}}\Pr(W_{\Bar{p},\Bar{q}})W_{\Bar{p},\Bar{q}}\rho W_{\Bar{p},\Bar{q}}^\dag ~,
\end{equation}
where $\rho$ is an $n$-qudit state, $W_{\Bar{p},\Bar{q}} = \otimes_{i=1}^n W_{q_{k_i},p_{k_i}}$ is a tensor product of single-qudit operators in the $D$-dimensional Weyl-Heisenberg group, and $\Pr(W_{\Bar{p},\Bar{q}})$ is the probability of the Weyl error $W_{\Bar{p},\Bar{q}}$ occurring. An important note about twirling in higher dimensions is that it is just as efficient as twirling in $D = 2$, as shown in \fig\ref{fig:qudit-twirling}, where we demonstrate the numerical results of twirling away off-diagonal elements (e.g., coherent errors) in qudit transfer matrices using Weyl twirling for different qudit dimensions. In fact, this is guaranteed by Hoeffding's inequality \cite{H63}, and in the context of QCVV it means that no additional sampling is required relative to qubit-based methods to achieve the same degree of noise tailoring \cite{goss2023extending}.

%%%%%%%%%%%%%%%%%%%%%%% Qudit Benchmarking %%%%%%%%%%%%%%%%%%%%%%% 
\section{QCVV for Qudits}\label{sec:qudit_benchmarking}

While the focus of this tutorial has been on characterizing and benchmarking the performance of qubit-based quantum computers, in recent years there have been significant efforts in realizing qudit-based quantum processors on platforms including superconducting circuits \cite{goss2022high, blok2021quantum, PhysRevX.13.021028, cao2023emulating}, trapped ions \cite{ringbauer2021universal, native-qudit-ion}, and photonic circuits \cite{lanyon2008manipulating, photonic-qudit}. Building a quantum computer based on qudits can yield significant advantages, such as improved quantum error correction \cite{PhysRevA.87.062338, qudit-toric-codes-threshold, muralidharan_zou_li_wen_jiang_2017, PhysRevX.2.041021}, more efficient quantum algorithms \cite{10.1145/3307650.3322253}, and naturally tailored quantum simulations \cite{sqed-simulation, blok2021quantum}. To benchmark a qudit-based quantum computer, it is first incumbent upon us to generalize much of the machinery that has already been developed for qubits. In Sec.~\ref{sec:weyl_gellman}, we generalized the qubit Pauli and Clifford operators for qudits. In this section, we introduce qudit transfer matrices, and then discuss tomographic reconstruction and randomized benchmarks generalized for qudits.

%%%%%%%%%%%%%%%%%%%%%%% Qudit transfer matrices %%%%%%%%%%%%%%%%%%%%%%%
\subsection{Qudit Transfer Matrices}\label{sec:qudit_proc_matrix}

Having constructed suitable bases for describing the unitary operations of qudits in Sec.~\ref{sec:weyl_gellman}, we can turn our attention to generalizing transfer matrices for qudits as well (see Sec.~\ref{sec:rep_quant_proc} for more information). Unsurprisingly, the requirements for quantum operations describing real, physical processes do not change for qudits, i.e., they must be \ac{CPTP} maps. First, we can expand a qudit density matrix $\rho$ in terms of the $n$-qudit Gell-Mann basis $\mathbb{G}_{D,n}$, 
\begin{equation}
    \rho = \sum_{G \in \mathbb{G}_{D,n}} \rho_G G
\end{equation}
where $\rho_G = \langle \braket{G | \rho} \rangle / D^n$ are the expansion coefficients, which can be \emph{vectorized} into a $D^{2n} \times 1$ column vector. Now, any map $\ket{\rho^\prime \rangle} = \Lambda \ket{\rho \rangle}$ can be completely described by a $D^{2n} \times D^{2n}$ transfer matrix with elements
\begin{equation}
    \Lambda_{ij} = \frac{1}{D^{n}} \Tr[G_i \E(G_j)] ~,
\end{equation}
where $\E$ denotes the Kraus map defined by \eq\ref{eq:kraus_pauli}.

%%%%%%%%%%%%%%%%%%%%%%% Qudit Tomography %%%%%%%%%%%%%%%%%%%%%%%
\subsection{Qudit Tomography}\label{sec:qudit_tomography}

%%%%%%%%%%%%%%%%%%%%%%% Qudit State Tomography %%%%%%%%%%%%%%%%%%%%%%%
\subsubsection{Qudit State Tomography}\label{sec:qudit_state_tomography}

\begin{figure}
    \centering
    \includegraphics[width = \columnwidth]{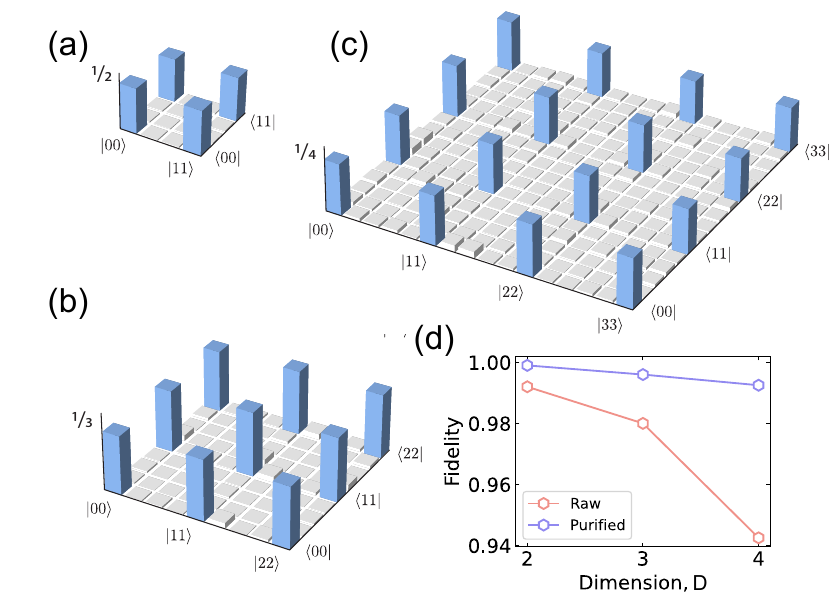}
    \caption{\textbf{Qudit State Tomography.} 
    The real component of an experimentally reconstructed density matrices for a 
    \textbf{(a)} two-qubit Bell state $\ket{\psi} = \frac{1}{\sqrt{2}}(\ket{00} + \ket{11})$, 
    \textbf{(b)} two-qutrit Bell state $\ket{\psi} = \frac{1}{\sqrt{3}}(\ket{00} + \ket{11} + \ket{22})$, and 
    \textbf{(c)} two-ququart Bell state $\ket{\psi} = \frac{1}{\sqrt{4}}(\ket{00} + \ket{11} + \ket{22} + \ket{33})$. 
    \textbf{(d)} The state fidelities for the raw and purified \cite{purification} density matrices. 
    (Figure reproduced with permission from Ref.~\cite{nguyen2023empowering}.)
    }
    \label{fig:qudit-state-tomo}
\end{figure}

Similar to qubits, the tomographic reconstruction of a qudit density matrix $\rho$ requires an informationally complete set of qudit basis measurements (e.g., Gell-Mann or Weyl-Heisenberg bases). This requires $D^{D^n} - 1$ independent experiments, from which we can reconstruct the density matrix:
\begin{equation}
    \rho = \frac{1}{D^n} \sum_{G \in \mathbb{G}} \braket{G} G ~.
    % \rho = \frac{1}{D^n} \sum_{\Vec{\sigma}}\expval{\Vec{\sigma_1}\otimes \Vec{\sigma_2} \dots \otimes \Vec{\sigma_n}}\Vec{\sigma_1}\otimes \Vec{\sigma_2} \dots \otimes \Vec{\sigma_n} ~
\end{equation}
From these measurements, as described in Sec.~\ref{sec:state_tomography}, the density matrix $\rho$ can be estimated using \ac{MLE} (see Sec.~\ref{sec:mle}). In practice, the single-qudit operations necessary to reconstruct an arbitrary qudit density matrix can be considered as the local projections onto all the computational states $\ket{0},\dots,\ket{D}$, as well as the $\sqrt{X^{jk}},\sqrt{Y^{jk}}$ projections over all local two-level subspaces of the qudit. The results of experimentally reconstructed qudit Bell states ($\ket{\psi} = \frac{1}{\sqrt{D}}\sum_{i=0}^{D-1} \ket{ii}$) for qudit dimension $D=2,3,4$ can be seen in \fig\ref{fig:qudit-state-tomo}.

%%%%%%%%%%%%%%%%%%%%%%% Qudit Process Tomography %%%%%%%%%%%%%%%%%%%%%%%
\subsubsection{Qudit Process Tomography}\label{sec:qudit_qpt}

\begin{figure*}
    \centering
    \includegraphics[width = \textwidth]{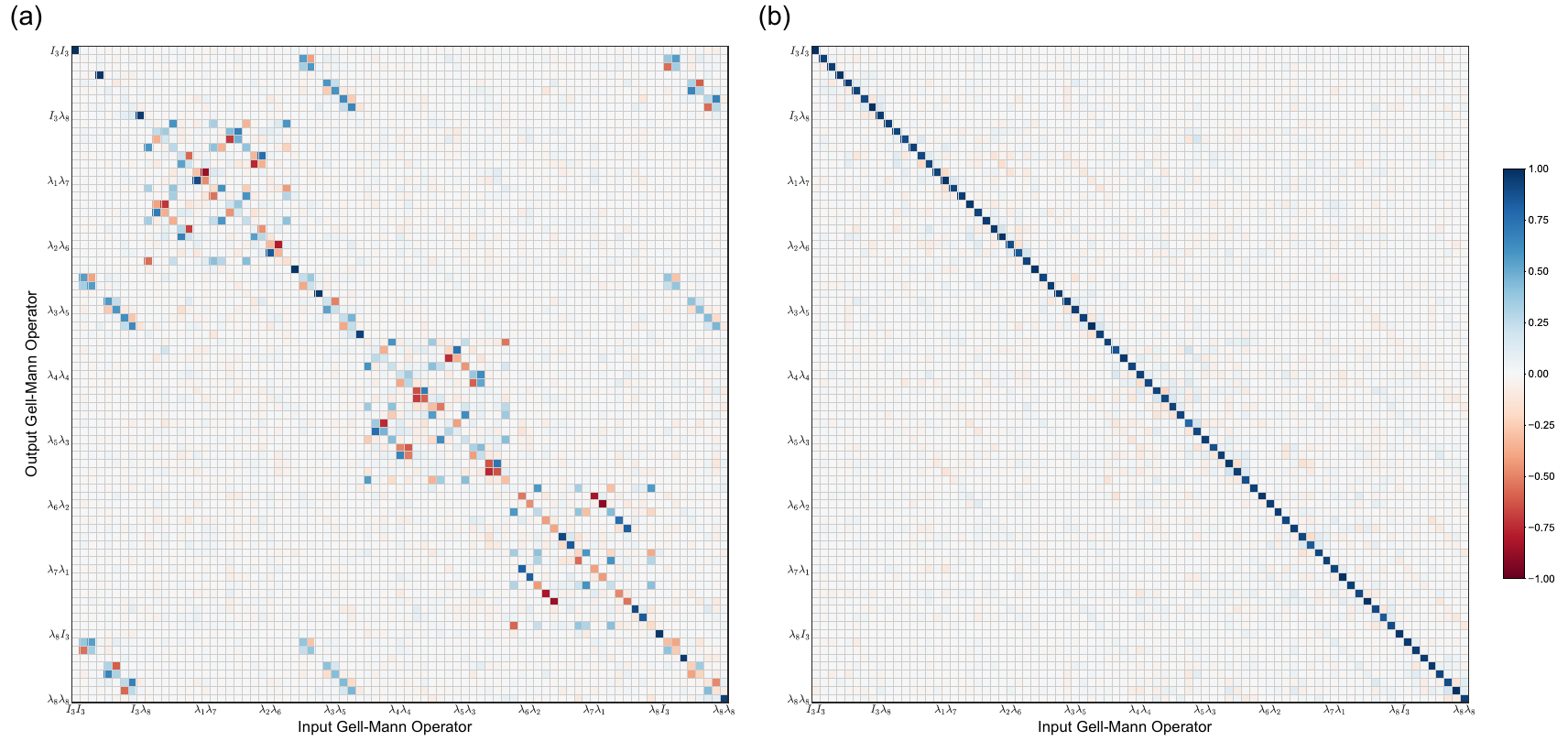}
    \caption{\textbf{Qudit Process Tomography.} 
    The transfer matrix of an experimentally realized two-qutrit CZ$^\dag$ gate \cite{goss2022high}. 
    \textbf{(a)} The transfer matrix ($\Lambda_{exp}$) in the qutrit Gell-Mann basis. 
    \textbf{(b)} The error matrix $\Lambda_{exp}^\dag \Lambda_{ideal}$, with corresponding process fidelity of 93.2\%. 
    (Figure reproduced with permission from \cite{goss2022high}.)}
    \label{fig:qudit-qpt}
\end{figure*}

As with state tomography, quantum process tomography (\ac{QPT}) can also be readily generalized to describe how a qudit operation maps input qudit states to output qudit states. Following the procedure described in Sec.~\ref{sec:qpt}, we can use the set of informationally complete operations outlined in Sec.~\ref{sec:qudit_state_tomography} for both the state preparations and measurement bases to tomographically reconstruct the qudit transfer matrix $\Lambda$. For example, \fig\ref{fig:qudit-qpt} shows the results of performing QPT on a two-qutrit CZ$^\dag$ gate in the Gell-Mann basis \cite{goss2022high}, where $U_{CZ^\dag} = \sum_{i,j \in \mathbb{Z}_3} \omega^{-ij}\ket{i,j}\bra{i,j}$. This required 81 different two-qutrit input states prepared using the following set of native gates on each qutrit: 
$\{ I$, $\sqrt{X^{01}}$, $\sqrt{Y^{01}}$, $X^{01}$, $X^{12}X^{01}$, $Y^{12}\sqrt{X^{01}}$, $\sqrt{X^{12}}X^{01}$, $\sqrt{Y^{12}}X^{01}$, $X^{12}\sqrt{X^{01}} \}$.
The same set of native gates is then used to perform state tomography on each input state, and the qudit transfer matrix $\Lambda$ is reconstructed using MLE. In \fig\ref{fig:qudit-qpt}, the tomographically reconstructed transfer matrix $\Lambda_{exp}$ and the error matrix $\Lambda_{ideal}^\dag \Lambda_{exp}$ are displayed in the qutrit Gell-Mann basis. From these results, the process fidelity is calculated as $F_e = \Tr[\Lambda_{ideal}^\dag \Lambda_{exp}]/d^2 = 93.2\% $. We note that the discrepancy between the process fidelity of the CZ$^\dag$ calculated via QPT and randomized benchmarks (introduced in the following section) can be attributed to \ac{SPAM} errors.

%%%%%%%%%%%%%%%%%%%%%%% Qudit Gate Set Tomography %%%%%%%%%%%%%%%%%%%%%%%
\subsubsection{Qudit Gate Set Tomography}\label{sec:qudit_gst}

Finally, we note that \ac{GST} (see Sec.~\ref{sec:gst}) can also be generalized for qudits, and has also been applied to study single qutrit gates in Ref.~\cite{cao2022efficient}, where it demonstrated good agreement between other SPAM-free characterization methods such as qutrit randomized benchmarking. Additionally, qudit-based GST methods can provide insight into non-Markovian errors and fine-grained error budgets for qudit gates, which are difficult to extract from lighter-weight methods such as randomized benchmarks.

%%%%%%%%%%%%%%%%%%%%%%% Qudit Randomized Benchmarking Toolbox %%%%%%%%%%%%%%%%%%%%%%%
\subsection{Qudit Randomized Benchmarks}\label{sec:qrb}

As with qubit-based randomized benchmarks (see Sec.~\ref{sec:randomized_benchmarks}), qudit-based randomized benchmarking techniques employ random circuits to enable the efficient characterization of quantum gate sets. Broadly speaking, all of these methods leverage twirling (see Sec.~\ref{sec:twirling_qudits}) to tailor noise into Pauli channels or a global depolarizing channel, yielding efficient estimates of process fidelities. In this section, we describe the generalizations required for performing randomized benchmarking, cycle benchmarking, and cross-entropy benchmarking on a qudit-based quantum computer. 

\begin{figure*}
    \centering
    \includegraphics[width = 0.9\textwidth]{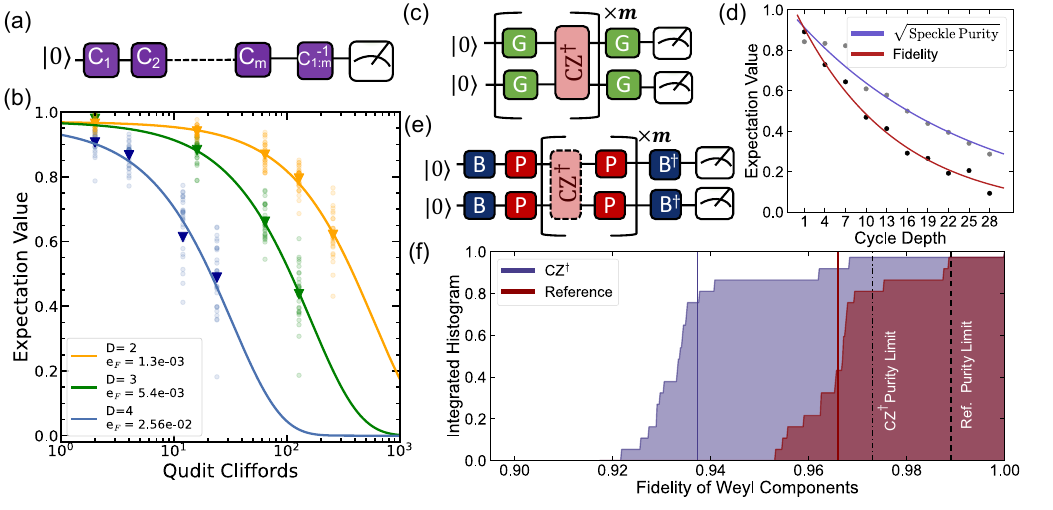}
    \caption{\textbf{Qudit Randomized Benchmarks.} 
    \textbf{(a)} Single-qudit Clifford RB circuits where $m$ different Clifford gates $C_i$ (purple) are selected from $\mathbb{C}_D$, after which the entire circuit is inverted with a single additional Clifford $C^{-1}_{1:m}$. 
    \textbf{(b)} The experimental results of qudit RB on a superconducting qudit for $D = \{2, 3, 4\}$ \cite{nguyen2023empowering}. The process infidelity $e_F$ is listed in the legend for each dimension $D$.
    \textbf{(c)} Circuit schematic of XEB. A CZ$^\dag$ gate is interleaved between $m$ cycles of random $\mathsf{SU}(3)$ gates (green).
    \textbf{(d)} A dressed CZ$^\dag$ fidelity of 0.933(3) was measured from the exponential decay of the XEB results \cite{goss2022high}. Additionally, the speckle-purity limited fidelity of the dressed cycle was estimated to be 0.961(3).
    \textbf{(e)} Circuit schematic of CB. The system is prepared in a Weyl basis state $B$ (blue), after which the CZ$^\dag$ gate is interleaved between $m$ cycles of random Weyl gates (red). Finally, the system is rotated back to the original Weyl basis with an additional final cycle of inverting Weyl gates $B^\dagger$.
    \textbf{(f)} An integrated histogram of CB for both the CZ$^\dag$ gate and a reference cycle from Ref.~\cite{goss2022high}, with the solid vertical lines giving the process fidelities of 0.936(1) and 0.966(1), respectively. Together, these yield an estimated gate fidelity of 97.3(1)\% for the CZ$^\dag$ gate. Moreover, one can extract an error budget directly from CB, giving purity-limited fidelities of 0.973(9) and 0.989 (with negligible error) for the dressed CZ$^\dag$ and reference cycles, respectively; together, this gives an estimated purity limit 0.986(9) for the isolated CZ$^\dag$ gate. 
    (Figures reproduced with permission from \cite{goss2022high, nguyen2023empowering}.)
    }
    \label{fig:qudit-rb}
\end{figure*}

%%%%%%%%%%%%%%%%%%%%%%% Qudit Randomized Benchmarking %%%%%%%%%%%%%%%%%%%%%%%
\subsubsection{Qudit Randomized Benchmarking}\label{sec:qudit_rb}

Having already defined the qudit Weyl-Heisenberg group and Clifford group in Sec.~\ref{sec:weyl_gellman}, it is now possible to describe qudit randomized benchmarking (\ac{RB}; see Sec.~\ref{sec:randomized_benchmarks} for a background on qubit RB). The procedure for qudit randomized benchmarking follows exactly as in the qubit case, where now the random Clifford gates are sampled uniformly from $\mathbb{C}_{D,n}$, with the final gate in any RB sequence chosen to decompose the entire circuit to the identity (or up to a random Weyl operator). In general, the ground state is prepared, and the success probability is fit to an exponential decay of the qudit $Z$ expectation value, calculated as
\begin{equation}
    \expval{Z}_D = \sum_{i=0}^{D-1} p(\ket{i}) \omega^i ~,
\end{equation}
where $\omega$ is again the $D$-th root of unity. The average Clifford gate fidelity is then calculated from fitting the exponential $\expval{Z}_D(m) = Af^m$ by measuring different circuit depths $m$ and converting the process polarization $f$ to an average gate fidelity or process fidelity (see Table \ref{tab:table_rel_fid_dep}). We note that although the qudit $Z$ operator is in general non-Hermitian, its phase does not change under depolarization. Therefore, the imaginary component --- which is initially zero due to preparing in the ground state --- remains zero throughout. 

Initial demonstrations of qudit randomized benchmarking have been performed for $D=3$ in Refs.~\cite{qutritrb, goss2023extending} and $D=4$ in Refs.~\cite{PhysRevX.13.021028, seifert2023exploring}. In \fig\ref{fig:qudit-rb}(a) -- (b), the circuits and results of performing RB on the same qudit operating in $D=\{2,3,4\}$ is shown \cite{nguyen2023empowering}, yielding average Clifford process fidelities of $F_e = \{0.99872(1)$, $ 0.9946(2), $ and $0.974(2)\}$. Since there are in general $\{2, 6, 12\}$ native gates (excluding software defined virtual $Z$ gates) needed to decompose Clifford gates in $D = \{2, 3, 4\}$, from the Clifford fidelities one can calculate the average native gate process fidelities, yielding $F_e = \{0.99936(3), 0.99909(4), 0.9978(2)\}$ for the results in \fig\ref{fig:qudit-rb}. We further note that interleaved RB and simultaneous RB can also be performed to obtain additional insight into specific qudit gate performance as well as effects from undesired crosstalk interactions \cite{qutritrb}.

%%%%%%%%%%%%%%%%%%%%%%% Qudit Cycle Benchmarking %%%%%%%%%%%%%%%%%%%%%%%
\subsubsection{Qudit Cycle Benchmarking}\label{sec:qudit_cb}

Cycle Benchmarking (\ac{CB}) is useful for qudit-based QCVV, specifically in the context of multi-qudit gates, as performing randomized benchmarking on a multi-qudit gate requires sampling and decomposing multi-qudit Clifford gates, often requiring many native multi-qudit gates. In contrast, CB can be performed with significantly fewer multi-qudit gates per circuit and can be scaled to studying larger systems (see Sec.~\ref{sec:cb}). For qudit CB, the eigenstates and twirling gates are chosen from the Weyl-Heisenberg group. Notably, each $W \in \mathbb{W_D}$ commutes with its Hermitian conjugate $W^\dag = W^{D-1}$, which also belongs to the Weyl-Heisenberg group. This implies that these operators share the same eigenbasis, such that $\expval{W^\dag} = \Bar{\expval{W}}$, which allow us to reduce the number of required measurements needed to characterize the Weyl decays \cite{qutritrb}.

Qutrit CB was first described and demonstrated in Ref.~\cite{qutritrb}, and was later used to benchmark two-qutrit CZ and CZ$^\dagger$ gates with interleaved gate fidelities as high as 95.2(3)\% and 97.3(1)\%, respectively \cite{goss2022high}. The circuits and integrated histogram results of the CB protocol for a CZ$^\dag$ gate are shown in \fig\ref{fig:qudit-rb}(e) -- (f).

%%%%%%%%%%%%%%%%%%%%%%% Qudit Cross-Entropy Benchmarking %%%%%%%%%%%%%%%%%%%%%%%
\subsubsection{Qudit Cross-Entropy Benchmarking}\label{sec:qudit_xeb}

The cross-entropy benchmarking (\ac{XEB}) protocol (see Sec.~\ref{sec:xeb}) can likewise be straightforwardly generalized to qudits. In the case of qudit XEB, the local twirling is now performed via Haar-random $\mathsf{SU}(d)$ gates, and the analysis is performed on the output ditstring (rather than bitstring) results. Qutrit XEB has been described and performed in Ref.~\cite{goss2022high}, where it was used to benchmark a two-qutrit CZ$^\dag$ gate. Those results can also be found in \fig\ref{fig:qudit-rb}(c) -- (d), where we also provide a circuit schematic for XEB sequences. We note here that the dressed gate fidelities for the CZ$^\dag$ estimated from qutrit CB and XEB agree to within error bars, which is expected due to the fact that twirling does not change the average gate fidelity or process fidelity of a gate (see Sec.~\ref{sec:twirling_channels}).

%%%%%%%%%%%%%%%%%%%%%%% Gauge Uncertainty %%%%%%%%%%%%%%%%%%%%%%% 
\section{Gauge Ambiguity in Pauli Noise Learning}\label{sec:pauli_gauge}

One prominent advantage of cycle benchmarking (CB; see Sec.~\ref{sec:cb}) or cycle error reconstruction (CER; see Sec.~\ref{sec:cer}) is the intrinsic robustness to \ac{SPAM} errors, which is a general feature of randomized benchmarking-like protocols. However, when benchmarking cycles containing multi-qubit Clifford gates, CER generally cannot resolve every Pauli fidelity (or Pauli error rate) individually. Instead, for certain subset of Pauli operators, only the geometric mean of Pauli fidelities can be estimated. See \fig\ref{fig:cer} for an example. This issue of ``degeneracy'' is not a drawback of any specific method, but is related to the fundamental notion of gauge ambiguity~\cite{nielsen2021gate} (see Sec.~\ref{sec:gauge}). That is, when taking unknown SPAM noise into account, there exist certain gauge degrees of freedom in the noise model that cannot be resolved. In this section, we discuss a theory~\cite{chen2023learnability} that fully characterizes the gauge-consistently learnable information in Pauli noise learning.

\begin{figure*}[!htp]
    \centering
    \includegraphics[width=0.95\textwidth]{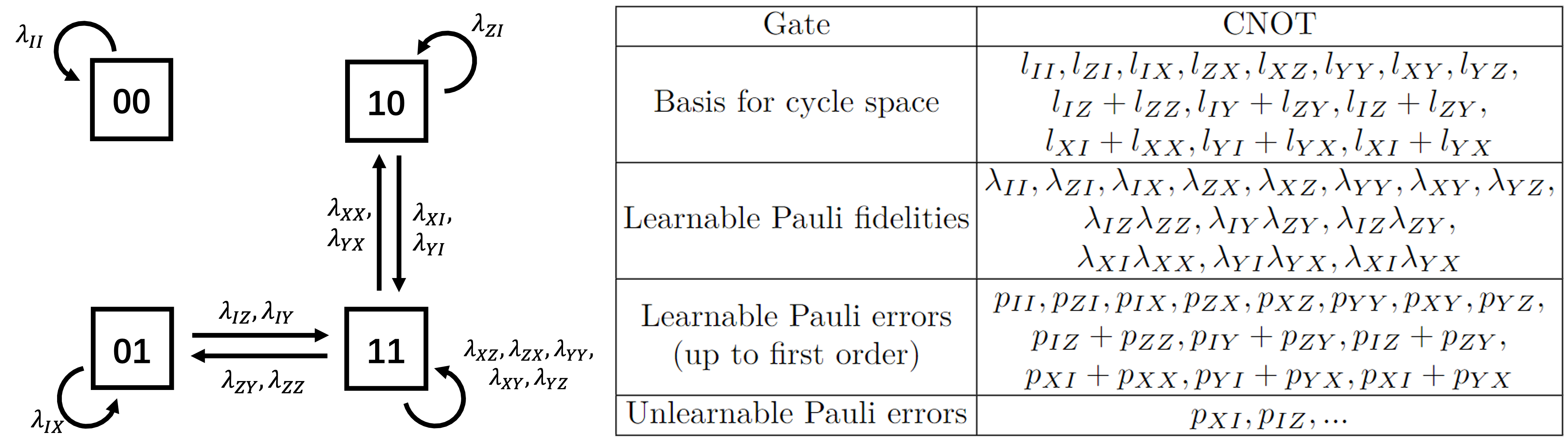}
    \caption{\textbf{Learnability of Pauli Noise.}
    (Left) Pattern transform graph of CNOT. (Right) Learnable and unlearnable information of the CNOT gate.}
    \label{fig:ptg}
\end{figure*}

We start with four assumptions about the noise: (1) any single-qubit unitary gate can be implemented perfectly; (2) a set of multi-qubit Clifford gates $\{G\}$ can be implemented with gate-dependent Pauli noise, i.e., $\widetilde{G} = G \circ \Lambda_{G}$, where $\widetilde{G}$ denotes the noisy gate and $\Lambda_{G}$ the transfer matrix capturing the Pauli noise; (3) any state preparation and \ac{POVM} measurement can be implemented subject to an unknown Pauli noise channel; (4) the Pauli fidelities of all Pauli channels are strictly positive. 
For the first condition, we can allow single-qubit gate cycles to have gate-independent noise, which are standard assumptions of CB and CER, but in that case one can simply absorb the noise into the multi-qubit Clifford gate (known as dressed cycles~\cite{carignan2023error}). 
The second and third assumptions can be guaranteed via randomized compiling~\cite{wallman2016noise, hashim2021randomized}. 
The last one is for regularization and should hold for any reasonable gate set.
Now, we ask the following question: what information of $\{\Lambda_{G}\}$ can be learned in a SPAM-robust manner despite the existence of unknown SPAM noise?

For the task of Pauli channel estimation, this question can be answered step-by-step: firstly, any individual Pauli fidelity, $\lambda_P^{G} \equiv \Tr[P\Lambda_{G}(P)]/2^n$, can be learned SPAM-robustly iff $G$ preserves the \emph{pattern} of $P$. The pattern of an $n$-qubit Pauli (with sign ignored) is defined as an $n$-bit string whose $i$th bit is 0 if $P_i = I$ and $1$ otherwise (e.g., $XZIYI\mapsto11010$). Take the CNOT gate as an example, which maps Paulis into Paulis (see Table \ref{tab:CNOT_conjugation}). Since 
$$
\begin{aligned}
    \mathrm{CNOT}:\quad ZI&\mapsto ZI~, \quad YY\mapsto XZ~, \quad XI\mapsto XX ~,\\
    \mathrm{pattern}:\quad01&\mapsto01~, \quad11\mapsto11~, \quad\quad10\mapsto11 ~,
\end{aligned}
$$
we have that $\lambda_{ZI}$ and $\lambda_{YY}$ are SPAM-robustly learnable, while $\lambda_{XI}$ is not. We can summarize how the CNOT changes the pattern of all 2-qubit Pauli operators in a \emph{pattern transform graph}, with 4 nodes and 16 edges, shown in  \fig\ref{fig:ptg}. Only those Pauli fidelities which lie on a self-loop are individually learnable.

Next, a product of Pauli fidelities $(\lambda_{P_1}\cdots\lambda_{P_M})$ is SPAM-robustly learnable if the corresponding edges form a cycle in the pattern transform graph. For the example of a CNOT gate, though $\lambda_{XI}$ and $\lambda_{XX}$ are individually unlearnable, their product $\lambda_{XI}\lambda_{XX}$ is learnable. 
A more rigorous statement goes as follows: define the log-Pauli fidelities $l_P^{G}\equiv\log\lambda_P^{G}$ for all $P$. The space of linear functions of  $\{l_P\}$ has a natural isomorphism to the edge space of the pattern transform graph. 
The result states that the learnable linear functions form a subspace corresponding to the cycle space of the graph, a notion from algebraic graph theory~\cite{bollobas1998modern}. 
The number of learnable/unlearnable degrees of freedom can also be inferred using graph-theoretical tools.
More details are presented in \cite{chen2023learnability,chen2024efficient}.

Here, we briefly sketch the proof of these results. To see that everything in the cycle space is learnable, one just needs to construct a proper CB-type experiment. Take the CNOT as an example: since the CNOT preserves $ZI$, to learn $\lambda_{ZI}$ we first prepare an eigenstate of $ZI$, repeat the CNOT $m$ times (under randomized compiling), and then measure the expectation value of $ZI$, from which we can fit an exponential decay of the form $A_{ZI} \lambda_{ZI}^m$, where $A_{ZI}$ is some SPAM-dependent coefficient. Similarly, since the CNOT preserves the pattern of $YY$, to learn $\lambda_{YY}$ we perform the same protocol; however, now we must interleave certain single-qubit gates (e.g., $\sqrt Z\otimes\sqrt X$) following each application of the CNOT, from which one can obtain $A_{YY} \lambda_{YY}^m$ (similar techniques are mentioned in~\cite{van2023probabilistic}). Finally, to learn products like $\lambda_{XI}\lambda_{XX}$, one can simply repeat the CNOT $2m$ times and measure in the eigenstate of $XX$, yielding $A_{XX}(\lambda_{XI}\lambda_{XX})^m$. To see that everything outside the cycle space is unlearnable, one can show that every cut in the pattern transform graph induces a gauge transformation. For example, for a CNOT gate, a cut between $10$ and the other nodes induces the following gauge transform 
$$
\begin{aligned}
    \lambda_{XI},\lambda_{YI}&\mapsto \kappa\lambda_{XI},\kappa\lambda_{XI} ~,  \\
    \quad\lambda_{XX},\lambda_{YX}&\mapsto \kappa^{-1}\lambda_{XX},\kappa^{-1}\lambda_{YX} ~,
\end{aligned}
$$
for a real number $\kappa$ sufficiently close to $1$.
The SPAM noise needs to change correspondingly and is omitted here. One can show that this is indeed a gauge transformation that preserve all assumptions of the Pauli noise model. Since any learnable function must be orthogonal to all gauge transformations, and the cut space is orthogonal complement to the cycle space, this implies any learnable function must be inside the cycle space.

In practice, any noise channel $\Lambda_{G}$ should be sufficiently close to identity, i.e., $\lambda_P\rightarrow 1$. 
In this regime, any function of $\Lambda_{G}$  can be approximated to first order by an affine function of $\{l_b^{G}\}_b$, and the above result can thus be used to infer the learnability of a general function to first order, including the Pauli error rates. Interestingly, it can be shown that the first-order learnable Pauli error rates are also isomorphic to the cycle space (implicit from~\cite[Lemma 3]{carignan2023error} and rigorously proven in~\cite[Appendix D]{zhang2024generalized}). In other words, the cycle space of the pattern transform graph is invariant under the Walsh-Hadamard transform. As a concrete example, in \fig\ref{fig:ptg}, we list the cycle basis, learnable fidelities, first-order learnable error rates, and a possible choice of gauge parameters for the CNOT.

We end this section with some relevant observations: 
\begin{itemize}
    \item The unlearnablity is rooted in the gauge ambiguity of SPAM noise. If SPAM noise is very small compared to the gate noise, one can expect the ambiguity for gate noise characterization to be negligible. However, there are experiments suggesting this might not be the case for state-of-the-art quantum computing platforms~\cite{chen2023learnability}. Nevertheless, one can always try to bound the unlearnable parameters using physicality constraints (e.g., CPTP conditions). 
    \item If we treat quantum circuits as a black box, any observable properties should, by definition, be learnable functions. Therefore, by properly characterizing all learnable degrees of freedom, one should in principle be able to perform error mitigation (see, e.g.,~\cite{endo2018practical}). However, such gauge-consistent error mitigation techniques within the Pauli noise model have yet to be developed. On the other hand, there exist error mitigation experiments based on Pauli noise learning that avoid the learnability issue by introducing additional assumptions~\cite{ferracin2022efficiently, van2023probabilistic}. It is an interesting direction to better understand the relation between noise learnability and quantum error mitigation. 
    \item It is common to make additional locality assumptions about the Pauli noise channels, so that the number of noise parameters becomes manageable. Ref.~\cite{chen2024efficient} studies how to reconcile the locality assumptions with the learnable of Pauli noise, giving a similar graph-algebraic characterization of learnable/unlearnable degrees of freedom as described here.
    \item Another practical issue for noise characterization is whether one can estimate a certain noise parameter to relative precision (using a small number of experiments). Achieving this usually requires multiple repetitions of specific gate sequence (known as ``germs'' in \ac{GST}) to amplify the noise parameter. It has been proven that any product of Pauli fidelities corresponding to a cycle in the pattern transform graph can be learned to relative precision by repeating certain sequence of gates~\cite{chen2024efficient, carignan2023error}. In contrast, an unlearnable function as discussed above cannot be estimated to arbitrarily small precision, no matter if it is additive precision or relative precision.
\end{itemize}

\end{document}